\pdfoutput=1
\documentclass[11pt,twoside,a4paper,cmspaper,final,collab]{cms-tdr}
\def\svnVersion{9ac7821}\def\svnDate{2025/02/18}\def\cmsCernNoTag{CERN-EP-2023-197}\def\cmsCernDate{\today}\def\cmsMessage{Published in the Journal of High Energy Physics as \href{http://dx.doi.org/10.1007/JHEP02(2025)064}{\doi{10.1007/JHEP02(2025)064}.}}
\begin{document}\cmsNoteHeader{TOP-20-006}

\newcommand{\ttother}{\ensuremath{\ttbar \, \text{other}\xspace}}
\newcommand{\Wjets}{\ensuremath{\PW\text{+jets}}\xspace}
\newcommand{\Zjets}{\ensuremath{\PZ\text{+jets}}\xspace}
\newcommand{\mumu}{\ensuremath{\PGmp\PGmm}\xspace}
\newcommand{\ee}{\ensuremath{\Pep\Pem}\xspace}
\newcommand{\emu}{\ensuremath{\Pepm\PGmmp}\xspace}
\DeclareRobustCommand{\PX}{{\HepParticle{X}{}{}}\Xspace}
\newcommand{\MadSpin}{{\textsc{MadSpin}}} 
\newcommand{\MGaMCatNLO}{MG5\_a\textsc{mc@nlo}[FxFx]}
\newcommand{\Matrix}{\textsc{Matrix} (NNLO)\xspace}
\newcommand{\MiNNLOPS}{\textsc{MiNNLOPS} (NNLOPS)\xspace}
\newcommand{\Stripper}{\textsc{Stripper} (NNLO)\xspace}
\newcommand{\appNTLO}{\ensuremath{a\text{N}^{3}\text{LO}}\xspace}
\newcommand{\MatrixOnly}{\textsc{Matrix}\xspace}
\newcommand{\MiNNLOPSOnly}{\textsc{MiNNLOPS}\xspace}
\newcommand{\StripperOnly}{\textsc{Stripper}\xspace}
\newcommand{\appNTLOOnly}{\ensuremath{a\text{N}^{3}\text{LO}}\xspace}
\newcommand{\mt}{\ensuremath{m_{\PQt}}\xspace}
\newcommand{\lumivalueSix}{36.3\fbinv}
\newcommand{\lumivalueSeven}{41.5\fbinv}
\newcommand{\lumivalueEight}{59.7\fbinv}
\newcommand{\lumivalue}{138\fbinv} 
\newcommand{\xsectheo}{\ensuremath{ 831 \pm^{20}_{30} \text{(scale)} \pm 35 \, (\text{PDF}+\alpha_s)\,\text{pb}}\xspace}
\newcommand{\xsectheoc}{831\unit{pb}}
\providecommand{\PYTHIAviii}{\PYTHIA\!8\xspace}
\providecommand{\HERWIGvii}{\HERWIG\!7\xspace}
\newcommand{\PowPyt}{\POWHEG{}+\PYTHIAviii\xspace}
\newcommand{\PowHer}{\POWHEG{}+\HERWIGvii\xspace}
\newcommand{\aMCPyt}{\MGaMCatNLO+\PYTHIAviii\xspace}
\newcommand{\chisq}{\ensuremath{\chi^2}\xspace}
\newcommand{\ndf}{dof\xspace}
\newcommand{\Irel}{\ensuremath{I_{\text{rel}}}\xspace}
\newcommand{\hdamp}{\ensuremath{h_{\text{damp}}}\xspace}
\newcommand{\mur}{\ensuremath{\mu_{\text{r}}}\xspace}
\newcommand{\muf}{\ensuremath{\mu_{\text{f}}}\xspace}

\newcommand{\mtt}{\ensuremath{m(\ttbar)}\xspace}
\newcommand{\ptt}{\ensuremath{\pt(\PQt)}\xspace}
\newcommand{\ptat}{\ensuremath{\pt(\PAQt)}\xspace}
\newcommand{\yt}{\ensuremath{y(\PQt)}\xspace}
\newcommand{\absyt}{\ensuremath{\abs{y(\PQt)}}\xspace}
\newcommand{\yat}{\ensuremath{y(\PAQt)}\xspace}
\newcommand{\ytt}{\ensuremath{y(\ttbar)}\xspace}
\newcommand{\absytt}{\ensuremath{\abs{y(\ttbar)}}\xspace}
\newcommand{\detatt}{\ensuremath{\Delta \eta(\PQt,\PAQt)}\xspace}
\newcommand{\absdetatt}{\ensuremath{\abs{\Delta \eta(\PQt,\PAQt)}}\xspace}
\newcommand{\dytt}{\ensuremath{\abs{y(\PQt)}-\abs{y(\PAQt)}}\xspace}
\newcommand{\absdytt}{\ensuremath{\abs{\abs{y(\PQt)}-\abs{y(\PAQt)}}}\xspace}
\newcommand{\dphitt}{\ensuremath{\abs{\Delta \phi(\PQt,\PAQt)}}\xspace} 
\newcommand{\pttt}{\ensuremath{\pt(\ttbar)}\xspace}
\newcommand{\rpttmtt}{\ensuremath{\ptt/\mtt}\xspace}
\newcommand{\rptttmtt}{\ensuremath{\pttt/\mtt}\xspace} 
\newcommand{\rptbsptts}{\ensuremath{(\pt(\PQb) + \pt(\PAQb))/(\ptt + \ptat)}\xspace}
\newcommand{\logxone}{\ensuremath{\log(\xi_{1})}\xspace}
\newcommand{\logxtwo}{\ensuremath{\log(\xi_{2})}\xspace}

\newcommand{\ptb}{\ensuremath{\pt(\PQb)}\xspace}
\newcommand{\ptll}{\ensuremath{\pt(\Pell\PAell)}\xspace}
\newcommand{\absetall}{\ensuremath{\abs{\eta(\Pell\PAell)}}\xspace}
\newcommand{\ptlep}{\ensuremath{\pt(\Pell)}\xspace}
\newcommand{\mll}{\ensuremath{m(\Pell\PAell)}\xspace}
\newcommand{\mbb}{\ensuremath{m(\PQb\PAQb)}\xspace}
\newcommand{\mlb}{\ensuremath{m(\Pell\PQb)}\xspace}
\newcommand{\mllbb}{\ensuremath{m(\Pell\PAell\PQb\PAQb)}\xspace}
\newcommand{\nj}{\ensuremath{N_{\text{jet}}}\xspace}

\newcommand{\ytptt}{\ensuremath{[\absyt,\, \ptt]}\xspace}
\newcommand{\yttpttt}{\ensuremath{[\absytt,\, \pttt]}\xspace}
\newcommand{\pttpttt}{\ensuremath{[\ptt,\, \pttt]}\xspace}
\newcommand{\mttyt}{\ensuremath{[\mtt,\, \absyt]}\xspace}
\newcommand{\mttytt}{\ensuremath{[\mtt,\, \absytt]}\xspace}
\newcommand{\mttdetatt}{\ensuremath{[\mtt,\, \absdetatt]}\xspace} 
\newcommand{\mttdphitt}{\ensuremath{[\mtt,\, \dphitt]}\xspace} 
\newcommand{\mttpttt}{\ensuremath{[\mtt,\, \pttt]}\xspace}
\newcommand{\mttptt}{\ensuremath{[\mtt,\, \ptt]}\xspace}
\newcommand{\etallmll}{\ensuremath{[\absetall,\, \mll]}\xspace}
\newcommand{\etallptll}{\ensuremath{[\absetall,\, \ptll]}\xspace}
\newcommand{\ptllmll}{\ensuremath{[\ptll,\, \mll]}\xspace}
\newcommand{\njforty}{\ensuremath{N_{\text{jet}} (\pt > 40\GeV) }\xspace}
\newcommand{\njhundred}{\ensuremath{N_{\text{jet}} (\pt > 100\GeV) }\xspace}
\newcommand{\njptt}{\ensuremath{[\nj,\, \ptt]}\xspace}
\newcommand{\njpttt}{\ensuremath{[\nj,\, \pttt]}\xspace}
\newcommand{\njyt}{\ensuremath{[\nj,\, \absyt]}\xspace}
\newcommand{\njytt}{\ensuremath{[\nj,\, \absytt]}\xspace}
\newcommand{\njmtt}{\ensuremath{[\nj,\, \mtt]}\xspace}
\newcommand{\njdetatt}{\ensuremath{[\nj,\, \absdetatt]}\xspace}

\newcommand{\ptttmttytt}{\ensuremath{[\pttt,\, \mtt,\, \absytt]}\xspace}
\newcommand{\njmttytttwo}{\ensuremath{[N^{0,1+}_{\text{jet}},\, \mtt,\, \absytt]}\xspace}
\newcommand{\njmttyttthree}{\ensuremath{[N^{0,1,2+}_{\text{jet}},\, \mtt,\, \absytt]}\xspace}
\newcommand{\njmttyttfour}{\ensuremath{[N^{0,1,2,3+}_{\text{jet}},\, \mtt,\, \absytt]}\xspace}

\newcommand{\pp}{\ensuremath{\Pp\Pp}\xspace}
\newcommand{\mb}{\unit{mb}}
\newcommand{\tW}{\ensuremath{\PQt\PW}\xspace}
\newcommand{\PowPytSh}{POW+PYT\xspace}
\newcommand{\PowHerSh}{POW+HER\xspace}
\newcommand{\aMCPytSh}{FxFx+PYT\xspace}
\newcommand{\mtmc}{\ensuremath{m_{\PQt}^{\text{MC}}}\xspace}
\newcommand{\smperc}{\ensuremath{<}1}

\ifthenelse{\boolean{cms@external}}{\providecommand{\cmsTable}[1]{\resizebox{\linewidth}{!}{#1}}}{\providecommand{\cmsTable}[1]{#1}}

\cmsNoteHeader{TOP-20-006} 
\title{Differential cross section measurements for the production of top quark pairs and of additional jets using dilepton events from \texorpdfstring{$\Pp\Pp$}{pp} collisions at \texorpdfstring{$\sqrt{s} = 13\TeV$}{sqrt(s) = 13 TeV}}
\titlerunning{Top quark pair production at 13\TeV}

\date{\today}

\abstract{
Differential cross sections for top quark pair (\ttbar) production are measured in proton-proton collisions at a center-of-mass energy of 13 TeV using a sample of events containing two oppositely charged leptons. The data were recorded with the CMS detector at the CERN Large Hadron Collider and correspond to an integrated luminosity of \lumivalue. The differential cross sections are measured as functions of kinematic observables of the \ttbar system, the top quark and antiquark and their decay products, as well as of the number of additional jets in the event. The results are presented as functions of up to three variables and are corrected to the parton and particle levels. When compared to standard model predictions based on quantum chromodynamics at different levels of accuracy, it is found that the calculations do not always describe the observed data. The deviations are found to be largest for the multi-differential cross sections.
}

\hypersetup{pdfauthor={CMS Collaboration},
pdftitle={Measurement of differential cross sections for the production of top quark pairs and of additional jets using dilepton events from pp collisions atsqrt(s) = 13 TeV},
pdfsubject={CMS},pdfkeywords={CMS, top, dilepton, unfolding, QCD, jets}}

\maketitle 

\maketitle \section{Introduction}
\begin{figure}[h]
    \centering
    \includegraphics[width=0.60\textwidth]{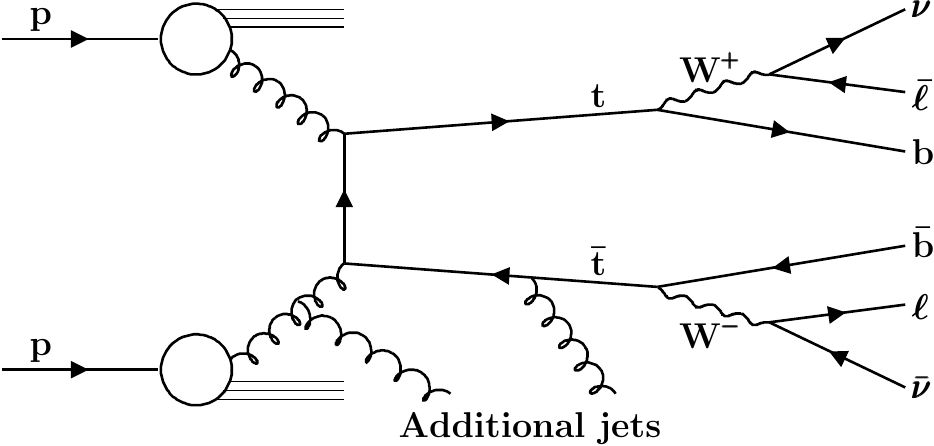}
    \caption{Illustration of a \pp collision with \ttbar plus additional jet production and
     subsequent dilepton decay of the \ttbar system.}
    \label{fig:intro_feyn}
\end{figure}

Measurements of top quark pair (\ttbar) production play a crucial role in
testing the validity of the standard model (SM) and in
searching for new phenomena~\cite{ParticleDataGroup:2022pth}.
The large data set of proton-proton (\pp) collisions
delivered during Run 2 at the CERN Large Hadron Collider (LHC)
in the years 2016 through 2018 enables
precision studies of \ttbar differential production cross sections as functions of
kinematic variables of various objects produced in the events.
The differential measurements provide
sensitivity to many new physics scenarios~\cite{Atwood:1994vm,Langacker:2008yv,Englert:2012by,ATLAS:2014abv,Albert:2016osu,CMS:2018ysw,CMS:2019pzc,Biekotter:2021qbc,Anuar:2024myn} for
which the \ttbar event topologies and kinematical distributions
are different from those in the SM.
The present analysis focuses on events in the \ttbar dilepton decay channel,
shown in Fig.~\ref{fig:intro_feyn}, where both \PW bosons
decay into a charged lepton and a neutrino.
Kinematic observables of the following objects are studied in the analysis:
the \ttbar system, the top quark (\PQt) and antiquark (\PAQt),
the charged leptons ($\Pell$ and $\PAell$)
and bottom quarks (\PQb and \PAQb) produced in the decay chain, and the additional jets in the event.
Electrons and muons produced directly in the \PW boson decays are considered as signal,
while \PGt leptons are not.
The bottom quarks are experimentally accessible through the associated \PQb jets.

In the SM context, the measured cross sections can be used to check predictions
of perturbative quantum chromodynamics (pQCD) and electroweak theory.
During the last decade, a variety of next-to-next-to-leading order (NNLO)
predictions~\cite{bib:kidonakis_13TeV,bib:Kidonakis:2019yji,bib:mitov,bib:Czakon:2017wor,Czakon:2018nun,bib:Catani:2019hip}
have become available for kinematic observables of the \ttbar system, top quark and antiquark, and 
recently also of the final-state leptons and \PQb jets~\cite{Czakon:2020qbd}.   
The situation is different for \ttbar events with additional energetic jets in
the final state, which, at LHC energies, contribute a large fraction to the total \ttbar cross section.
The NNLO corrections are not yet established for these high multiplicity radiative processes;
nevertheless, a comparison of the available pQCD models to data provides an important
benchmark test.

Differential cross sections for \ttbar production
have been measured previously in \pp collisions at the LHC
at $\sqrt{s} = 7\TeV$~\cite{bib:ATLAS,bib:TOP-11-013_paper,bib:ATLASnew,Aad:2015eia,Aaboud:2016iot},
8\TeV~\cite{Aaboud:2016iot,Khachatryan:2015oqa,Aad:2015mbv,Aad:2015hna,Khachatryan:2015fwh,Khachatryan:2149620,Sirunyan:2017azo,Aad:2019mkw},
and 13\TeV~\cite{Khachatryan:2016mnb,Aaboud:2016xii,Aaboud:2016syx,ATLAS:2017cez,TOP-16-007,ATLAS:2018orx,ATLAS:2018acq,Sirunyan:2018wem,Sirunyan:2018ucr,Sirunyan:2019zvx,Aad:2019ntk,Aad:2019hzw,Aad:2020nsf,CMS:2020tvq,CMS:2021vhb,ATLAS:2022xfj,ATLAS:2022mlu,ATLAS:2023gsl},
in the channels with either both, one, or neither of the \PW bosons
emitted in the decays of \PQt and \PAQt decaying leptonically.
The nominal predictions of modern pQCD calculations
generally fail to describe several kinematic distributions.
For instance, most theoretical models predict a spectrum for the transverse momentum \pt
of the top quark, that is harder than observed~\cite{Sirunyan:2018ucr,Sirunyan:2018wem}.

We present measurements of differential \ttbar cross sections in \pp collisions at $\sqrt{s} = 13\TeV$
using data taken with the CMS detector during the Run-2 operation of the LHC.
The analysis is based on an integrated luminosity of \lumivalue, where \lumivalueSix
were recorded in 2016, \lumivalueSeven in 2017, and \lumivalueEight in 2018.
The dilepton decay channel has a relatively small branching fraction and significantly lower background compared to other \ttbar decay channels.
As a consequence of the excellent lepton energy resolution, the precise measurement of kinematic observables based on 
lepton pairs is unique to the dilepton channel. However, because of the presence of two neutrinos in the final state, the measurement of top quark kinematic observables in the dilepton channel requires specialized kinematic 
reconstruction techniques.

The analysis follows the procedures and strategies of Refs.~\cite{Sirunyan:2018ucr,Sirunyan:2019zvx}
for which only the 2016 data were used.
All measurements are compared to predictions from Monte Carlo (MC) generators 
with next-to-leading order (NLO) accuracy in QCD at the matrix element level interfaced to parton shower 
simulations. 
Selected cross section measurements are also compared to a variety of predictions with precision 
beyond NLO.

The distributions studied in this paper are of basic kinematic observables including the \pt,
and pseudorapidities $\eta$ or rapidities $y$ of single objects,
\eg \ptt and \yt, or of compound systems, \eg \pttt and \ytt.
Distributions of invariant masses of compound objects are also investigated, \eg \mtt, as well as the 
azimuthal or rapidity differences
between two objects, \eg \dphitt, \dytt, and \absdetatt, where $\phi$ is the azimuthal angle in radians.
Cross sections are measured at the particle level in a fiducial phase space that is close to that of the detector
acceptance.
In addition, we extract cross sections for kinematic observables of the top quark and antiquark and
the \ttbar system
defined at the parton level in the full phase space, which allows a comparison to
a larger set of higher-order pQCD calculations.
Both absolute and normalized differential cross sections are presented.
The latter are obtained
by dividing the former by the sum of the cross sections measured in the differential bins,
leading to a reduction of systematic uncertainties.
Cross sections are measured as functions of one kinematic variable (single-differential),
or multi-differentially as functions of two or three variables (double- or triple-differential).
The improvements of this analysis compared to Refs.~\cite{Sirunyan:2018ucr,Sirunyan:2019zvx} fall into two 
categories:

\begin{enumerate}
\item Measurements are expanded by considering new kinematic observables, using refined binnings and extending the
phase space range. An example of a new kinematic observable is
the ratio \rpttmtt, revealing interesting details of the \ttbar production dynamics.
A finer binning and extended phase space range is used, in particular for the
kinematic distributions of leptons and \PQb jets.
For measurements with additional jets in the events, a systematic survey
of the correlations of the top quark and \ttbar kinematic variables 
with the number of additional jets in the events is
performed in the dilepton channel for the first time.

\item The statistical and systematic uncertainties of the measurements are generally 
reduced by a factor of about two, 
the latter profiting from the following improvements: 
using refined procedures and algorithms, such as for identifying \PQb jets and measuring their
\pt; applying precise calibrations determined separately for each year of data taking, such as for
the jet
energy scale; having better estimates for some important background process contributions using in situ 
constraints from data;
exploiting MC simulated samples with reduced statistical uncertainties to correct the data for detector 
effects and for assessing systematic uncertainties.

\end{enumerate}

The paper is structured as follows:
Section~\ref{sec:cms} provides a brief description of the CMS detector.
Details of the event simulation are given in Section~\ref{sec:simulation}.
The event selection is detailed in Section~\ref{sec:sel},
followed by a description of the kinematic reconstruction in Section~\ref{sec:kinrec}
where comparisons between data and simulations are shown.
The signal extraction and unfolding procedure are explained in Section~\ref{sec:unfold},
together with the definitions of the parton and particle level phase spaces.
The method to extract the differential cross sections is discussed in Section~\ref{sec:xsec},
and the assessment of the systematic uncertainties is presented in Section~\ref{sec:systematics}.
Results and comparisons to theoretical predictions are shown in Section~\ref{sec:res}.
Finally, Section~\ref{sec:concl} provides a summary. Tabulated
results can be found in HEPData~\cite{hepdata}.

\section{The CMS detector}
\label{sec:cms}
The central feature of the CMS apparatus is a superconducting solenoid of 6\unit{m} internal diameter, providing a
magnetic field of 3.8\unit{T}.
Within the solenoid volume are a silicon pixel and strip tracker, a lead tungstate crystal electromagnetic
calorimeter (ECAL), and a brass and scintillator hadron calorimeter (HCAL),
each composed of a barrel and two endcap sections. Forward calorimeters extend the $\eta$ coverage provided by the
barrel and endcap detectors.
Muons are measured in gas-ionization detectors embedded in the steel flux-return yoke outside the solenoid.
Events of interest are selected using a two-tiered trigger system~\cite{Khachatryan:2016bia}.
The first level (L1), composed of custom hardware processors, uses information from the 
calorimeters and muon detectors
to select events at a rate of around 100\unit{kHz} within a time interval of less than 4\mus.
The second level, known as the high-level trigger (HLT), consists of a farm of processors running a
 version
of the full event reconstruction software optimized for fast processing, and reduces the event rate to 
around
1\unit{kHz} before data storage.
A more detailed description of the CMS detector, together with a definition of the coordinate system used and
the relevant kinematic variables,
can be found in Ref.~\cite{bib:Chatrchyan:2008zzk}.

\section{Event simulation}
\label{sec:simulation}
Simulations of physics processes are performed with MC event generators for
three main purposes.
First, to obtain representative predictions of \ttbar production cross sections
to be compared to the measurements.
Second, to determine corrections for the effects of hadronization, reconstruction, and selection 
efficiencies, as well as resolutions. These corrections are obtained by passing generated \ttbar 
signal events through a detector simulation, and are applied for the unfolding of the data. Last, to 
obtain predictions for the backgrounds by 
passing generated background events through the detector simulation.
All MC programs used in this analysis perform the event generation in several stages:
matrix element (ME) level, parton showering matched to ME, hadronization,
and the underlying event (UE), including multi-parton interactions (MPIs).
For all simulations, the proton structure is described by the NNPDF3.1 NNLO 
set~\cite{bib:NNPDF,Ball:2017nwa} 
of parton distribution functions (PDFs), unless stated otherwise.  
For all simulations with top quark production, 
the value of the top quark mass parameter is fixed to $\mtmc=172.5\GeV$. 
The \ttbar signal process is simulated with ME calculations at NLO in QCD.
For the nominal signal simulation, the \POWHEG 
(version~2)~\cite{Nason:2004rx,bib:powheg,bib:powheg2,Frixione:2007nw} 
generator is taken.
The \hdamp parameter of \POWHEG, which regulates the damping of real emissions in the NLO calculation
when matching to the parton shower, is set to $\hdamp = 1.379\, \mtmc$~\cite{Sirunyan:2019dfx}.
The \PYTHIAviii program (version~8.230)~\cite{Sjostrand:2007gs} with the CP5 tune~\cite{Sirunyan:2019dfx} is used
to model parton showering, hadronization, and the UE.
This setup, referred to as \PowPyt, is used for the detector corrections of the \ttbar signal process in the
data, with appropriate variations for assessing the theoretical model uncertainties, described in
Section~\ref{sec:sys_theo}.
The generator-level cross sections of \PowPyt are also used as theoretical predictions that are
compared in Section~\ref{sec:res} to the measured \ttbar cross sections, as well as the predictions from two other
models.
The first alternative model is based on the \MGvATNLO
(version 2.4.2)~\cite{Alwall:2014hca} generator, interfaced with \PYTHIAviii using the CP5 tune.
At the ME level, up to two extra partons are included at NLO.
The events are matched to \PYTHIAviii using the FxFx prescription~\cite{Frederix:2012ps,bib:CMS:2016kle},
and \MadSpin~\cite{bib:madspin} is used to model the decays of the top quarks, while preserving their spin 
correlation.
The whole setup is referred to as \aMCPyt.
The second alternative model is \POWHEG interfaced to \HERWIGvii~\cite{Bellm_2016}
using the CH3 tune~\cite{cmscollaboration2020development} and is referred to as \PowHer.

The main background contributions originate from single top quarks produced in association
with a {\PW} boson (\tW),\, $\PZ/\PGg^{*}$ bosons produced with additional jets (\Zjets), {\PW} boson production
with additional jets (\Wjets),
and diboson ($\PW\PW$, $\PW\PZ$, and $\PZ\PZ$) events.
Other backgrounds are negligible.

For all background samples, parton showering, hadronization,
and the UE are simulated with \PYTHIAviii.
For single top quark production, the
$t$-channel and \tW processes are
simulated at NLO with \POWHEG~\cite{bib:powheg1,Frederix:2012dh,bib:powheg3}, 
and the $s$-channel process at LO with \MGvATNLO.
In all three cases the \PYTHIAviii CP5 tune is used.
For all other background samples discussed in the following, 
the CP5 tune is applied for the 2017 and 2018 samples,
and the CUETP8M1 tune~\cite{bib:CMS:2016kle,bib:CUETP8tune,Skands:2014pea} for the 2016 samples.
For the latter, the PDF set NNPDF3.0~\cite{NNPDF:2014otw} is used
with the order (\eg NLO) of the respective simulation.
The \Zjets process is simulated at NLO using \MGaMCatNLO\ with up to two additional partons at the ME level.
The \Wjets process is simulated at leading order (LO) using \MGvATNLO
with up to four additional partons at the ME level and matched to \PYTHIAviii using
the MLM prescription~\cite{bib:MLM,Mrenna:2003if}.
Diboson events are simulated at LO with \PYTHIAviii.

Predictions are normalized based on their inclusive
theoretical cross sections and the integrated luminosity of the data sample.
For $s$- and $t$-channel single top quark production, the
cross sections are calculated at NLO
with \textsc{Hathor} (version 2.1)~\cite{Kant:2014oha}.
For single top quark production in the \tW channel, the
approximate NNLO calculations from Ref.~\cite{bib:twchan} are used.
For \Zjets and \Wjets processes, NNLO predictions obtained with \FEWZ~\cite{Li:2012wna} are 
taken, and for diboson production the NLO calculations from Ref.~\cite{bib:mcfm:diboson} are applied.
The \ttbar simulation is normalized to a cross section of \xsectheo
calculated with the \TOPpp (version~2.0) program~\cite{Czakon:2011xx}
at NNLO, including resummation of next-to-next-to-leading logarithm (NNLL)
soft-gluon terms~\cite{bib:xs1,bib:xs2,bib:xs3,bib:xs4,bib:xs7,Czakon:2013goa}, and assuming
a top quark pole mass of 172.5\GeV.

The CMS detector response is simulated using \GEANTfour~\cite{bib:geant}.
The effect of additional \pp interactions within the same or nearby bunch crossings (pileup) is taken 
into account by adding
simulated minimum-bias interactions to the simulated data.
Events in the simulation are then weighted to reproduce the pileup distribution in the data,
which is estimated from the measured bunch-to-bunch instantaneous
luminosity and assuming a total inelastic \pp cross section of 69.2\mb~\cite{Aaboud:2016mmw}.
Separate simulations are employed for the data taken in the three years 2016--2018,
in order to match the varying detector performance and data-taking conditions.
At every step of the analysis, the simulated samples from different years are added together
and used as one single sample, both at the detector and the generator levels.  

Correction factors described in Sections~\ref{sec:unfold} and~\ref{sec:systematics}, 
subsequently referred to as scale factors, are used to reconcile the number of expected events from 
simulation with what
is observed in data. They are applied, for example, to correct a detector efficiency in the 
simulation to match the one observed in data, or to scale a background prediction.

\section{Event selection}
\label{sec:sel}
The event selection closely follows that of Ref.~\cite{Sirunyan:2018ucr}.
Events are selected corresponding to the decay chain
where both top quarks decay into a {\PW} boson and a bottom quark,
and each of the {\PW} bosons decays directly into an electron or a muon and a neutrino.
This specification defines the signal process, while
all other \ttbar events, including those with at least one electron or muon originating from the decay
of a \PGt lepton, are treated as background which is taken into account,
as detailed in Section~\ref{sec:unfold}.
The signal includes three distinct channels: two same-flavor channels corresponding
to two electrons (\ee) or two muons (\mumu),
and the different-flavor channel corresponding to one electron and one muon (\emu).
Results are obtained by combining the three channels and adding, at every step of the analysis,  
the samples from the years 2016--2018.

At the HLT level, events are selected either by 
single-lepton or dilepton triggers.
The former require the presence of at least one electron or muon,
and the latter the presence of either two electrons, two muons, or an electron and a muon.
For all employed triggers, the leptons are required to fulfill isolation criteria that are looser
than those applied later in the offline analysis.
For the single-electron triggers, a \pt threshold of 27 (32)\GeV is applied in 2016 (2017--2018),
while for single-muon triggers the \pt threshold is 24 (27)\GeV in 2016/2018 (2017).
The dilepton triggers select events based on the leptons with the highest (leading)
and second-highest (trailing) \pt in the event.
The same-flavor dilepton triggers require either an electron pair with $\pt > 23$ (12)\GeV for the leading
(trailing)
electron or a muon pair with $\pt > 17$ (8)\GeV for the leading (trailing) muon.
The different-flavor dilepton triggers require either an electron with  $\pt > 23$\GeV
and a muon with $\pt > 8$\GeV, or a muon with $\pt > 23$\GeV and an electron with $\pt > 12$\GeV.
The analysis mainly relies on the dilepton triggers, while the single-lepton triggers
help to improve the overall trigger efficiency by about 10\%.

The particle-flow (PF) algorithm~\cite{bib:Sirunyan:2017ulk} aims to reconstruct and identify each individual particle
with an optimized combination of information from the various elements of the CMS detector.
The energy of muons is obtained from the curvature of the corresponding track.
The energy of electrons is inferred from a combination of the electron momentum at the primary interaction vertex
as determined by the tracker, the energy of the corresponding ECAL cluster,
and the energy sum of all bremsstrahlung photons spatially compatible with originating from the electron track.
The energy of photons is directly obtained from the ECAL measurement.
The energy of charged hadrons is determined from a combination of their momentum
measured in the tracker and the matching ECAL and HCAL energy deposits,
corrected for the response function of the calorimeters to hadronic showers.
Finally, the energy of neutral hadrons is obtained from the corresponding corrected ECAL and HCAL energies.

The measurements presented in this paper depend on the reconstruction and identification of electrons, muons, jets,
and missing transverse momentum \ptvecmiss associated with neutrinos. Electrons and muons
are selected if they are compatible with originating from the primary \pp interaction vertex.
The primary vertex (PV) is taken to be the vertex corresponding to the hardest scattering in the
event, evaluated using tracking information alone, as described in Section 9.4.1 of Ref.~\cite{CMS-TDR-15-02}.

For both electrons and muons, the ``tight'' identification criteria as described
in Refs.~\cite{CMS:2020uim,Sirunyan:2018} are applied.
Reconstructed electrons~\cite{CMS:2020uim} are required to have $\pt > 25$ (20)\GeV
for the leading (trailing) candidate
and $\abs{\eta} < 2.4$.
Electron candidates with ECAL clusters in the transition region
between the ECAL barrel and endcap, $1.44 <\abs{\eta_{\mathrm{cluster}}}< 1.57$, are excluded 
since the reconstruction of an electron object in this region is not optimal.
A relative isolation
\Irel is defined as the \pt sum of all neutral and charged hadrons, and photon 
candidates
within a distance of 0.3 from the electron in $\eta$-$\phi$~space,
divided by the \pt of the electron candidate.
A maximum value of \Irel is allowed, in the range 0.05--0.10, depending on the
\pt and $\eta$ of the electron candidate.
Further electron identification requirements are applied to reject misidentified electron candidates
and candidates originating from photon conversions. Reconstructed muons~\cite{Sirunyan:2018} are
required to have $\pt > 25$ (20)\GeV for the leading (trailing) candidate and $\abs{\eta}<2.4$.
An isolation requirement of $\Irel < 0.15$ is applied, including particles
within a distance of 0.4 in $\eta$-$\phi$~space from the muon candidate.
Furthermore, muon identification requirements are applied to reject misidentified muon candidates
and muons originating from in-flight decays.
For both electron and muon candidates, \Irel is corrected for residual pileup effects. Finally, for the targeted prompt leptons in 
the \ttbar dilepton decay channel, the total selection efficiencies
are about 90\% for muons and 70\% for electrons.

Jets are reconstructed by clustering the PF candidates using the anti-\kt jet
algorithm~\cite{Cacciari:2008gp,Cacciari:2011ma}
with a distance parameter of 0.4.
The jet energies are corrected following the procedures described in Ref.~\cite{Khachatryan:2016kdb}.
After correcting for all residual energy deposits from charged and neutral
particles from pileup, \pt- and $\eta$-dependent jet energy adjustments are applied to correct for the 
detector response.
Jets are required to have $\pt > 30\GeV$ and $\abs{\eta} < 2.4$ and a
distance in $\eta$-$\phi$~space of at least 0.4 to the closest selected lepton.

The \PQb jets are identified with
the deep neural network algorithm 
\textsc{DeepCSV}~\cite{Sirunyan:2017ezt},
based on tracking and secondary vertex information.
The chosen working point of the network discriminator has a {\PQb}-jet tagging efficiency of 
${\approx \,}80$--90\%
and a mistagging efficiency of  ${\approx \,}1\%$ for jets originating from gluons, as well as \PQu, 
\PQd, or \PQs quarks, and ${\approx \,}30$--40\% for jets originating from {\PQc} quarks.
The energy measurement of the {\PQb}-tagged jets is improved using a deep neural
network estimator~\cite{Sirunyan:2019wwa} that performs a regression after all other jet energy
corrections have been applied.

The \ptvecmiss is computed as the negative vector sum
of the transverse momenta of all the PF candidates in an event,
and its magnitude is denoted as \ptmiss~\cite{Sirunyan:2019kia}.
The \ptvecmiss is updated when accounting for corrections to the energy scale of reconstructed jets in the event.
The pileup per particle identification algorithm~\cite{Bertolini:2014bba}
is applied to reduce the pileup dependence of the \ptvecmiss observable.
The \ptvecmiss is computed from the PF candidates weighted
by their probability to originate from the primary interaction vertex~\cite{Sirunyan:2019kia}.

Events are selected offline if they contain exactly two isolated,
oppositely charged electrons or muons, (\ee, \mumu, \emu)
and at least two jets, with at least one of these jets being {\PQb} tagged.
Events with an invariant mass of the lepton pair \mll below 20\GeV
are removed in order to suppress contributions from resonance decays and low-mass Drell--Yan processes.
Backgrounds from \Zjets processes in the \ee and \mumu channels are further reduced
by requiring $\mll <76\GeV$ or $\mll >106\GeV$, and $\ptmiss > 40\GeV$.

In this analysis, the \ttbar production cross section is also measured as a function of the extra jet 
multiplicity \nj.
Extra jets (also referred to as additional jets) are jets arising primarily from hard QCD radiation
and not from the top quark decays.
At the reconstruction level, the extra jets in dilepton \ttbar candidate events
are defined as jets with $\pt > 40\GeV$ and $\abs{\eta} < 2.4$ that
are isolated from the leptons and from the {\PQb} jets originating from the top quark decays.
These two {\PQb} jets are identified by the \ttbar kinematic reconstruction
algorithms discussed in Section~\ref{sec:kinrec}.
The extra jet is considered isolated when having a distance to the
leptons in $\eta$-$\phi$~space of at least 0.4 
and a distance to the {\PQb} jets 
from top quark decays of at least 0.8.
The requirements on \pt and isolation of extra jets largely eliminate
the expected contributions from gluons radiated off \PQb quarks produced in the top quark decays.

\section{Kinematic reconstruction of the \texorpdfstring{\ttbar}{ttbar} system}
\label{sec:kinrec}

The four-momenta of the top quark and antiquark are determined
from the four-momenta of their decay products using a
kinematic reconstruction method referred to as the ``full kinematic reconstruction''~\cite{Khachatryan:2015oqa}.
In this reconstruction, the two pairs consisting of a lepton and a \PQb jet from the decay 
are identified, and the top quark (antiquark) is associated with the pair containing the lepton with 
positive (negative) charge.
The three-momenta of the neutrino ($\PGn$) and antineutrino (${\PAGn}$) are
reconstructed using algebraic equations deduced from the following six kinematic constraints: the conservation of the total \pt in the event,
and the masses of the {\PW} bosons and of the top quark and antiquark.
The mass values used in the constraints are 172.5\GeV for the top quark and antiquark,
and 80.4\GeV for the {\PW} bosons.
The \ptvecmiss in the event is assumed to originate solely from the two neutrinos.
An ambiguity can arise due to multiple algebraic solutions of the constraint equations for the neutrino 
momenta,
which is resolved by taking the solution with the smallest invariant mass of the \ttbar system~\cite{Korol:2016wzq}.
The reconstruction is performed 100 times.
Each time, the measured energies and directions
of the reconstructed leptons and jets are randomly smeared in accordance with their resolutions.
This procedure recovers events that initially yield no solution because of measurement fluctuations.
The three-momenta of the two neutrinos are determined as a weighted average over all smeared solutions.
For each solution, the weight is calculated based on the expected true spectrum of the invariant mass
of the lepton and the \PQb jet originating from the decay of a top quark \mlb and taking
the product of the two weights for the top quark and antiquark decay chains.
All possible lepton-jet combinations in the event that satisfy the requirement $\mlb < 180\GeV$ are 
considered.
Combinations are ranked based on the presence of \PQb-tagged jets in the assignments, \ie a combination 
with
both leptons assigned to \PQb-tagged jets is preferred over those with one or no \PQb-tagged jet.
Among assignments with an equal number of \PQb-tagged jets, the one with the highest sum of weights is 
taken.
Events with no solution after smearing are rejected.
The efficiency of the full kinematic reconstruction is defined as the number of events for 
which a solution
is found divided by the total number of selected \ttbar events after the full event selection 
described in Section~\ref{sec:sel}.
Consistent results are observed in data and simulation.
The efficiency for signal events is about 90\%.
Performing the reconstruction more than
100 times does not significantly 
alter the efficiency and kinematic resolutions.
After applying the complete event selection and the full kinematic reconstruction,
about 1.2 million events are observed, with
shares of 56, 14, and 30\%
for the \emu, \ee, and \mumu channels, respectively.
Combining all decay channels, the estimated signal fraction in data is about 80\%.

\begin{figure*}
    \centering
    \includegraphics[width=0.48\textwidth]{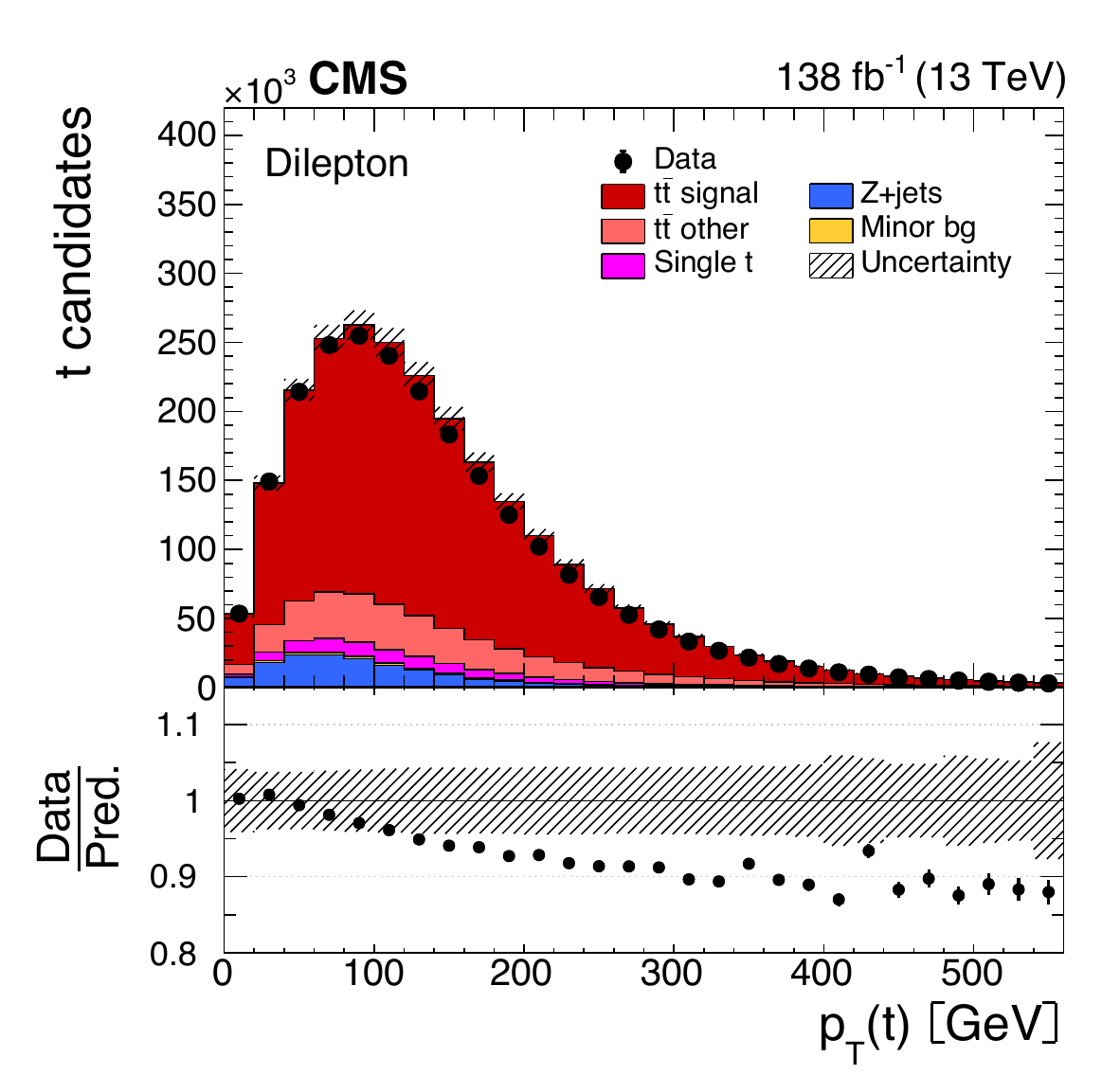}
    \includegraphics[width=0.48\textwidth]{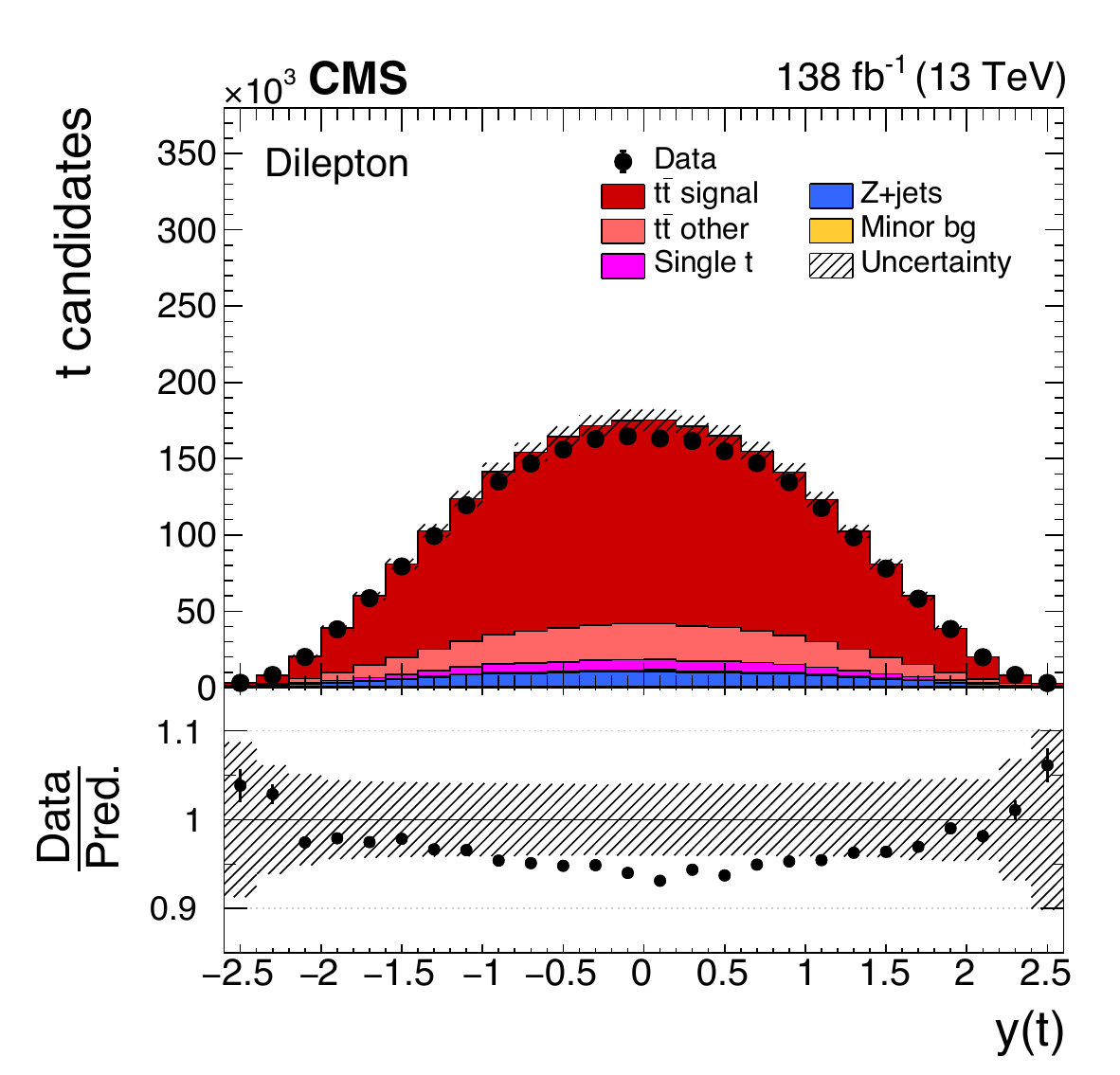}
    \includegraphics[width=0.48\textwidth]{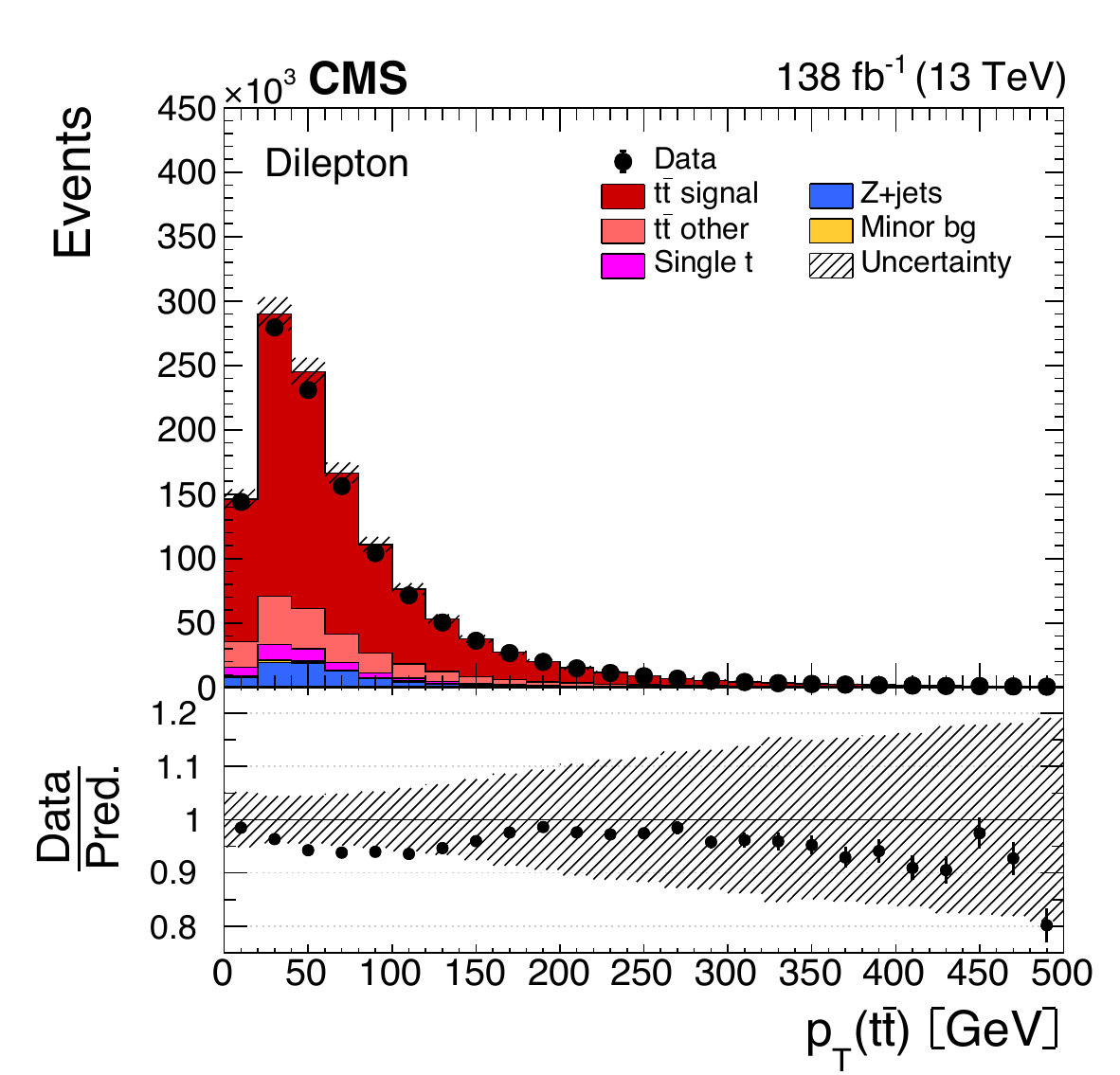}
    \includegraphics[width=0.48\textwidth]{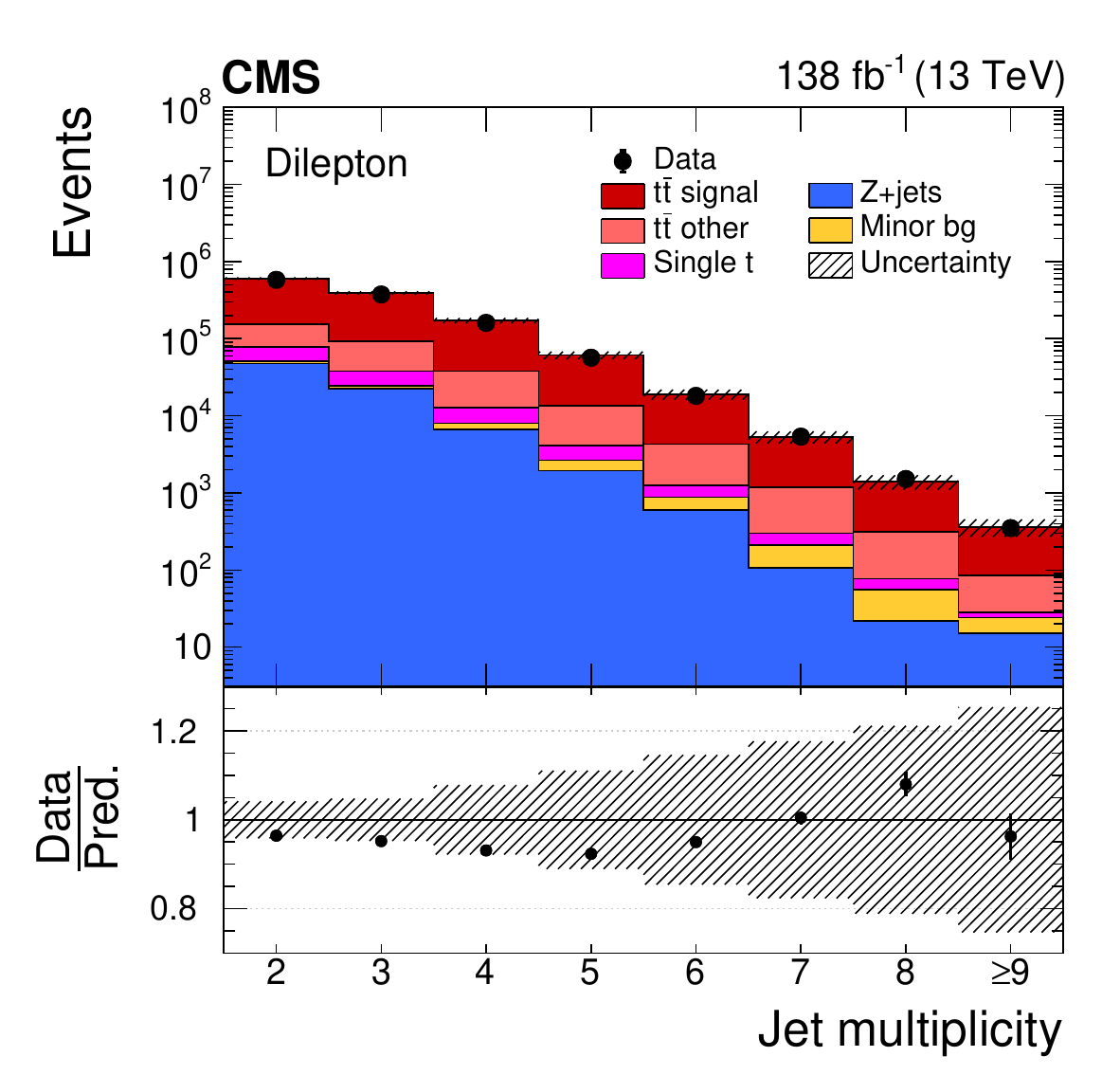}
    \caption{Distributions of \ptt (upper left), \yt (upper right), \pttt (lower left),
    and jet multiplicity (lower right) obtained in selected events with the full kinematic 
reconstruction. For the first two distributions, ``\PQt'' refers to both top quark and antiquark.
    The three dilepton channels (\ee, \mumu, and \emu) are added together.
    The data with vertical bars corresponding to their statistical uncertainties
    are plotted together with distributions of simulated signal and background processes.
    The hatched regions depict the systematic shape uncertainties in the signal and backgrounds
 (as detailed in
Section~\ref{sec:systematics}).
    The lower panel in each plot shows the ratio of the observed data event yields to those expected in the
 simulation.}
    \label{fig:cp}
\end{figure*}

\begin{figure*}
    \centering
    \includegraphics[width=0.48\textwidth]{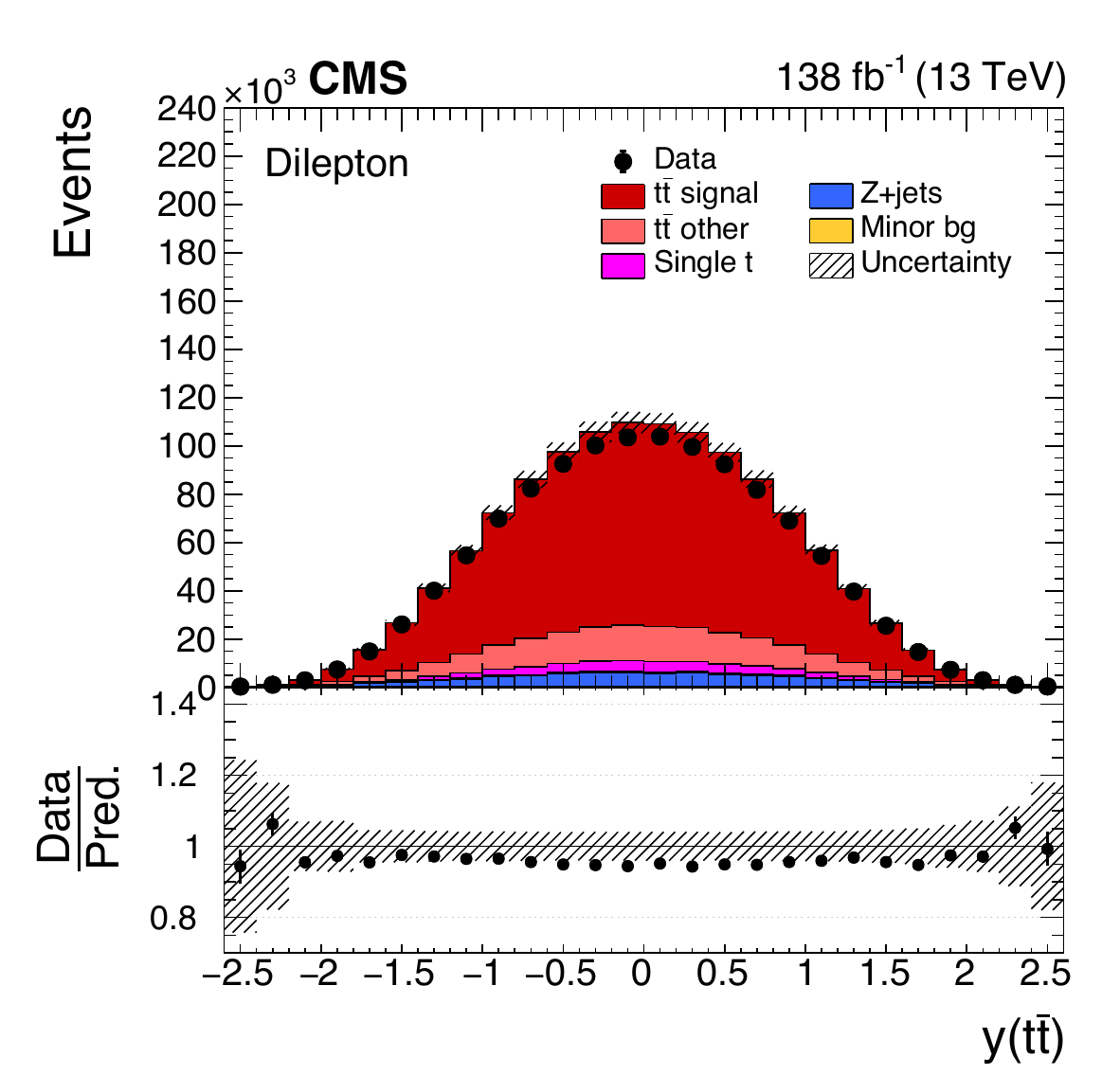}
    \includegraphics[width=0.48\textwidth]{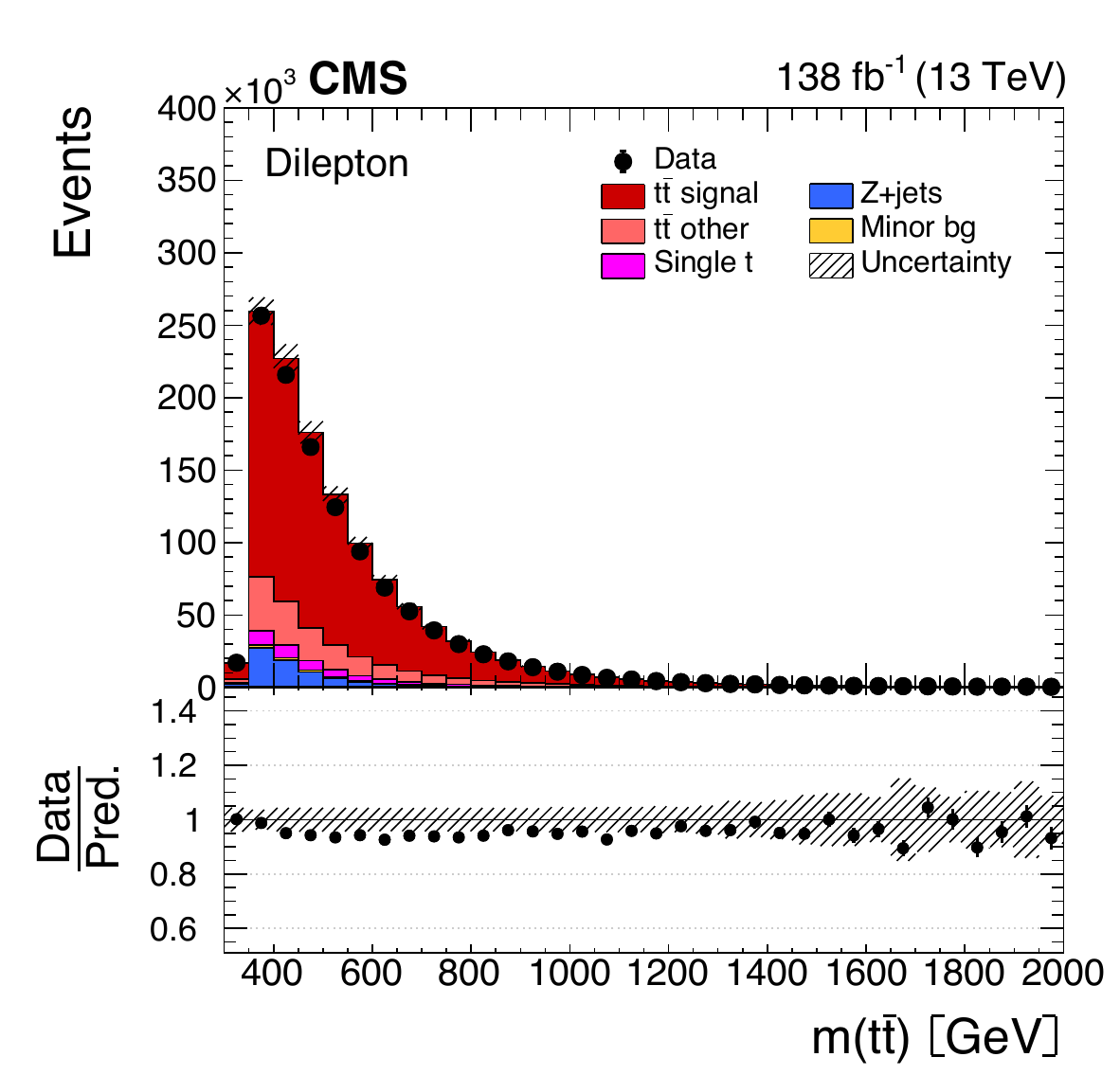}
    \includegraphics[width=0.49\textwidth]{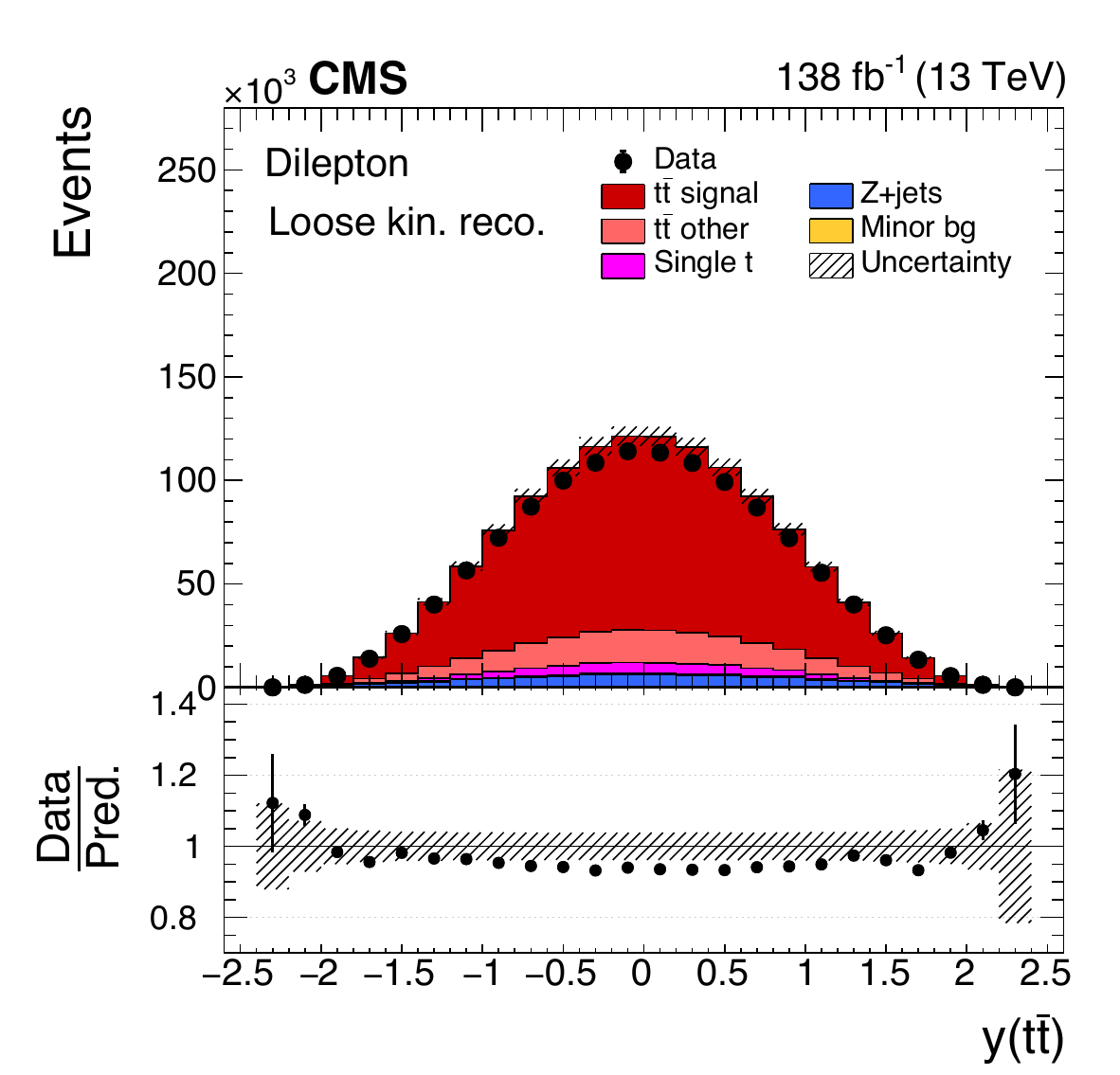}
    \includegraphics[width=0.49\textwidth]{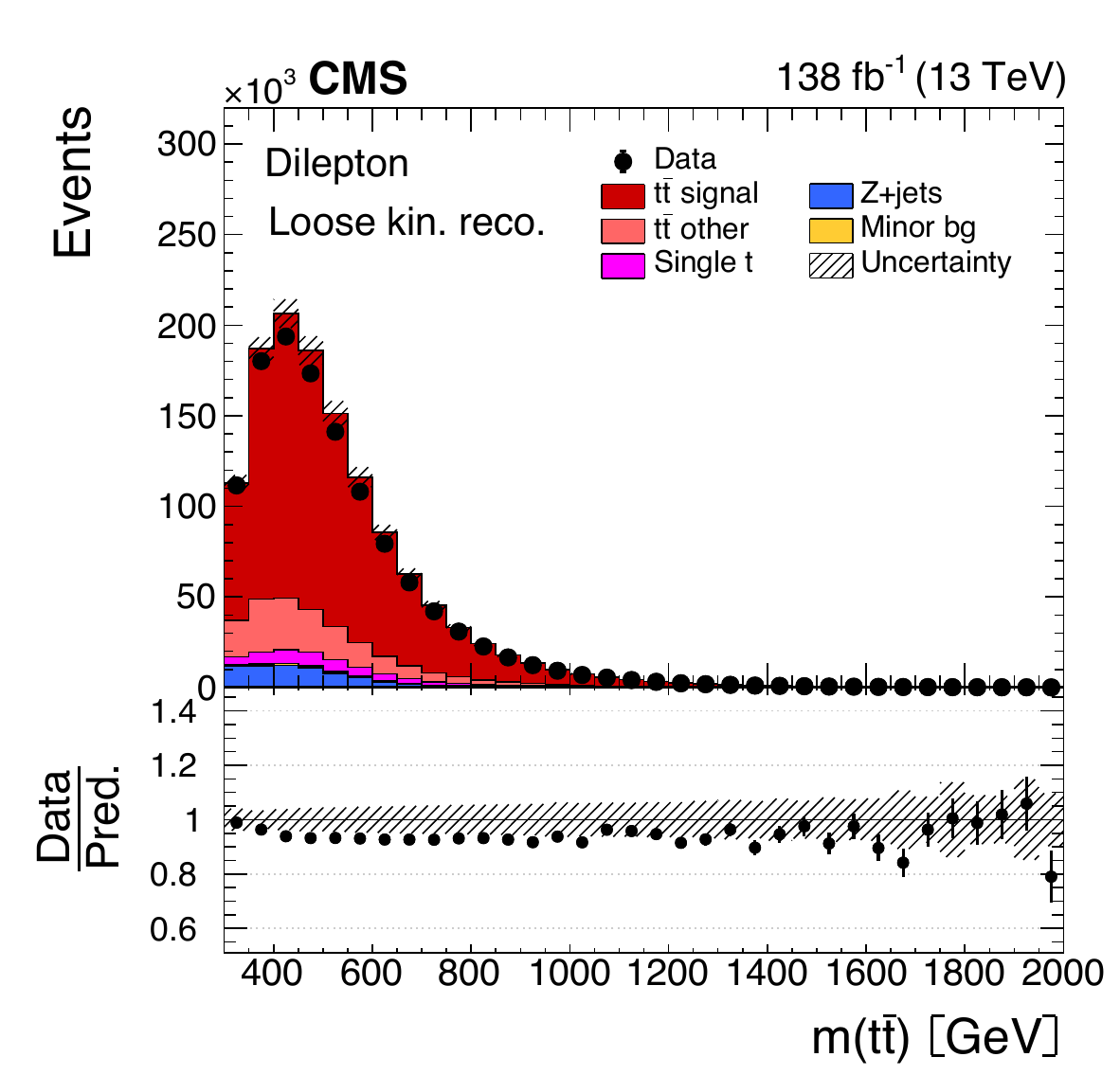}
    \caption{Distributions of \ytt (left) and \mtt (right)
    obtained in selected events with the full (upper) and the loose kinematic reconstruction (lower).
    Further details can be found in the caption of Fig.~\ref{fig:cp}.
    }
    \label{fig:cp:loosekr}
\end{figure*}

Distributions of the reconstructed top quark and \ttbar kinematic variables,
obtained with the full kinematic reconstruction, are shown in Fig.~\ref{fig:cp} and the upper plots 
in Fig.~\ref{fig:cp:loosekr}. Furthermore, the multiplicity of jets in the events is presented in 
Fig.~\ref{fig:cp}. The \ptt and \yt spectra include contributions from both the top quark
and antiquark.
The expected signal and background contributions are estimated
as described in Section~\ref{sec:unfold}, using the MC simulations for the various processes introduced
in Section~\ref{sec:simulation}.
The events labeled as ``$\ttother$'' show the expected contributions from
\ttbar final states other than the signal,
dominated by events
with one or both of the {\PW} bosons decaying into \PGt leptons with subsequent decay into
electrons or muons.
The expected events labeled as ``Minor bg'' constitute minor background contributions from 
diboson and \Wjets processes.
In general, the data are reasonably well described by the simulation that overestimates, however, 
the total number of events by about 5\%.
Some trends are visible, in particular for \ptt, where the simulation
predicts a somewhat harder spectrum than that observed in data, as seen in previous differential \ttbar 
cross section
measurements~\cite{Khachatryan:2015oqa,Khachatryan:2015fwh,Sirunyan:2017azo,Khachatryan:2016mnb,Sirunyan:2018wem,TOP-16-007,Sirunyan:2018ucr}. 
The mismodeling of the data by the simulation is accounted for by
the relevant systematic uncertainties described in 
Section~\ref{sec:systematics}. The checks discussed in Section~\ref{sec:xsec}
confirm that the residual mismodeling of the \ptt distribution
does not introduce a significant bias in the measurement.

The \mtt value obtained using the full kinematic reconstruction
described above is sensitive to the value of the top quark mass used as a kinematic constraint.
An alternative algorithm is employed that reconstructs the \ttbar kinematic variables without using the 
top quark mass constraint.
This algorithm is referred to as the ``loose kinematic reconstruction''.
It is used in this analysis for measuring differential \ttbar cross sections as a function of
\mtt, in order to preserve the sensitivity of the data for a future top quark mass extraction, as 
performed in Ref.~\cite{Sirunyan:2019zvx}.
In contrast to the full kinematic reconstruction, this algorithm tackles the reconstruction of 
the $\PGn\PAGn$ system as a whole.
The $\ell$\PQb pairs are selected and ranked as described for the full kinematic reconstruction. Among combinations with equal number of \PQb-tagged jets, the ones with the leading and trailing \pt jets are chosen.
The kinematic variables of the $\PGn\PAGn$ system are obtained as follows:
its \ptvec is set equal to \ptvecmiss and its
unknown longitudinal momentum and energy are set equal to
the values of the lepton pair.
Additional constraints are applied on the invariant mass of the neutrino pair, $m(\PGn\PAGn) \ge 0$, and on the invariant mass of the {\PW} bosons, $m(\PWp\PWm) \ge 2m_{\PW}$.
These constraints have only minor impact on the performance of the reconstruction.
This method achieves an average reconstruction efficiency  
of approximately 96\% for signal events and provides
\ttbar kinematic resolutions 
comparable to those obtained with the full kinematic reconstruction.
As in the case of the full kinematic reconstruction, events with no valid solution for the loose
kinematic reconstruction are excluded from further analysis.
The presence of a solution for the full kinematic reconstruction in an event is also required for
the cross section measurements as functions of lepton and \PQb jet kinematic variables
discussed in Section~\ref{sec:res_2}.
Figure~\ref{fig:cp:loosekr} (lower row) displays the distributions of the reconstructed \ttbar 
invariant mass
and rapidity using the loose kinematic reconstruction.
These distributions are similar to the ones obtained 
with the full kinematic reconstruction, shown in the upper row, except for \mtt 
in the threshold region that is smeared out due the omission of the top quark mass constraint.

The loose kinematic reconstruction is used for a subset of measurements presented in 
Section~\ref{sec:res}, namely, for single-differential cross sections as functions of \mtt, 
and for all multi-differential cross sections composed of combinations of \mtt with \pttt, \ytt, or \nj.
For all other cross sections as functions of kinematic observables
of the \ttbar system or of the top quark and antiquark,
the full kinematic reconstruction is used.

\section{Signal extraction and unfolding}
\label{sec:unfold}
The number of signal events
is obtained for each histogram bin by subtracting the expected number of background events from
the observed number of events.

The expected
background contribution from single top quark, \Wjets, and diboson processes are taken directly from the MC simulations.
The \Zjets contribution is also estimated from the MC simulation, but corrected by
global normalization scale factors,
which are determined from a binned template fit to the data, using the method described in Ref.~\cite{Barlow:1993dm},
as implemented in the \textsc{TFractionFitter} class in \textsc{ROOT}~\cite{Brun:1997pa}.
In this procedure, the event fraction of \Zjets process and of the sum of all other contributions
are fitted to the \mll distributions in the data, within the \PZ boson peak signal region 
$76 < \mll <106\GeV$.
The template distributions are obtained from the MC-simulated samples.
Separate normalization scale factors are fitted for the simulations of the \Zjets process in
the \ee and \mumu channels and the scale factor for the \emu channel is calculated as the geometric mean
of these factors.
The nominal scale factors are determined for a data selection with relaxed requirements,
omitting those on \ptmiss, the number of \PQb-tagged jets, and the presence of a 
solution for the kinematic
reconstruction, in order to have a clean sample of \Zjets events.
The scale factors are compatible, within a few percent, with unity.
The \mll window chosen for the fits ensures that there is no overlap
with the sample used for the analysis and is still large enough to provide a good
separation of \Zjets from the other processes.
As a cross-check, the template fits are performed over a wider \mll
range, starting at 20\GeV and extending to values much above the \PZ boson peak signal region,
and only small scale factor variations are observed of the order of one percent.

After the subtraction of non-\ttbar backgrounds, the resulting
event yields are corrected for ``$\ttother$'' contributions, introduced in Section~\ref{sec:kinrec}.
These events arise from the same \ttbar production process as the signal and thus the normalization of 
this background is fixed to that of the signal.
For each bin, the number of events obtained after the subtraction of all other background sources
is multiplied by the ratio of the number of selected \ttbar signal events to the total number of 
selected \ttbar events
(\ie the signal and all other \ttbar events) in simulation.

The numbers of signal events are obtained by adding together the event yields in the \ee, \mumu,
and \emu channels, 
subtracting the background, and correcting for detector effects using the \textsc{TUnfold} package~\cite{Schmitt:2012kp}.
The addition of the three channels before applying detector corrections
is justified by the fact
that background levels are small in all channels and kinematic resolutions are comparable.
The measurements in the separate channels yield consistent results.

\begin{figure*}
    \centering
    \includegraphics[width=0.48\textwidth]{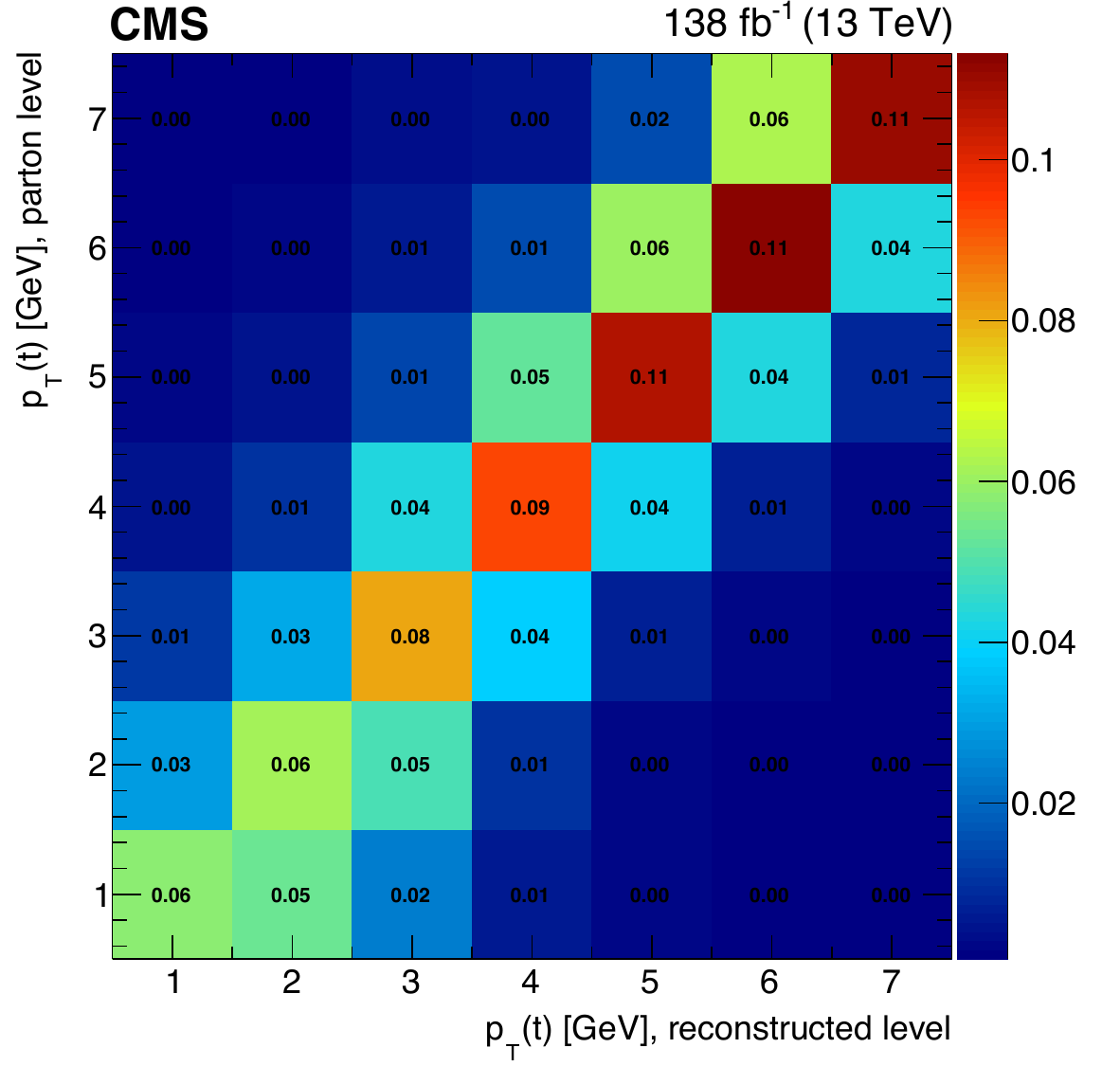}
    \includegraphics[width=0.48\textwidth]{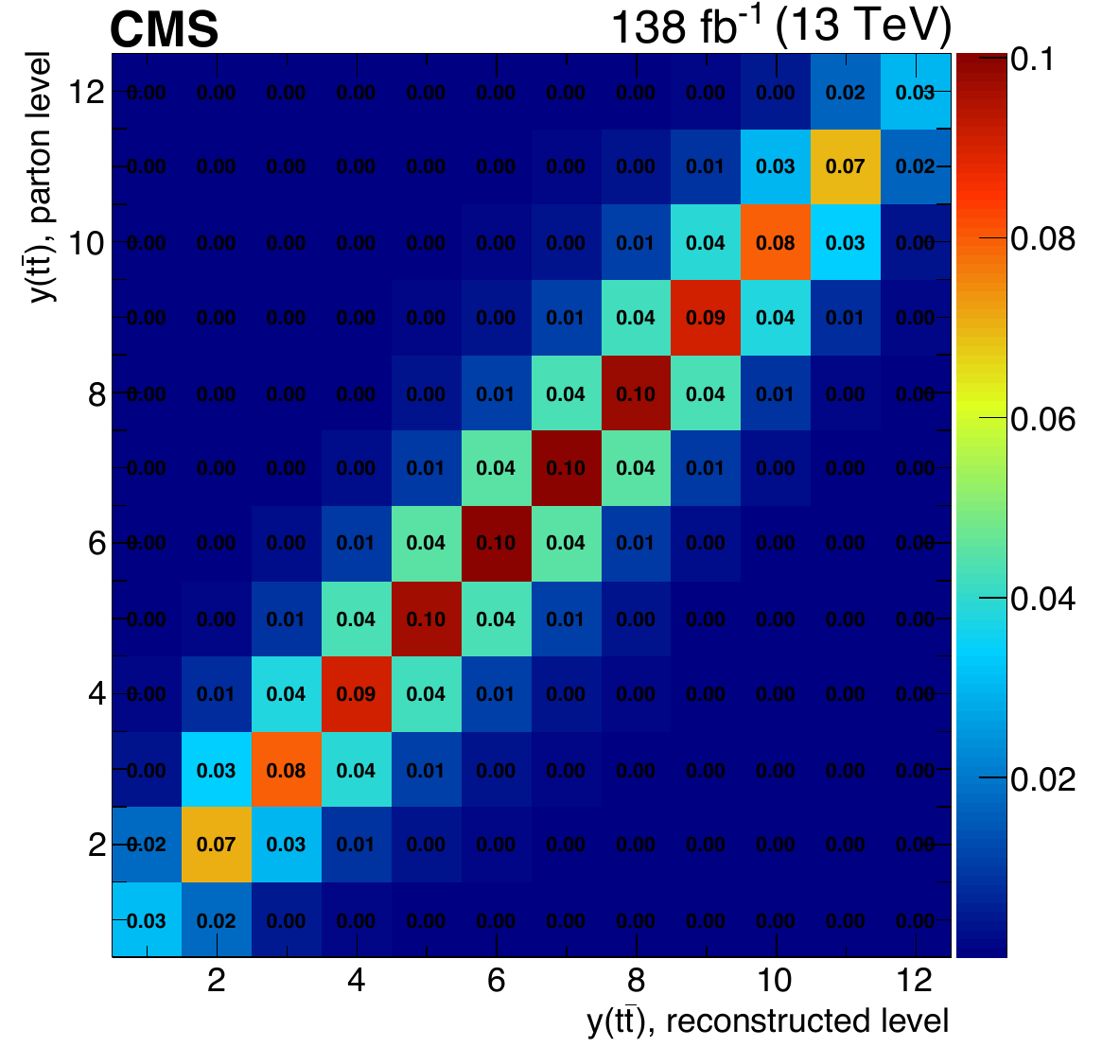}
    \caption{Response matrices for the unfolding
      of the \ptt (left) and \ytt (right)
      distributions at the parton level, as extracted from the nominal \PowPyt\ \ttbar signal simulation. The ranges of the observables for a given bin number 
can be read off from the corresponding cross section distributions in 
Figs.~\ref{fig:res_ptt} and~\ref{fig:res_pttt}.
    }
    \label{fig:respm}
\end{figure*}

In the unfolding procedure the response matrix plays an important role.
An element of this matrix specifies the probability for an event originating
from one bin of the true distribution to be observed in a specific bin of the reconstructed
observables.
The response matrix models the effects of acceptance,
detector efficiencies, and resolutions in a given phase space.
It is calculated for each measured distribution using the \ttbar signal simulation, 
and is defined either at the particle level in a fiducial phase space or at the parton level in the full
phase space, using the corresponding generator-level definitions discussed in 
Section~\ref{sec:truthlevel}.
At the detector level, the number of bins used per kinematic variable 
is typically two times larger than the number of bins used at the generator level.
Figure~\ref{fig:respm} shows example 
response matrices obtained for the \ptt and \ytt distributions
at the parton level.
For illustrative reasons, they are displayed with the
same binnings at both the parton and detector levels.

In \textsc{TUnfold}, the distribution of unfolded event numbers is extracted
from the measured signal distribution at the detector level by performing a \chisq fit. The fit model consists of the sum of template distributions, with one distribution per cross section bin, 
constructed from the product of the unfolded event number in that bin
and the respective column of the response matrix.
An additional \chisq term is included representing Tikhonov regularization~\cite{Tikhonov:1963},
based on second-order derivatives and using the nominal \ttbar simulation as the bias vector.
The regularization reduces unphysically large high-frequency components of the unfolded spectrum.
The regularization strength that minimizes the global correlation coefficient~\cite{Schmitt:2016orm} is 
chosen.
The statistical uncertainties of the simulated \ttbar signal samples
used to derive the response matrices introduce 
small additional uncertainties in the unfolded event yields.
These are accounted for in the \textsc{TUnfold}
procedure through error propagation.

\subsection{Generator-level definitions}
\label{sec:truthlevel}

The definitions of the generator level that are used in the construction of 
the response matrices follow, to a large extent, those applied in 
Refs.~\cite{Sirunyan:2018ucr,Sirunyan:2019zvx,bib:CMS-NOTE-2017-004}.

For the parton-level results, the momenta of the parton-level top quarks are defined after QCD radiation, but before the top quark decays. The unfolded signal event numbers are
corrected for the branching fractions $\PW \to \Pell \PGn$. The parton-level results are 
extrapolated to the full
phase space using the default \ttbar simulation. The extrapolation is implicitly performed by counting
each simulated event in the response
matrix that enters a specific parton-level bin of a differential cross section.

For the particle-level results, the generator-level objects are defined by the stable particles 
(\ie those with lifetime $\tau >0.3 \times 10^{-10}$ s)
in the simulation. The selection of these objects is intended to match as closely as possible 
the detector-level requirements used to select \ttbar events. 
It is described together with the generator-level top quark kinematic reconstruction procedure 
in Refs.~\cite{Sirunyan:2018ucr,bib:CMS-NOTE-2017-004}, and is summarized below.

\begin{itemize}
\item All simulated electrons and muons, including those from \PGt lepton decays, but not originating 
from the decay of a hadron, are
corrected (``dressed'') for bremsstrahlung effects 
by adding the momentum of each photon to that of the
closest lepton if their
separation in $\eta$-$\phi$~space is $< 0.1$. Leptons are required to have $\pt > 20\GeV$ and 
$\abs{\eta} < 2.4$.

\item Only neutrinos originating from nonhadronic decays (\ie prompt neutrinos) are used.

\item Jets are clustered using the anti-\kt jet algorithm~\cite{Cacciari:2008gp, Cacciari:2011ma} with a
distance parameter of 0.4.
All stable particles, with the exception of the
dressed leptons and prompt neutrinos, are clustered.
Jets with $\pt > 30\GeV$ and $\abs{\eta} < 2.4$ are selected if there is no electron or muon, as defined
above, within a distance of 0.4 in $\eta$-$\phi$~space.

\item $\PQb$ jets are defined as those jets that contain a $\PQb$ hadron using the ghost-matching
technique~\cite{bib:PUSubtraction}: as a result of the short lifetime of $\PQb$ hadrons, only their decay
products are considered for the
jet clustering. However, to allow their association with a jet, the $\PQb$ hadrons are also included with
their momenta scaled down to a
negligible value. This preserves the information of their directions, but removes their impact on the jet
clustering.

\item The following additional event-level requirements are applied to define the fiducial phase space 
region in which the particle-level
cross sections are measured: we require that the \PW bosons produced from decays of the top quark and
antiquark in a \ttbar event themselves decay to an electron or a muon; 
we reject events where these \PW bosons decay to \PGt leptons; we require exactly two selected 
lepton candidates with opposite charges, $ \mll > 20\GeV$, and at 
least two \PQb jets.

\item The top quark reconstruction at the particle level proceeds as follows. Prompt neutrinos
 are combined
with the dressed leptons to form \PW boson candidates. We take the permutation of neutrinos and leptons that minimizes the sum of the absolute values of the differences between the mass of each 
neutrino-lepton pair
and the nominal \PW boson mass of 80.4\GeV. 
Subsequently, the \PW boson candidates are
combined with \PQb jets to form particle-level top quark candidates by minimizing the sum of the absolute
values of the differences between the mass of each pair and the nominal top quark mass of 172.5\GeV. 
\end{itemize}

Due to the finite detector resolution, events
that are outside the fiducial phase space region at the generator level
can be measured inside the accepted region at the detector level.
These events are subtracted, before the unfolding, by a fractional
correction of the observed number of events after subtracting all other backgrounds.
The correction, performed separately for each
detector-level bin, is defined as the number of events in the \ttbar
signal simulation that pass both the detector- and particle-level selection criteria,
divided by the number of all events fulfilling the detector-level requirements.

When measuring the differential cross sections as functions of \nj 
(see Section~\ref{sec:res_3}), the top quark and antiquark are measured
either at the parton level in the
full phase space or at the particle level in a fiducial phase space, as described above, while the 
additional jets are measured
at the particle level only. The definition of these extra jets differs from the one given above only in one aspect: they
have to be isolated from the
charged leptons (\Pe or \PGm) and \PQb quarks originating from the top quark decays, as represented
by the corresponding particle level leptons and \PQb jets, with a minimal
distance to these leptons
(\PQb quarks) of 0.4 (0.8) in $\eta$-$\phi$~space. The larger distance to the \PQb quarks is required
to avoid selecting jets coming from gluon radiation from the \PQb quarks as additional jets. In addition, 
the additional jets are required
to have $\pt > 40\GeV$ and $\abs{\eta} < 2.4$.
Specifically for the measurements of the top quark and antiquark at the parton level, two more 
differences in the definition of the extra jets are introduced~\cite{Sirunyan:2019zvx}. The neutrinos
from decays of hadrons are excluded in the clustering of these jets, and the charged leptons and \PQb quarks
used in the jet isolation
are taken directly after the \PW boson and top quark decays, respectively.

\section{Cross section measurement}
\label{sec:xsec}

For a given variable $X$, the absolute differential \ttbar cross section ${\rd\sigma_{i}}/{\rd X}$
is extracted via the relation
\begin{equation}
\label{eq:diffXsec}
\frac{\rd\sigma_{i}}{\rd X}=\frac{1}{\lumi} \frac{x_{i}}{\Delta^{X}_{i}},
\end{equation}
where \lumi is the integrated luminosity, $x_{i}$ is the number of unfolded signal events observed
in bin $i$,
and $\Delta_{i}^{X}$ is the bin width.
The numbers $x_{i}$ are calculated with respect to the \ttbar parton or particle generator levels defined in Section~\ref{sec:truthlevel}.
The normalized differential cross section is obtained by dividing the absolute differential cross section
by the measured total cross section $\sigma$ in the same phase space,
which is evaluated by summing the binned cross section measurements over all bins of the observable $X$.
For differential cross sections measured simultaneously as functions of two or three variables,
the following criteria are adopted for optimizing their display.
The measured cross sections are divided by the bin width of the variable that is chosen as the last one.
They present single-differential
cross sections as functions of the last variable in different ranges of the first, or
first and second variables, and are referred to as double- or triple-differential
cross sections, respectively.

The bin widths 
at generator level
are chosen based on the resolutions of the kinematic variables,
such that the purity and the stability of each bin are generally above 30\% for single-differential
cross sections and above 20\% for double- or triple-differential cross sections.
For assessing the purity and stability, the binning at detector level is adjusted to be equal to the 
binning at generator level.
For a given bin, the purity is defined as the fraction of events in the \ttbar signal simulation that are
generated and reconstructed in the same bin
with respect to the total number of events reconstructed in that bin.
To evaluate the stability, the number of events in the \ttbar signal simulation that are generated and
reconstructed in a given bin are divided
by the total number of reconstructed events generated in the bin.

\begin{figure*}[!phtb]
\centering
\includegraphics[width=0.49\textwidth]{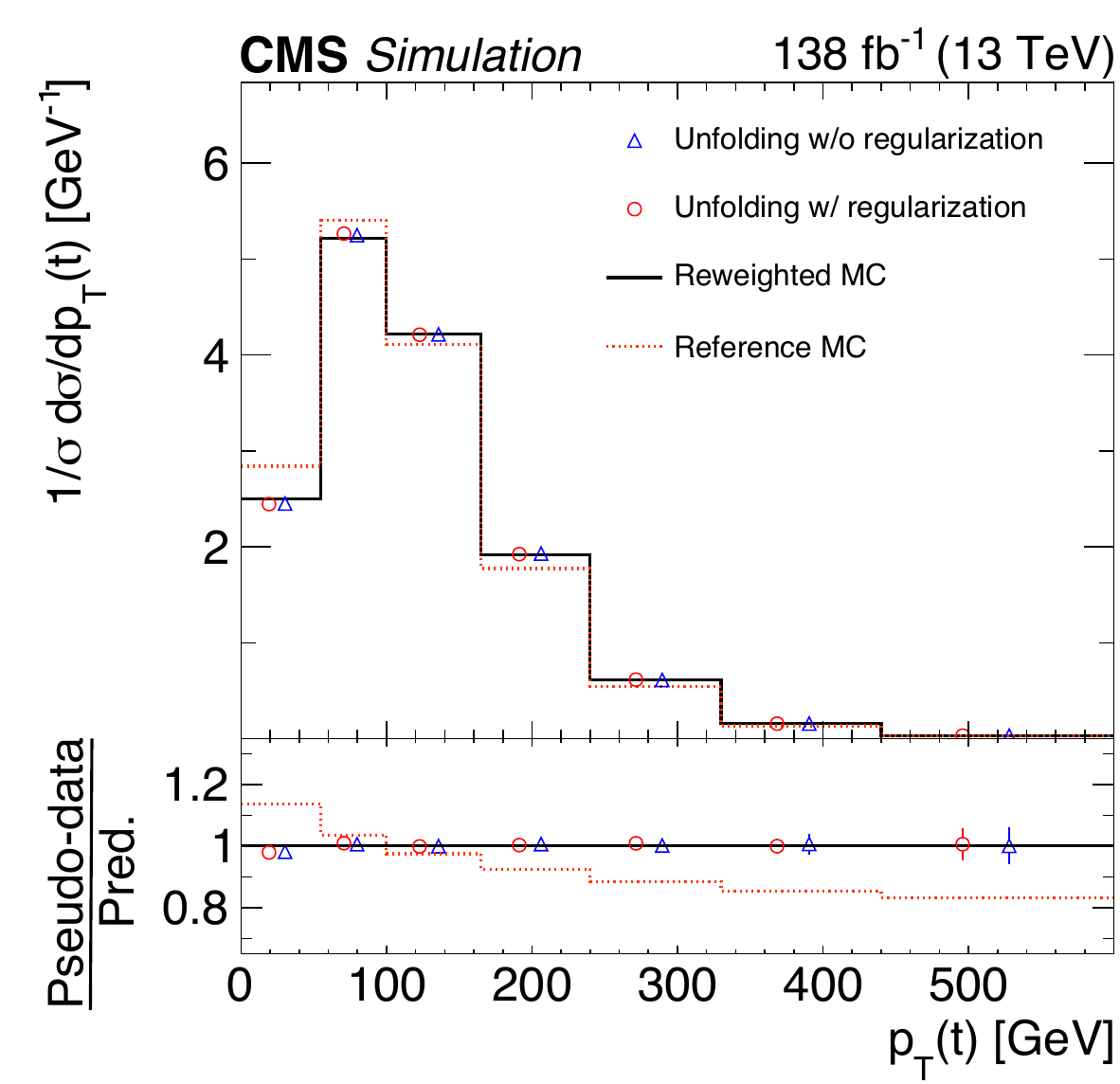}
\includegraphics[width=0.49\textwidth]{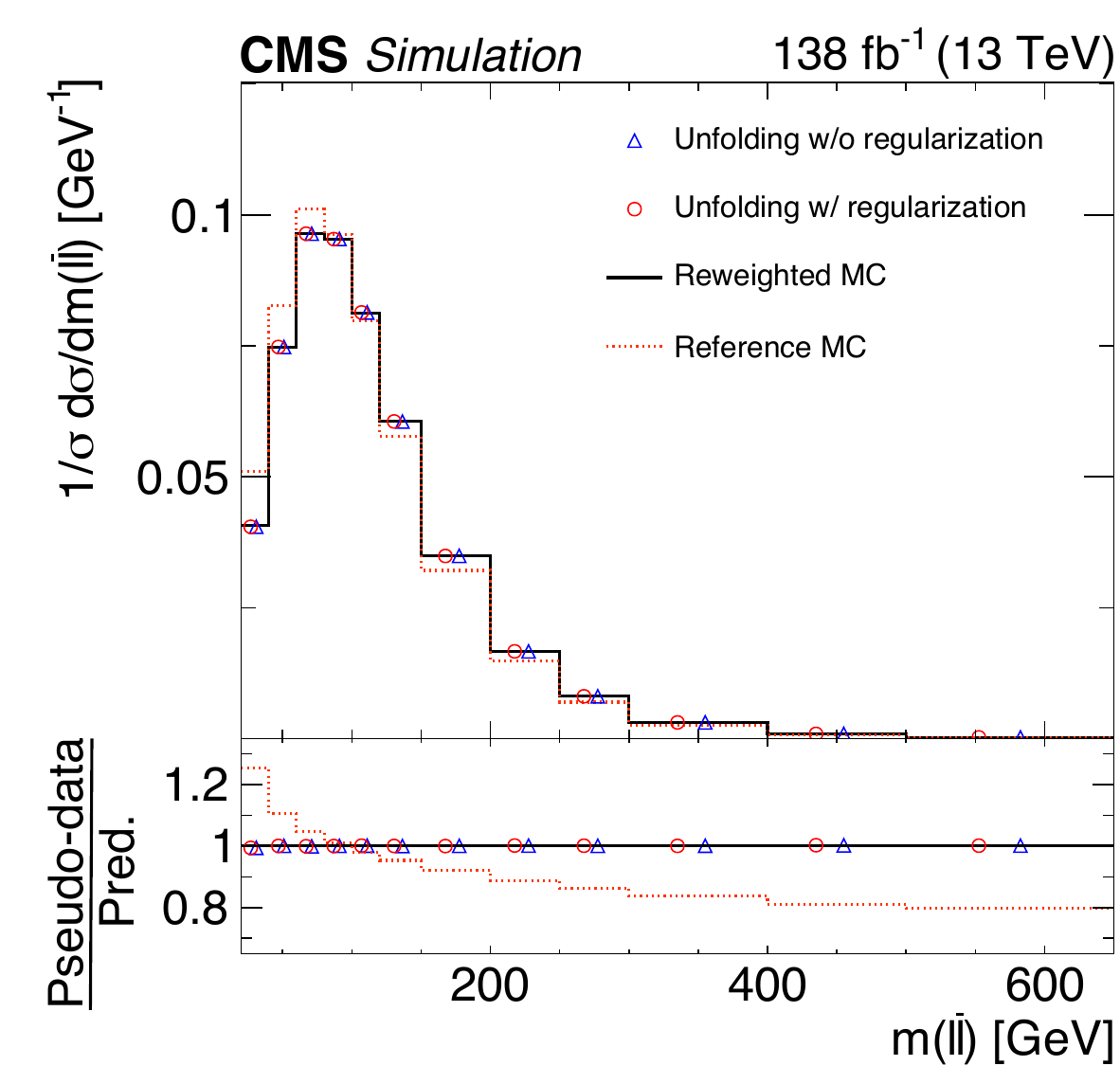}
\caption {Reweighting test for the extraction of the normalized differential cross sections
as functions of \ptt (left) and \mll (right).
The former cross section is  measured at the parton level in the full phase space
and the latter at the particle level in a fiducial phase space.
The nominal \ttbar signal MC spectra are shown as dotted red histograms and the assumed
true spectra, obtained from reweighting, as solid black histograms.
The unfolded spectra, using pseudo-data based on the true spectra but using the nominal spectra
for the detector corrections and bias vector in the regularization, are presented as open
red circles. The unfolded spectra with the regularization switched off are also shown (open blue triangles).
The statistical uncertainties in the unfolded cross sections are represented by a vertical bar
on the corresponding points.
The lower panel in each plot shows the ratios of the pseudo-data to the predicted spectra.}
\label{fig:xsec_closure}
\end{figure*}

The cross section measurement based on the signal extraction and unfolding procedure
described in Section~\ref{sec:unfold} is validated with closure tests.
Large numbers ($\sim$ 1000) of pseudo-data sets are generated 
from the \ttbar signal MC simulations and analyzed as if
they were real data.
The unfolded differential cross sections are found to be unbiased and to provide
proper 68\% confidence intervals within $\pm 1$ estimated standard deviation uncertainties.
A second test is performed by unfolding pseudo-data sets, generated
using reweighted \ttbar signal simulations, with the response matrix and bias vector taken from
the nominal simulation.
The reweighting is performed as a function of the differential cross section kinematic observables 
at the generator level
and is used to introduce controlled shape variations, \eg, making the \ptt spectrum harder or softer.
This test probes the robustness of the unfolding procedure with respect to the underlying physics model in
the simulation,
which impacts both the response matrix and the regularization bias vector.
Figure~\ref{fig:xsec_closure} shows two examples of these tests, performed
for cross sections as functions of \ptt and \mll.
The applied reweightings follow approximately parabolic functions and lead to shape
distortions of about $\pm 20\%$ at thresholds and end points of the kinematic spectra, comprising
the differences observed between data and the nominal simulation (as shown by figures in Section~\ref{sec:res}).
The unfolding is performed with the standard regularization
procedure and alternatively with switching it off; the obtained cross sections are found to deviate
at most by a few permille, showing that biases induced by the regularization are small.
The unfolded cross sections vary visibly from the true values only in the kinematic threshold regions, 
with maximum differences of the order of 1\%.
Effects of similar size are also seen for other single- and multi-differential cross sections
and demonstrate an overall good robustness of the unfolding procedure.
The simple reweighting approach discussed here is not suitable for quantifying 
the measurement uncertainty from the underlying physics model; this is done, 
instead, by the dedicated set of variations applied to the \ttbar signal 
simulation detailed in Section~\ref{sec:sys_theo}.

\section{Systematic uncertainties}
\label{sec:systematics}
The systematic uncertainties in the measured differential cross sections are grouped into three categories:
experimental uncertainties from the imperfect modeling of the detector response, theoretical uncertainties
from the modeling of the signal, and the uncertainties in the numbers of events from background processes.
The systematic uncertainty is assessed source by source 
largely following the prescriptions used in Refs.~\cite{Sirunyan:2018ucr,Sirunyan:2019zvx}.
For each change made, the cross section is recalculated,
which for most sources involves a repetition of the full analysis.
The difference with respect to the nominal result in each bin is taken as the systematic uncertainty.

Separate simulations are employed for the data taken in the three years 2016, 2017, and 2018,
in order to match the varying detector performance and data-taking conditions.
The correlations of systematic uncertainties among the measurements in the different periods 
must be
specified.
For theoretical uncertainties we assume a 100\% correlation, as
the same theoretical models and variations are used for all periods.
For experimental uncertainties we use either uncorrelated, partial, or full correlations.
For the case of partial correlations we make use of the
varied and nominal simulations for each year.
The correlated part of the uncertainty is assessed by using
the varied simulations for all three years, but rescaling
the resulting systematic uncertainty by a factor $\sqrt{\rho}$,
where $\rho$ specifies the level of correlation, \eg 50\%.
For the uncorrelated part, we separately vary
the simulation for each year, keeping the nominal simulation for
the other two years, and add the resulting uncertainties
in quadrature, after rescaling them by a factor $\sqrt{1-\rho}$.

\subsection{Experimental uncertainties}
Most experimental uncertainties are assessed by variations
that are simultaneously applied to the signal and background simulations.
The following sources are considered:
\begin{itemize}
\item
The uncertainties in the integrated luminosities of the 2016, 2017, and 2018 data samples
are 1.2, 2.3, and 2.5\%, respectively, and are 30\% correlated between the years~\cite{CMS:2021xjt,LUMI17,LUMI18}.
The resulting total normalization uncertainty on the absolute cross sections is 1.6\%.
\item
The uncertainty in the amount of pileup is assessed by varying the value of the total \pp inelastic cross section,
which is used to estimate the mean number of additional \pp interactions, by its measurement uncertainty of
$\pm4.6\%$~\cite{Aaboud:2016mmw}, leading to a 100\% correlated uncertainty among the three years.
\item
The efficiencies for the dilepton and single-lepton triggers are measured with independent
triggers based on a \ptmiss requirement.
Scale factors are calculated in bins of lepton \pt, independently for the years 2016, 2017, and 2018.
They agree with unity typically within 1\%.
The scale factors are varied within their uncertainties.
The uncertainties are assumed to be uncorrelated among the years.
\item
Lepton identification and isolation efficiencies are determined using the ``tag-and-probe'' method
with \Zjets event samples~\cite{Chatrchyan:2011cm,bib:TOP-15-003_paper}.
The efficiencies are assessed in two-dimensional bins of lepton $\eta$ and \pt.
The corresponding scale factors typically agree with unity within 10\% for electrons and 3\% for muons.
The scale factors are varied within their calibration uncertainties and
a 100\% correlation among the years is assumed.
An implicit assumption made in the analysis is that the scale factors derived from the \Zjets sample
are applicable to the \ttbar samples, where the efficiency for lepton isolation is reduced
due to the typically larger number of jets present in the events.
This assumption is validated through studies of \ttbar-enriched samples using an event selection similar to the current analysis~\cite{CMS:2020mpn}.
In these studies, the lepton isolation criteria are relaxed for one lepton, and the efficiency of passing
the criteria is measured in both data and simulation.
The observed efficiencies for electrons (muons) differ by at most 1\% (0.5\%) 
from the nominal values. These maximum variations
are taken as additional uncertainties.
\item
  The uncertainty arising from the jet energy scale (JES)
  is assessed by varying the energy scale correction for all
  jets in the signal and background simulations by one
  standard deviation, in bins of \pt
  and $\eta$. This uncertainty is divided into seven independent
  sources, including those from extrapolating between
  samples with different jet-flavor compositions
  and the impact of pileup collisions on the corrections derivation~\cite{Khachatryan:2016kdb}.  
  Most sources
  are treated as uncorrelated across
  data-taking periods. However, sources related to theoretical predictions
  in the MC simulation, such as those used to extrapolate between different
  jet-flavor compositions, are treated as correlated
  across data-taking periods.
  The JES variations are also propagated to the uncertainties in \ptvecmiss.
\item
The uncertainty from the jet energy resolution (JER) is evaluated by the variation
of the simulated JER by $\pm 1$ standard deviation in different $\eta$ regions~\cite{Khachatryan:2016kdb}.
An additional uncertainty in the calculation of \ptvecmiss is estimated by varying the energies of reconstructed particles not clustered into jets.
Since both sources of uncertainty are primarily affected by
varying detector conditions that are not time-correlated,
they are treated as uncorrelated across the different years.
\item
During the 2016--2017 data-taking periods, a gradual shift in the timing of the inputs of
the ECAL L1 trigger in the forward endcap region ($\abs{\eta} > 2.4$) led to a particular 
trigger inefficiency.
A correction for this effect was determined using an unbiased data sample and is found to be
relevant in events containing high-\pt jets 
pointing to the most forward ECAL region ($2.4 < \abs{\eta} < 3.0$).
While no reconstructed objects at
this pseudorapidity enter the measurements, the systematic variation of 20\% in this correction
for affected events nevertheless leads to a small measurement uncertainty.
\item
Scale factors for the \PQb tagging efficiency of individual \PQb jets and the mistagging efficiencies
of \PQc quark, and light-flavor and gluon jets are measured using
dedicated calibration samples~\cite{Sirunyan:2017ezt}.
The factors are parameterized as functions of jet \pt and $\eta$.
For the systematic uncertainty evaluation the scale factors are varied
within their estimated uncertainties~\cite{Sirunyan:2017ezt}.
The variations for \PQb and \PQc jets are treated as fully correlated,
while independent variations are applied to the light jets.
The statistical uncertainties of the scale factors, arising from
the limited size of the calibration samples, are
uncorrelated among the years. These uncertainties play a significant
role for the light jets, and hence, the light jet
uncertainties are treated as uncorrelated across the years.
All other \PQb tagging uncertainties are treated as fully correlated
across the years.
\end{itemize}

\subsection{Theoretical uncertainties}
\label{sec:sys_theo}
The uncertainties in the modeling of the \ttbar events, comprising signal and other final states,
are assessed with appropriate variations of the nominal
simulation based on \PowPyt and the CP5 tune, introduced in Section~\ref{sec:simulation}:
\begin{itemize}
\item
The uncertainty arising from missing higher-order terms in the simulation of the signal process at the ME level
is estimated
by varying the renormalization and factorization scales (denoted as \mur and \muf,
respectively)
in the \POWHEG simulation up and down by factors of two with respect to the nominal values.
The nominal scales are defined in the \POWHEG sample as
${\mur = \muf =} \sqrt{\smash[b]{{(\mtmc)}^2 + p^2_{\mathrm{T,\PQt}}}}$.
Here, $p_{\mathrm{T,\PQt}}$ denotes the \pt of the top quark in the \ttbar rest frame.
In total, six variations are applied: two with \mur fixed, two with \muf fixed,
and two with both scales varied in the same direction (up or down).
For each measurement bin, the envelope of the resulting measurement variations is taken as the final uncertainty.
\item
For the parton shower simulation, uncertainties are separately assessed for initial- and final-state radiation,
by varying the respective shower
scales up and down by factors of two.
\item
The uncertainty in the damping of real emissions
in the NLO calcululation when matching to the parton shower 
is derived by varying the \hdamp parameter in \POWHEG from its nominal value of 1.379\,\mtmc
to 0.8738\,\mtmc and 2.305\,\mtmc, according to the tuning results of Ref.~\cite{Sirunyan:2019dfx}.
\item
The uncertainty from the choice of PDFs is assessed by reweighting the \ttbar signal simulation
according to the $\pm 1$ standard deviation along the directions of the 100 eigenvectors of the NNPDF3.1 error
PDF sets~\cite{Ball:2017nwa}
and adding the resulting measurement variations in quadrature.
In addition, the value of the strong coupling \alpS is independently varied within its uncertainty in the PDF set.
\item
The dependence of the measurement on the assumed top quark mass is estimated by varying \mtmc
in the \ttbar simulation by $\pm 1\GeV$ around the nominal value of 172.5\GeV.
\item
The uncertainty related to modeling of the UE is estimated by varying the CP5 tune parameters
within their uncertainties determined in the tuning process~\cite{Sirunyan:2019dfx}.
\item
The nominal \PYTHIAviii setup includes a model of color reconnection (CR) based on MPIs with early
resonance decays switched off.
The analysis is repeated with three other CR models within \PYTHIAviii: the MPI-based scheme
with early resonance decays
switched on, a gluon-move scheme~\cite{Argyropoulos:2014zoa}, and a QCD-inspired scheme~\cite{Christiansen:2015yqa}.
The total CR related uncertainty is taken to be the maximum deviation from the nominal result.
\item
The \PQb jet energy response is different for semileptonic decays of {\PQb} hadrons  
and thus their branching fractions are varied within world average uncertainties~\cite{ParticleDataGroup:2022pth}.
The parton level cross sections are corrected for the branching fractions for $\PW\to \Pell\PGn$ 
which are taken from Ref.~\cite{ParticleDataGroup:2022pth} and varied according to their uncertainty of 1.5\%.
\end{itemize}
The uncertainties associated with the values of \hdamp, \mtmc, and the CR treatment
are evaluated using separate \ttbar simulations incorporating the varied values,
while all other theoretical uncertainties are assessed by applying appropriate event weights in the nominal simulations.

\subsection{Background uncertainties}
The contributions from non-\ttbar background processes are overall at the level of a few percent,
and the uncertainties are treated as global normalization uncertainties in
the MC simulated processes.
The uncertainty in the \Zjets background normalization is assessed
by repeating the template fits to \mll distributions, described in Section~\ref{sec:unfold},
with varied event selections.
The nominal fits are performed with a selection dropping all requirements on \ptmiss,
the number of \PQb-tagged jets, and the \ttbar kinematic reconstruction.
In the first varied scenario, the $\ptmiss > 40\GeV$ requirement is switched on again for the \ee and \mumu channels,
in the next one it is additionally required to have at least one \PQb-tagged jet in the event,
and in the last one the criterion of finding a solution to the full kinematic reconstruction is imposed.
The variations are studied separately for each channel (\ee and \mumu) and each year.
The maximum scale factor variation observed among all channels, years, and event selections is used
to derive a $\pm 20\%$ systematic uncertainty in the \Zjets normalization.

For the single top quark, \Wjets, and diboson backgrounds,
a normalization uncertainty of $\pm 30\%$ is taken, following the prescription from our previous
analyses~\cite{Sirunyan:2018ucr,Sirunyan:2019zvx}. This value is confirmed for the \tW background (the dominant contribution among these processes) by analyzing the ratio of the numbers of events with one and two \PQb-tagged jets in the \emu sample. This ratio is higher for the \tW process than for \ttbar events. The observed ratio in data agrees well with the nominal simulations but deteriorates significantly when the assumed \tW cross section is varied by more than 30\%.

\subsection{Summary of uncertainties}
The total systematic uncertainty in each measurement bin is assessed by adding all contributions 
described
above in quadrature, separately for positive and negative cross section variations.
If a systematic uncertainty leads to two cross section variations of the same sign, the largest one is 
taken
and the opposite variation is set to zero.
These procedures are applied only to the cross section plots while
the full information on all measurement variations is
preserved in HEPdata~\cite{hepdata} and also contributes
to all theory-to-data \chisq comparisons presented
in this paper, following the \chisq definition in
Appendix~\ref{sec:app}.
The total uncertainties for the measured cross sections range 2--20\%, depending on the observable and the bin.
They are dominated by the systematic uncertainties.

\begin{figure}[htp]
    \centering
    \includegraphics[width=0.6\textwidth]{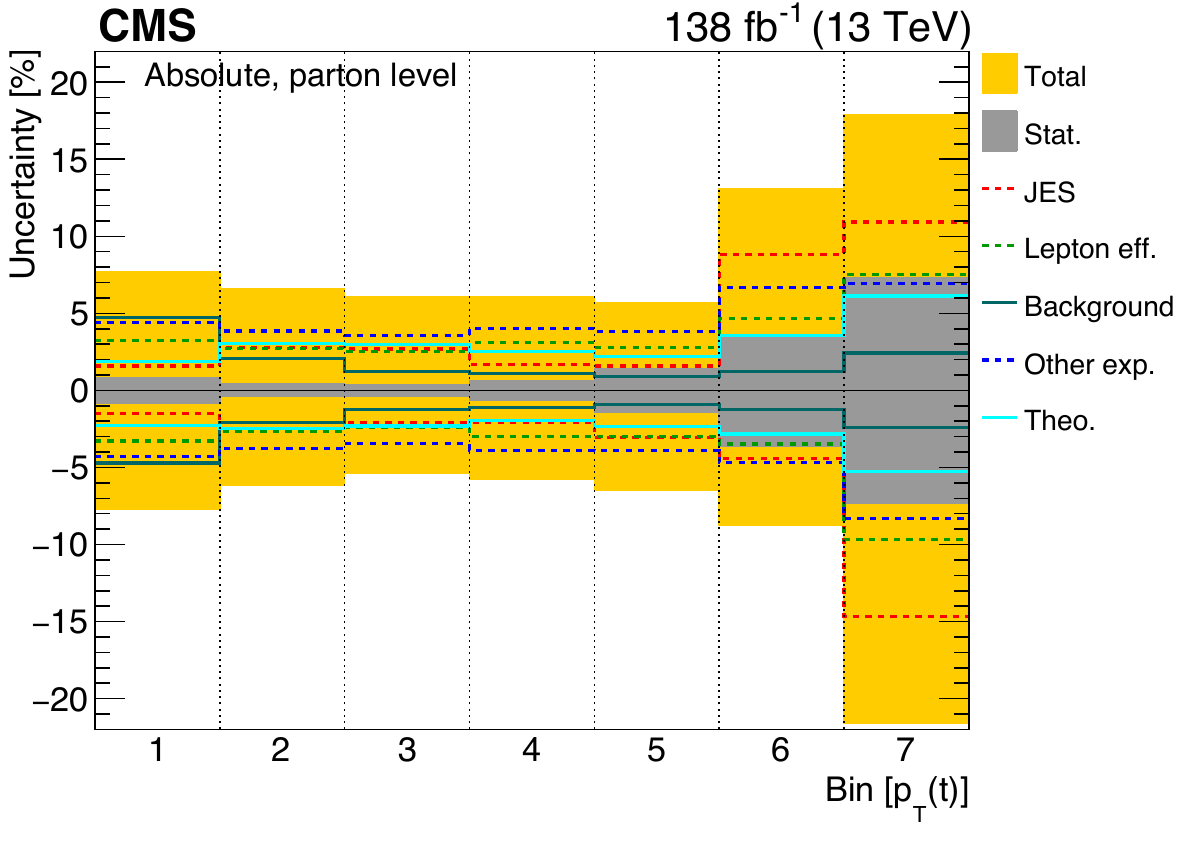}
    \includegraphics[width=0.6\textwidth]{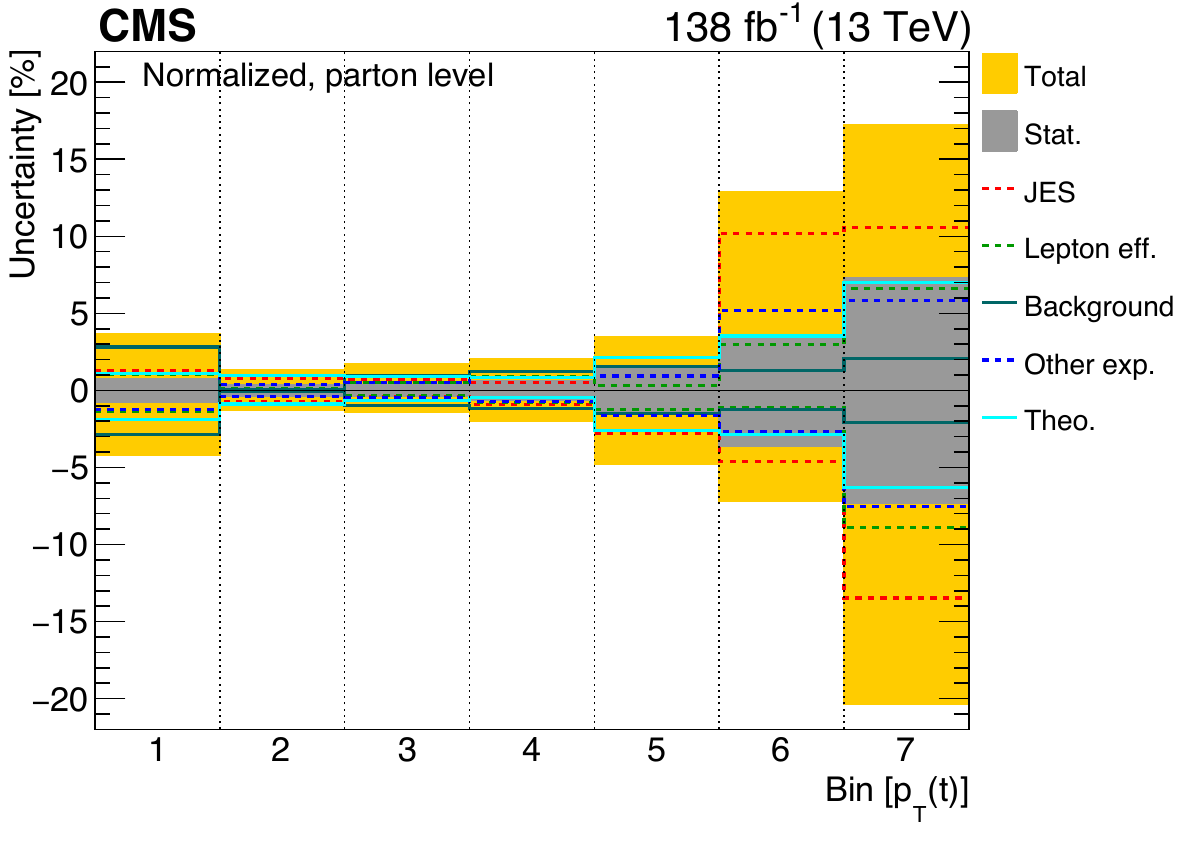}
    \includegraphics[width=0.6\textwidth]{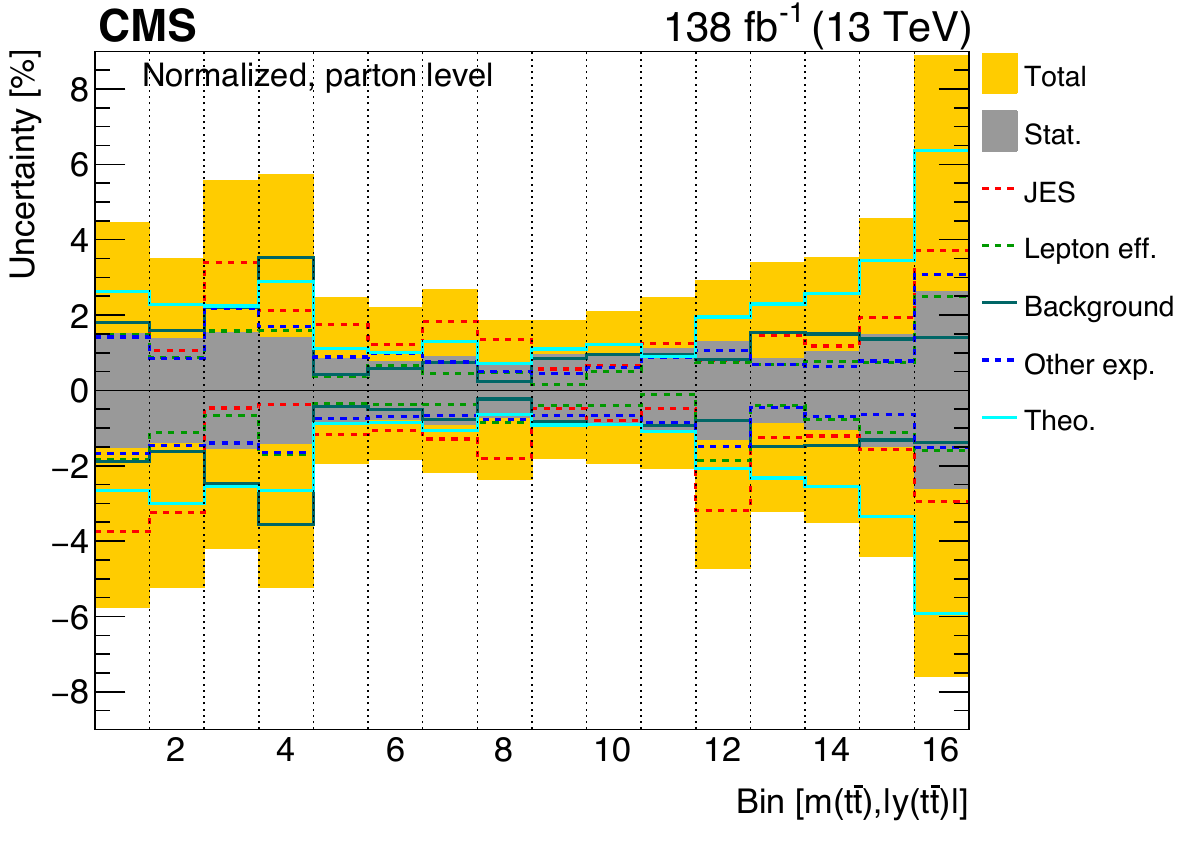}
    \caption{The various sources of systematic uncertainty and their relative contributions to the 
overall uncertainty are shown for several parton-level 
measurements: absolute \ptt (upper), normalized \ptt (middle), and normalized \mttytt (lower).  
The statistical uncertainties and the total uncertainties (statistical and systematic 
uncertainties added in quadrature) are shown as grey and yellow bands, respectively.
The ranges of the observables for a given bin number 
can be read off from the corresponding cross section distributions in 
Figs.~\ref{fig:res_ptt} and~\ref{fig:res_mttytt}.
}
    \label{fig:unc-comp1}
\end{figure}

\begin{figure}[htp]
    \centering
    \includegraphics[width=0.6\textwidth]{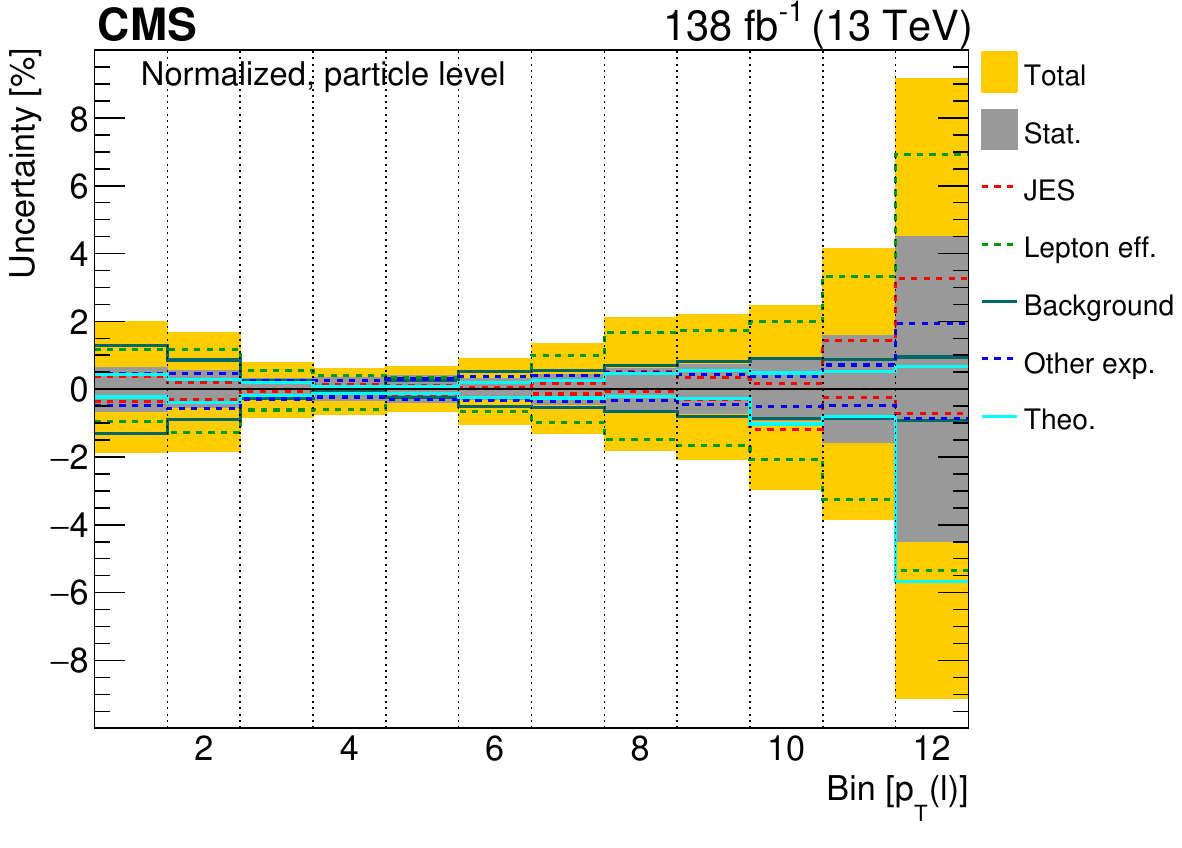}
    \includegraphics[width=0.6\textwidth]{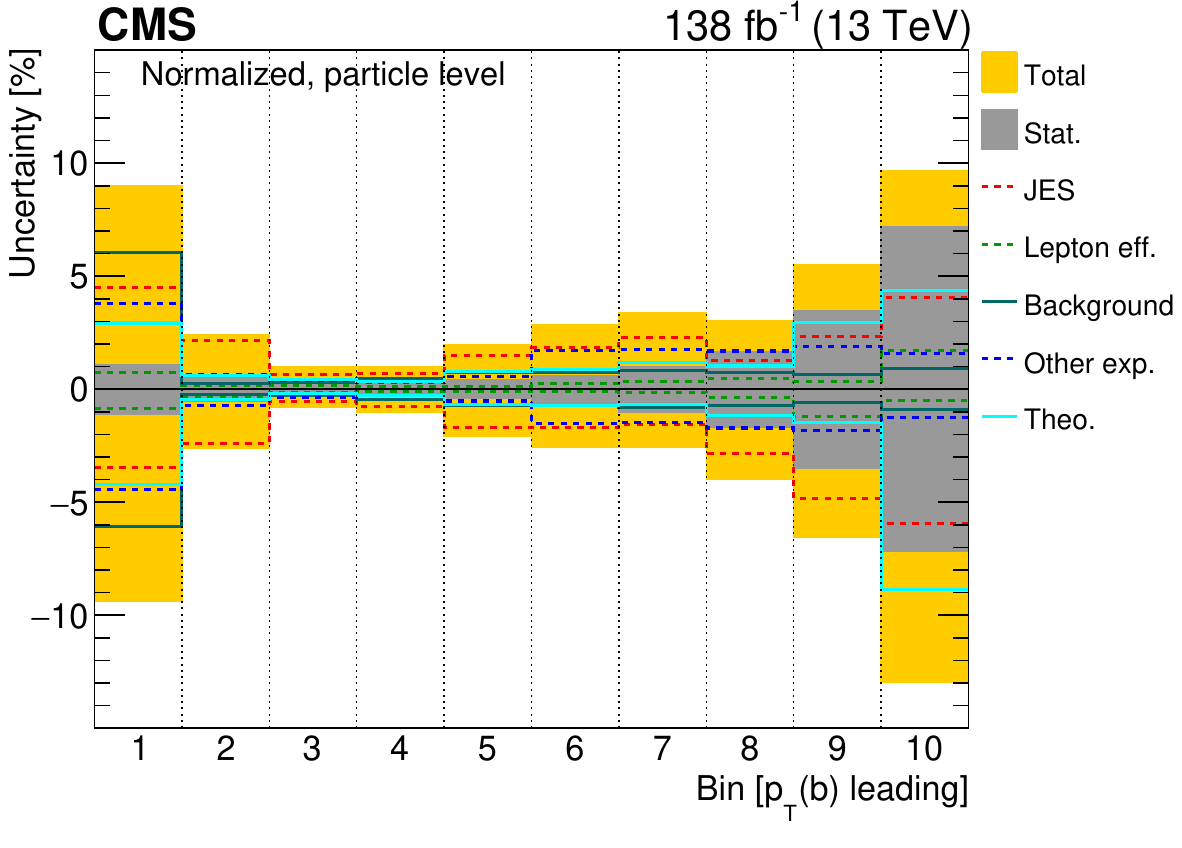}
    \includegraphics[width=0.6\textwidth]{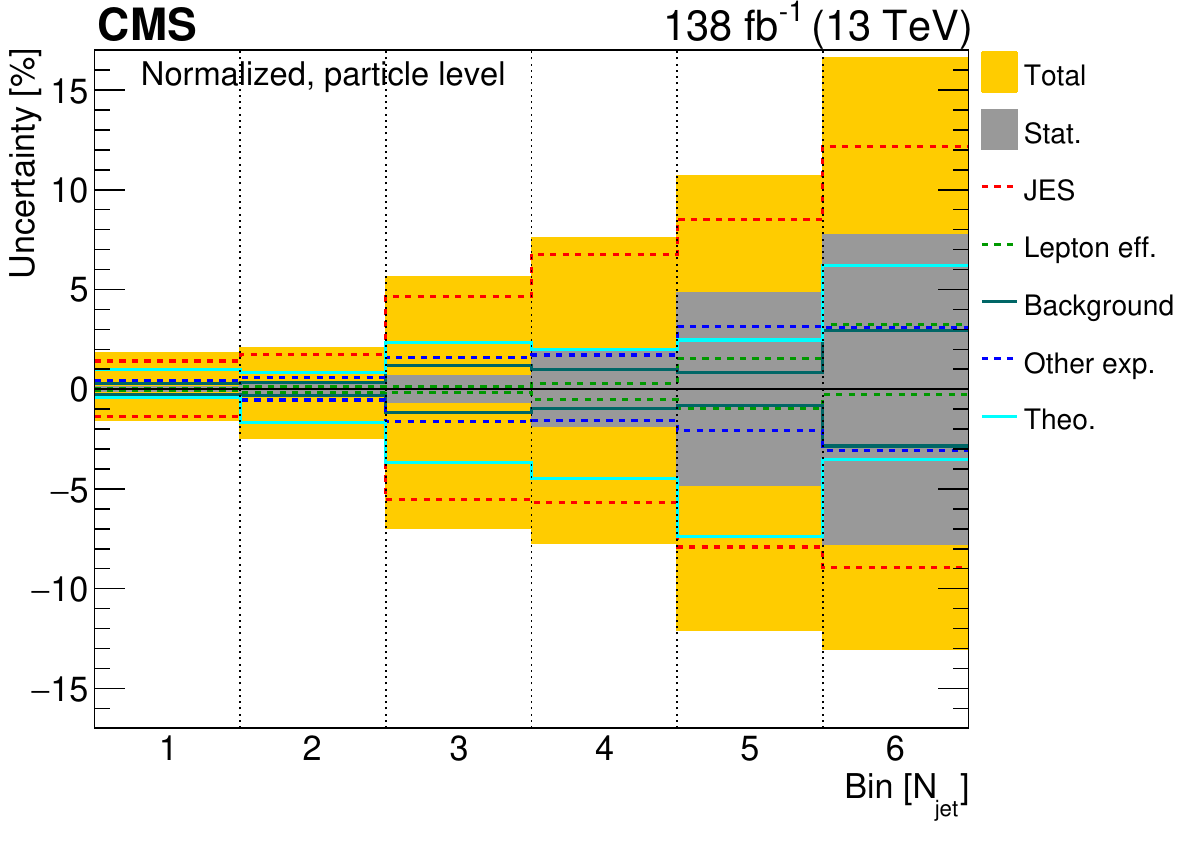}
    \caption{The various sources of systematic uncertainty and their relative contributions to the
overall uncertainty are shown for several normalized 
particle-level measurements: \pt of the lepton (upper), 
\pt of the leading \PQb jet (middle), and \nj (lower).
The statistical uncertainties and the total uncertainties (statistical and systematic
uncertainties added in quadrature) are shown as grey and yellow bands, respectively.
The ranges of the observables for a given bin number 
can be read off from the corresponding cross section distributions in  
Figs.~\ref{fig:res_ptlep},~\ref{fig:res_ptb}, and~\ref{fig:res_nj40}.
}
    \label{fig:unc-comp2}
\end{figure}

The uncertainties are illustrated in Figs.~\ref{fig:unc-comp1}--\ref{fig:unc-comp2},
showing the relative contributions from the various sources for selected differential cross sections.
Individual sources affecting a particular uncertainty (\eg JES) are added in quadrature and shown as a
 single
component. Additional experimental systematic uncertainties and all contributions from theoretical
uncertainties are also added in quadrature, respectively, and shown as single components.
For most bins in a majority of the distributions, the JES is the dominant systematic uncertainty.
The evaluation of this uncertainty is also affected by the limited number of simulated \ttbar signal events, particularly
at high transverse momenta or invariant masses.
Important contributions arise from other experimental sources, as well as from
theoretical and background uncertainties. Among the significant experimental sources of uncertainties
are the lepton and \PQb tagging efficiencies, and for measurements of absolute cross sections, the 
integrated luminosity.
For the theoretical uncertainties, the following sources contribute significantly, with
relative magnitudes depending on the observables and phase space region: ME level and final-state 
radiation scales, \hdamp parameter, top quark mass, underlying event, and CR.

\clearpage

\section{Results}
\label{sec:res}
Cross sections for \ttbar and top quark kinematic variables
are discussed in Section~\ref{sec:res_1}, for lepton and \PQb jet
variables in Section~\ref{sec:res_2}, and for events with additional jets in
Section~\ref{sec:res_3}.
The three predictions based on MC simulations introduced in Section~\ref{sec:simulation}: 
\PowPyt (`POW+PYT'), \PowHer (`POW+HER'), and \aMCPyt (`FxFx+PYT'), are used for comparisons 
to data.
The \PowPytSh and \PowHerSh theoretical predictions differ by the parton-shower method, 
hadronization and event tune (\pt-ordered parton showering, string hadronization model, and CP5 tune in \PowPytSh, 
or angular ordered parton showering, cluster hadronization model, and CH3 tune in \PowHerSh),
 while the \PowPytSh and \aMCPytSh predictions adopt different matrix elements 
(inclusive \ttbar production at NLO in \PowPytSh, or \ttbar 
with up to two extra partons at NLO in \aMCPytSh) and different methods for matching with parton shower 
(correcting the first parton shower emission to the NLO result in \PowPytSh, 
or subtracting from the exact NLO result its parton shower approximation in \aMCPytSh).
In Section~\ref{sec:res_theory_comp}, several theoretical calculations with beyond-NLO precision are introduced
and their predictions compared to the data for a subset of the measured cross sections.
Finally, a study of the sensitivity of the normalized cross sections to the PDFs is presented
in Section~\ref{sec:res_pdf_comp}.
For each data-to-theory comparison, a \chisq statistic
and the number of degrees of freedom (\ndf) are reported.
One statistic, denoted in the following as ``standard \chisq'', probes directly the quality of the 
nominal predictions. It takes all measurement uncertainties into account,
including bin-to-bin correlations, but neglects the uncertainties in the theoretical predictions. 
For \PowPytSh, additional \chisq values that include
prediction uncertainties are also provided,
applying the full set of 
uncertainties discussed in Section~\ref{sec:sys_theo}
to the generator-level predictions.
The exact definition of the \chisq values is given in Appendix~\ref{sec:app}.
Normalized cross sections and tables providing the \chisq values are presented in this section, while absolute cross sections and corresponding \chisq tables are summarized in Appendix~\ref{sec:res_abs}. The $p$-values denoting the probability for finding a \chisq of equal or larger size than observed are tabulated, for all results, in Appendix~\ref{sec:res_pval}.

\subsection{Results for top quark and \texorpdfstring{\ttbar kinematic observables at the parton and particle levels}{}}
\label{sec:res_1}
The studies presented in this subsection aim to provide
a comprehensive survey of the kinematic spectra of the top quark and antiquark,
the \ttbar system, and their correlations.
\subsubsection{Single-differential cross sections}
\begin{figure*}[!phtb]
\centering
\includegraphics[width=0.49\textwidth]{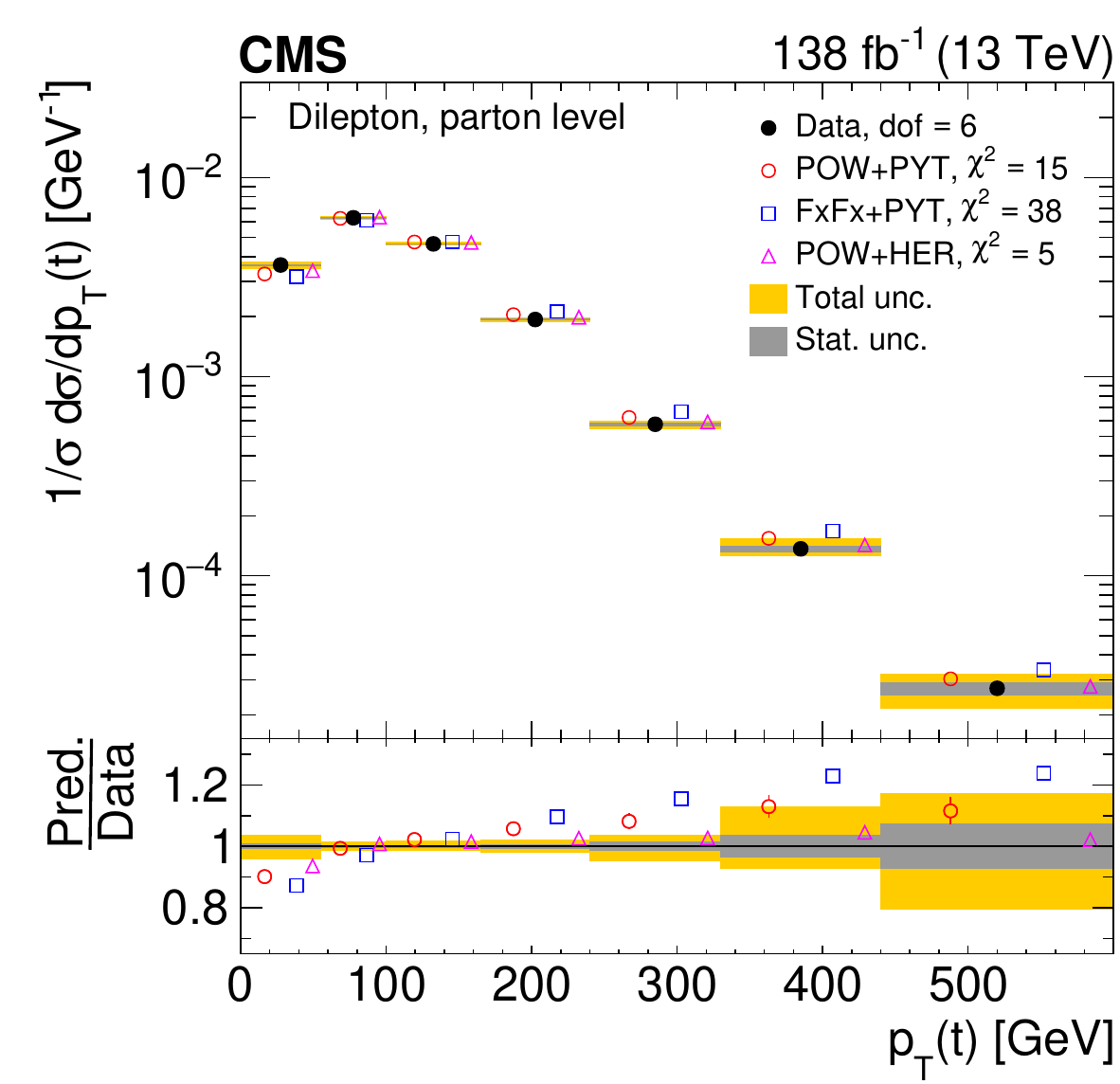}
\includegraphics[width=0.49\textwidth]{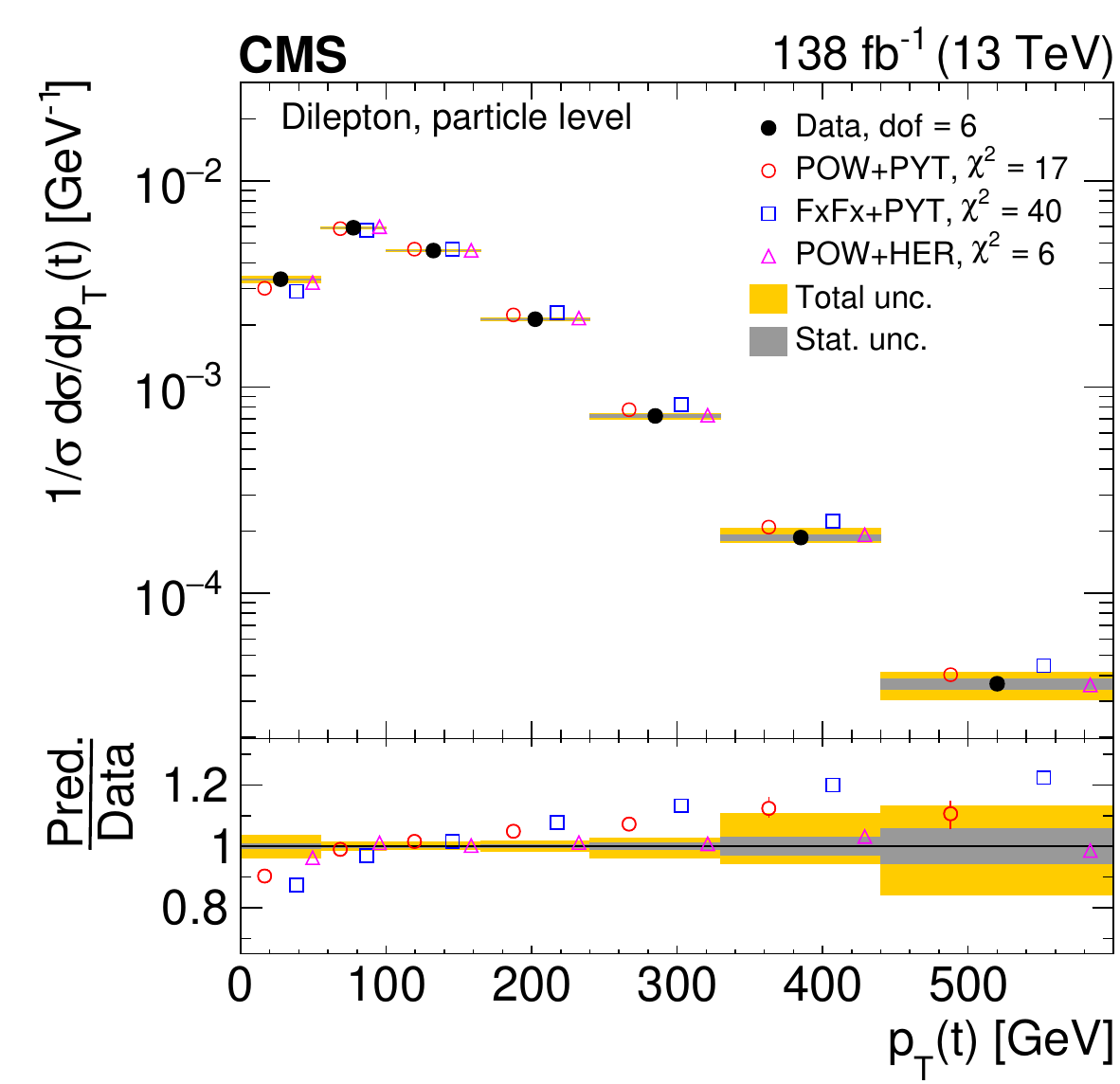}
\includegraphics[width=0.49\textwidth]{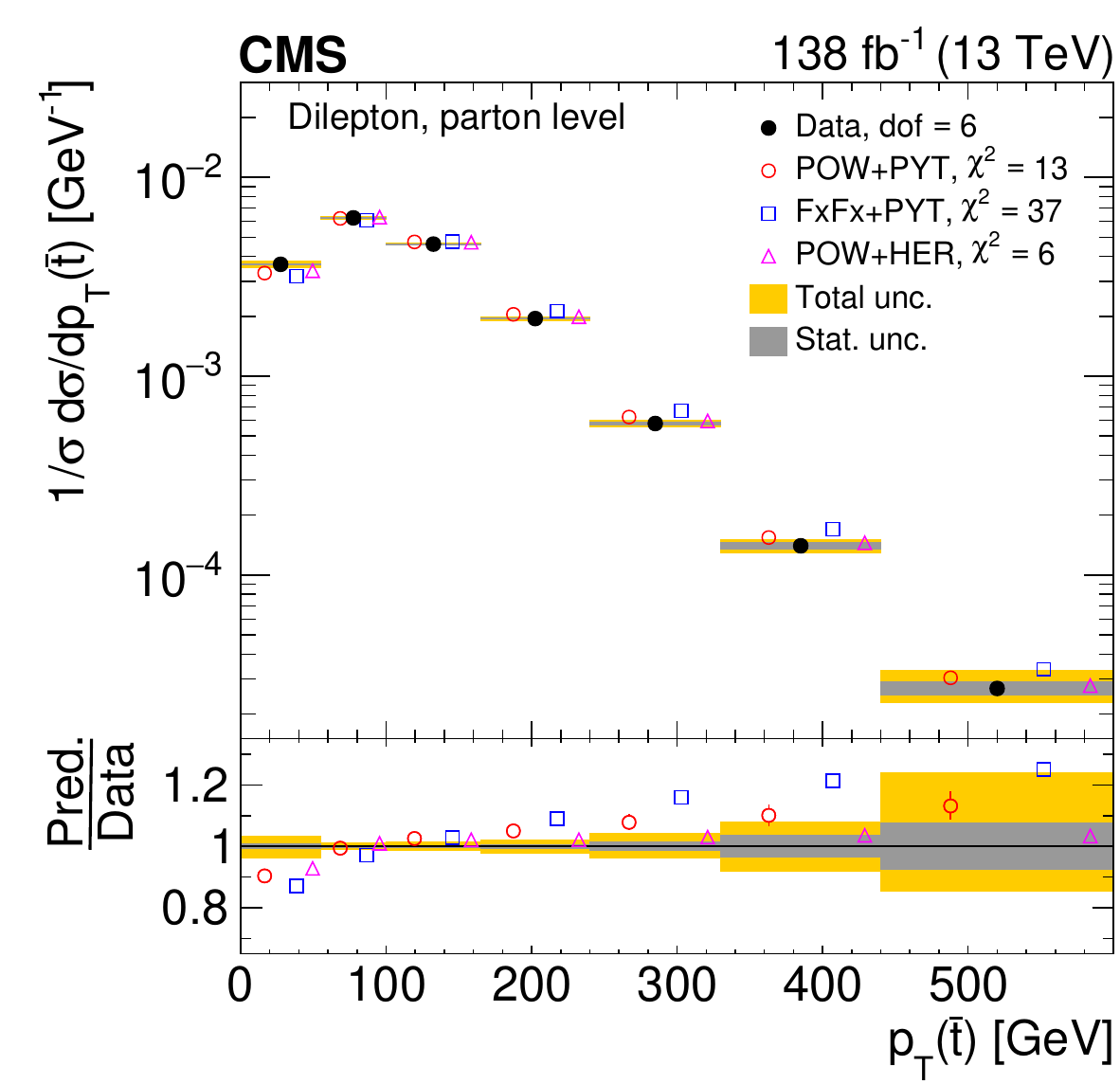}
\includegraphics[width=0.49\textwidth]{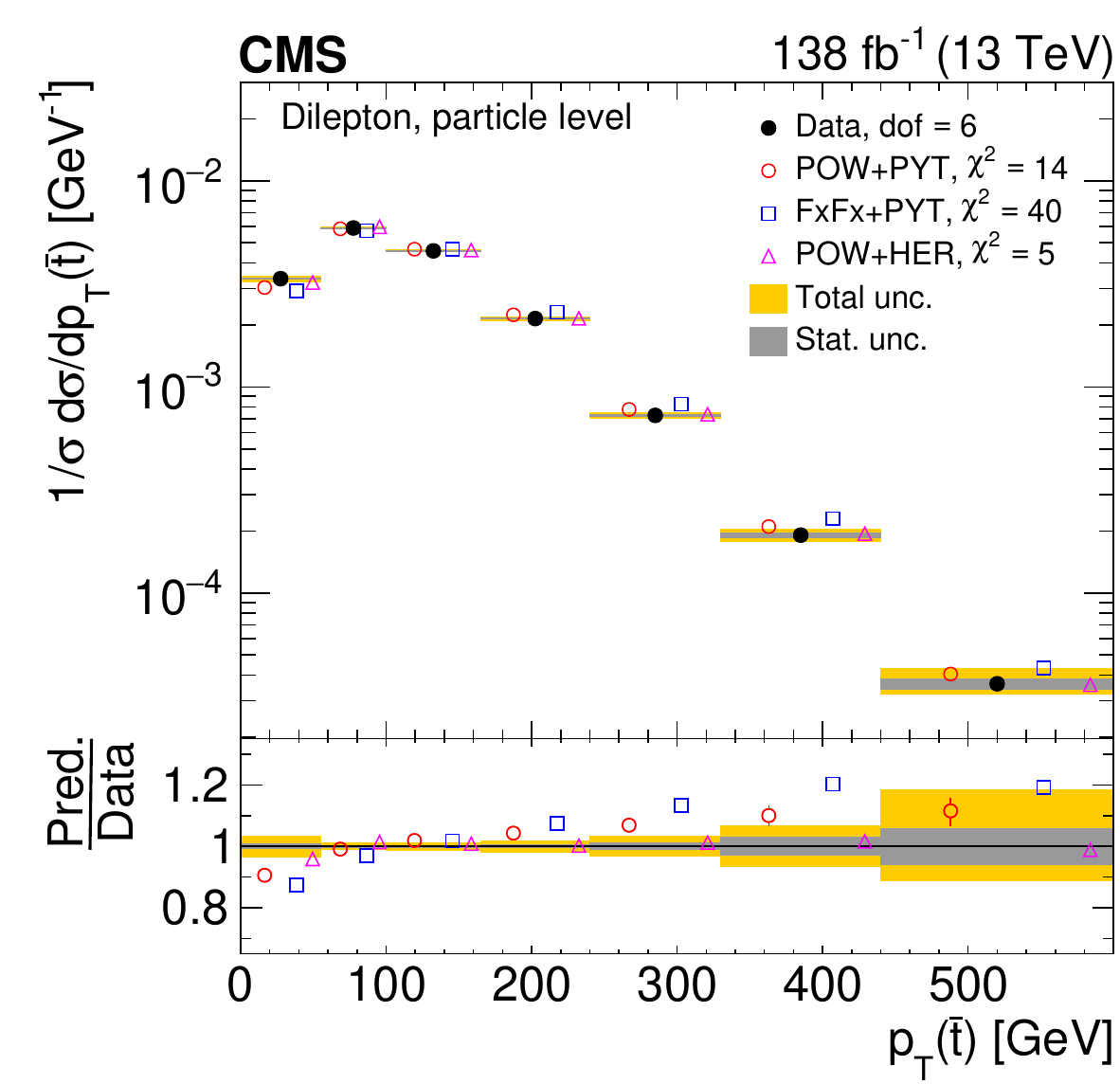}
\caption {Normalized differential \ttbar production cross sections as functions of \ptt (upper) and \ptat (lower),
measured
at the parton level in the full phase space (left) and at the particle level in a fiducial phase space (right).
The data are shown as filled circles with grey and yellow bands indicating the statistical and total uncertainties
(statistical and systematic uncertainties added in quadrature), respectively.
For each distribution, the number of degrees of freedom (dof) is also provided.
The cross sections are compared to various MC predictions (other points).
The estimated uncertainties in the \PowPyt (`POW-PYT') simulation are represented by vertical bars on the
corresponding points.
For each MC model, a value of \chisq is reported that takes into account the measurement uncertainties.
The lower panel in each plot shows the ratios of the predictions to the data.}
\label{fig:res_ptt}
\end{figure*}

\begin{figure*}[!phtb]
\centering
\includegraphics[width=0.49\textwidth]{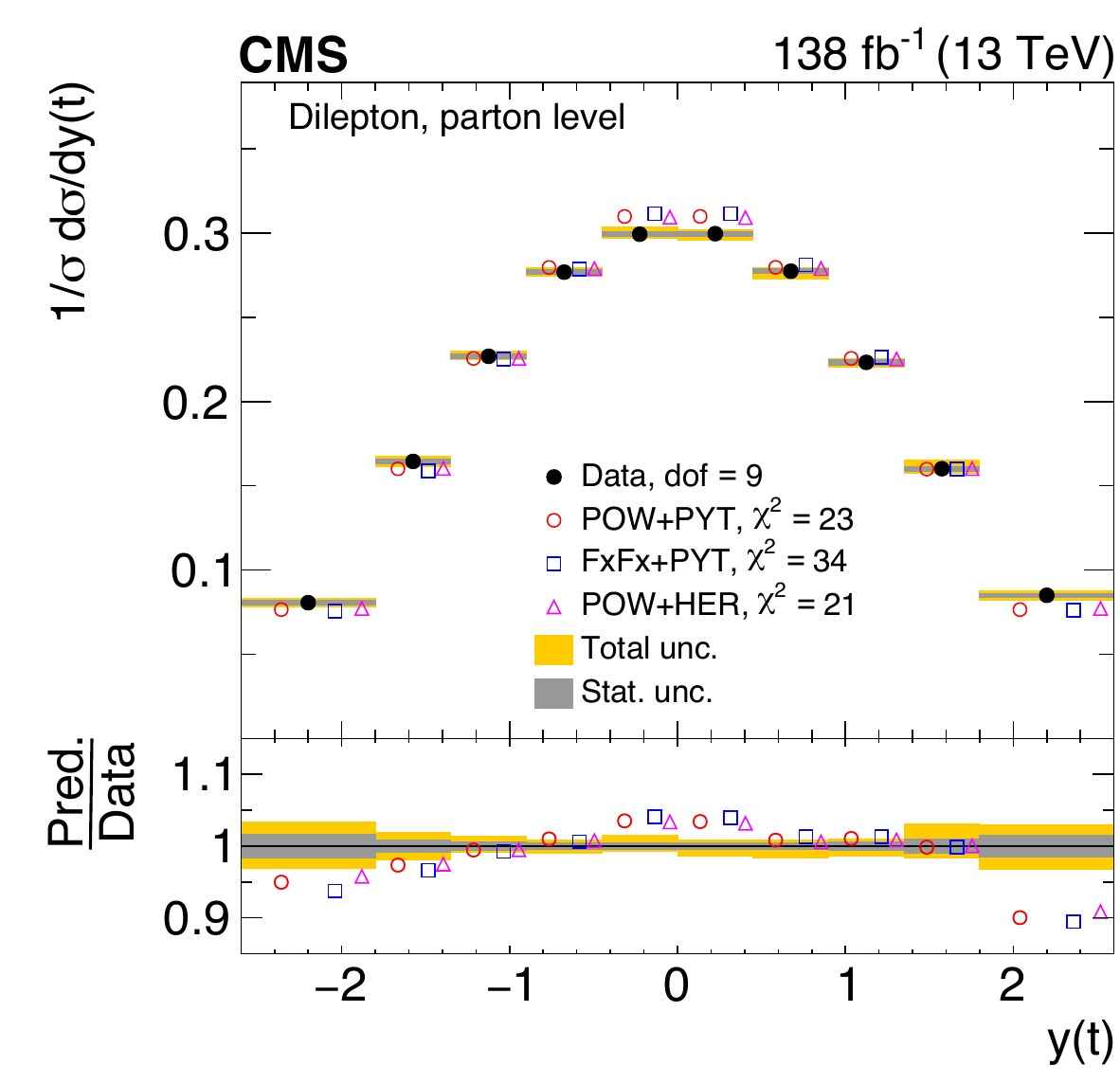}
\includegraphics[width=0.49\textwidth]{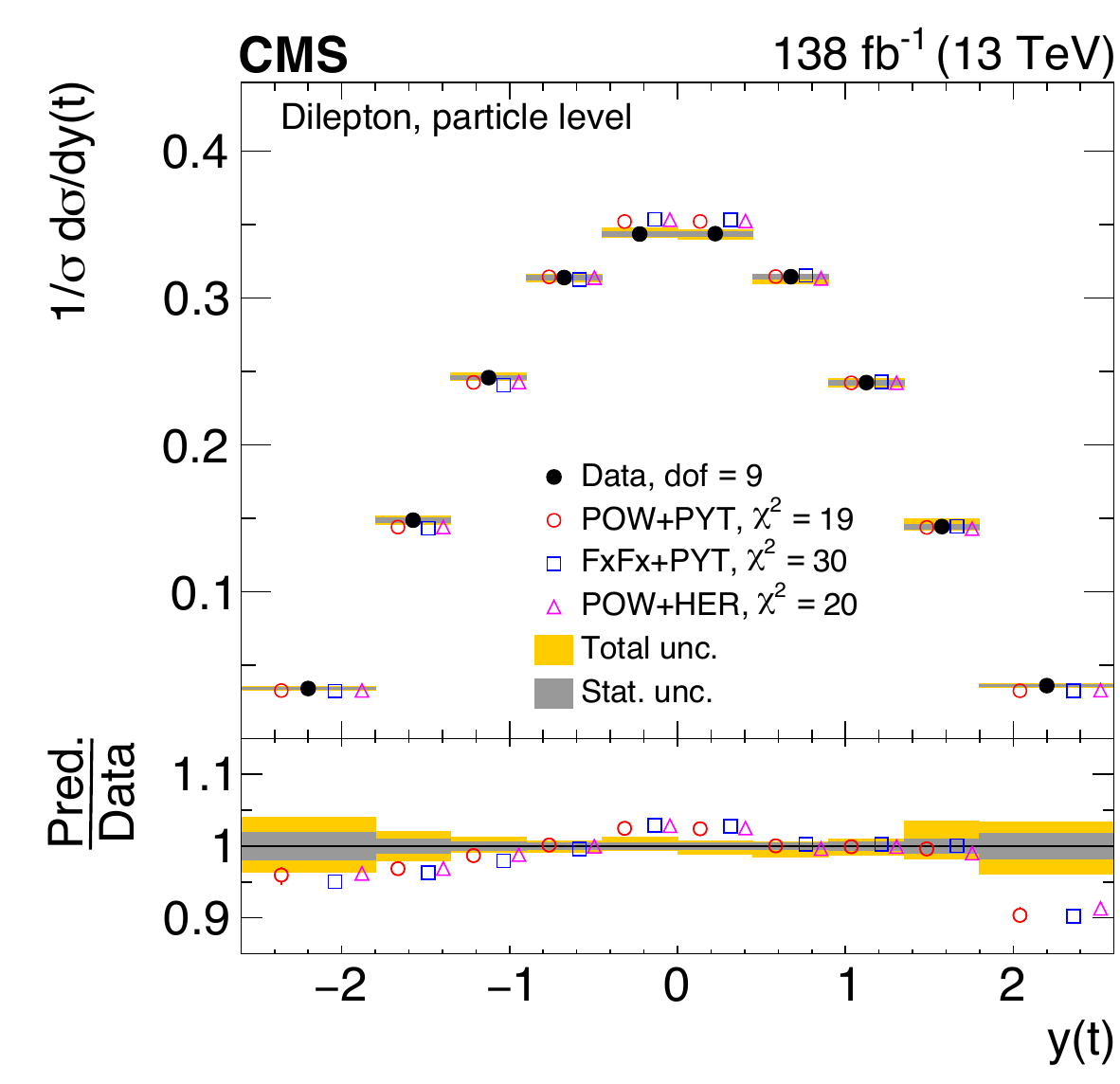}
\includegraphics[width=0.49\textwidth]{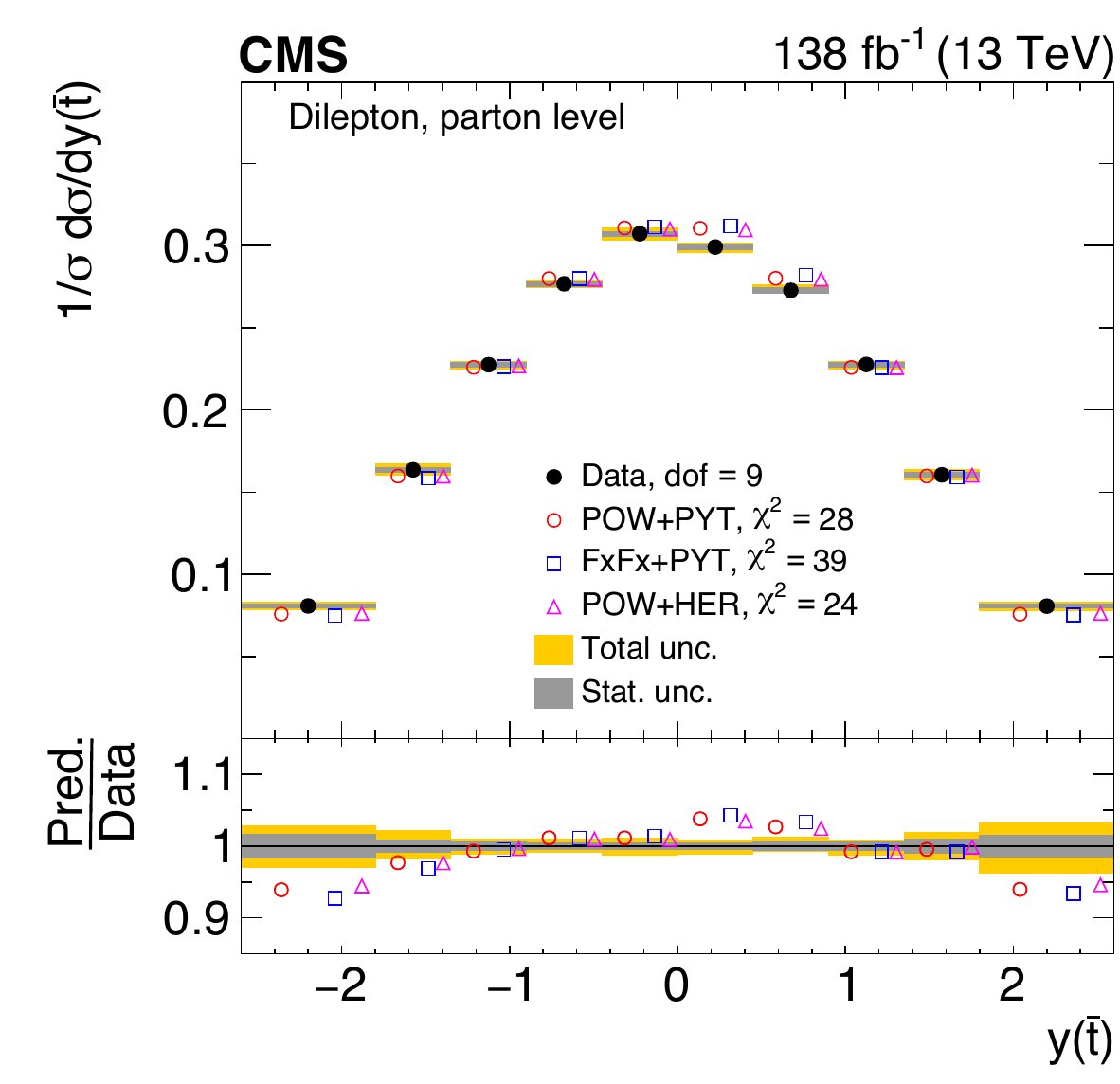}
\includegraphics[width=0.49\textwidth]{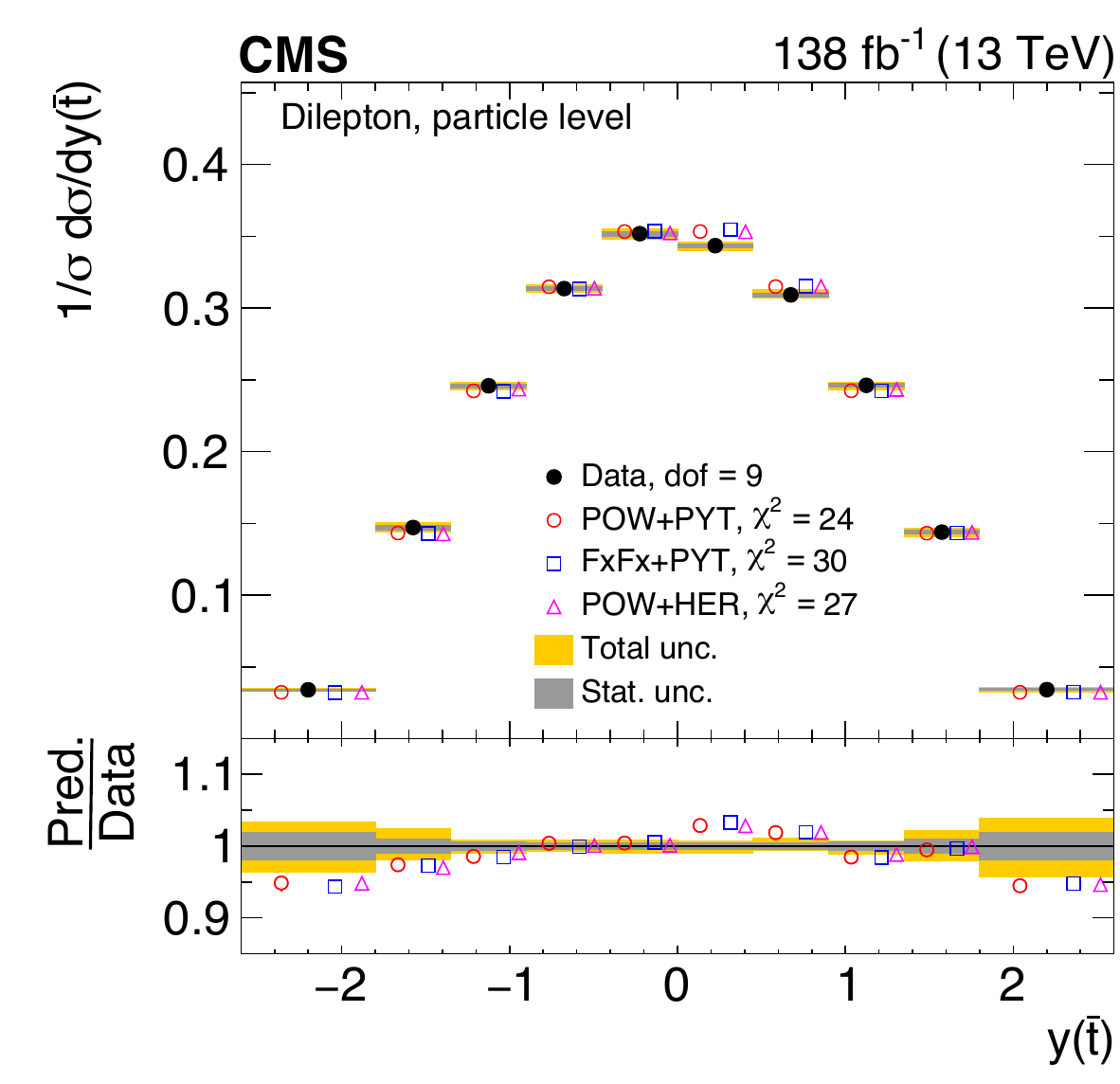}
\caption{Normalized differential \ttbar production cross sections as functions of \yt (upper) and \yat (lower)
are shown for data (filled circles) and
various MC predictions (other points).
Further details can be found in the caption of Fig.~\ref{fig:res_ptt}.}
\label{fig:res_yt}
\end{figure*}

\begin{figure*}[!phtb]
\centering
\includegraphics[width=0.49\textwidth]{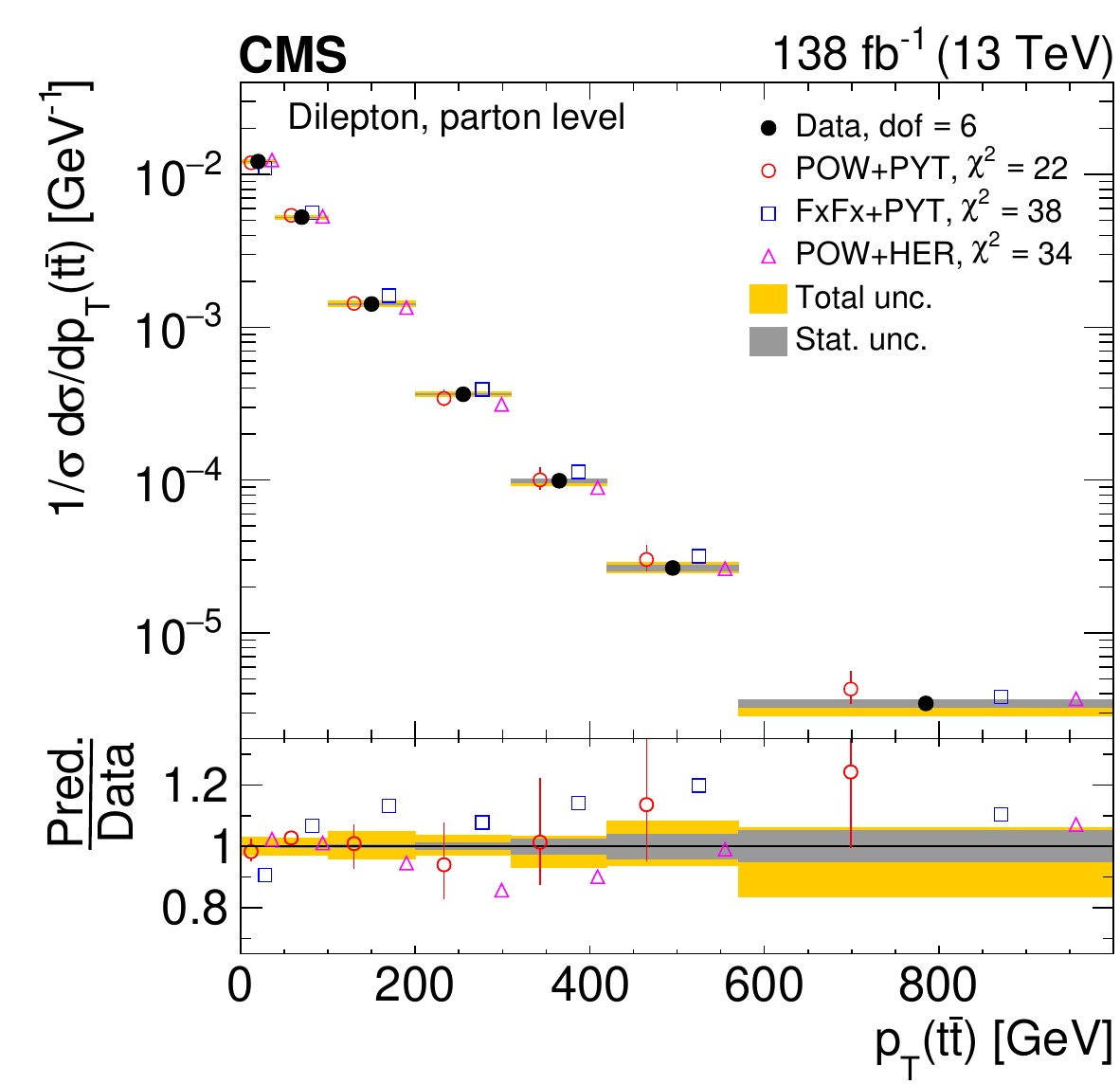}
\includegraphics[width=0.49\textwidth]{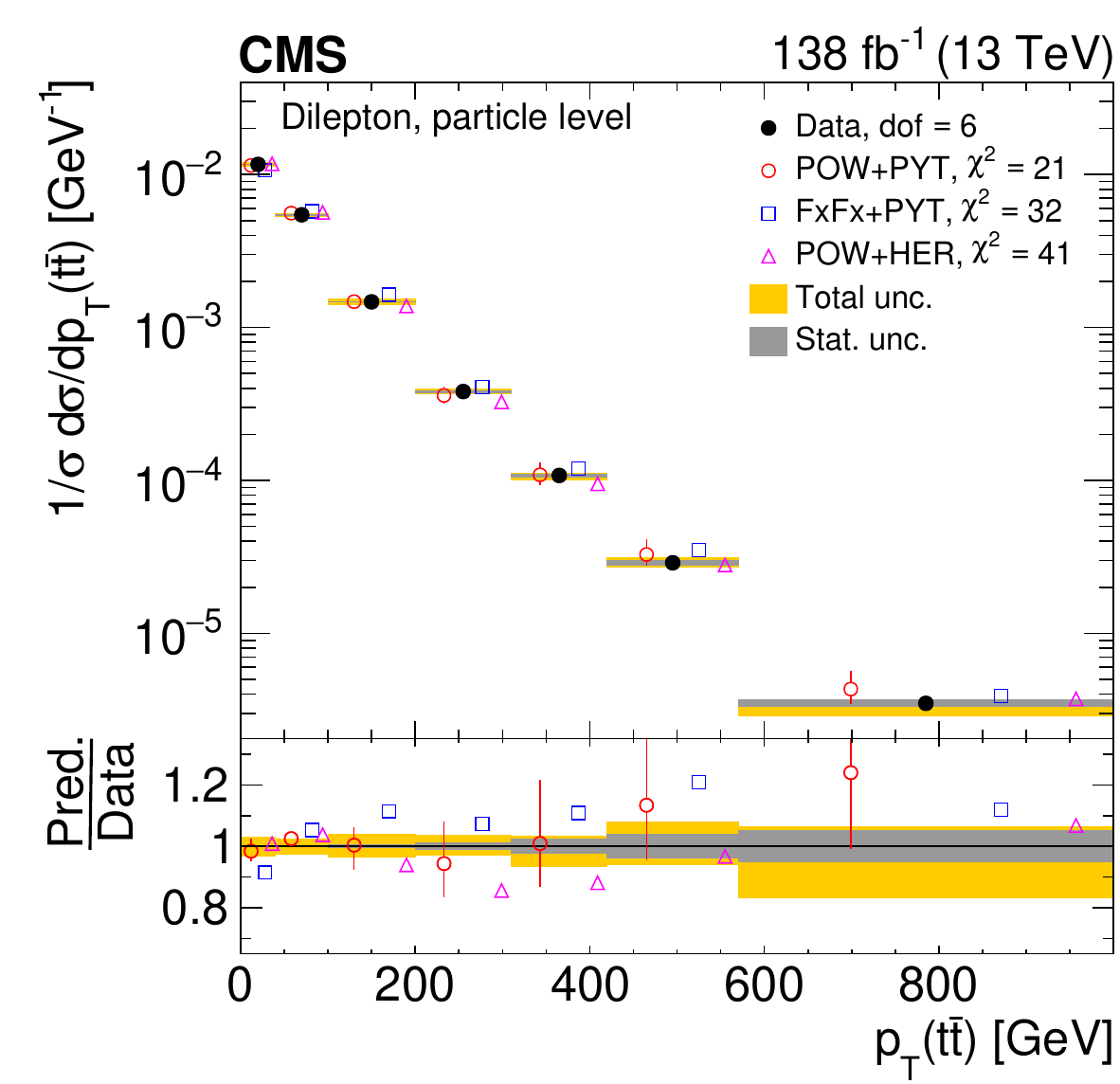}
\includegraphics[width=0.49\textwidth]{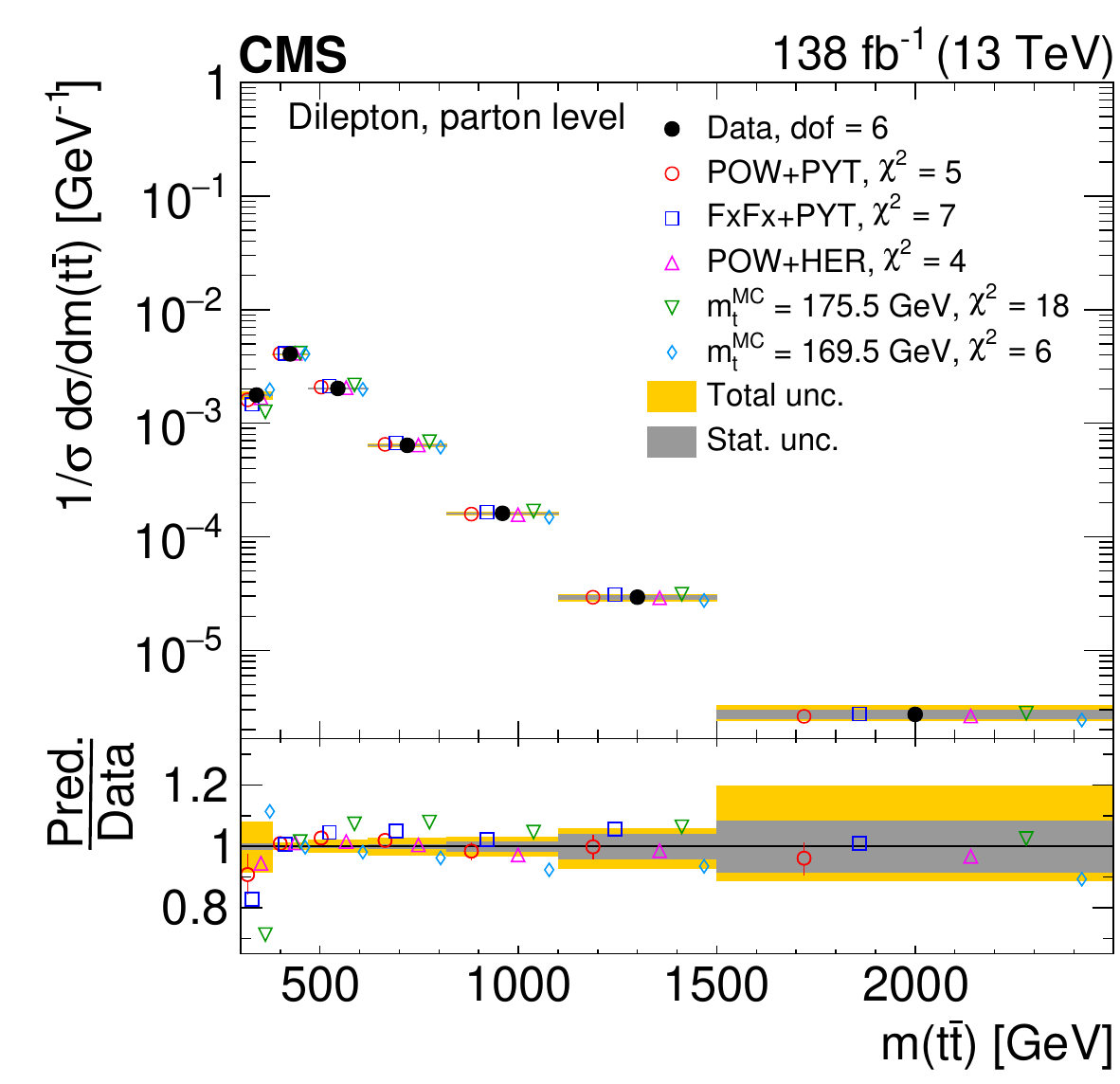}
\includegraphics[width=0.49\textwidth]{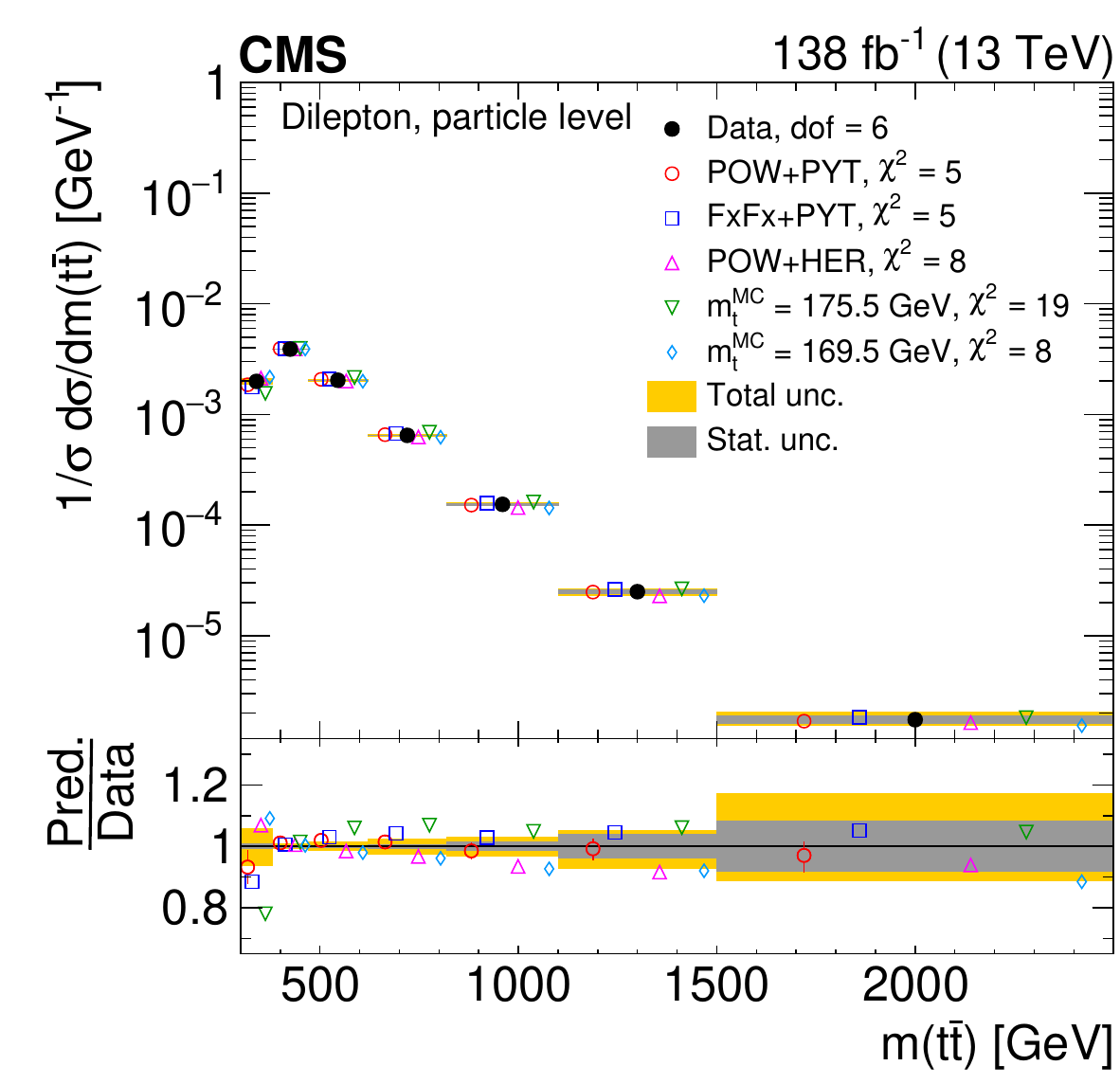}
\includegraphics[width=0.49\textwidth]{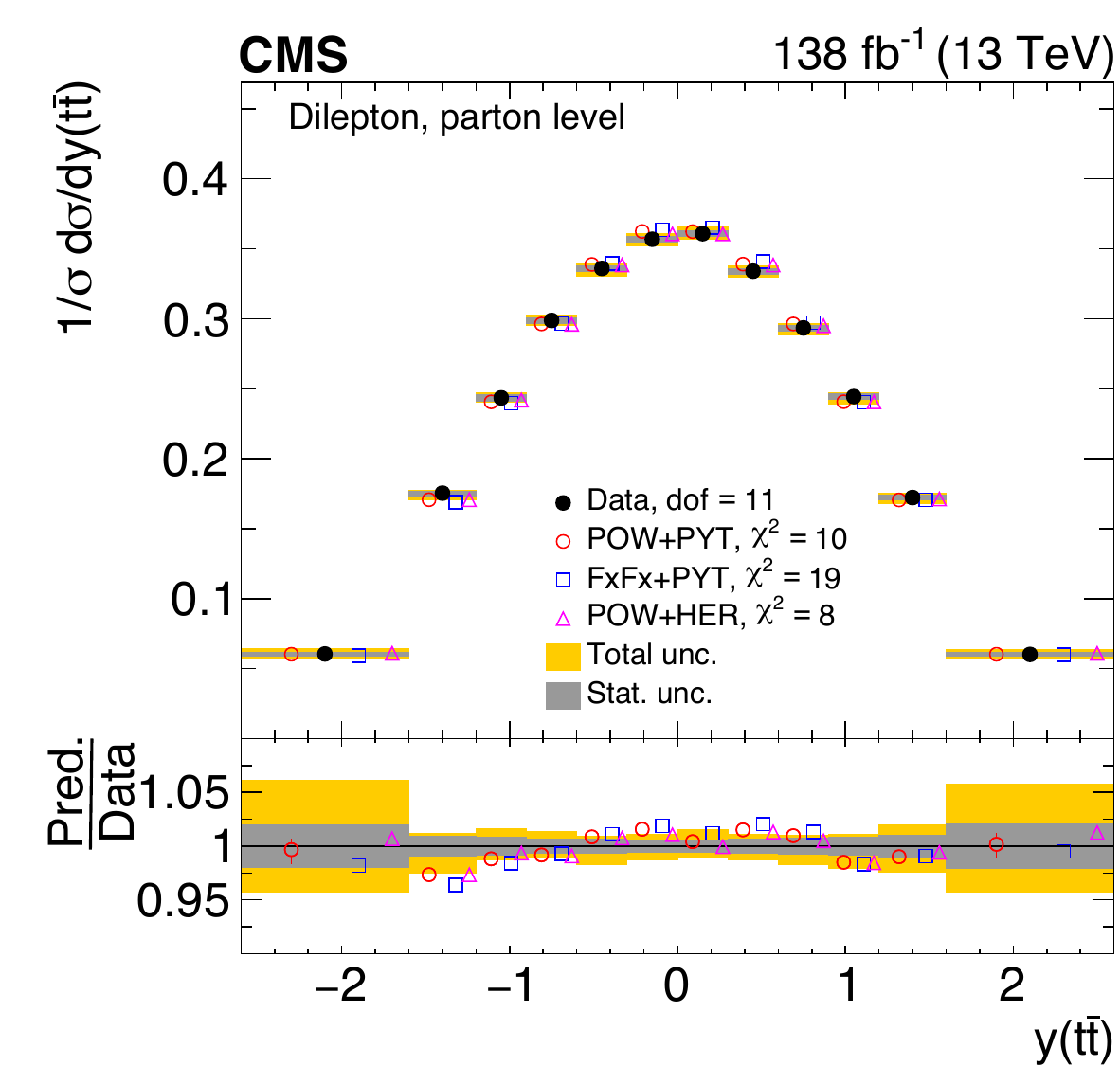}
\includegraphics[width=0.49\textwidth]{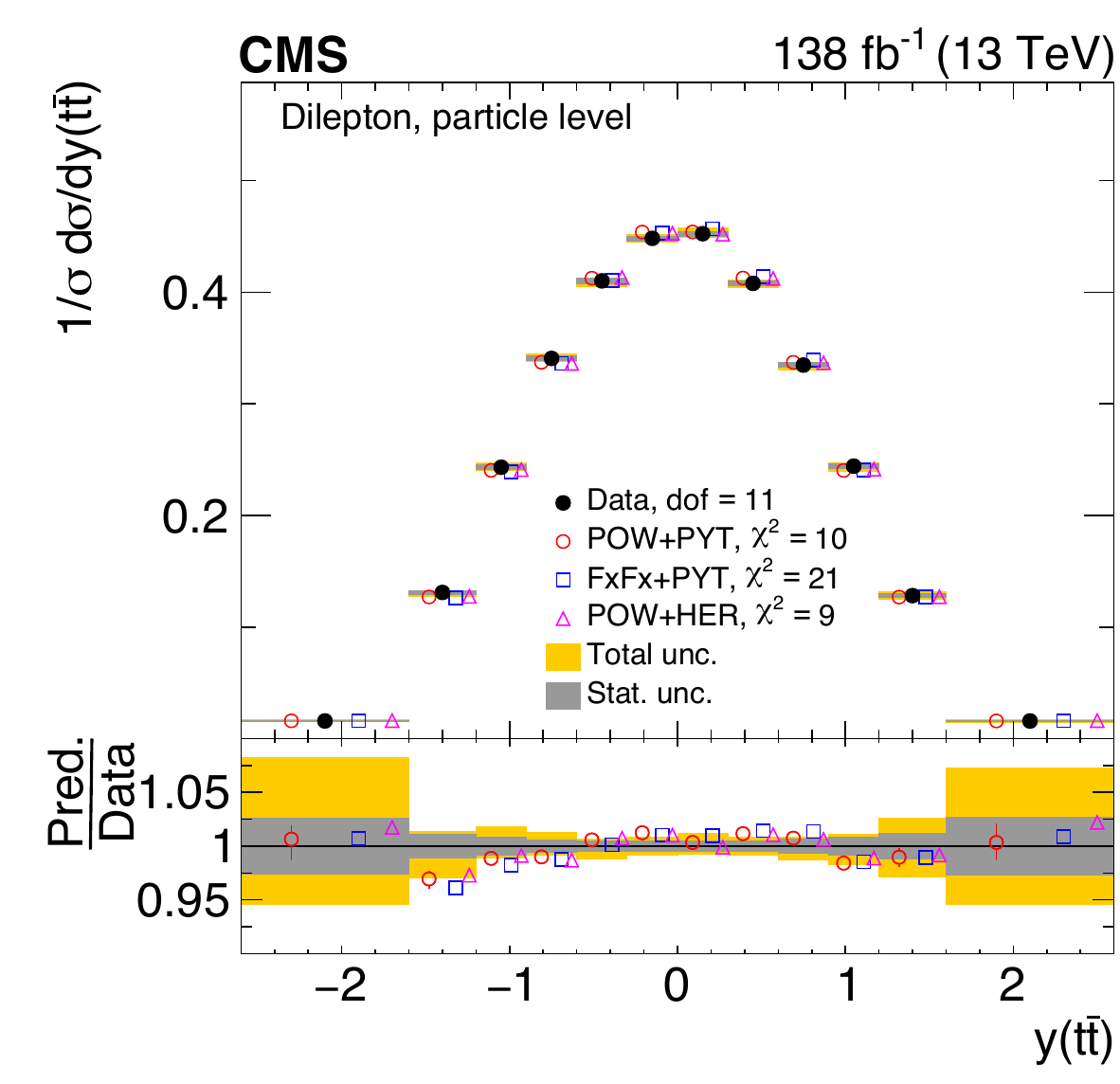}
\caption{Normalized differential \ttbar production cross sections as functions of \pttt (upper), \mtt (middle)
and \ytt (lower)
are shown for data (filled circles) and various MC predictions (other points). The \mtt distributions are also
compared to  \PowPyt (`POW-PYT') simulations with different values of \mtmc.
Further details can be found in the caption of Fig.~\ref{fig:res_ptt}.}
\label{fig:res_pttt}
\end{figure*}

\begin{figure*}[!phtb]
\centering
\includegraphics[width=0.49\textwidth]{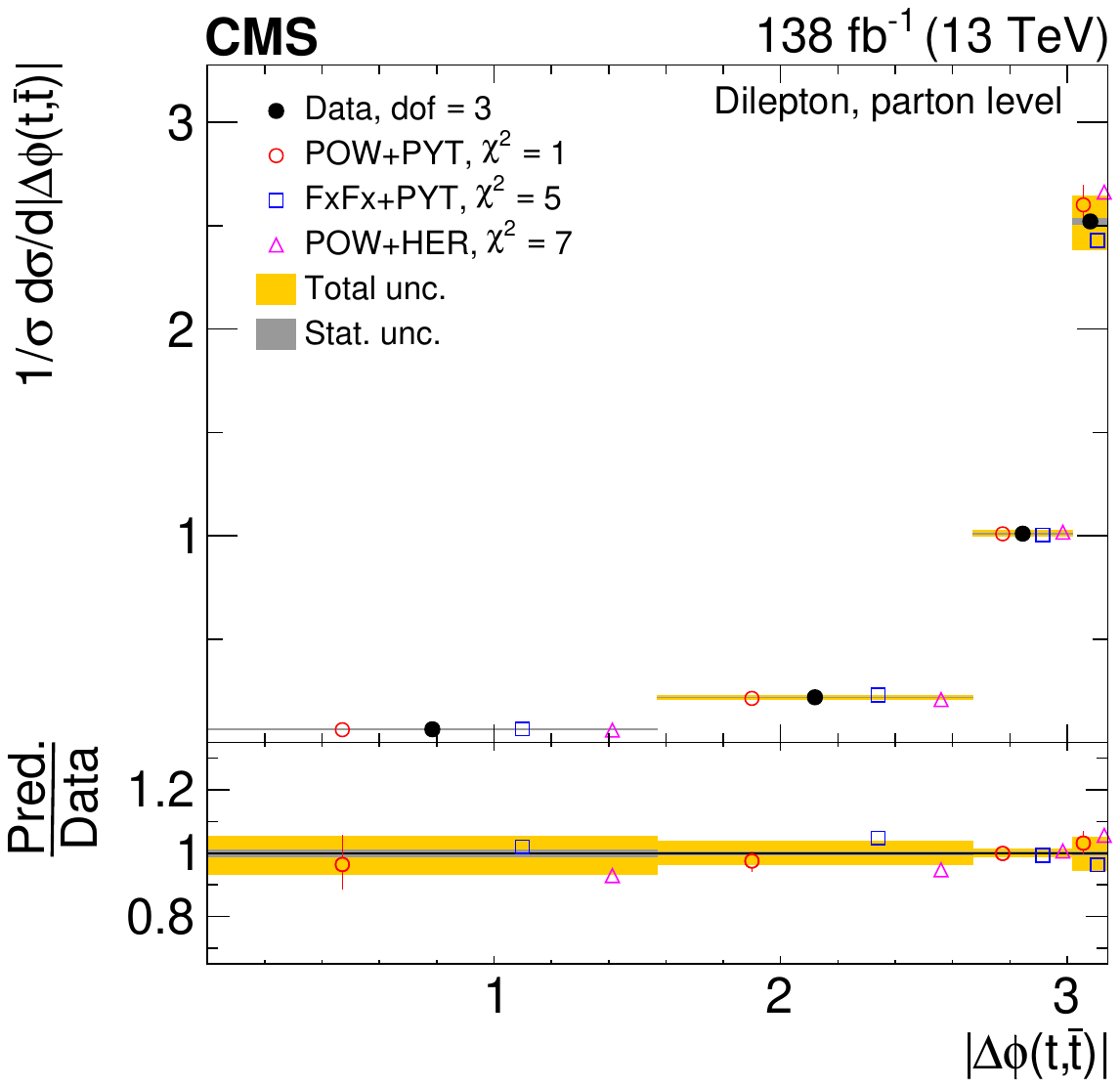}
\includegraphics[width=0.49\textwidth]{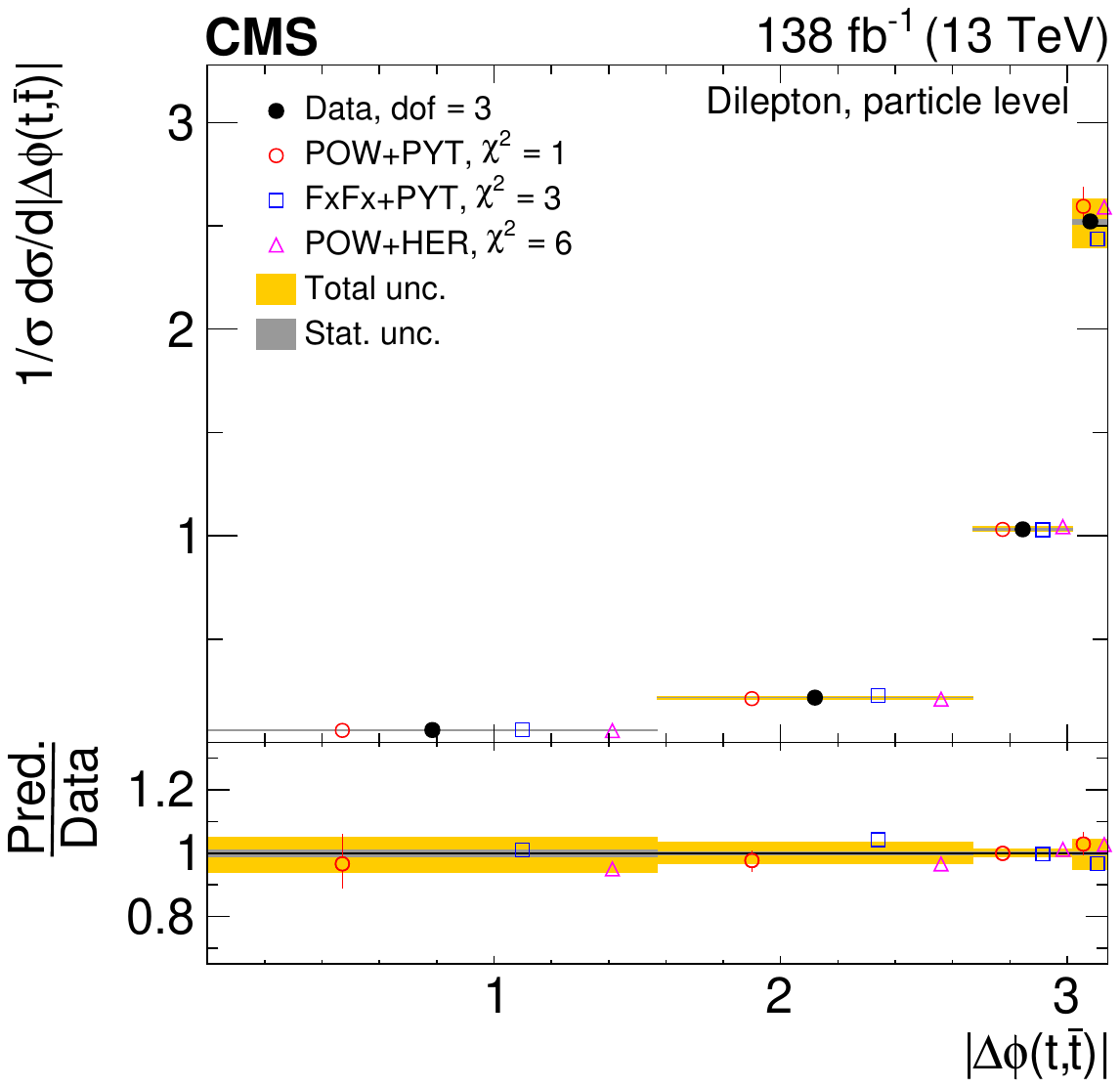}
\includegraphics[width=0.49\textwidth]{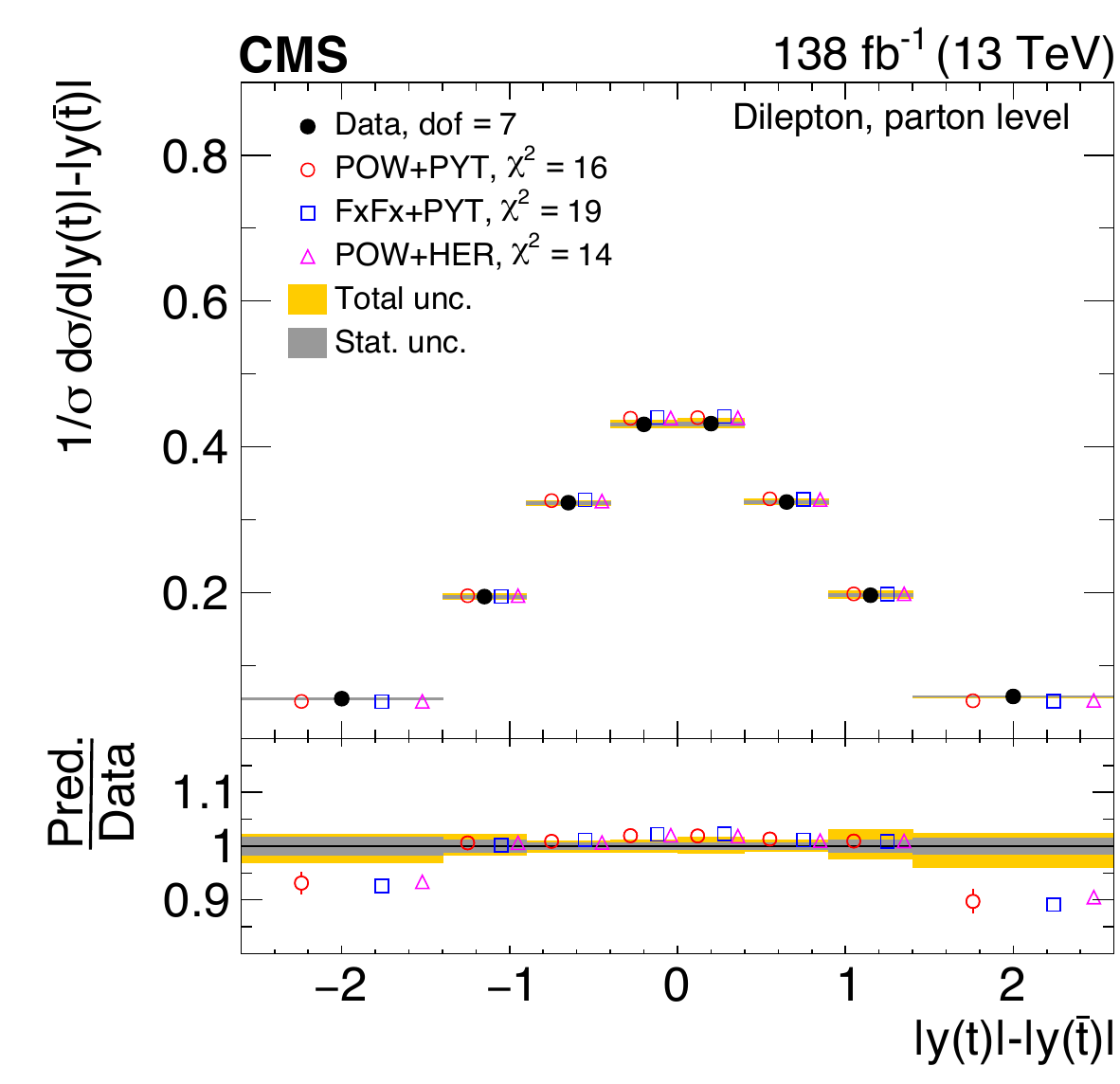}
\includegraphics[width=0.49\textwidth]{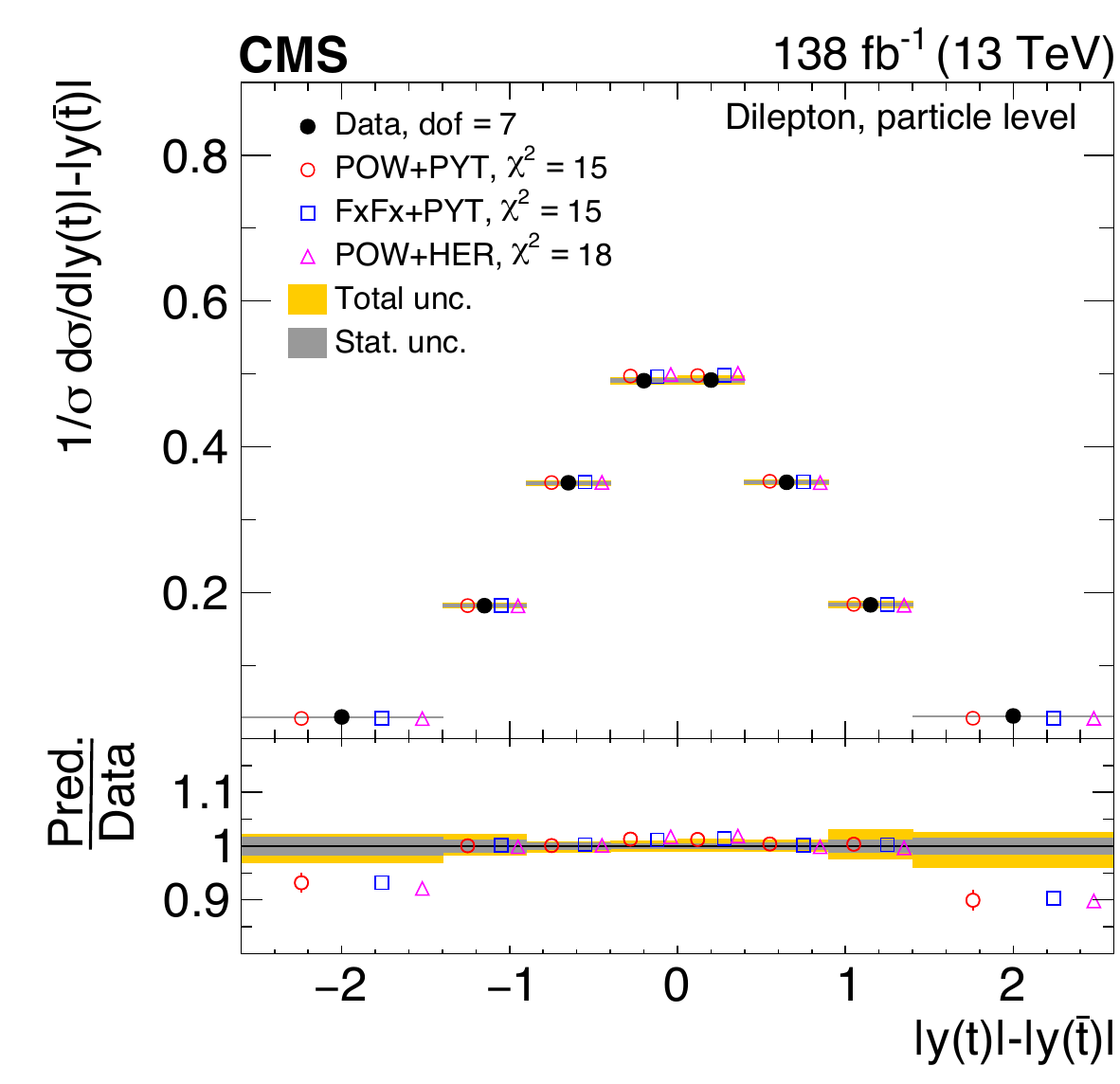}
\caption{Normalized differential \ttbar production cross sections as functions of \dphitt (upper) and \dytt (lower)
are shown for data (filled circles) and various MC predictions (other points).
Further details can be found in the caption of Fig.~\ref{fig:res_ptt}.}
\label{fig:res_dphitt}
\end{figure*}

\begin{figure*}[!phtb]
\centering
\includegraphics[width=0.49\textwidth]{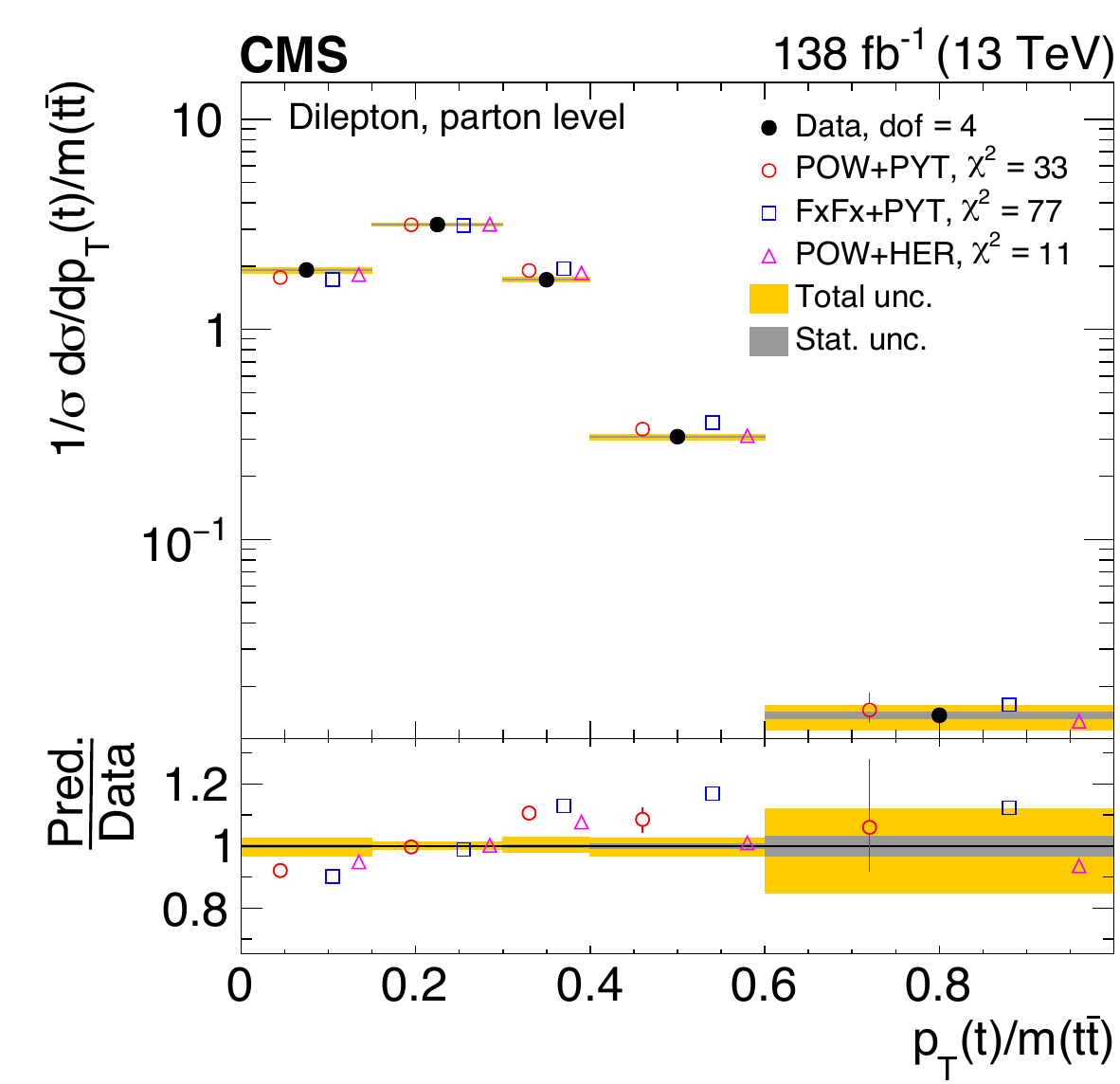}
\includegraphics[width=0.49\textwidth]{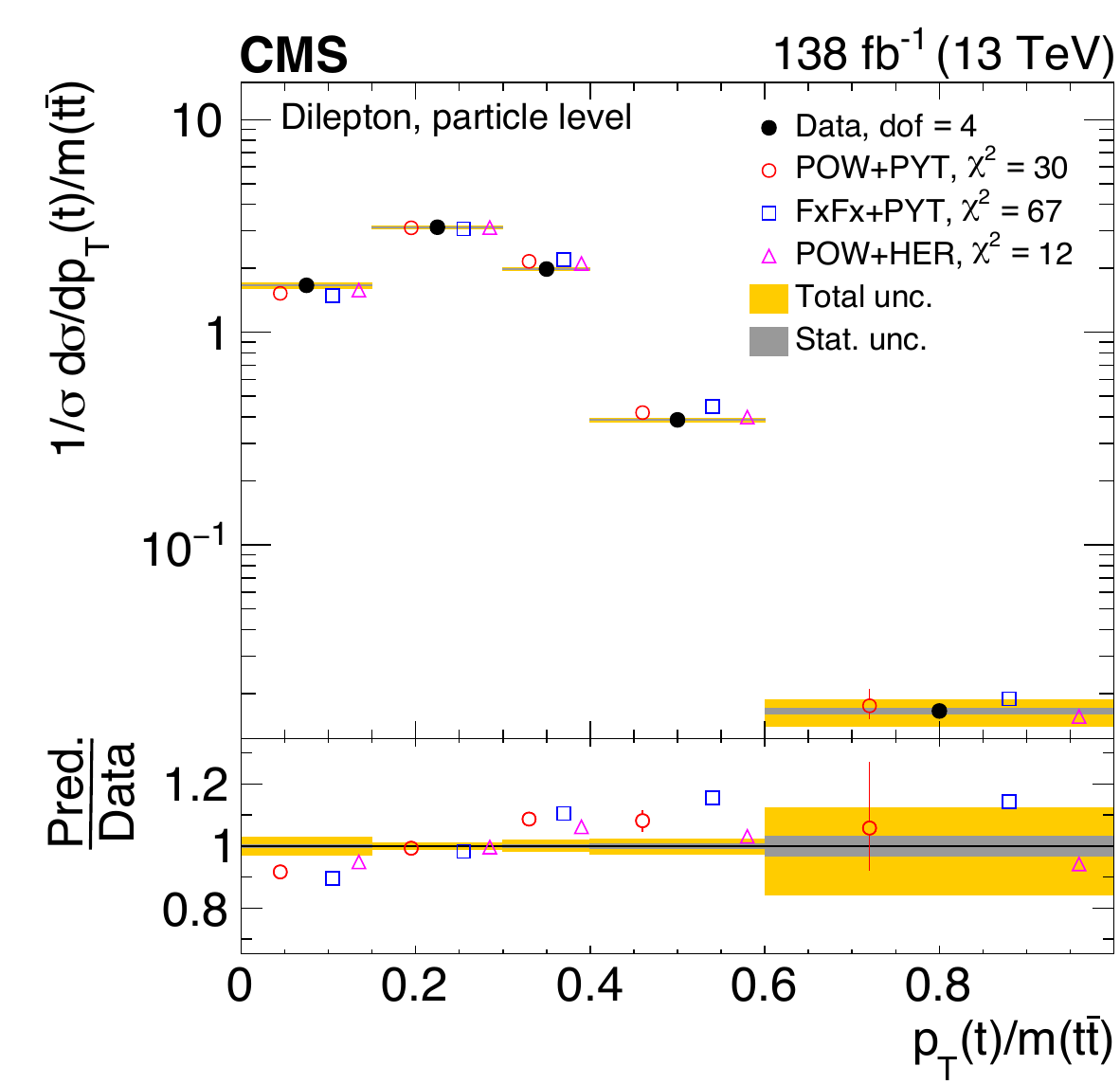}
\includegraphics[width=0.49\textwidth]{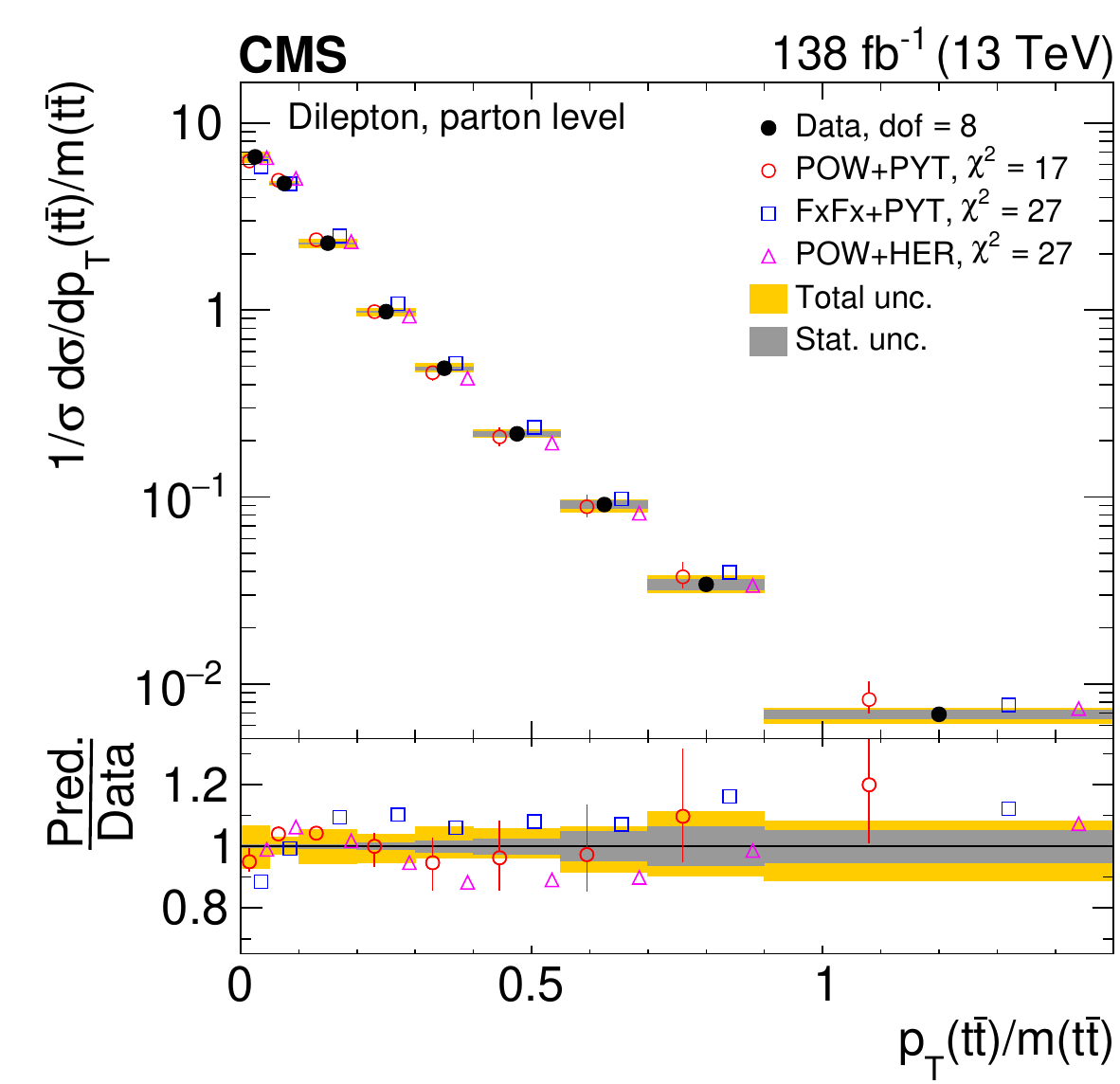}
\includegraphics[width=0.49\textwidth]{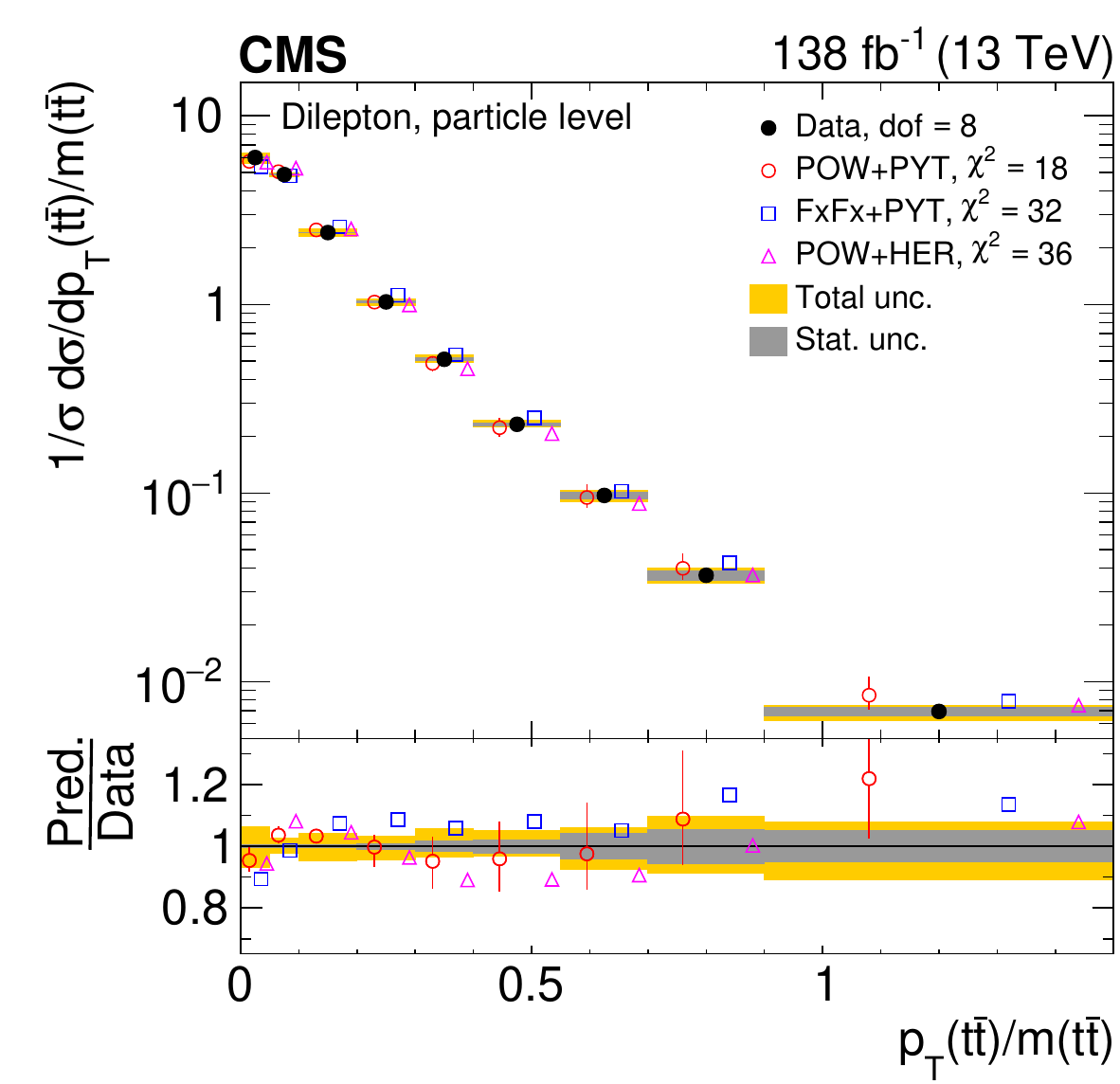}
\caption{Normalized differential \ttbar production cross sections as functions of \rpttmtt (upper) and
\rptttmtt (lower) are shown for data (filled circles) and various MC predictions (other points).
Further details can be found in the caption of Fig.~\ref{fig:res_ptt}.}
\label{fig:res_rpttmtt}
\end{figure*}

\begin{figure*}[!phtb]
\centering
\includegraphics[width=0.49\textwidth]{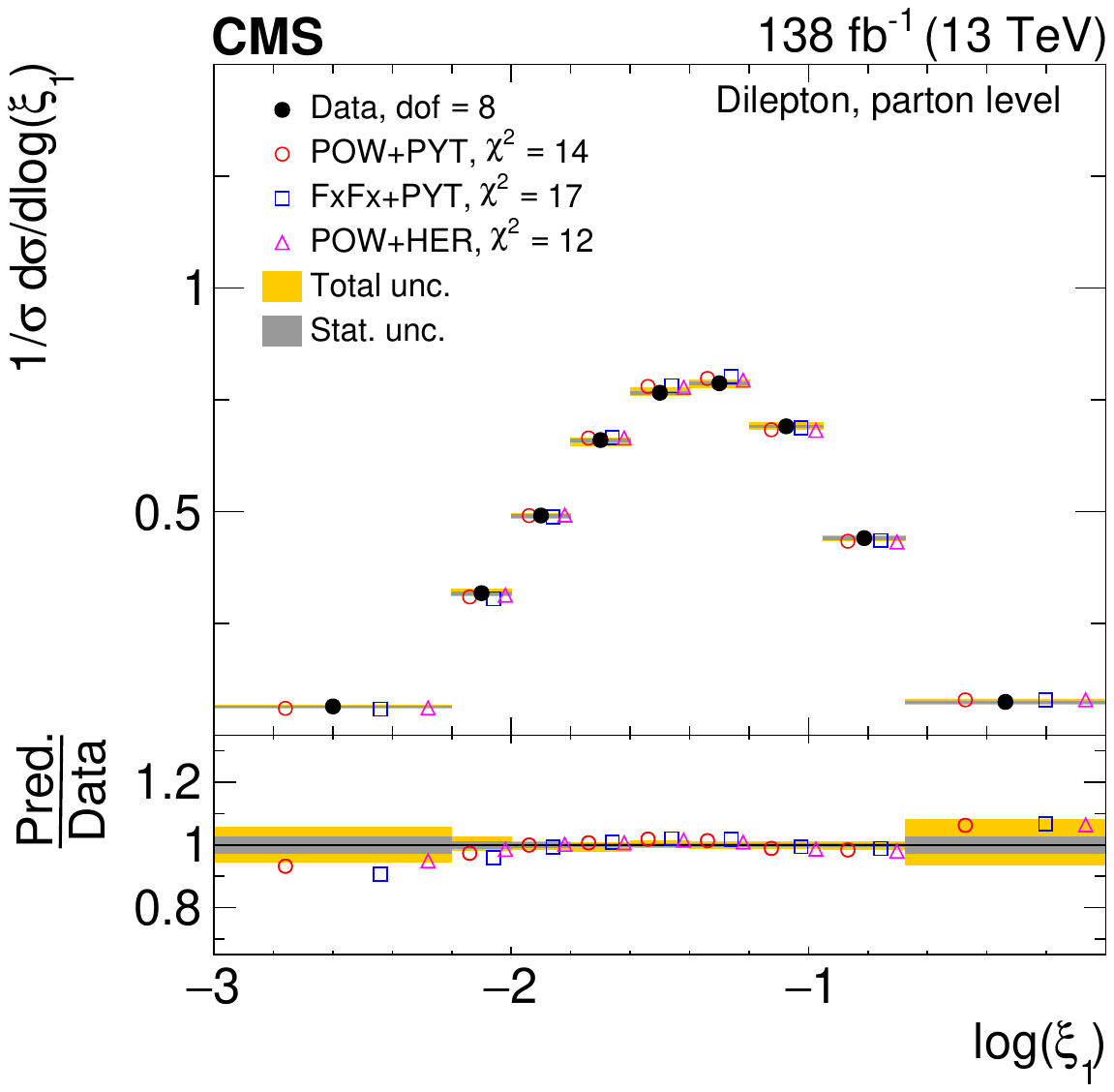}
\includegraphics[width=0.49\textwidth]{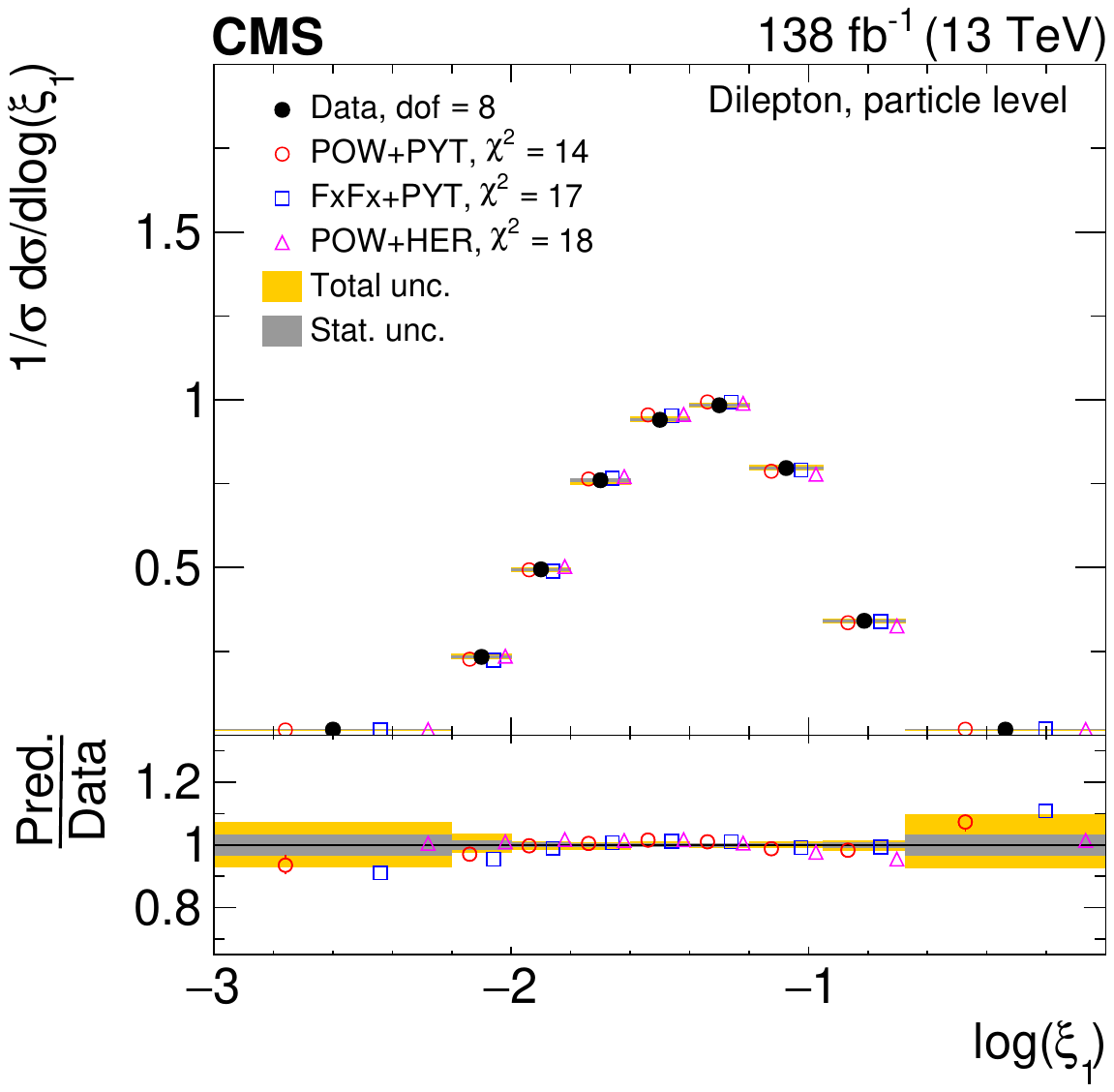}
\includegraphics[width=0.49\textwidth]{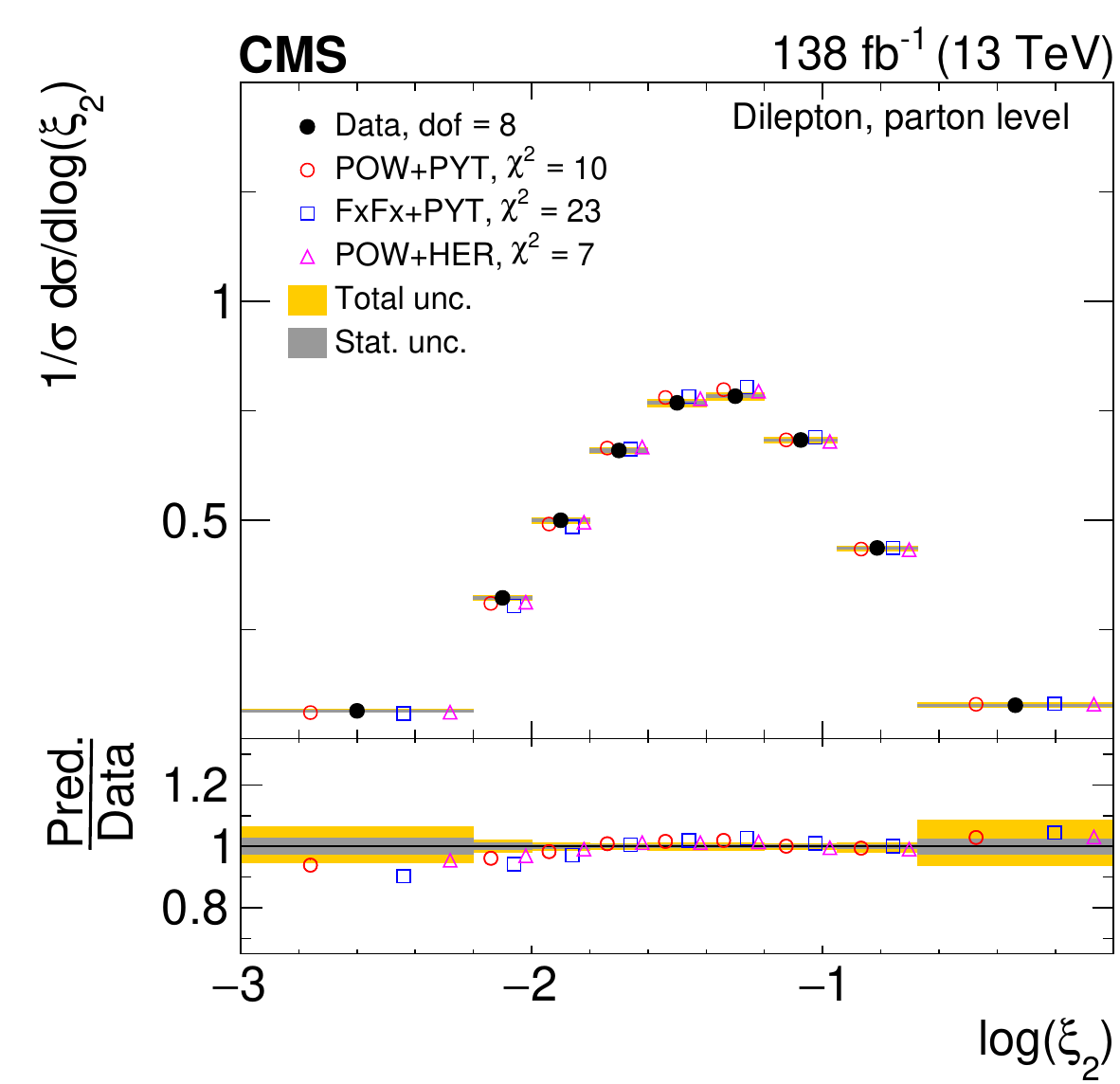}
\includegraphics[width=0.49\textwidth]{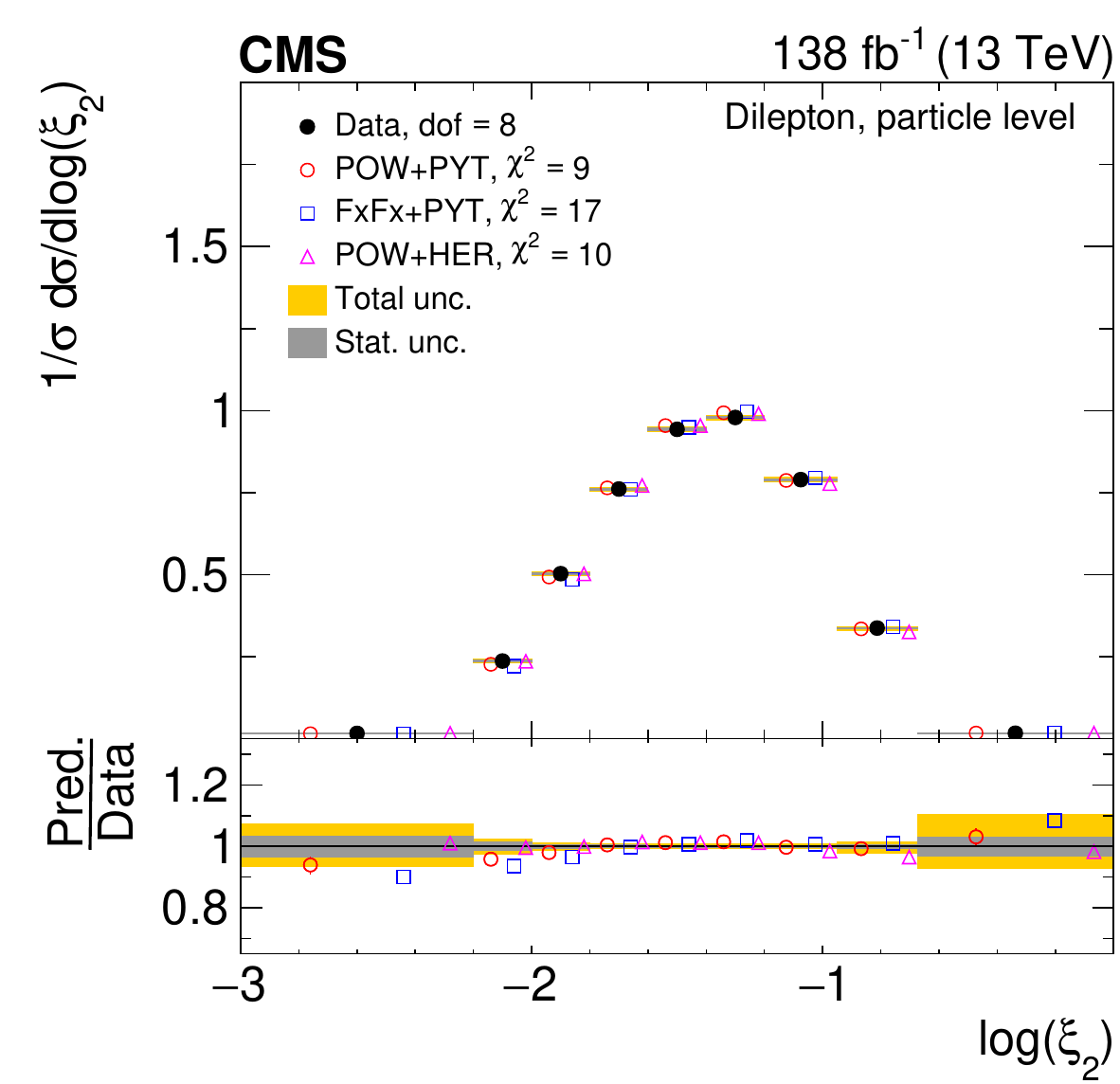}
\caption{Normalized differential \ttbar production cross sections as functions of \logxone (upper) and
\logxtwo (lower)
are shown for data (filled circles) and various MC predictions (other points).
Further details can be found in the caption of Fig.~\ref{fig:res_ptt}.}
\label{fig:res_logxone}
\end{figure*}

The single-differential cross sections
are shown in Figs.~\ref{fig:res_ptt}--\ref{fig:res_logxone}. The \chisq values of the model-to-data comparisons are listed in Tables~\ref{tab:chi2mc_1d_nor_parton}--\ref{tab:chi2mc_1d_nor_particle_ttbar}, and the corresponding $p$-values in Tables~\ref{tab:pvaluemc_1d_nor_parton}--\ref{tab:pvaluemc_1d_nor_particle_ttbar}.
First, we present measurements where the top quark and antiquark kinematics are studied.
Figure~\ref{fig:res_ptt} illustrates the distributions of \ptt and \ptat.
Both the \PowPytSh and, in particular, the \aMCPytSh models
predict harder spectra than observed, while \PowHerSh
provides a reasonable description of the data, as supported by the $p$-value of
the \chisq test.
The discrepancy between \PowPytSh and the data is smaller
than what was observed in our previous analysis~\cite{Sirunyan:2018ucr}.
This reduction can be attributed to the use of the more-recent tune CP5~\cite{Sirunyan:2019dfx} for the
\PYTHIAviii part of the calculation, whereas in the previous analysis the
CUETP8M2T4 tune~\cite{bib:CMS:2016kle,bib:CUETP8tune,Skands:2014pea} was applied. The \aMCPytSh model provides the poorest description of the data, 
as indicated by the particularly large \chisq values.
Figure~\ref{fig:res_yt} depicts the \yt and \yat distributions.
All models predict a slightly more central distribution than observed
in data.
One general observation can be made at this point: the comparisons of predictions and data for the
cross sections at the parton and particle levels show similar patterns.
This is also the case for most other cross sections presented in this paper;
consequently, a separate discussion of the parton- and particle-level cross sections
will be only given in case of significant differences.

Figure~\ref{fig:res_pttt} shows the distributions of the complete set of \ttbar
kinematic observables: \pttt, \mtt, and \ytt.
For \pttt, the phase space covered in our previous analysis~\cite{Sirunyan:2018ucr} is extended up to 1\TeV.
The description of the \pttt distribution is particularly
difficult, since nonzero values indicate the presence of extra QCD radiation in the event recoiling
against the \ttbar system, which directly probes higher-order processes in the calculation.
This effect is also characterized by large theory uncertainties as 
shown for the \PowPytSh prediction.
The three MC models differ in the predicted shape of the \pttt distribution
and none of them is able to describe the data accurately.
The best description is provided by \PowPytSh,
which tends to overshoot the data only in the higher-\pttt range.
The \aMCPytSh model predicts too many events in the intermediate \pttt ranges,
while \PowHerSh predicts too few.
The description by \PowHerSh is somewhat improved compared
to what was observed in our previous analysis~\cite{Sirunyan:2018ucr}, 
which can also be attributed to using a newer version of the \HERWIG MC generator, \HERWIGvii~\cite{Bellm_2016},
instead of \HERWIGpp~\cite{bib:herwigpp}.
It is interesting to note that also in two recent measurements~\cite{Aad:2019ntk,Sirunyan:2018wem} by the ATLAS and 
CMS Collaborations, the \pttt distribution was found to be rather poorly described by several models.
In the present analysis, the \mtt spectrum is
overall reasonably
well described by the models, with the exception
of the first bin near the threshold where the predictions lie somewhat below
the data.
This region is known to be particularly sensitive to the value of the top quark
mass assumed in the calculations.
To investigate this discrepancy in more detail, the \PowPytSh
prediction is also shown for two other values of the top quark mass parameter, $\mtmc=169.5\GeV$ and
 $\mtmc=175.5\GeV$,
compared to the value of 172.5\GeV taken in the nominal simulation.
For the lower value, the cross section prediction moves up in the lowest \mtt bin by about 20\%
and the whole mass spectrum is a bit softer than in the data, however still providing a 
\chisq with a good $p$-value.
Using the higher value, the resulting predicted \mtt spectrum is clearly harder than what
is observed in data. This discrepancy is also
reflected by an increase of the \chisq value of about 13 units compared to using the central mass value,
proving the high sensitivity of the \mtt distribution to the top quark mass value, 
as explored in the CMS analysis~\cite{CMS:2019jul}.  
The \ytt distribution of the data
is described reasonably well by all models.

Moving on to studies of kinematic correlations between top quark and antiquark,
we show in Fig.~\ref{fig:res_dphitt} the distribution of the absolute value of the azimuthal angle
difference between the top quark and antiquark \dphitt, and
the difference of the absolute values of the top quark and antiquark rapidities \dytt,
related to the charge asymmetry~\cite{Chwalek:2022nbk}. For \dphitt, angles smaller than $\pi$ are directly sensitive to additional
QCD radiation in the event.
The models provide a good description of the data.
The \dytt spectrum of the data
is not perfectly described by the MC models, as they predict
more events at small rapidity separations and fewer at larger values
$\absdytt \approx 2$.

Figure~\ref{fig:res_rpttmtt} presents the distributions of two ratios: \rpttmtt and \rptttmtt.
The study of the first quantity is inspired by the observation in Ref.~\cite{Sirunyan:2019zvx}
that at large \mtt the effect of the models predicting harder \ptt spectra is enhanced.
Indeed, all three tested models predict a significantly harder spectrum for the ratio of the two variables
and the \chisq values indicate a poor data description.
The distribution of the second ratio, \rptttmtt,
is expected to be sensitive to \pt resummation
effects. The description by the three models follows largely
the trends observed for \pttt.

Finally, we study two observables $\xi_{1}$ and $\xi_{2}$, defined as
$\xi_{1} = [ E(\ttbar) - p_z(\ttbar) ] / 2 E_{\mathrm{p}}$
and
$\xi_{2} = [ E(\ttbar) + p_z(\ttbar) ] / 2 E_{\mathrm{p}}$,
with $E_{\mathrm{p}}$ denoting the proton beam energy, and $p_z(\ttbar)$ the magnitude 
of the \ttbar momentum along the beam axis.
In the leading order pQCD picture
of the $\pp \to \ttbar$  process, these variables represent the
proton momentum fractions
of the two partons entering the hard interaction.
Figure~\ref{fig:res_logxone} depicts
the \logxone and \logxtwo distributions.
Overall, these PDF-sensitive distributions are reasonably
well described by all the models.
Of special interest is the last bin
covering proton momentum fractions above $\approx$0.2, where the
uncertainties in the PDFs, in particular for the gluon distribution,
start to grow.

The description of the kinematic distributions of top quark, antiquark, and \ttbar system
by the three MC models can be summarized as follows.
The models tend to predict, for the individual quarks, harder
\pt spectra and slightly more-central rapidity distributions than those observed in data.
A reasonable description is provided for the \ttbar rapidity and invariant mass spectra.
The latter is described less well by the \PowPytSh simulation that uses a larger value of \mtmc, as it
predicts a harder spectrum than that observed in data.
The \dphitt distribution is modeled well,
and smaller rapidity separations are predicted than observed in data on average.
The \aMCPytSh model provides overall the least accurate description of the data, in particular by
predicting harder \pt spectra for the top quark, antiquark, and the \ttbar system.
Among all the models, \PowHerSh predicts the softest \pt spectra for the top quark, antiquark, and
the \ttbar system,
which matches the data well for the former two, but are too soft for the latter.
The \PowPytSh model
provides the best description of the \pt distribution of the \ttbar system.
The standard \chisq values indicate a rather poor description of several distributions
by some of the models.
For \PowPytSh, the \chisq values that include
the prediction uncertainties
(see Tables~\ref{tab:chi2mc_1d_nor_parton}--\ref{tab:chi2mc_1d_nor_particle_ttbar})
are significantly smaller than the standard
\chisq values. However, for a few distributions, such as \rpttmtt, the $p$-values
of these additional \chisq tests remain too low for a reasonable 
description of the data. This is typically the case for the distributions with the
largest visible discrepancies.

\clearpage

\begin{table*}
\centering
 \topcaption{The \chisq values and \ndf of the measured normalized single-differential cross sections 
for \ttbar and top quark kinematic observables at the parton level are shown with respect to the predictions of various MC generators. 
The \chisq values are calculated taking only measurement uncertainties into account and excluding theory uncertainties.  
For \PowPytSh, the \chisq values including theory uncertainties are indicated with the brackets (w. unc.).}
\label{tab:chi2mc_1d_nor_parton}
\renewcommand{\arraystretch}{1.4}
\centering
\begin{tabular}{lccccc}
\multirow{1}{*}{Cross section} & \hspace*{0.3 cm} \multirow{2}{*}{\ndf} \hspace*{0.3 cm} & \multicolumn{3}{c}{\chisq} \\
\cline{3-5}
{variables} && \PowPytSh (w. unc.)  & \aMCPytSh  & \PowHerSh   \\
\hline
\ptt& 6 & 15 \: (11) & 38 & 5 \\
\ptat& 6 & 13 \: (9) & 37 & 6 \\
\yt& 9 & 23 \: (20) & 34 & 21 \\
\yat& 9 & 28 \: (25) & 39 & 24 \\
\pttt& 6 & 22 \: (7) & 38 & 34 \\
\ytt& 11 & 10 \: (8) & 19 & 8 \\
\mtt  & 6 & 5 \: (3) & 7 & 4 \\
\dphitt& 3 & 1 \: (0) & 5 & 7 \\
\dytt& 7 & 16 \: (9) & 19 & 14 \\
\rpttmtt& 4 & 33 \: (20) & 77 & 11 \\
\rptttmtt& 8 & 17 \: (6) & 27 & 27 \\
\logxone& 8 & 14 \: (10) & 17 & 12 \\
\logxtwo& 8 & 10 \: (7) & 23 & 7 \\
 \end{tabular}
\end{table*}

\begin{table*}
\centering
 \topcaption{The \chisq values and \ndf of the measured normalized single-differential cross sections for \ttbar and top quark kinematic observables at the particle level are shown with respect to the predictions of various MC generators. 
The \chisq values are calculated taking only measurement uncertainties into account and excluding theory uncertainties.  
For \PowPytSh, the \chisq values including theory uncertainties are indicated with the brackets (w. unc.).}
 \label{tab:chi2mc_1d_nor_particle_ttbar}
 \renewcommand{\arraystretch}{1.4}
 \centering
 \begin{tabular}{lccccc}
 \multirow{1}{*}{Cross section} & \hspace*{0.3 cm} \multirow{2}{*}{\ndf} \hspace*{0.3 cm} & \multicolumn{3}{c}{\chisq} \\
 \cline{3-5}
{variables} && \PowPytSh (w. unc.)  & \aMCPytSh  & \PowHerSh   \\
\hline
\ptt& 6 & 17 \: (12) & 40 & 6 \\
\ptat& 6 & 14 \: (9) & 40 & 5 \\
\yt& 9 & 19 \: (16) & 30 & 20 \\
\yat& 9 & 24 \: (20) & 30 & 27 \\
\pttt& 6 & 21 \: (7) & 32 & 41 \\
\ytt& 11 & 10 \: (7) & 21 & 9 \\
\mtt  & 6 & 5 \: (3) & 5 & 8 \\
\dphitt& 3 & 1 \: (0) & 3 & 6 \\
\dytt& 7 & 15 \: (10) & 15 & 18 \\
\rpttmtt& 4 & 30 \: (20) & 67 & 12 \\
\rptttmtt& 8 & 18 \: (7) & 32 & 36 \\
\logxone& 8 & 14 \: (9) & 17 & 18 \\
\logxtwo& 8 & 9 \: (6) & 17 & 10 \\
 \end{tabular}
\end{table*}

\clearpage

\subsubsection{Multi-differential cross section measurements}
\begin{figure}
\centering
\includegraphics[width=0.99\textwidth]{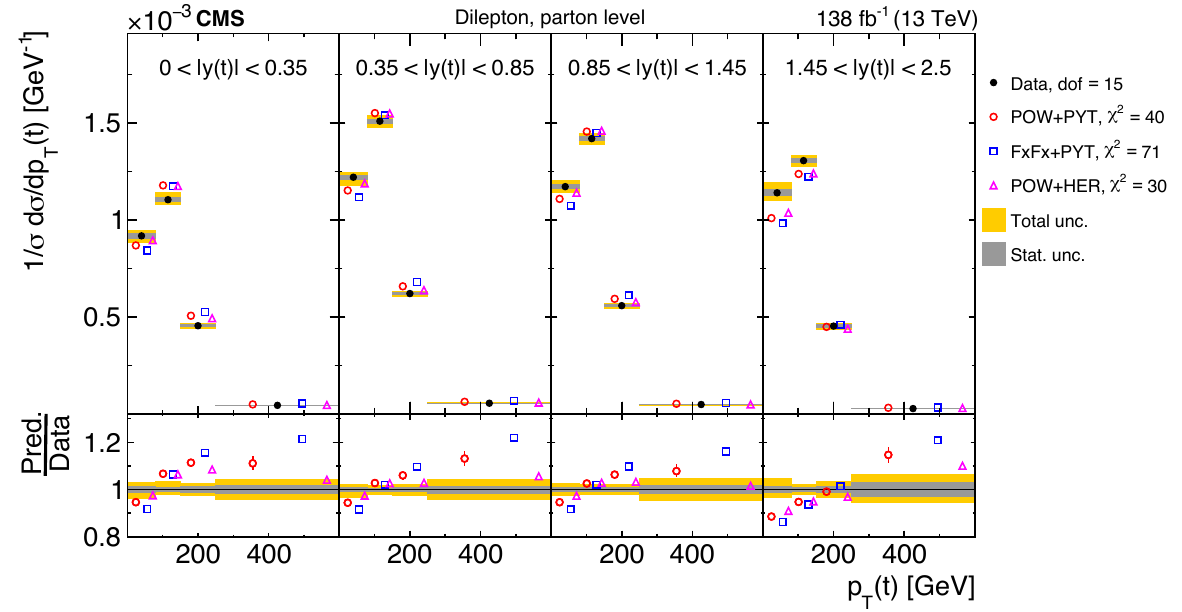}
\includegraphics[width=0.99\textwidth]{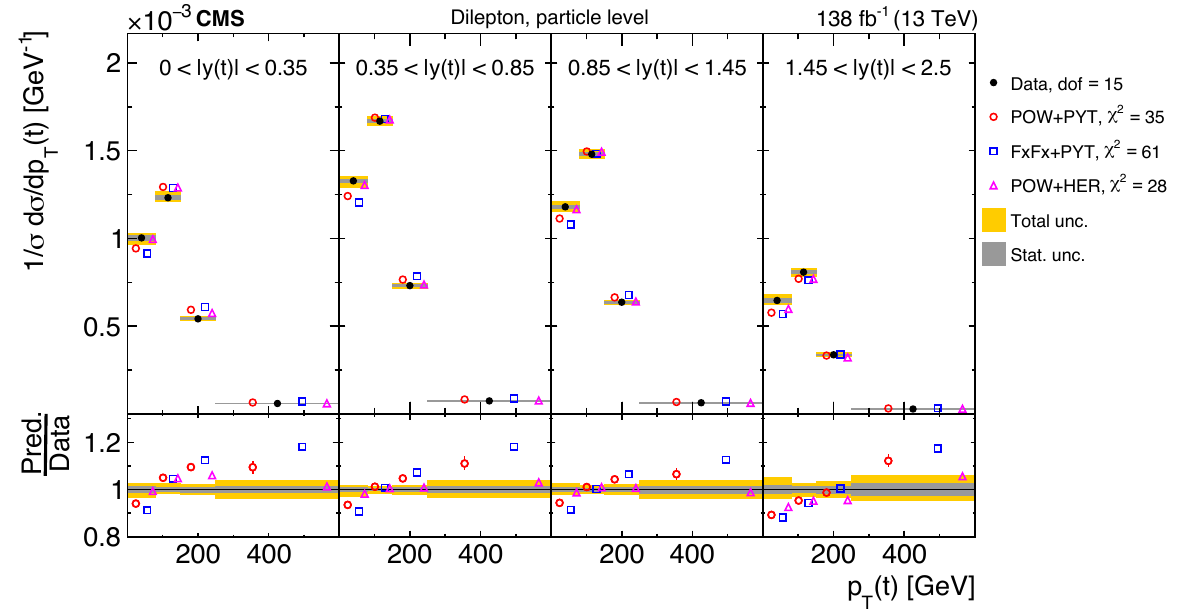}
\caption{Normalized \ytptt cross sections measured at the parton level in the full phase space (upper) and at
the particle
level in a fiducial phase space (lower). The data are shown as filled circles with grey and yellow bands indicating the
statistical and total uncertainties (statistical and systematic uncertainties added in quadrature), respectively.
For each distribution, the number of degrees of freedom (dof) is also provided.
The cross
sections are compared to various MC predictions (other points). The estimated uncertainties in the \PowPyt (`POW-PYT')
simulation are represented by vertical bars on the corresponding points. For each MC model, a value of \chisq
is reported
that takes into account the measurement uncertainties. The lower panel in each plot shows the ratios of the
predictions to the data.}
    \label{fig:res_ytptt}
\end{figure}

\begin{figure}
\centering
\includegraphics[width=0.99\textwidth]{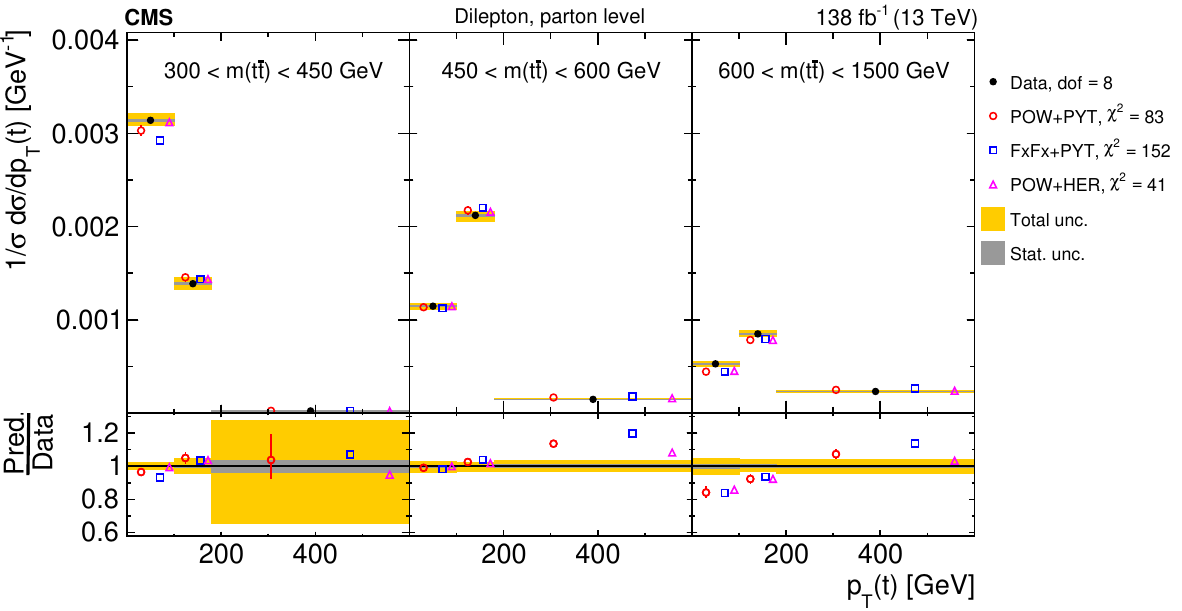}
\includegraphics[width=0.99\textwidth]{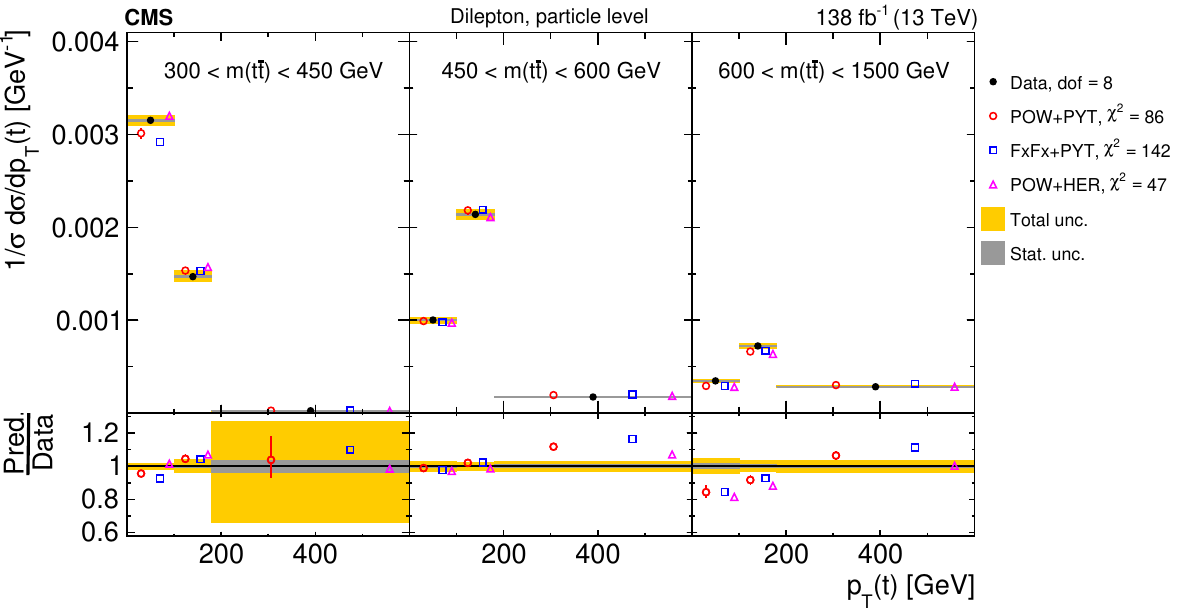}
\caption{Normalized \mttptt cross sections are shown for data (filled circles) and various MC predictions
(other points).
    Further details can be found in the caption of Fig.~\ref{fig:res_ytptt}.}
    \label{fig:res_mttptt}
\end{figure}

\begin{figure}
\centering
\includegraphics[width=0.99\textwidth]{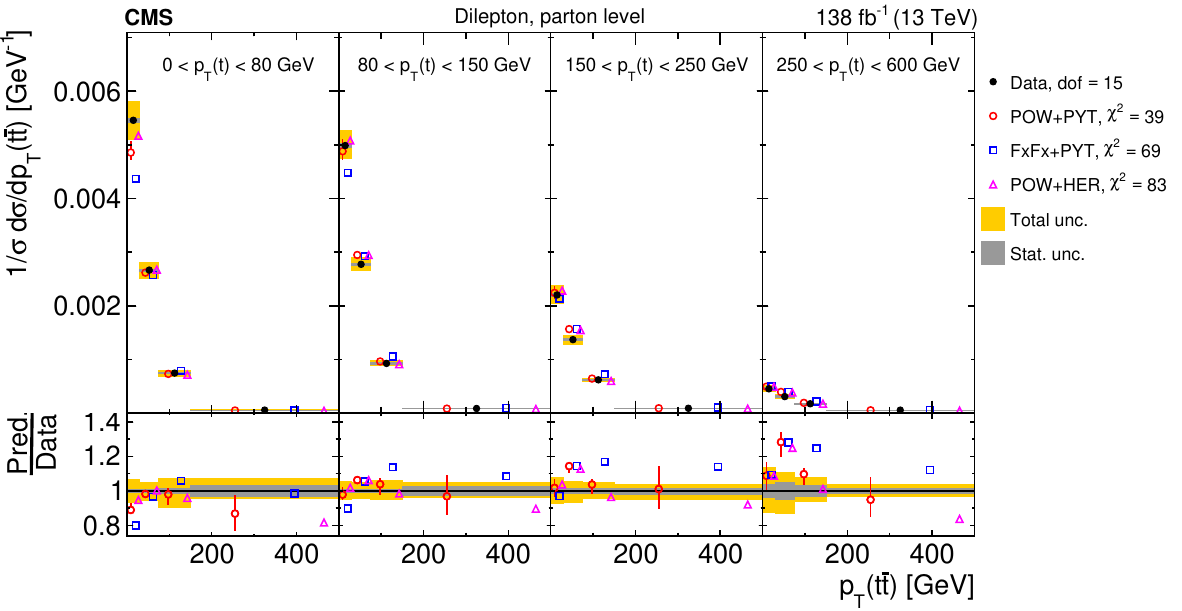}
\includegraphics[width=0.99\textwidth]{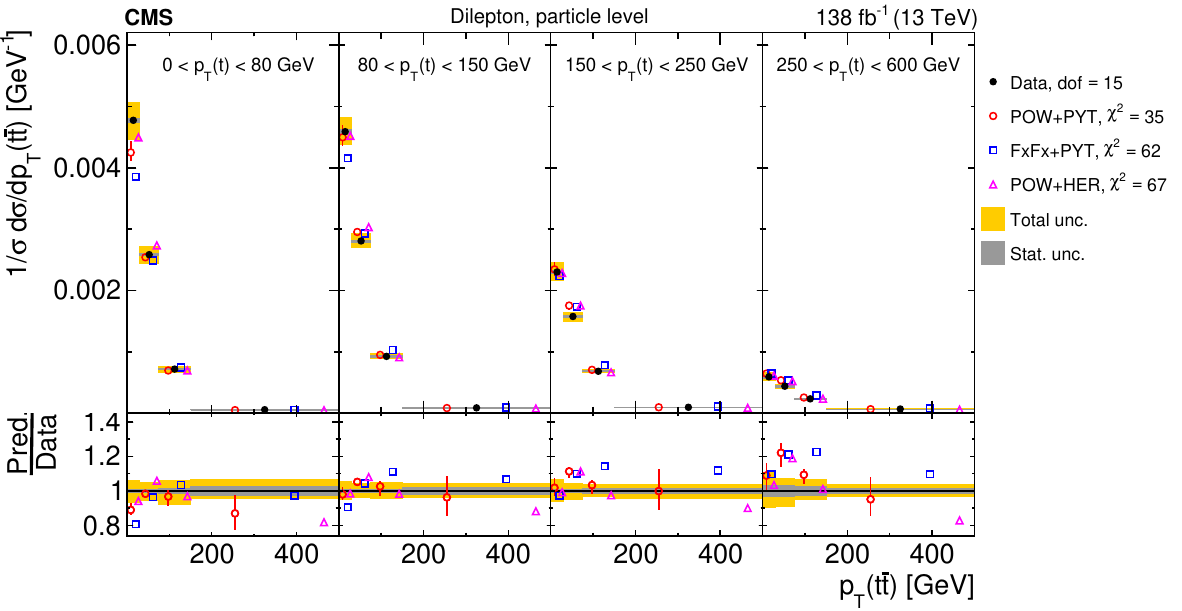}
\caption{Normalized \pttpttt cross sections are shown for data (filled circles) and various MC predictions
(other points).
    Further details can be found in the caption of Fig.~\ref{fig:res_ytptt}.}
    \label{fig:res_pttpttt}
\end{figure}

\begin{figure}
\centering
\includegraphics[width=0.99\textwidth]{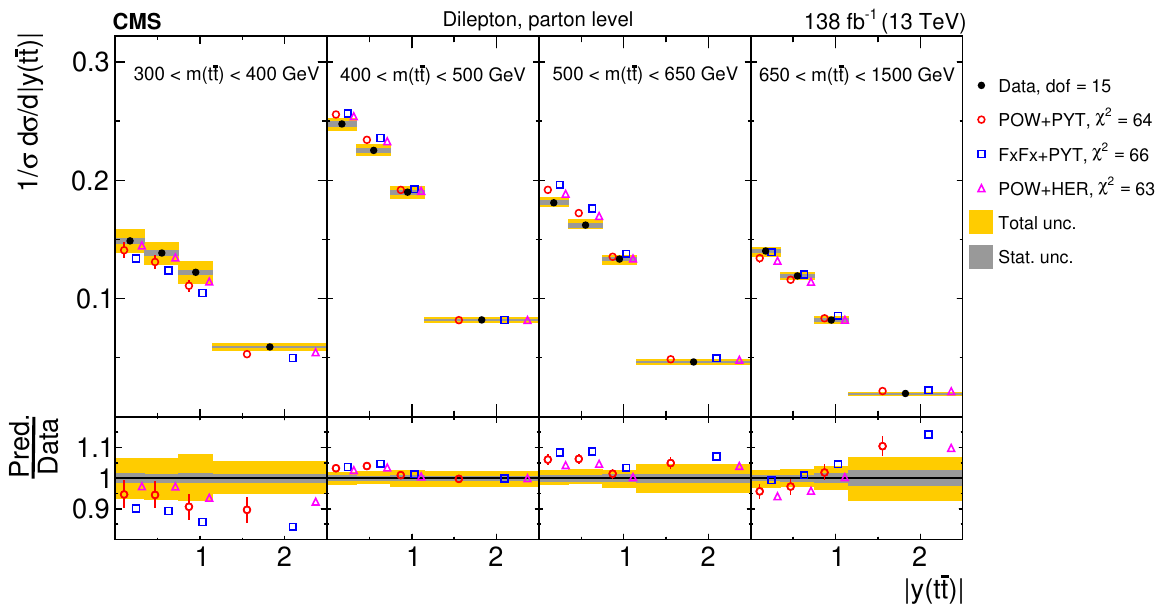}
\includegraphics[width=0.99\textwidth]{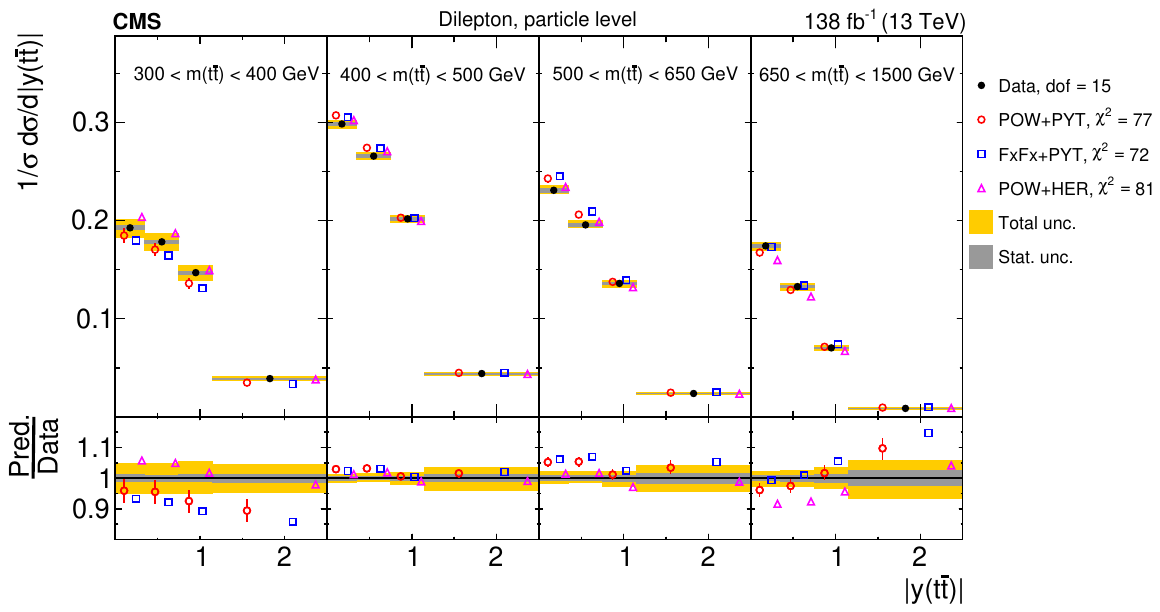}
\caption{Normalized \mttytt cross sections are shown for data (filled circles) and various MC predictions 
(other points).
    Further details can be found in the caption of Fig.~\ref{fig:res_ytptt}.}
    \label{fig:res_mttytt}
\end{figure}

\begin{figure}
\centering
\includegraphics[width=0.99\textwidth]{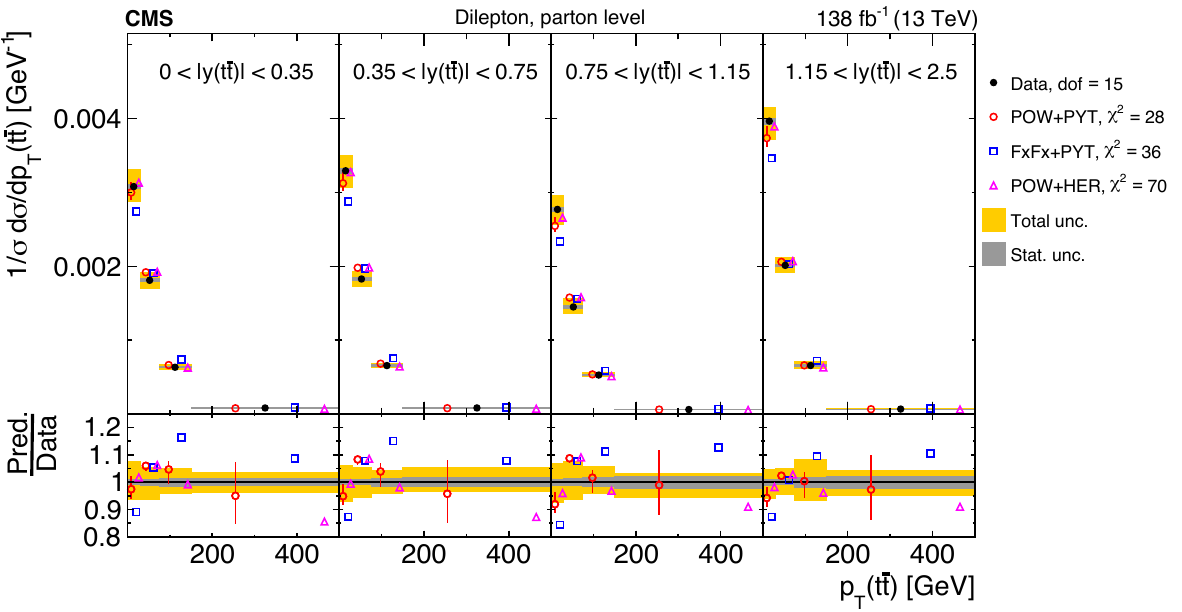}
\includegraphics[width=0.99\textwidth]{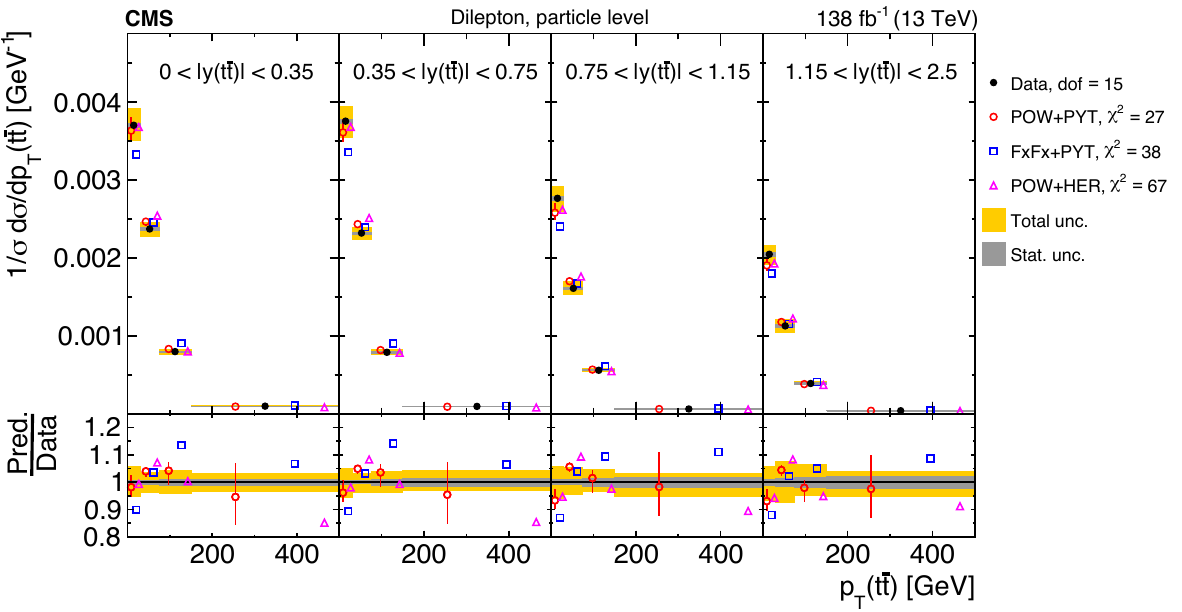}
\caption{Normalized \yttpttt cross sections are shown for data (filled circles) and various MC predictions
(other points).
    Further details can be found in the caption of Fig.~\ref{fig:res_ytptt}.}
    \label{fig:res_yttpttt}
\end{figure}

\begin{figure}
\centering
\includegraphics[width=0.99\textwidth]{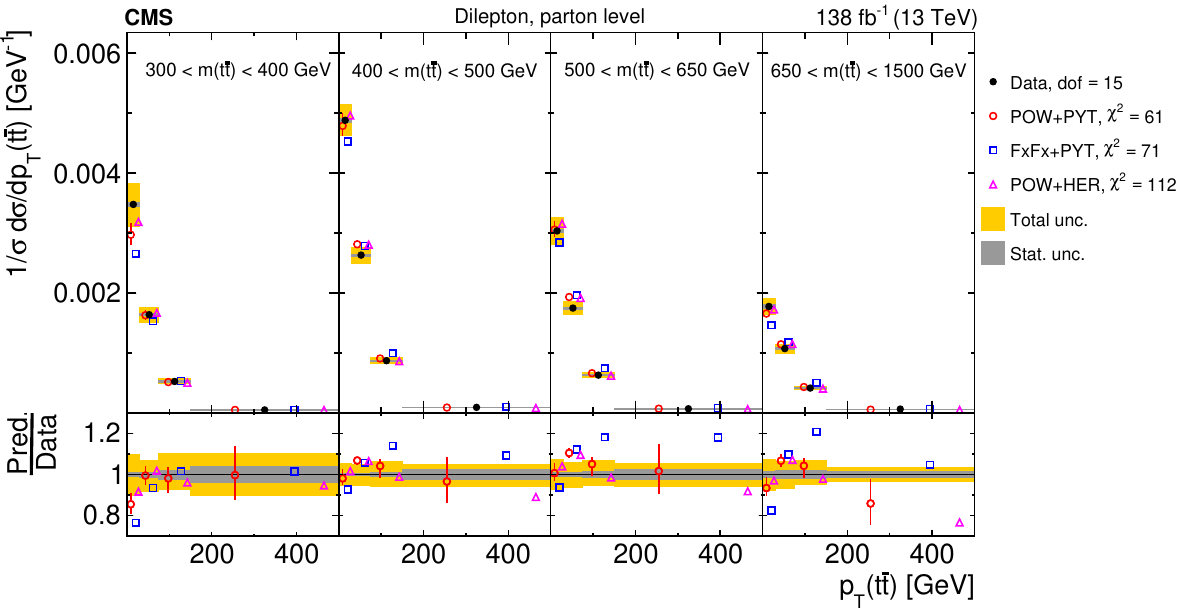}
\includegraphics[width=0.99\textwidth]{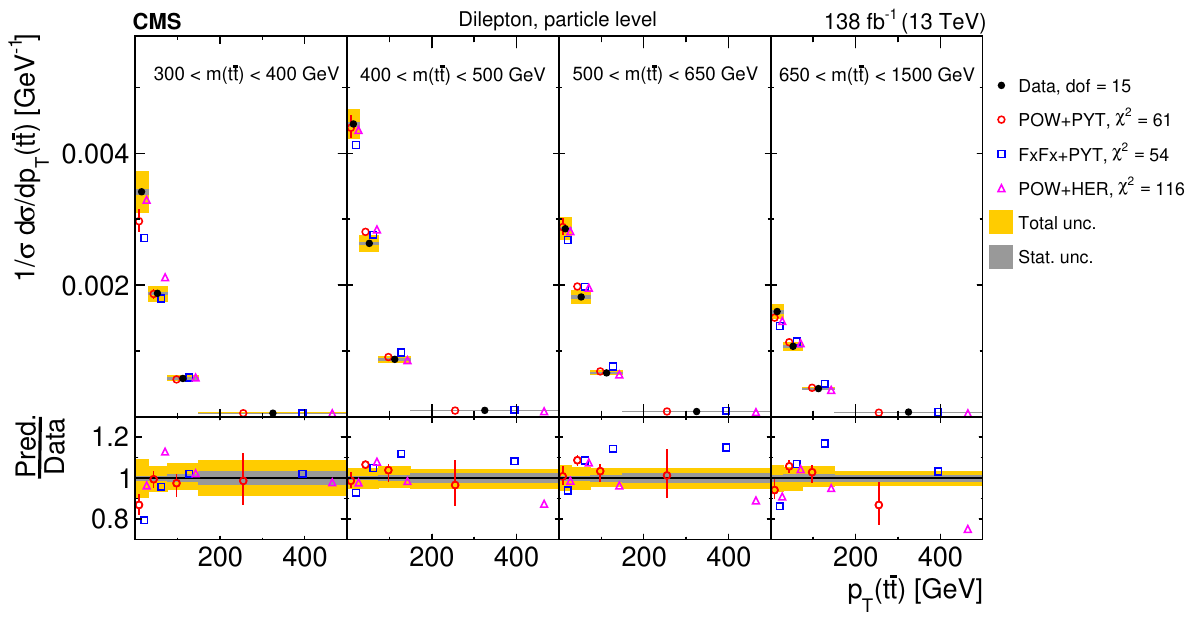}
\caption{Normalized \mttpttt cross sections are shown for data (filled circles) and various MC predictions
(other points).
    Further details can be found in the caption of Fig.~\ref{fig:res_ytptt}.}
    \label{fig:res_mttpttt}
\end{figure}

\begin{figure}
\centering
\includegraphics[width=0.99\textwidth]{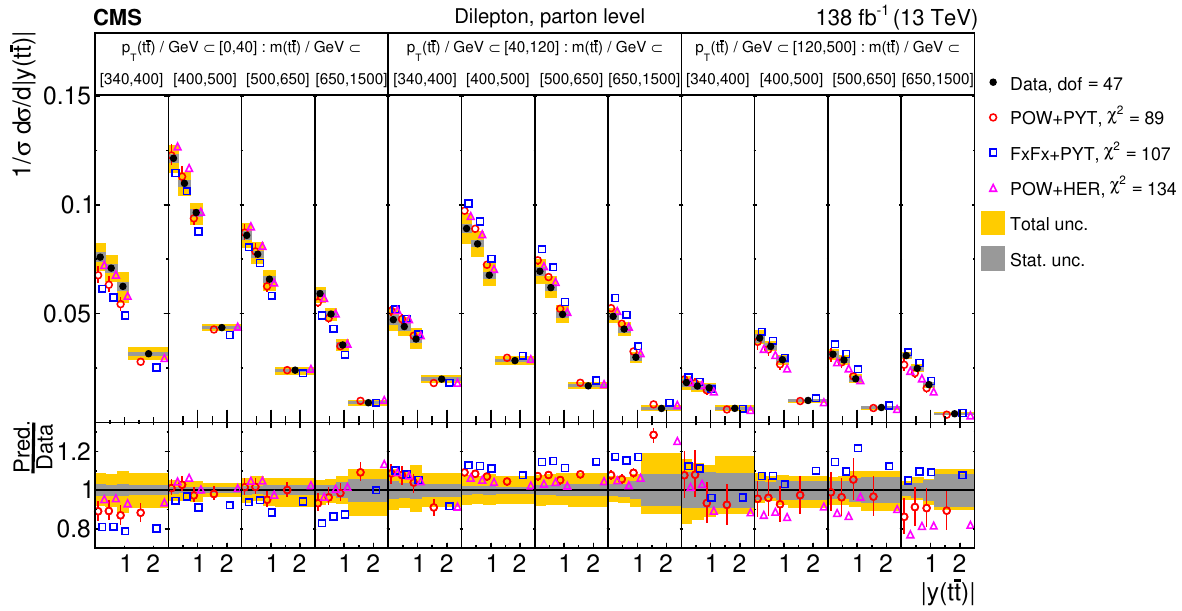}
\includegraphics[width=0.99\textwidth]{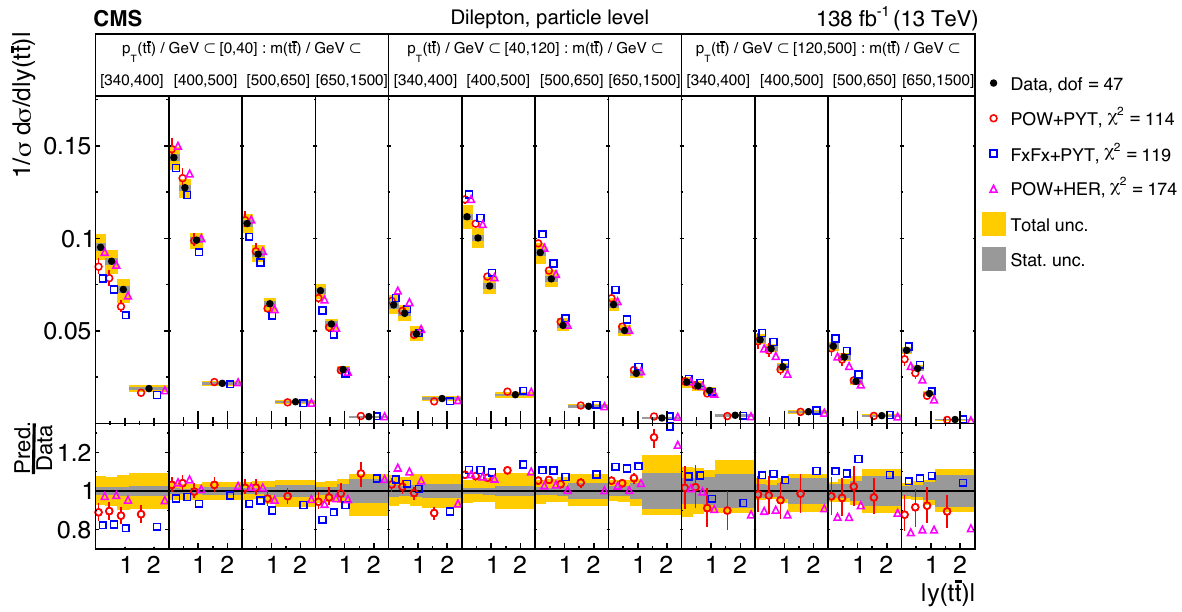}
\caption{Normalized \ptttmttytt cross sections are shown for data (filled circles) and various MC predictions
(other points).
     Further details can be found in the caption of Fig.~\ref{fig:res_ytptt}.}
    \label{fig:res_ptttmttytt}
\end{figure}

\begin{figure}
\centering
\includegraphics[width=0.99\textwidth]{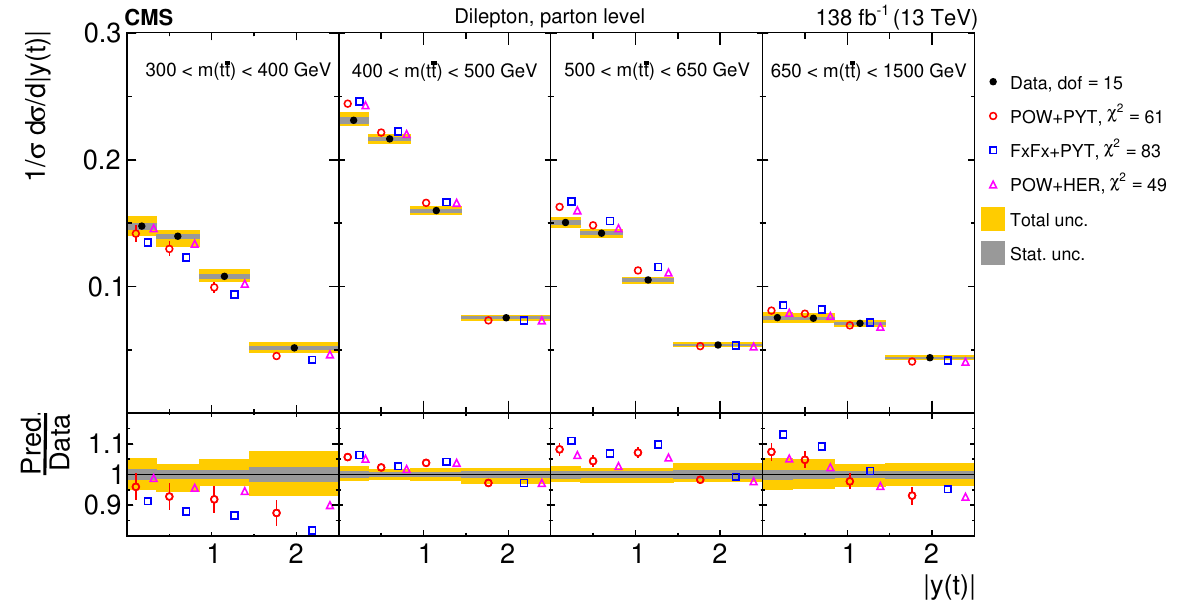}
\includegraphics[width=0.99\textwidth]{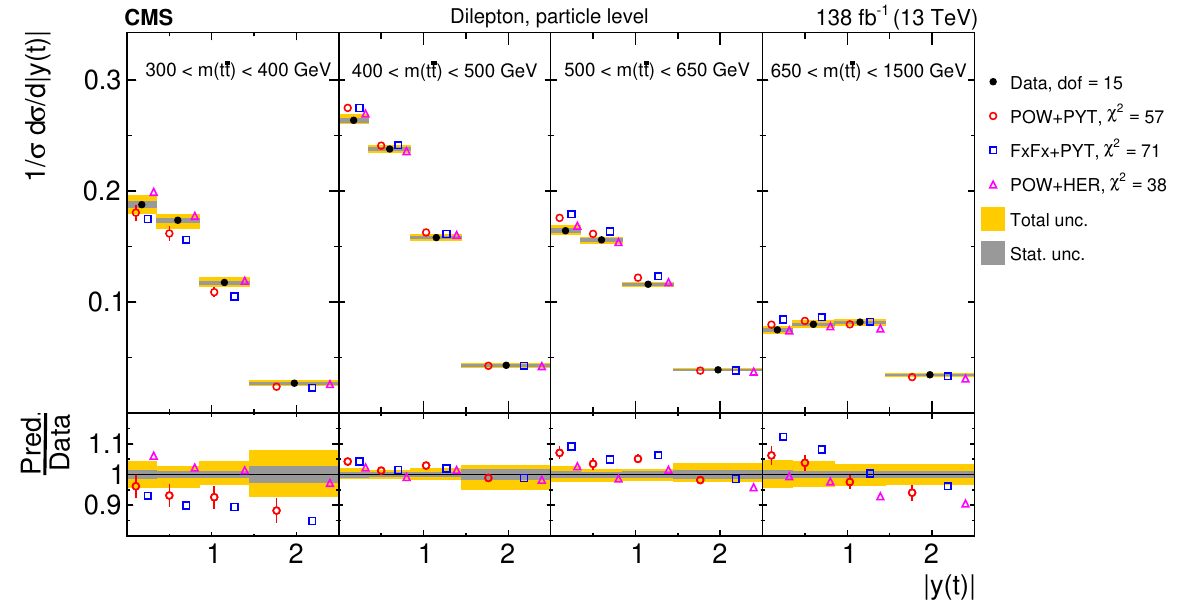}
\caption{Normalized \mttyt cross sections are shown for data (filled circles) and various MC predictions
(other points).
    Further details can be found in the caption of Fig.~\ref{fig:res_ytptt}.}
    \label{fig:res_mttyt}
\end{figure}

\begin{figure}
\centering
\includegraphics[width=0.99\textwidth]{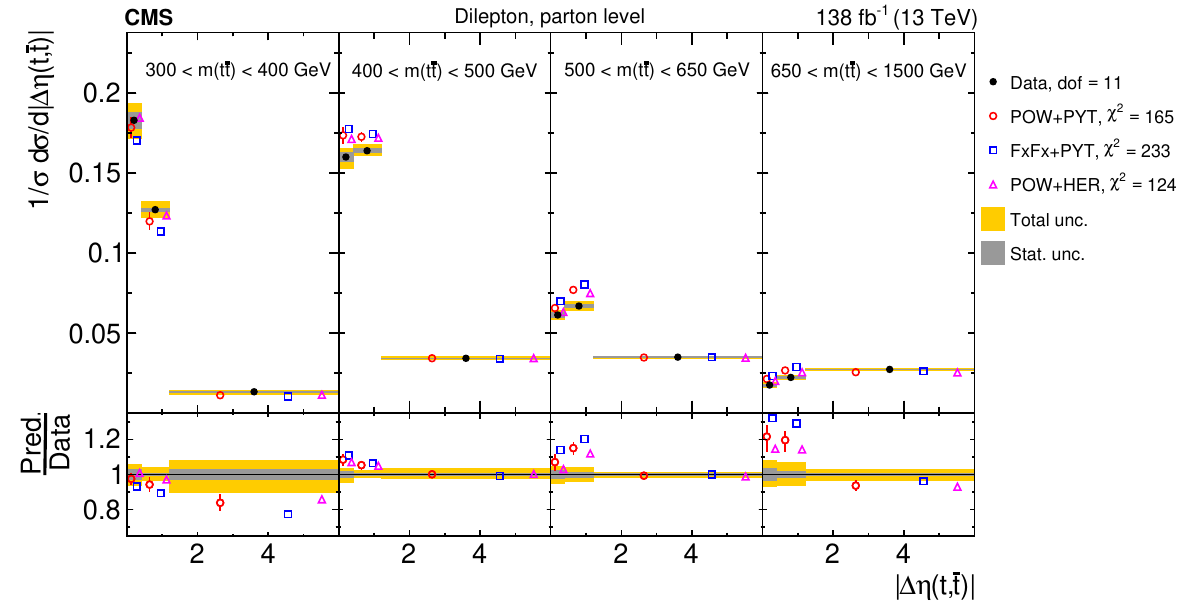}
\includegraphics[width=0.99\textwidth]{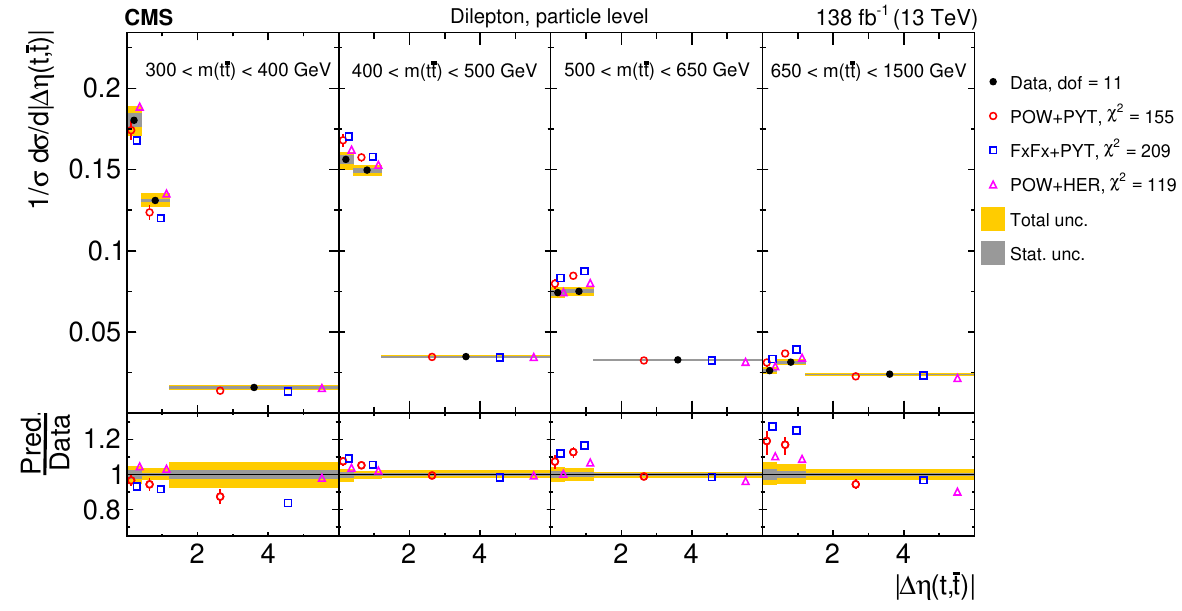}
\caption{Normalized \mttdetatt cross sections are shown for data (filled circles) and various MC predictions
(other points).
    Further details can be found in the caption of Fig.~\ref{fig:res_ytptt}.}
    \label{fig:res_mttdetatt}
\end{figure}

\begin{figure}
\centering
\includegraphics[width=0.99\textwidth]{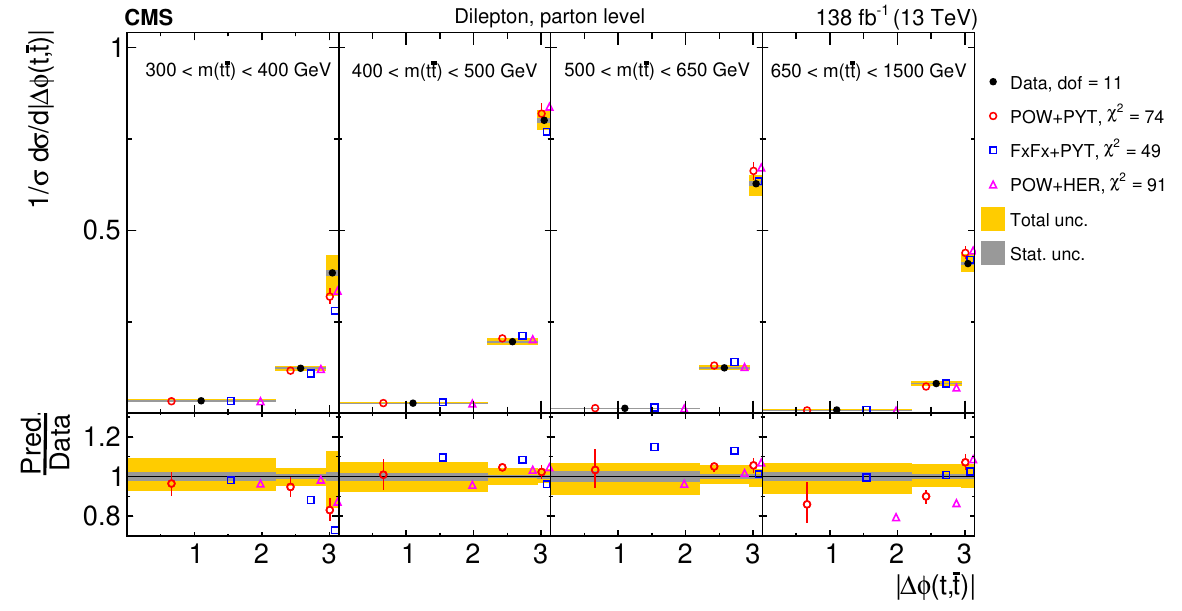}
\includegraphics[width=0.99\textwidth]{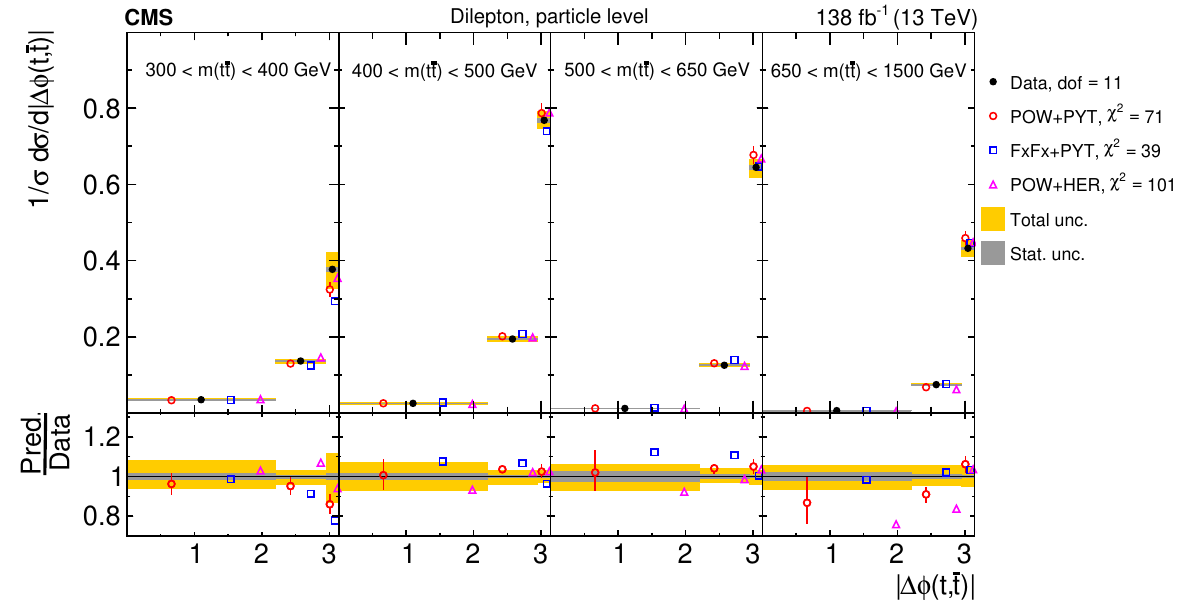}
\caption{Normalized \mttdphitt cross sections are shown for data (filled circles) and various MC predictions
(other points).
    Further details can be found in the caption of Fig.~\ref{fig:res_ytptt}.}
    \label{fig:res_mttdphitt}
\end{figure}

Studies of differential \ttbar cross sections performed
as functions of several kinematic observables shed light on the details of the
\ttbar production dynamics and can help to better understand the origin of
model-to-data discrepancies seen in single-differential cross sections.
The double-differential measurements performed in this analysis, \eg as functions of \absyt and \ptt,
are denoted in the following as \ytptt, \etc,
and an analogous notation is adopted for triple-differential cross sections.
The cross section results are
shown in Figs.~\ref{fig:res_ytptt}--\ref{fig:res_mttdphitt}.
The upper and lower plots show the cross sections at the parton and particle levels, respectively.
The \chisq values of model-to-data comparisons are listed
in Tables~\ref{tab:chi2mc_nor_parton_ttbar}--\ref{tab:chi2mc_nor_particle_ttbar}, and the corresponding $p$-values in Tables~\ref{tab:pvaluemc_nor_parton_ttbar}--\ref{tab:pvaluemc_nor_particle_ttbar}.

In the first set of studies,
we investigate how the \ptt distribution is correlated
with other event kinematic observables.
In Fig.~\ref{fig:res_ytptt}, the \ptt distribution is compared in
different ranges of \absyt to predictions from the
same three MC models discussed above.
The data distribution is softer than that of the predictions
over the entire \yt range.
The \PowHerSh prediction that gave a reasonable description
of the single-differential \ptt cross section (Fig.~\ref{fig:res_ptt}),
still provides the best description, but exhibits a stronger positive \ptt
slope with respect to the data in the lowest and highest \absyt ranges.
Figure~\ref{fig:res_mttptt} shows the \ptt distributions
in different \mtt ranges.
Similar to the result of the previous analysis~\cite{Sirunyan:2019zvx},
this is among the double-differential cross sections poorly described by the models.
While the \PowHerSh model describes the \ptt distribution
in the lowest \mtt range near threshold reasonably well, it joins
\PowPytSh and \aMCPytSh in a trend that grows with increasing \mtt
to predict a \ptt spectrum that is harder than observed
in the data.
Figure~\ref{fig:res_pttpttt} illustrates the \pttt spectrum in different \ptt ranges.
Larger \ptt values can be kinematically correlated
with higher \pttt values when the \ttbar system is recoiling against additional QCD radiation in the event.
The \aMCPytSh model predicts a harder \pttt spectrum
than observed in the data for all \ptt ranges.
The \PowPytSh and \PowHerSh predictions tend to overshoot
the data in the higher \ptt ranges at the lower \pttt values.

Next we investigate the
\ttbar kinematic observables.
Figure~\ref{fig:res_mttytt}
illustrates the \mttytt distributions.
Both variables are kinematically correlated
with the observables $\xi_{1}$ and $\xi_{2}$ introduced above,
and their combination is known to provide optimal information
for constraining the PDFs~\cite{Sirunyan:2017azo}.
For low- and medium-\mtt regions, the predictions are slightly more central than the data,
though in the highest \mtt range the opposite effect is observed.
In Fig.~\ref{fig:res_yttpttt}, the spectrum of \pttt is shown in bins of
\absytt. These two observables are rather uncorrelated, and
thus it is not surprising that the description
of the \pttt distribution by the models is similar in all \absytt ranges.
Figure~\ref{fig:res_mttpttt} presents the \mttpttt distributions.
This is an interesting observable combination since the phase space
for QCD radiation that is driving the \pttt observable is increasing with higher \mtt.
The trends observed in the single-differential \pttt distribution 
(see Fig.~\ref{fig:res_pttt}), namely, that \aMCPytSh
(\PowHerSh) predicts a too hard (soft) spectrum, are somewhat enhanced in the higher
\mtt ranges.
Figure~\ref{fig:res_ptttmttytt} shows the first simultaneous study
of all three \ttbar kinematic observables: \pt, mass,
and rapidity. Overall, \PowPytSh provides a fairly reasonable description, though
\aMCPytSh and \PowHerSh show deficiencies in specific
\mtt and \pttt phase space regions, with mostly small or moderate dependencies on \ytt.

Finally, we perform several studies of \mtt, investigating
its correlation to several other kinematic observables.
Figure~\ref{fig:res_mttyt} illustrates the distributions of \absyt.
The trend that the predictions exhibit
a more-central rapidity distribution than the data increases slightly with higher \mtt.
In Fig.~\ref{fig:res_mttdetatt}, the \absdetatt distributions are shown, which are 
sensitive to the hard scattering dynamics.
The data favor larger rapidity separations than predicted by the models,
and the significance of this effect increases at larger \mtt.
The disagreement is the strongest for \aMCPytSh.
For a given \mtt value, larger \absdetatt are correlated with lower \ptt values on average.
This gives a hint that the effects of the models predicting harder \ptt spectra and smaller
\absdetatt distributions,
which are enhanced in the higher \mtt ranges (Figs.~\ref{fig:res_mttptt} and
~\ref{fig:res_mttdetatt}), are related.
Figure~\ref{fig:res_mttdphitt} depicts the \mttdphitt distributions.
At low \mtt, the data prefer a slightly more back-to-back distribution
of the top quark and antiquark compared to the models, but in the highest
\mtt range, the trend reverses.
Of all models considered, the \aMCPytSh simulation provides the best description and \PowHerSh the worst.

The observations made with the multi-differential cross sections
can be summarized as follows.
The \pt of the top quark and \ttbar, and the \ttbar invariant mass
are, in general, mildly correlated with rapidity of the same objects,
and also the quality of their description by the MC models is nearly independent
of the rapidity.
As expected, larger kinematical correlations are observed
between \pt and the mass observables.
The trends of harder top quark \pt spectra and smaller rapidity separations
between top quark and antiquark, when comparing the models to the data, is clearly
enhanced at higher \mtt.
The tendency for \aMCPytSh (\PowHerSh) to predict \pt spectra for the \ttbar system that are too hard (soft)
is also stronger at higher \mtt.
In general, the $p$-values associated 
with the standard \chisq values of the model-to-data comparisons
are often much lower for the multi-differential \ttbar cross sections
than for the single-differential results presented in the previous subsection.
This result is in line with the observations in recent comparable \ttbar differential
cross section papers from the ATLAS and CMS Collaborations~\cite{Sirunyan:2018wem,Aad:2019ntk}.
Another interesting observation is that the \chisq values for both the single- and multi-differential 
\ttbar
cross sections are, on average, significantly higher than those observed in the corresponding
previous measurements~\cite{Sirunyan:2018ucr,Sirunyan:2019zvx} based on
the 2016 data set only, which can be attributed to a substantially improved
measurement precision. The \chisq values that include the prediction uncertainties for the 
\PowPytSh model (see Tables~\ref{tab:chi2mc_nor_parton_ttbar}--\ref{tab:chi2mc_nor_particle_ttbar})
are, in general, significantly lower than the standard ones. However, for several 
distributions such as \mttdphitt, the values remain too high to indicate a good description of the 
data.

\clearpage

\begin{table*}
 \centering
 \topcaption{The \chisq values and \ndf of the measured normalized multi-differential cross sections for \ttbar and top quark kinematic observables at the parton level are shown with respect to the predictions of various MC generators. 
The \chisq values are calculated taking only measurement uncertainties into account and excluding theory uncertainties.  
For \PowPytSh, the \chisq values including theory uncertainties are indicated with the brackets (w. unc.).}
 \label{tab:chi2mc_nor_parton_ttbar}
 \renewcommand{\arraystretch}{1.4}
 \centering
 \begin{tabular}{lccccc}
 \multirow{1}{*}{Cross section} & \hspace*{0.3 cm} \multirow{2}{*}{\ndf} \hspace*{0.3 cm} & \multicolumn{3}{c}{\chisq} \\
 \cline{3-5}
{variables} && \PowPytSh (w. unc.)  & \aMCPytSh  & \PowHerSh   \\
\hline
\ytptt& 15 & 40 \: (31) & 71 & 30 \\
\mttptt& 8 & 83 \: (35) & 152 & 41 \\
\pttpttt& 15 & 39 \: (21) & 69 & 83 \\
\mttytt  & 15 & 64 \: (42) & 66 & 63 \\
\yttpttt& 15 & 28 \: (15) & 36 & 70 \\
\mttpttt  & 15 & 61 \: (43) & 71 & 112 \\
\ptttmttytt  & 47 & 89 \: (64) & 107 & 134 \\
\mttyt& 15 & 61 \: (37) & 83 & 49 \\
\mttdetatt& 11 & 165 \: (31) & 233 & 124 \\
\mttdphitt& 11 & 74 \: (47) & 49 & 91 \\
 \end{tabular}
\end{table*}

\begin{table*}
 \centering
 \topcaption{The \chisq values and \ndf of the measured normalized multi-differential cross sections for \ttbar and top quark kinematic observables at the particle level are shown with respect to the predictions of various MC generators. 
The \chisq values are calculated taking only measurement uncertainties into account and excluding theory uncertainties.  
For \PowPytSh, the \chisq values including theory uncertainties are indicated with the brackets (w. unc.).}
 \label{tab:chi2mc_nor_particle_ttbar}
 \renewcommand{\arraystretch}{1.4}
 \centering
 \begin{tabular}{lccccc}
 \multirow{1}{*}{Cross section} & \hspace*{0.3 cm} \multirow{2}{*}{\ndf} \hspace*{0.3 cm} & \multicolumn{3}{c}{\chisq} \\
 \cline{3-5}
{variables} && \PowPytSh (w. unc.)  & \aMCPytSh  & \PowHerSh   \\
\hline
\ytptt& 15 & 35 \: (25) & 61 & 28 \\
\mttptt& 8 & 86 \: (36) & 142 & 47 \\
\pttpttt& 15 & 35 \: (19) & 62 & 67 \\
\mttytt  & 15 & 77 \: (40) & 72 & 81 \\
\yttpttt  & 15 & 27 \: (18) & 38 & 67 \\
\mttpttt   & 15 & 61 \: (36) & 54 & 116 \\
\ptttmttytt  & 47 & 114 \: (68) & 119 & 174 \\
\mttyt& 15 & 57 \: (26) & 71 & 38 \\
\mttdetatt& 11 & 155 \: (30) & 209 & 119 \\
\mttdphitt& 11 & 71 \: (42) & 39 & 101 \\
 \end{tabular}
\end{table*}

\clearpage

\subsection{Results for lepton and \texorpdfstring{\PQb jet kinematic variables at the particle level}{}}
\label{sec:res_2}
\begin{figure*}[!phtb]
\centering
\includegraphics[width=0.49\textwidth]{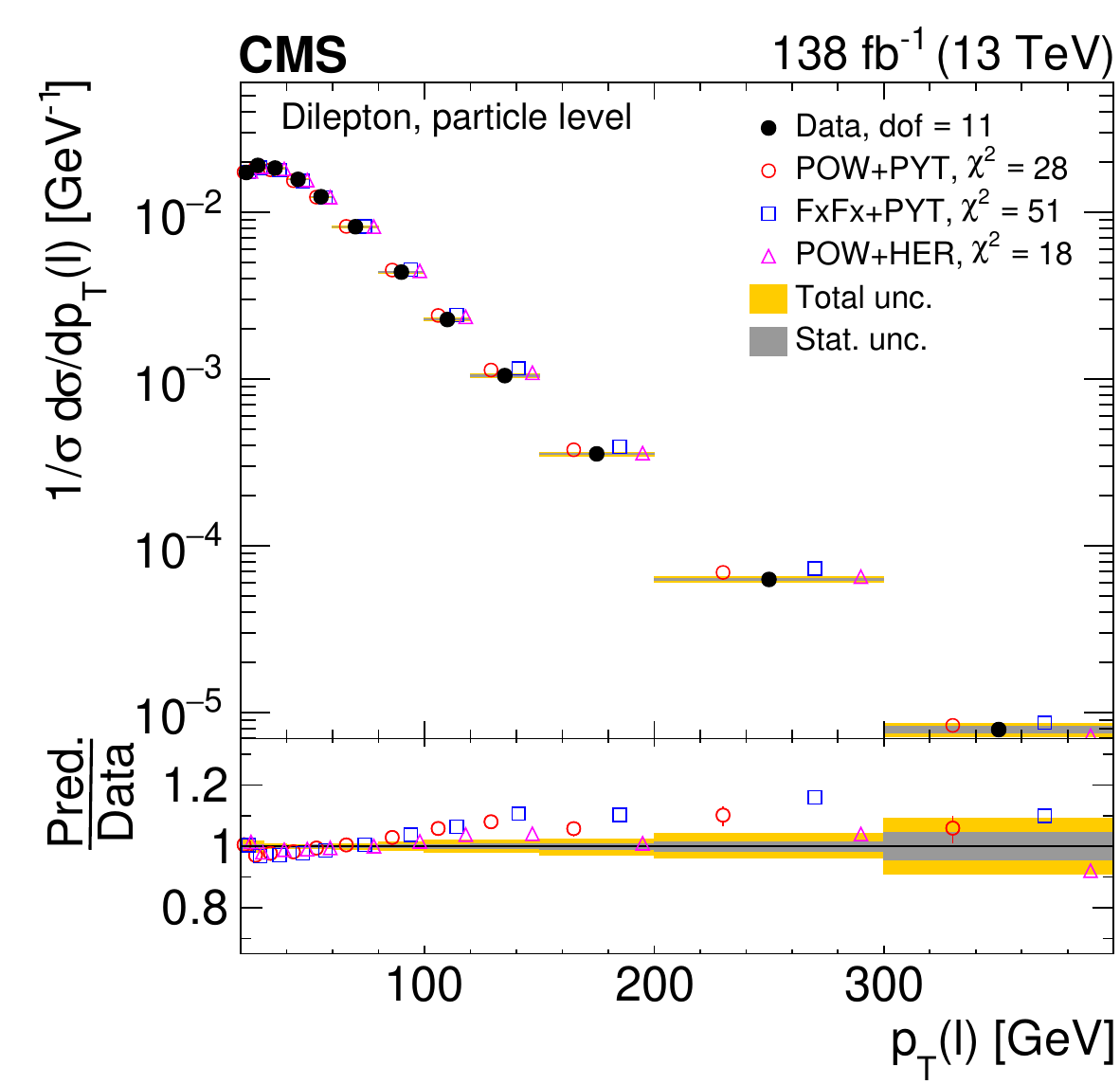}
\includegraphics[width=0.49\textwidth]{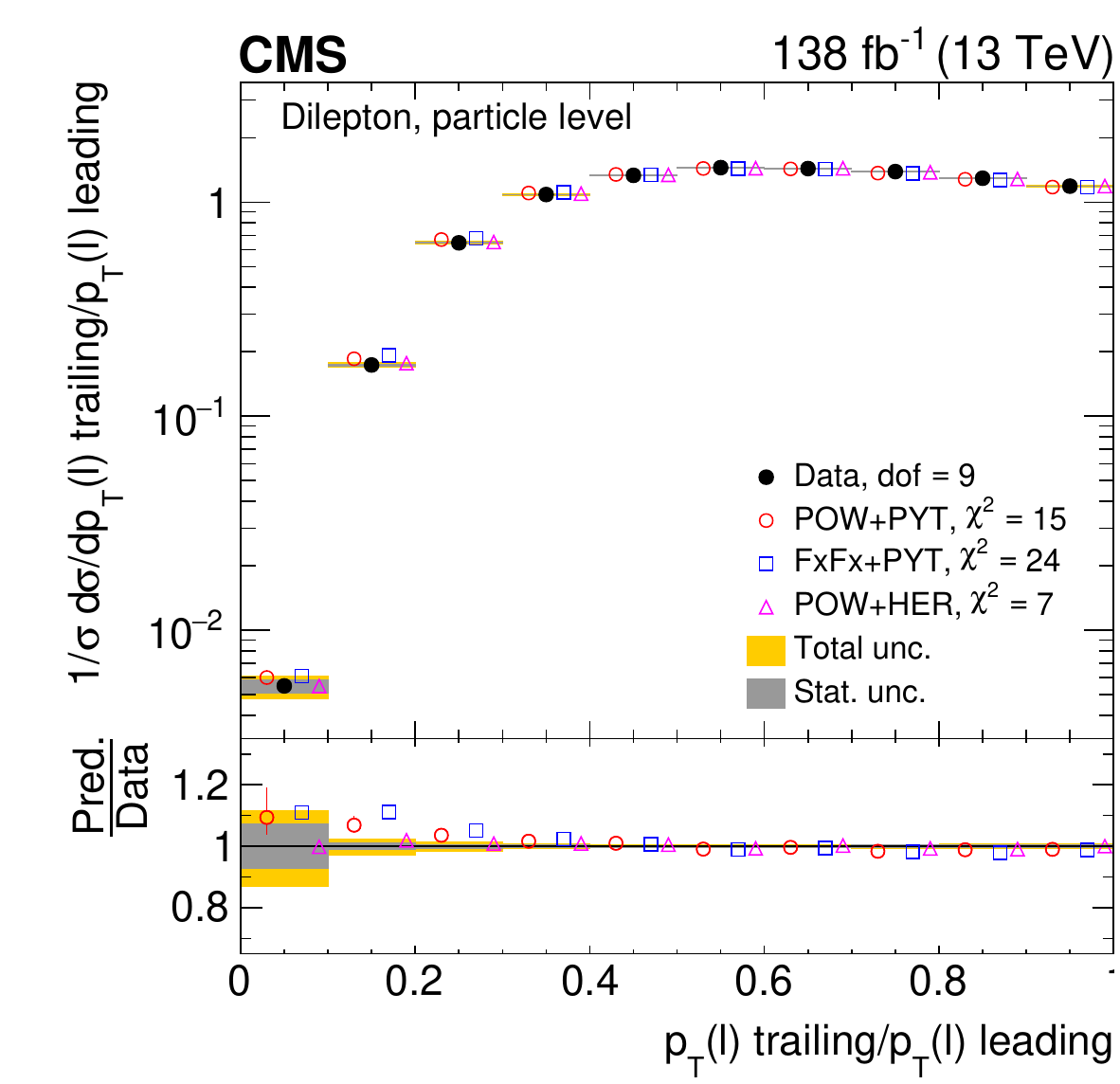}
\includegraphics[width=0.49\textwidth]{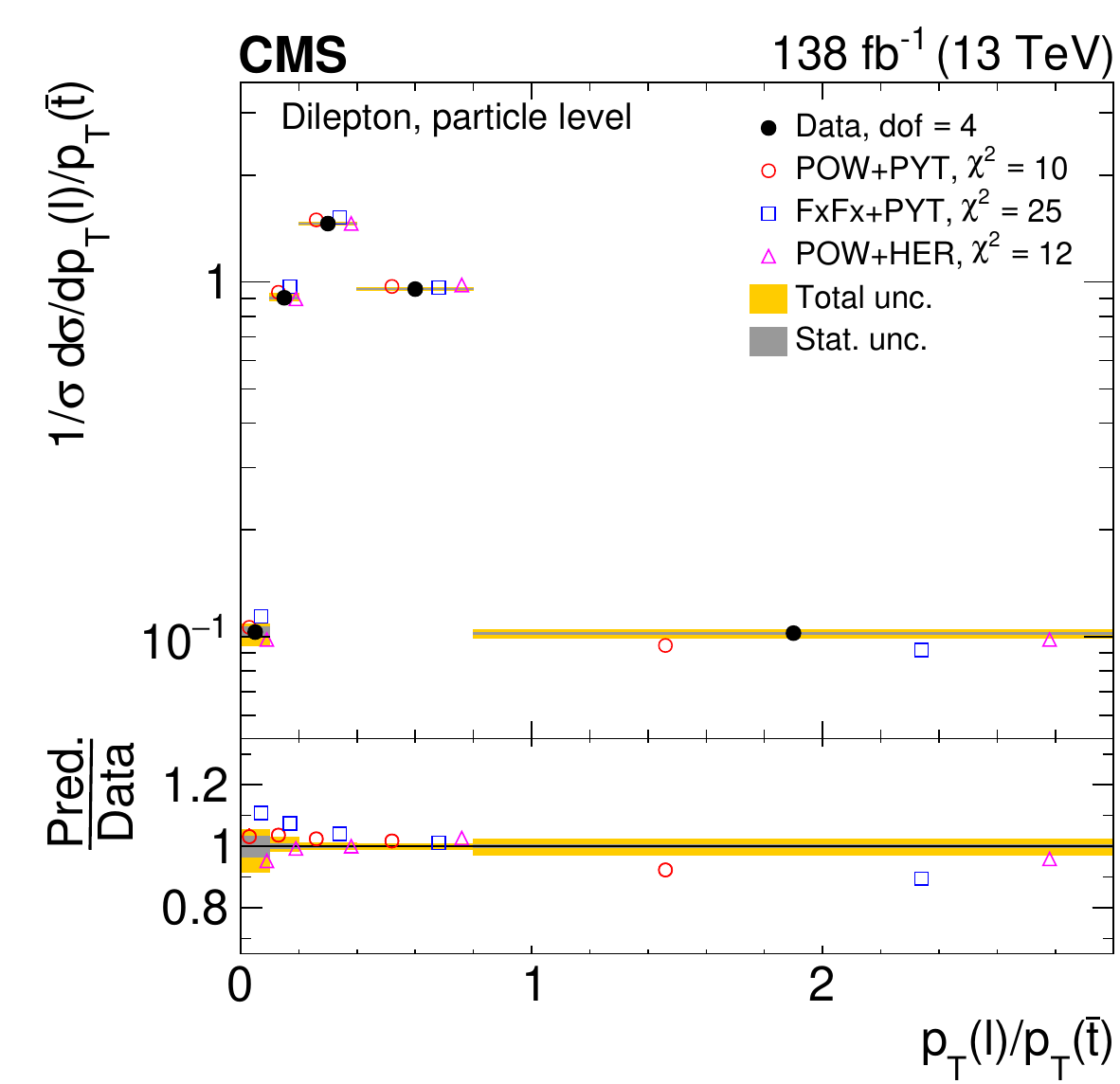}
\caption{Normalized differential \ttbar production cross sections as functions of \pt of the lepton (upper left),
of the ratio of the trailing and leading lepton \pt
(upper right), and of the ratio of lepton and top antiquark \pt (lower),
measured at the particle level in a fiducial phase space.
The data are shown as filled circles with grey and yellow bands indicating the statistical and total uncertainties
(statistical and systematic uncertainties added in quadrature), respectively.
For each distribution, the number of degrees of freedom (dof) is also provided.
The cross sections are compared to various MC predictions (other points).
The estimated uncertainties in the \PowPyt (`POW-PYT') simulation are represented by vertical bars on the
corresponding points.
For each MC model, a value of \chisq is reported that takes into account the measurement uncertainties.
The lower panel in each plot shows the ratios of the predictions to the data.}
\label{fig:res_ptlep}
\end{figure*}

\begin{figure*}[!phtb]
\centering
\includegraphics[width=0.49\textwidth]{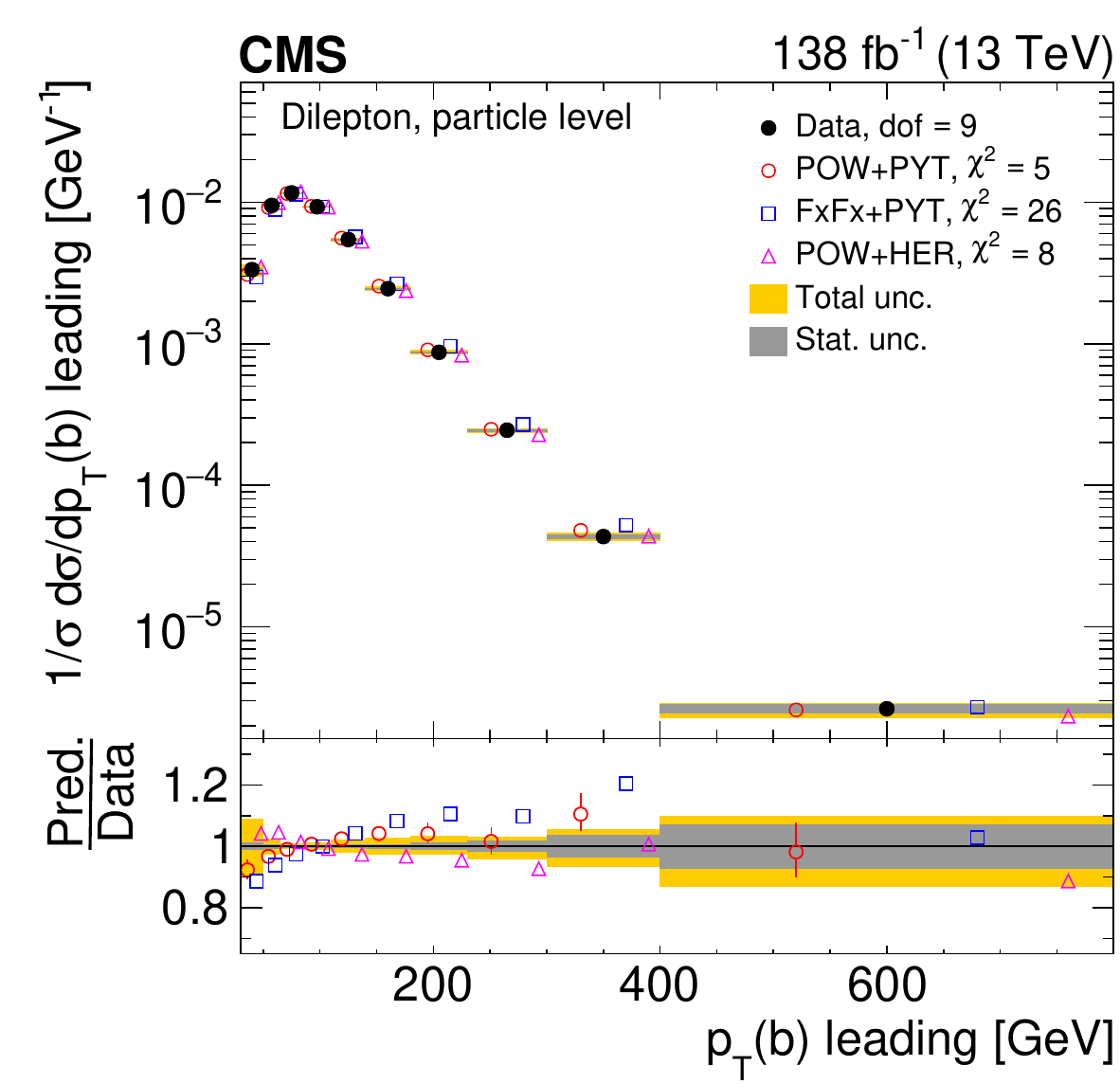}
\includegraphics[width=0.49\textwidth]{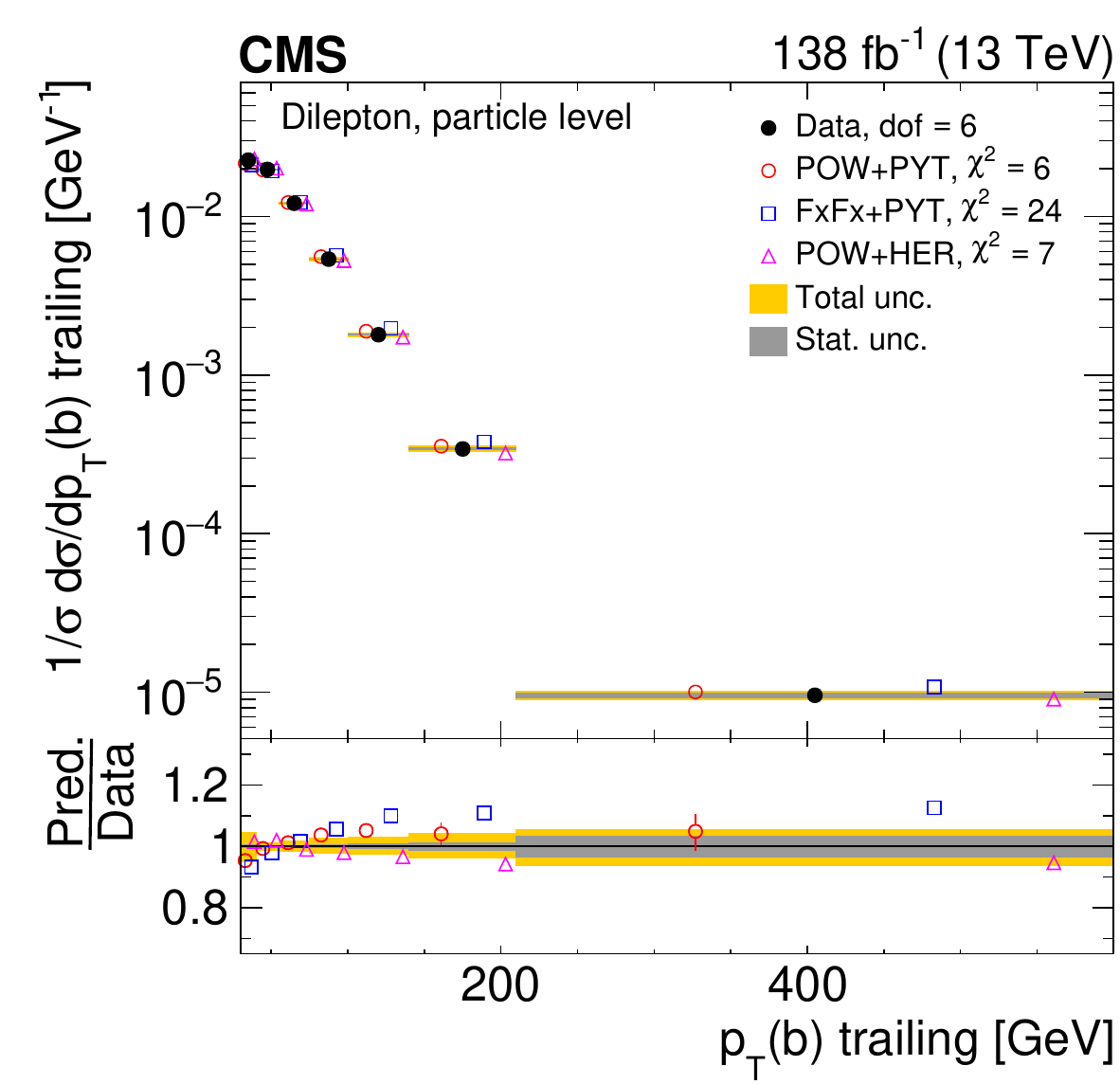}
\includegraphics[width=0.49\textwidth]{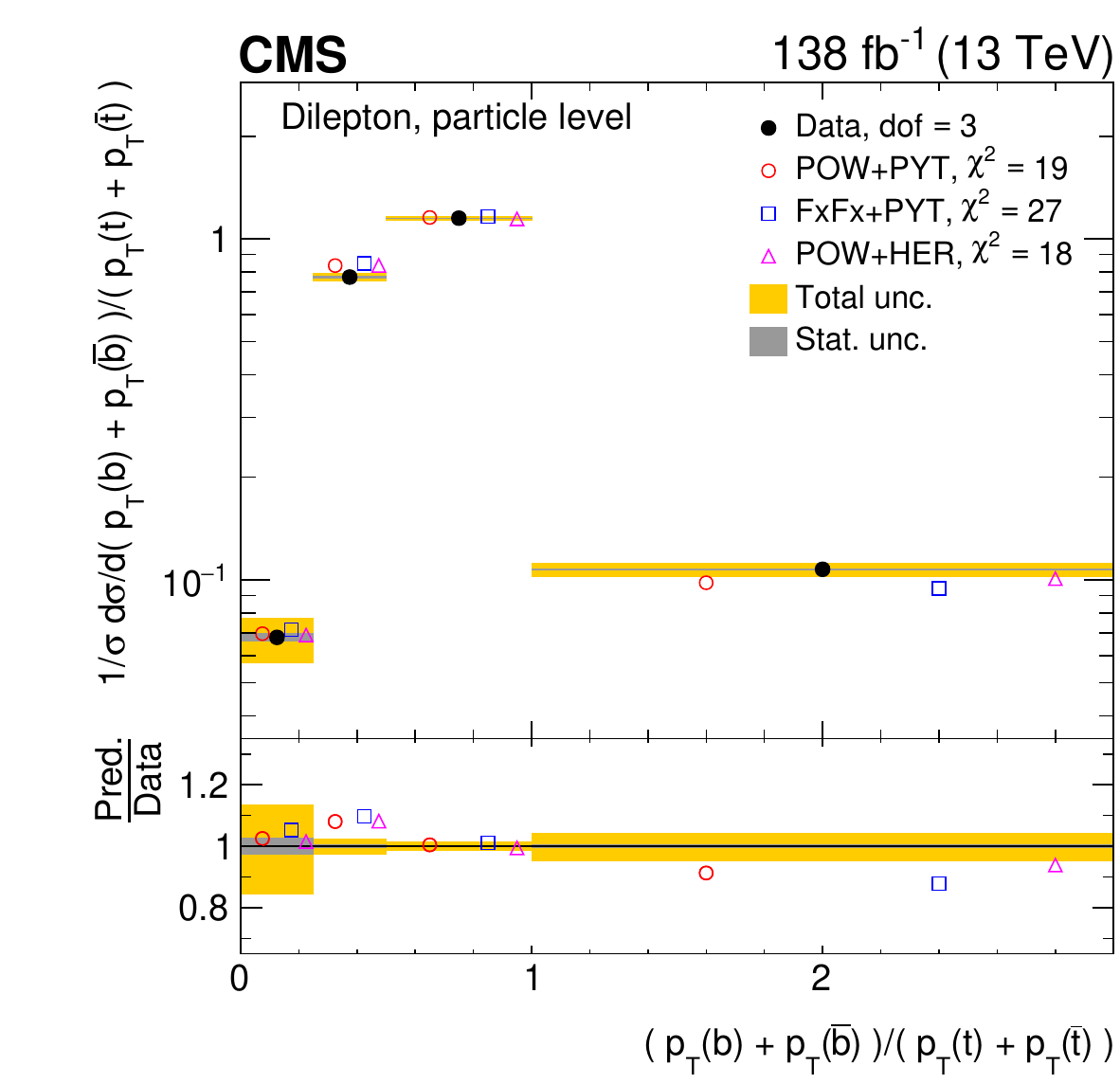}
\caption{Normalized differential \ttbar production cross sections as functions of the \pt of
the leading (upper left) and trailing (upper right) \PQb jet, and \rptbsptts (lower).
Further details can be found in the caption of Fig.~\ref{fig:res_ptlep}.}
\label{fig:res_ptb}
\end{figure*}

\begin{figure*}[!phtb]
\centering
\includegraphics[width=0.49\textwidth]{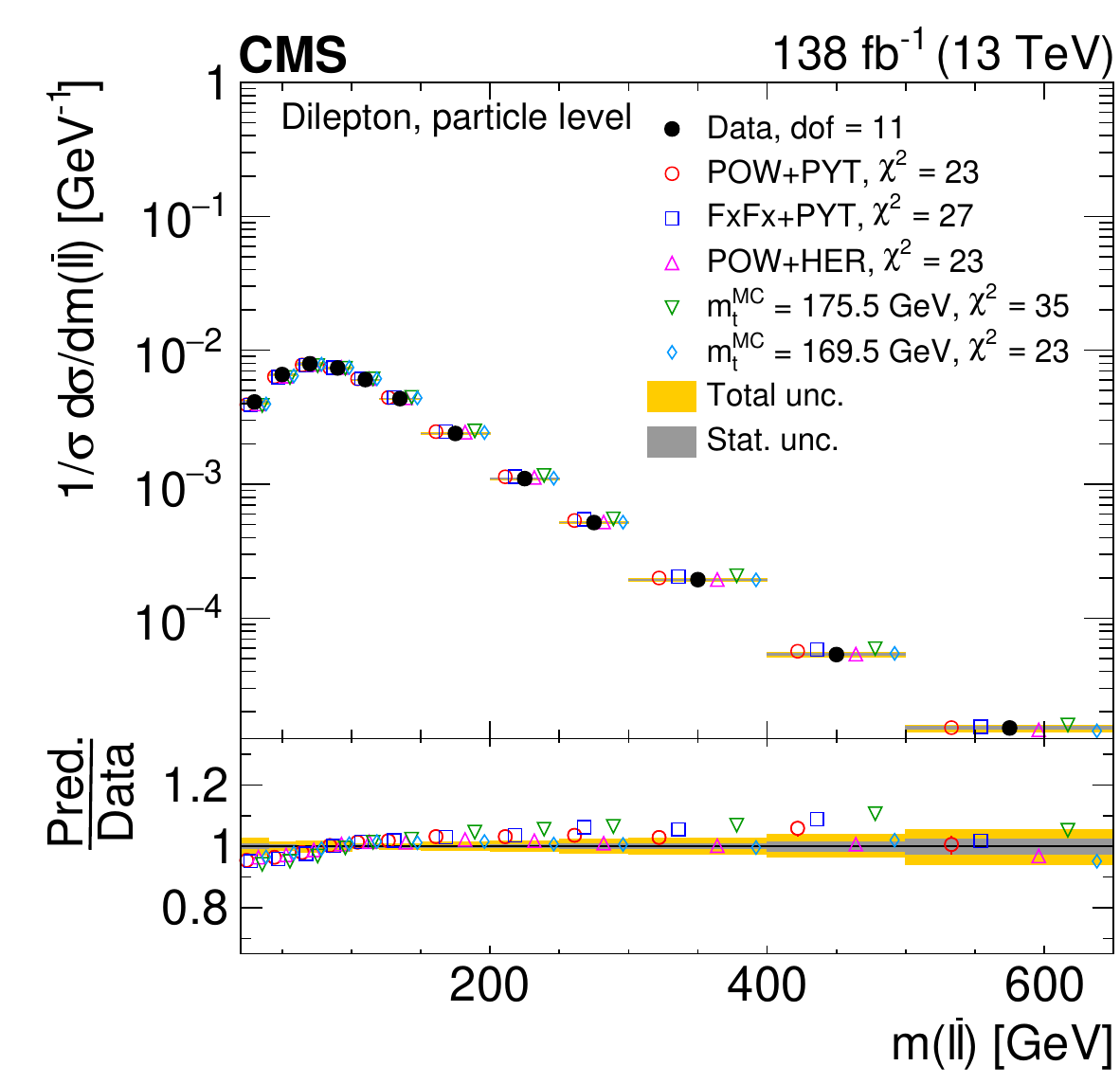}
\includegraphics[width=0.49\textwidth]{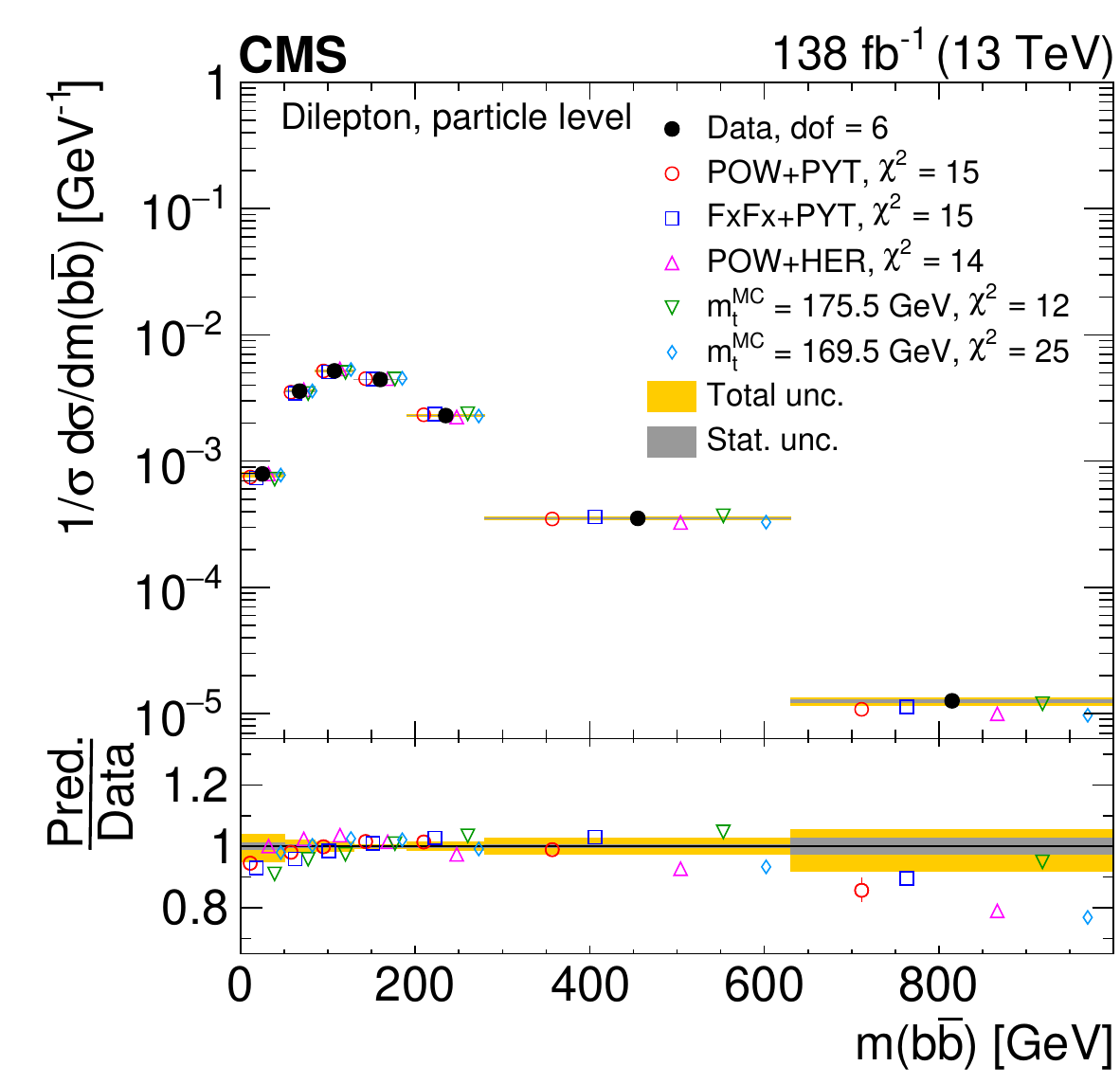}
\includegraphics[width=0.49\textwidth]{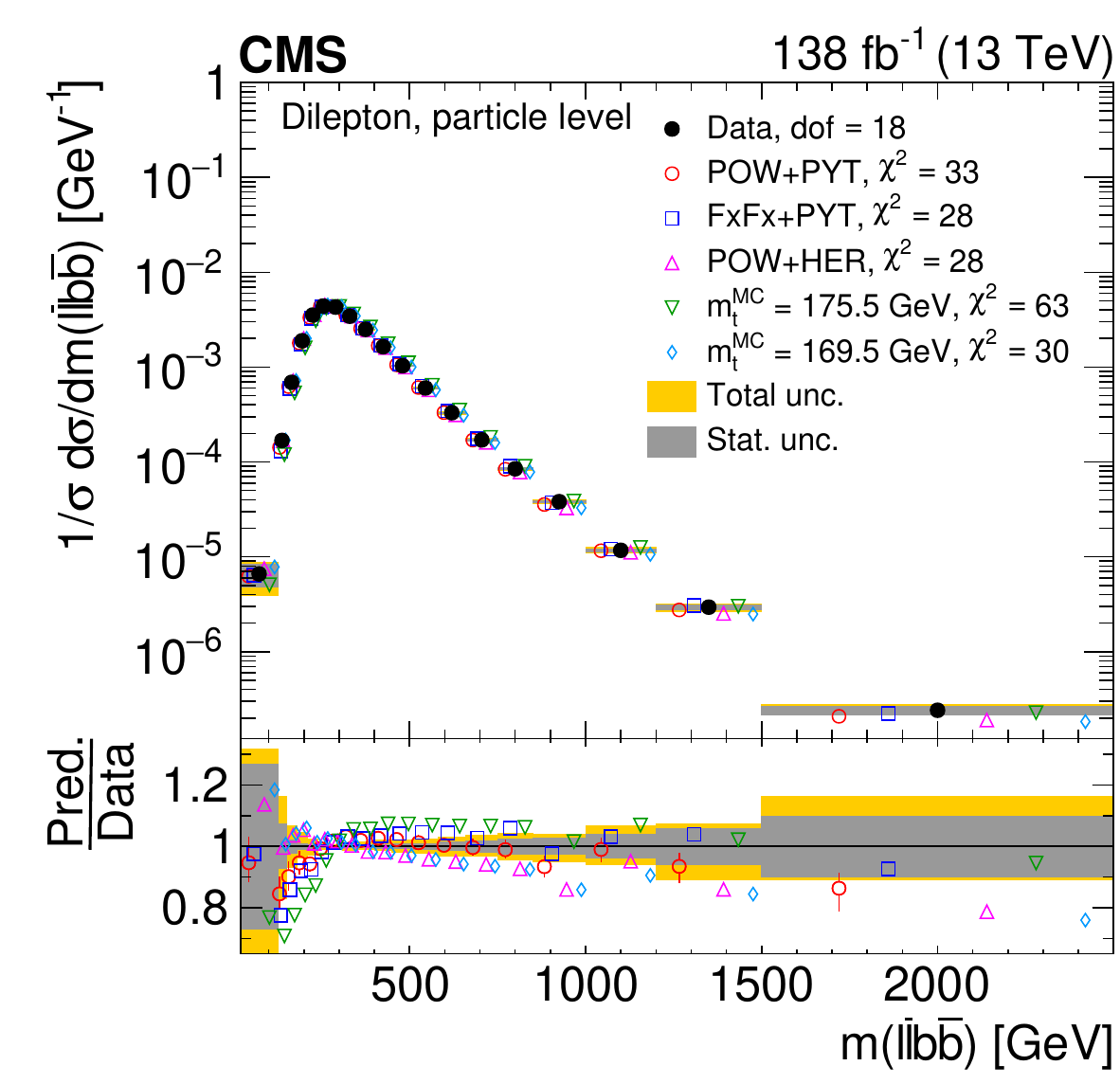}
\caption{Normalized differential \ttbar production cross sections as functions of \mll (upper left), \mbb
(upper right), and \mllbb (lower)
are shown for data (filled circles) and various MC predictions (other points).
The distributions are also compared to \PowPyt (`POW-PYT') simulations with different values of \mtmc.
Further details can be found in the caption of Fig.~\ref{fig:res_ptlep}.}
\label{fig:res_mll}
\end{figure*}

In this subsection we present
selected kinematic distributions of the leptons and \PQb jets
produced in the decays of the top quark and antiquark, at the particle level
in a fiducial phase space.
These distributions are sensitive to both the dynamics
of the \ttbar production and its decay.
The kinematic observables of these
objects are measured very precisely with the CMS detector.
The single-differential cross sections studied at the particle level are
shown in Figs.~\ref{fig:res_ptlep}--\ref{fig:res_mll}. The model-to-data \chisq and corresponding $p$-values are listed in Tables~\ref{tab:chi2mc_1d_nor_particle_lepb} and~\ref{tab:pvaluemc_1d_nor_particle_lepb}, respectively. 

First we investigate the lepton kinematic observables.
Figure~\ref{fig:res_ptlep} shows the distributions of the
\pt of the lepton (with negative charge), the ratio of the trailing and leading lepton \pt,
and the ratio of the lepton and top antiquark \pt.
As for the top antiquark (Fig.~\ref{fig:res_ptt}), the MC models predict
a harder \pt spectrum for the leptons
than observed, with \aMCPytSh exhibiting a stronger
discrepancy.
For the ratio of the trailing and leading lepton \pt,
the models predict distributions that are slightly too soft, with
\aMCPytSh again showing a more significant deviation than
\PowPytSh and \PowHerSh.
The distribution of the ratio of the lepton and top quark \pt
shows an interesting excess of data over the predictions
for ratios above 0.8, which is an indication of a failure of the models
to describe the dynamics of the top quark decay.

Next, we study three \PQb jet observables, as shown in Fig.~\ref{fig:res_ptb}.
The leading (trailing) \PQb jet is defined as the \PQb jet from the
decay of the \ttbar system with the higher (lower) \pt.
The distributions of the leading and trailing \PQb jet \pt
are reasonably well described by the \PowPytSh and \PowHerSh models,
while \aMCPytSh predicts a spectrum that is too hard.
The third studied variable is the ratio between the sum of the \pt of the \PQb and \PAQb jets over the sum
of the \pt of
the top quark and antiquark \rptbsptts.
All models predict a distribution that is somewhat too soft.

In the next set of studies, shown in Fig.~\ref{fig:res_mll},
we analyze the invariant mass spectra of the lepton pair \mll,
of the \PQb jet pair \mbb, and of the combined system \mllbb.
We investigate whether the model descriptions for
the mass spectra of these partial decay systems follow the good description observed
for the full \ttbar system, \mtt (Fig.~\ref{fig:res_pttt}, middle plot).
The \mll distributions show a clear trend towards
a somewhat harder spectrum in the model predictions compared to the data.
The \mbb spectra are reasonably well described overall by all three predictions, with
a small trend of the data overshooting the predictions near threshold and at large
mass values.
The \mllbb distributions show a similar trend.
Furthermore, we investigate the sensitivity of the different mass spectra to the value of
the top quark mass assumed in the \PowPytSh calculation, by showing the predictions
for \mtmc= 169.5 and 175.5\GeV, compared to the nominal prediction using
a value of 172.5\GeV.
It is clear that the predicted \mll and \mbb spectra become
harder with increasing \mtmc, although the effects
are diluted in the regions of small invariant masses compared to the \mtt distribution
shown in Fig.~\ref{fig:res_pttt}.
The \mllbb distribution clearly exhibits a better sensitivity to the \mtmc value in
the small invariant mass region, comparable to the one observed for the \mtt spectrum.

\begin{figure}
    \centering
\includegraphics[width=0.49\textwidth]{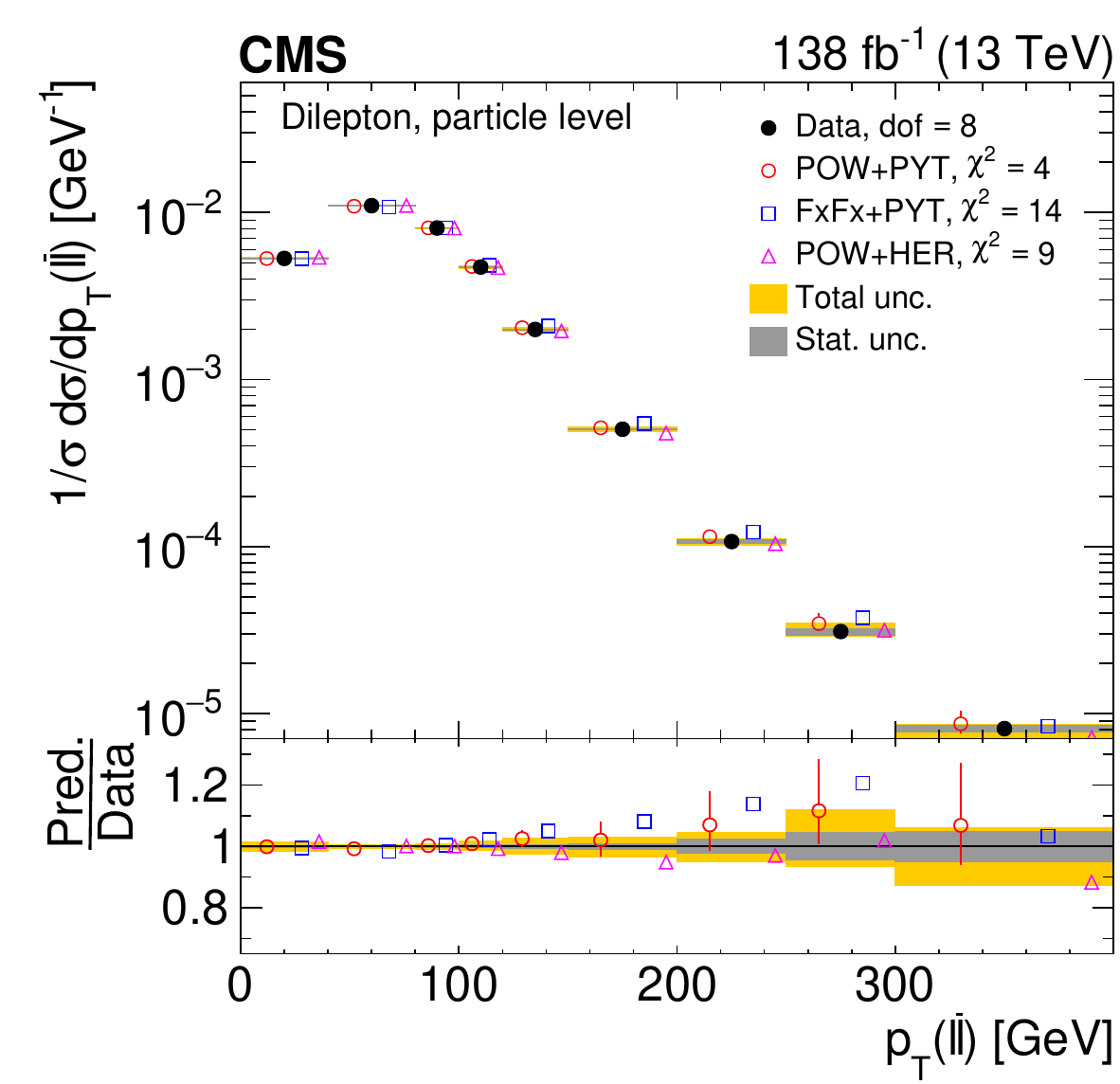}
\includegraphics[width=0.49\textwidth]{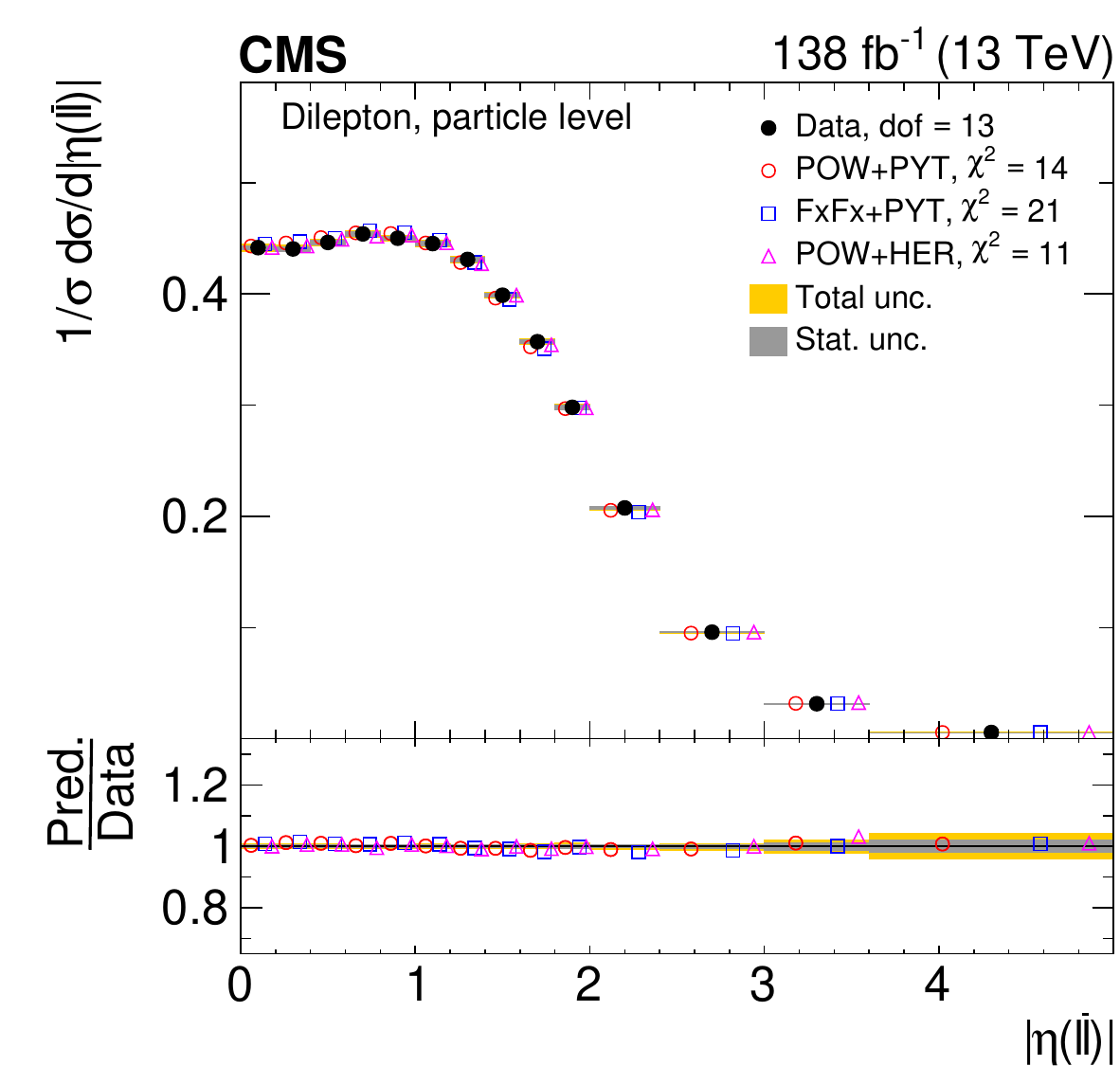}
    \caption{Normalized differential \ttbar production cross sections as functions of \ptll (left) and \absetall
(right)
are shown for data (filled circles) and various MC predictions (other points).
Further details can be found in the caption of Fig.~\ref{fig:res_ptlep}.}
\label{fig:res_ptll}
\end{figure}

\begin{figure}
    \centering
\includegraphics[width=1.00\textwidth]{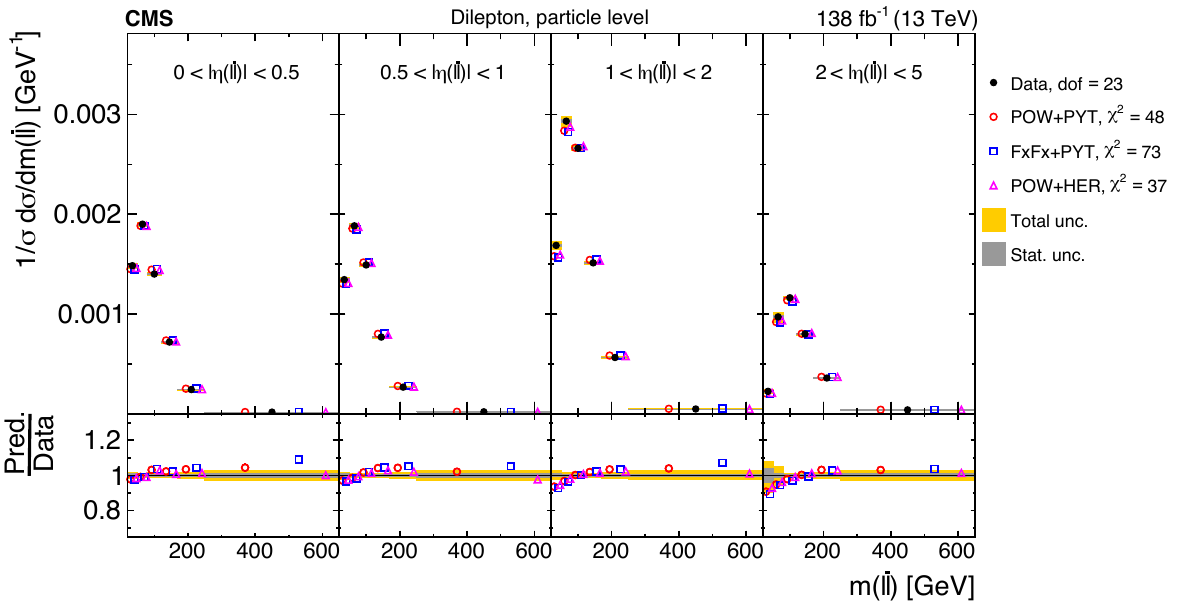}
\caption{Normalized \etallmll cross sections are shown for data (filled circles) and various MC predictions
(other points).
    Further details can be found in the caption of Fig.~\ref{fig:res_ptlep}.}
    \label{fig:res-etallmll}
\end{figure}

\begin{figure}
\centering
\includegraphics[width=1.00\textwidth]{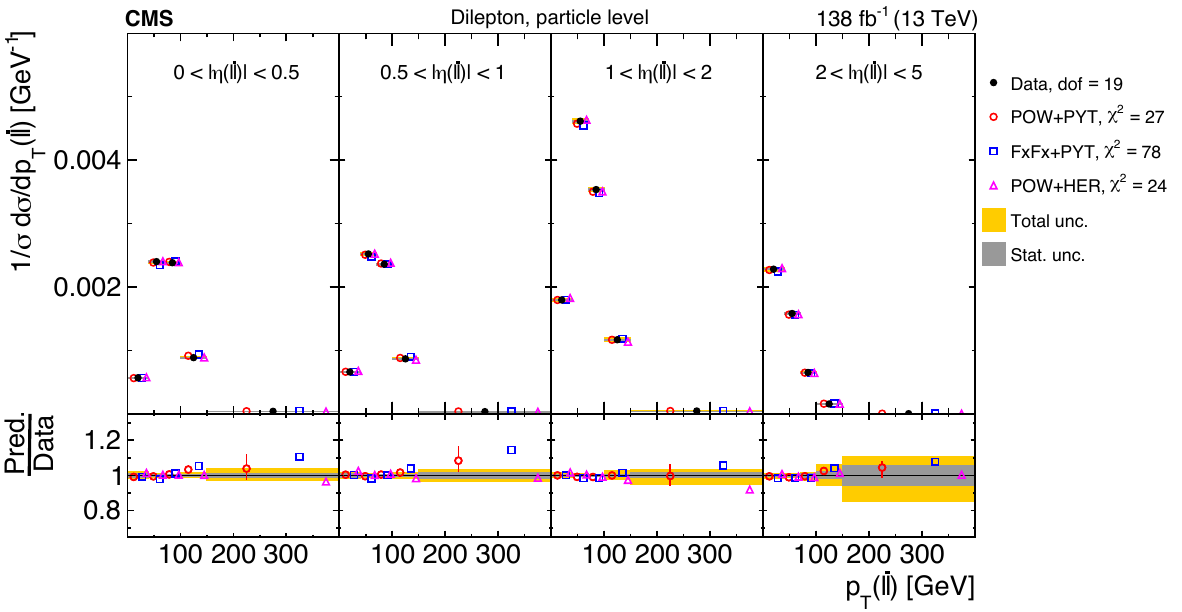}
\caption{Normalized \etallptll cross sections are shown for data (filled circles) and various MC predictions
(other points).
    Further details can be found in the caption of Fig.~\ref{fig:res_ptlep}.}
    \label{fig:res-etallptll}
\end{figure}

\begin{figure}
\centering
\includegraphics[width=1.00\textwidth]{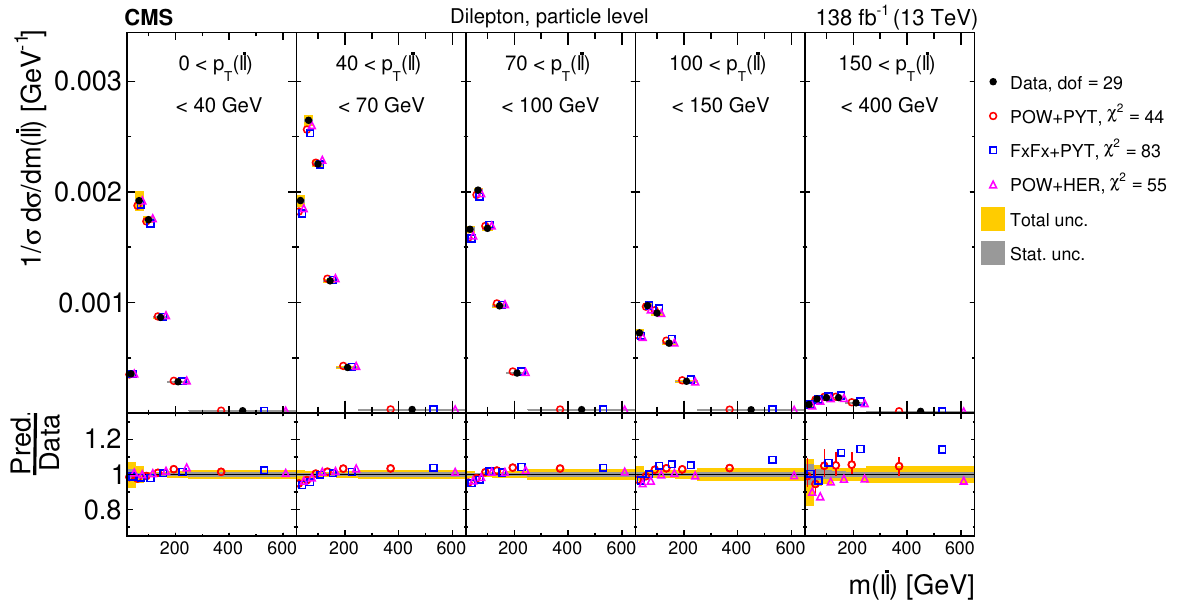}
\caption{Normalized \ptllmll cross sections are shown for data (filled circles) and various MC predictions
(other points).
    Further details can be found in the caption of Fig.~\ref{fig:res_ptlep}.}
    \label{fig:res-ptllmll}
\end{figure}

Finally, we study additional dilepton distributions,
whose kinematic obervables are among those in the present analysis
that are reconstructed with highest precision.
Figure~\ref{fig:res_ptll} shows the \ptll and \absetall distributions.
Overall, they are reasonably well described by the three MC models, with
\aMCPytSh predicting a \ptll spectrum that is slightly too hard and an \absetall
distribution that is a bit too central.
Figures~\ref{fig:res-etallmll}-\ref{fig:res-ptllmll}
show double-differential cross sections, illustrating
the correlations between the dilepton kinematic observables.
The \etallmll distributions show that the tendency of the MC predictions to provide
a \mll spectrum that is harder than in data is a bit enhanced towards high values of \absetall.
The \etallptll spectra are well described by the \PowPytSh and \PowHerSh models,
while \aMCPytSh predicts \ptll distributions that are somewhat too hard at small \absetall
values.
For the \ptllmll distributions the \mll
spectra clearly become harder towards larger values of \ptll and this
effect is a bit more pronounced in the MC models than in the data.

The observations made with the differential \ttbar cross sections as functions
of charged lepton and \PQb jet kinematic observables can be summarized as follows.
These measurements are the most accurate of all results presented in this document.
Overall, the predictions from the MC models agree rather well with each other.
The models predict harder distributions for the lepton \pt and the invariant
mass of the lepton pair.
The distributions of \pt of leading and trailing \PQb jets,
and of \mbb are reasonably well described,
except for \aMCPytSh, which predicts harder \pt spectra.
Among the invariant mass spectra,
the \mllbb distribution clearly shows the strongest sensitivity to the value
of the top quark mass used in the \PowPytSh calculation.
The single-differential \pt and $\eta$ distributions of the dilepton system
are reasonably well described by the models overall,
although for double-differential distributions, including also \mll as a possible second 
variable,
some tensions are visible, in particular for \aMCPytSh.
The standard \chisq values indicate, in general, a poor quality in the description of the data
by the nominal predictions. 
The inclusion of the prediction uncertainties for the \PowPytSh model (see Table~\ref{tab:chi2mc_1d_nor_particle_lepb}) results in \chisq tests with reasonable $p$-values in most cases.

\clearpage

\clearpage

\clearpage

\begin{table*}
\centering
 \topcaption{The \chisq values and \ndf of the measured normalized single-differential cross sections for lepton and \PQb-jet kinematic observables at the particle level are shown with respect to the predictions of various MC generators. 
The \chisq values are calculated taking only measurement uncertainties into account and excluding theory uncertainties.  
For \PowPytSh, the \chisq values including theory uncertainties are indicated with the brackets (w. unc.).}
 \label{tab:chi2mc_1d_nor_particle_lepb}
 \renewcommand{\arraystretch}{1.4}
 \centering
 \begin{tabular}{lccccc}
 \multirow{1}{*}{Cross section} & \hspace*{0.3 cm} \multirow{2}{*}{\ndf} \hspace*{0.3 cm} & \multicolumn{3}{c}{\chisq} \\
 \cline{3-5}
{variables} && \PowPytSh (w. unc.)  & \aMCPytSh  & \PowHerSh   \\
\hline
\ptlep& 11 & 28 \: (18) & 51 & 18 \\
\ptlep trailing/\ptlep leading& 9 & 15 \: (11) & 24 & 7 \\
\ptlep/\ptat& 4 & 10 \: (9) & 25 & 12 \\
\ptb leading& 9 & 5 \: (4) & 26 & 8 \\
\ptb trailing& 6 & 6 \: (4) & 24 & 7 \\
\rptbsptts& 3 & 19 \: (15) & 27 & 18 \\
\mll& 11 & 23 \: (20) & 27 & 23 \\
\mbb& 6 & 15 \: (12) & 15 & 14 \\
\mllbb& 18 & 33 \: (18) & 28 & 28 \\
\ptll& 8 & 4 \: (3) & 14 & 9 \\
\absetall& 13 & 14 \: (9) & 21 & 11 \\
\etallmll& 23 & 48 \: (28) & 73 & 37 \\
\etallptll& 19 & 27 \: (14) & 78 & 24 \\
\ptllmll& 29 & 44 \: (37) & 83 & 55 \\
 \end{tabular} 
\end{table*}
\clearpage

\subsection{Results as a function of additional-jet multiplicity}
\label{sec:res_3}
\begin{figure}
\centering
\includegraphics[width=0.49\textwidth]{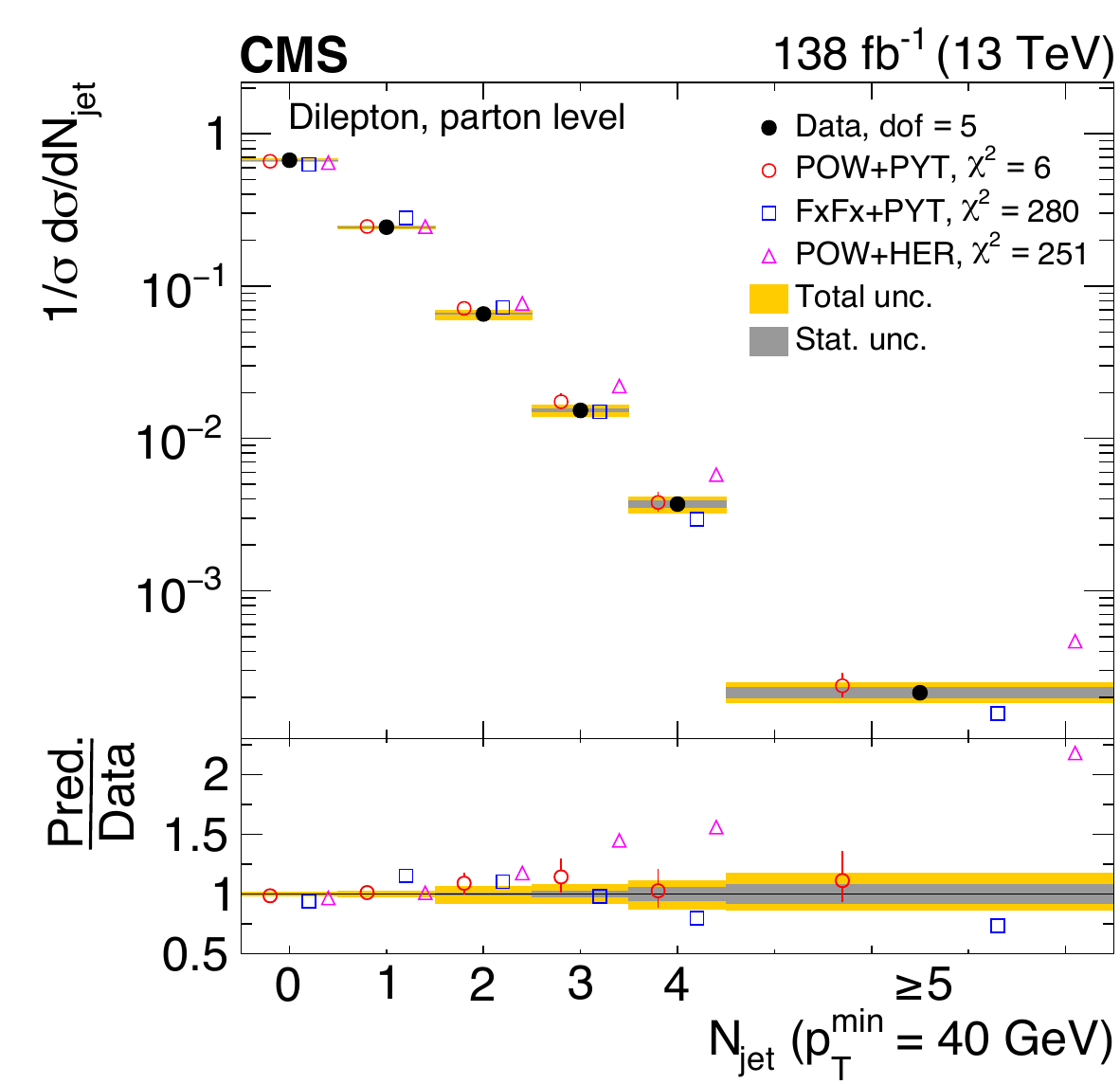}
\includegraphics[width=0.49\textwidth]{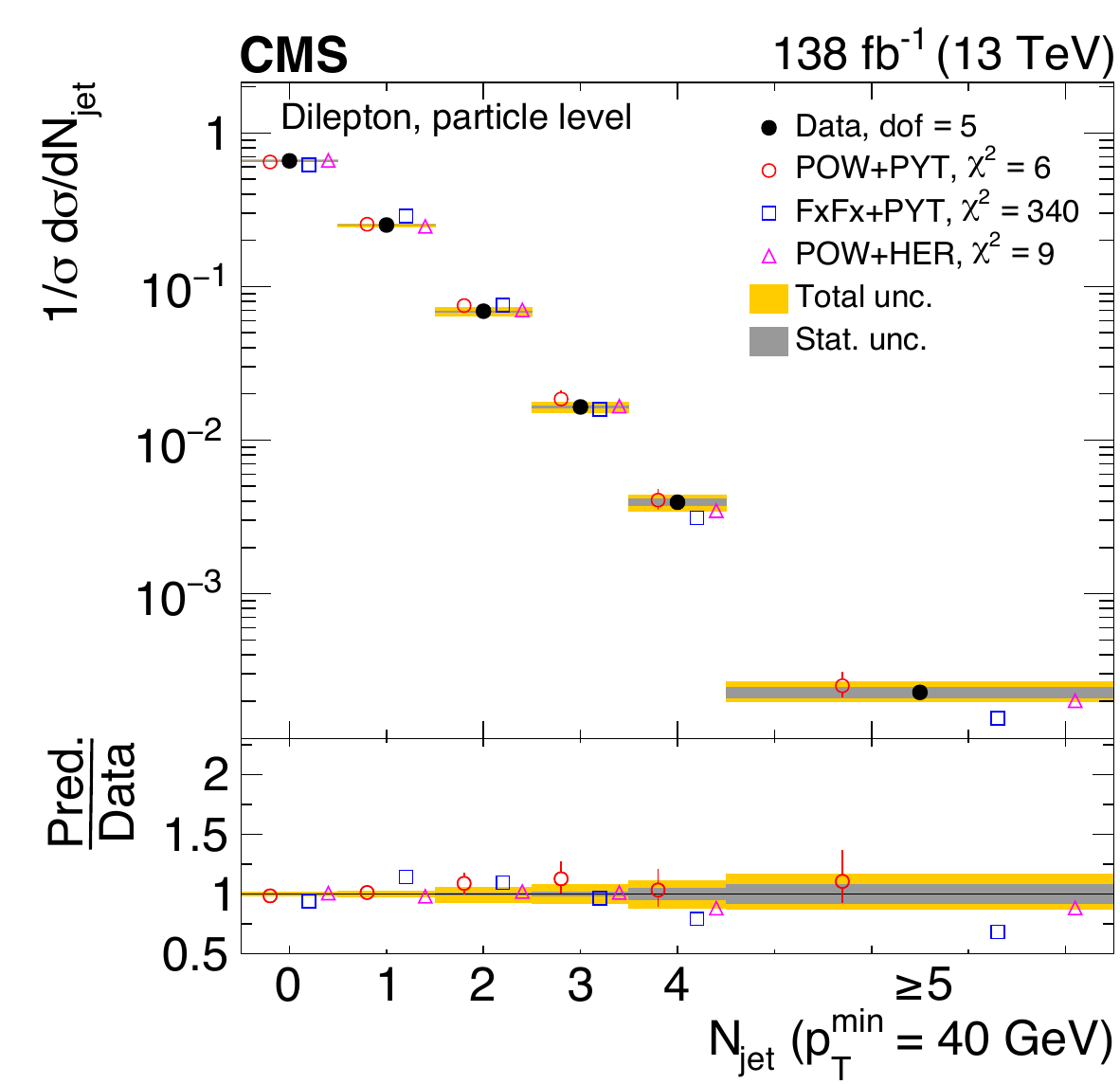}
\includegraphics[width=0.49\textwidth]{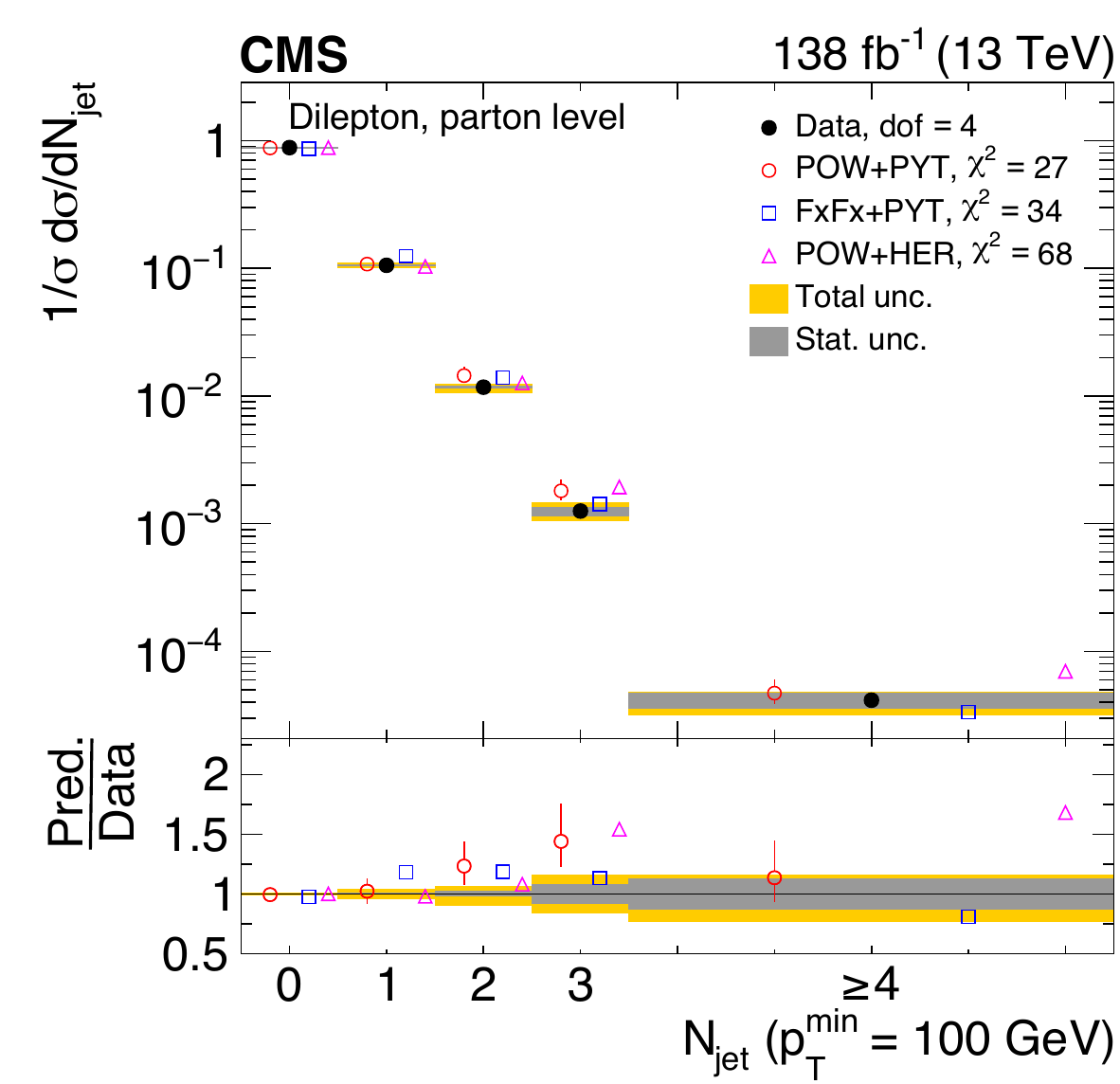}
\includegraphics[width=0.49\textwidth]{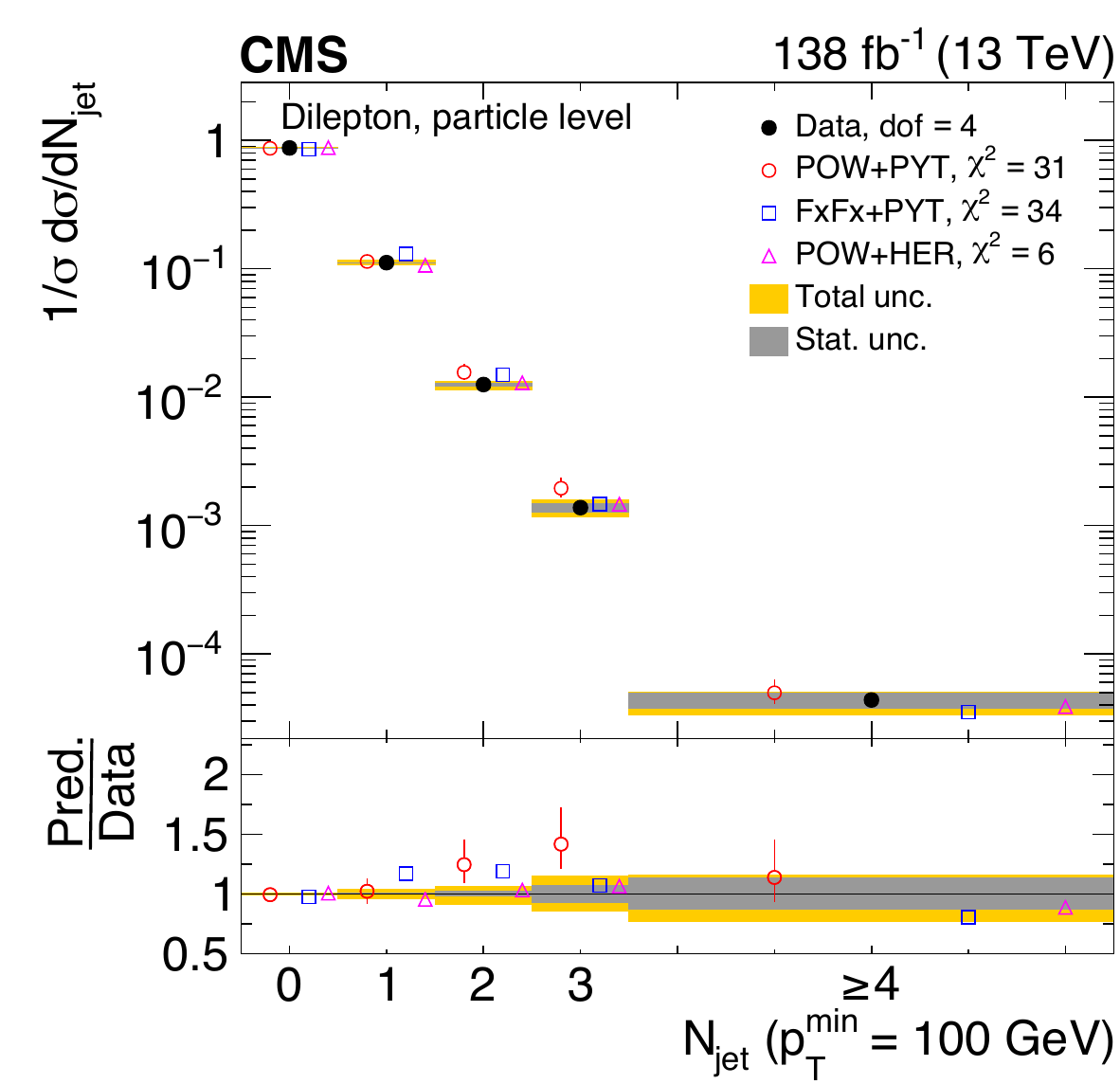}
\caption {Normalized differential \ttbar production cross sections as a function of \nj, for a minimum
jet \pt of 40\GeV (upper) and 100\GeV (lower), measured at the parton level in the full phase space (left) and at the 
particle level in a fiducial phase space (right).
The data are shown as filled circles with grey and yellow bands indicating the statistical and total uncertainties
(statistical and systematic uncertainties added in quadrature), respectively.
For each distribution, the number of degrees of freedom (dof) is also provided.
The cross sections are compared to various MC predictions (other points).
The estimated uncertainties in the \PowPyt (`POW-PYT') simulation are represented by vertical bars on the
corresponding points.
For each MC model, a value of \chisq is reported that takes into account the measurement uncertainties.
The lower panel in each plot shows the ratios of the predictions to the data.}
    \label{fig:res_nj40}
\end{figure}

\begin{figure}
\centering
\includegraphics[width=0.99\textwidth]{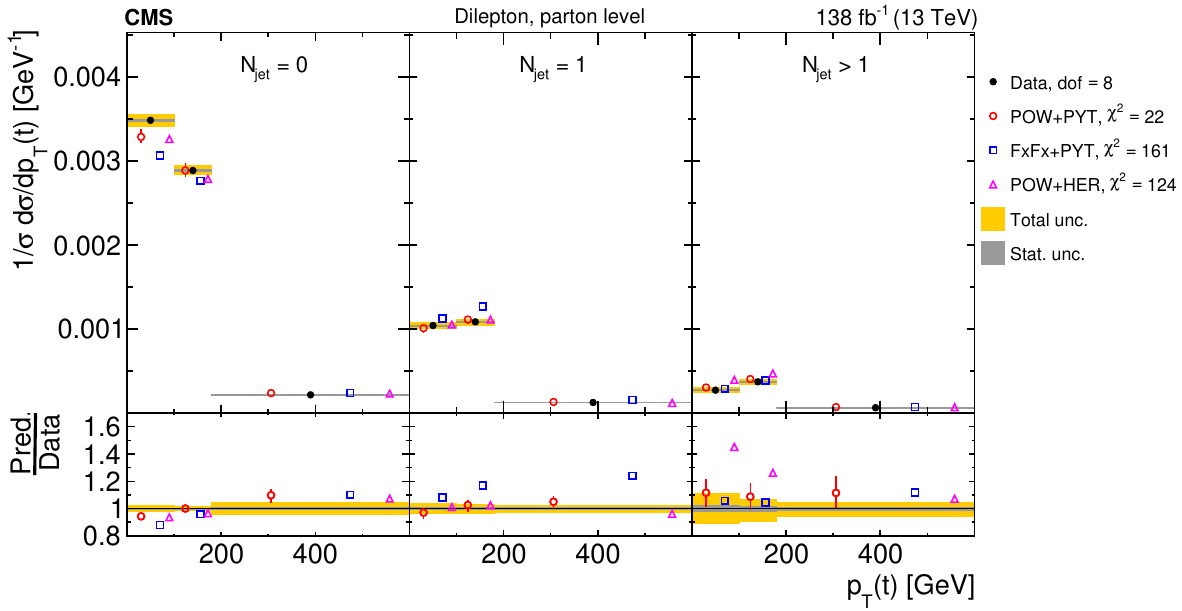}
\includegraphics[width=0.99\textwidth]{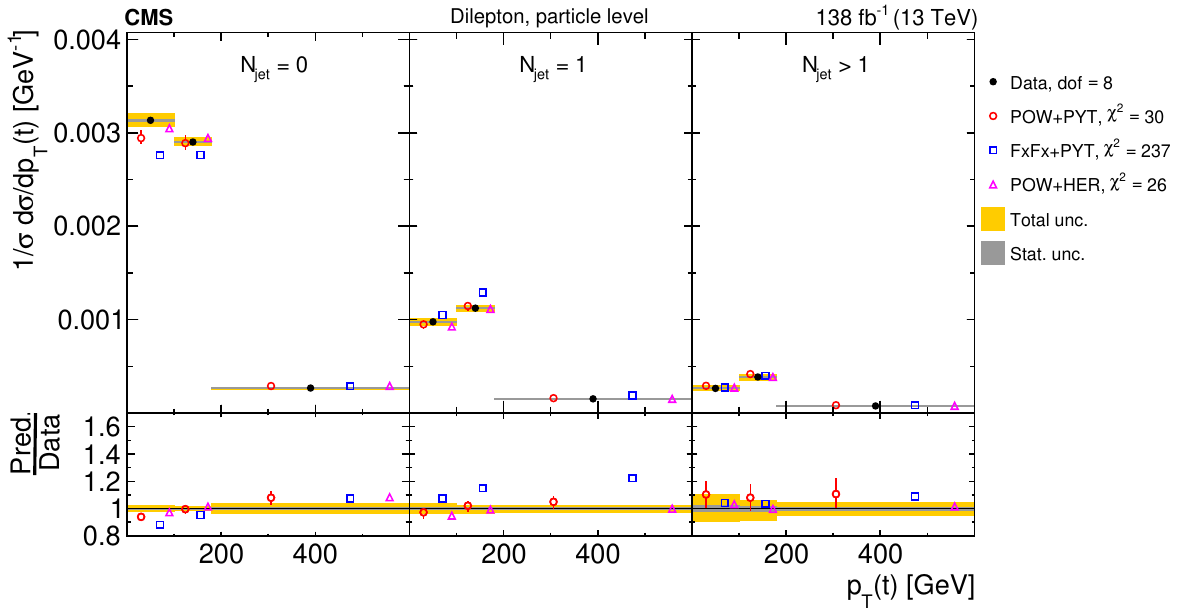}
\caption{Normalized \njptt cross sections measured at the parton level in the full phase space (upper) and
at the particle level in a fiducial phase space (lower). The data are shown as filled circles with grey and
yellow bands indicating the statistical and total uncertainties (statistical and systematic uncertainties
added in quadrature), respectively.
For each distribution, the number of degrees of freedom (dof) is also provided.
The cross sections are compared to various MC predictions (other points). The estimated uncertainties in the
\PowPyt (`POW-PYT') simulation
are represented by vertical bars on the corresponding points.
For each MC model, a value of \chisq is reported that takes into account the measurement uncertainties.
The lower panel in each
plot shows the ratios of the predictions to the data.}
      \label{fig:res_njptt}
\end{figure}

\begin{figure}
\centering
\includegraphics[width=0.99\textwidth]{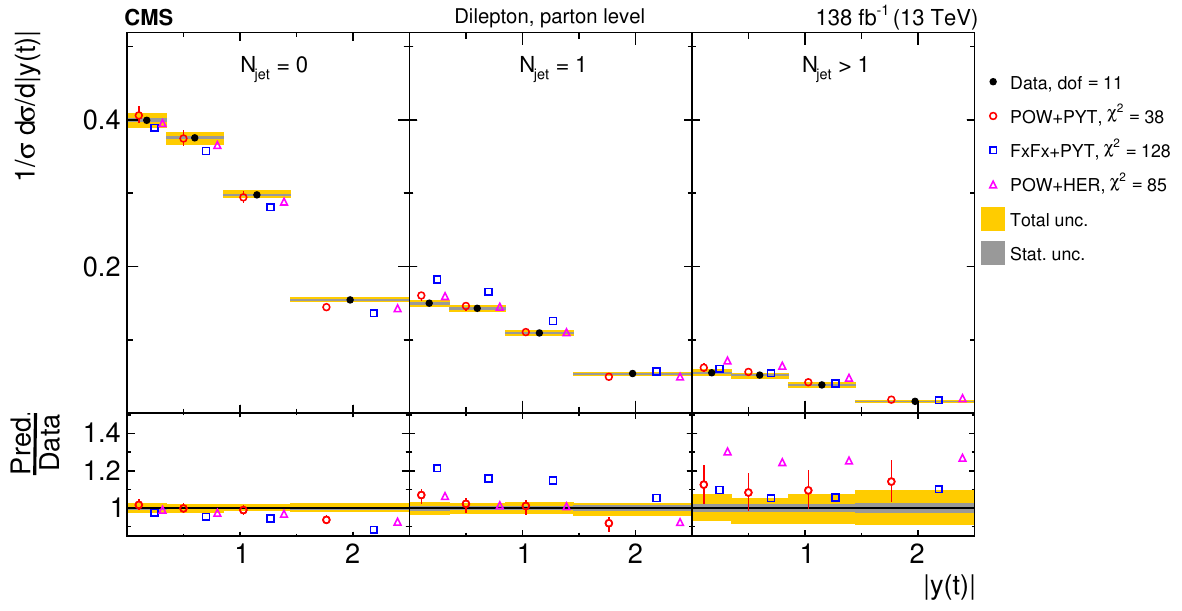}
\includegraphics[width=0.99\textwidth]{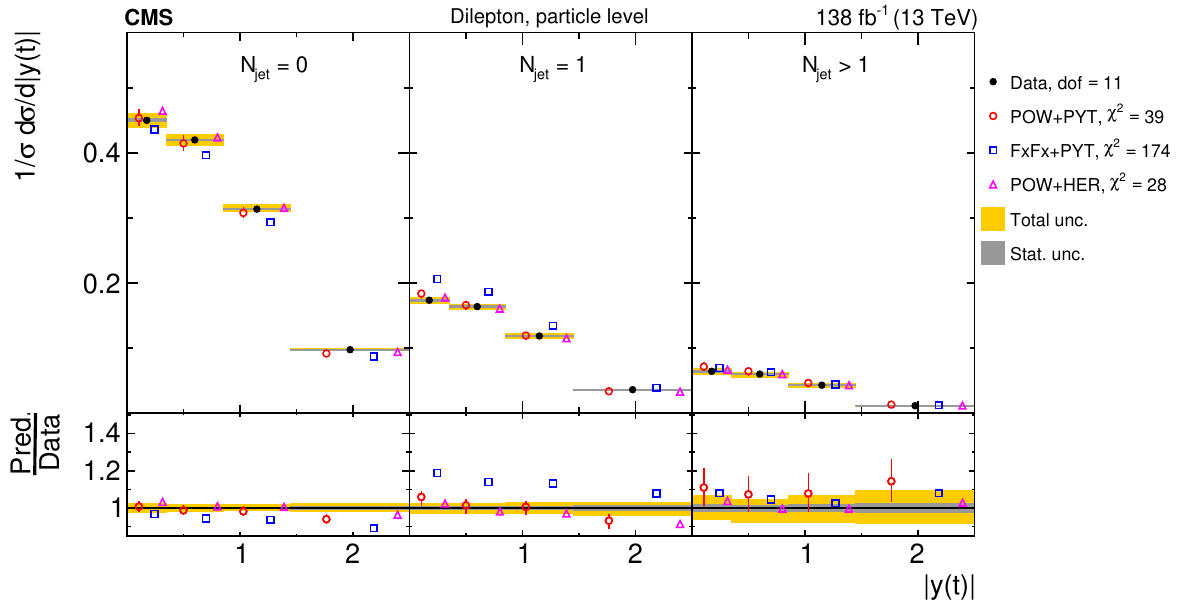}
\caption{Normalized \njyt cross sections are shown for data (filled circles) and various MC predictions
(other points).
    Further details can be found in the caption of Fig.~\ref{fig:res_njptt}.}
    \label{fig:res_njyt}
\end{figure}

\begin{figure}
\centering
\includegraphics[width=0.99\textwidth]{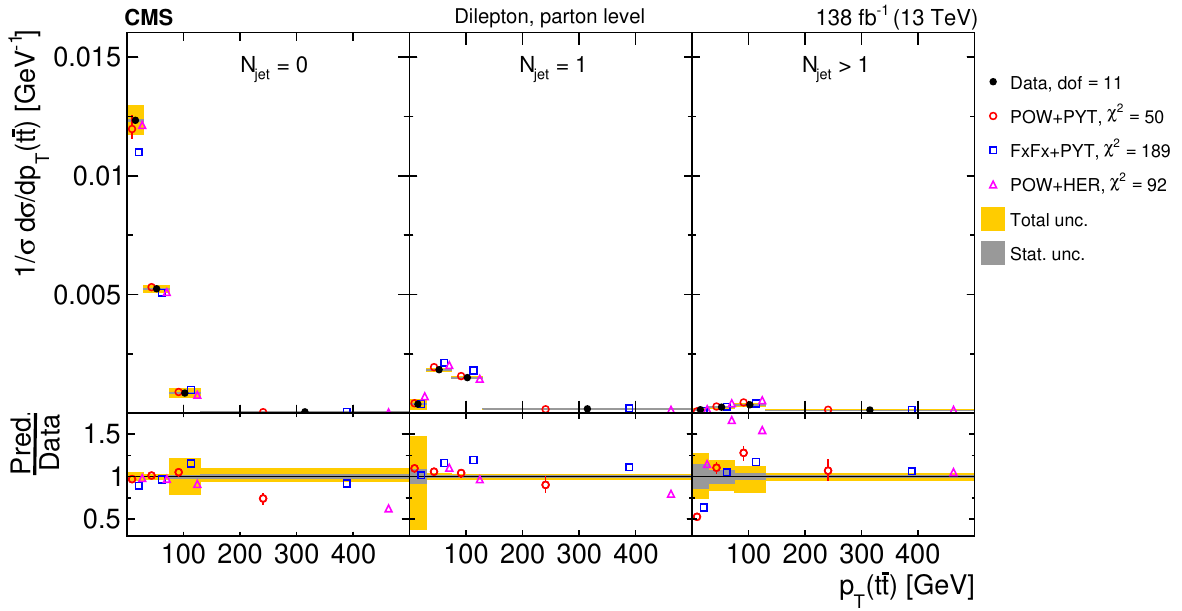}
\includegraphics[width=0.99\textwidth]{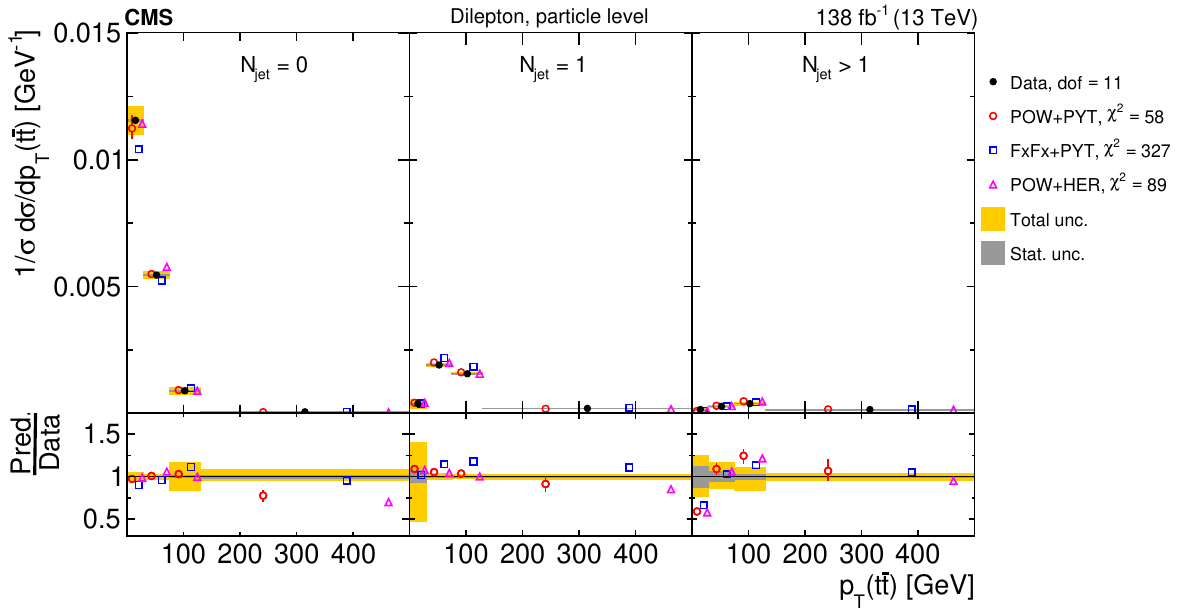}
\caption{Normalized \njpttt cross sections are shown for data (filled circles) and various MC predictions
(other points).
    Further details can be found in the caption of Fig.~\ref{fig:res_njptt}.}
    \label{fig:res_njpttt}
\end{figure}

\begin{figure}
\centering
\includegraphics[width=0.99\textwidth]{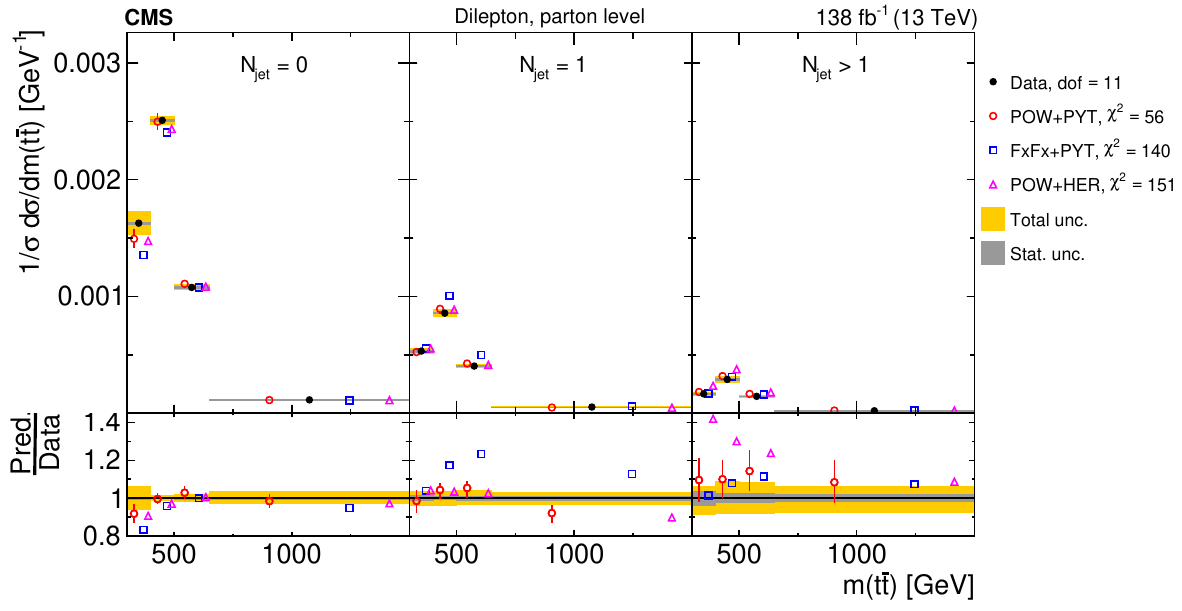}
\includegraphics[width=0.99\textwidth]{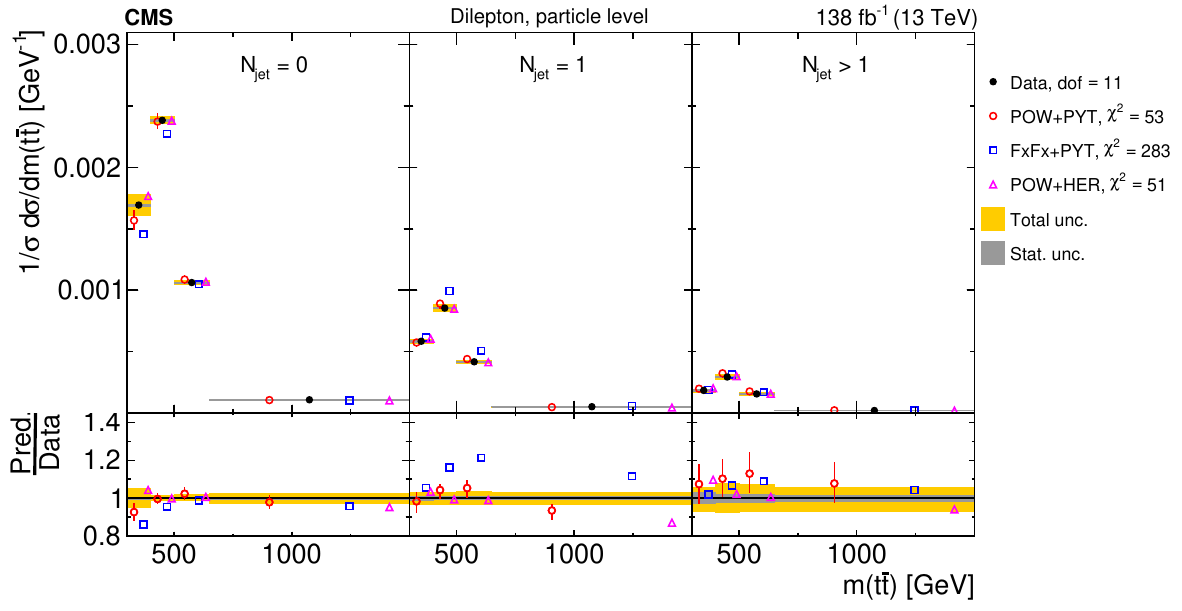}
\caption{Normalized \njmtt cross sections are shown for data (filled circles) and various MC predictions
(other points).
    Further details can be found in the caption of Fig.~\ref{fig:res_njptt}.}
    \label{fig:res_njmtt}
\end{figure}

\begin{figure}
\centering
\includegraphics[width=0.99\textwidth]{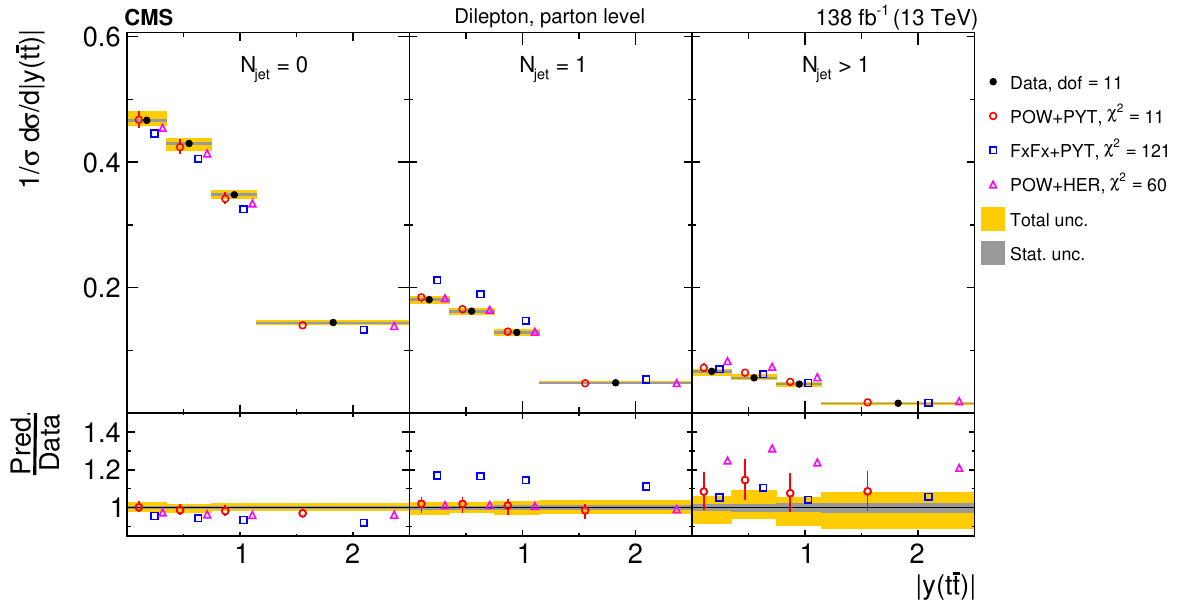}
\includegraphics[width=0.99\textwidth]{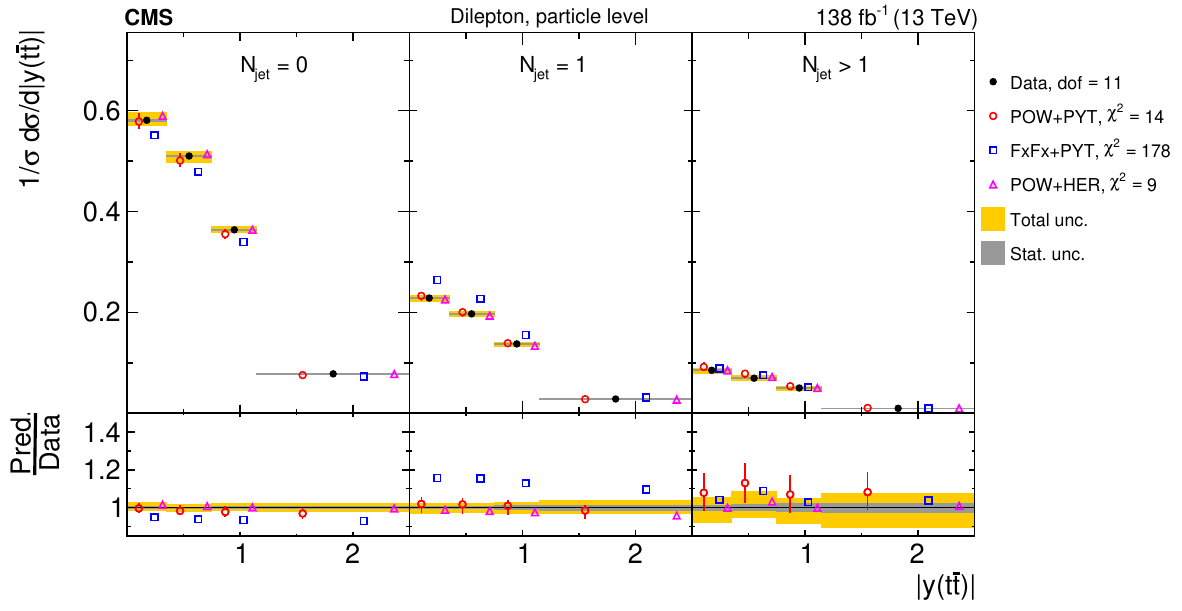}
\caption{Normalized \njytt cross sections are shown for data (filled circles) and various MC predictions
(other points).
    Further details can be found in the caption of Fig.~\ref{fig:res_njptt}.
     }
    \label{fig:res_njytt}
\end{figure}

\begin{figure}
\centering
\includegraphics[width=0.99\textwidth]{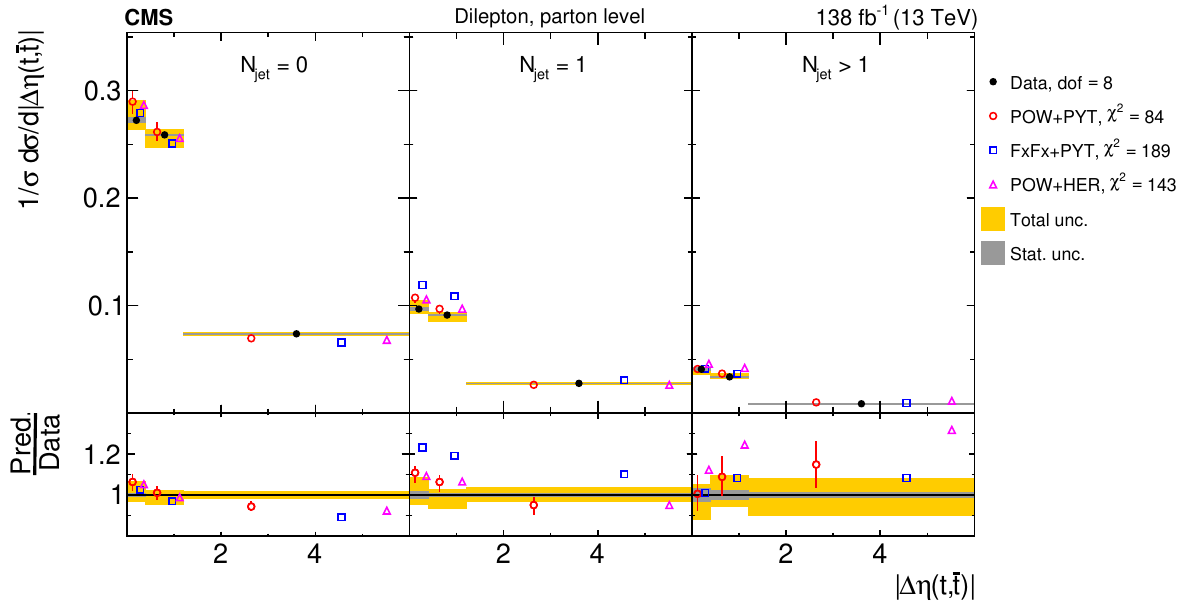}
\includegraphics[width=0.99\textwidth]{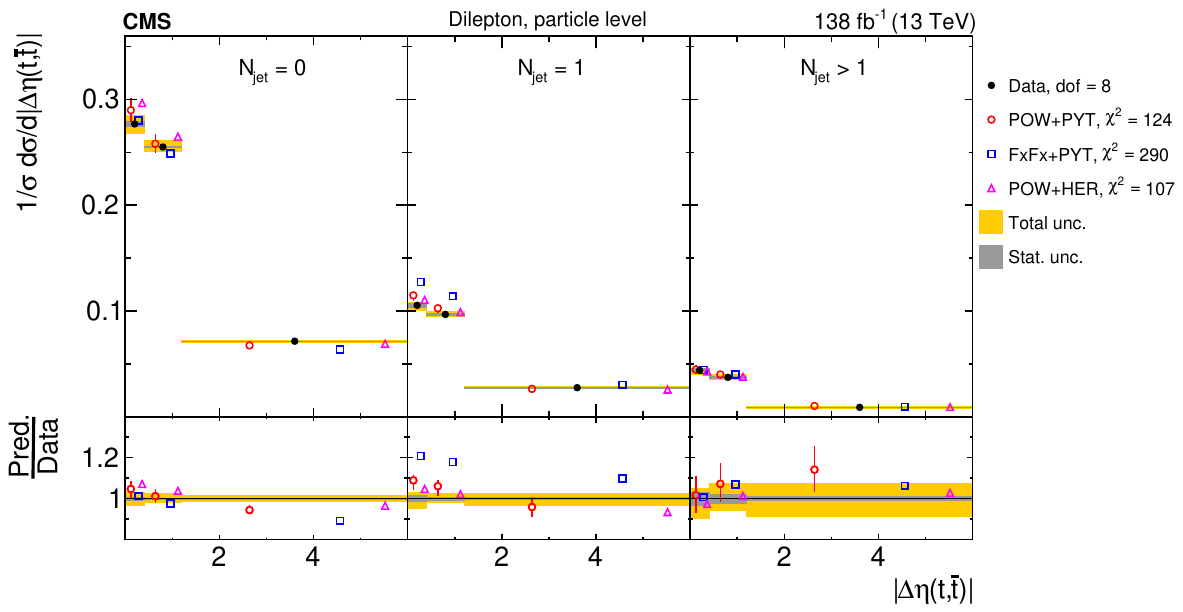}
\caption{Normalized \njdetatt cross sections are shown for data (filled circles) and various MC predictions
(other points).
    Further details can be found in the caption of Fig.~\ref{fig:res_njptt}.
     }
    \label{fig:res_njdeta}
\end{figure}

\begin{figure}
\centering
\includegraphics[width=1.00\textwidth]{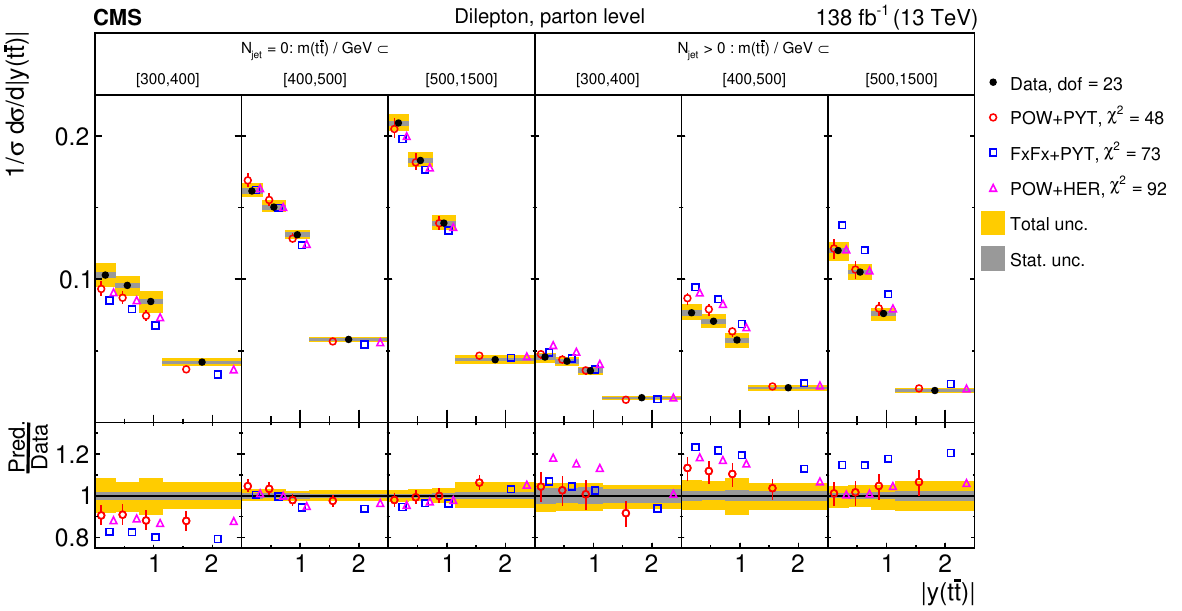}
\includegraphics[width=1.00\textwidth]{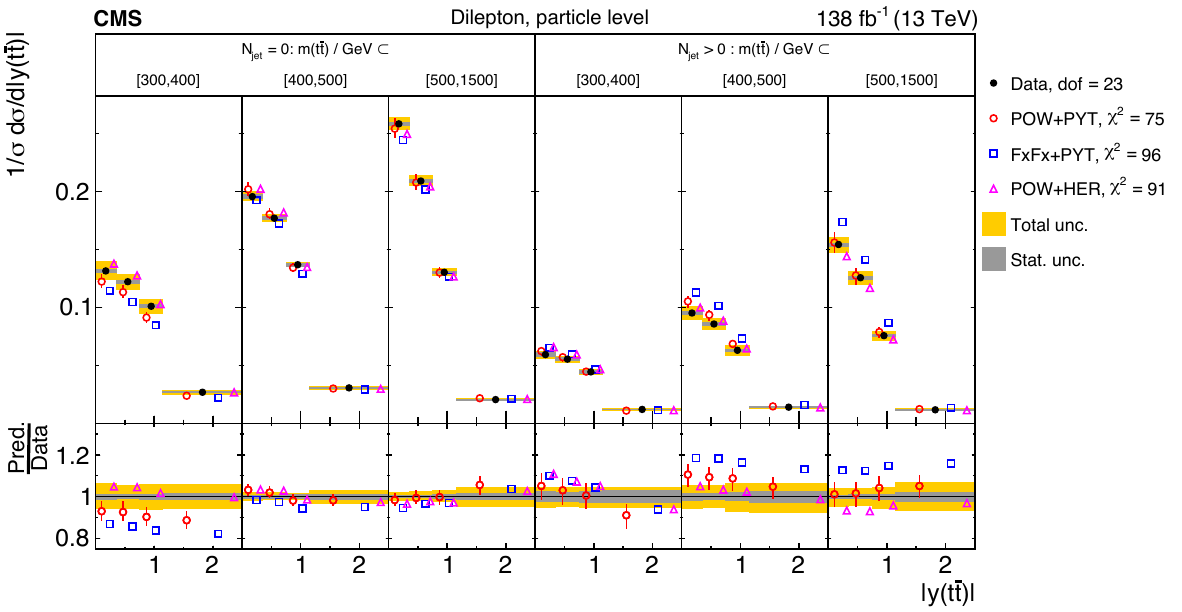}
\caption{Normalized \njmttytttwo cross sections are shown for data (filled circles) and various MC predictions
(other points).
    Further details can be found in the caption of Fig.~\ref{fig:res_njptt}.
     }
    \label{fig:res_nj2mttytt}
\end{figure}

\begin{figure}
\centering
\includegraphics[width=1.00\textwidth]{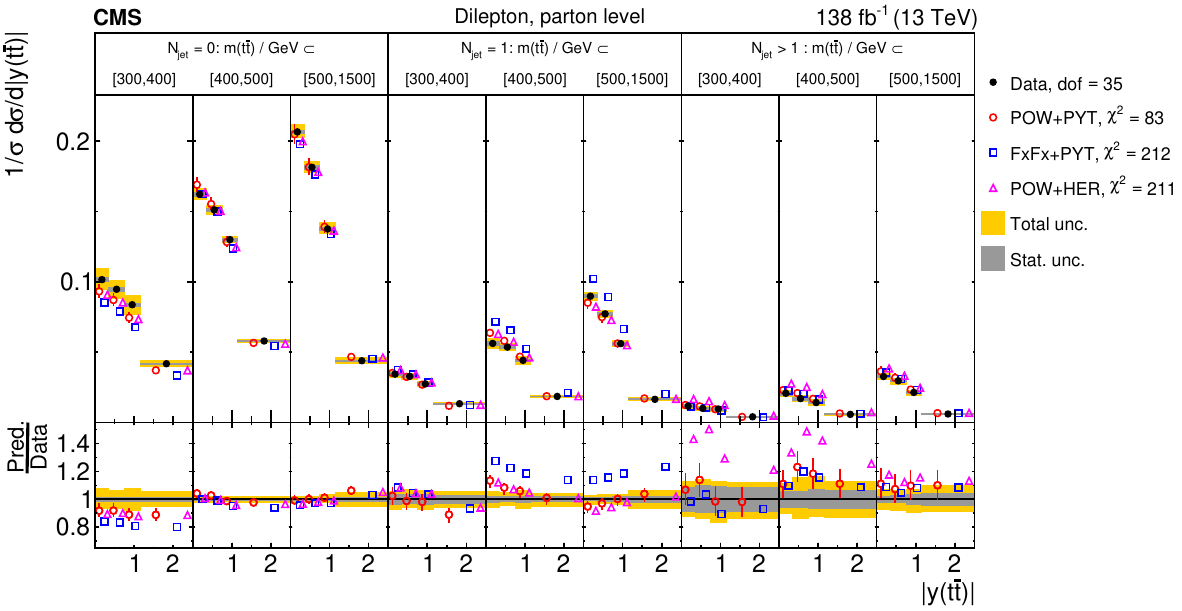}
\includegraphics[width=1.00\textwidth]{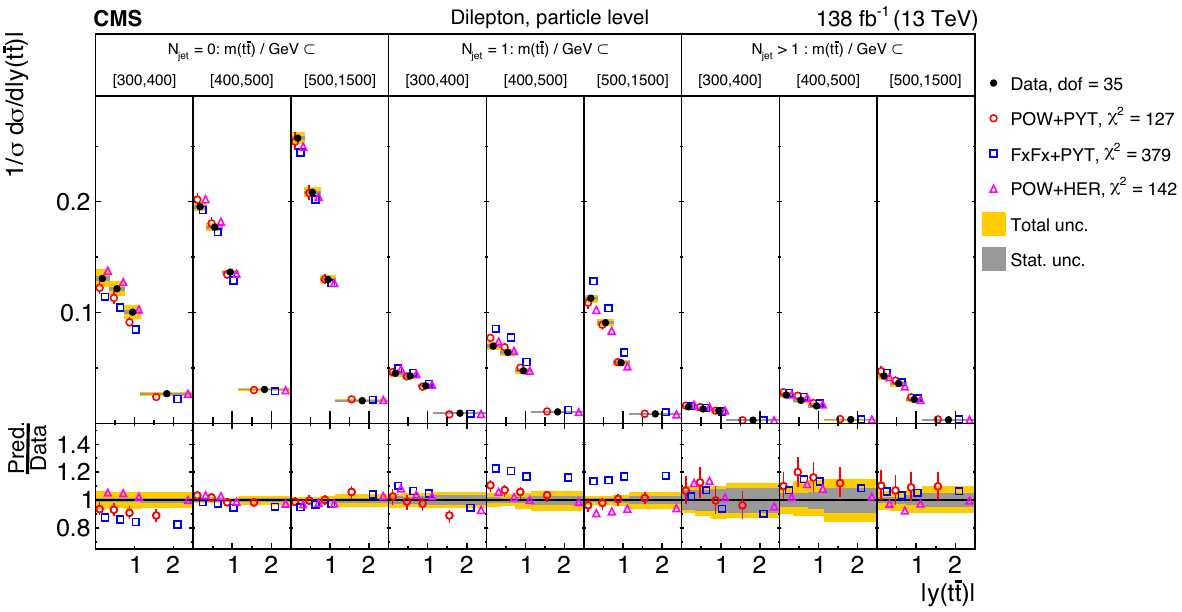}
\caption{Normalized \njmttyttthree cross sections are shown for data (filled circles) and various MC predictions
(other points).
    Further details can be found in the caption of Fig.~\ref{fig:res_njptt}.
     }
    \label{fig:res_nj3mttytt}
\end{figure}

\begin{figure}
\centering
\includegraphics[width=1.00\textwidth]{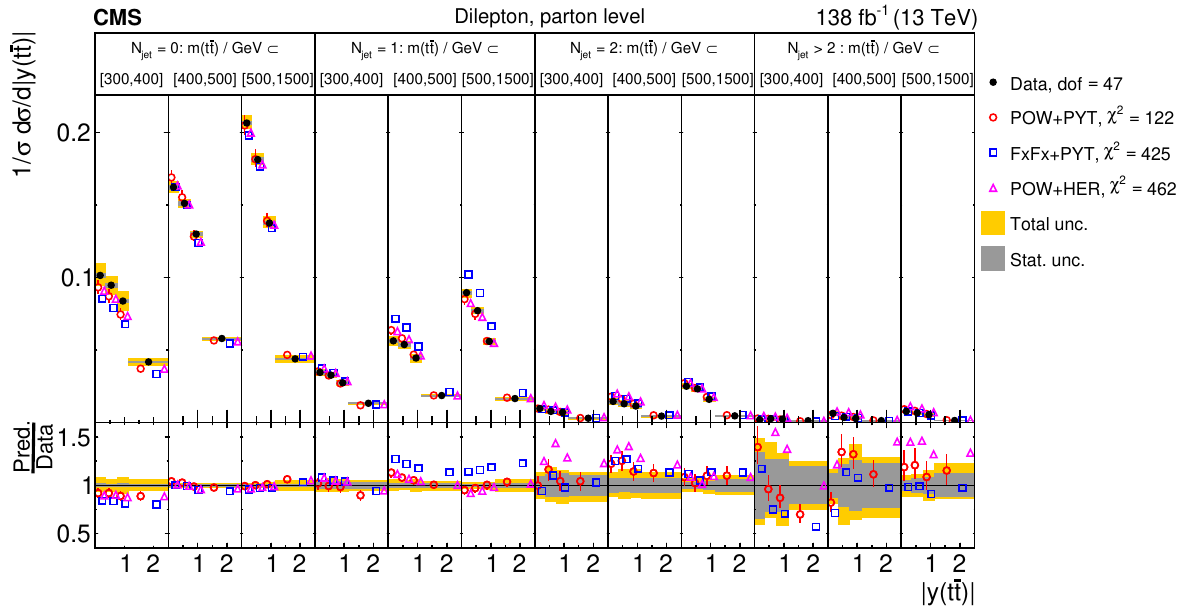}
\includegraphics[width=1.00\textwidth]{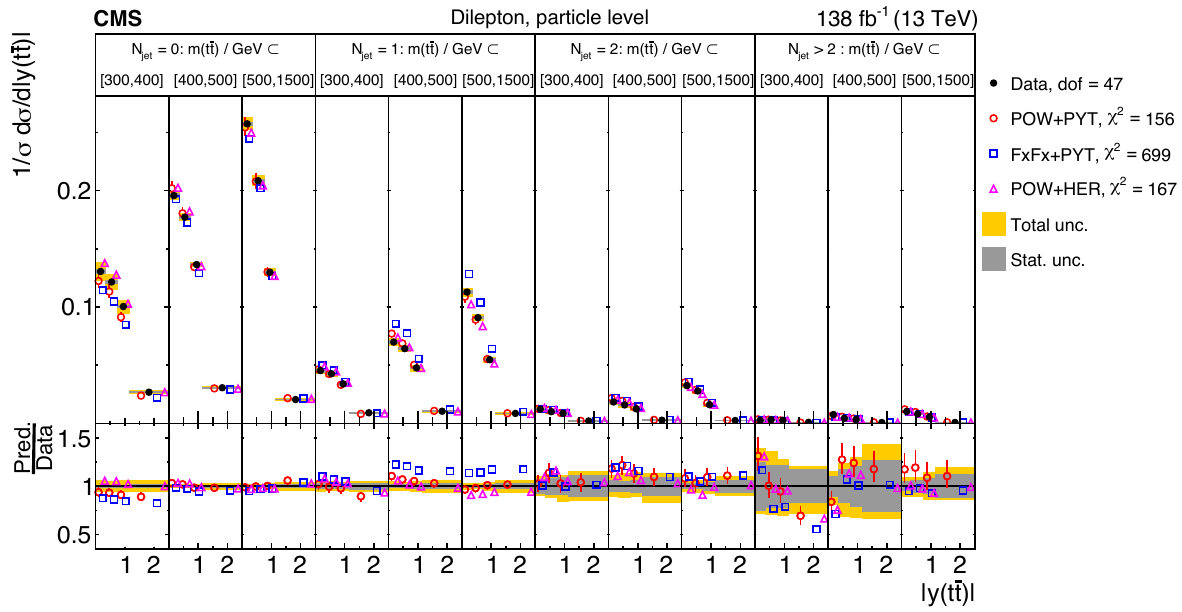}
\caption{Normalized \njmttyttfour cross sections are shown for data (filled circles) and various MC predictions
(other points).
    Further details can be found in the caption of Fig.~\ref{fig:res_njptt}.
     }
    \label{fig:res_nj4mttytt}
\end{figure}

In the final set of studies, we measure differential \ttbar production cross sections
as a function of the multiplicity of additional jets in the events.
These investigations provide an exemplary testing ground for the understanding
of perturbative QCD.
In particular, additional jets provide a second hard kinematic scale in the events,
competing with the \ttbar invariant mass, and thus give rise to a multiscale situation
that provides a challenge for the perturbative expansion~\cite{Diehl:mla}.
The definitions of the cross sections are given in Section~\ref{sec:truthlevel}.
Additional jets are measured at the particle level, and
the top quark and antiquark are either measured at the parton level
in the full phase space or at the particle level in a fiducial phase space.
The cross sections are
shown in Figs.~\ref{fig:res_nj40}--\ref{fig:res_nj4mttytt}.
The upper and lower plots in the figures depict the cross sections at the parton
and particle levels, respectively.
The \chisq values of model-to-data comparisons
are listed in Tables~\ref{tab:chi2mc_nor_parton_addjets}--\ref{tab:chi2mc_nor_particle_addjets} and the corresponding $p$-values in Tables~\ref{tab:pvaluemc_nor_parton_addjets}--\ref{tab:pvaluemc_nor_particle_addjets}.

We first discuss the additional-jet multiplicity \nj distribution
and its dependence on the minimum jet \pt requirement.
This distribution provides a direct view of the amount of higher-order QCD radiation
in \ttbar events.
Figure~\ref{fig:res_nj40}
shows the \nj cross section distributions for minimum
\pt values of 40 and 100\GeV, respectively.
The data are compared to the same three MC models as discussed above.
The \PowPytSh MC provides a good description of the \nj
distribution for the lower \pt value, though for the higher \pt value
it starts to overshoot the data at larger jet multiplicities $\nj \ge 2$.
The \aMCPytSh model exhibits a low accuracy, nearly independent of the \pt requirement.
Its cross section prediction is too low for $\nj=0$ and too high for $\nj = 1$,
though it is reasonable for larger \nj.
The effect is underlined by large \chisq values.
The description of the data by \PowPytSh and
\aMCPytSh is consistent for the parton- and particle-level cross sections,
but for \PowHerSh a different picture emerges.
This model clearly predicts too many
extra jets for the parton-level cross section,
but describes the data well at the particle level in a fiducial phase space.
The parton level is defined for the full phase space. However,
there might be a larger contribution from extra jets in \PowHerSh from events
that are predominantly outside the particle-level fiducial phase space for the
top quark and antiquark.

Next we investigate
the top quark and \ttbar kinematics as a function of \nj.
This allows us to map the kinematic correlations
to additional jets and to check whether description deficiencies seen, \eg
for the top quark transverse momentum \ptt spectrum (Fig.~\ref{fig:res_ptt}), are associated with 
specific \nj values.
Figure~\ref{fig:res_njptt} shows the \ptt distributions.
For \PowPytSh, the effect of a harder \ptt spectrum compared to that observed in data
seems to be slightly enhanced in the lower two jet multiplicity bins $\nj=0,1$,
though a reasonable description is seen for higher multiplicities $\nj>1$.
The \aMCPytSh model predicts harder \ptt spectra, rather independent of \nj.
Among all considered predictions, \PowHerSh provides the best description of the \ptt distributions
at the particle level, but fails at the parton level, where
its large excess at $\nj>1$ is accompanied by a \ptt spectrum that falls too steeply.
The distributions of \absyt are illustrated in Fig.~\ref{fig:res_njyt}.
The data prefer slightly less-central \absyt distributions
than the models, with weak dependence on \nj.
Figure~\ref{fig:res_njpttt} shows the \njpttt distributions.
The two variables are correlated since
additional QCD radiation in the event leads to nonzero values of \pttt,
as well as to values $\nj \ge 1$.
For low jet multiplicities, $\nj=0,1$, the \PowPytSh model provides
a reasonable description of the measured \pttt spectrum, though
for $\nj>1$ it predicts a rise of the cross section that is too steep over the first three \pttt
ranges from 0 to 100\GeV.
The ratio of the \aMCPytSh model to the data
always exhibits a positive slope over the first three \pttt ranges, irrespectively
of \nj.
The \PowHerSh prediction mostly follows the \PowPytSh model, except
at high \pttt where it is a bit lower.

The distributions of \mtt are shown in Fig.~\ref{fig:res_njmtt}.
The shapes of these distributions are fairly well modeled
by \PowPytSh, for all \nj ranges.
The \aMCPytSh calculation clearly predicts much harder \mtt spectra for $\nj=1$.
The \PowHerSh provides a fair description at the \ttbar particle level,
but fails again at the parton level, where it predicts
a much softer \mtt distribution for $\nj>1$.
Figure~\ref{fig:res_njytt} displays the \absytt spectra.
The shapes of these distributions are reasonably well described by the models,
irrespectively of the \nj range, but specific \nj-dependent normalization problems
remain for \aMCPytSh and \PowHerSh at the particle level.
Figure~\ref{fig:res_njdeta} depicts the \detatt distributions. 
The description by the models is poor, which is also indicated by the large \chisq values
with very small $p$-values. 
For the first two \nj bins there is a clear trend
for the models to predict rapidity separations
between top quark and antiquark that are too small, while in the last bin
the trend is slightly reversed.

Figures~\ref{fig:res_nj2mttytt}--\ref{fig:res_nj4mttytt} show
triple-differential cross sections
as functions of \nj, \mtt, and \absytt.
These cross sections were one of the highlights of our previous analysis~\cite{Sirunyan:2019zvx},
where it was demonstrated that they can be used for a simultaneous extraction of the top quark pole mass,
\alpS, and the PDFs, with good precision.
We measure the cross sections separately using two ($\nj = 0$ and $\nj \ge 1$), three ($\nj = 0$, $\nj = 1$,
and $\nj \ge 2$),
and four ($\nj = 0$, $\nj = 1$, $\nj=2$ and $\nj \ge 3$) bins of \nj for the particle-level jets.
{These cross sections are denoted as \njmttytttwo, \njmttyttthree, and \njmttyttfour, respectively.}
The \njmttyttfour cross sections are measured for the first time.
A striking feature of the comparisons of the MC models to the data is
a growing discrepancy when going from two to three and four bins of \nj.
The \njmttyttfour cross sections clearly exhibit
the best power for distinguishing the models.
Among all tested predictions, the \PowPytSh model provides the overall best description.
The \aMCPytSh calculation exhibits the same normalization problems versus \nj that are visible in the \nj spectrum
(Fig.~\ref{fig:res_nj40}), coupled with a small trend to predict cross sections towards large \mtt that are too
high, though the \absytt shapes are described reasonably well.
The \PowHerSh model delivers a description at the particle level that is comparable to that of \PowPytSh,
but fails at the parton level where it overshoots the data for the $\nj=2$ and $\nj>2$ bins.

The comparisons of MC models and data
can be summarized as follows.
The \PowPytSh calculation clearly provides the best description
of the additional-jet multiplicity \nj in \ttbar events.
The \aMCPytSh and \PowHerSh models fail, the former predicting too many events with $\nj=1$
and the latter too many events at the parton level in the full phase space with $\nj>1$.
There are rather weak kinematic correlations of the top quark and \ttbar rapidity spectra with \nj,
and the quality of the descriptions of the rapidity spectra by the models is rather
independent of \nj.
As expected, there are larger kinematic correlations of top quark and \ttbar transverse momenta
or \ttbar invariant mass with \nj, showing harder spectra for larger \nj.
All models exhibit different level of discrepancies for these distributions that depend on the jet 
multiplicity.
For instance, there is an indication that the problem of harder \ptt distributions in the models
is localized at small jet multiplicities.
The theory-to-data \chisq values indicate a rather poor description of the data by the nominal 
model predictions for many of the discussed distributions.
The \chisq values that include the prediction uncertainties for \PowPytSh 
(see Tables~\ref{tab:chi2mc_nor_parton_addjets}--\ref{tab:chi2mc_nor_particle_addjets}) are significantly 
reduced, but the corresponding $p$-values remain too small for a reasonable description of several measured observables.

\clearpage
\begin{table*}
 \centering
 \topcaption{The \chisq values and \ndf of the measured normalized differential cross sections as a function of the additional-jet multiplicity in the events, at the parton level of the top quark and antiquark, are shown with respect to the predictions of various MC generators. 
The \chisq values are calculated taking only measurement uncertainties into account and excluding theory uncertainties.  
For \PowPytSh, the \chisq values including theory uncertainties are indicated with the brackets (w. unc.).}
 \label{tab:chi2mc_nor_parton_addjets}
 \renewcommand{\arraystretch}{1.4}
 \centering
 \begin{tabular}{lccccc}
 \multirow{1}{*}{Cross section} & \hspace*{0.3 cm} \multirow{2}{*}{\ndf} \hspace*{0.3 cm} & \multicolumn{3}{c}{\chisq} \\
 \cline{3-5}
{variables} && \PowPytSh (w. unc.)  & \aMCPytSh  & \PowHerSh   \\
\hline
\njforty & 5 & 6 \: (3) & 280 & 251 \\
\njhundred & 4 & 27 \: (8) & 34 & 68 \\
\njptt& 8 & 22 \: (12) & 161 & 124 \\
\njyt& 11 & 38 \: (29) & 128 & 85 \\
\njpttt& 11 & 50 \: (37) & 189 & 92 \\
\njmtt  & 11 & 56 \: (41) & 140 & 151 \\
\njytt& 11 & 11 \: (5) & 121 & 60 \\
\njdetatt& 8 & 84 \: (37) & 189 & 143 \\
\njmttytttwo  & 23 & 48 \: (35) & 73 & 92 \\
\njmttyttthree  & 35 & 83 \: (57) & 212 & 211 \\
\njmttyttfour  & 47 & 122 \: (84) & 425 & 462 \\
 \end{tabular} 
\end{table*}

\begin{table*}
 \centering
 \topcaption{The \chisq values and \ndf of the measured normalized differential cross sections as a function of the additional-jet multiplicity in the events, at the particle level of the top quark and antiquark, are shown with respect to the predictions of various MC generators. 
The \chisq values are calculated taking only measurement uncertainties into account and excluding theory uncertainties.  
For \PowPytSh, the \chisq values including theory uncertainties are indicated with the brackets (w. unc.).}
 \label{tab:chi2mc_nor_particle_addjets}
 \renewcommand{\arraystretch}{1.4}
 \centering
 \begin{tabular}{lccccc}
 \multirow{1}{*}{Cross section} & \hspace*{0.3 cm} \multirow{2}{*}{\ndf} \hspace*{0.3 cm} & \multicolumn{3}{c}{\chisq} \\
 \cline{3-5}
{variables} && \PowPytSh (w. unc.)  & \aMCPytSh  & \PowHerSh   \\
\hline
\njforty & 5 & 6 \: (3) & 340 & 9 \\
\njhundred & 4 & 31 \: (8) & 34 & 6 \\
\njptt& 8 & 30 \: (13) & 237 & 26 \\
\njyt& 11 & 39 \: (24) & 174 & 28 \\
\njpttt& 11 & 58 \: (37) & 327 & 89 \\
\njmtt  & 11 & 53 \: (35) & 283 & 51 \\
\njytt& 11 & 14 \: (5) & 178 & 9 \\
\njdetatt& 8 & 124 \: (41) & 290 & 107 \\
\njmttytttwo  & 23 & 75 \: (45) & 96 & 91 \\
\njmttyttthree  & 35 & 127 \: (69) & 379 & 142 \\
\njmttyttfour  & 47 & 156 \: (94) & 699 & 167 \\
 \end{tabular}
\end{table*}

\clearpage

\subsection{Comparisons to higher-order theoretical predictions}
\label{sec:res_theory_comp}
In this Section, the measured cross sections are compared to the following calculations
of beyond-NLO precision in QCD:
\begin{itemize}
        \item \textbf{\appNTLO}:
An approximate next-to-NNLO calculation \cite{bib:kidonakis_13TeV,bib:Kidonakis:2019yji},
based on the resummation of soft-gluon contributions at NNLL accuracy in the moment-space approach.
This prediction is only available for the \ptt, \yt, and \ytptt cross sections.
The renormalization and factorization scales are set to \mt for the \yt distributions
and to \mT for \ptt and \ytptt. Here, \mt denotes the top quark mass
and \mT is defined as $\mT = \sqrt{\mt^2 + \ptt^2}$.

\item \textbf{\Matrix}: A prediction with full NNLO accuracy in QCD obtained with
 the \textsc{Matrix} package~\cite{bib:Grazzini:2017mhc, bib:Catani:2019hip, bib:Catani:2019iny, bib:Buccioni:2019sur, bib:Cascioli:2011va, bib:Denner:2016kdg, bib:Catani:2012qa, bib:Catani:2007vq}.
The dynamic scales are set to $\HT/4$, where \HT denotes the sum of the top quark and antiquark \mT values.

\item \textbf{\Stripper}: A calculation with full NNLO precision in QCD using the
\textsc{Stripper} framework~\cite{bib:Czakon:2010td, Czakon:2020qbd, bib:Czakon:2014oma, bib:Czakon:2011ve, bib:Czakon:2017wor}.
The dynamic scales are set to $\HT/4$.
For parton level cross sections 
as functions of the \ttbar and top quark kinematic observables, the \textsc{Stripper} predictions are expected 
to be identical to the results obtained with \textsc{Matrix}. 
The \textsc{Stripper} calculation also provides cross sections at NNLO accuracy for the process 
 {\ensuremath{\pp \to \ttbar \to \PQb \PAQb \Pell\PAell\PGn\PAGn + \PX}}~\cite{Czakon:2020qbd},  
that can be compared to the particle level measurements obtained 
in this analysis as functions of the \ttbar, top quark, and lepton and \PQb jet kinematic variables.

\item \textbf{\MiNNLOPS}: A prediction with full NNLO precision in QCD, complemented with parton showers and
computed using the \POWHEG-BOX-V2 \cite{bib:powheg2}.
These calculations are obtained using the \textsc{MiNNLOPS} 
method~\cite{bib:Monni:2019whf,bib:Monni:2020nks,bib:Mazzitelli:2021mmm},
which supplements the \textsc{MiNLO} prescription \cite{bib:Hamilton:2012np, bib:Hamilton:2012rf}
with the missing pieces to reach NNLO accuracy for inclusive observables.
The \mur for the two powers of \alpS is set to $\HT/4$,
and the scale of the modified logarithms is set to half of this value.
Parton showering is simulated with \PYTHIAviii, and includes the effects of underlying event and hadronization.
\end{itemize}

In all predictions, the top quark mass is set to $\mt = 172.5\GeV$
and the NNPDF3.1 NNLO PDF set~\cite{Ball:2017nwa} is used.
In the following, the calculations are
collectively referred to as theoretical predictions.

\begin{figure*}[!phtb]
\centering
\includegraphics[width=0.49\textwidth]{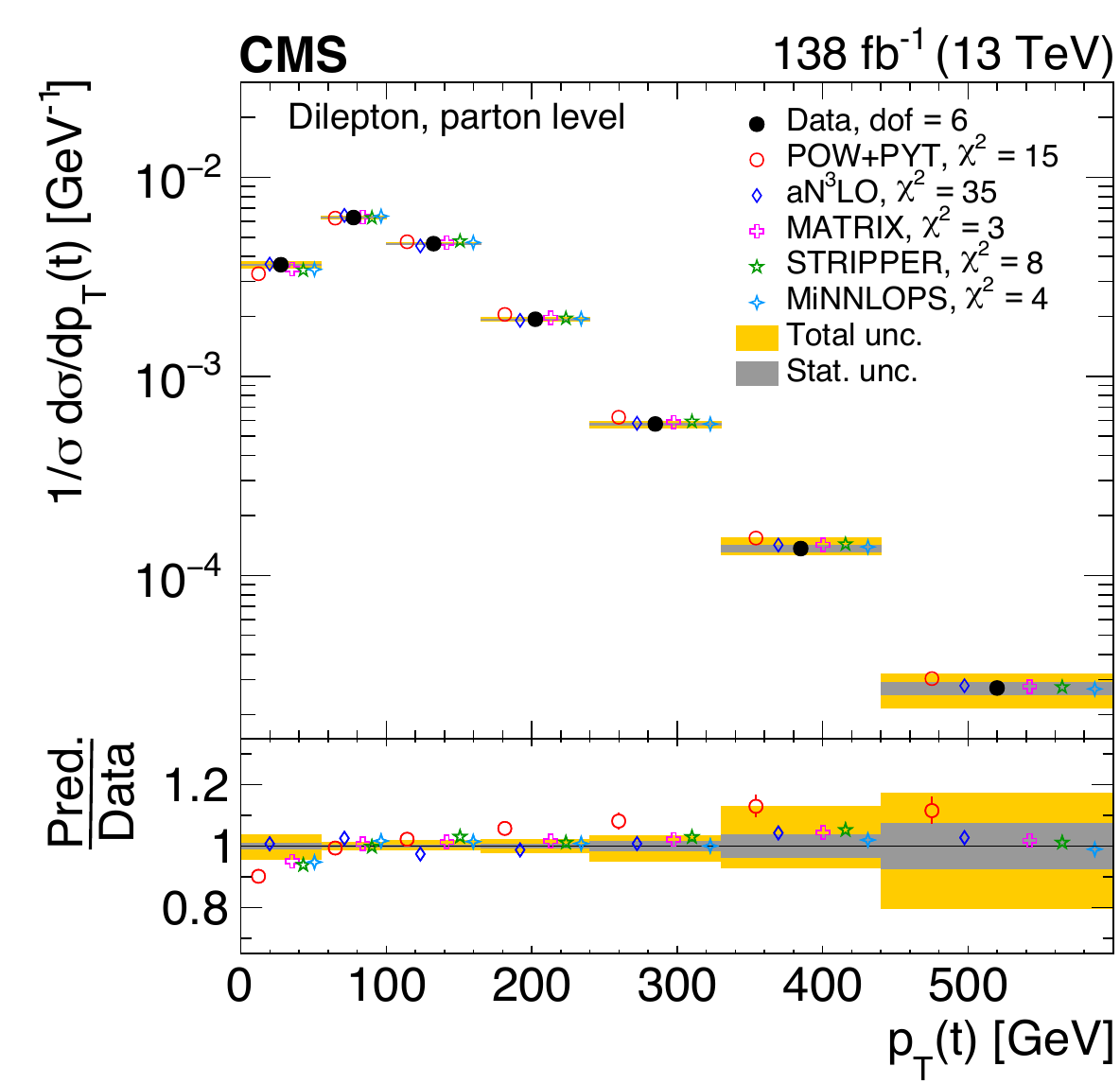}
\includegraphics[width=0.49\textwidth]{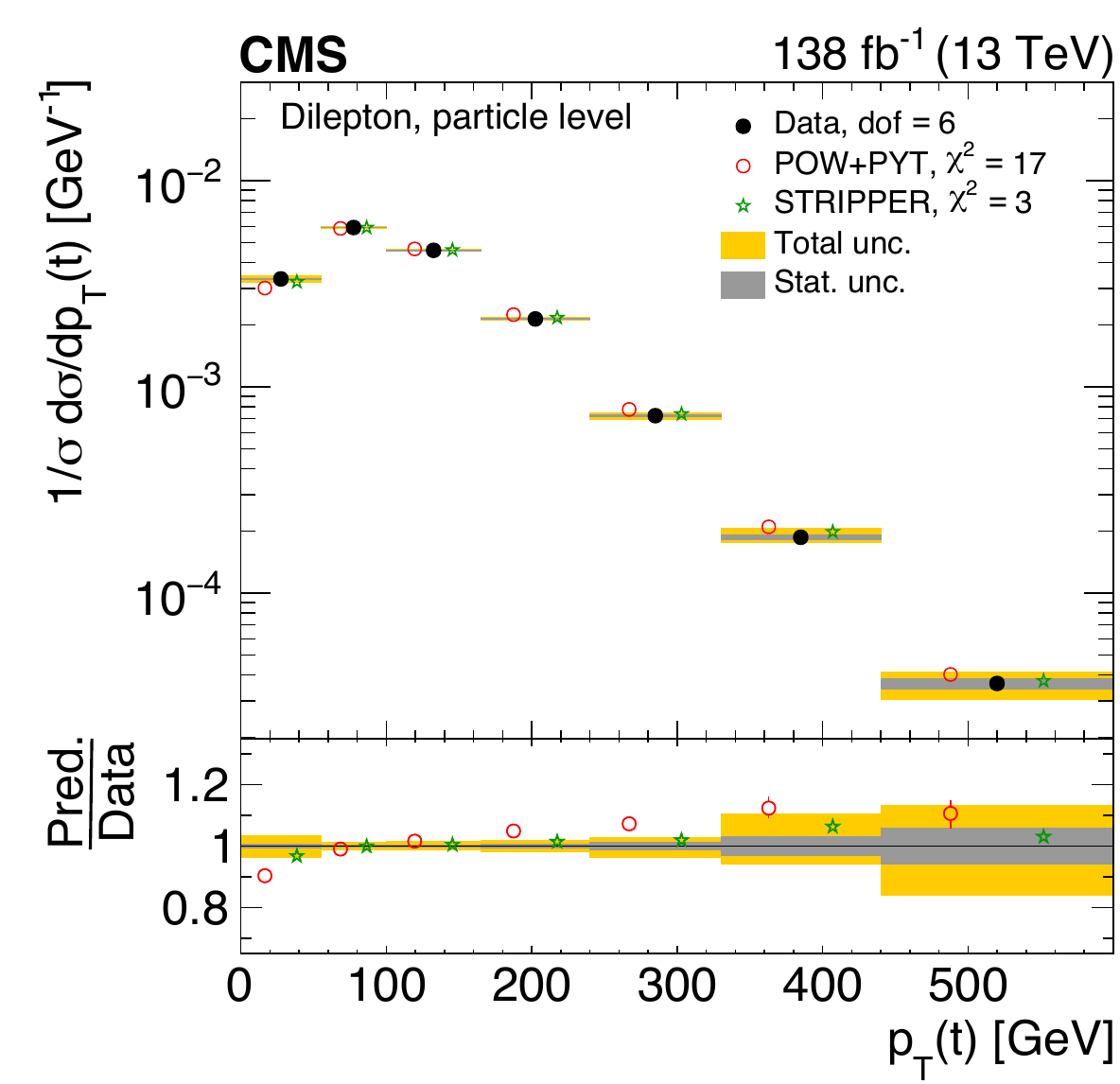}
\includegraphics[width=0.49\textwidth]{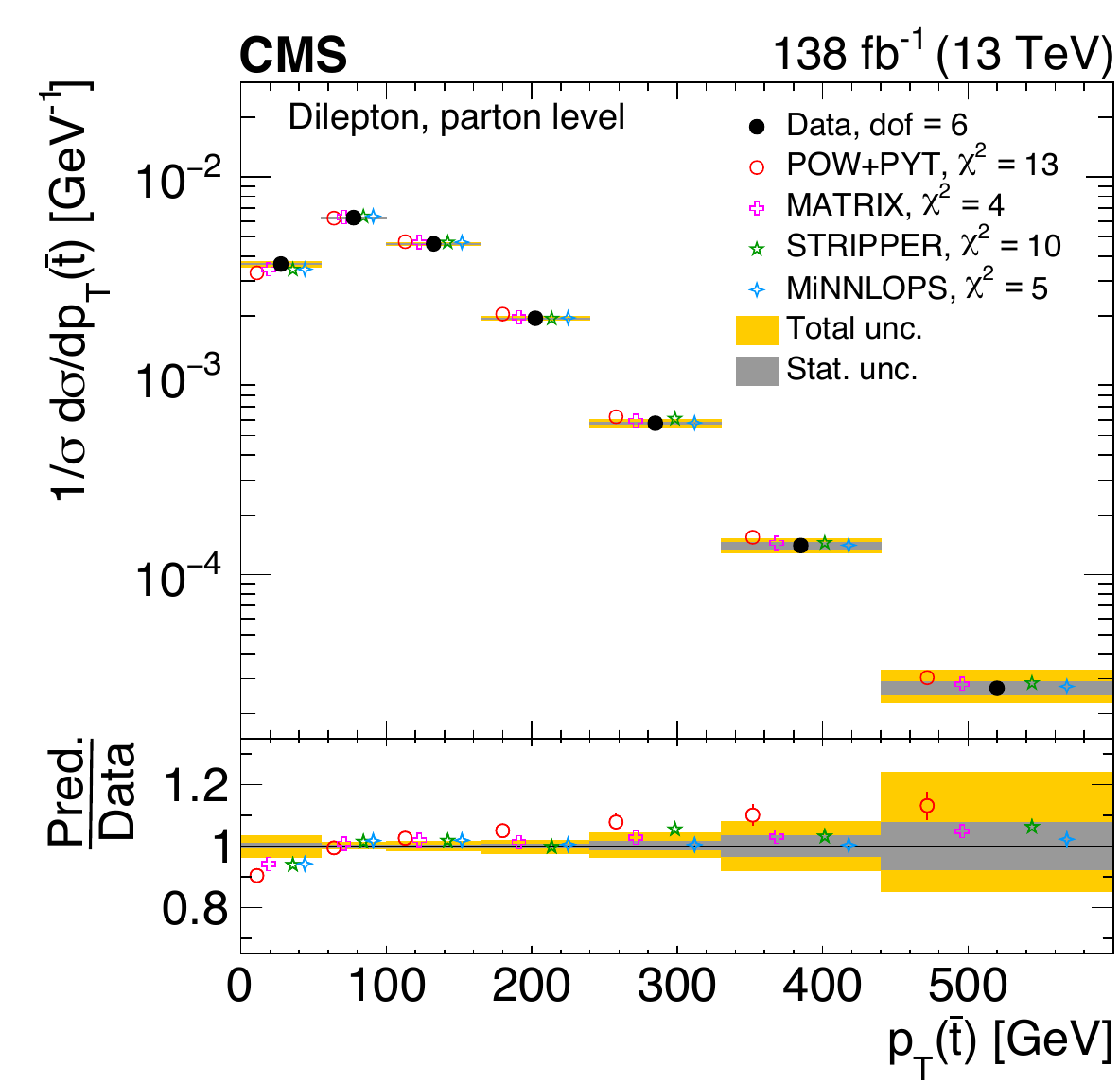}
\includegraphics[width=0.49\textwidth]{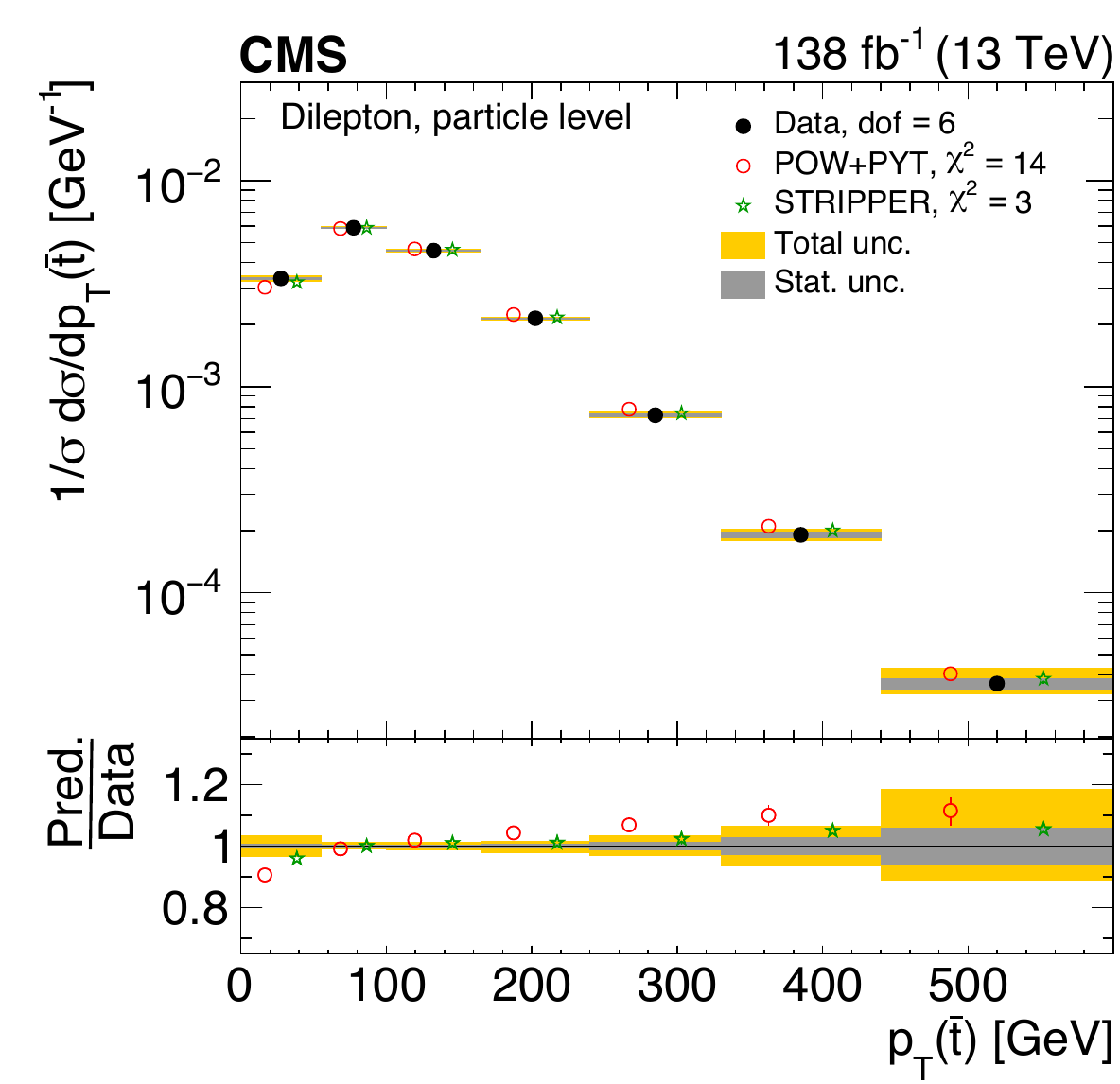}
\caption {Normalized differential \ttbar production cross sections as functions of \ptt (upper) and \ptat (lower),
 measured
at the parton level in the full phase space (left) and at the particle level in a fiducial phase space (right).
The data are shown as filled circles with grey and yellow bands indicating the statistical and total uncertainties
(statistical and systematic uncertainties added in quadrature), respectively.
For each distribution, the number of degrees of freedom (dof) is also provided.
The cross sections are compared to predictions from the \PowPyt (`POW-PYT', open circles) simulation and various
theoretical predictions with beyond-NLO precision (other points).
The estimated uncertainties in the \PowPytSh model are represented by vertical bars on the corresponding points.
For each model, a value of \chisq is reported that takes into account the measurement uncertainties.
The lower panel in each plot shows the ratios of the predictions to the data.}
\label{fig:xsec-1d-theory-nor-ptt-ptat}
\end{figure*}

\begin{figure*}[!phtb]
\centering
\includegraphics[width=0.49\textwidth]{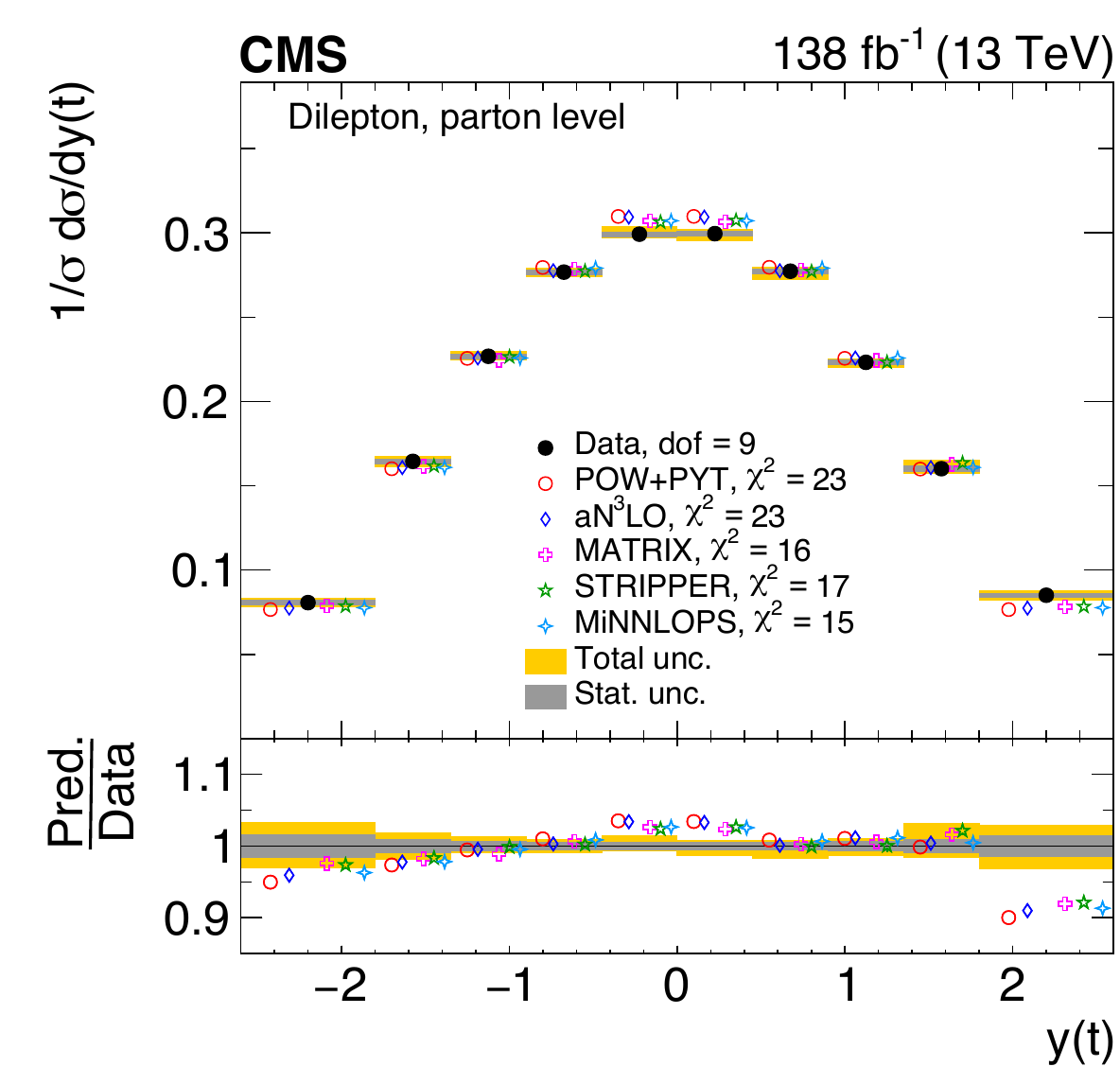}
\includegraphics[width=0.49\textwidth]{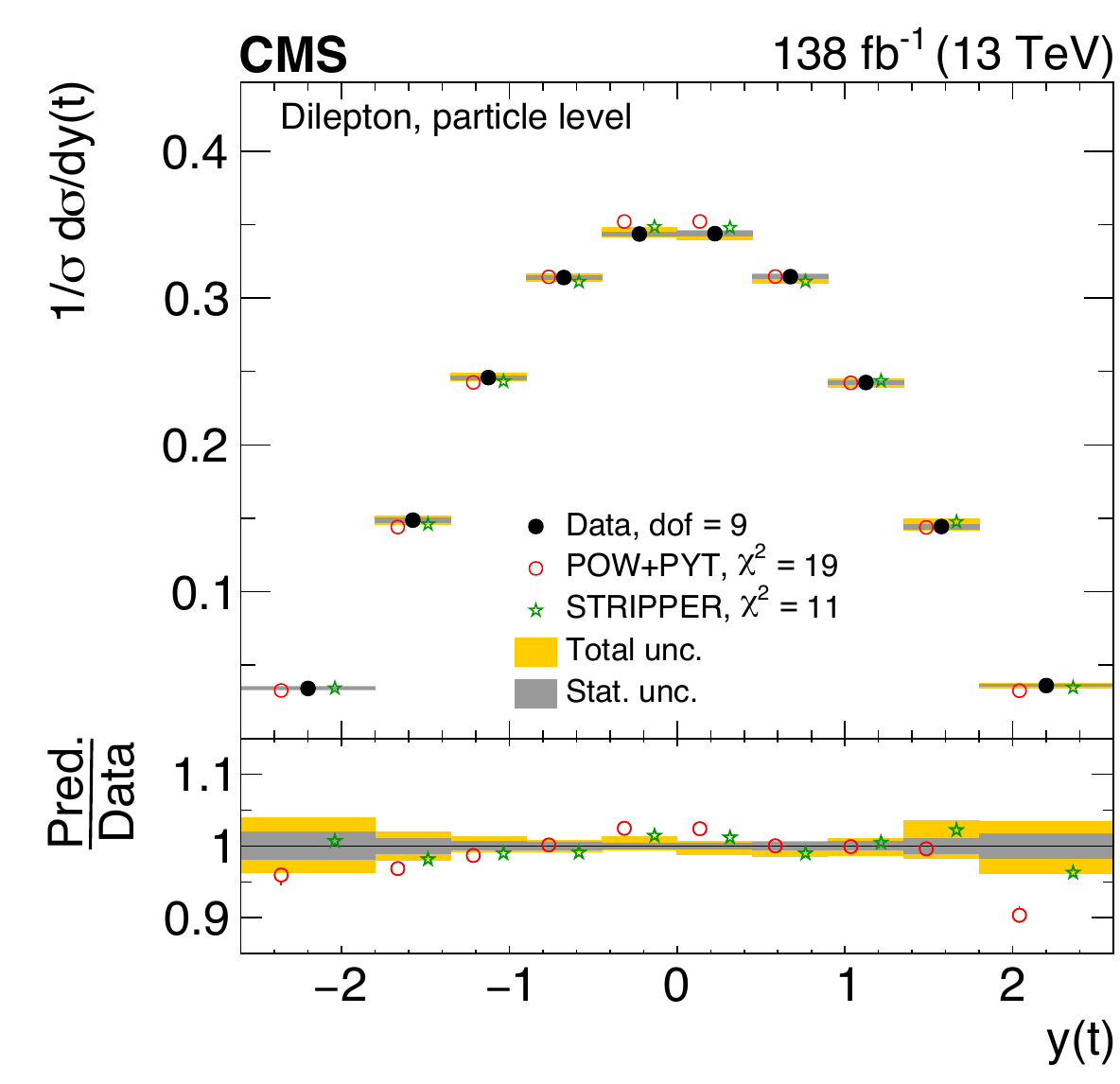}
\includegraphics[width=0.49\textwidth]{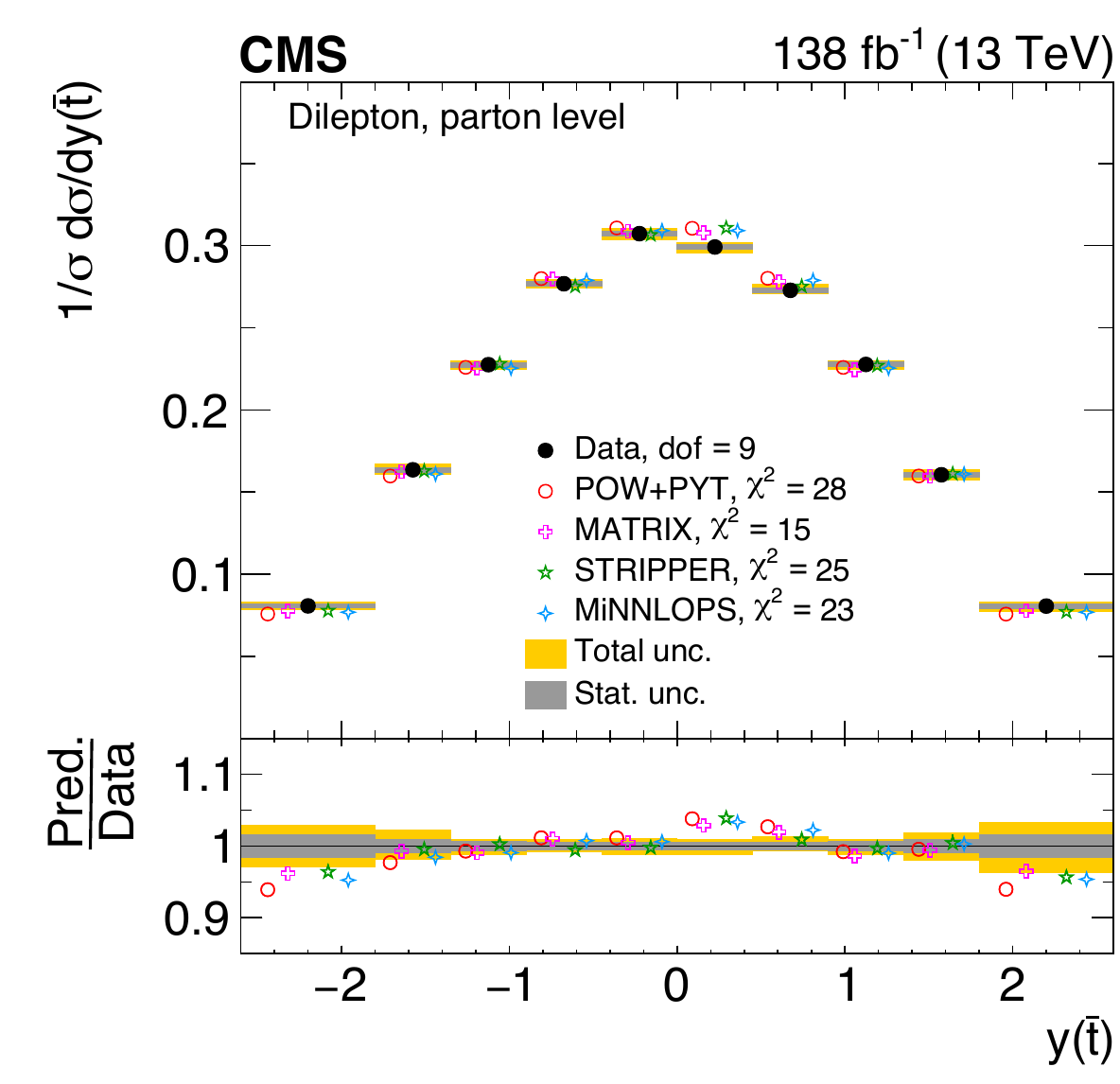}
\includegraphics[width=0.49\textwidth]{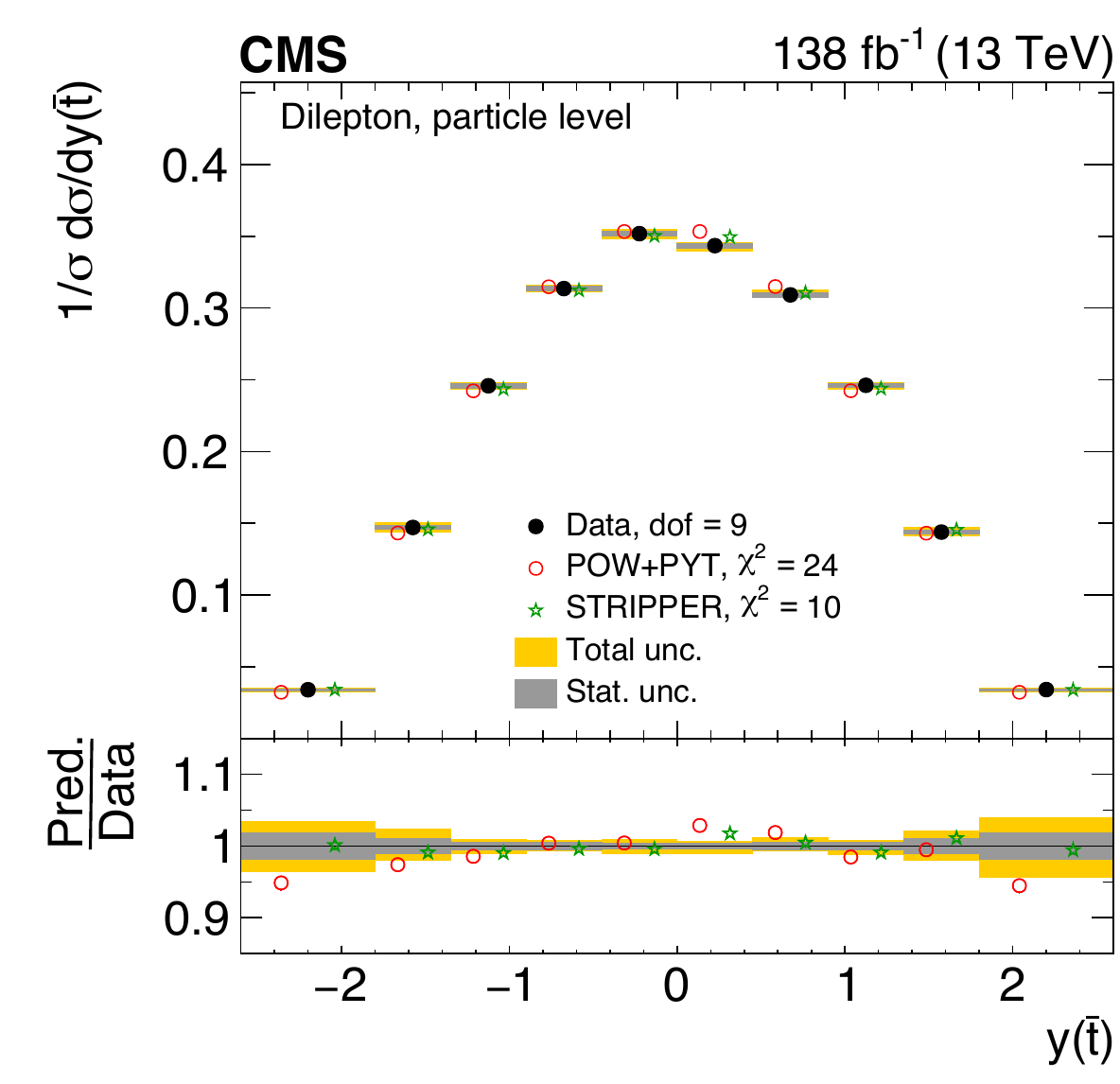}
\caption{Normalized differential \ttbar production cross sections as functions of \yt (upper) and \yat (lower)
are shown for data (filled circles), \PowPyt (`POW-PYT', open circles) simulation, and various theoretical
predictions with beyond-NLO precision (other points).
Further details can be found in the caption of Fig.~\ref{fig:xsec-1d-theory-nor-ptt-ptat}.}
\label{fig:xsec-1d-theory-nor-yt-yat}
\end{figure*}

\begin{figure*}[!phtb]
\centering
\includegraphics[width=0.49\textwidth]{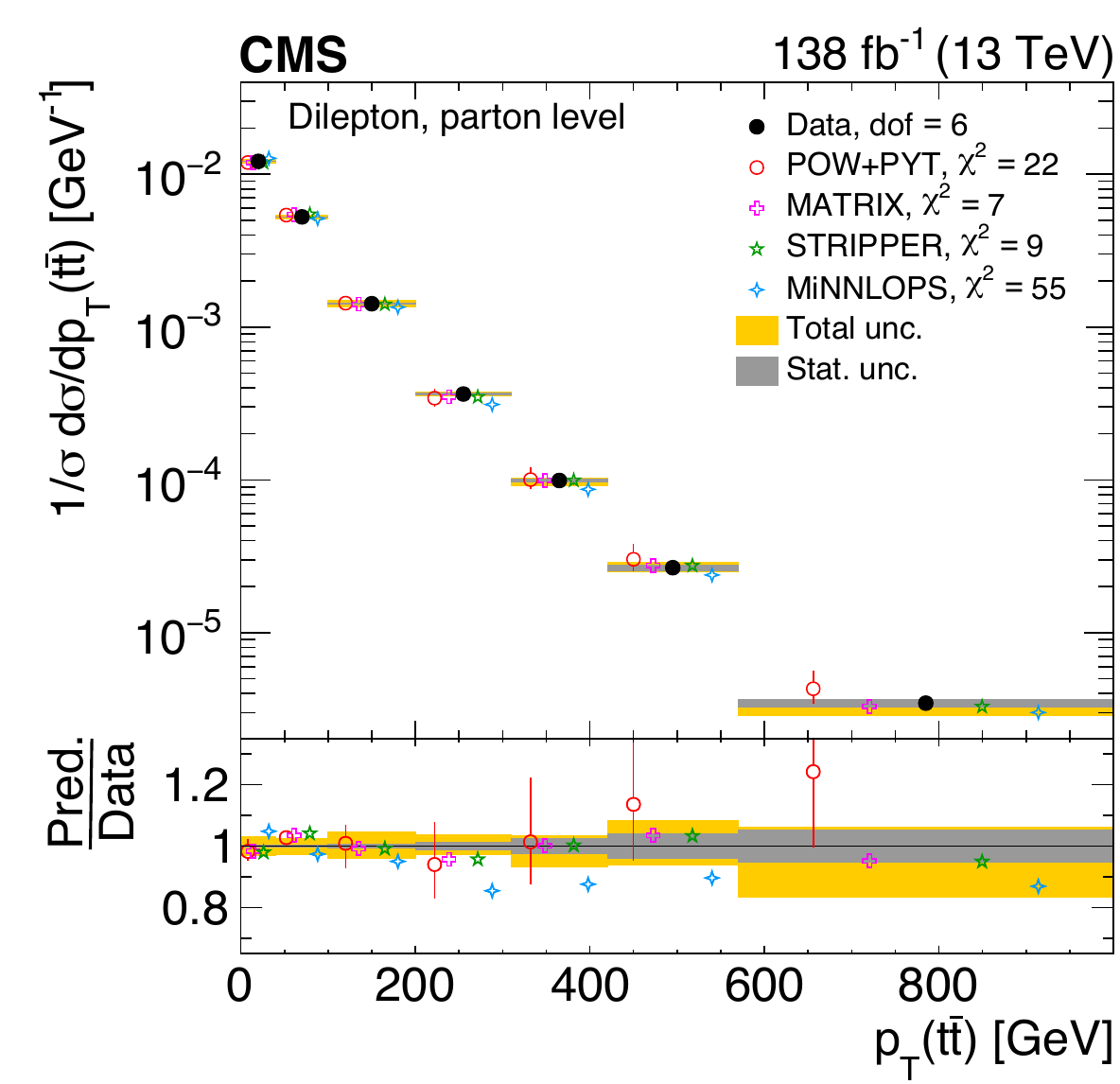}
\includegraphics[width=0.49\textwidth]{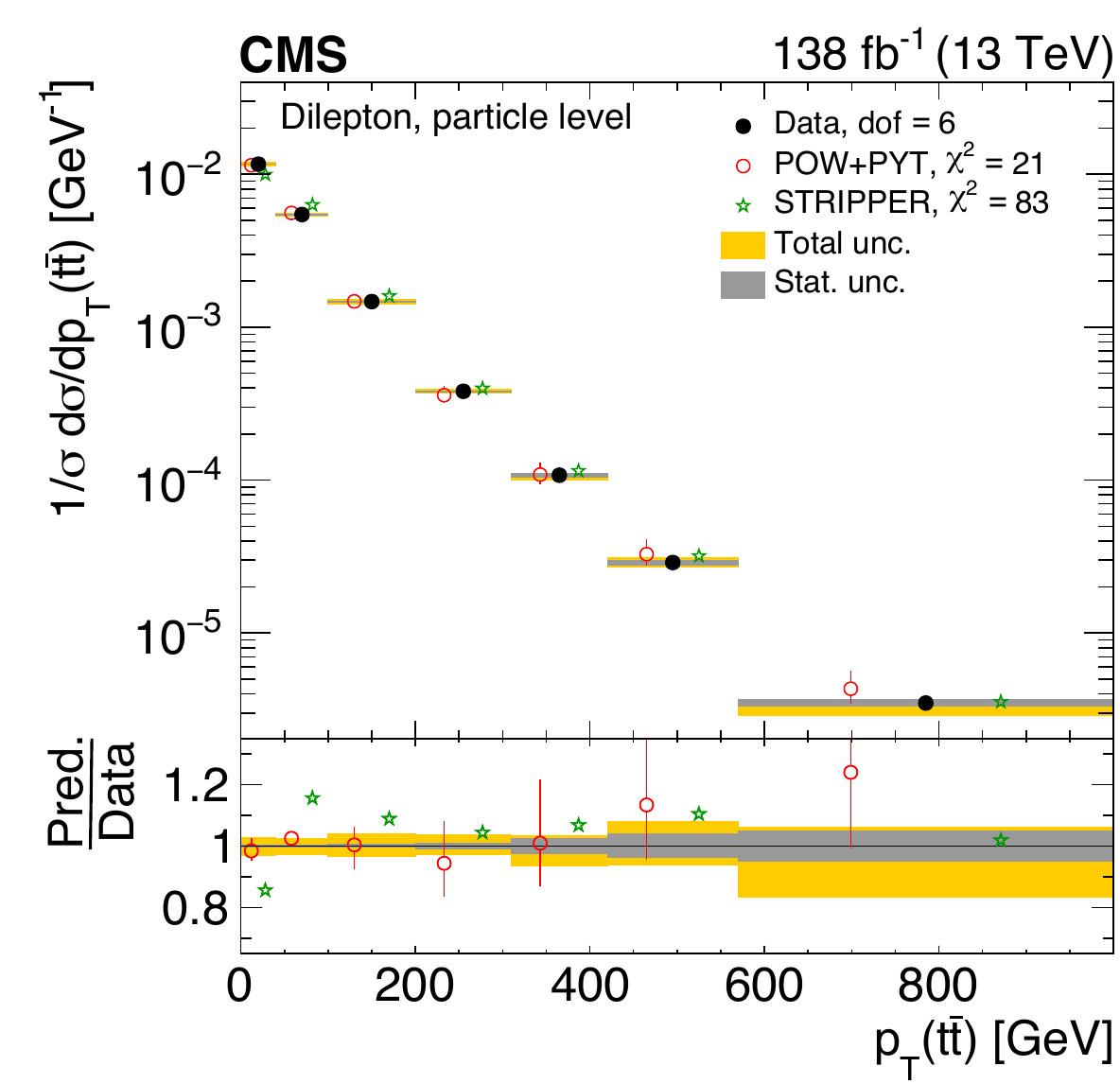}
\includegraphics[width=0.49\textwidth]{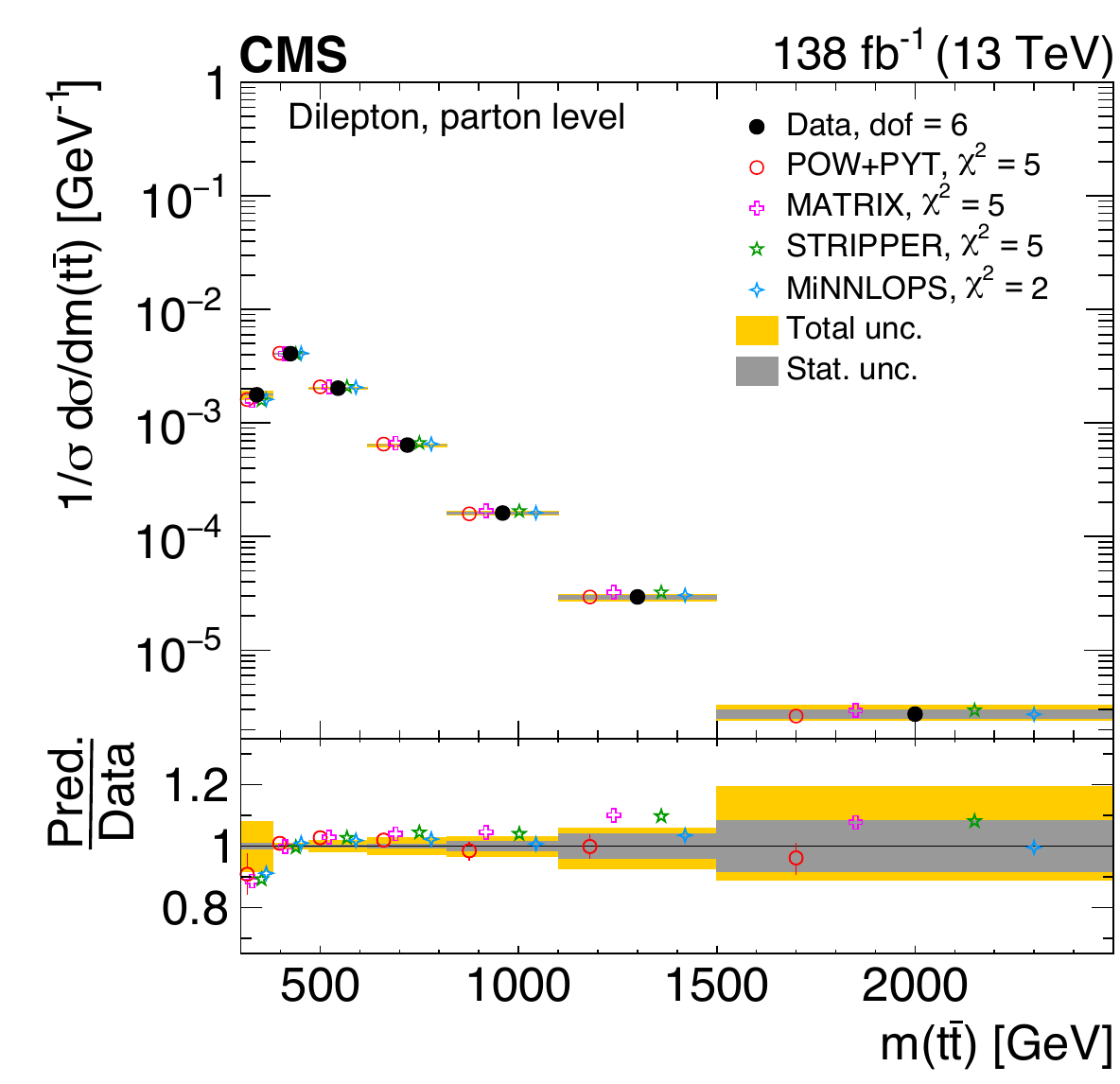}
\includegraphics[width=0.49\textwidth]{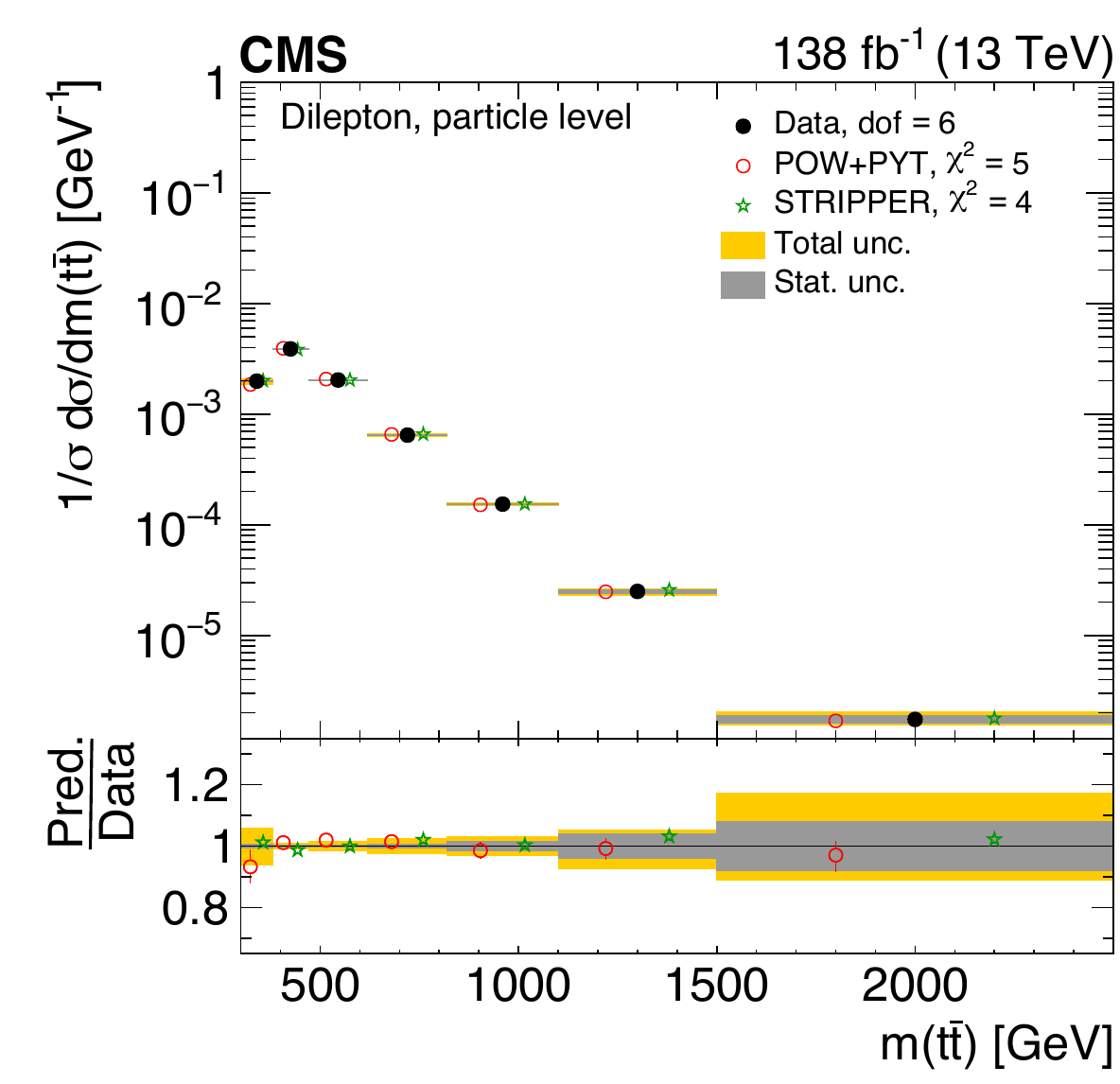}
\includegraphics[width=0.49\textwidth]{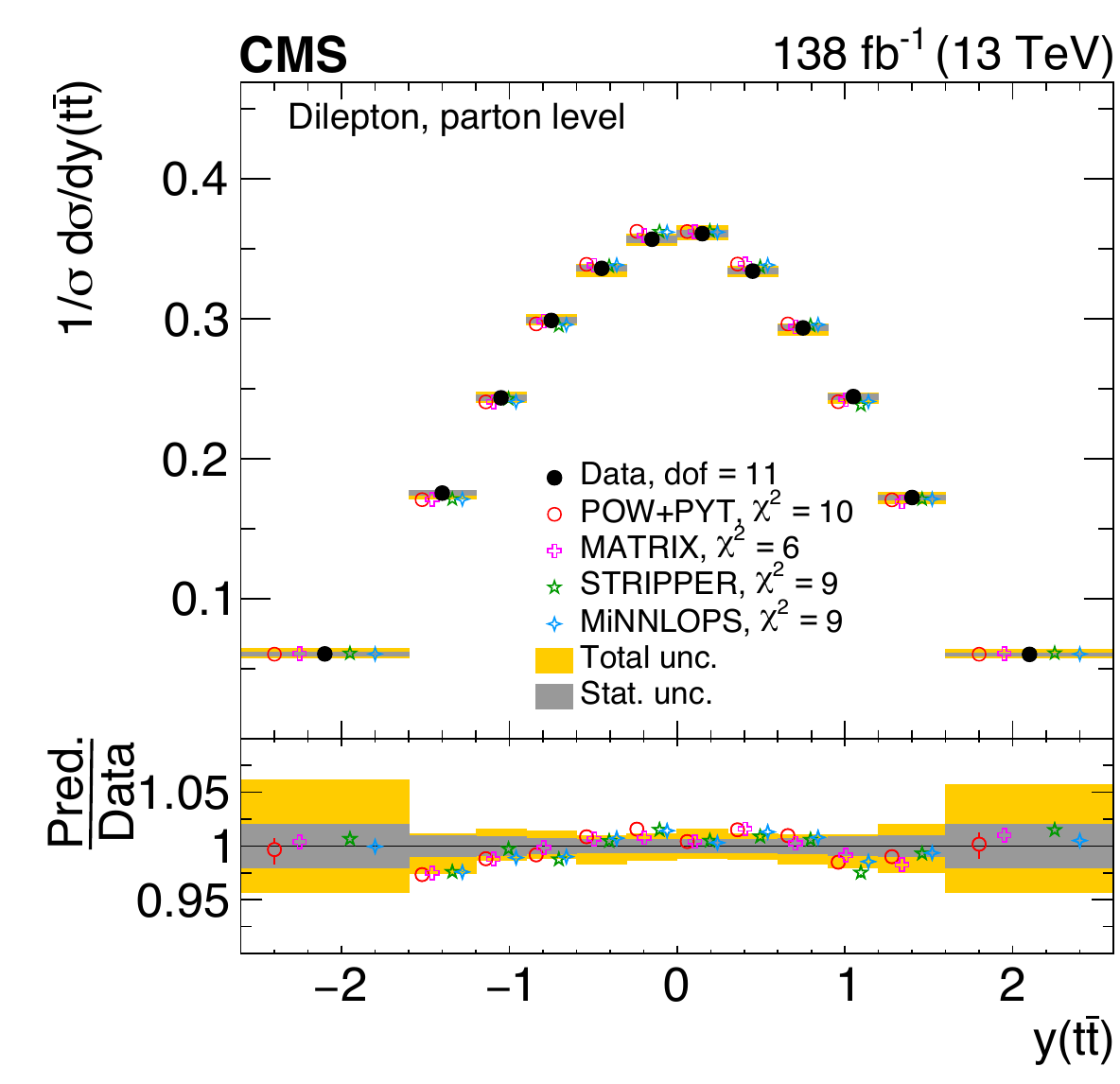}
\includegraphics[width=0.49\textwidth]{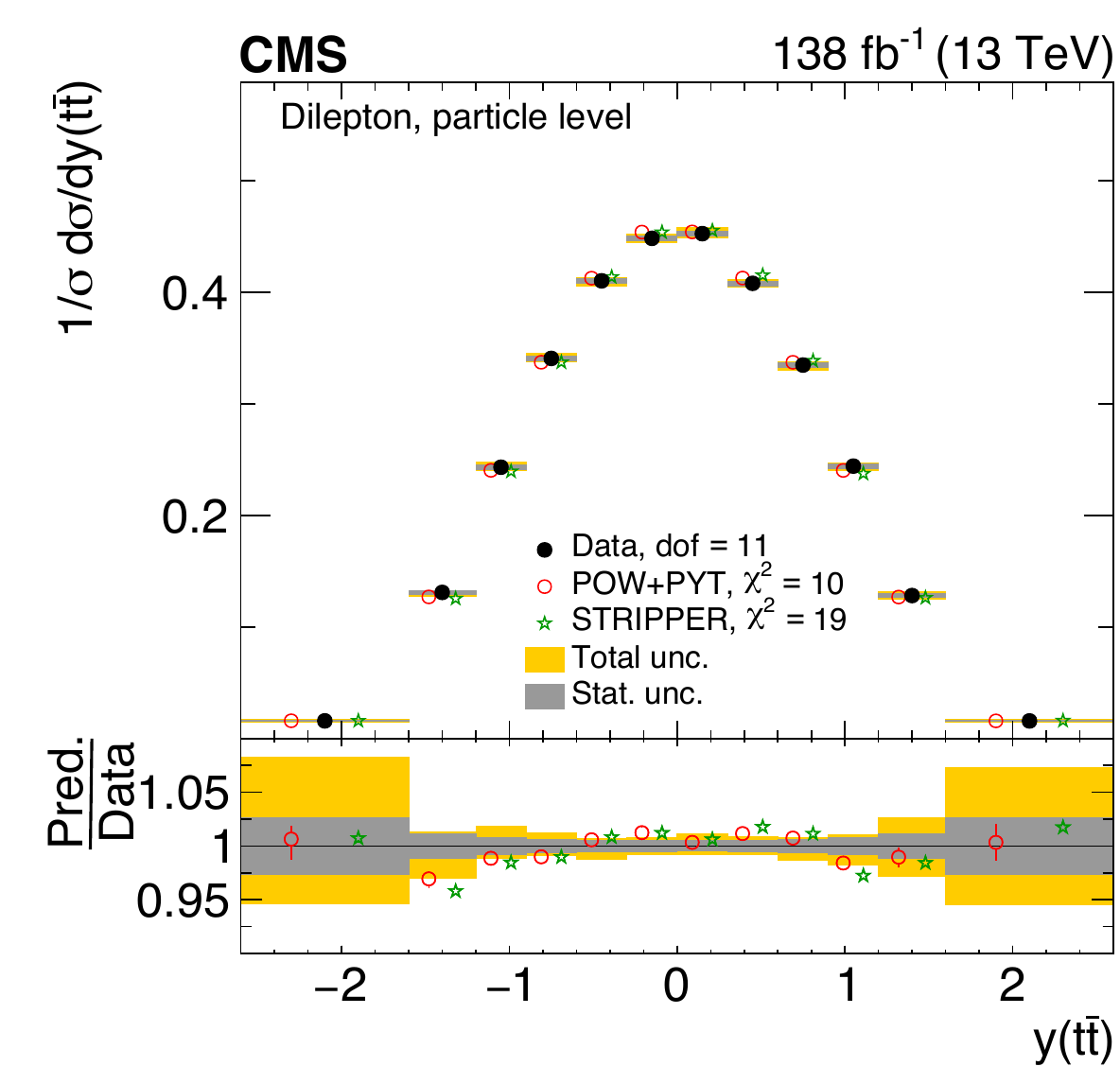}
\caption{Normalized differential \ttbar production cross sections as functions of \pttt (upper), \mtt (middle),
and \ytt (lower)
are shown for data (filled circles), \PowPyt (`POW-PYT', open circles) simulation, and various theoretical
predictions with beyond-NLO precision (other points).
Further details can be found in the caption of Fig.~\ref{fig:xsec-1d-theory-nor-ptt-ptat}.}
\label{fig:xsec-1d-theory-nor-pttt-mtt-ytt}
\end{figure*}

\begin{figure*}[!phtb]
\centering
\includegraphics[width=0.49\textwidth]{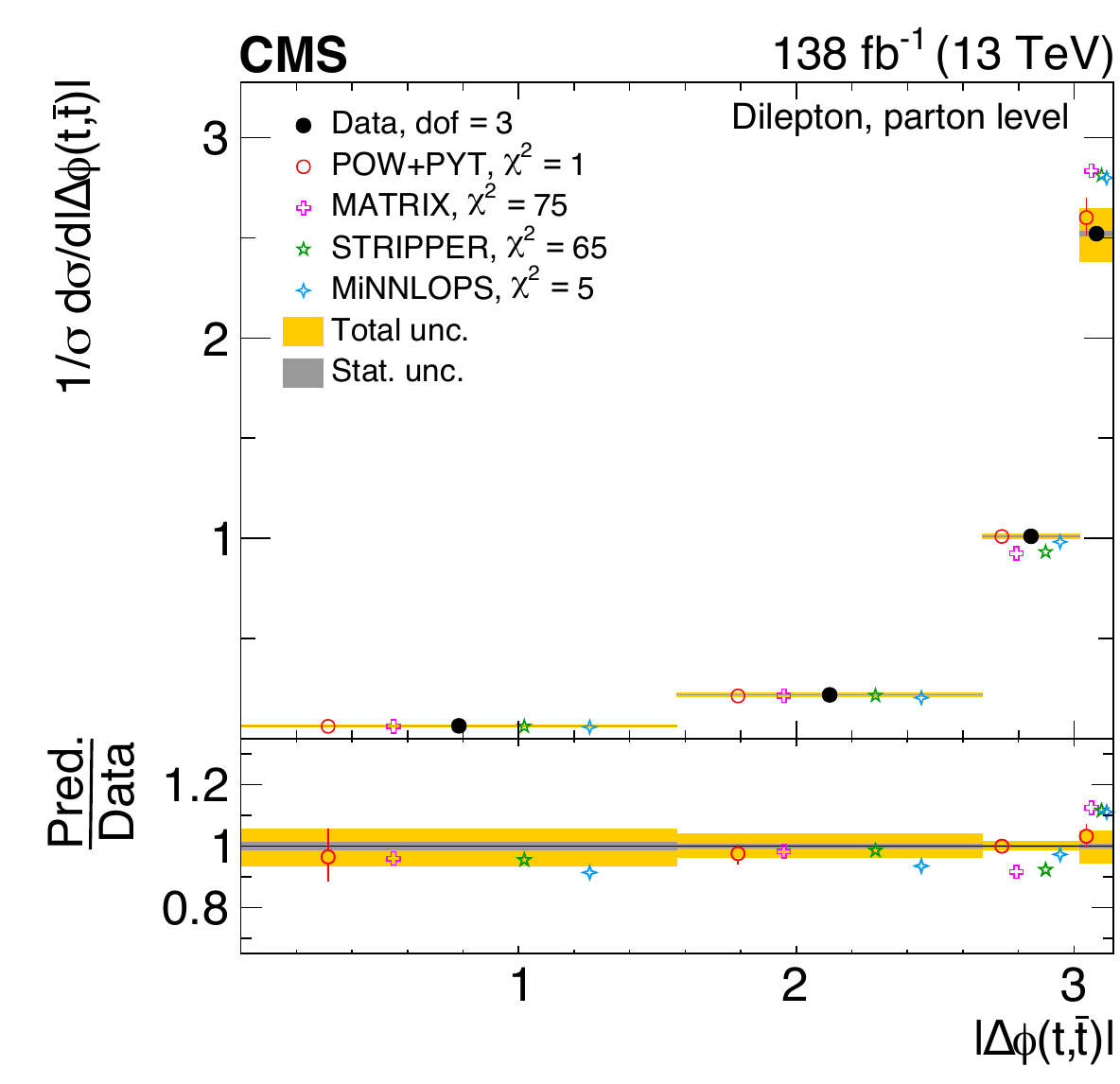}
\includegraphics[width=0.49\textwidth]{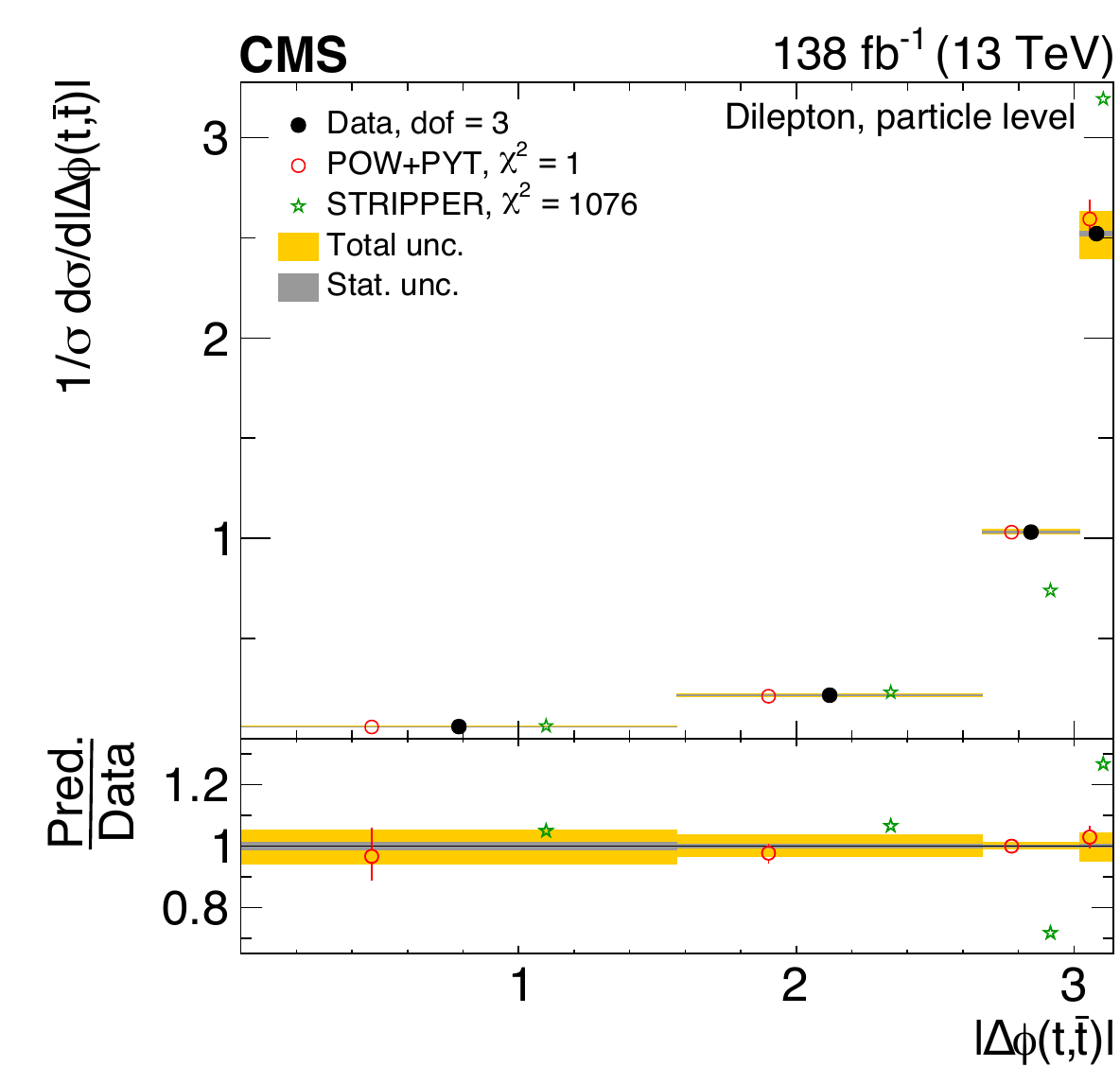}
\includegraphics[width=0.49\textwidth]{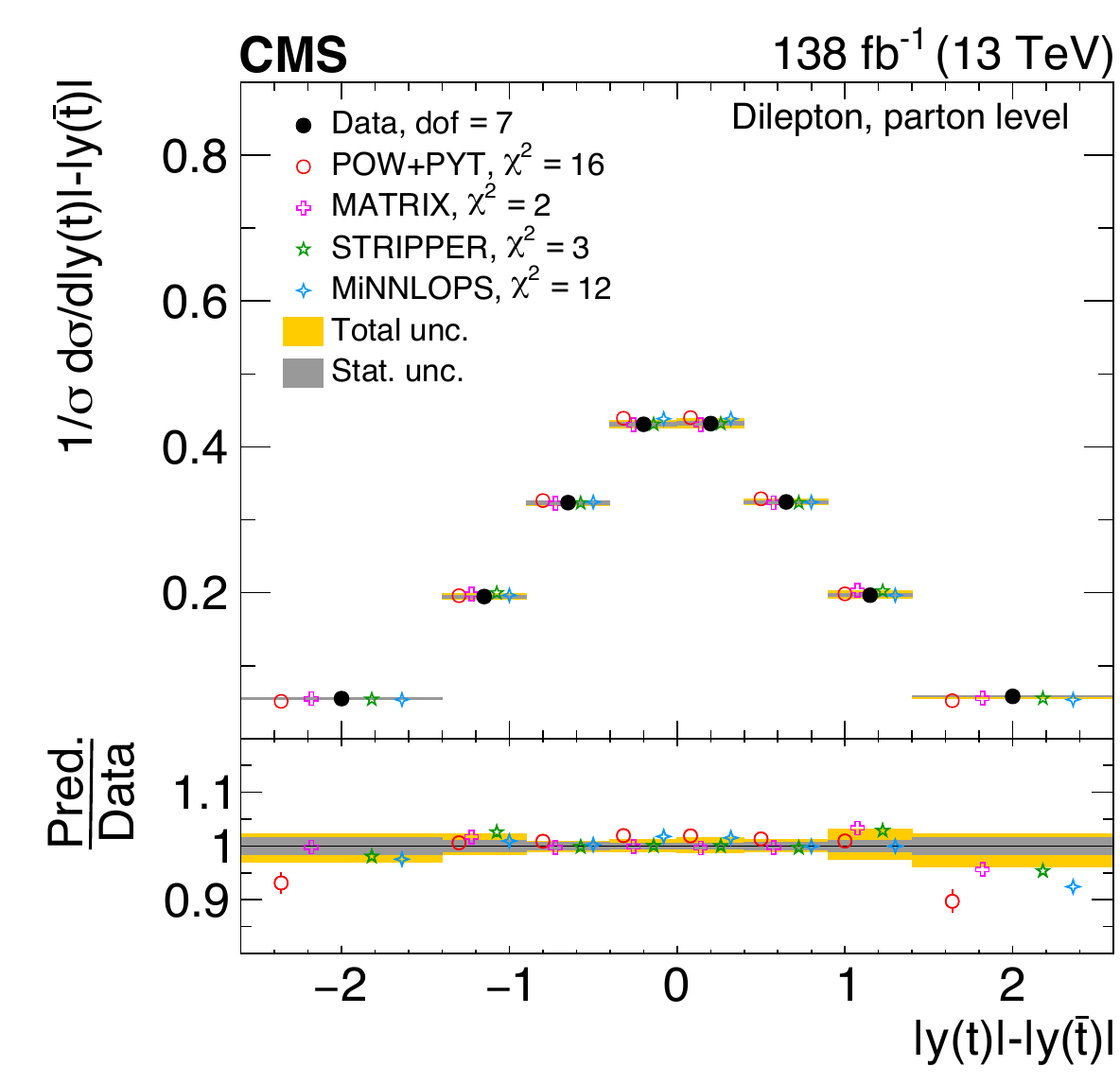}
\includegraphics[width=0.49\textwidth]{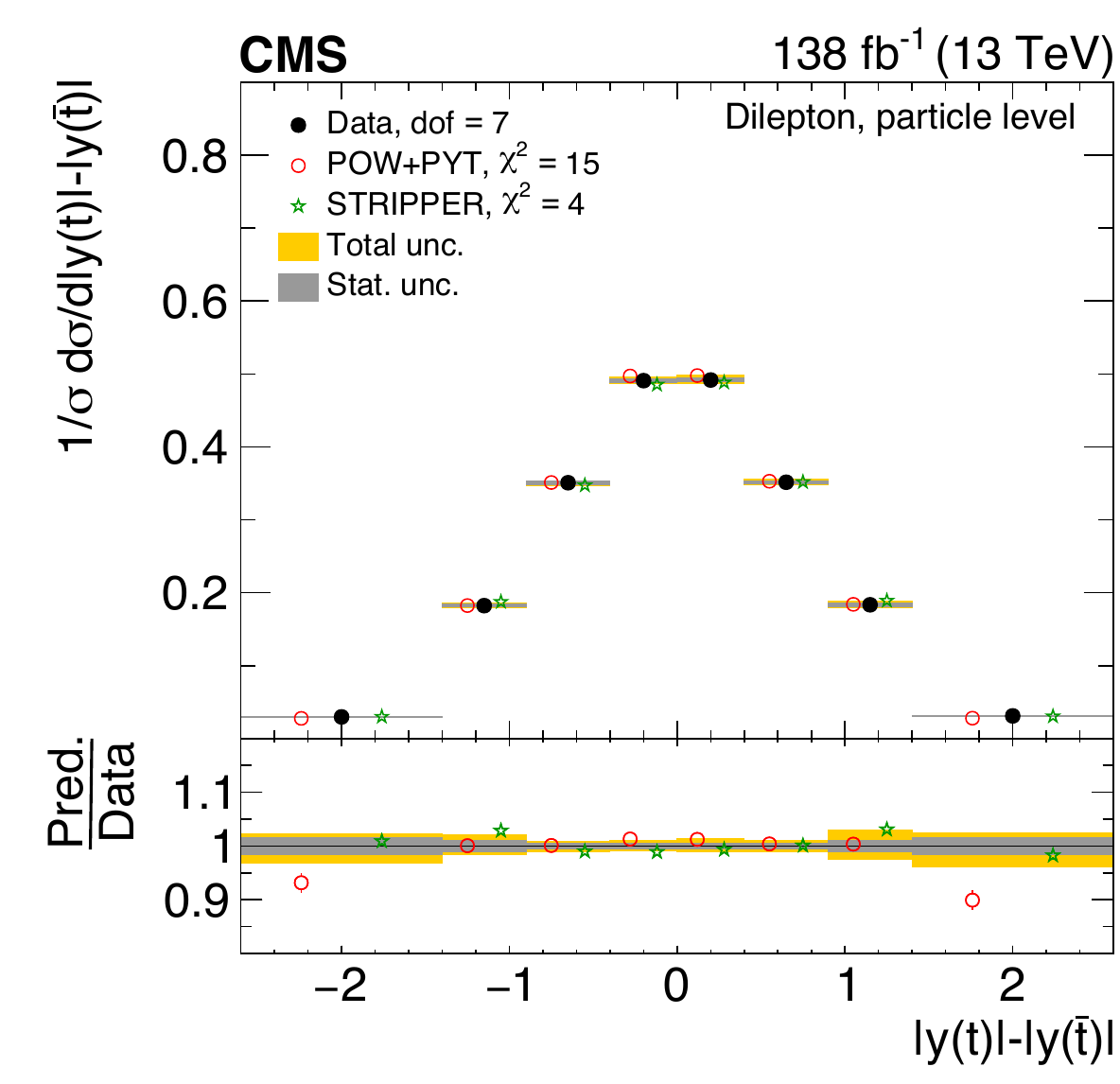}
\caption{Normalized differential \ttbar production cross sections as functions of \dphitt (upper) and \dytt
(lower)
are shown for data (filled circles), \PowPyt (`POW-PYT', open circles) simulation, and various theoretical
predictions with beyond-NLO precision (other points).
Further details can be found in the caption of Fig.~\ref{fig:xsec-1d-theory-nor-ptt-ptat}.}
\label{fig:xsec-1d-theory-nor-dphitt-dytt}
\end{figure*}

\begin{figure*}[!phtb]
\centering
\includegraphics[width=0.49\textwidth]{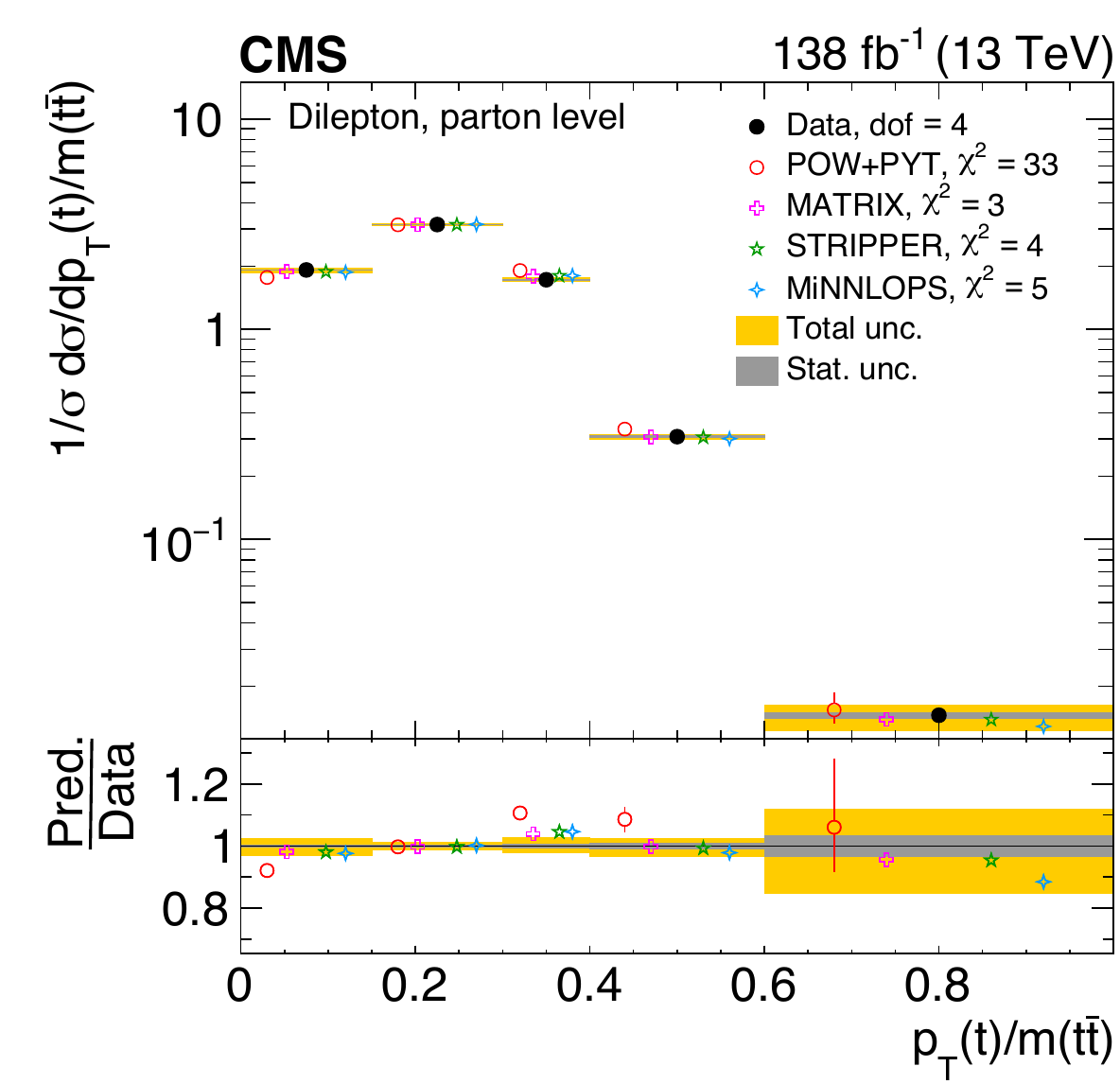}
\includegraphics[width=0.49\textwidth]{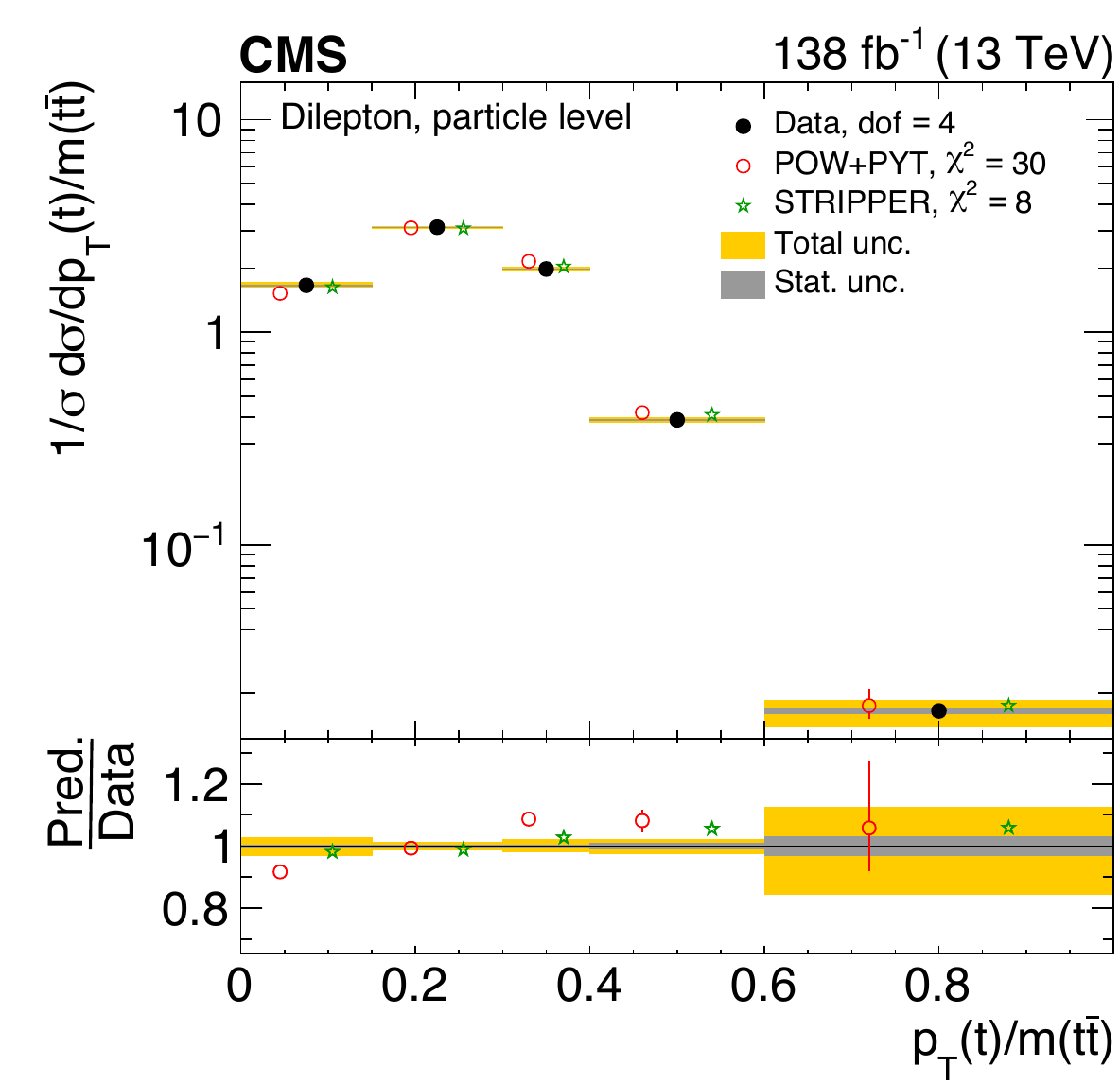}
\includegraphics[width=0.49\textwidth]{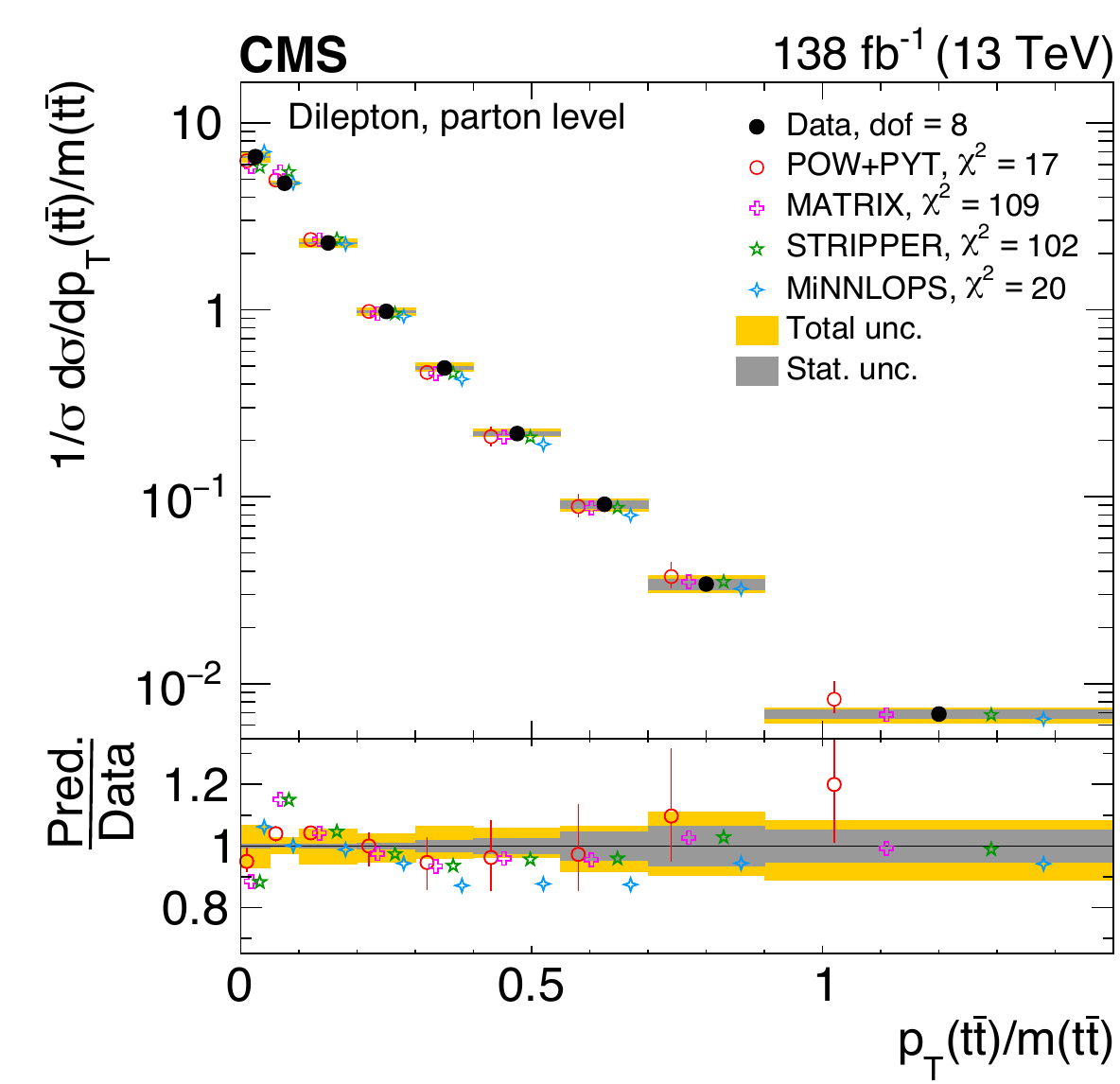}
\includegraphics[width=0.49\textwidth]{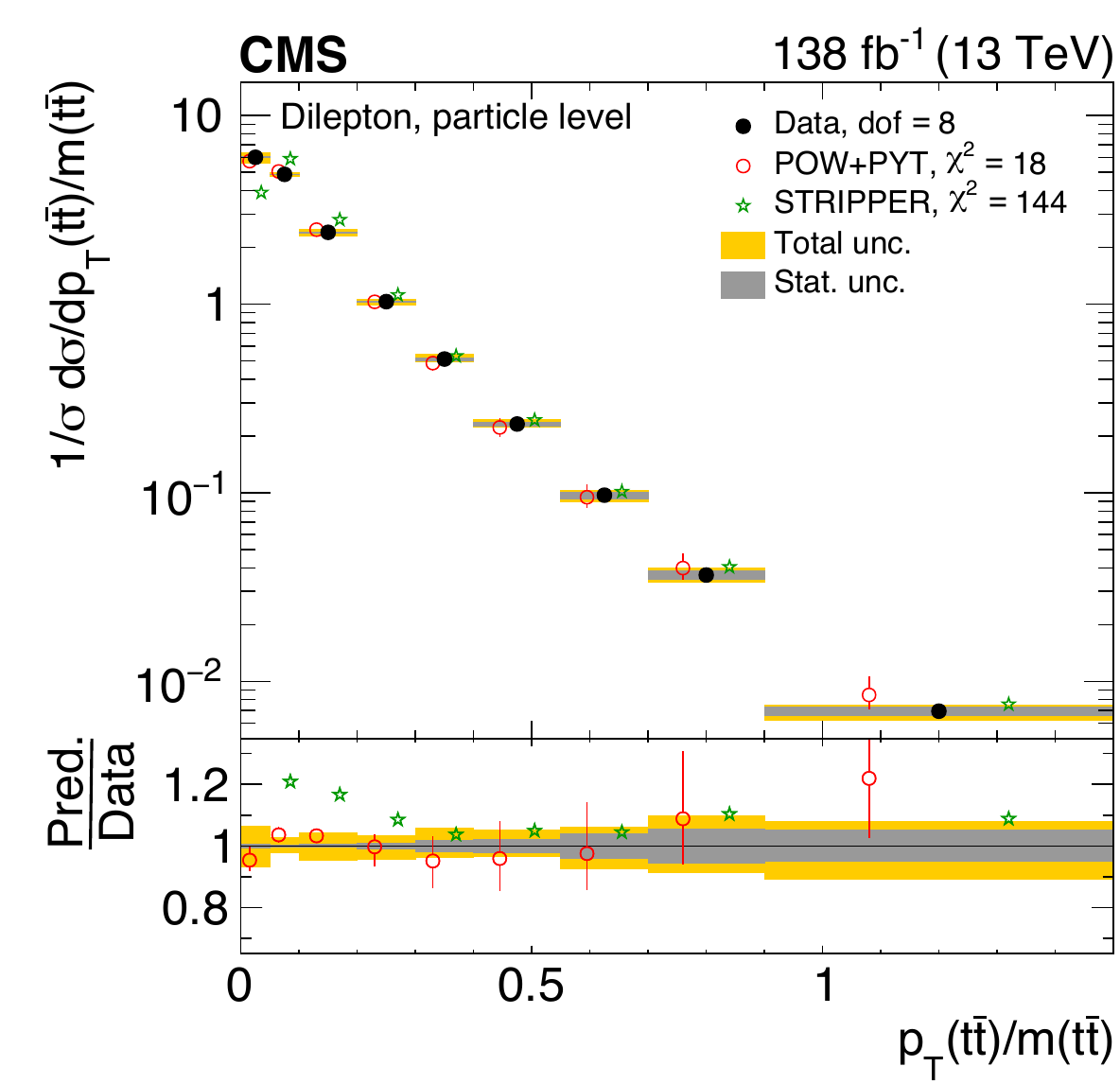}
\caption{Normalized differential \ttbar production cross sections as functions of \rpttmtt (upper) and \rptttmtt
(lower)
are shown for data (filled circles), \PowPyt (`POW-PYT', open circles) simulation, and various theoretical
predictions with beyond-NLO precision (other points).
Further details can be found in the caption of Fig.~\ref{fig:xsec-1d-theory-nor-ptt-ptat}.}
\label{fig:xsec-1d-theory-nor-rpttmtt-rptttmtt}
\end{figure*}

\begin{figure*}[!phtb]
\centering
\includegraphics[width=0.49\textwidth]{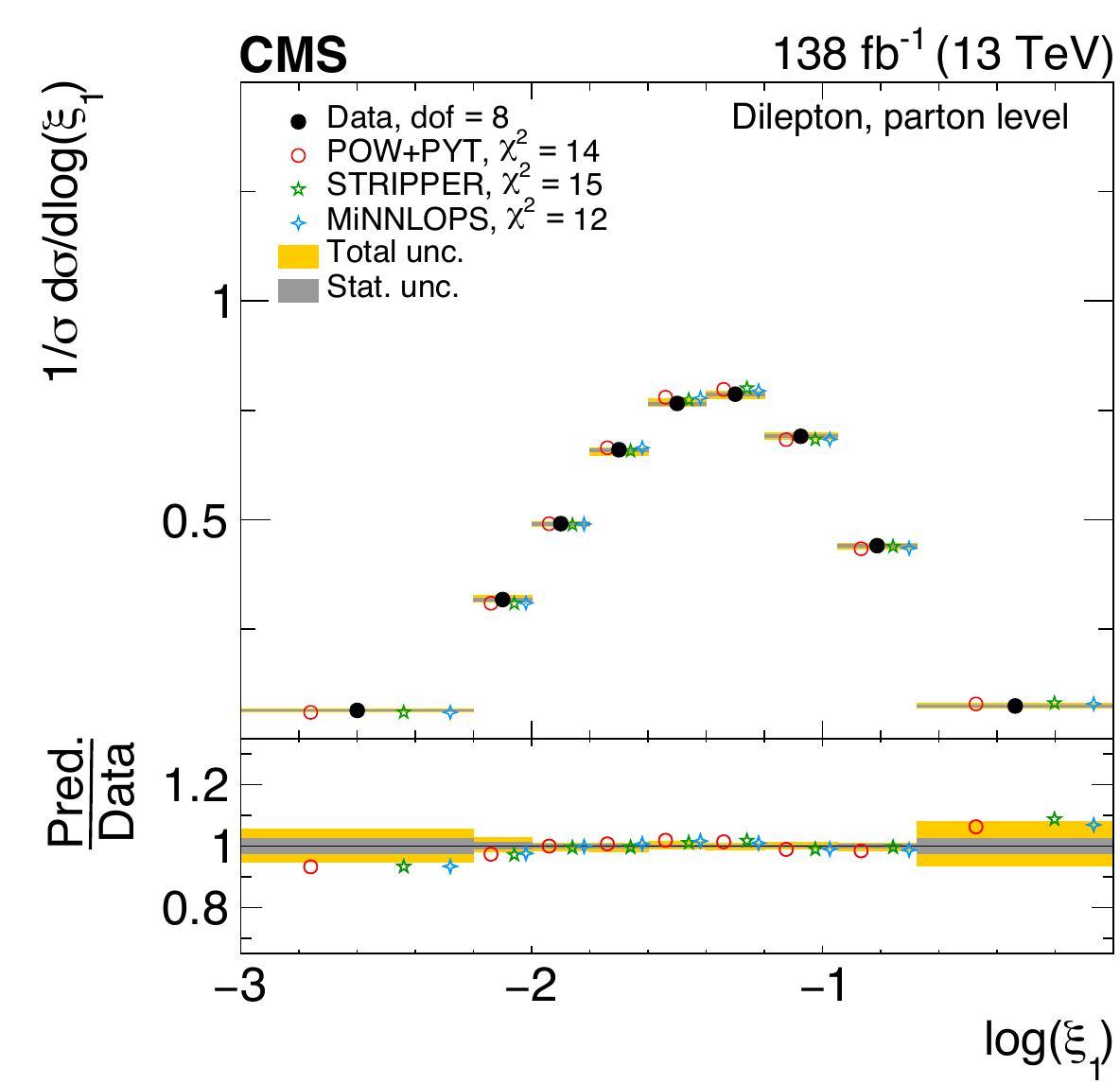}
\includegraphics[width=0.49\textwidth]{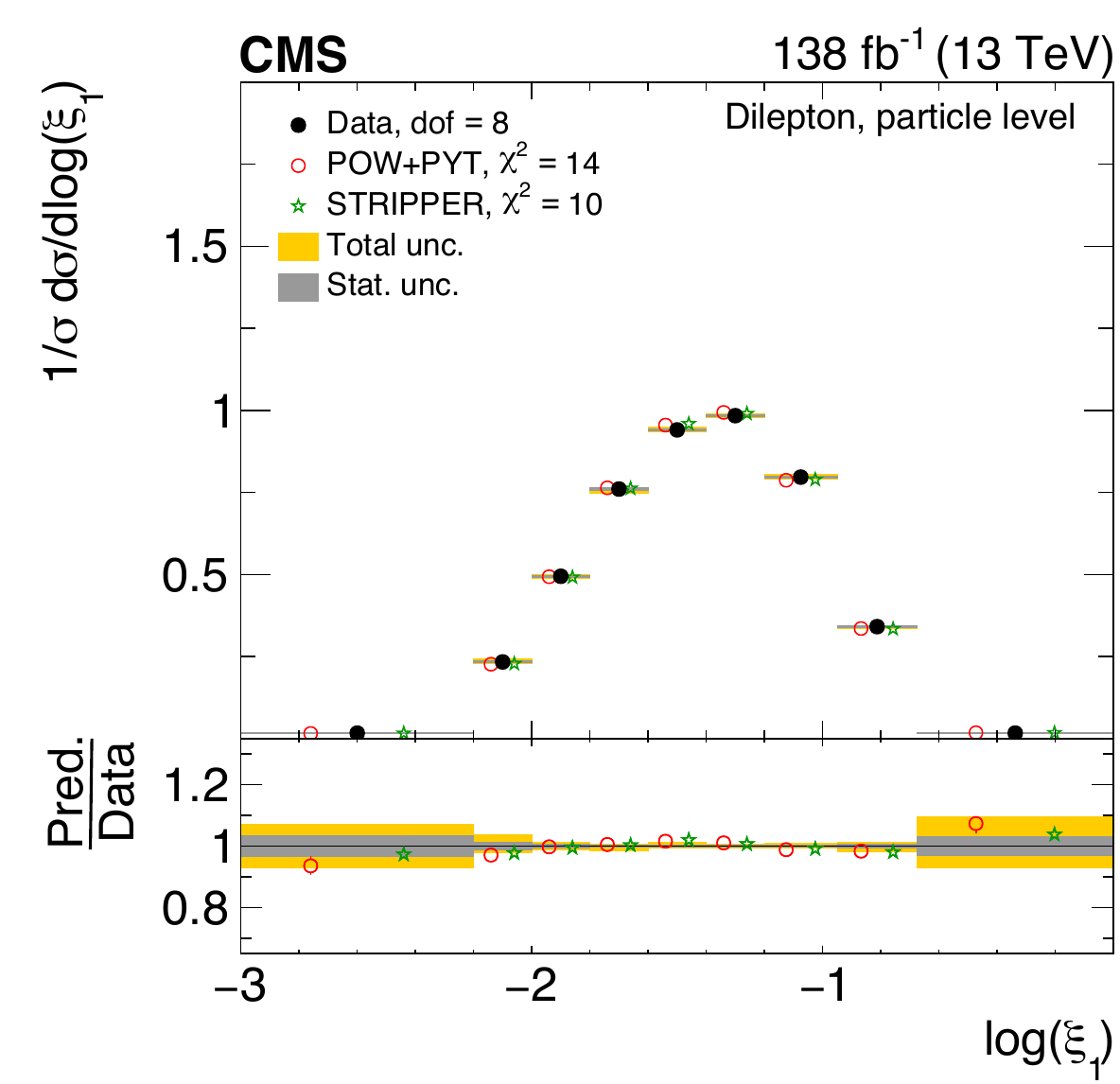}
\includegraphics[width=0.49\textwidth]{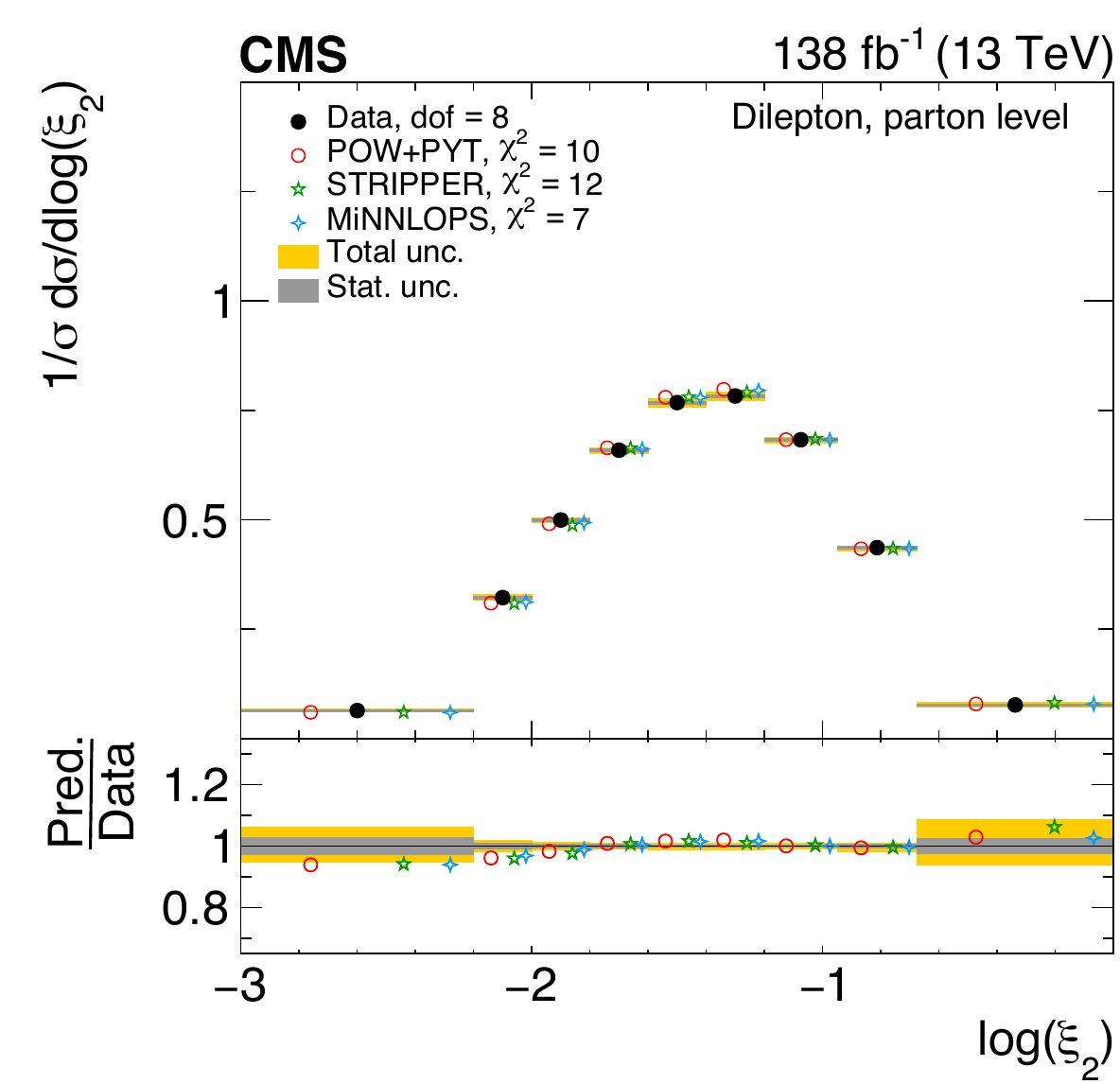}
\includegraphics[width=0.49\textwidth]{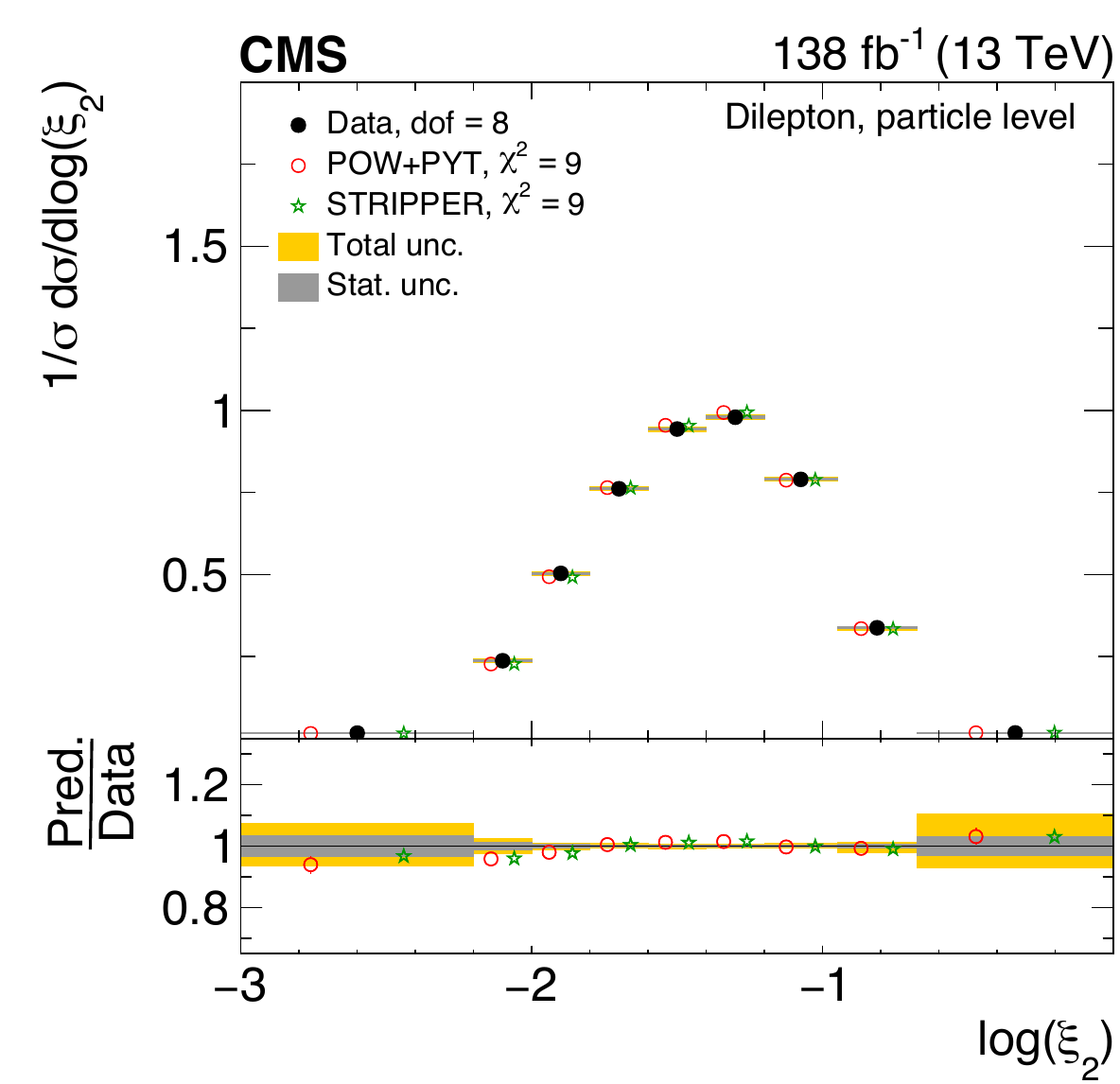}
\caption{Normalized differential \ttbar production cross sections as functions of \logxone (upper) and \logxtwo (lower)
are shown for data (filled circles), \PowPyt (`POW-PYT', open circles) simulation, and \StripperOnly
NNLO calculation (stars).
Further details can be found in the caption of Fig.~\ref{fig:xsec-1d-theory-nor-ptt-ptat}.}
\label{fig:xsec-1d-theory-nor-logxone-logxtwo}
\end{figure*}

\begin{figure}
\centering
\includegraphics[width=0.99\textwidth]{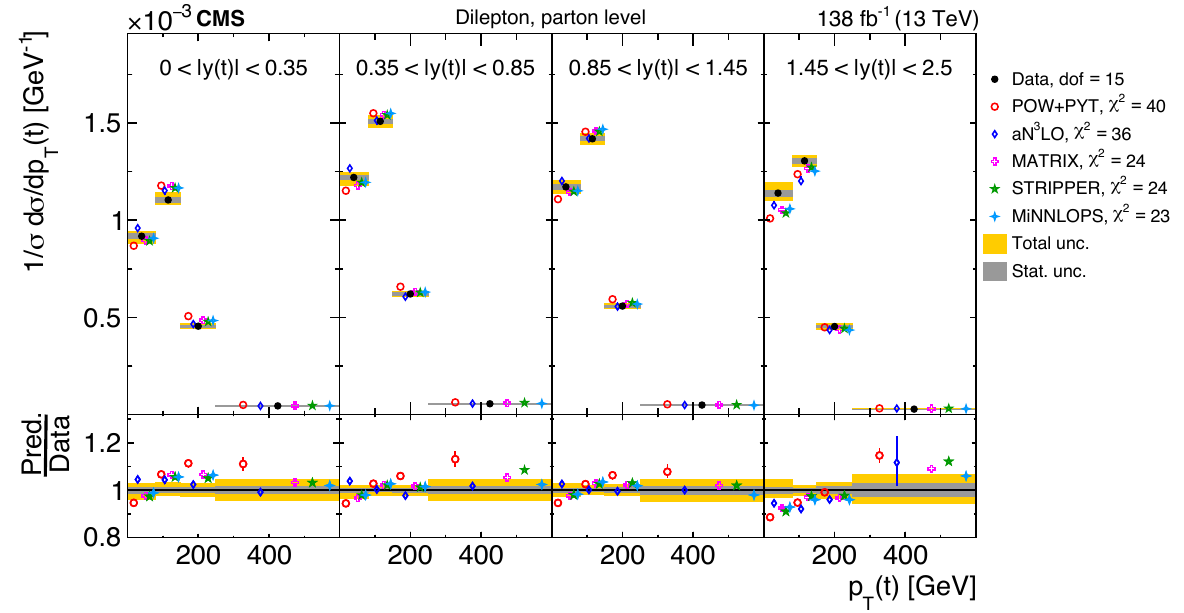}
\includegraphics[width=0.99\textwidth]{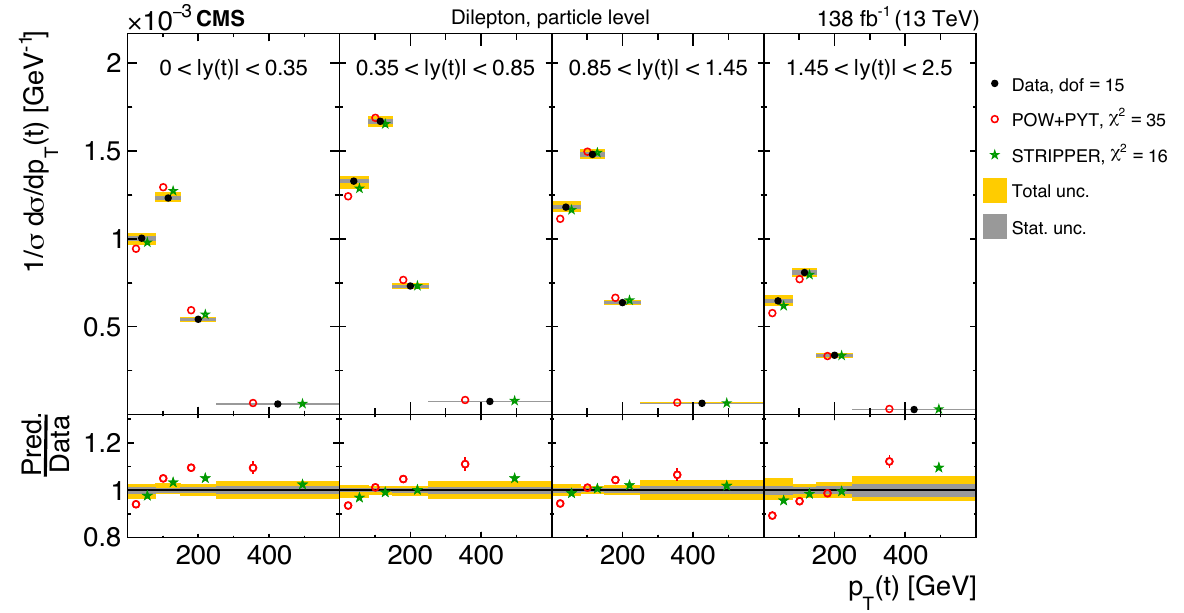}
\caption{Normalized \ytptt cross sections measured at the parton level in the full phase space (upper) and at
the particle level in a fiducial phase space (lower). The data are shown as filled circles with grey and yellow
bands indicating the statistical and total uncertainties (statistical and systematic uncertainties added in
quadrature), respectively. For each distribution, the number of degrees of freedom (dof) is also provided.
The cross sections are compared to predictions from the \PowPyt (`POW-PYT', open
circles) simulation and various theoretical predictions with beyond-NLO precision (other points). The estimated uncertainties in the \PowPytSh model are represented by vertical bars on the corresponding points.
For each model, a value of \chisq is reported that takes into account the measurement uncertainties. The lower panel in each plot
shows the ratios of the predictions to the data.}
    \label{fig:xsec-md-theory-nor-ytptt}
\end{figure}

\begin{figure}
\centering
\includegraphics[width=0.99\textwidth]{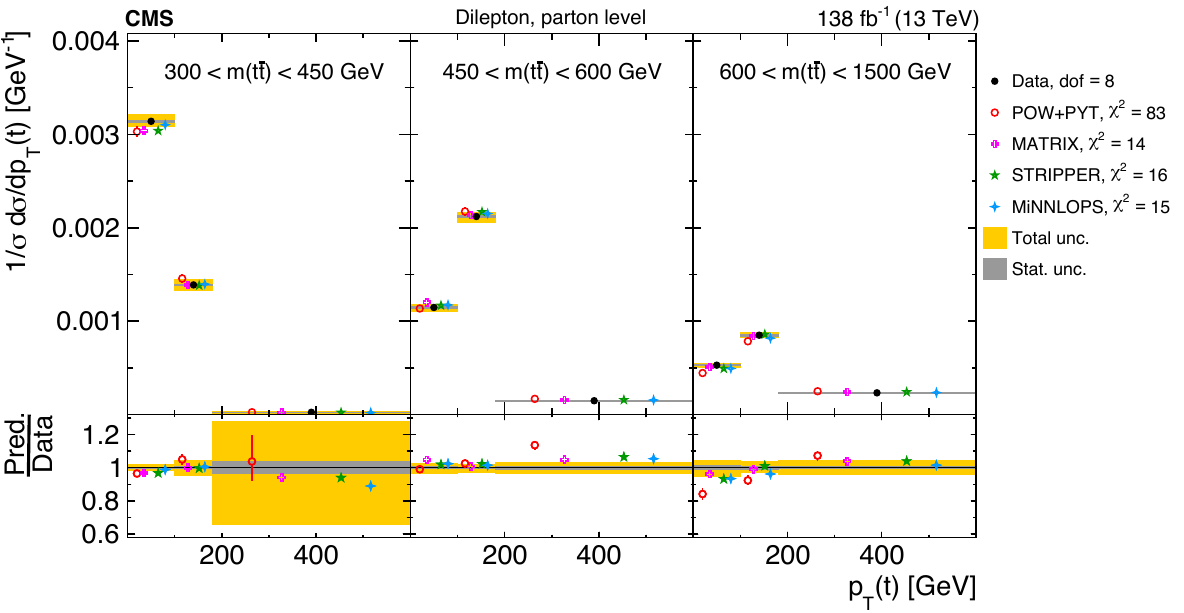}
\includegraphics[width=0.99\textwidth]{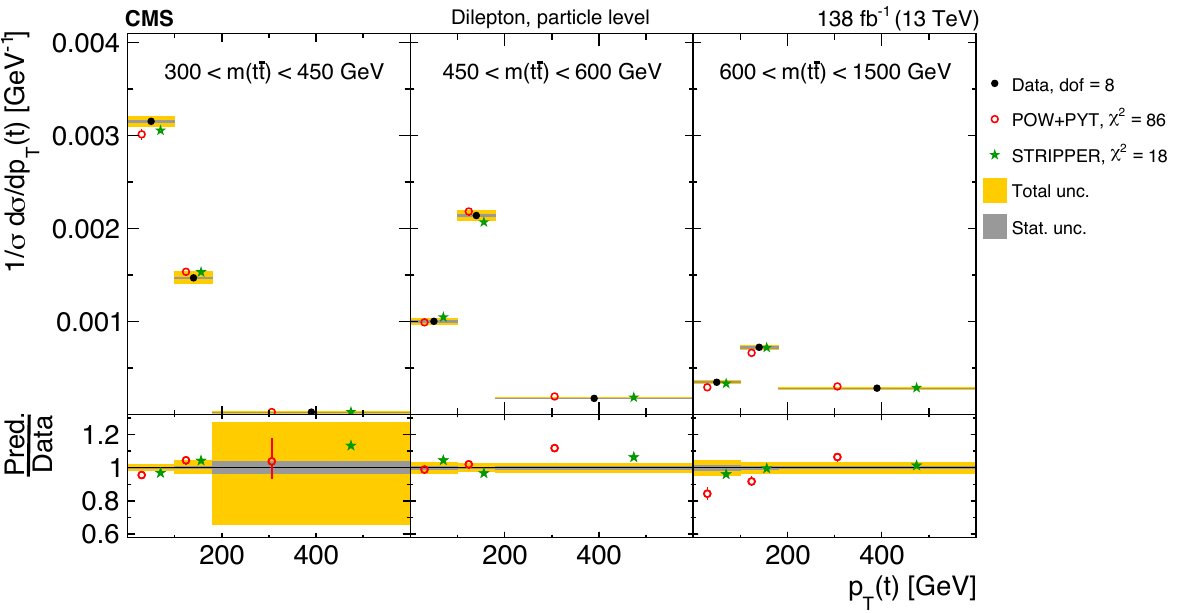}
\caption{Normalized \mttptt cross sections are shown for data (filled circles), \PowPyt (`POW-PYT', open
circles) simulation, and various theoretical predictions with beyond-NLO precision (other points).
    Further details can be found in the caption of Fig.~\ref{fig:xsec-md-theory-nor-ytptt}.}
    \label{fig:xsec-md-theory-nor-mttptt}
\end{figure}

\begin{figure}
\centering
\includegraphics[width=0.99\textwidth]{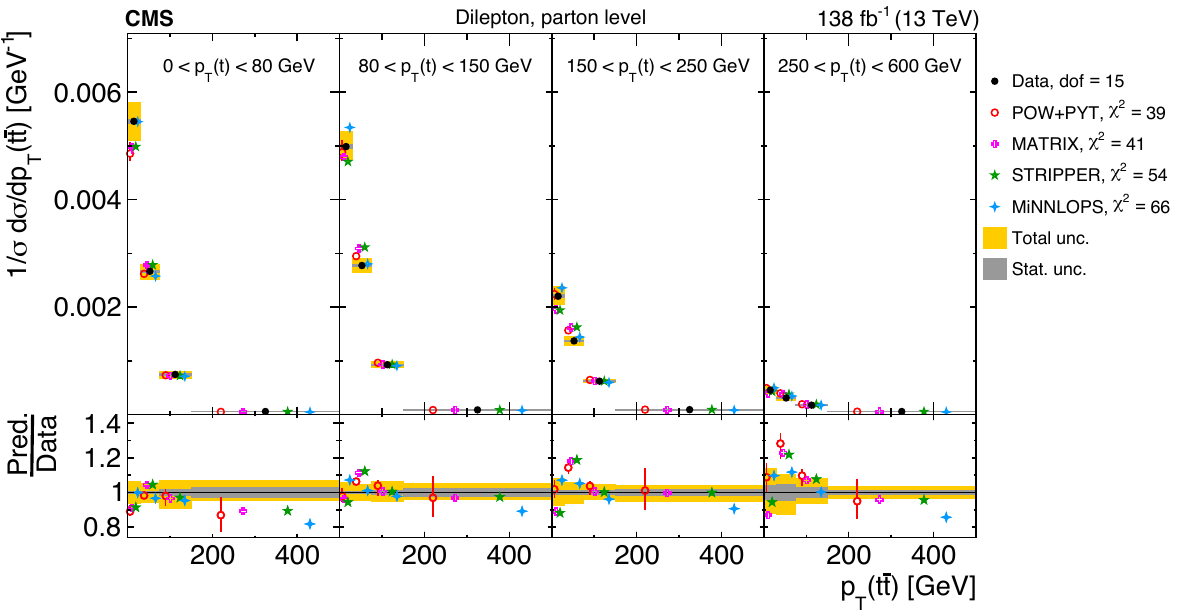}
\includegraphics[width=0.99\textwidth]{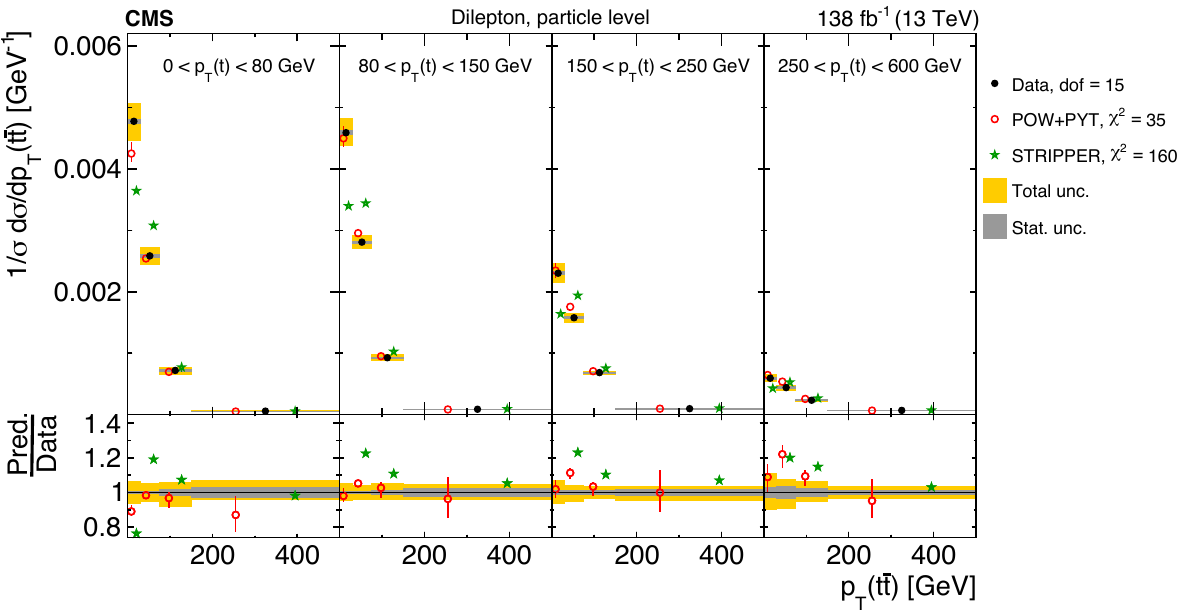}
\caption{Normalized \pttpttt cross sections are shown for data (filled circles), \PowPyt (`POW-PYT', open
circles) simulation, and various theoretical predictions with beyond-NLO precision (other points).
    Further details can be found in the caption of Fig.~\ref{fig:xsec-md-theory-nor-ytptt}.}
    \label{fig:xsec-md-theory-nor-pttpttt}
\end{figure}

\begin{figure}
\centering
\includegraphics[width=0.99\textwidth]{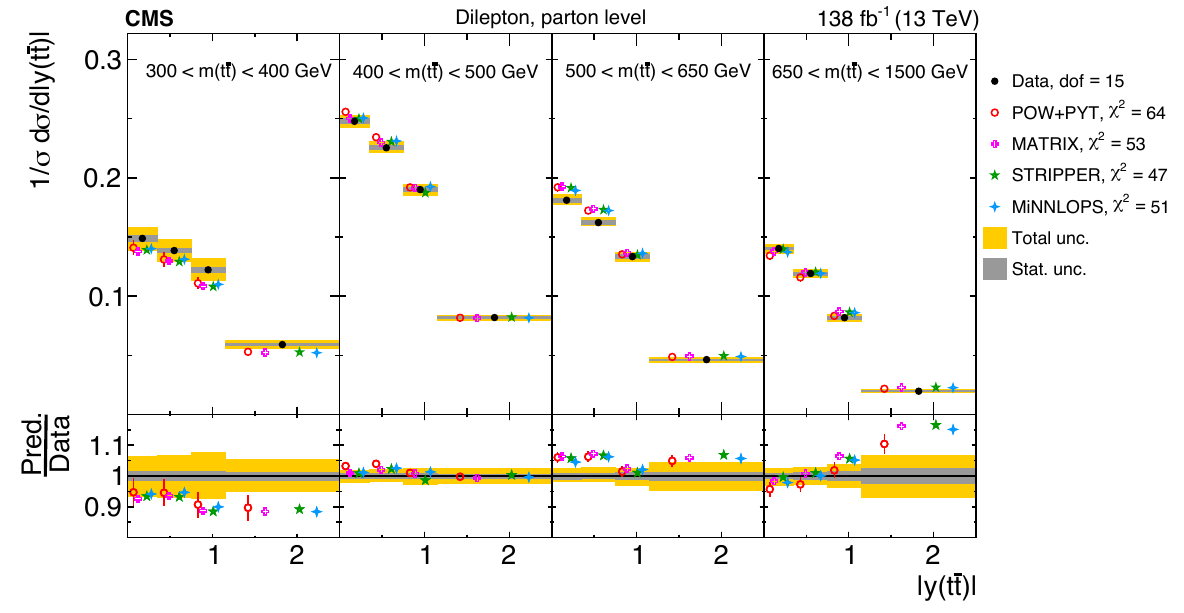}
\includegraphics[width=0.99\textwidth]{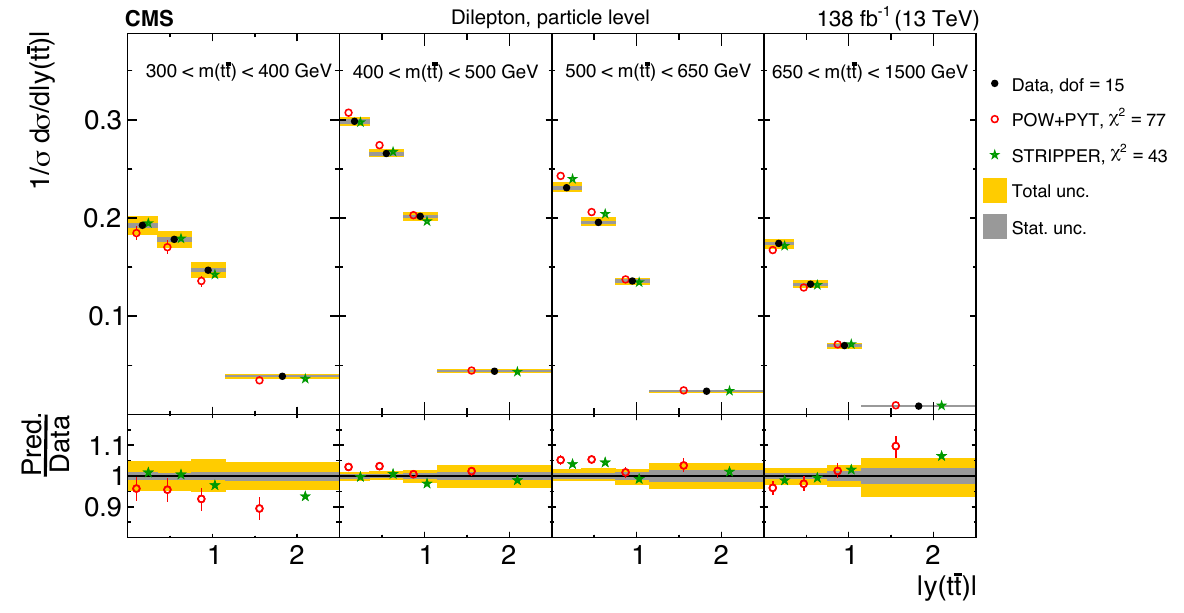}
\caption{Normalized \mttytt cross sections are shown for data (filled circles), \PowPyt (`POW-PYT', open
circles) simulation, and various theoretical predictions with beyond-NLO precision (other points).
    Further details can be found in the caption of Fig.~\ref{fig:xsec-md-theory-nor-ytptt}.}
    \label{fig:xsec-md-theory-nor-mttytt}
\end{figure}

\begin{figure}
\centering
\includegraphics[width=0.99\textwidth]{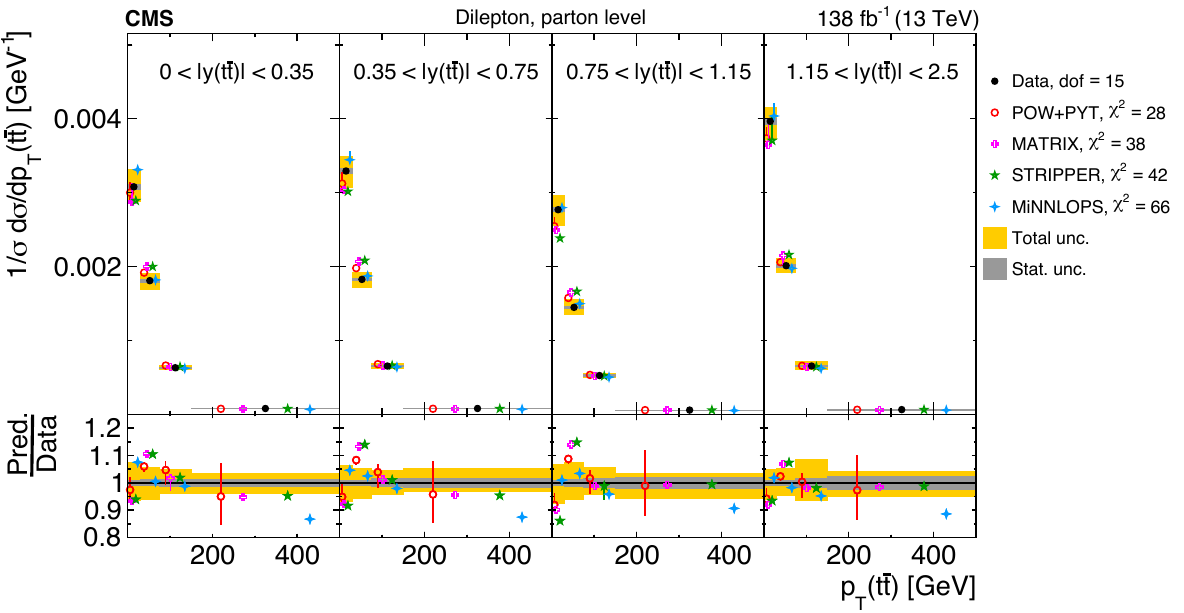}
\includegraphics[width=0.99\textwidth]{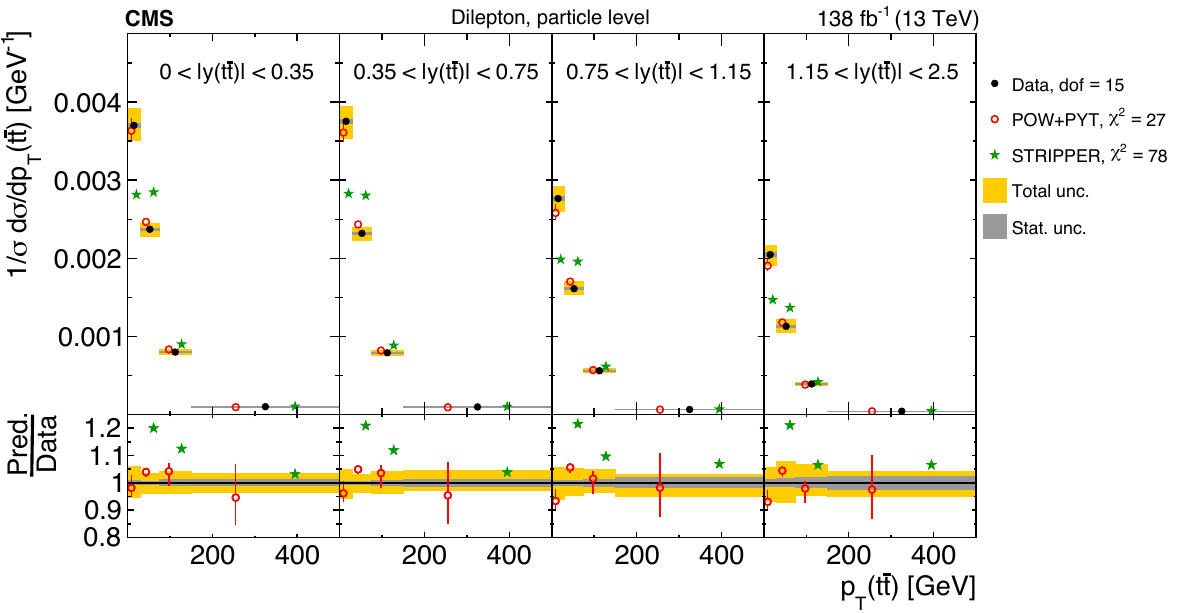}
    \caption{Normalized \yttpttt cross sections are shown for data (filled circles), \PowPyt (`POW-PYT', open
circles) simulation, and various theoretical predictions with beyond-NLO precision (other points).
    Further details can be found in the caption of Fig.~\ref{fig:xsec-md-theory-nor-ytptt}.}
    \label{fig:xsec-md-theory-nor-yttpttt}
\end{figure}

\begin{figure}
\centering
\includegraphics[width=0.99\textwidth]{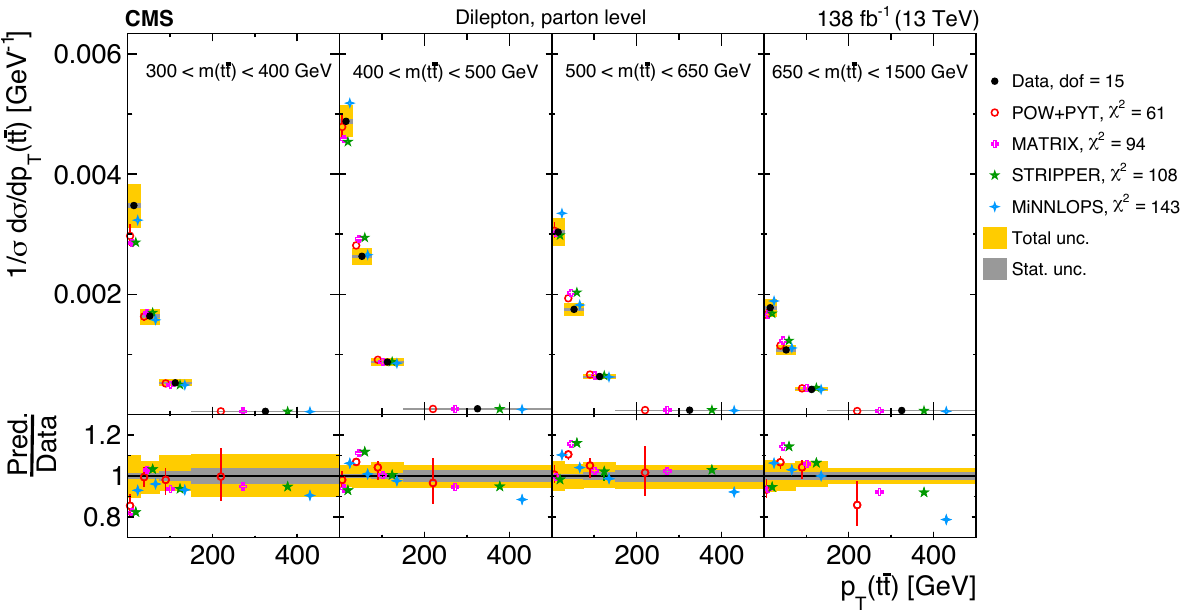}
\includegraphics[width=0.99\textwidth]{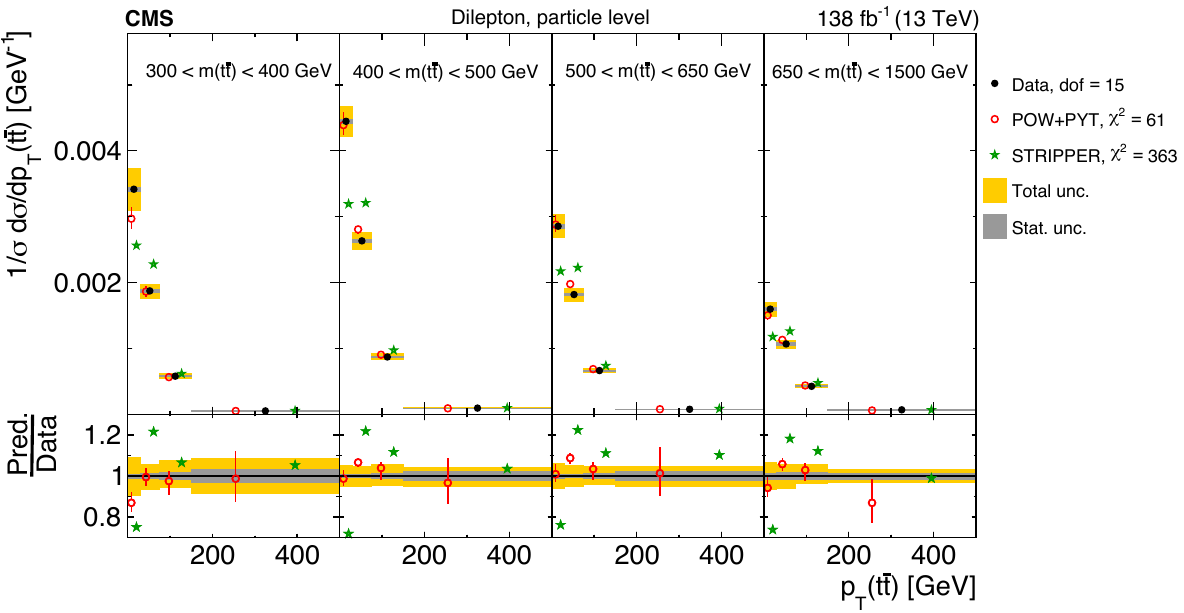}
\caption{Normalized \mttpttt cross sections are shown for data (filled circles), \PowPyt (`POW-PYT', open
circles) simulation, and various theoretical predictions with beyond-NLO precision (other points).
    Further details can be found in the caption of Fig.~\ref{fig:xsec-md-theory-nor-ytptt}.}
    \label{fig:xsec-md-theory-nor-mttpttt}
\end{figure}

\begin{figure}
\centering
\includegraphics[width=0.99\textwidth]{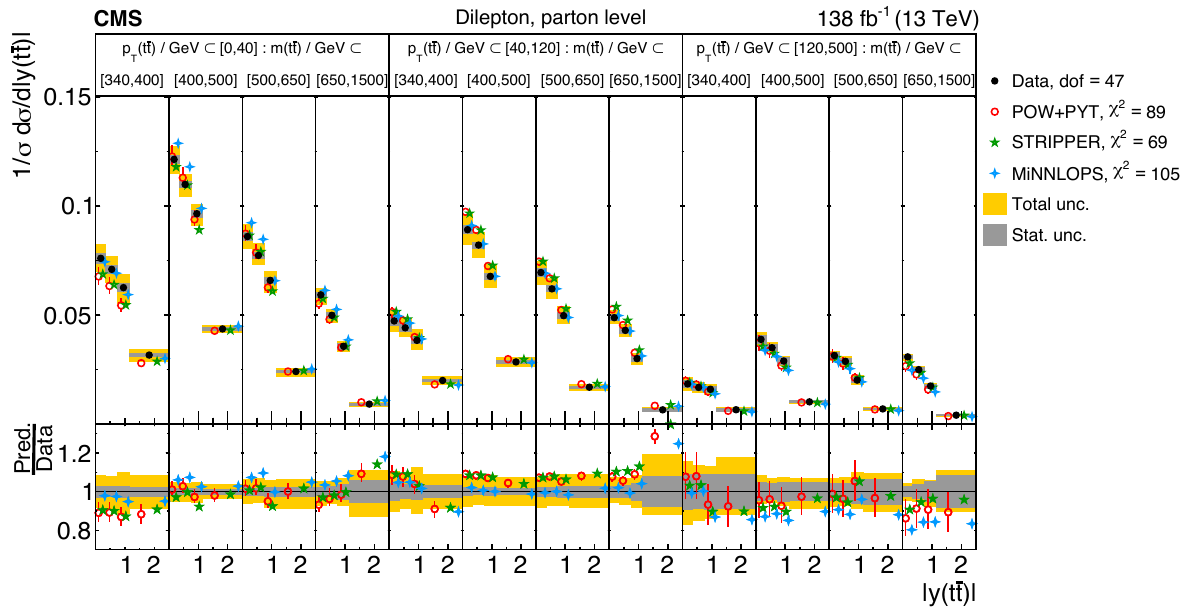}
\includegraphics[width=0.99\textwidth]{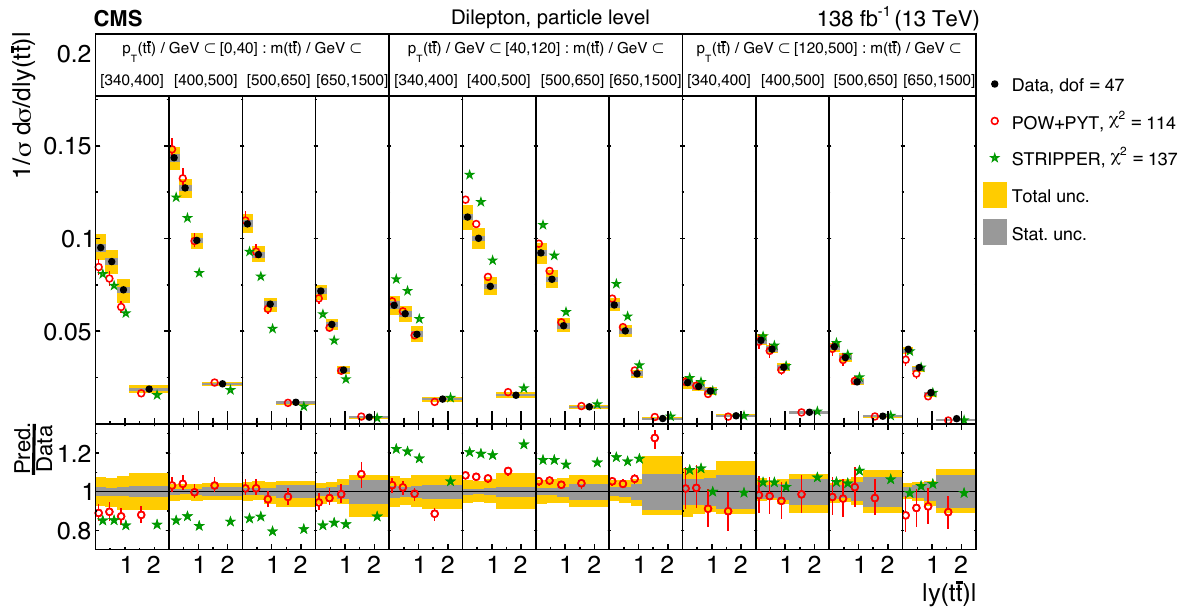}
\caption{Normalized \ptttmttytt cross sections are shown for data (filled circles),
             \PowPyt (`POW-PYT', open circles) simulation, and various theoretical predictions with beyond-NLO precision (other points).
    Further details can be found in the caption of Fig.~\ref{fig:xsec-md-theory-nor-ytptt}.}
    \label{fig:xsec-md-theory-nor-ptttmttytt}
\end{figure}

\begin{figure}
\centering
\includegraphics[width=0.99\textwidth]{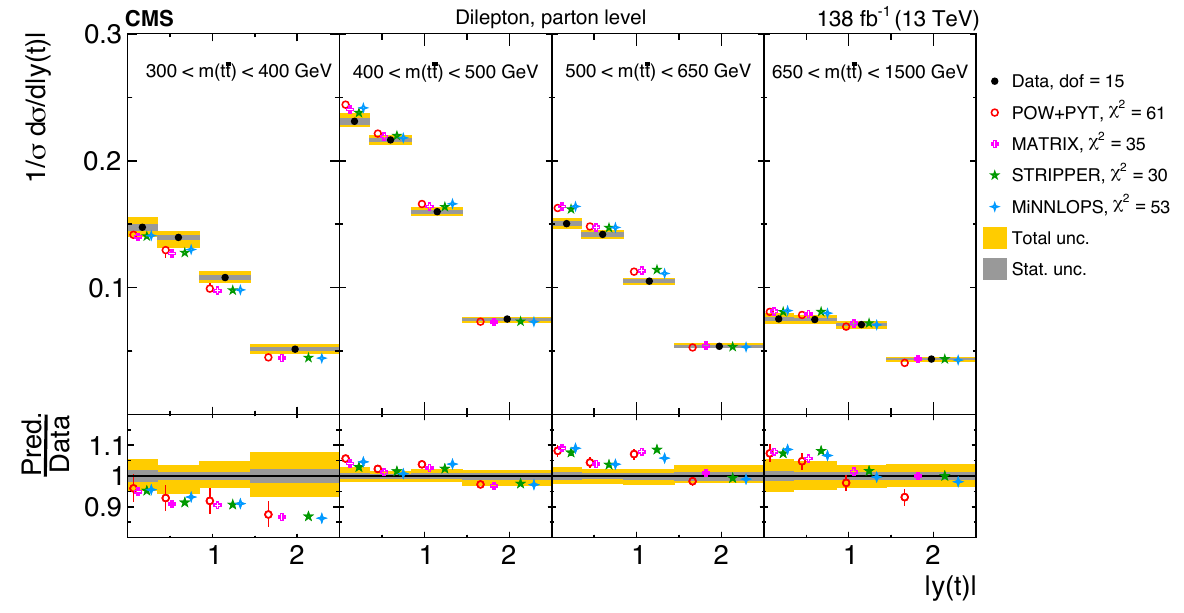}
\includegraphics[width=0.99\textwidth]{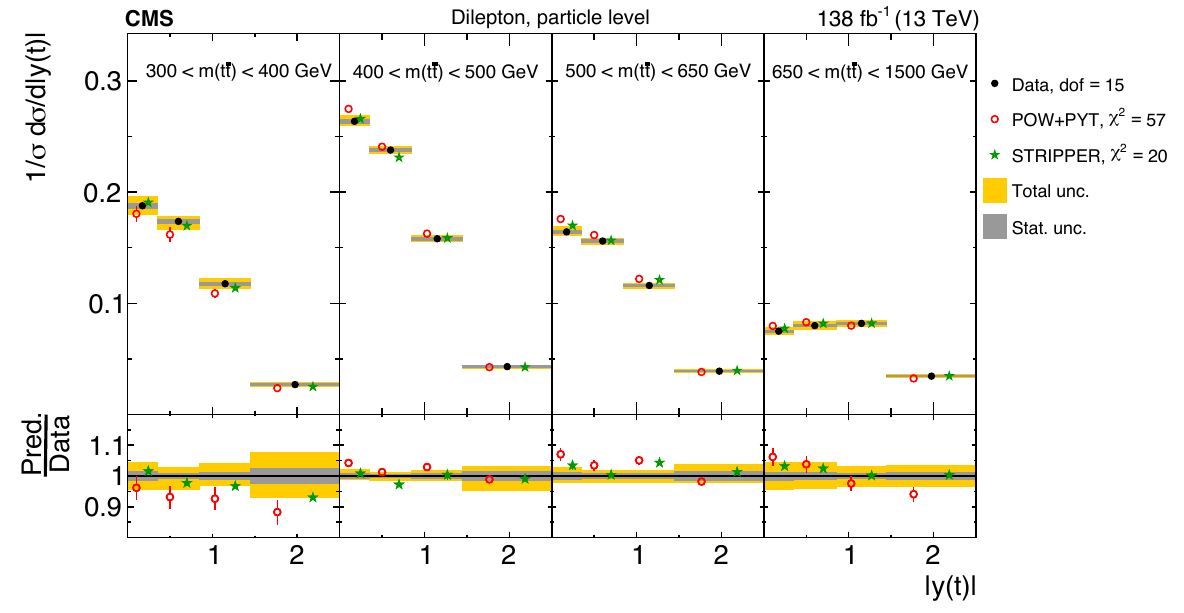}
\caption{Normalized \mttyt cross sections are shown for data (filled circles), \PowPyt (`POW-PYT', open
circles) simulation, and various theoretical predictions with beyond-NLO precision (other points).
    Further details can be found in the caption of Fig.~\ref{fig:xsec-md-theory-nor-ytptt}.}
    \label{fig:xsec-md-theory-nor-mttyt}
\end{figure}

\begin{figure}
\centering
\includegraphics[width=0.99\textwidth]{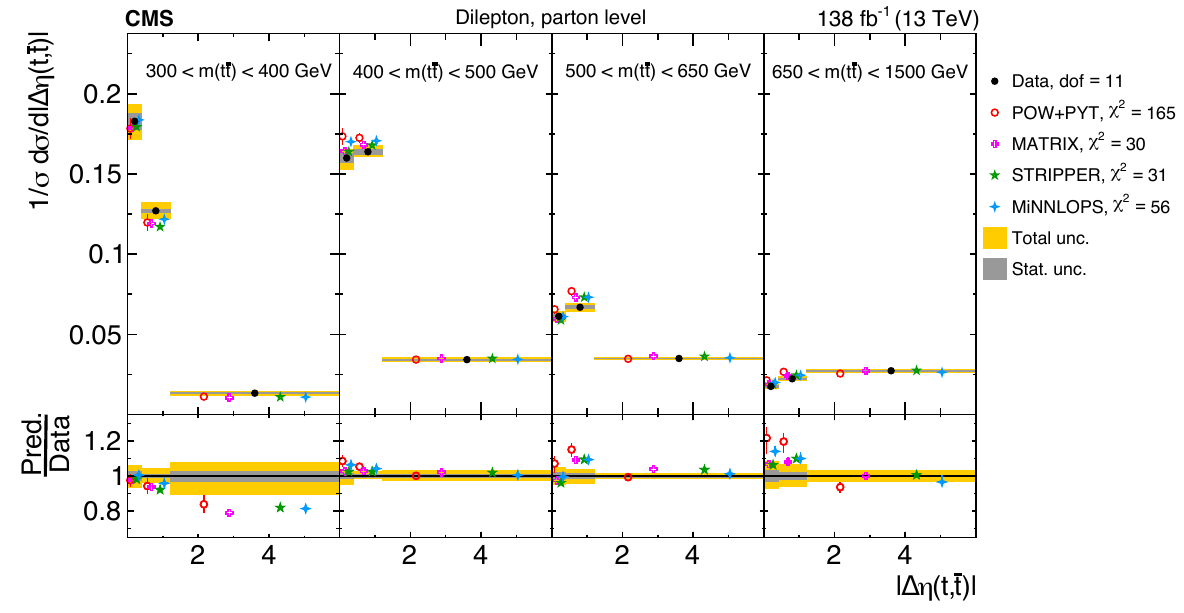}
\includegraphics[width=0.99\textwidth]{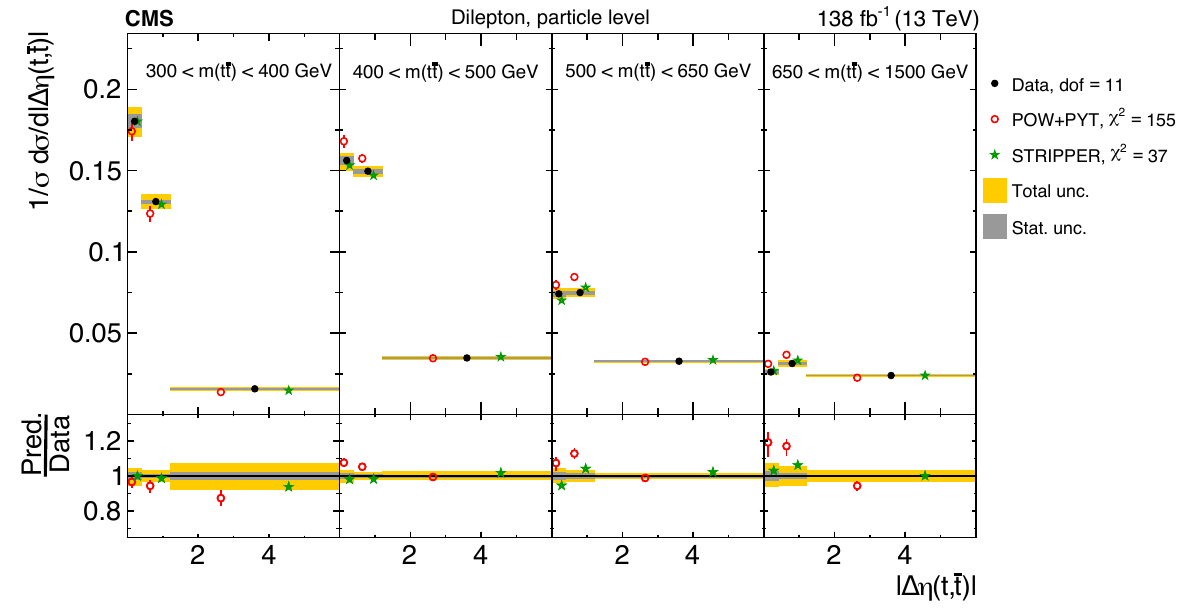}
\caption{Normalized \mttdetatt cross sections are shown for data (filled circles), \PowPyt (`POW-PYT', open
circles) simulation, and various theoretical predictions with beyond-NLO precision (other points).
    Further details can be found in the caption of Fig.~\ref{fig:xsec-md-theory-nor-ytptt}.}
    \label{fig:xsec-md-theory-nor-mttdetatt}
\end{figure}

\begin{figure}
\centering
\includegraphics[width=0.99\textwidth]{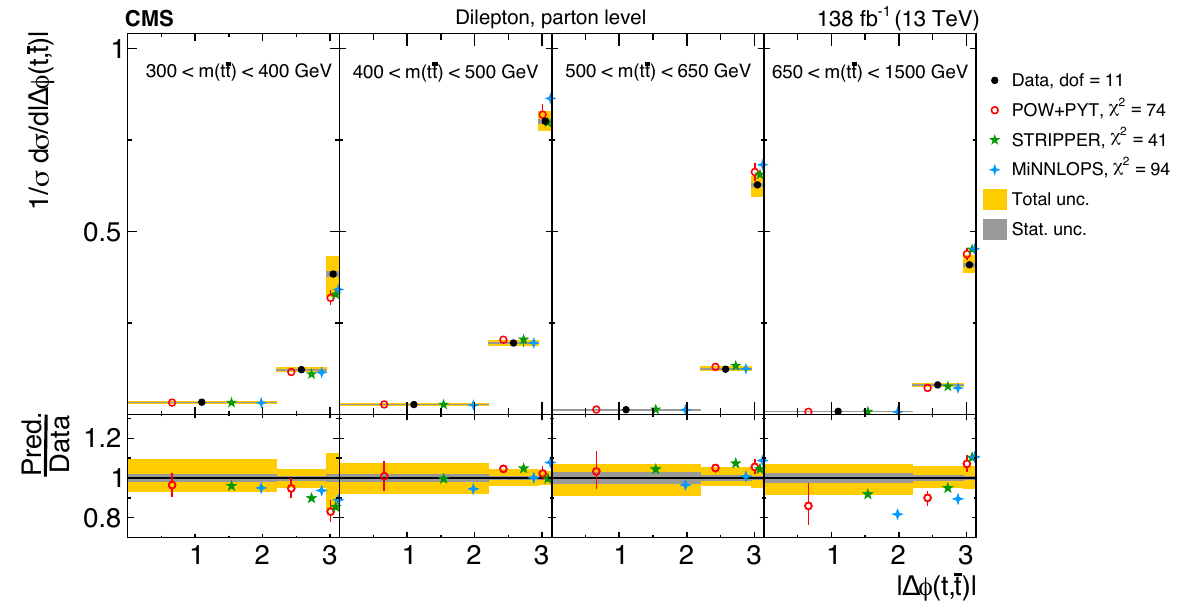}
\includegraphics[width=0.99\textwidth]{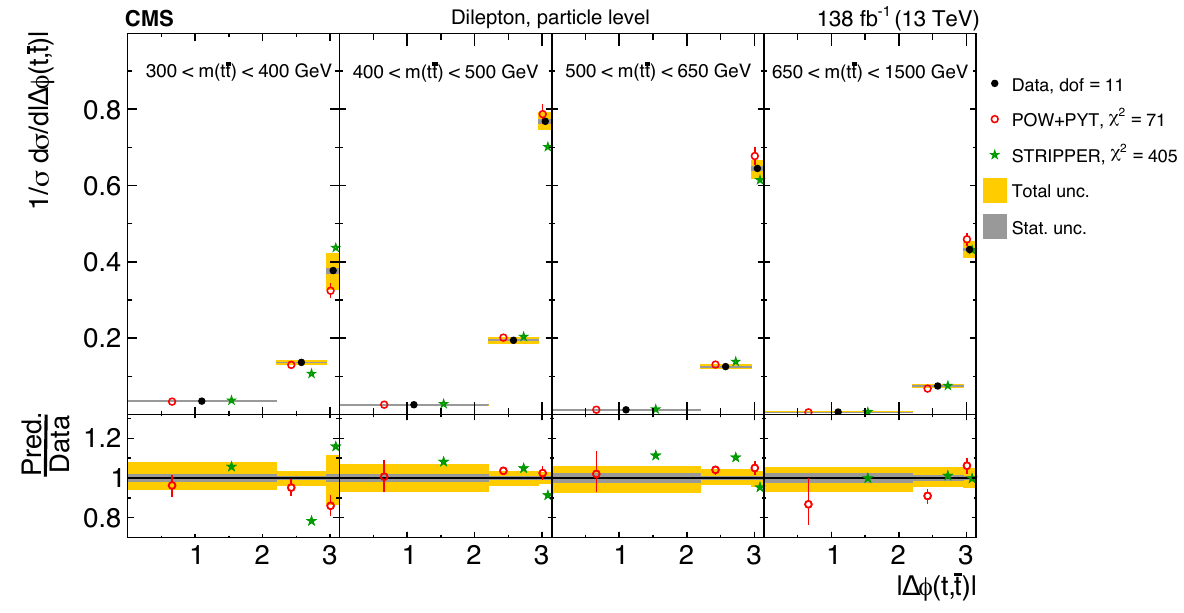}
\caption{Normalized \mttdphitt cross sections are shown for data (filled circles), \PowPyt (`POW-PYT', open
circles) simulation, and various theoretical predictions with beyond-NLO precision (other points).
    Further details can be found in the caption of Fig.~\ref{fig:xsec-md-theory-nor-ytptt}.}
    \label{fig:xsec-md-theory-nor-mttdphitt}
\end{figure}

\begin{figure*}[!phtb]
\centering
\includegraphics[width=0.49\textwidth]{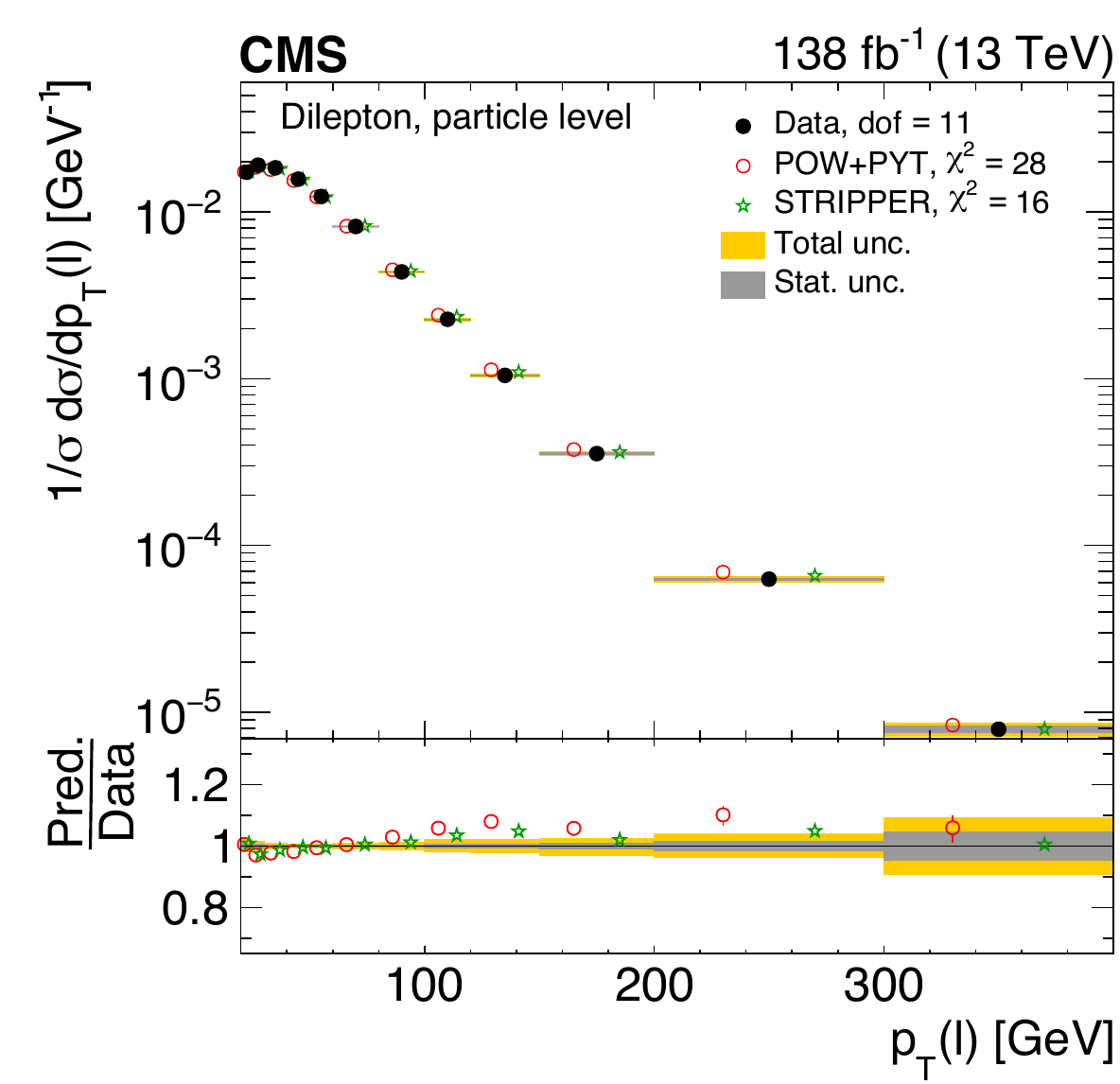}
\includegraphics[width=0.49\textwidth]{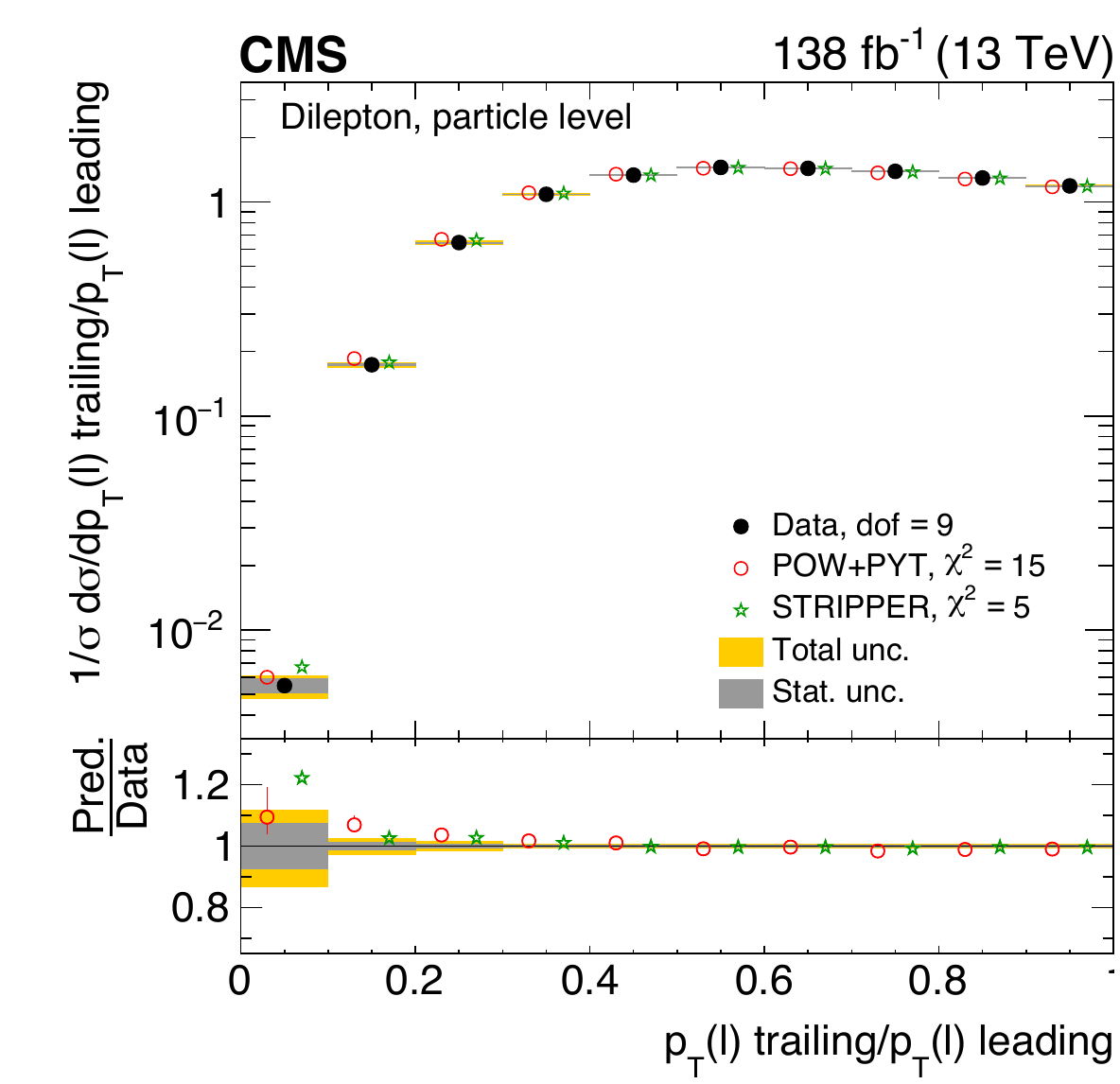}
\includegraphics[width=0.49\textwidth]{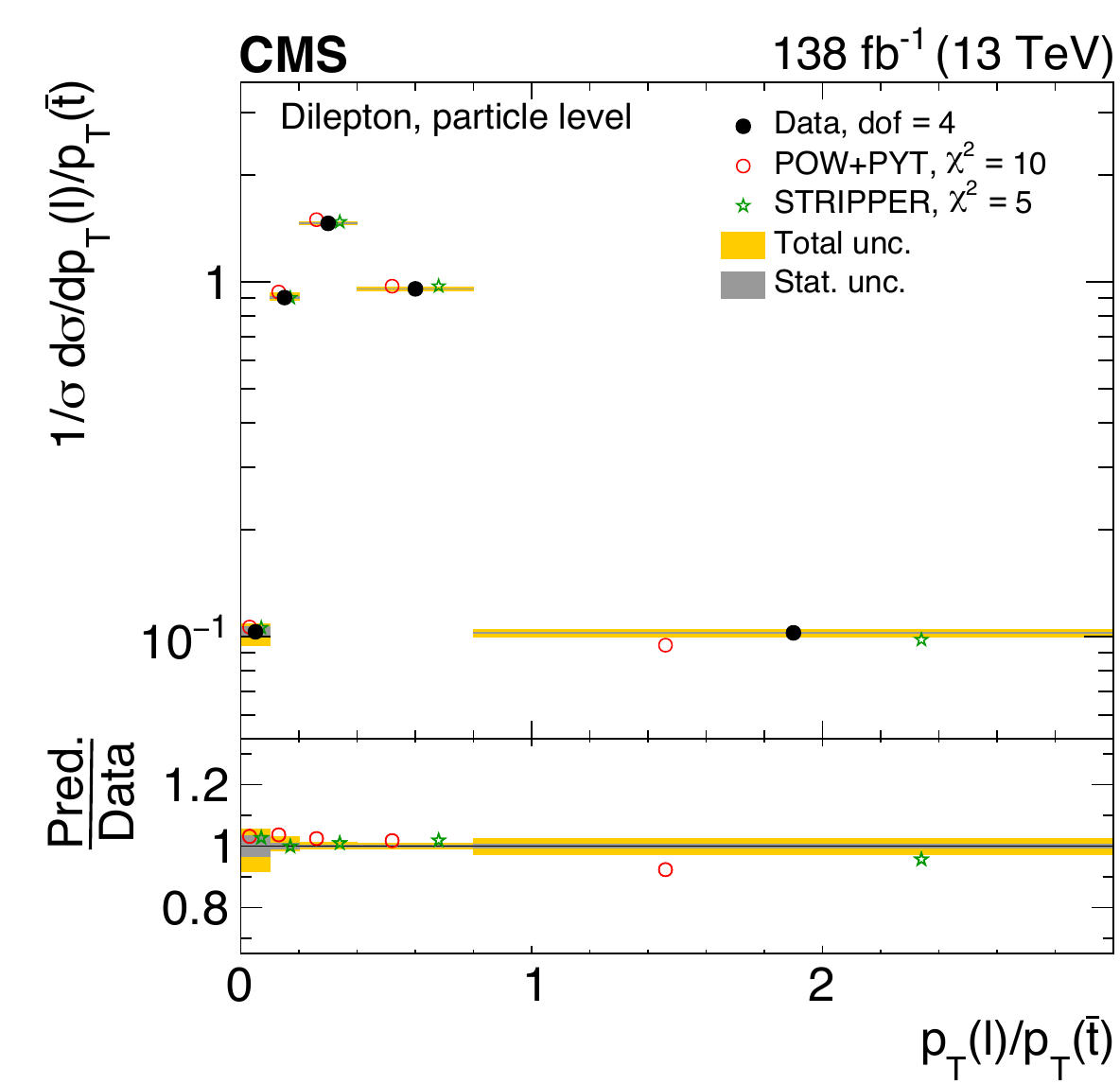}
\caption{Normalized differential \ttbar production cross sections as functions of \pt of the lepton (upper left),
 of the ratio of the trailing and leading lepton \pt
(upper right), and of the ratio of lepton and top antiquark \pt (lower middle), measured at the particle level in
a fiducial phase space.
The data are shown as filled circles with grey and yellow bands indicating the statistical and total uncertainties
(statistical and systematic uncertainties added in quadrature), respectively.
For each distribution, the number of degrees of freedom (dof) is also provided.
The cross sections are compared to predictions from the \PowPyt (`POW-PYT', open circles) simulation and
\StripperOnly NNLO calculation (stars).
The estimated uncertainties in the \PowPytSh model are represented by vertical bars on the corresponding points.
For each model, a value of \chisq is reported that takes into account the measurement uncertainties.
The lower panel in each plot shows the ratios of the predictions to the data.}
\label{fig:xsec-1d-theory-nor-ptlep-rptleptonic-rptlepptt}
\end{figure*}

\begin{figure*}[!phtb]
\centering
\includegraphics[width=0.49\textwidth]{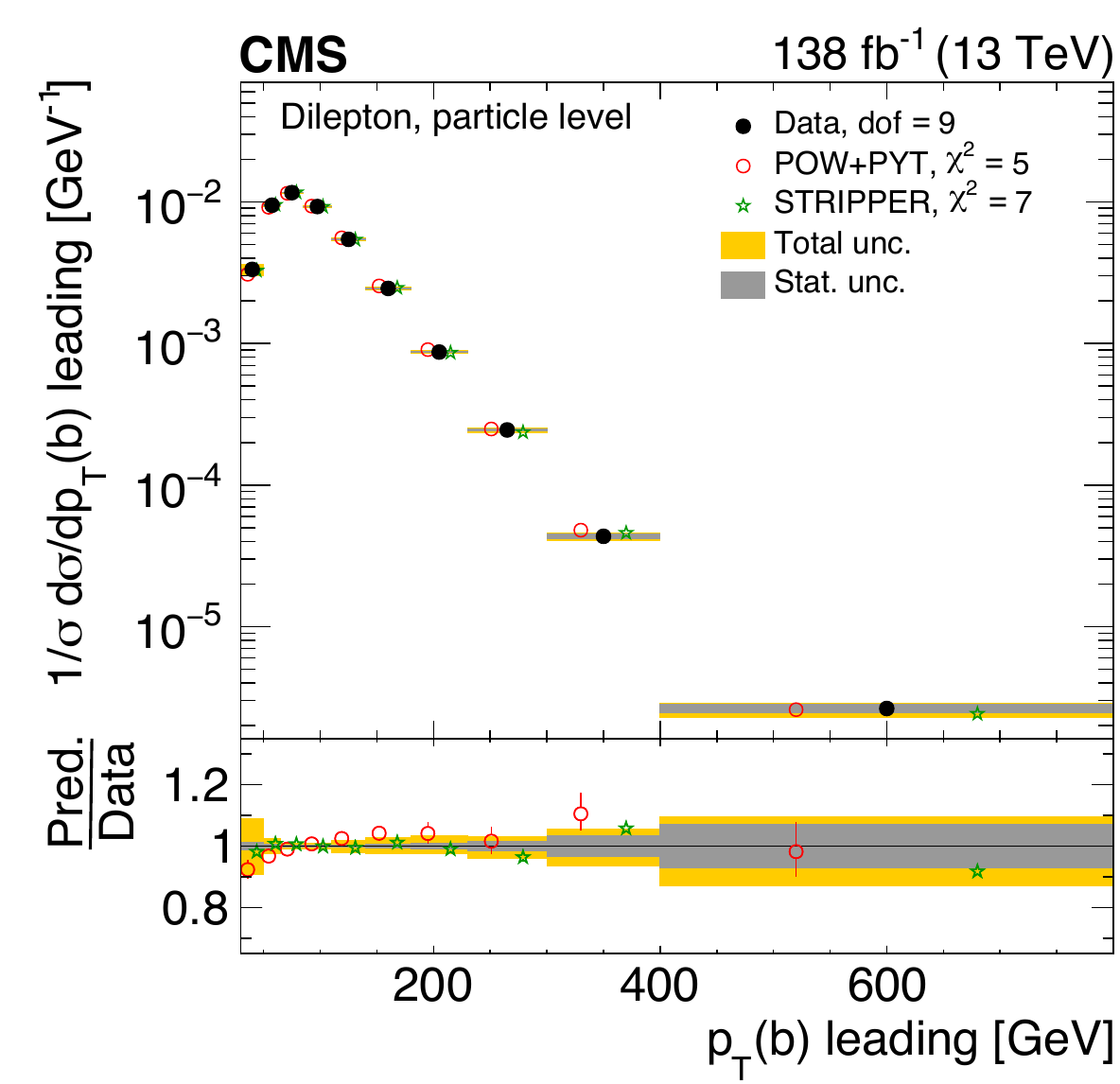}
\includegraphics[width=0.49\textwidth]{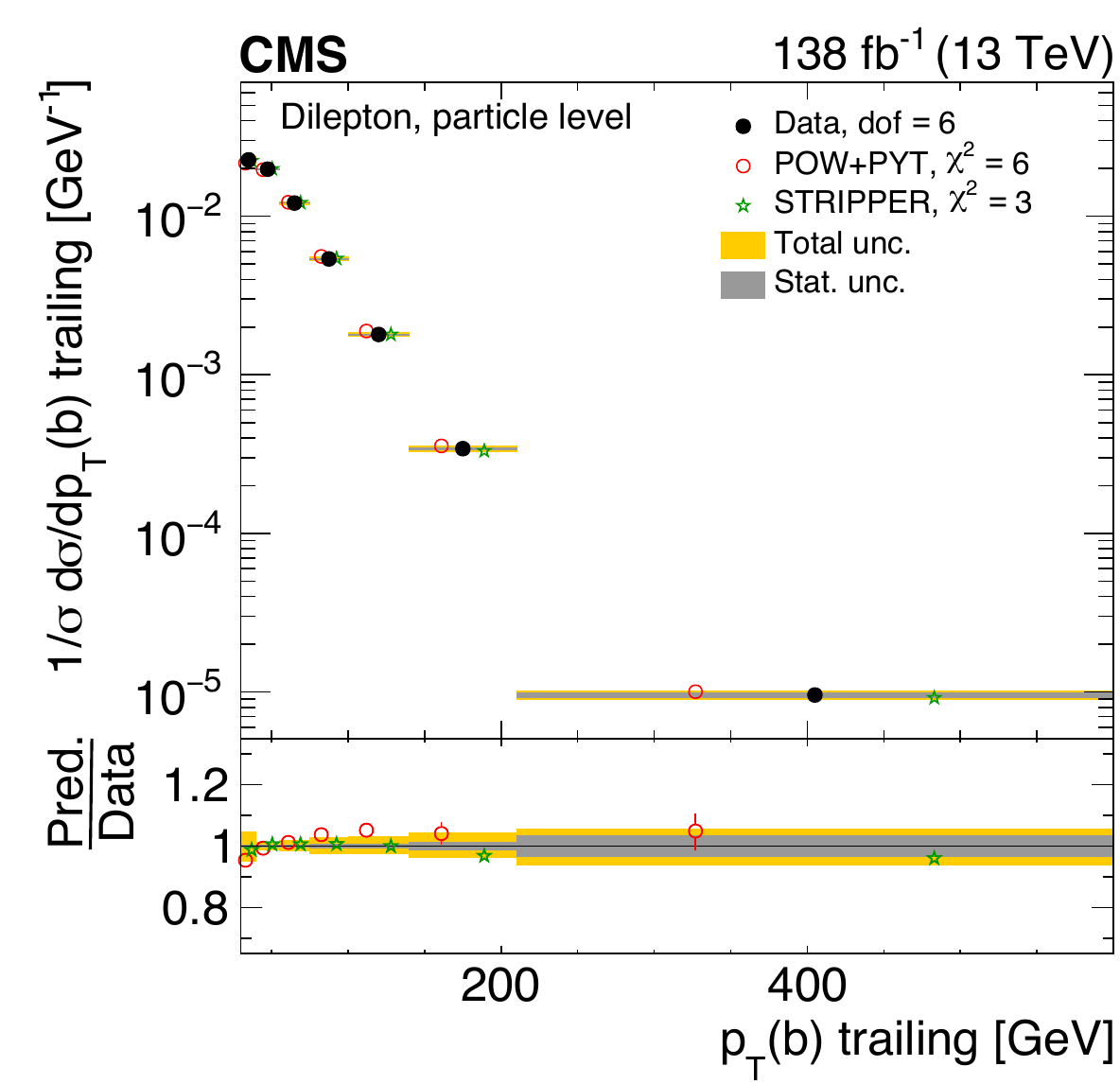}
\includegraphics[width=0.49\textwidth]{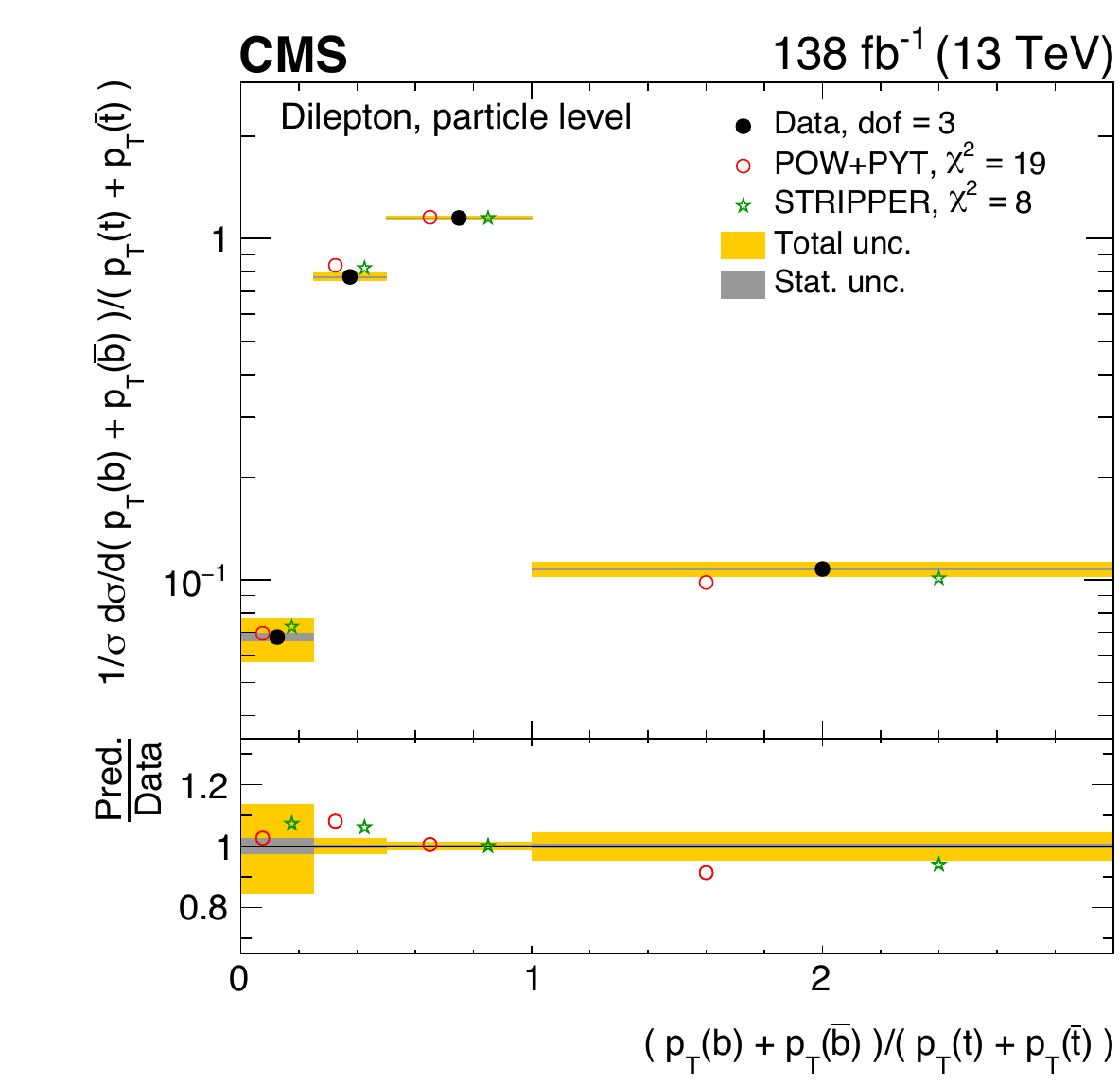}
\caption{Normalized differential \ttbar production cross sections as functions of the \pt of
the leading (upper left) and trailing (upper right) \PQb jet, and \rptbsptts (lower)
are shown for data (filled circles), \PowPyt (`POW-PYT', open circles) simulation, and \StripperOnly NNLO
calculation (stars).
Further details can be found in the caption of Fig.~\ref{fig:xsec-1d-theory-nor-ptlep-rptleptonic-rptlepptt}.}
\label{fig:xsec-1d-theory-nor-ptb-ptnb-rptbsptts}
\end{figure*}

\begin{figure*}[!phtb]
\centering
\includegraphics[width=0.49\textwidth]{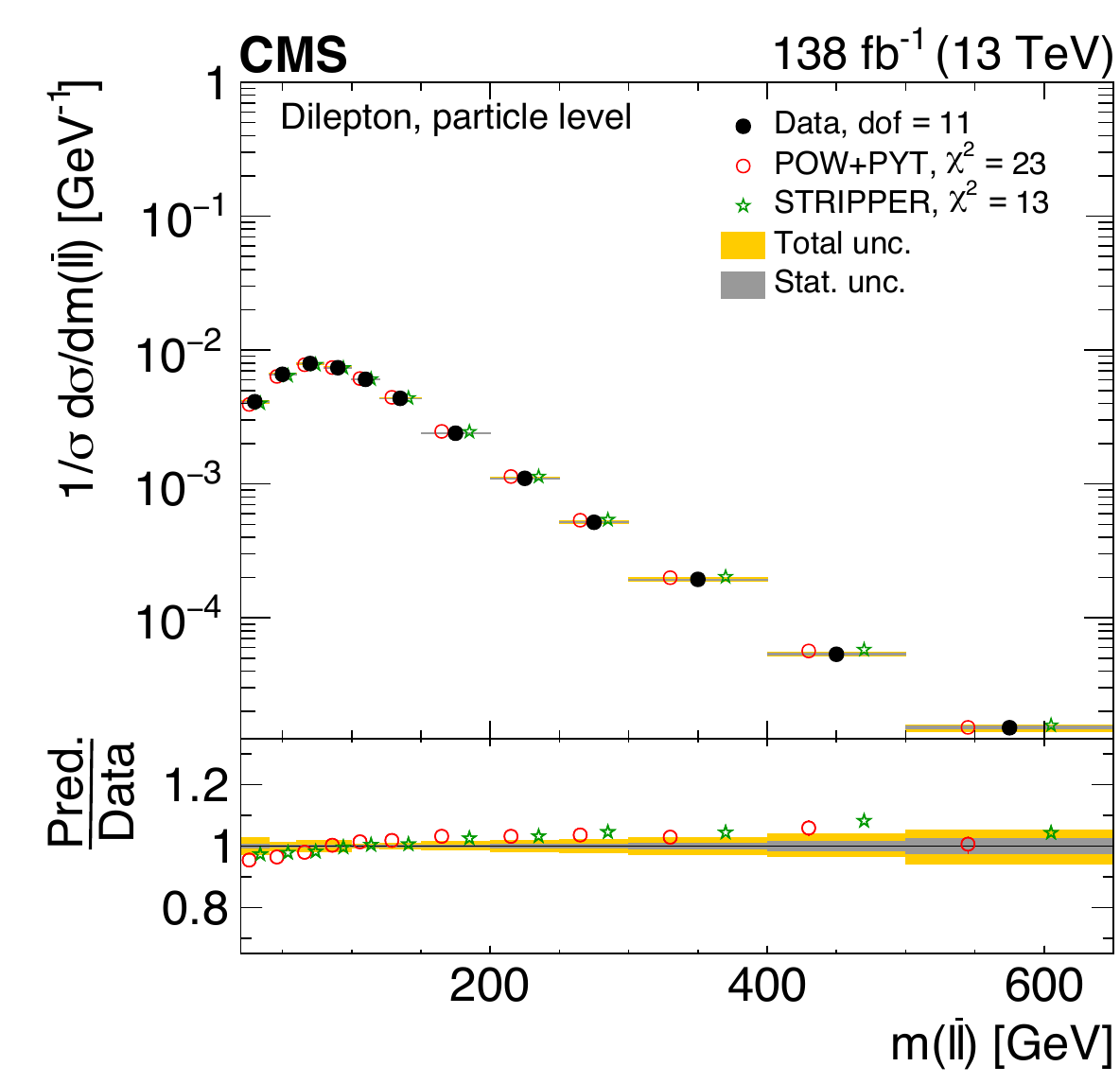}
\includegraphics[width=0.49\textwidth]{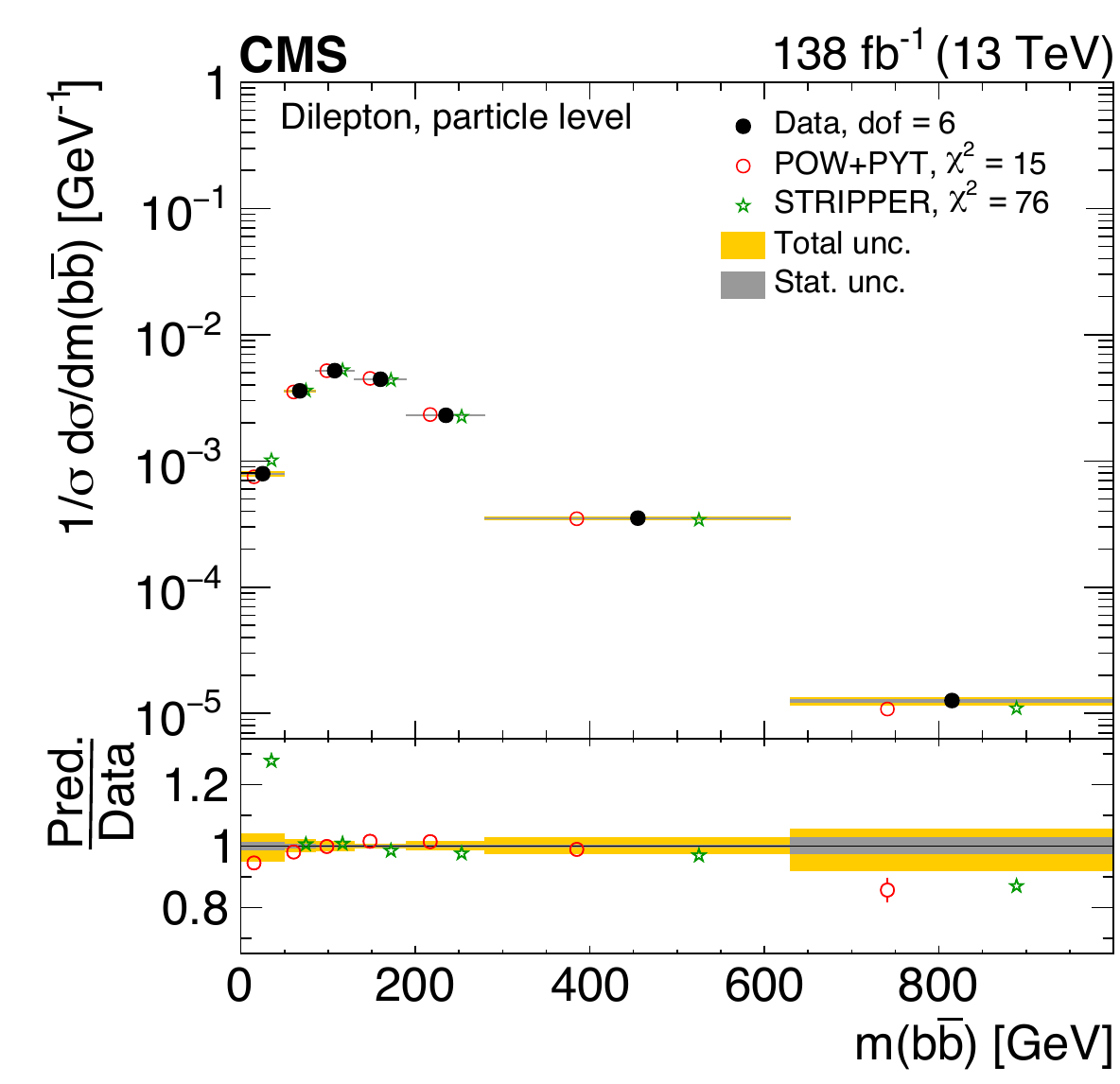}
\includegraphics[width=0.49\textwidth]{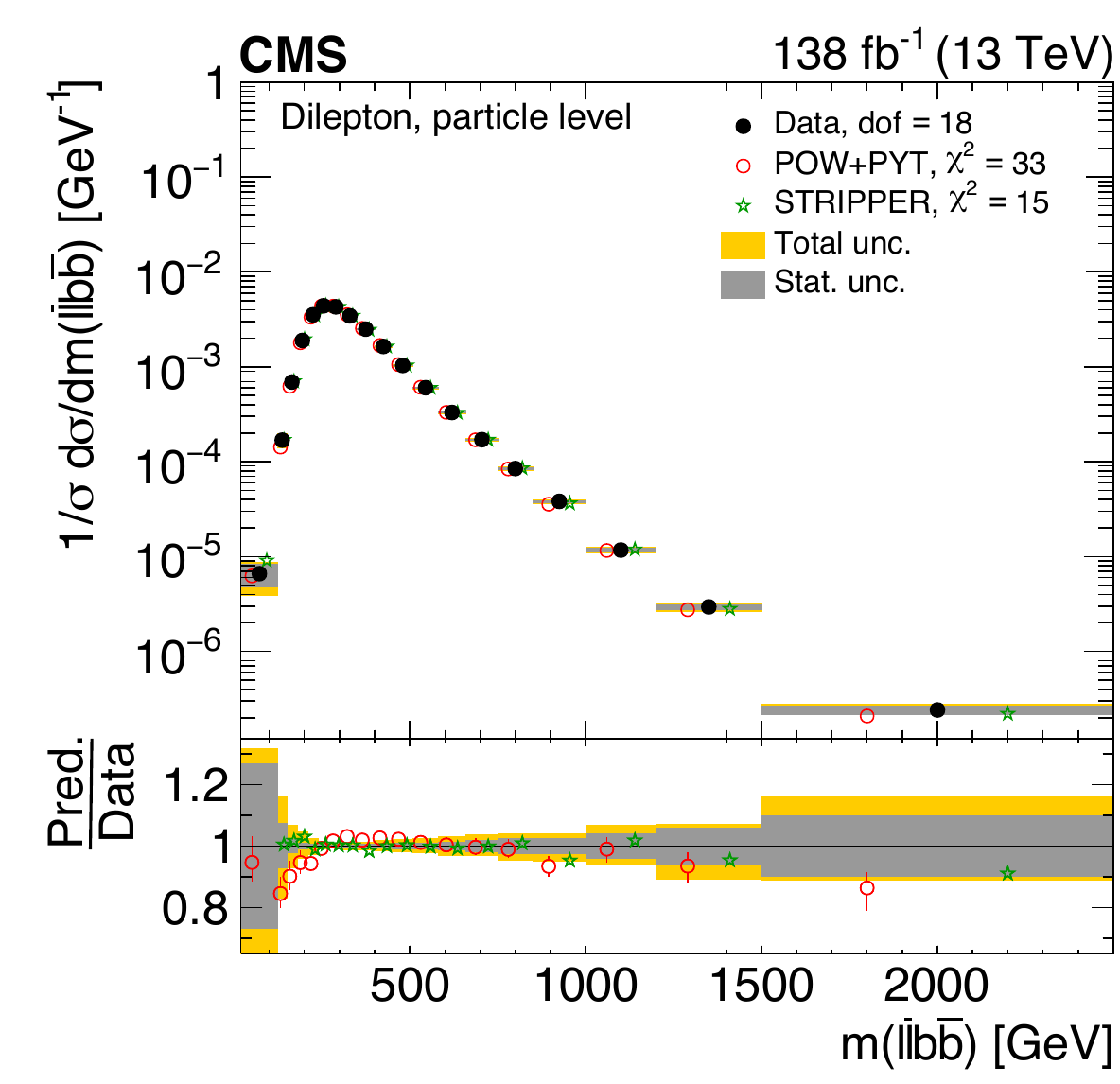}
\caption{Normalized differential \ttbar production cross sections as functions of \mll (upper left), \mbb
(upper right), and \mllbb (lower)
are shown for data (filled circles), \PowPyt (`POW-PYT', open circles) simulation, and \StripperOnly NNLO
calculation (stars).
Further details can be found in the caption of Fig.~\ref{fig:xsec-1d-theory-nor-ptlep-rptleptonic-rptlepptt}.}
\label{fig:xsec-1d-theory-nor-mll-mbb-mllbb}
\end{figure*}

\begin{figure}
\centering
\includegraphics[width=0.49\textwidth]{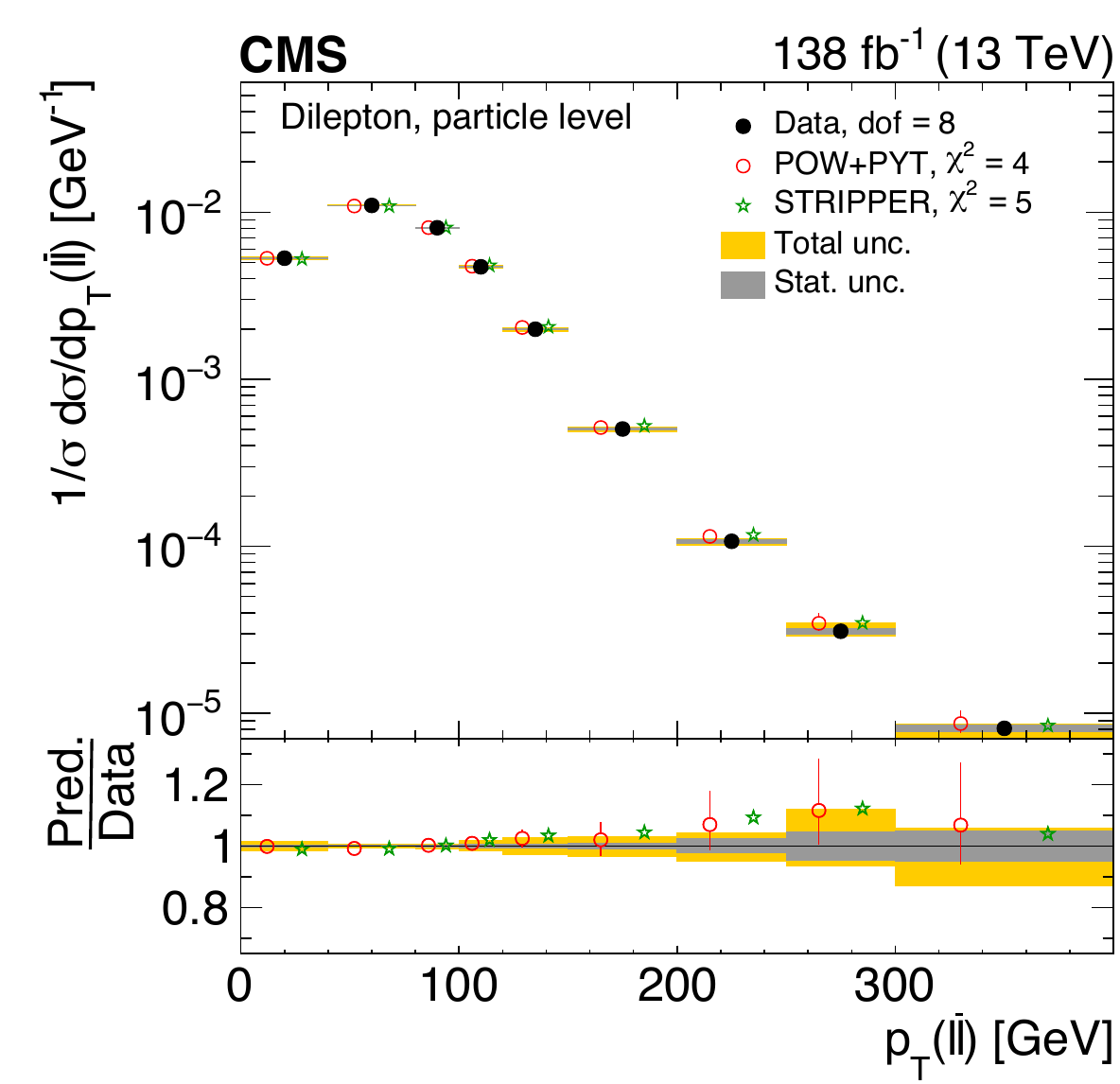}
\includegraphics[width=0.49\textwidth]{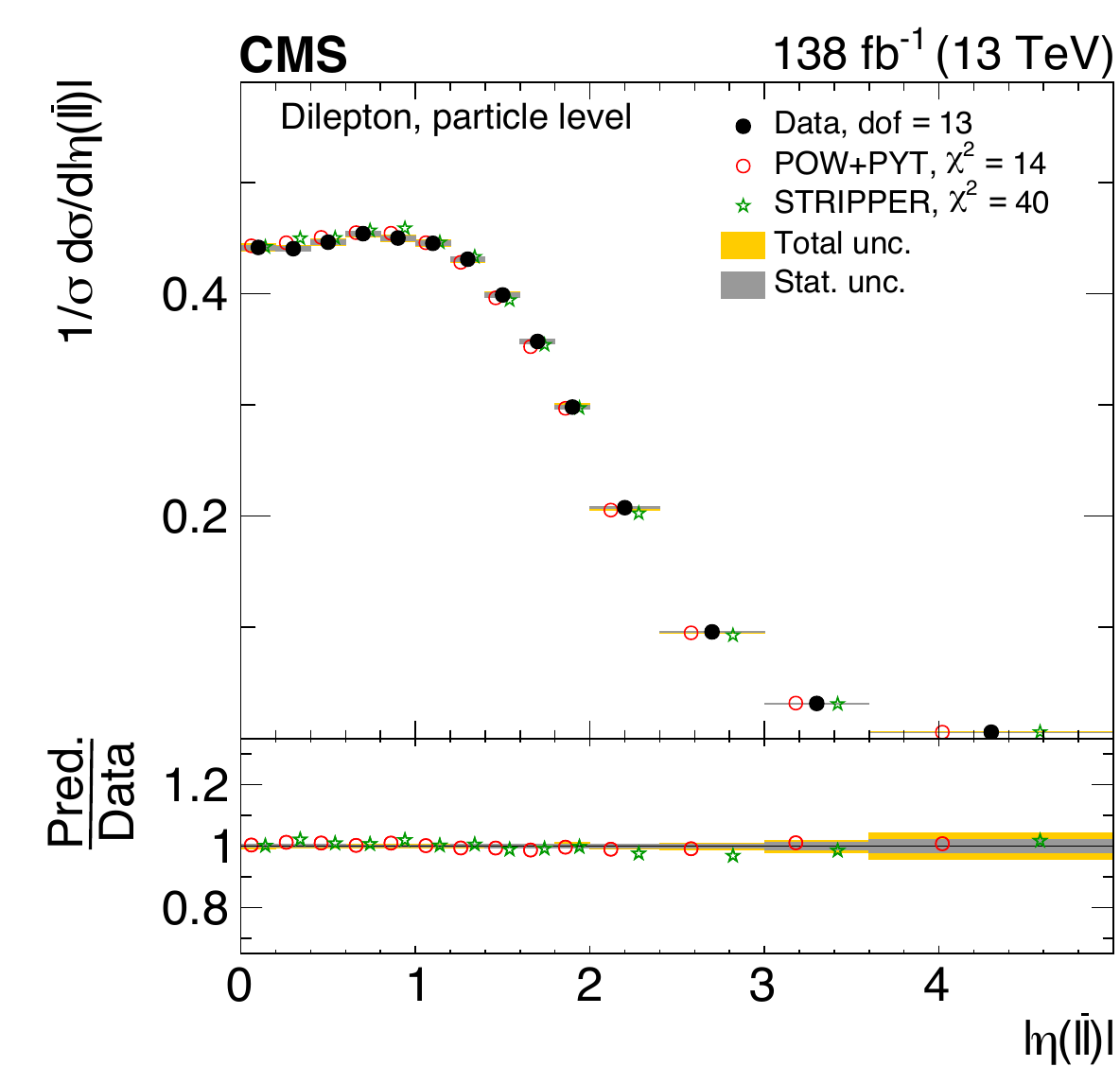}
\caption{Normalized differential \ttbar production cross sections as functions of \ptll (left) and \absetall (right)
are shown for data (filled circles), \PowPyt (`POW-PYT', open circles) simulation, and \StripperOnly NNLO calculation (stars).
Further details can be found in the caption of Fig.~\ref{fig:xsec-1d-theory-nor-ptlep-rptleptonic-rptlepptt}.}
\label{fig:xsec-1d-theory-nor-ptll-absetall}
\end{figure}

\begin{figure}
\centering
\includegraphics[width=1.00\textwidth]{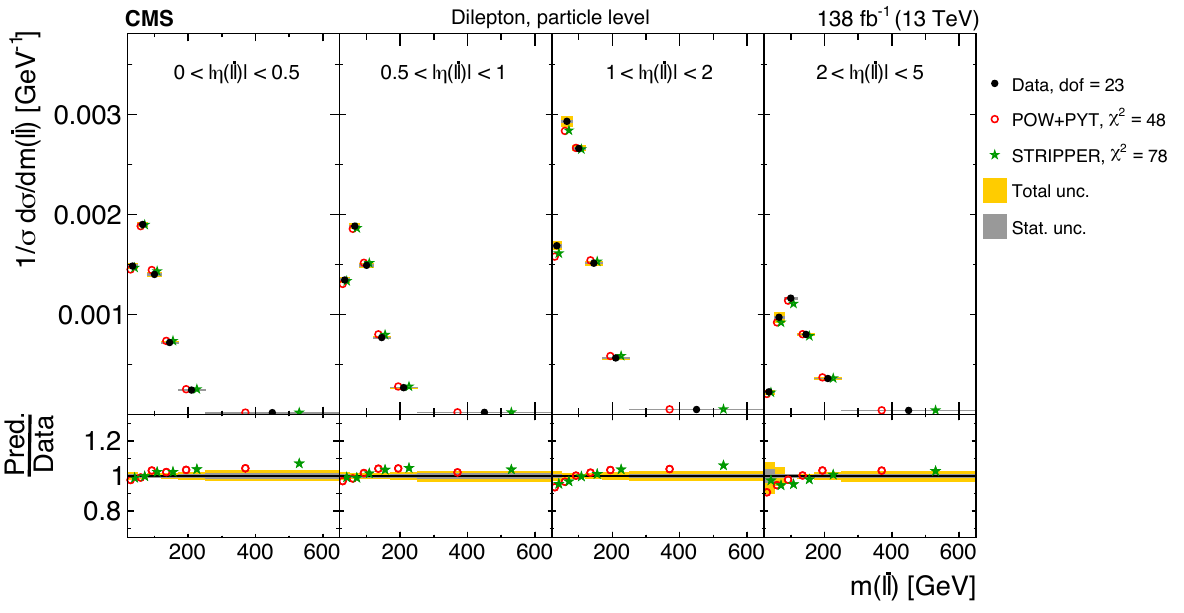}
\caption{Normalized \etallmll cross sections are shown for data (filled circles), \PowPyt (`POW-PYT', open circles) 
simulation, and \StripperOnly NNLO calculation (stars).
    Further details can be found in the caption of Fig.~\ref{fig:xsec-1d-theory-nor-ptlep-rptleptonic-rptlepptt}.}
    \label{fig:xsec-2d-theory-nor-etallmll}
\end{figure}

\begin{figure}
\centering
\includegraphics[width=1.00\textwidth]{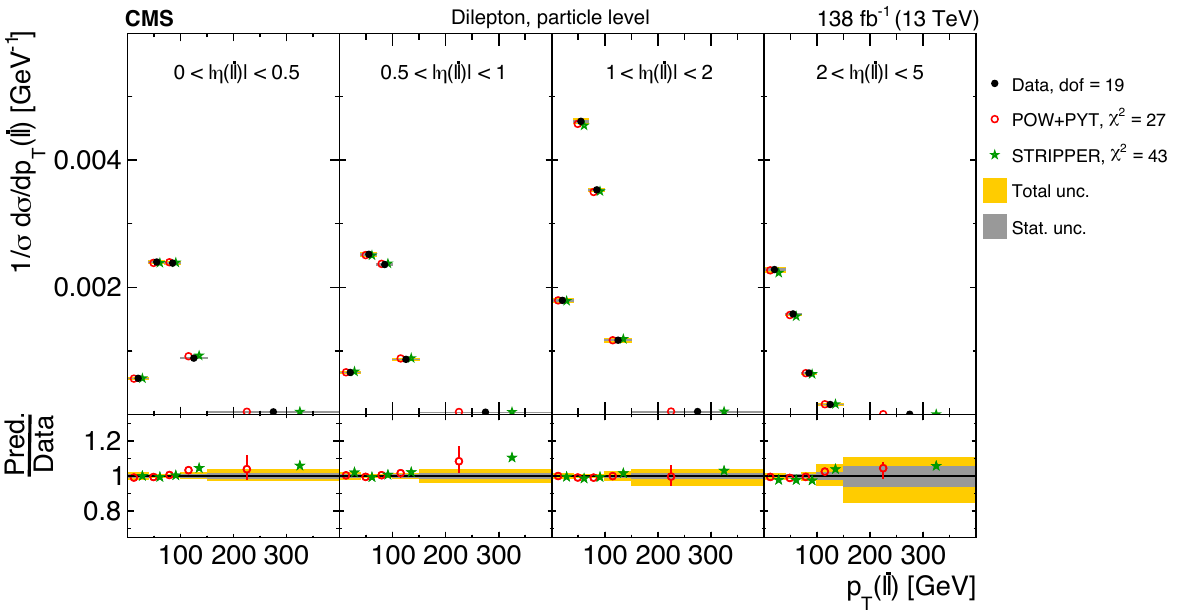}
\caption{Normalized \etallptll cross sections are shown for data (filled circles), \PowPyt (`POW-PYT', open
circles) simulation, and \StripperOnly NNLO calculation (stars).
    Further details can be found in the caption of Fig.~\ref{fig:xsec-1d-theory-nor-ptlep-rptleptonic-rptlepptt}.}
    \label{fig:xsec-2d-theory-nor-etallptll}
\end{figure}

\begin{figure}
\centering
\includegraphics[width=1.00\textwidth]{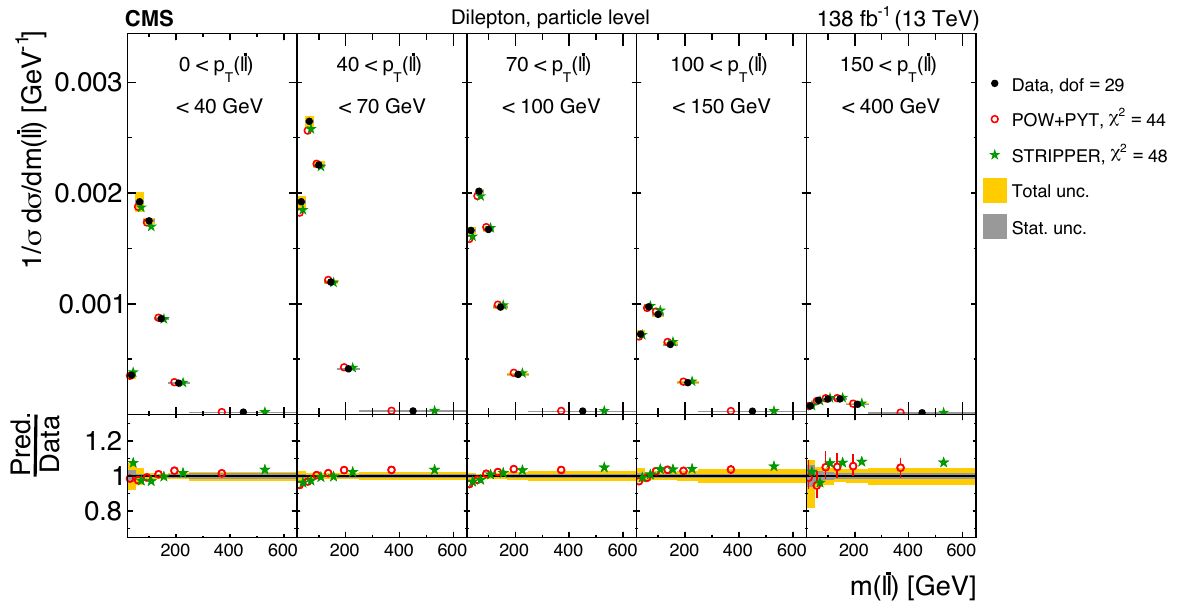}
\caption{Normalized \ptllmll cross sections are shown for data (filled circles), \PowPyt (`POW-PYT', open
circles) simulation, and \StripperOnly NNLO calculation (stars).
    Further details can be found in the caption of Fig.~\ref{fig:xsec-1d-theory-nor-ptlep-rptleptonic-rptlepptt}.}
    \label{fig:xsec-2d-theory-nor-ptllmll}
\end{figure}

The comparisons of the theoretical predictions, indicated
by different symbols,
to the measured normalized differential
cross sections are shown in Figs.~\ref{fig:xsec-1d-theory-nor-ptt-ptat}--\ref{fig:xsec-2d-theory-nor-ptllmll}.
The predictions from \PowPytSh are also shown, serving as a reference for the description by the MC models,
discussed in Sections~\ref{sec:res_1} and \ref{sec:res_2}.
The \chisq values of model-to-data comparisons
are listed in Tables~\ref{tab:chi2fixTheo_1d_nor_parton}--\ref{tab:chi2fixTheo_1d_nor_particle_lepb}, and the corresponding $p$-values can be found in Tables~\ref{tab:pvaluefixTheo_1d_nor_parton}--\ref{tab:pvaluefixTheo_1d_nor_particle_lepb}. The values are provided considering only measurement
uncertainties; for \PowPytSh, additional values are presented,
which include the full prediction uncertainties.
Comparisons of the predictions to the measured absolute cross sections are provided in Appendix~\ref{sec:res_th_abs}.
To illustrate the magnitude of the perturbative
uncertainties, the beyond-NLO calculations are displayed with vertical
bars constructed from the envelope of
six \mur and \muf variations, following the procedure outlined in Section~\ref{sec:sys_theo}.

For the cross sections as functions of the \ttbar and top quark kinematic observables
at the parton and particle levels, shown in Figs.~\ref{fig:xsec-1d-theory-nor-ptt-ptat}--\ref{fig:xsec-md-theory-nor-mttdphitt},
the theoretical models provide descriptions of the data that are similar
or improved in quality, compared to \PowPytSh, with a few exceptions.
For the \ptt and \ptat distributions shown in Fig.~\ref{fig:xsec-1d-theory-nor-ptt-ptat},
the \MatrixOnly, \StripperOnly, and \MiNNLOPSOnly models provide
a good description of the data, with no discernible trend towards a distribution that is harder, as exhibited by \PowPytSh.
For the \appNTLOOnly calculation, some wiggles are visible in the distribution of the
ratio to the data, leading to a rather poor \chisq.
The \StripperOnly model describes the data also at the particle level reasonably well.
The theoretical models also provide a reasonable description of other measured cross sections related to
top quark \pt, such as \rpttmtt (Fig. \ref{fig:xsec-1d-theory-nor-rpttmtt-rptttmtt}), and \mttptt
(Fig. \ref{fig:xsec-md-theory-nor-mttptt}),
clearly improving upon \PowPytSh.
For observables related to the top quark rapidity and the rapidity
and mass of the \ttbar system, the trends between data and theoretical calculations
are mostly similar to those of \PowPytSh.
Exemplary cases are the rapidity spectra \yt and \yat, depicted in Fig.~\ref{fig:xsec-1d-theory-nor-yt-yat},
where the models exhibit rapidity distributions that are more central than what is observed in data,
and the \mttytt cross sections (Fig.~\ref{fig:xsec-md-theory-nor-mttytt}), where the
predictions overshoot the data in the high-\mtt region at large \ytt.

Figures~\ref{fig:xsec-1d-theory-nor-pttt-mtt-ytt} and~\ref{fig:xsec-1d-theory-nor-dphitt-dytt} present the cross sections as functions of \pttt and \dphitt, which directly probe higher-order QCD effects.
These effects are reflected in the large uncertainties
of the theoretical predictions, as shown in the corresponding absolute
cross section measurements 
in Figs.~\ref{fig:xsec-1d-theory-abs-pttt-mtt-ytt} and~\ref{fig:xsec-1d-theory-abs-dphitt-dytt}.
The central predictions of these models fail to describe the data accurately.
Similar observations can be made for the multi-differential cross sections involving these two kinematic variables.

Summarizing, the beyond-NLO theoretical predictions provide descriptions
of the data that are of similar or improved quality, compared to \PowPytSh,
except for some of the kinematic distributions that are directly
sensitive to higher-order QCD corrections.

For the differential cross sections as functions of lepton and \PQb jet kinematic variables at the particle level,
shown in Figs.~\ref{fig:xsec-1d-theory-nor-ptlep-rptleptonic-rptlepptt}--\ref{fig:xsec-2d-theory-nor-ptllmll},
the \StripperOnly calculation provides descriptions
that are overall of similar quality compared to \PowPytSh.
One exception is the \mbb distribution, where \StripperOnly clearly
predicts too many events in the lowest mass bin.

\clearpage

\clearpage

\clearpage

\clearpage

\clearpage

\clearpage

\begin{sidewaystable*}
\centering
 \topcaption{The \chisq values and \ndf of the measured normalized single-differential cross sections for \ttbar and top quark kinematic observables at the parton level are shown with respect to the 
\PowPyt (`POW-PYT') simulation and various theoretical predictions with beyond-NLO precision.
The \chisq values are calculated taking only measurement uncertainties into account and excluding theory uncertainties.  
For \PowPytSh, the \chisq values including theory uncertainties are indicated with the brackets (w. unc.).}
 \label{tab:chi2fixTheo_1d_nor_parton}
 \renewcommand{\arraystretch}{1.4}
 \centering
 \begin{tabular}{lcccccc}
 \multirow{1}{*}{Cross section} & \hspace*{0.3 cm} \multirow{2}{*}{\ndf} \hspace*{0.3 cm} & \multicolumn{5}{c}{\chisq} \\
 \cline{3-7}
{variables} && \PowPytSh (w. unc.)  & \appNTLOOnly  & \MatrixOnly  & \StripperOnly  & \MiNNLOPSOnly  \\
\hline
\ptt& 6 & 15 \: (11) & 35  & 3  & 8  & 4  \\
\ptat& 6 & 13 \: (9) & \NA & 4  & 10  & 5  \\
\yt& 9 & 23 \: (20) & 23  & 16  & 17  & 15  \\
\yat& 9 & 28 \: (25) & \NA & 15 & 25  & 23  \\
\pttt& 6 & 22 \: (7) & \NA & 7  & 9  & 55  \\
\ytt& 11 & 10 \: (8) & \NA & 6  & 9  & 9 \\
\mtt  & 6 & 5 \: (3) & \NA & 5  & 5  & 2  \\
\dphitt& 3 & 1 \: (0) & \NA & 75  & 65  & 5  \\
\dytt& 7 & 16 \: (9) & \NA & 2 & 3 & 12  \\
\rpttmtt& 4 & 33 \: (20) & \NA & 3 & 4 & 5  \\
\rptttmtt& 8 & 17 \: (6) & \NA & 109 & 102  & 20  \\
\logxone& 8 & 14 \: (10) & \NA & \NA & 15 & 12  \\
\logxtwo& 8 & 10 \: (7) & \NA & \NA & 12  & 7  \\
 \end{tabular}
\end{sidewaystable*}

\begin{table*}
\centering
 \topcaption{The \chisq values and \ndf of the measured normalized single-differential cross sections for \ttbar and top quark kinematic observables at the particle level are shown with respect to the \PowPyt (`POW-PYT') simulation and the \StripperOnly NNLO calculation. 
The \chisq values are calculated taking only measurement uncertainties into account and excluding theory uncertainties.  
For \PowPytSh, the \chisq values including theory uncertainties are indicated with the brackets (w. unc.).}
 \label{tab:chi2fixTheo_1d_nor_particle_ttbar}
 \renewcommand{\arraystretch}{1.4}
 \centering
 \begin{tabular}{lccc}
 \multirow{1}{*}{Cross section} & \hspace*{0.3 cm} \multirow{2}{*}{\ndf} \hspace*{0.3 cm} & \multicolumn{2}{c}{\chisq} \\
 \cline{3-4}
{variables} && \PowPytSh (w. unc.)  & \StripperOnly  \\
\hline
\ptt& 6 & 17 \: (12) & 3 \\
\ptat& 6 & 14 \: (9) & 3 \\
\yt& 9 & 19 \: (16) & 11 \\
\yat& 9 & 24 \: (20) & 10 \\
\pttt& 6 & 21 \: (7) & 83 \\
\ytt& 11 & 10 \: (7) & 19 \\
\mtt  & 6 & 5 \: (3) & 4 \\
\dphitt& 3 & 1 \: (0) & 1076 \\
\dytt& 7 & 15 \: (10) & 4 \\
\rpttmtt& 4 & 30 \: (20) & 8 \\
\rptttmtt& 8 & 18 \: (7) & 144  \\
\logxone& 8 & 14 \: (9) & 10  \\
\logxtwo& 8 & 9 \: (6) & 9  \\
 \end{tabular}
\end{table*}

\begin{sidewaystable*}
 \centering
 \topcaption{The \chisq values and \ndf of the measured normalized multi-differential cross sections for \ttbar and top quark kinematic observables at the parton level are shown with respect to
the \PowPyt (`POW-PYT') simulation and various theoretical predictions with beyond-NLO precision. 
The \chisq values are calculated taking only measurement uncertainties into account and excluding theory uncertainties.  
For \PowPytSh, the \chisq values including theory uncertainties are indicated with the brackets (w. unc.).}
 \label{tab:chi2fixTheo_nor_parton_ttbar}
 \renewcommand{\arraystretch}{1.4}
 \centering
 \begin{tabular}{lcccccc}
 \multirow{1}{*}{Cross section} & \hspace*{0.3 cm} \multirow{2}{*}{\ndf} \hspace*{0.3 cm} & \multicolumn{5}{c}{\chisq} \\
 \cline{3-7}
{variables} && \PowPytSh (w. unc.)  & \appNTLOOnly  & \MatrixOnly  & \StripperOnly  & \MiNNLOPSOnly   \\
\hline
\ytptt& 15 & 40 \: (31) & 36 & 24 & 24 & 23  \\
\mttptt& 8 & 83 \: (35) & \NA & 14 & 16  & 15  \\
\pttpttt& 15 & 39 \: (21) & \NA & 41 & 54 & 66  \\
\mttytt  & 15 & 64 \: (42) & \NA & 53 & 47 & 51  \\
\yttpttt& 15 & 28 \: (15) & \NA & 38 & 42  & 66  \\
\mttpttt  & 15 & 61 \: (43) & \NA & 94 & 108  & 143  \\
\ptttmttytt  & 47 & 89 \: (64) & \NA & \NA & 69 & 105  \\
\mttyt& 15 & 61 \: (37) & \NA & 35 & 30 & 53  \\
\mttdetatt& 11 & 165 \: (31) & \NA & 30 & 31 & 56  \\
\mttdphitt& 11 & 74 \: (47) & \NA & \NA & 41  & 94  \\
 \end{tabular}
\end{sidewaystable*}

\begin{table*}
 \centering
 \topcaption{The \chisq values and \ndf of the measured normalized multi-differential cross sections for \ttbar and top quark kinematic observables at the particle level are shown with respect to the \PowPyt (`POW-PYT') simulation and the \StripperOnly NNLO calculation. 
The \chisq values are calculated taking only measurement uncertainties into account and excluding theory uncertainties.  
For \PowPytSh, the \chisq values including theory uncertainties are indicated with the brackets (w. unc.).}
 \label{tab:chi2fixTheo_nor_particle_ttbar}
 \renewcommand{\arraystretch}{1.4}
 \centering
 \begin{tabular}{lccc}
 \multirow{1}{*}{Cross section} & \hspace*{0.3 cm} \multirow{2}{*}{\ndf} \hspace*{0.3 cm} & \multicolumn{2}{c}{\chisq} \\
 \cline{3-4}
{variables} && \PowPytSh (w. unc.)  & \StripperOnly  \\
\hline
\ytptt& 15 & 35 \: (25) & 16 \\
\mttptt& 8 & 86 \: (36) & 18 \\
\pttpttt& 15 & 35 \: (19) & 160  \\
\mttytt  & 15 & 77 \: (40) & 43  \\
\yttpttt& 15 & 27 \: (18) & 78  \\
\mttpttt  & 15 & 61 \: (36) & 363  \\
\ptttmttytt  & 47 & 114 \: (68) & 137  \\
\mttyt& 15 & 57 \: (26) & 20 \\
\mttdetatt& 11 & 155 \: (30) & 37  \\
\mttdphitt& 11 & 71 \: (42) & 405  \\
 \end{tabular}
\end{table*}

\begin{table*}
\centering
 \topcaption{The \chisq values and \ndf of the measured normalized single-differential cross sections for lepton and \PQb-jet kinematic observables at the particle level are shown with respect to the \PowPyt (`POW-PYT') simulation and the \StripperOnly NNLO calculation. 
The \chisq values are calculated taking only measurement uncertainties into account and excluding theory uncertainties.  
For \PowPytSh, the \chisq values including theory uncertainties are indicated with the brackets (w. unc.).}
 \label{tab:chi2fixTheo_1d_nor_particle_lepb}
 \renewcommand{\arraystretch}{1.4}
 \centering
 \begin{tabular}{lccc}
 \multirow{1}{*}{Cross section} & \hspace*{0.3 cm} \multirow{2}{*}{\ndf} \hspace*{0.3 cm} & \multicolumn{2}{c}{\chisq} \\
 \cline{3-4}
{variables} && \PowPytSh (w. unc.)  & \StripperOnly \\
\hline
\ptlep& 11 & 28 \: (18) & 16\\
\ptlep trailing/\ptlep leading& 9 & 15 \: (11) & 5  \\
\ptlep/\ptat& 4 & 10 \: (9) & 5 \\
\ptb leading& 9 & 5 \: (4) & 7  \\
\ptb trailing& 6 & 6 \: (4) & 3  \\
\rptbsptts& 3 & 19 \: (15) & 8  \\
\mll& 11 & 23 \: (20) & 13  \\
\mbb& 6 & 15 \: (12) & 76 \\
\mllbb& 18 & 33 \: (18) & 15 \\
\ptll& 8 & 4 \: (3) & 5 \\
\absetall& 13 & 14 \: (9) & 40 \\
\etallmll& 23 & 48 \: (28) & 78 \\
\etallptll& 19 & 27 \: (14) & 43  \\
\ptllmll& 29 & 44 \: (37) & 48  \\
 \end{tabular}
\end{table*}

\clearpage

\subsection{Comparisons to \texorpdfstring{\PowPyt}{PowPyt} predictions using different PDFs}
\label{sec:res_pdf_comp}

In this subsection the
sensitivity of the measured differential \ttbar production cross sections
to the PDFs is assessed.
The \ttbar production at the LHC is known to be particularly sensitive
to the gluon density at higher proton momentum fractions $x$.
In the previous CMS analyses~\cite{Sirunyan:2017azo,Sirunyan:2019zvx}, systematic investigations were performed
in which different sets of \ttbar kinematic spectra were included in PDF fits and their impact evaluated.
A particularly large sensitivity was observed in Ref.~\cite{Sirunyan:2017azo}
for the \mttytt cross sections measured at $\sqrt{s}=8\TeV$, leading to a strong constraint on the gluon density
at $x$ values around 0.3.
In the present analysis, the PDF sensitivity is assessed
by comparing kinematic spectra of the top quark or the \ttbar system
with the \PowPyt predictions, evaluated with a number of different PDF sets:
NNPDF3.1~\cite{Ball:2017nwa}, CT14~\cite{Dulat:2015mca},
ABMP16~\cite{Alekhin:2018pai}, MMHT2014~\cite{Harland-Lang:2014zoa}, and HERAPDF2.0~\cite{Abramowicz:2015mha}.
For the NNPDF3.1 PDFs, both the NNLO (used in the nominal simulation) and NLO variants are studied, while 
only the NLO versions are taken for all other sets.
The PDFs differ in the input data and methodologies
that were used to extract them, as discussed elsewhere~\cite{Butterworth:2015oua,Accardi:2016ndt}.

\begin{figure*}[!phtb]
\centering
\includegraphics[width=0.49\textwidth]{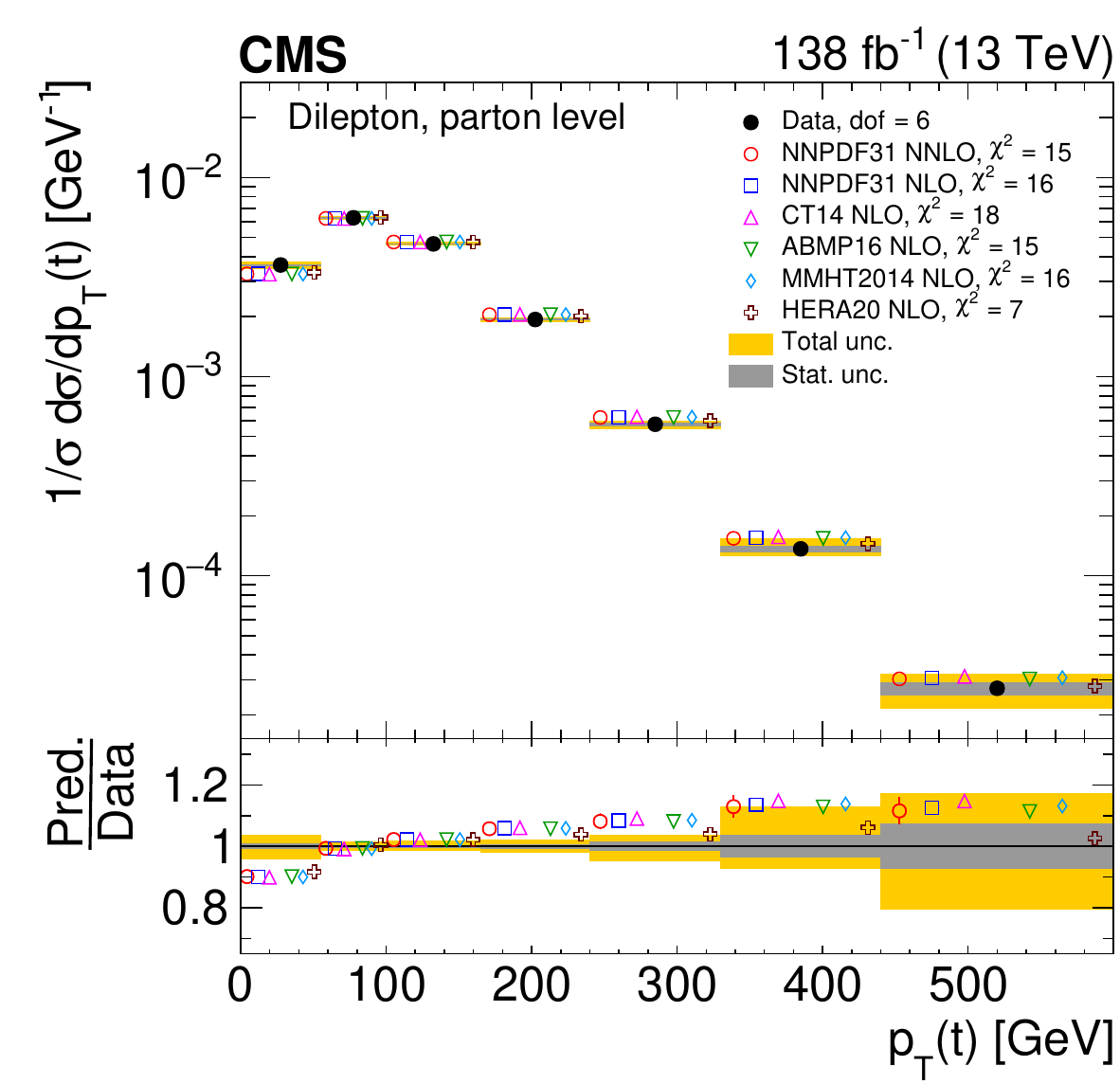}
\includegraphics[width=0.49\textwidth]{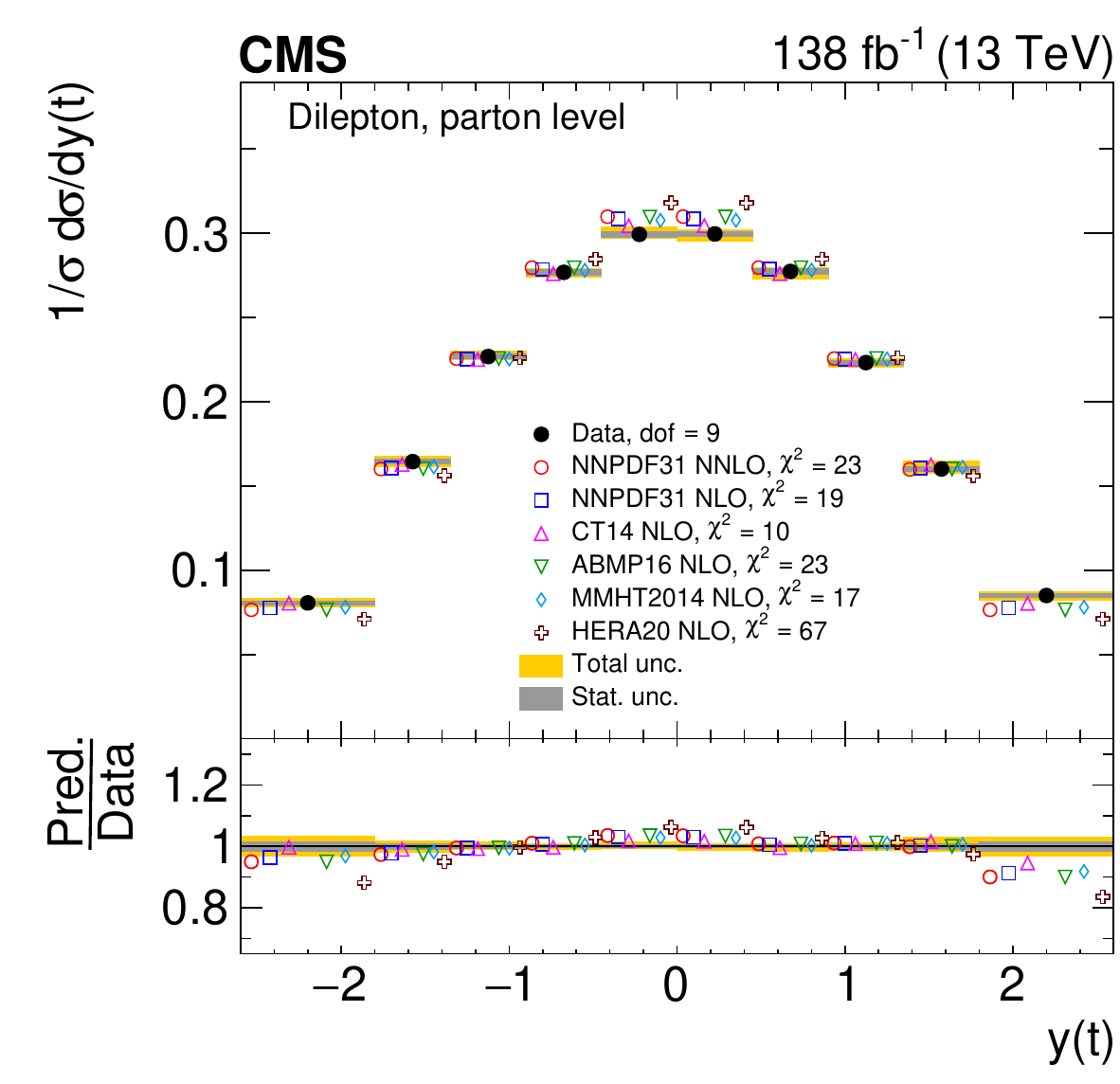}
\includegraphics[width=0.49\textwidth]{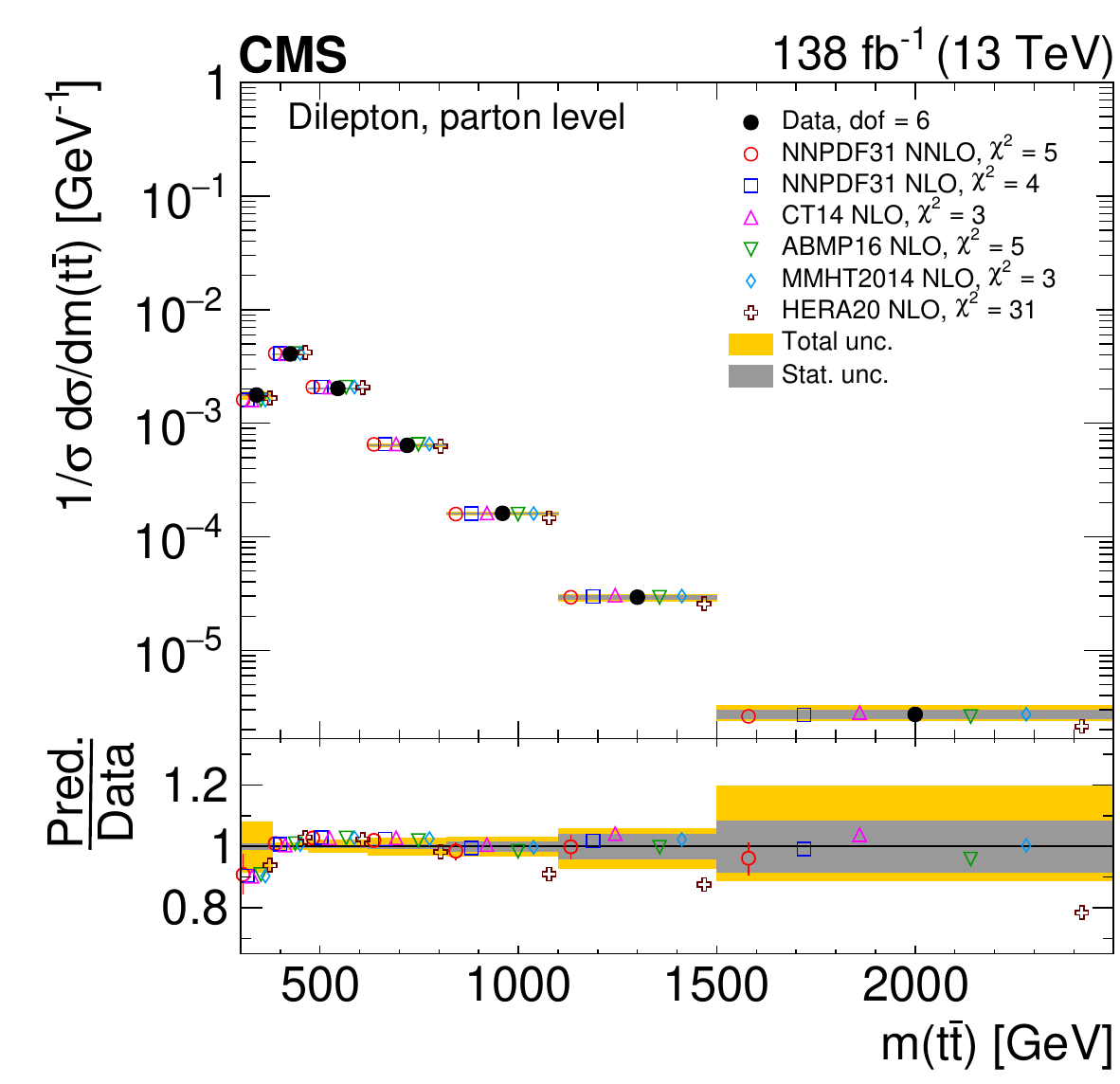}
\includegraphics[width=0.49\textwidth]{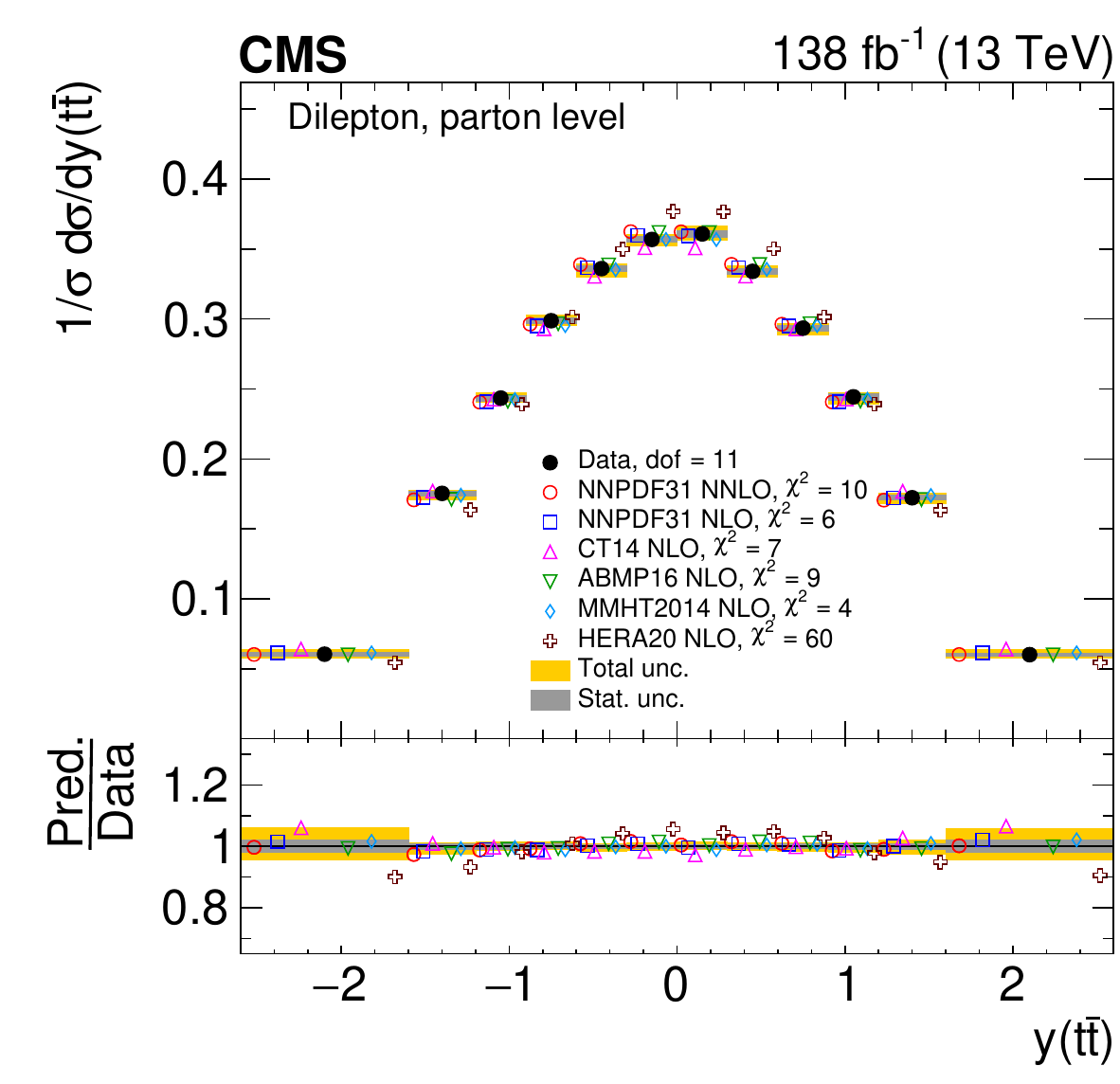}
\caption{Normalized differential \ttbar production cross sections as functions of \ptt (upper left), \yt
(upper right),
\mtt (lower left), and \ytt (lower right),
measured at the parton level in the full phase space.
The data are shown as filled circles with grey and yellow bands indicating the statistical and total uncertainties
(statistical and systematic uncertainties added in quadrature), respectively.
For each distribution, the number of degrees of freedom (dof) is also provided.
The cross sections are compared to predictions from the \PowPyt (`POW-PYT') simulation with various PDF sets.
The nominal prediction (open circles) uses the PDF set NNPDF3.1 at NNLO accuracy, assuming a top quark mass value
of 172.5\GeV and $\alpS = 0.118$.
The alternative PDF sets (other points) constitute NNPDF3.1, CT14, ABMP16, MMHT2014, and HERAPDF2.0
at NLO accuracy and assume the same values for the top quark mass and \alpS as the nominal NNPDF3.1 NNLO PDF
set.
The estimated uncertainties in the nominal prediction are represented by vertical bars on the corresponding
points.
For each PDF set, a value of \chisq is reported that takes into account the measurement uncertainties.
The lower panel in each plot shows the ratios of the predictions to the data.}
\label{fig:xsec-1d-PDF-nor-mtt-ytt-logxone-logxtwo}
\end{figure*}

\begin{figure}
\centering
\includegraphics[width=0.49\textwidth]{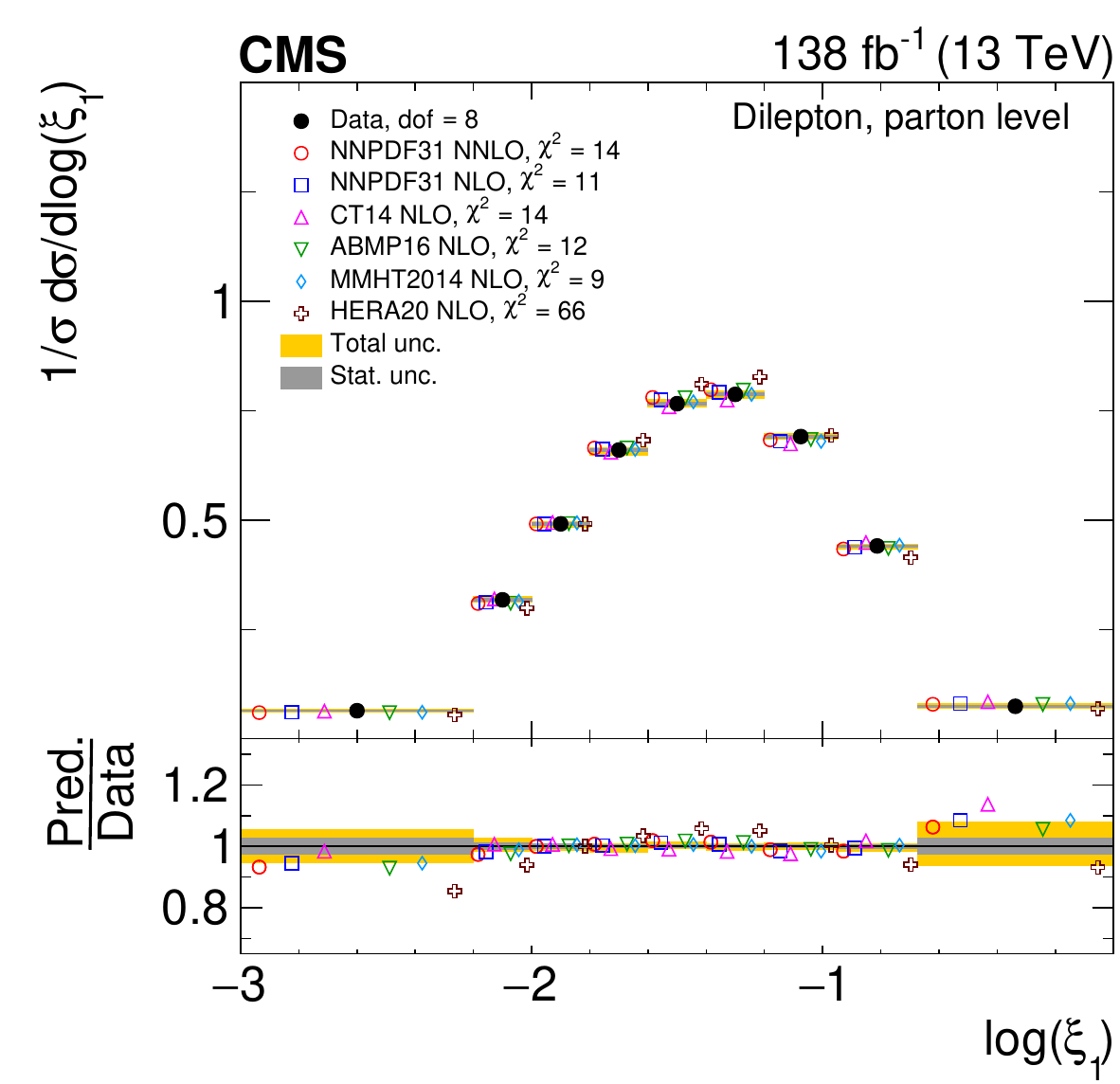}
\includegraphics[width=0.49\textwidth]{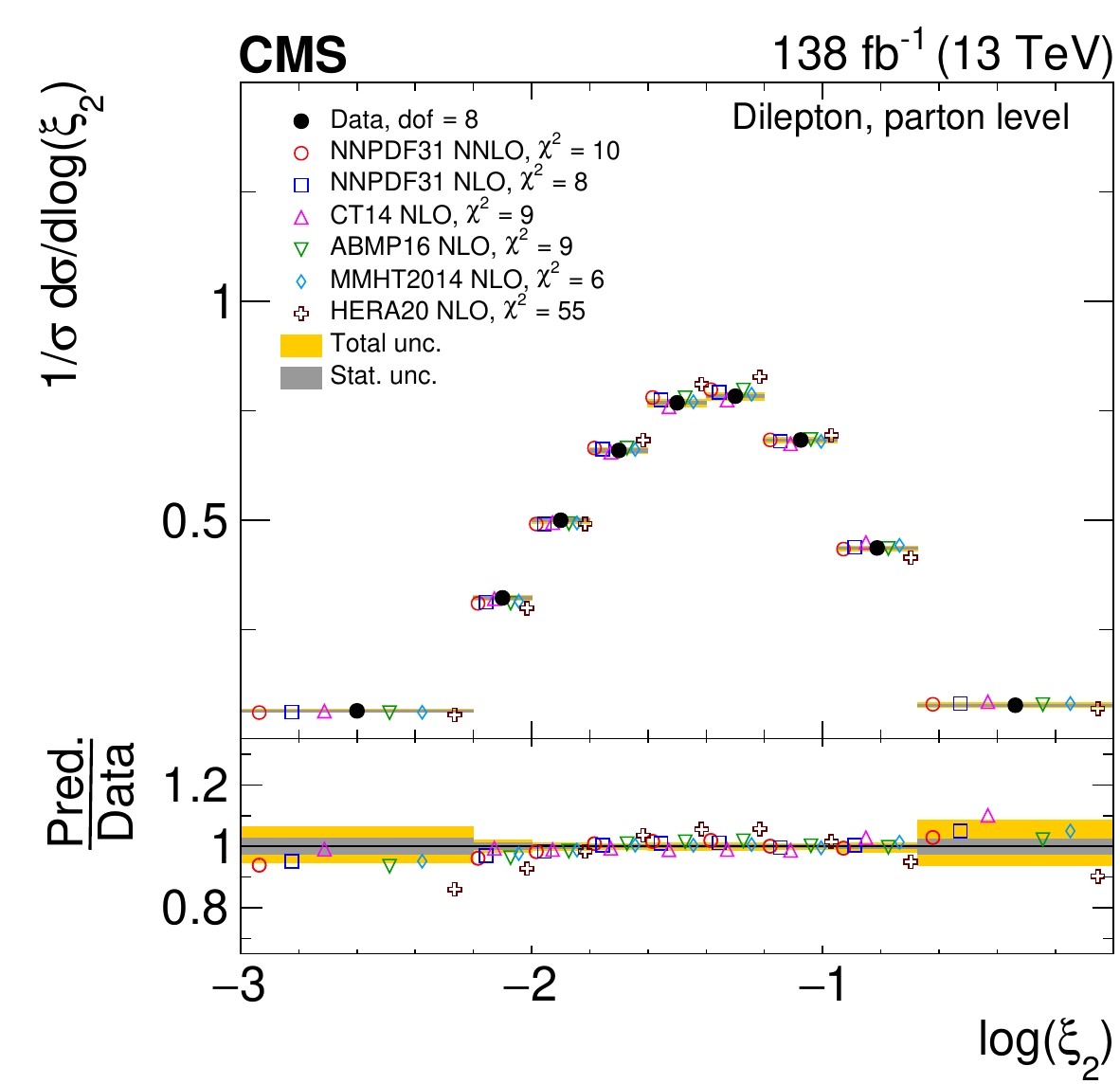}
\includegraphics[width=0.99\textwidth]{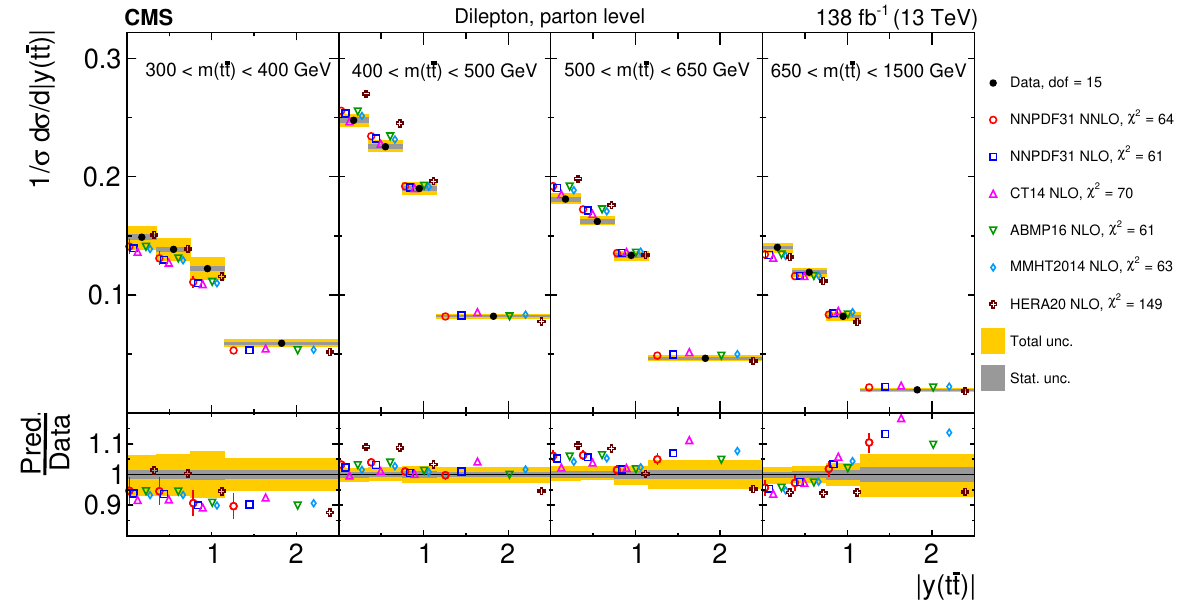}
\caption{Normalized \logxone (upper left), \logxtwo (upper right), and \mttytt (lower) cross sections are shown
for data (filled circles) and predictions from the \PowPyt (`POW-PYT') simulation with various PDF sets (other
points).
Further details can be found in the caption of Fig.~\ref{fig:xsec-1d-PDF-nor-mtt-ytt-logxone-logxtwo}.}
\label{fig:xsec-1d-PDF-nor-mttytt}
\end{figure}

Figure~\ref{fig:xsec-1d-PDF-nor-mtt-ytt-logxone-logxtwo} shows the measured normalized 
single-differential cross sections,
defined at the parton level, as functions of
\ptt, \yt, \mtt, and \ytt, and compared to the predictions obtained with different PDF sets.
Most of the PDFs provide a similar
description of the data, except HERAPDF2.0 NLO that describes the \ptt distribution well but exhibits
clear shape deviations for the other spectra, leading to large \chisq values.
In Fig.~\ref{fig:xsec-1d-PDF-nor-mttytt},
the \logxone, \logxtwo, and \mttytt cross section distributions are shown.
As discussed above, in the LO QCD picture the \logxone and \logxtwo variables
represent the proton momentum fractions
carried by the two partons entering the hard interaction.
The HERAPDF2.0 NLO prediction undershoots the data
near the smallest and largest values of these observables,
which may give a hint of a possibly wrong $x$ dependence of the
gluon density in this PDF set.
For \mttytt, all PDFs again provide data descriptions of similar quality, except HERAPDF2.0 NLO,
that predicts too few events towards large \mtt and \ytt values.
The particularly large \chisq for HERAPDF2.0 NLO for this distribution
indicates an enhanced PDF sensitivity
for this double-differential cross section.
Summarizing, the differential
spectra of the top quark and the \ttbar system show some sensitivity
to the PDFs and, therefore, it will be useful to include the corresponding data in future global
PDF fits.

\clearpage

\clearpage

\section{Summary}
\label{sec:concl}

A measurement of differential top quark pair (\ttbar) production cross sections in 
proton-proton collisions at $\sqrt{s}=13$\TeV was presented,
performed with events containing two oppositely charged leptons (electrons or muons).
The data used in this analysis were recorded in the years 2016 through 2018 with the CMS detector at the LHC
and correspond to an integrated luminosity of \lumivalue.
Differential cross sections are measured as functions of kinematical observables of the \ttbar system, the top
quark and antiquark and their decay products, and the total number of additional jets in the event not 
originating from the \ttbar decay.
The measurements are performed as functions of single observables, 
or simultaneously as functions of two or three
kinematic variables.
The differential cross sections are defined both with particle-level objects
in a fiducial phase space close to that of the detector acceptance and with
parton-level top quarks in the full phase space.
Overall, both the statistical and the systematic uncertainties in the measurements
are improved by a factor of about two compared to the previous
analyses~\cite{Sirunyan:2018ucr,Sirunyan:2019zvx} which are based on the 2016 data set.

Predictions of several next-to-leading-order (NLO) Monte Carlo (MC) event generators
that differ in the hard matrix element, parton shower, and hadronization models
were compared to the data.
The predictions of these MC models, without
taking theoretical uncertainties into account, generally fail
to describe many of the measured cross sections in their full kinematic range.
The predicted transverse momentum \pt distributions of the top quark 
and antiquark are harder than observed in the data,
and the rapidity distributions are more central.
The invariant mass and rapidity distributions of the \ttbar system are reasonably well described by the models overall.
The predictions for the \ttbar transverse momentum distribution differ from the data even more than the top quark and antiquark distributions do; none of them provides a good description of the data.
Double- and triple-differential cross sections show large 
model-to-data discrepancies, for instance the effect
of a harder top quark \pt spectrum \ptt in the models is pronounced at high \mtt.
Differential cross sections as functions of kinematic observables of the leptons
and \PQb jets originating from the decay of the top quark and antiquark are measured with
high precision.
Overall, the observed trends for these objects follow those for the top quarks and antiquarks, with the models
predicting
harder \pt spectra than seen in the data.
For the leptons, this effect is somewhat enhanced and furthermore the dilepton invariant mass spectrum
is harder in the models than in the data.
The distribution of the multiplicity of additional jets in \ttbar events 
shows varying level
of agreement between data and the models.
When considered as a function of jet multiplicity, the evolution of the shapes
of \ttbar, top quark and antiquark kinematic distributions is different
for the models and for data.
There is an indication that the trend of harder \ptt distributions in the models
is localized at small jet multiplicities.

Selected kinematic distributions 
were also compared to a variety of theoretical predictions beyond NLO precision.
For observables of the top quark and the \ttbar system,
these predictions provide descriptions of the data that are of similar or improved quality, compared
to the MC model best describing each variable, except for some of the kinematic spectra that are directly sensitive to higher-order QCD effects.
For observables associated with the leptons and \PQb jets, the quality of the tested next-to-NLO model is on average comparable to but not better than that of the NLO MC models.
Comparing several kinematic distributions of the top quark and the \ttbar system
to NLO MC models using various parton distribution function (PDF) sets,
clear differences are observed which indicate a sensitivity to PDFs that could be exploited in future PDF fits.

For each distribution, the quality of the description of the data 
by the models has been assessed with a \chisq test statistic.
When only the measurement uncertainties are taken into account in the calculation (\ie neglecting the 
uncertainties on the predictions), the $p$-values obtained from the \chisq tests are in general close to zero, pointing to a 
poor description of the data by the nominal models.
For the \PowPyt model, additional \chisq values have been evaluated
including the uncertainties on the prediction. This inclusion often
leads to substantially reduced \chisq values with reasonable $p$-values.
However, for several 
distributions, and in particular for a larger fraction of the multi-differential distributions, the observed differences between data and simulation
still remain significant, providing important input for future theoretical predictions.

\begin{acknowledgments}
We congratulate our colleagues in the CERN accelerator departments for the excellent performance of the LHC and thank the technical and administrative staffs at CERN and at other CMS institutes for their contributions to the success of the CMS effort. In addition, we gratefully acknowledge the computing centers and personnel of the Worldwide LHC Computing Grid and other centers for delivering so effectively the computing infrastructure essential to our analyses. Finally, we acknowledge the enduring support for the construction and operation of the LHC, the CMS detector, and the supporting computing infrastructure provided by the following funding agencies: SC (Armenia), BMBWF and FWF (Austria); FNRS and FWO (Belgium); CNPq, CAPES, FAPERJ, FAPERGS, and FAPESP (Brazil); MES and BNSF (Bulgaria); CERN; CAS, MoST, and NSFC (China); MINCIENCIAS (Colombia); MSES and CSF (Croatia); RIF (Cyprus); SENESCYT (Ecuador); ERC PRG, RVTT3 and TK202 (Estonia); Academy of Finland, MEC, and HIP (Finland); CEA and CNRS/IN2P3 (France); SRNSF (Georgia); BMBF, DFG, and HGF (Germany); GSRI (Greece); NKFIH (Hungary); DAE and DST (India); IPM (Iran); SFI (Ireland); INFN (Italy); MSIP and NRF (Republic of Korea); MES (Latvia); LAS (Lithuania); MOE and UM (Malaysia); BUAP, CINVESTAV, CONACYT, LNS, SEP, and UASLP-FAI (Mexico); MOS (Montenegro); MBIE (New Zealand); PAEC (Pakistan); MES and NSC (Poland); FCT (Portugal); MESTD (Serbia); MCIN/AEI and PCTI (Spain); MOSTR (Sri Lanka); Swiss Funding Agencies (Switzerland); MST (Taipei); MHESI and NSTDA (Thailand); TUBITAK and TENMAK (Turkey); NASU (Ukraine); STFC (United Kingdom); DOE and NSF (USA).

\hyphenation{Rachada-pisek} Individuals have received support from the Marie-Curie program and the European Research Council and Horizon 2020 Grant, contract Nos.\ 675440, 724704, 752730, 758316, 765710, 824093, and COST Action CA16108 (European Union); the Leventis Foundation; the Alfred P.\ Sloan Foundation; the Alexander von Humboldt Foundation; the Science Committee, project no. 22rl-037 (Armenia); the Belgian Federal Science Policy Office; the Fonds pour la Formation \`a la Recherche dans l'Industrie et dans l'Agriculture (FRIA-Belgium); the Agentschap voor Innovatie door Wetenschap en Technologie (IWT-Belgium); the F.R.S.-FNRS and FWO (Belgium) under the ``Excellence of Science -- EOS" -- be.h project n.\ 30820817; the Beijing Municipal Science \& Technology Commission, No. Z191100007219010 and Fundamental Research Funds for the Central Universities (China); the Ministry of Education, Youth and Sports (MEYS) of the Czech Republic; the Shota Rustaveli National Science Foundation, grant FR-22-985 (Georgia); the Deutsche Forschungsgemeinschaft (DFG), under Germany's Excellence Strategy -- EXC 2121 ``Quantum Universe" -- 390833306, and under project number 400140256 - GRK2497; the Hellenic Foundation for Research and Innovation (HFRI), Project Number 2288 (Greece); the Hungarian Academy of Sciences, the New National Excellence Program - \'UNKP, the NKFIH research grants K 124845, K 124850, K 128713, K 128786, K 129058, K 131991, K 133046, K 138136, K 143460, K 143477, 2020-2.2.1-ED-2021-00181, and TKP2021-NKTA-64 (Hungary); the Council of Science and Industrial Research, India; ICSC -- National Research Center for High Performance Computing, Big Data and Quantum Computing, funded by the EU NexGeneration program (Italy); the Latvian Council of Science; the Ministry of Education and Science, project no. 2022/WK/14, and the National Science Center, contracts Opus 2021/41/B/ST2/01369 and 2021/43/B/ST2/01552 (Poland); the Funda\c{c}\~ao para a Ci\^encia e a Tecnologia, grant CEECIND/01334/2018 (Portugal); the National Priorities Research Program by Qatar National Research Fund; MCIN/AEI/10.13039/501100011033, ERDF ``a way of making Europe", and the Programa Estatal de Fomento de la Investigaci{\'o}n Cient{\'i}fica y T{\'e}cnica de Excelencia Mar\'{\i}a de Maeztu, grant MDM-2017-0765 and Programa Severo Ochoa del Principado de Asturias (Spain); the Chulalongkorn Academic into Its 2nd Century Project Advancement Project, and the National Science, Research and Innovation Fund via the Program Management Unit for Human Resources \& Institutional Development, Research and Innovation, grant B37G660013 (Thailand); the Kavli Foundation; the Nvidia Corporation; the SuperMicro Corporation; the Welch Foundation, contract C-1845; and the Weston Havens Foundation (USA).

\end{acknowledgments}

\bibliography{auto_generated}   

\begin{appendix}

\section{Definition of \texorpdfstring{$\chi^2$}{} for theory-to-data comparisons}
\label{sec:app}

We define here the \chisq values for
the theory-to-data comparisons shown in Section~\ref{sec:res} and Appendix~\ref{sec:res_abs}.
The standard \chisq values are calculated taking the measurement uncertainties into account, but ignoring the uncertainties
in the predictions:
\begin{equation}
\label{eq:chi2nm1}
\chisq = \mathbf{R}^{T}_{N} \mathbf{Cov}^{-1}_{N} \mathbf{R}_{N},
\end{equation}
where $N$ denotes the number of bins of the respective cross section distribution and
$\mathbf{R}_{N}$ the vector of differences of the measured cross sections and the corresponding predictions.
The covariance matrix $\mathbf{Cov}$ is calculated as:
\begin{equation}
\label{eq:covmat}
\mathbf{Cov} = \mathbf{Cov}^\text{unf} + \mathbf{Cov}^\text{syst},
\end{equation}
where $\mathbf{Cov}^\text{unf}$ and $\mathbf{Cov}^\text{syst}$ are the covariance matrices representing
the statistical uncertainties from the unfolding,
and the systematic uncertainties, respectively.
The systematic covariance matrix $\mathbf{Cov}^\text{syst}$ is calculated as
\begin{equation}
    \mathbf{Cov}^\text{syst}_{ij} = \sum_{k,l} \frac{1}{N_k} C_{j,k,l}C_{i,k,l},\\ \quad 1 \le i \le N, \quad 1 \le j \le N.
    \label{eq:chi2}
    \end{equation}
Here, $C_{i,k,l}$ denotes the signed systematic shift of the measurement in the 
$i$th bin arising from variation $l$ of source $k$ and
$N_k$ is the number of variations for source $k$.
The sums run over all sources of systematic uncertainties and all their corresponding variations.
Most of the systematic uncertainty sources in this analysis consist of positive and negative variations
and thus have $N_k = 2$, though one model uncertainty
(namely, the model of color reconnection) consists of more than two variations,
a property that is accounted for in Eq.~(\ref{eq:chi2}).
For the \PowPytSh model,  
additional \chisq values are provided including
the uncertainties in the predictions.
This is achieved by adding the covariance matrix of the predictions, 
$\mathbf{Cov}^\text{pred}$, calculated analogously to Eq.~(\ref{eq:chi2}), to $\mathbf{Cov}$.

For normalized cross sections the \chisq values are calculated as 
\begin{equation}
\label{eq:chi2nm1_norm}
\chisq = \mathbf{R}^{T}_{N-1} \mathbf{Cov}^{-1}_{N-1} \mathbf{R}_{N-1},
\end{equation}
where $\mathbf{R}_{N-1}$ is the column vector of the residuals 
calculated as the difference of the measured normalized cross sections and the corresponding predictions
and discarding one of the $N$ bins, and $\mathbf{Cov}_{N-1}$ is the $(N-1)\times(N-1)$ 
submatrix obtained from the full covariance matrix by
discarding the corresponding row and column.
The matrix $\mathbf{Cov}_{N-1}$ obtained in this way is invertible, 
while the original covariance matrix $\mathbf{Cov}$ is singular 
because for normalized cross sections one degree of freedom is lost.

\clearpage

\section{Results for absolute cross sections}
\label{sec:res_abs}

Absolute differential cross sections corresponding to the normalized ones discussed in Section~\ref{sec:res}
are presented in the following, together with tables listing \chisq values of all prediction-to-data
comparisons.

\subsection{Comparisons to MC simulations}
\label{sec:res_mc_abs}

\begin{figure*}[!phtb]
\centering
\includegraphics[width=0.49\textwidth]{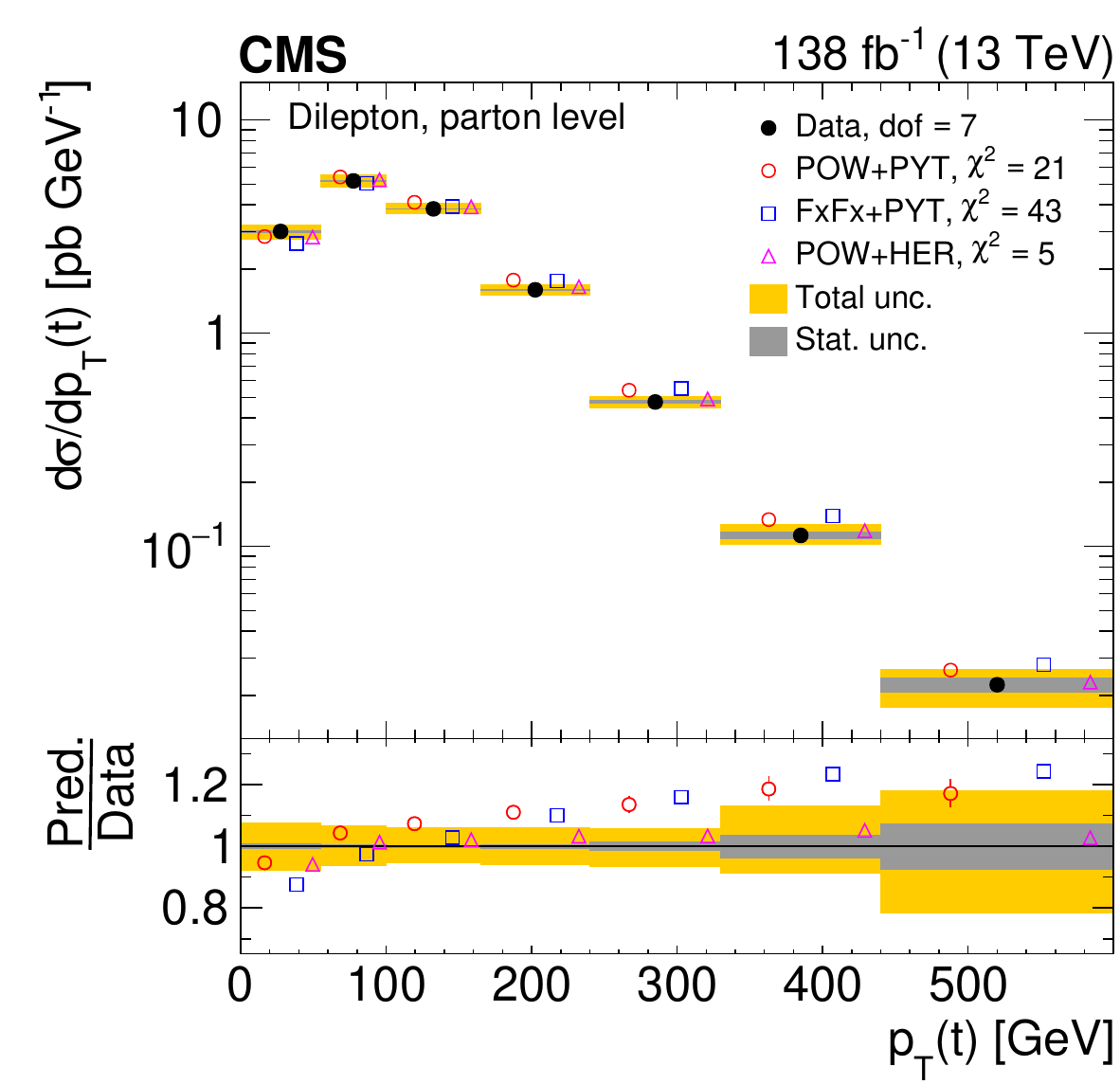}
\includegraphics[width=0.49\textwidth]{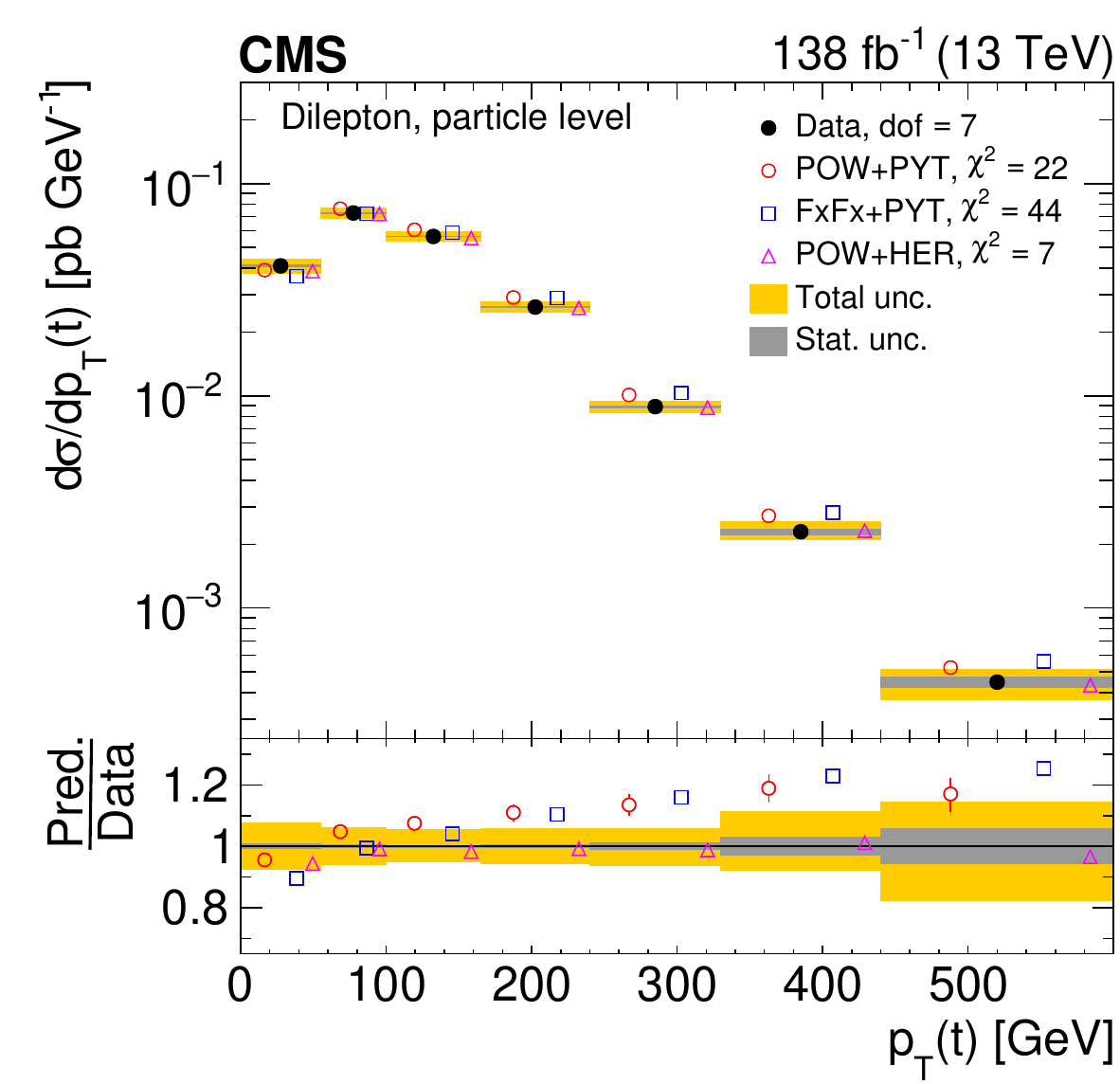}
\includegraphics[width=0.49\textwidth]{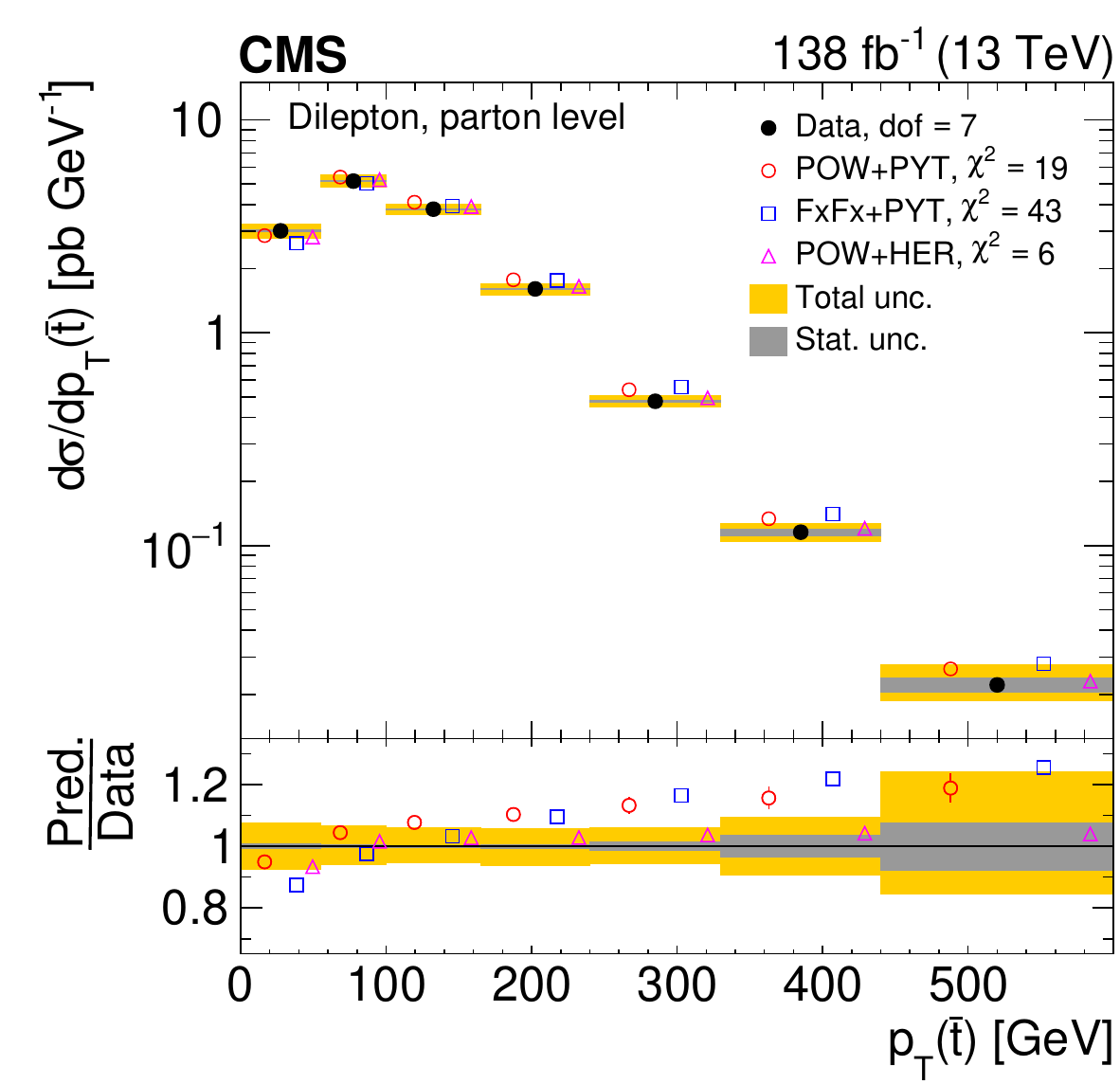}
\includegraphics[width=0.49\textwidth]{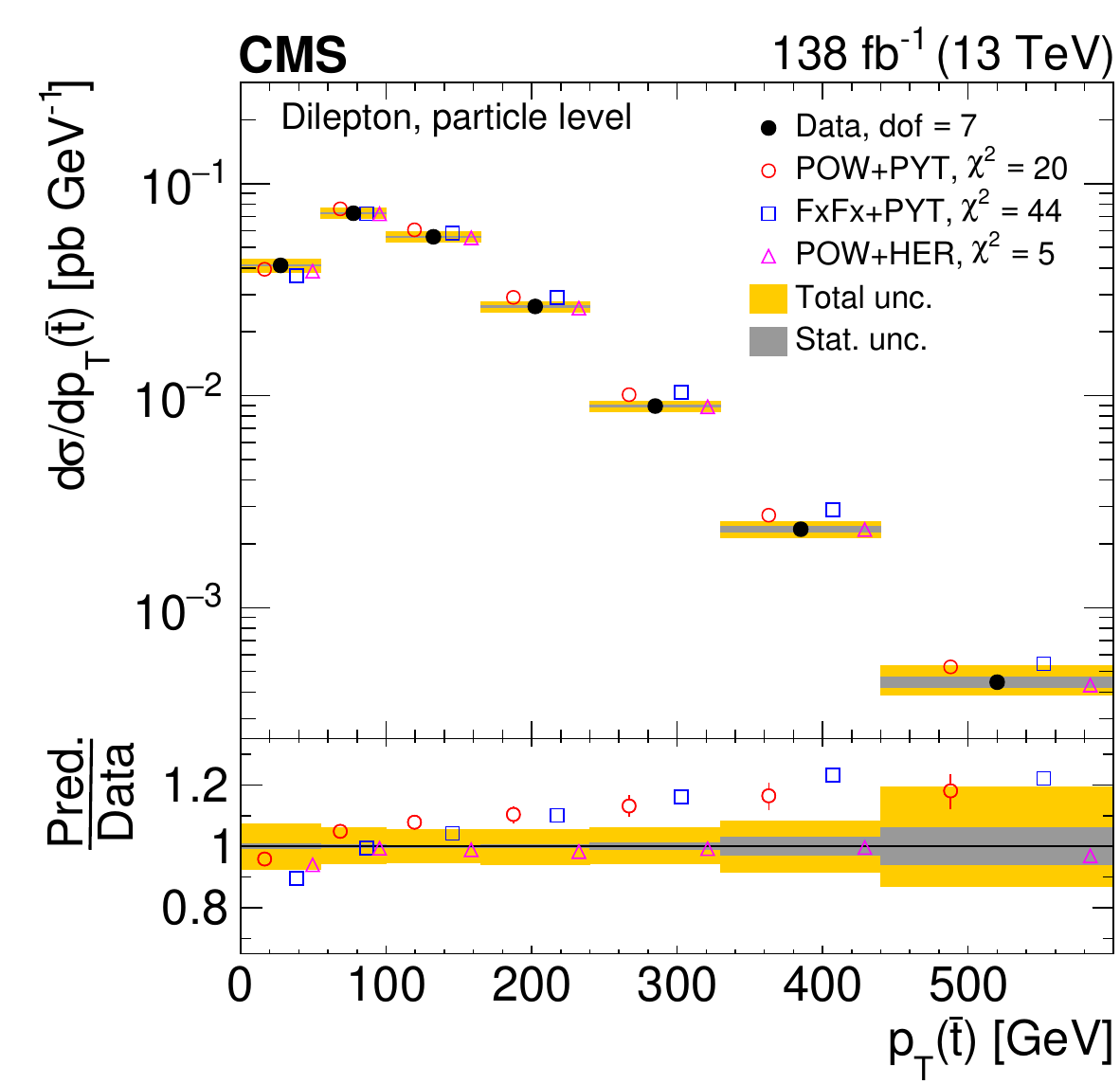}
\caption {Absolute differential \ttbar production cross sections as functions of \ptt (upper) and \ptat (lower),
measured
at the parton level in the full phase space (left) and at the particle level in a fiducial phase space (right).
The data are shown as filled circles with grey and yellow bands indicating the statistical and total uncertainties
(statistical
and systematic uncertainties added in quadrature), respectively.
For each distribution, the number of degrees of freedom (dof) is also provided.
The cross sections are compared to various MC predictions (other points).
The estimated uncertainties in the \PowPyt (`POW-PYT') simulation are represented by vertical bars on the
corresponding points.
For each MC model, a value of \chisq is reported that takes into account the measurement uncertainties.
The lower panel in each plot shows the ratios of the predictions to the data.}
\label{fig:res_ptt_abs}
\end{figure*}

\begin{figure*}[!phtb]
\centering
\includegraphics[width=0.49\textwidth]{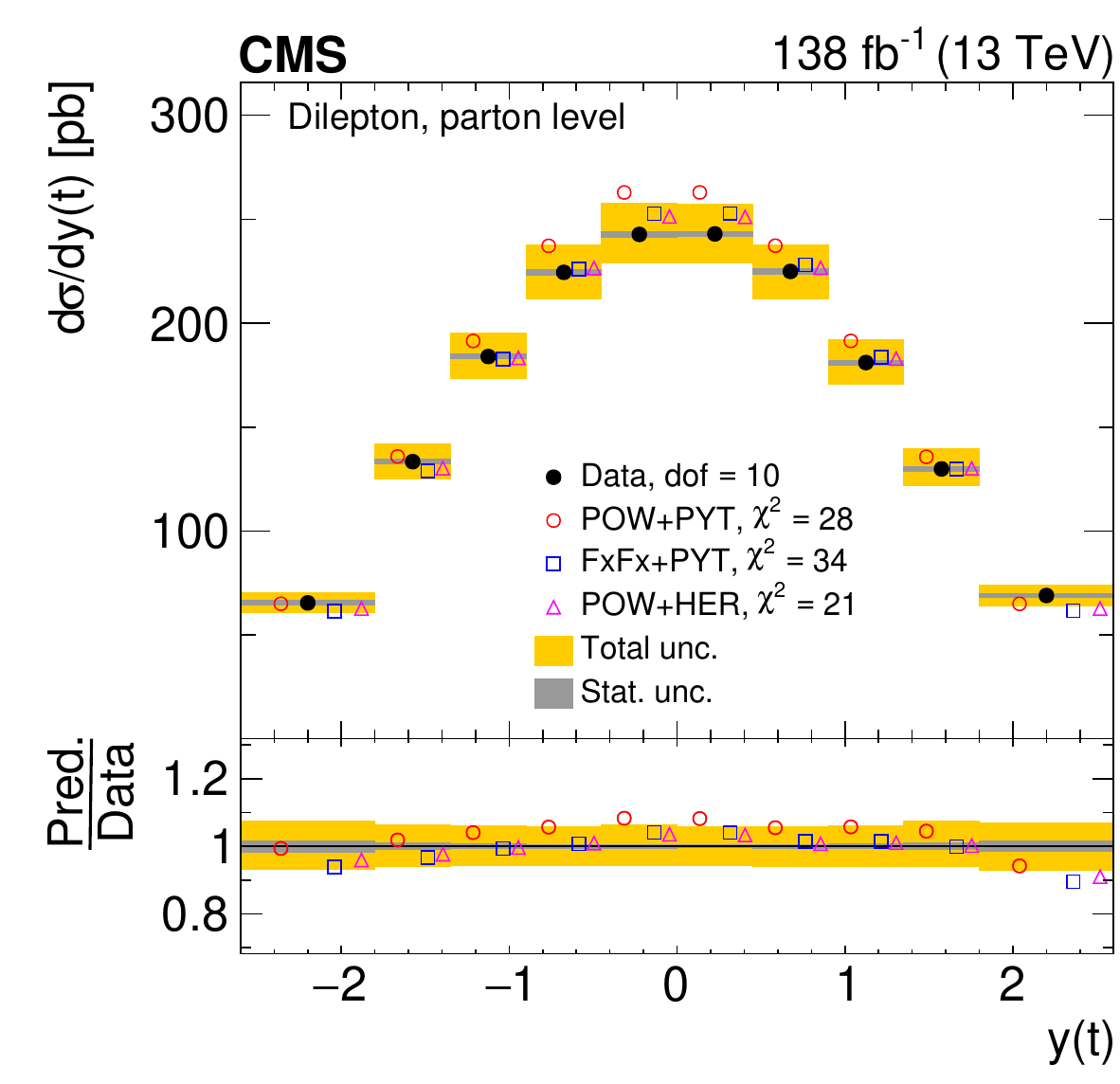}
\includegraphics[width=0.49\textwidth]{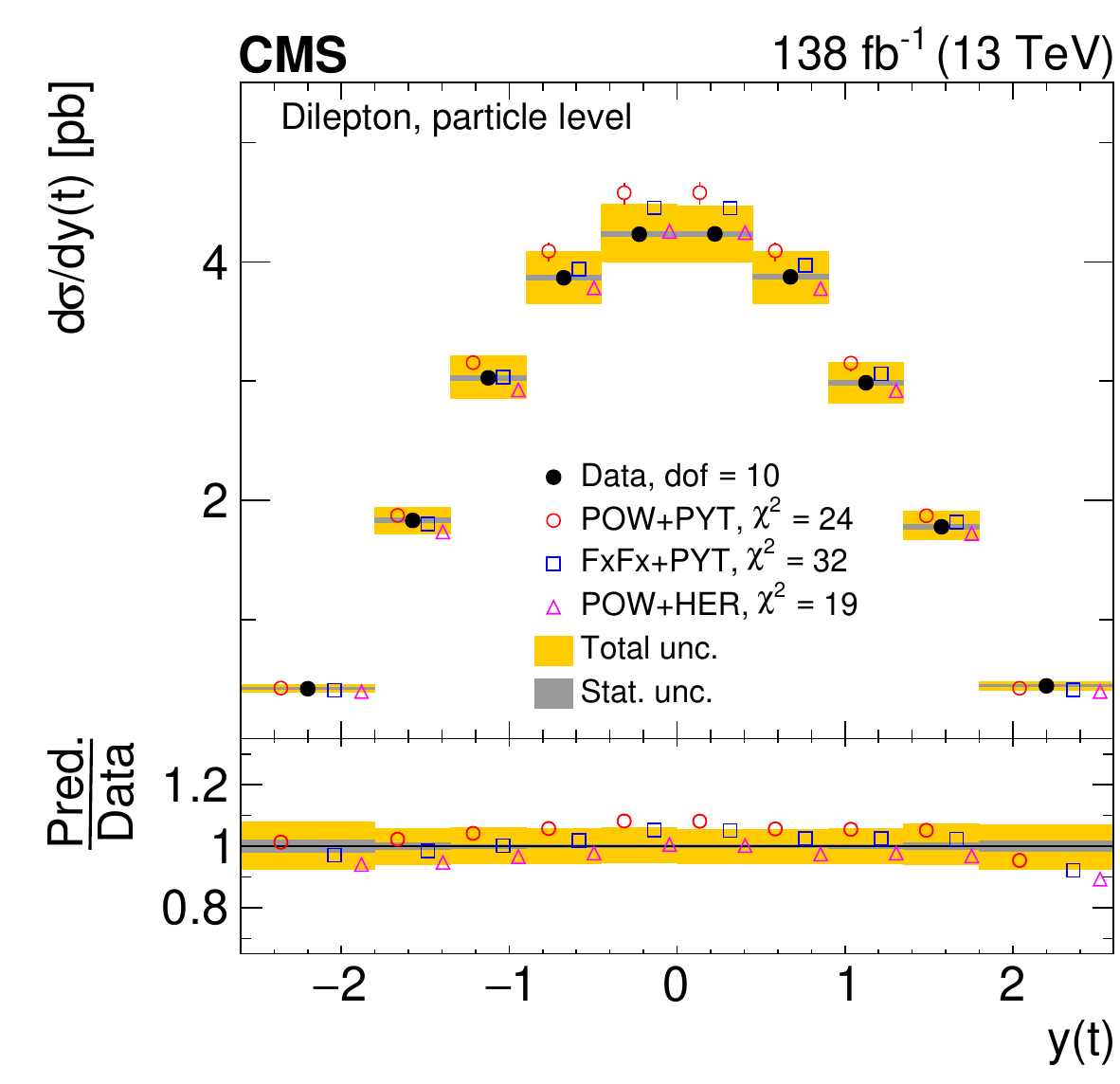}
\includegraphics[width=0.49\textwidth]{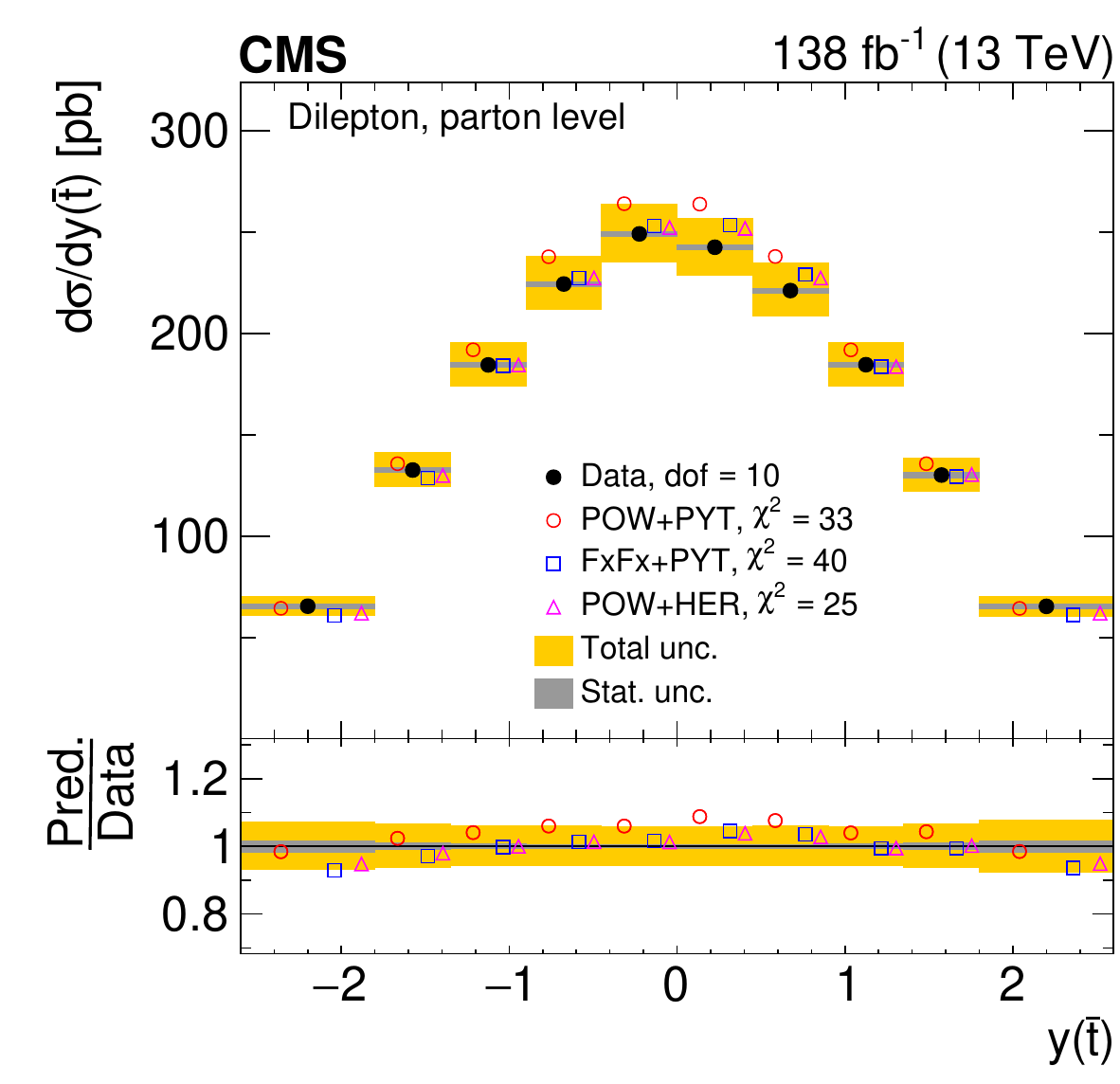}
\includegraphics[width=0.49\textwidth]{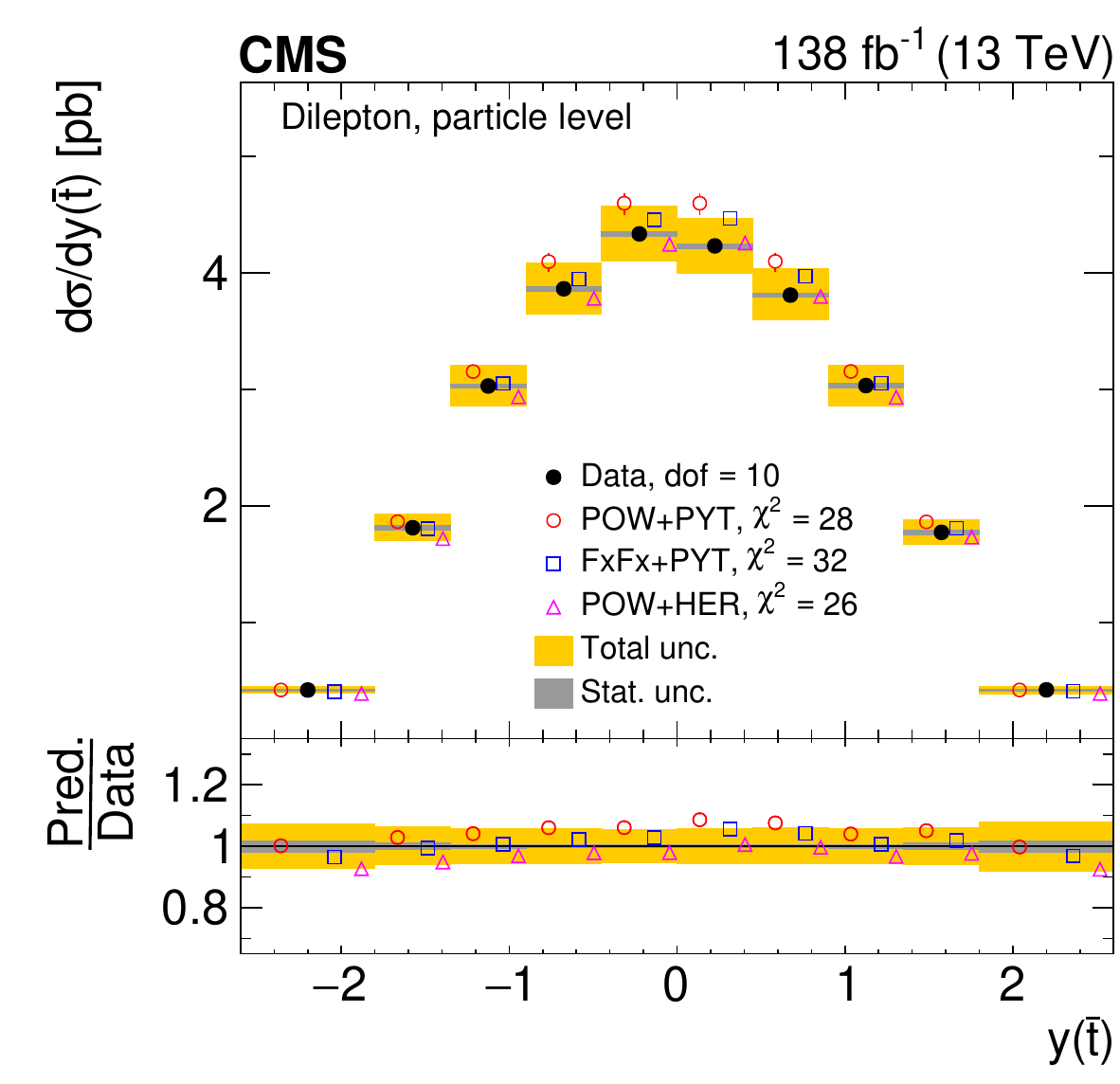}
\caption{Absolute differential \ttbar production cross sections as functions of \yt (upper) and \yat (lower) are
shown for data (filled circles) and
various MC predictions (other points).
Further details can be found in the caption of Fig.~\ref{fig:res_ptt_abs}.}
\label{fig:res_yt_abs}
\end{figure*}

\begin{figure*}[!phtb]
\centering
\includegraphics[width=0.49\textwidth]{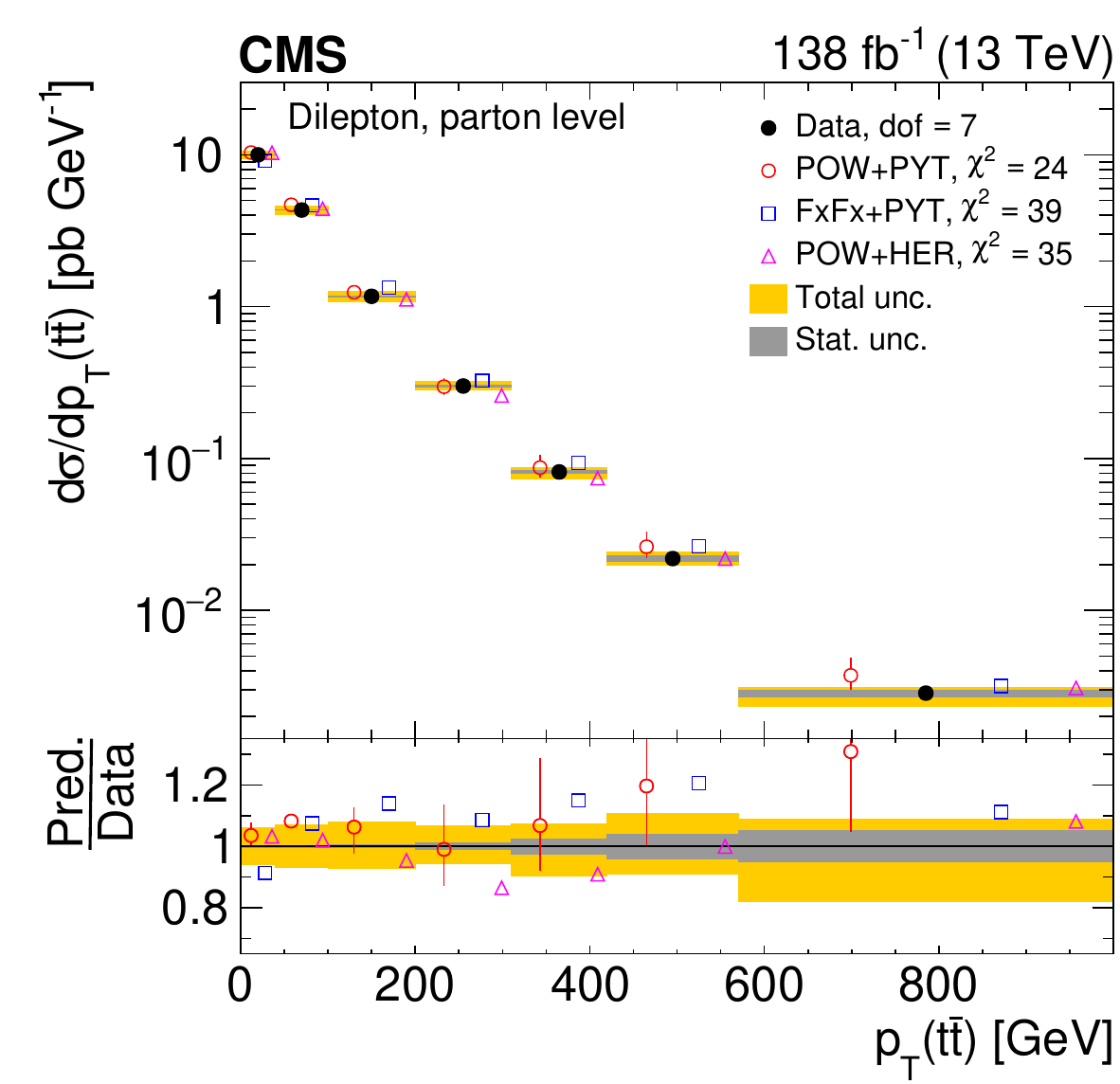}
\includegraphics[width=0.49\textwidth]{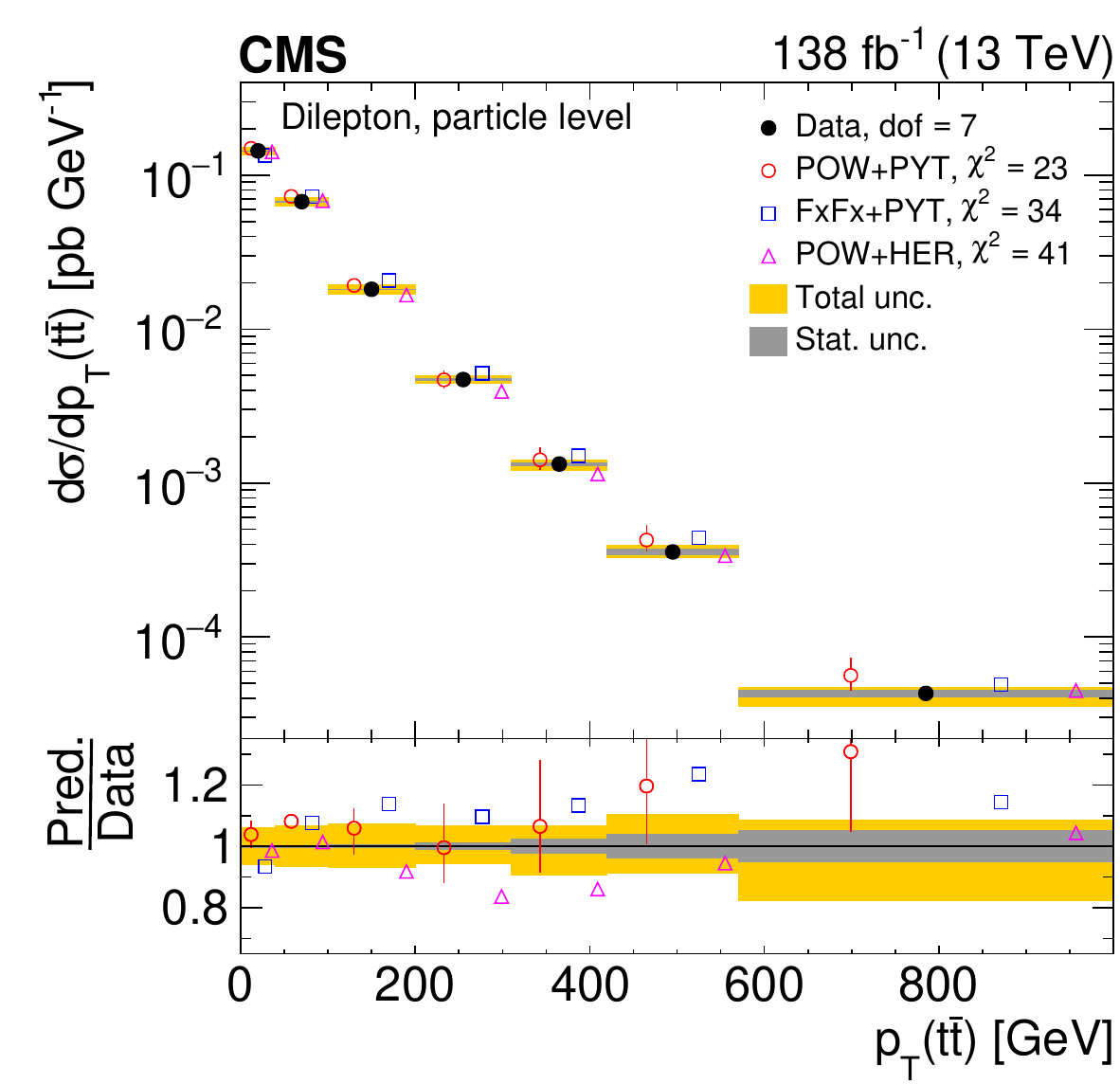}
\includegraphics[width=0.49\textwidth]{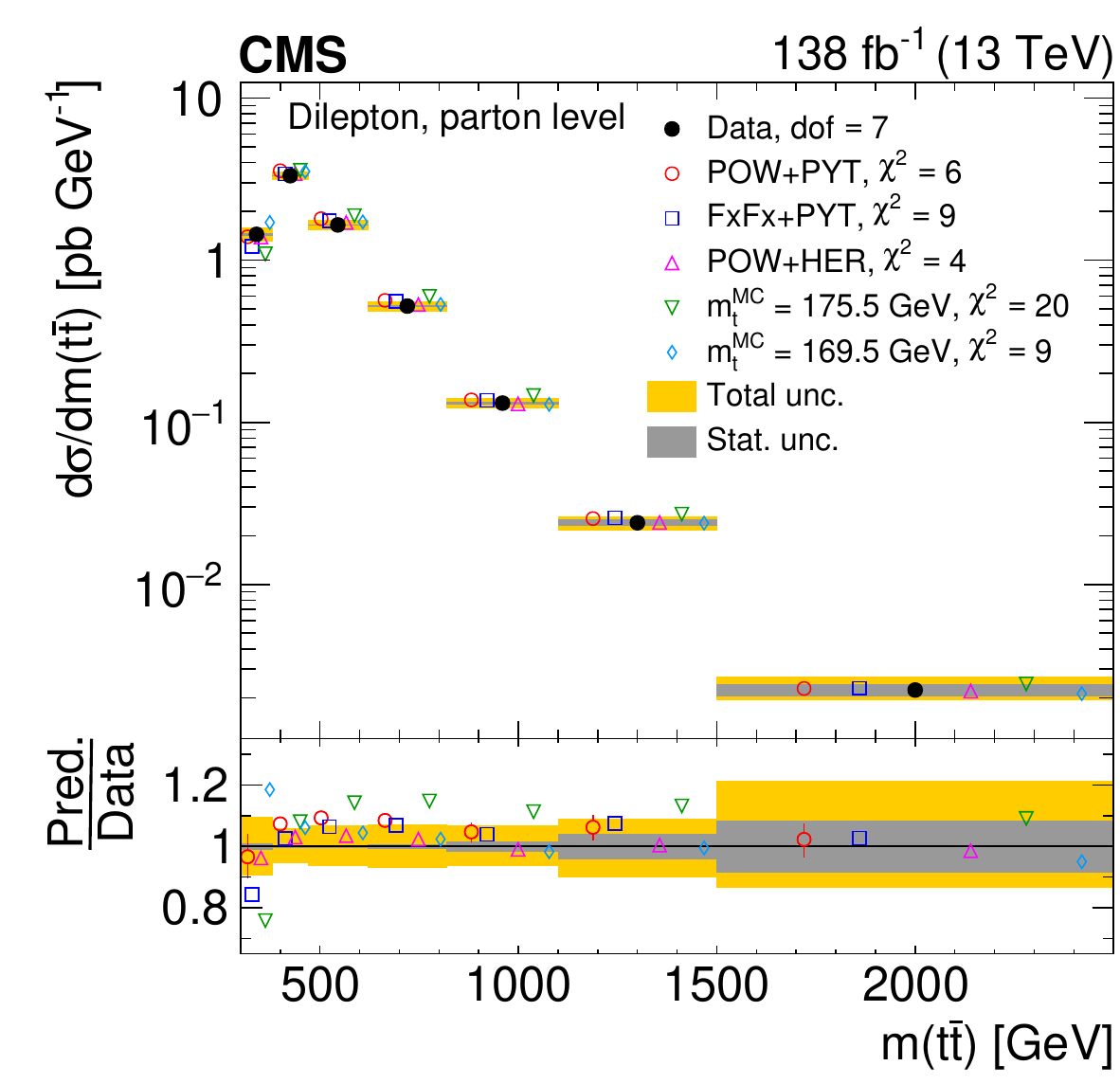}
\includegraphics[width=0.49\textwidth]{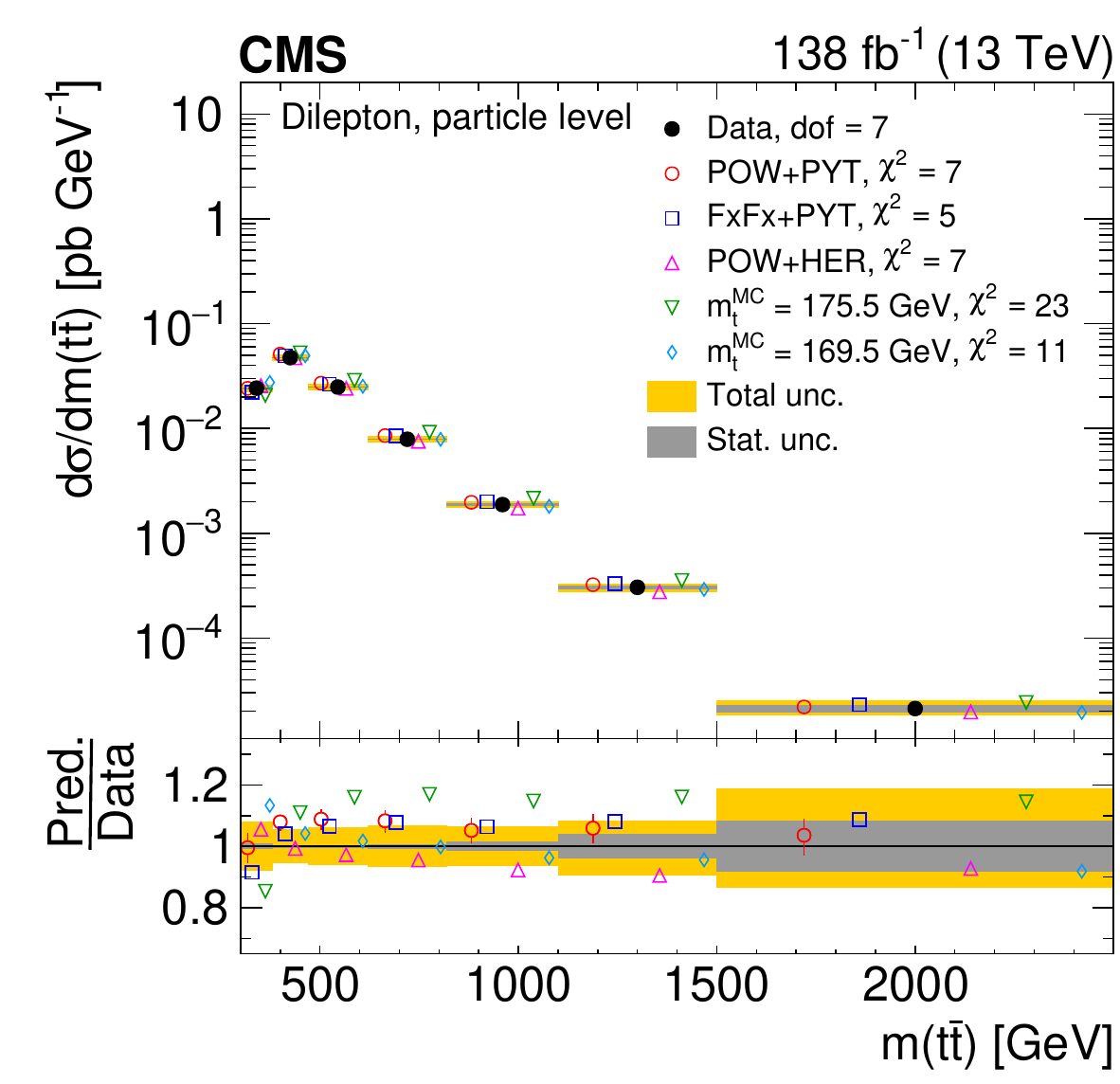}
\includegraphics[width=0.49\textwidth]{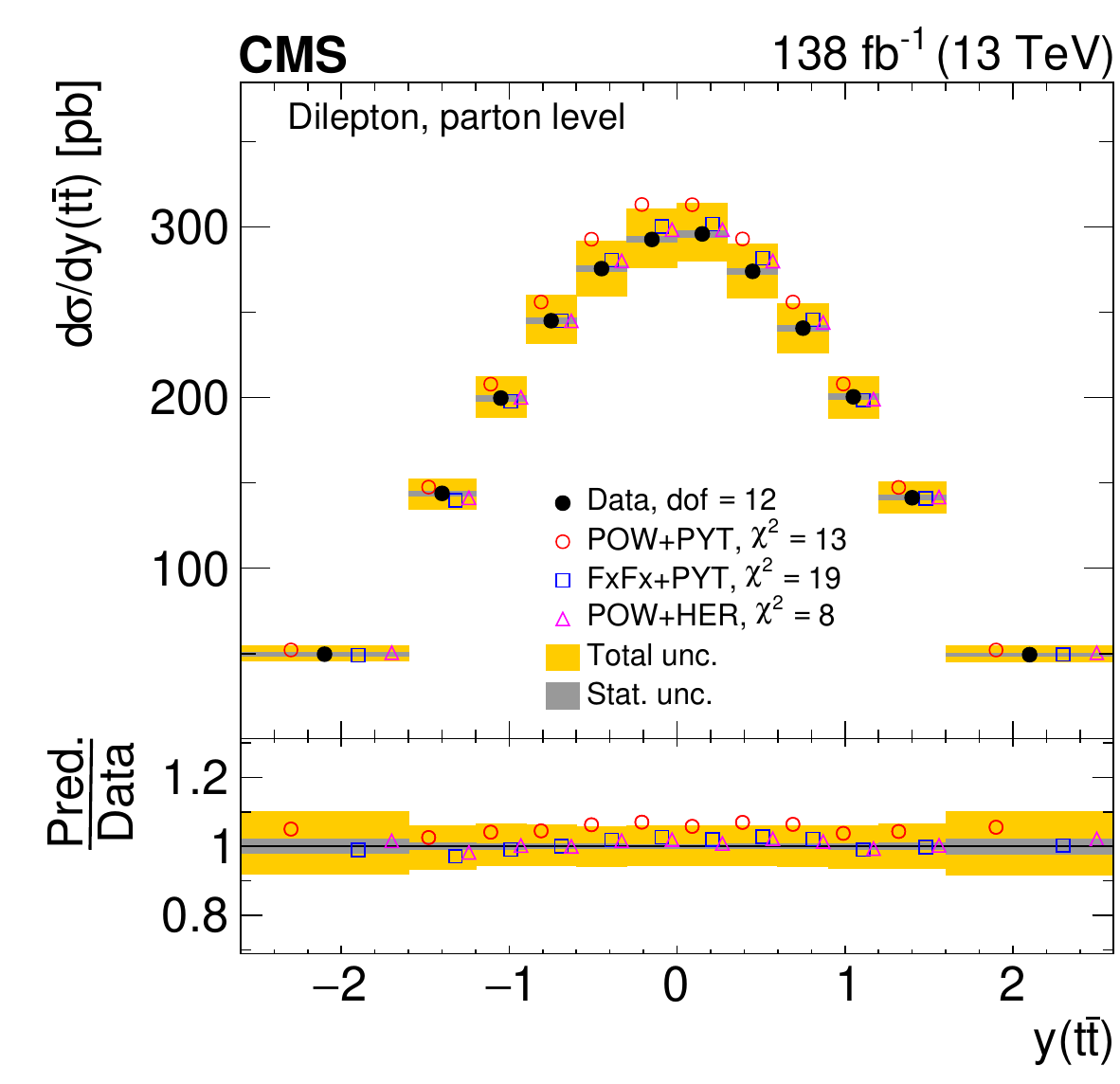}
\includegraphics[width=0.49\textwidth]{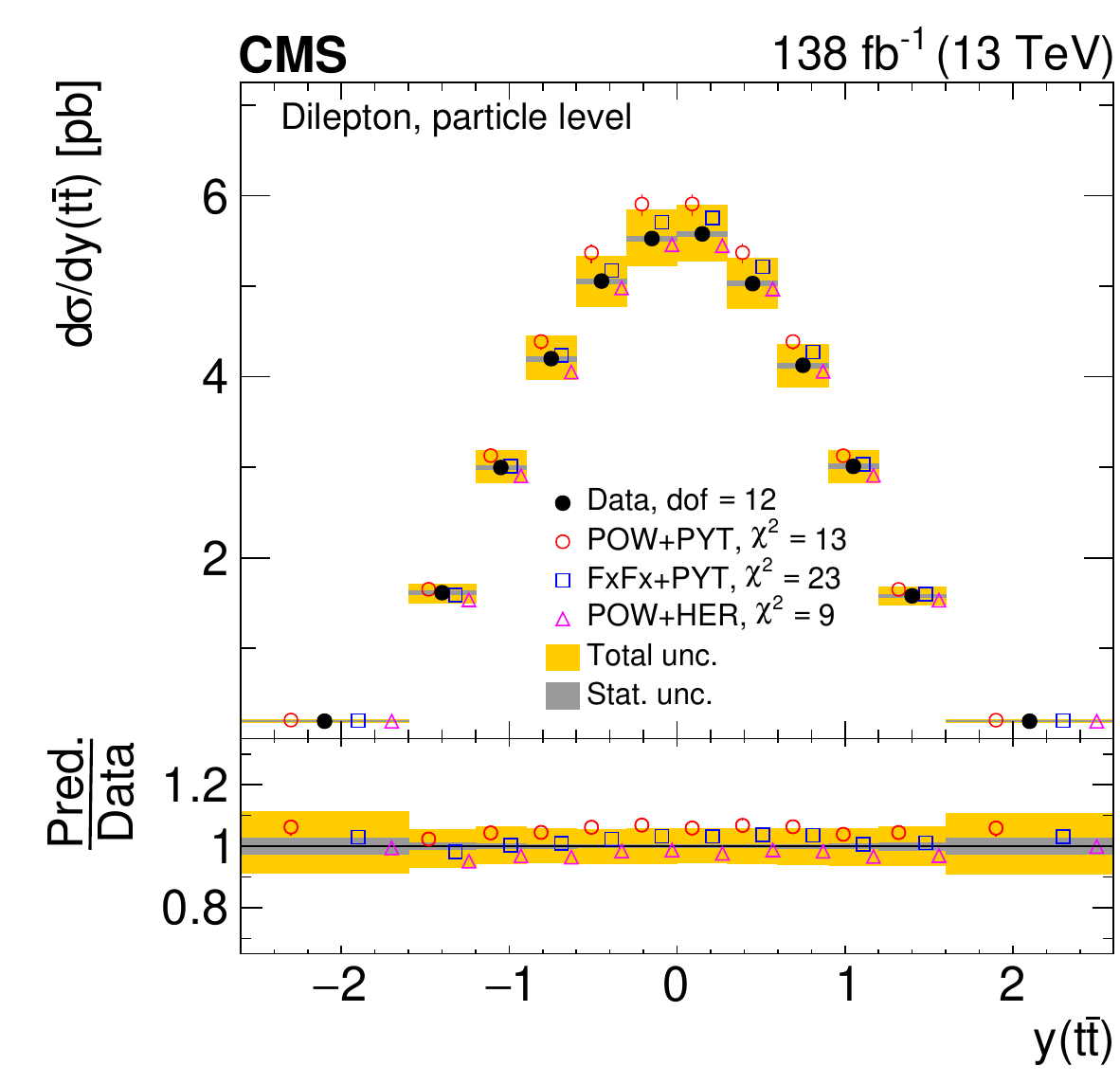}
\caption{Absolute differential \ttbar production cross sections as functions of \pttt (upper), \mtt (middle) and
\ytt (lower)
are shown for data (filled circles) and various MC predictions (other points).
Further details can be found in the caption of Fig.~\ref{fig:res_ptt_abs}.}
\label{fig:res_pttt_abs}
\end{figure*}

\begin{figure*}[!phtb]
\centering
\includegraphics[width=0.49\textwidth]{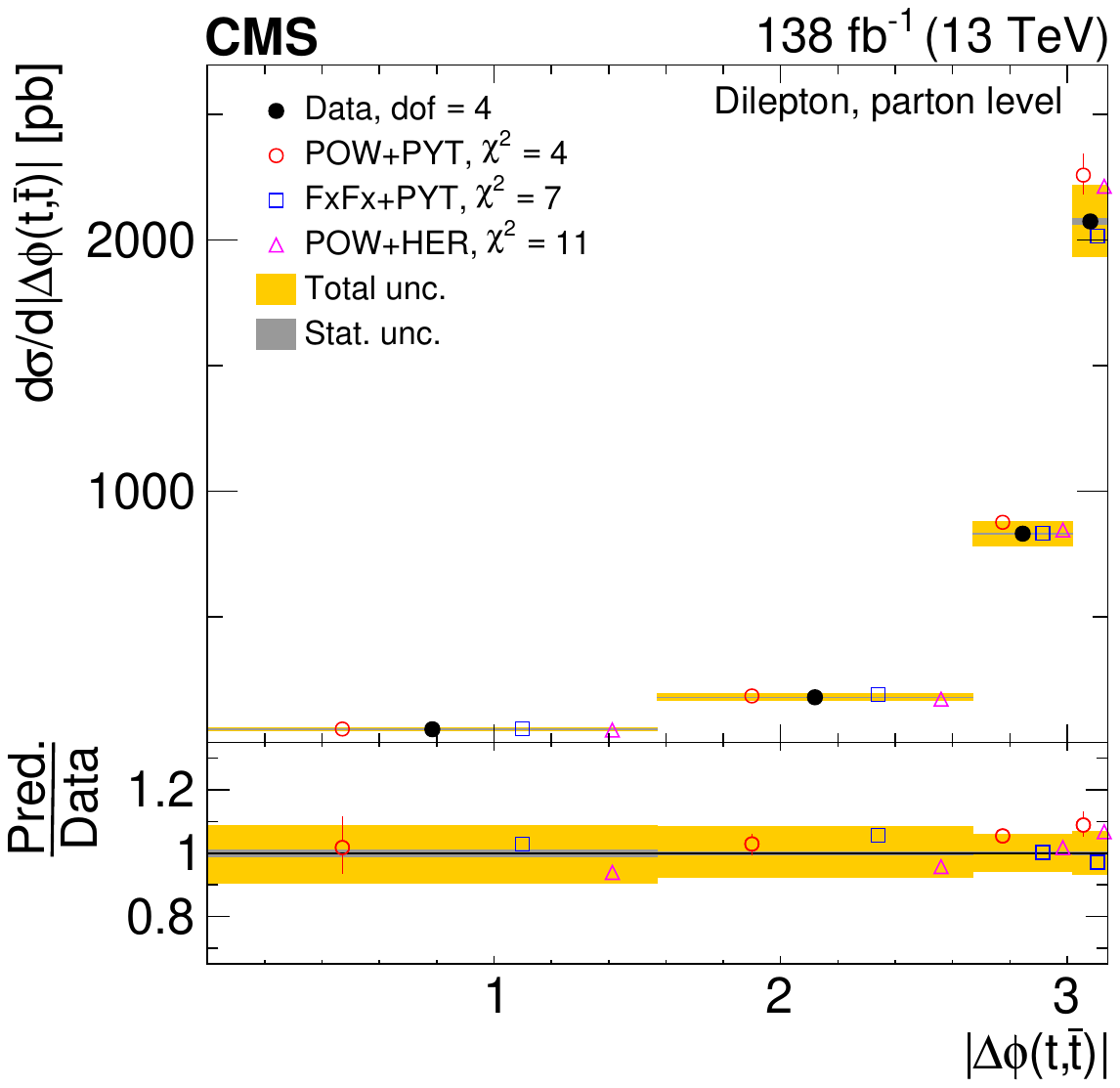}
\includegraphics[width=0.49\textwidth]{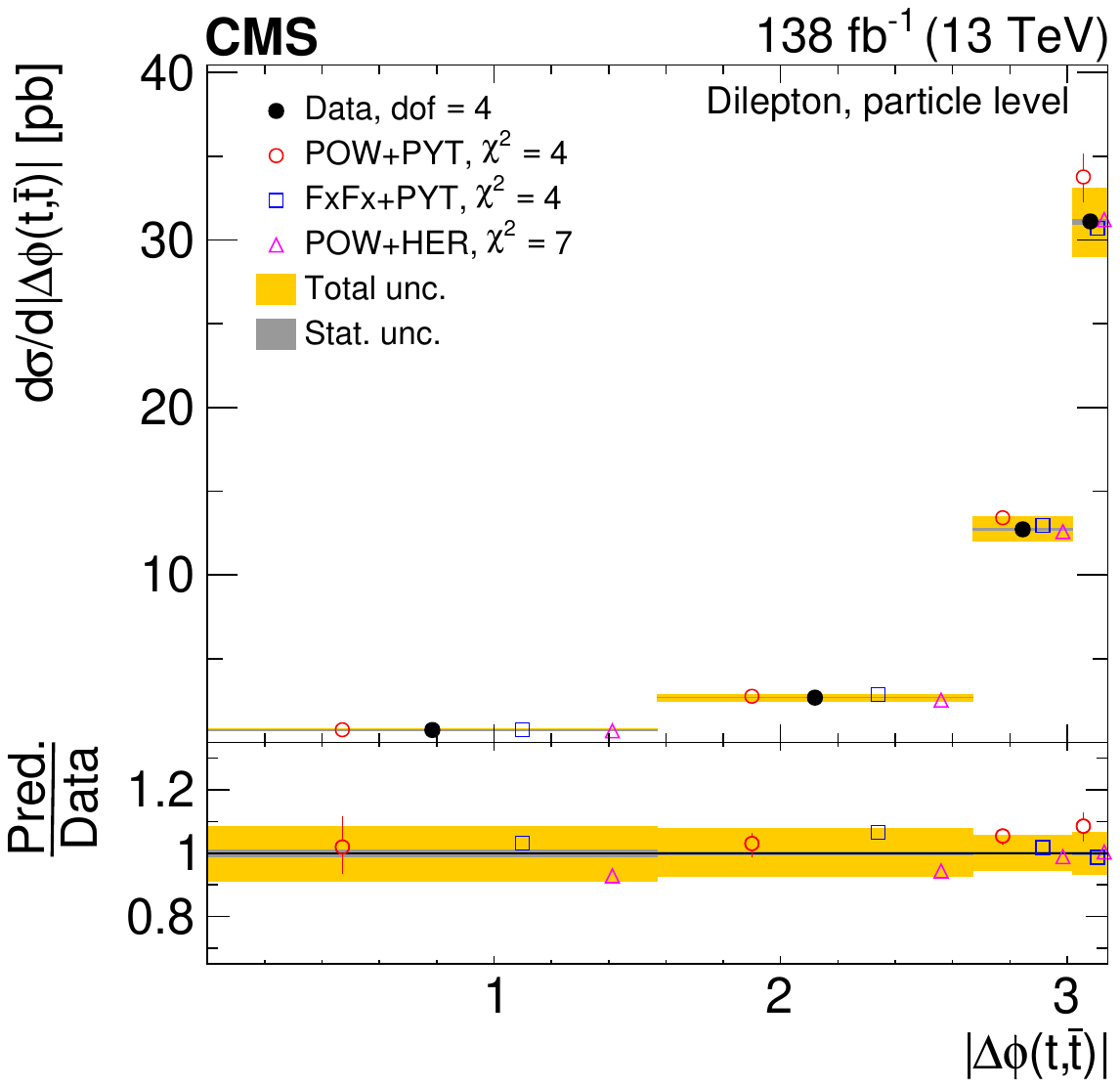}
\includegraphics[width=0.49\textwidth]{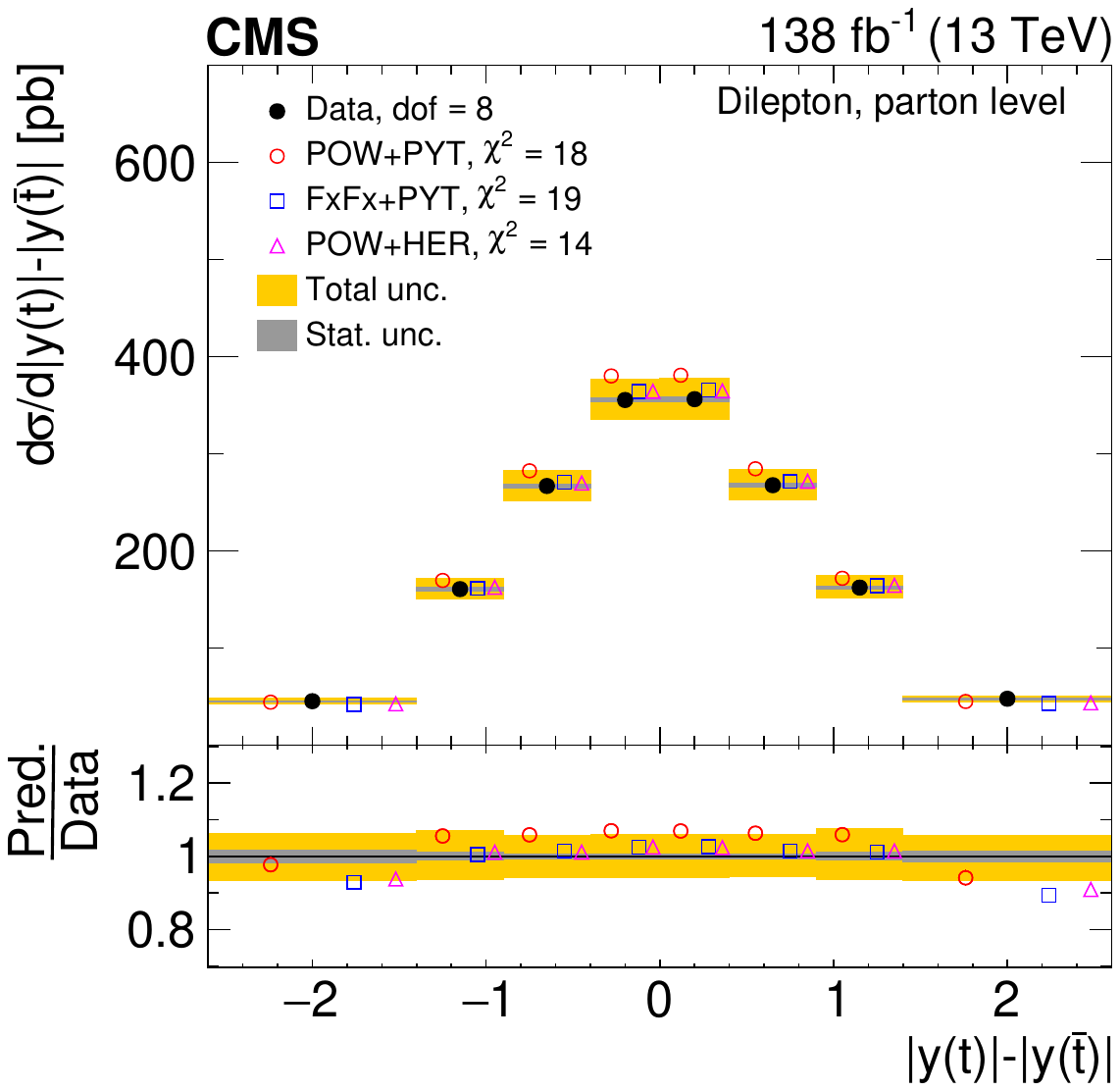}
\includegraphics[width=0.49\textwidth]{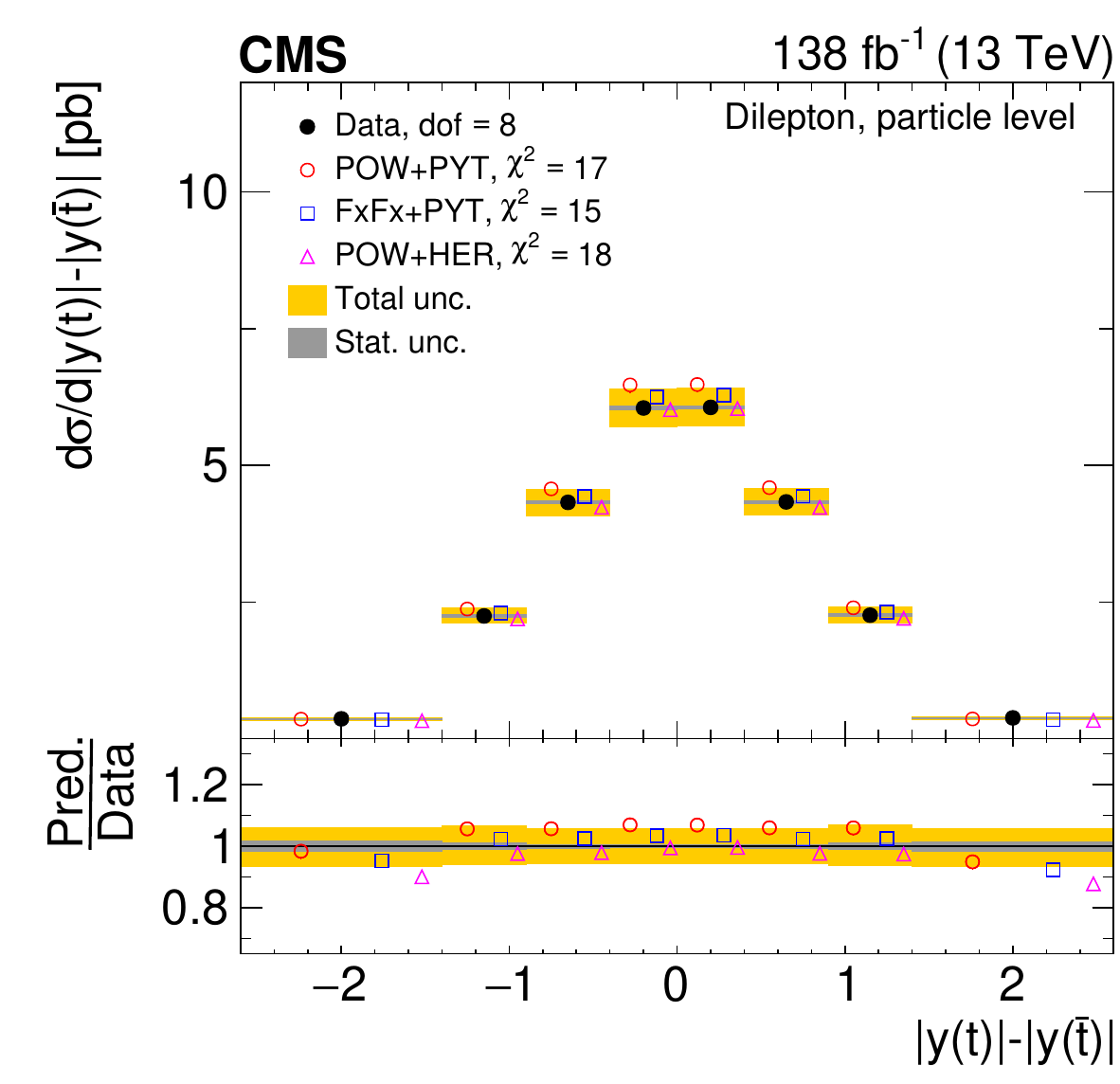}
\caption{Absolute differential \ttbar production cross sections as functions of \dphitt (upper) and \dytt (lower)
are shown for the data (filled circles) and various MC predictions (other points).
Further details can be found in the caption of Fig.~\ref{fig:res_ptt_abs}.}
\label{fig:res_dphitt_abs}
\end{figure*}

\begin{figure*}[!phtb]
\centering
\includegraphics[width=0.49\textwidth]{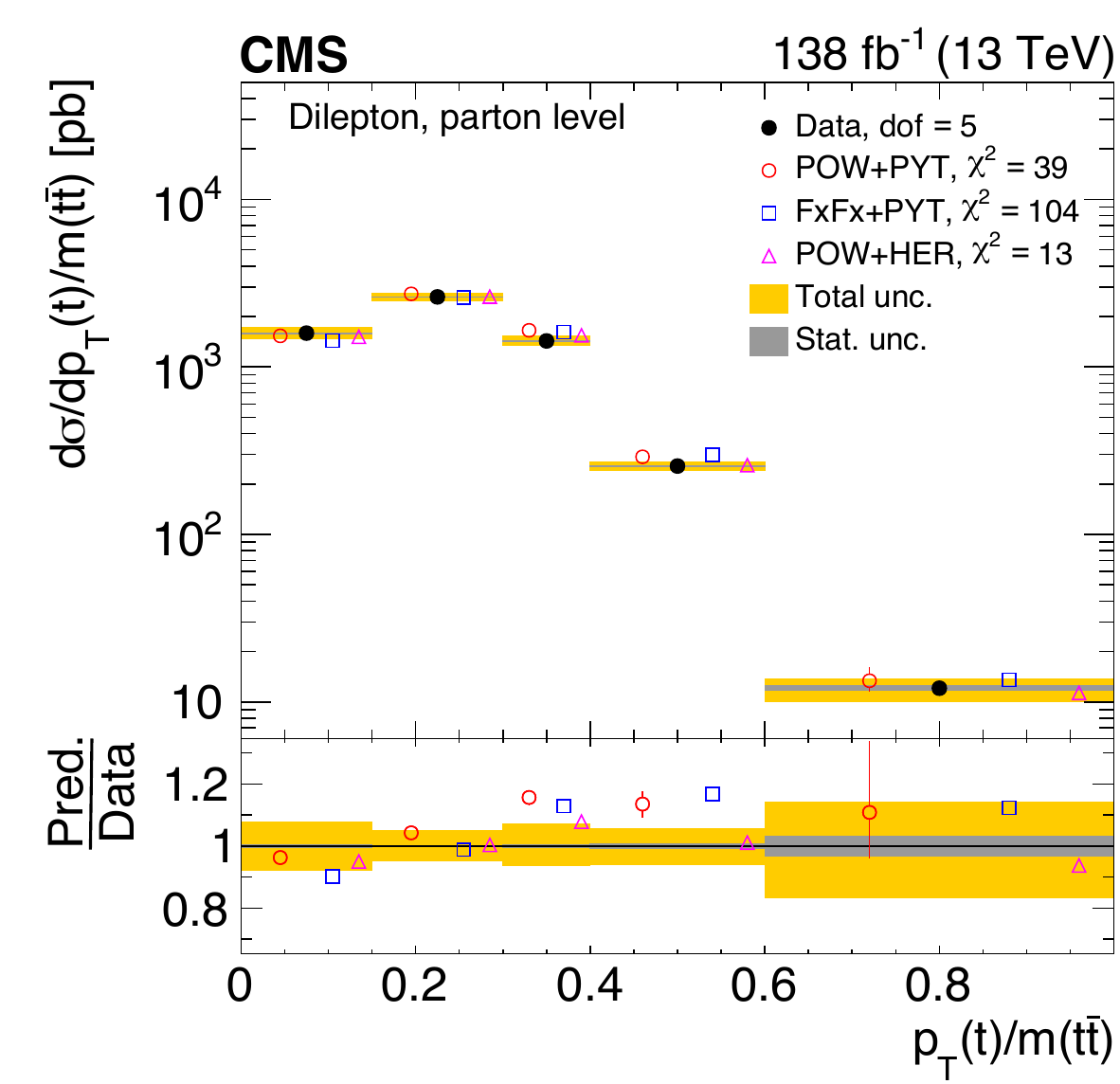}
\includegraphics[width=0.49\textwidth]{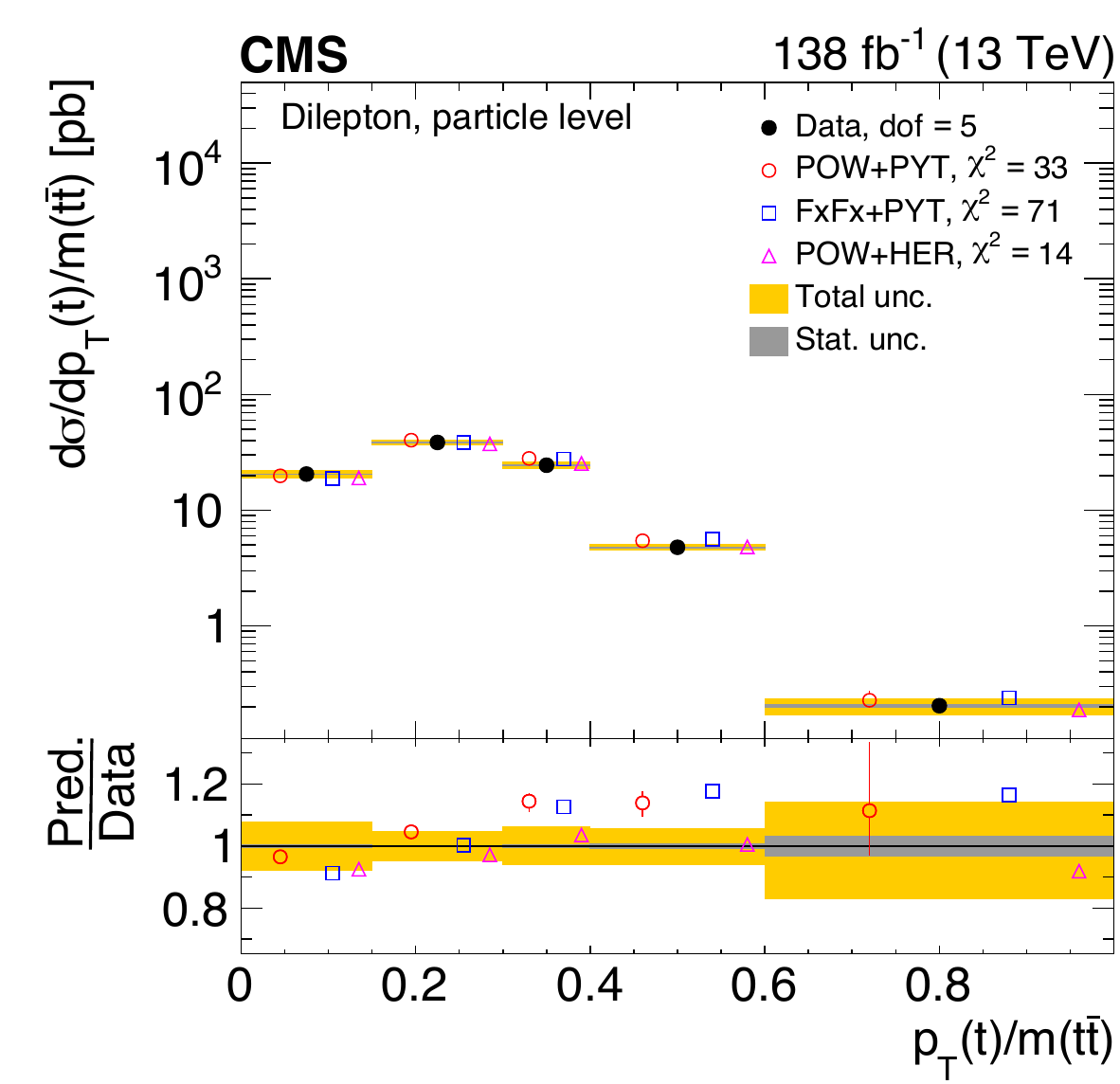}
\includegraphics[width=0.49\textwidth]{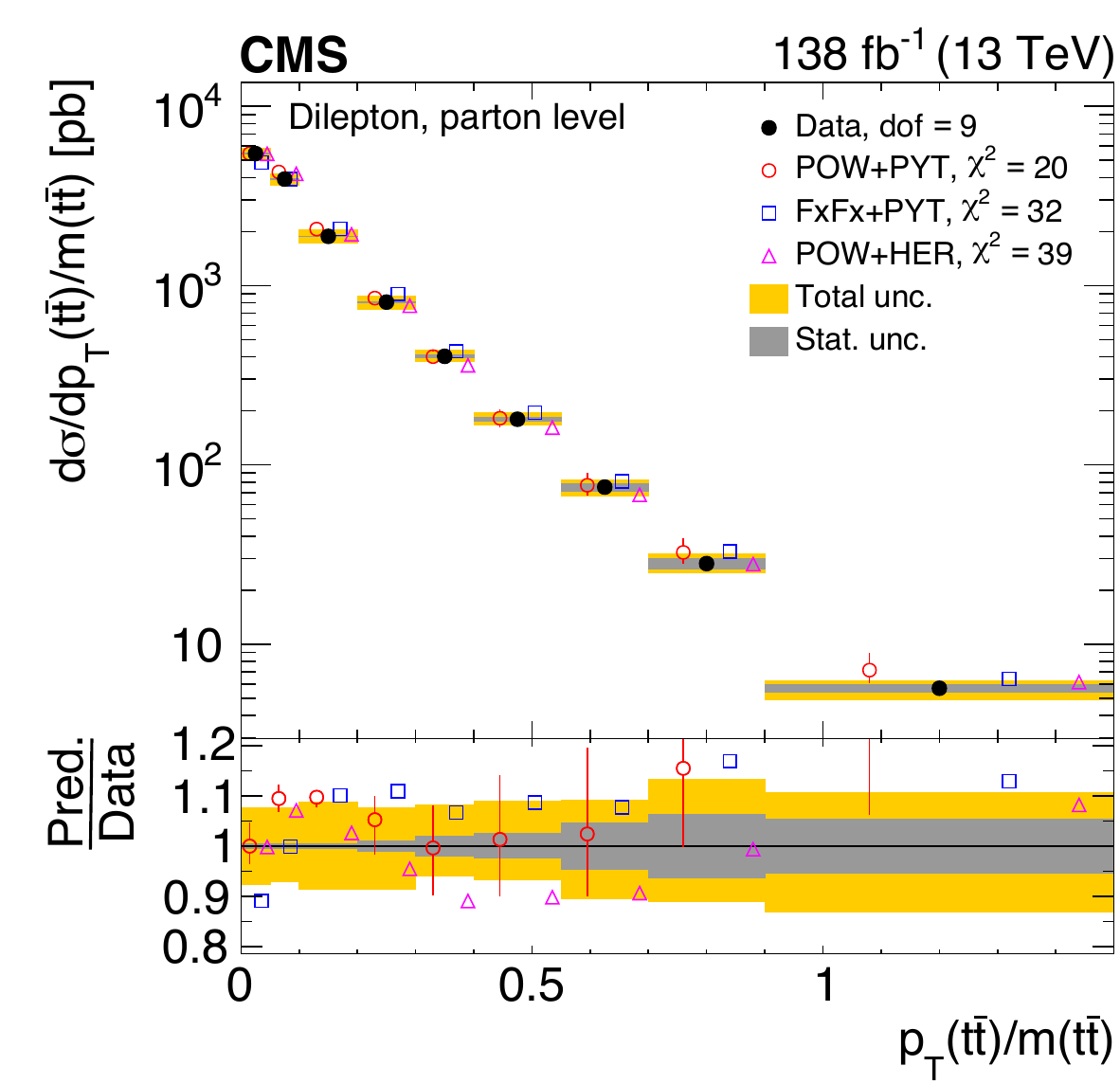}
\includegraphics[width=0.49\textwidth]{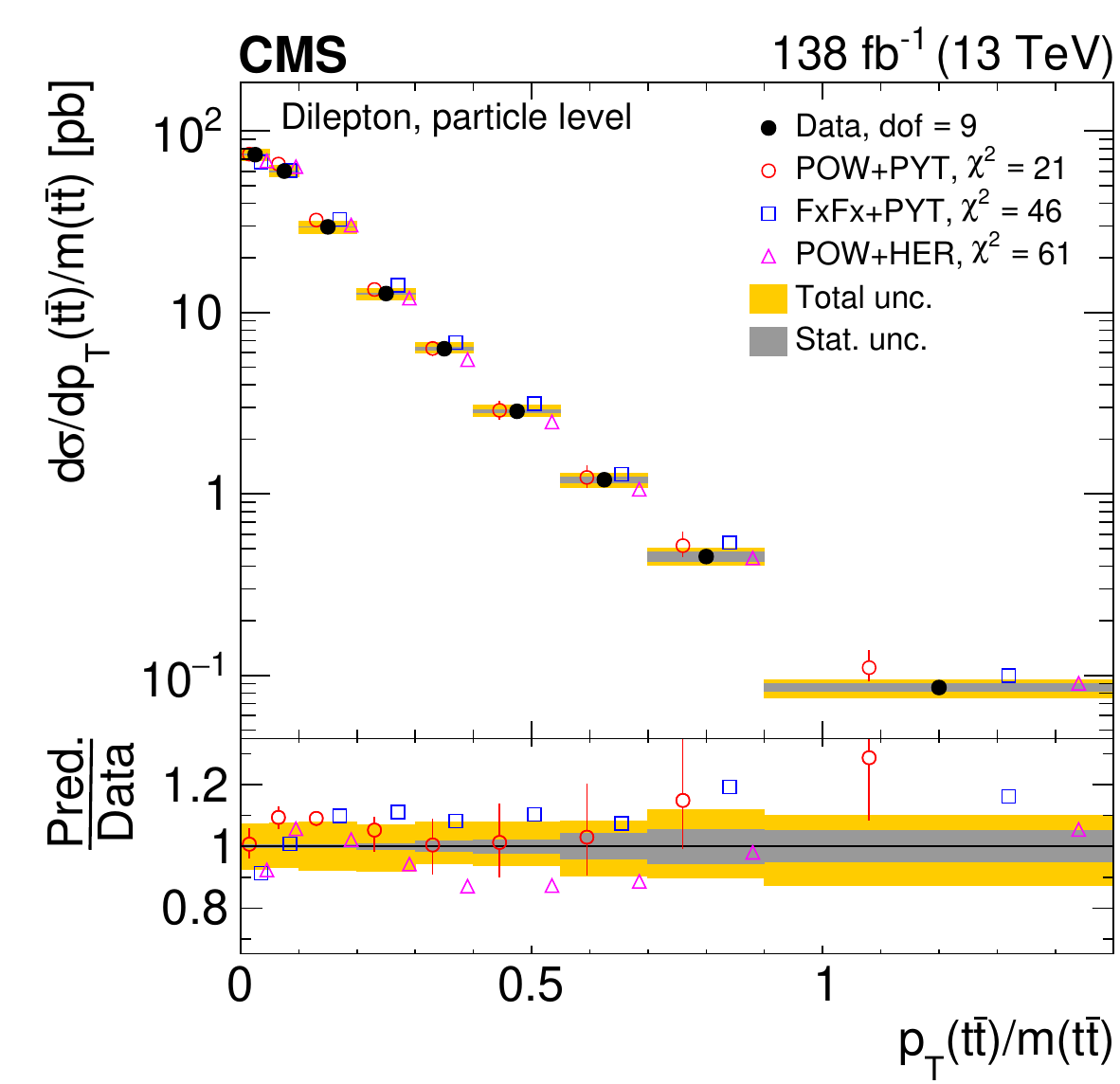}
\caption{Absolute differential \ttbar production cross sections as functions of \rpttmtt (upper) and
\rptttmtt (lower)
are shown for the data (filled circles) and various MC predictions (other points).
Further details can be found in the caption of Fig.~\ref{fig:res_ptt_abs}.}
\label{fig:res_rpttmtt_abs}
\end{figure*}

\begin{figure*}[!phtb]
\centering
\includegraphics[width=0.49\textwidth]{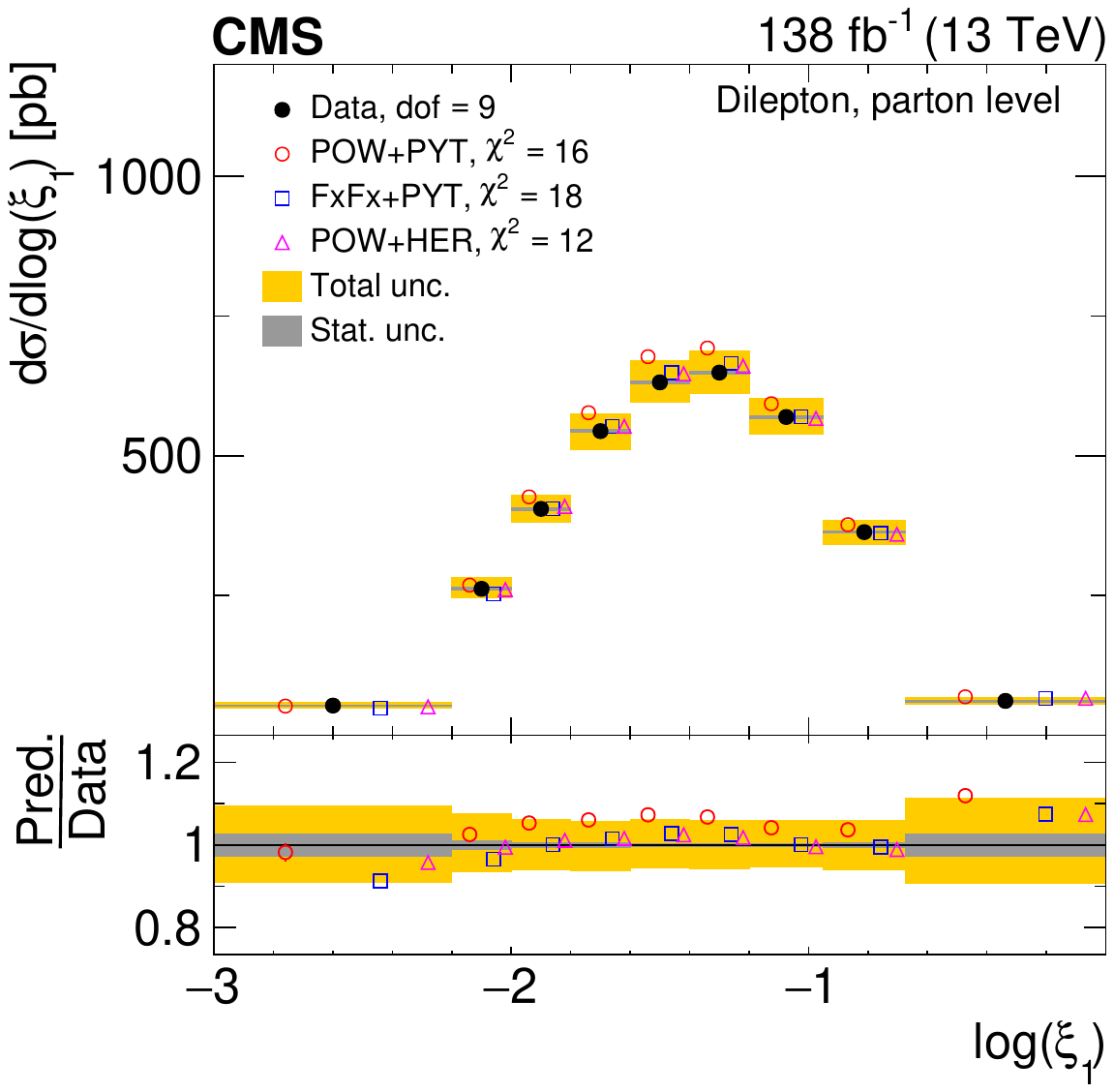}
\includegraphics[width=0.49\textwidth]{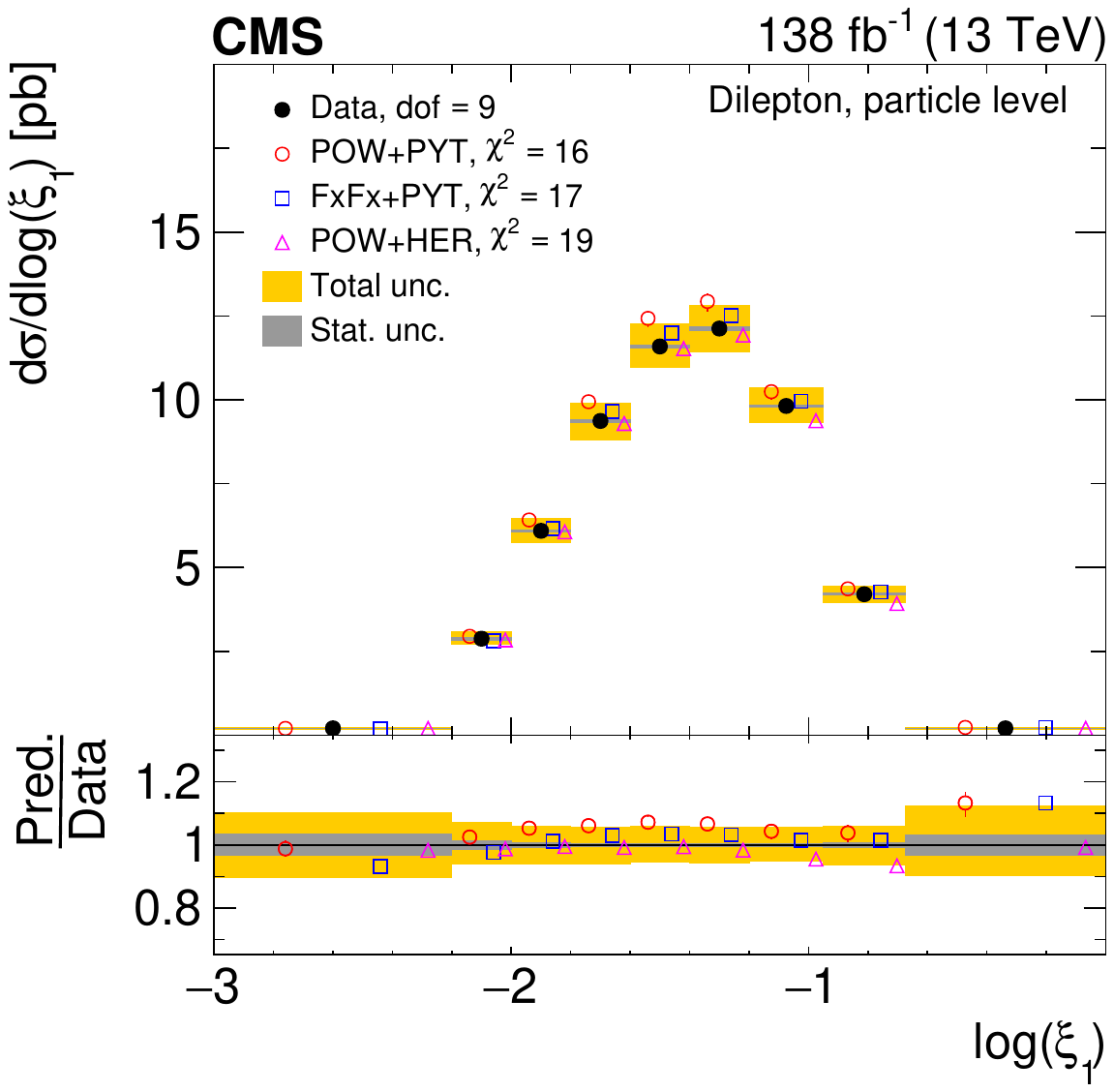}
\includegraphics[width=0.49\textwidth]{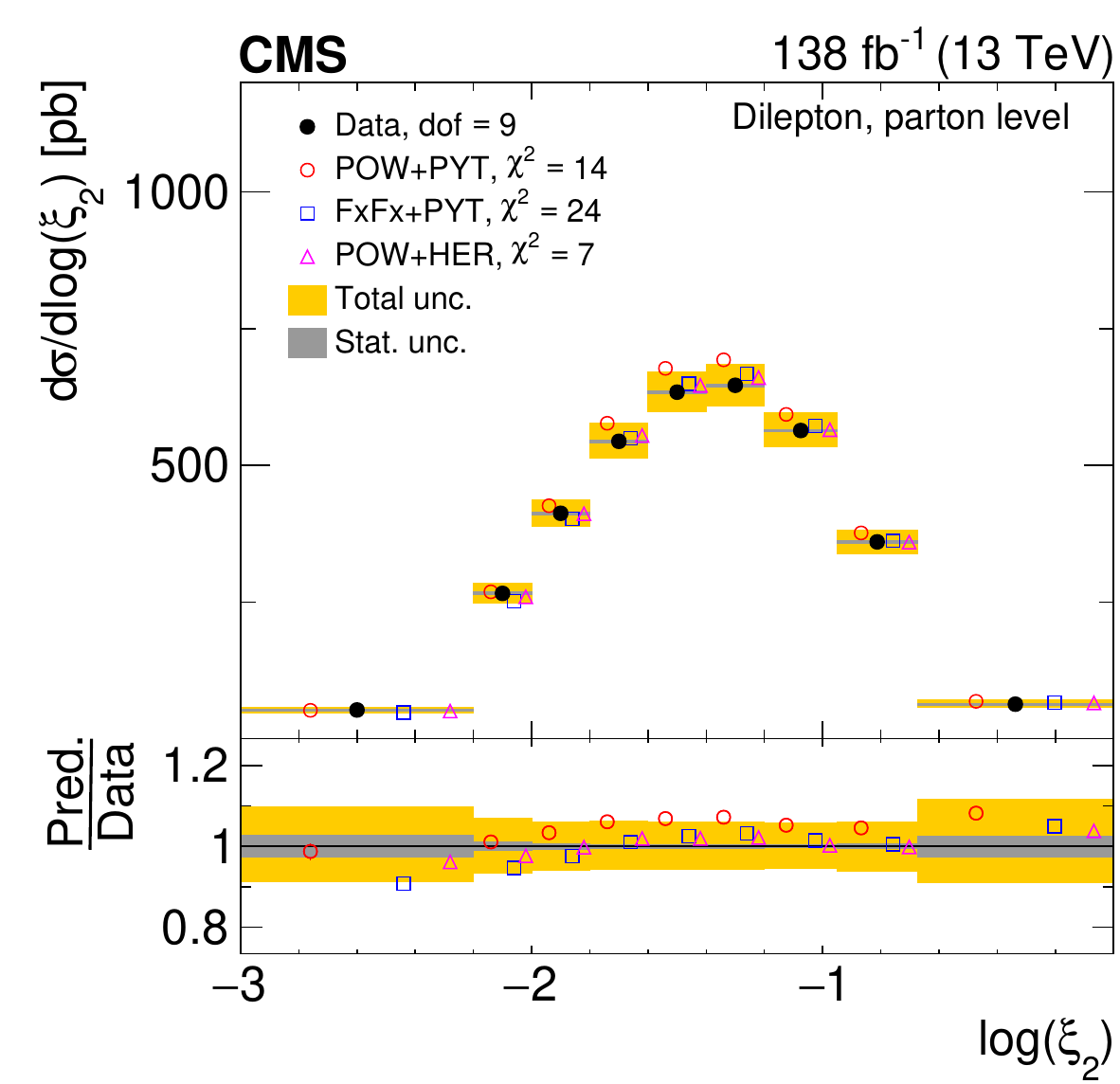}
\includegraphics[width=0.49\textwidth]{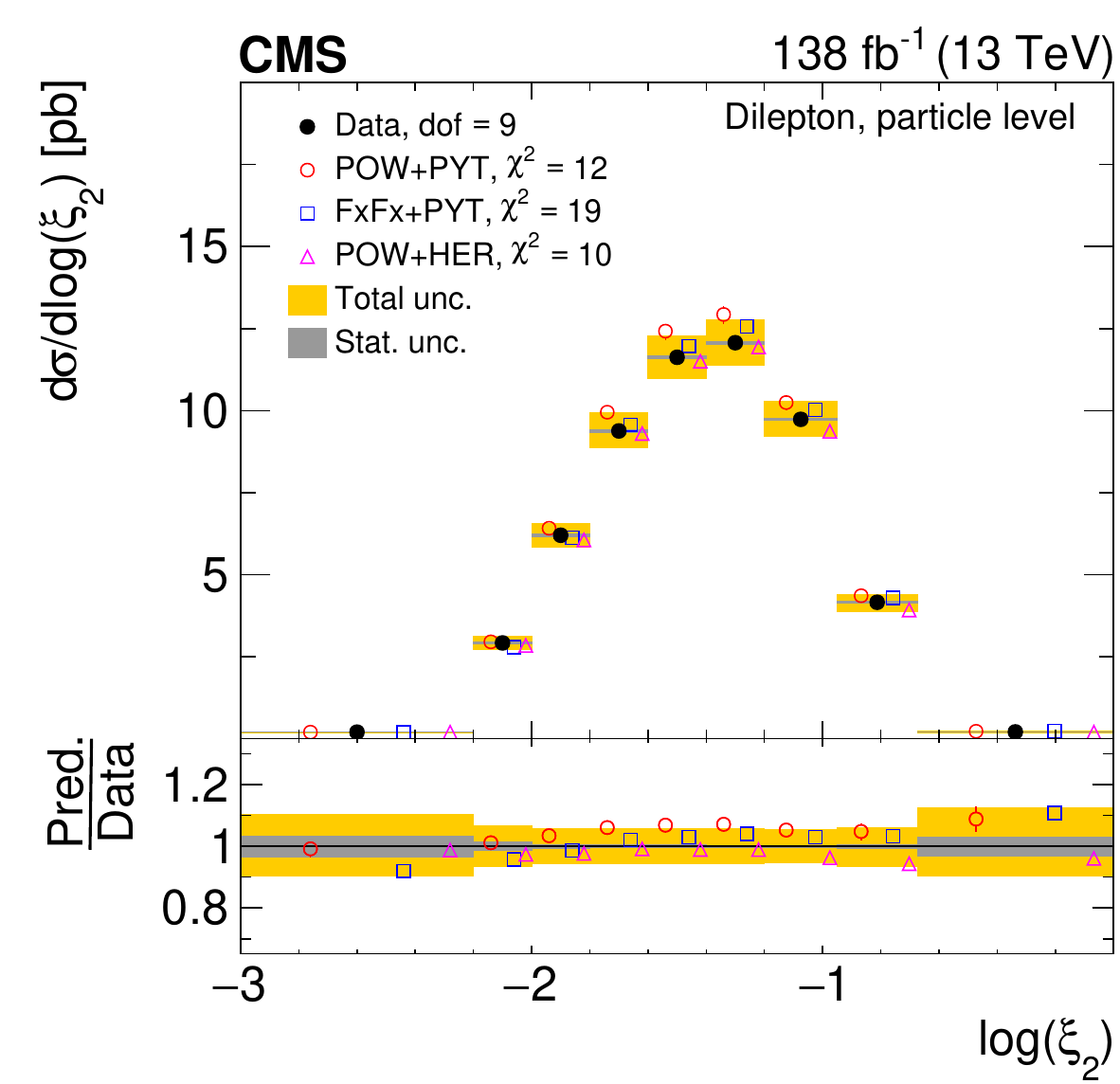}
\caption{Absolute differential \ttbar production cross sections as functions of \logxone (upper) and \logxtwo (lower)
are shown for the data (filled circles) and various MC predictions (other points).
Further details can be found in the caption of Fig.~\ref{fig:res_ptt_abs}.}
\label{fig:res_logxone_abs}
\end{figure*}

\begin{figure}
\centering
\includegraphics[width=0.99\textwidth]{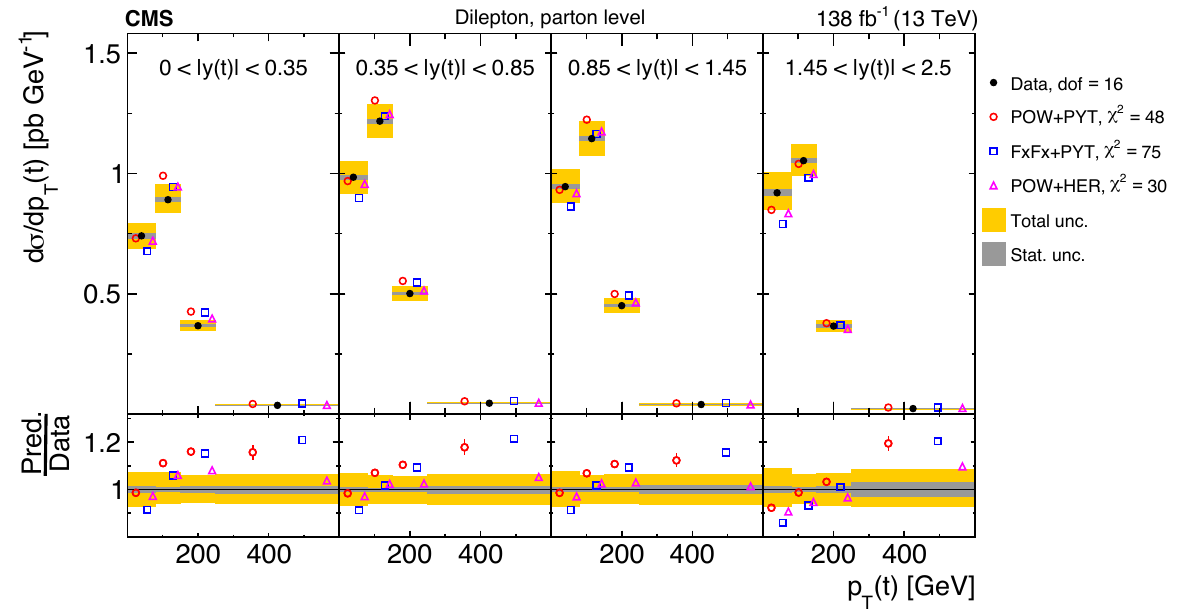}
\includegraphics[width=0.99\textwidth]{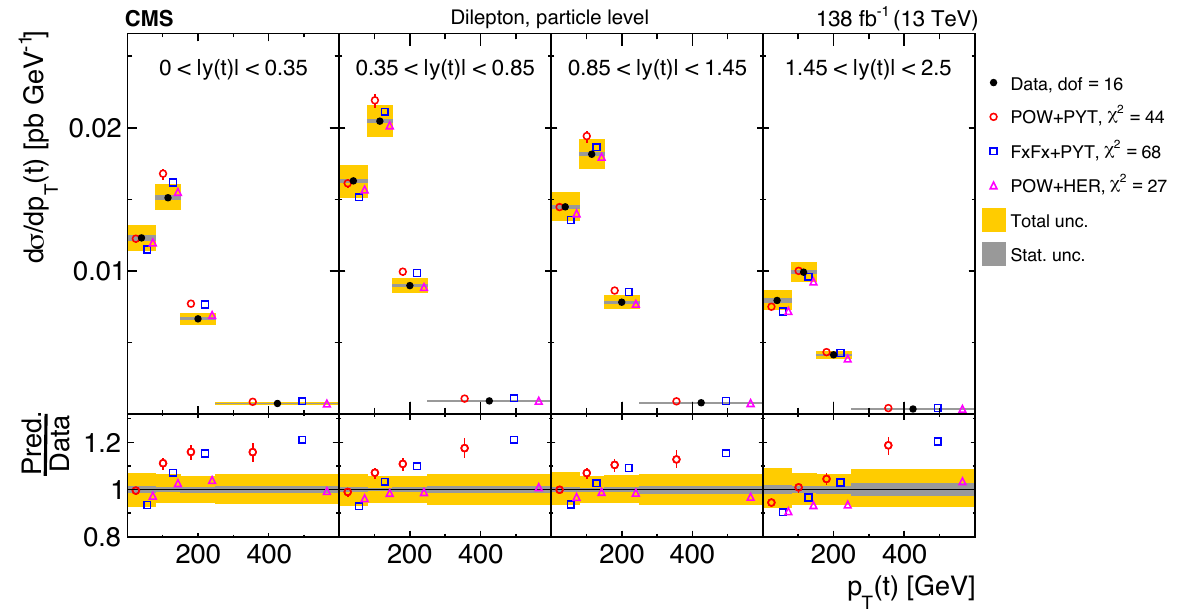}
\caption{Absolute \ytptt cross sections measured at the parton level in the full phase space (upper) and
at the particle level in a fiducial phase space (lower).
  The data are shown as filled circles with grey and yellow bands indicating the statistical and total uncertainties
(statistical and systematic uncertainties added in quadrature), respectively.
For each distribution, the number of degrees of freedom (dof) is also provided.
  The cross sections are compared to various MC predictions (other points). The estimated uncertainties in the \PowPyt
(`POW-PYT') simulation
  are represented by vertical bars on the corresponding points. For each MC model, a value of \chisq is
reported that takes into account the measurement uncertainties.
  The lower panel in each plot shows the ratios of the predictions to the data.}
    \label{fig:res_ytptt_abs}
\end{figure}

\begin{figure}
\centering
\includegraphics[width=0.99\textwidth]{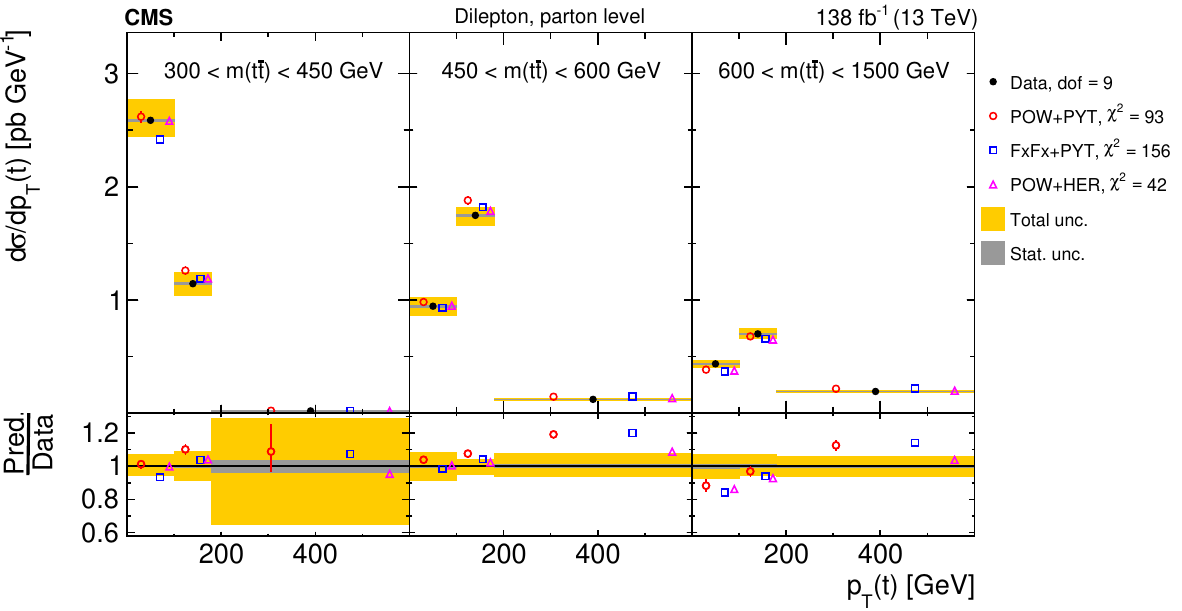}
\includegraphics[width=0.99\textwidth]{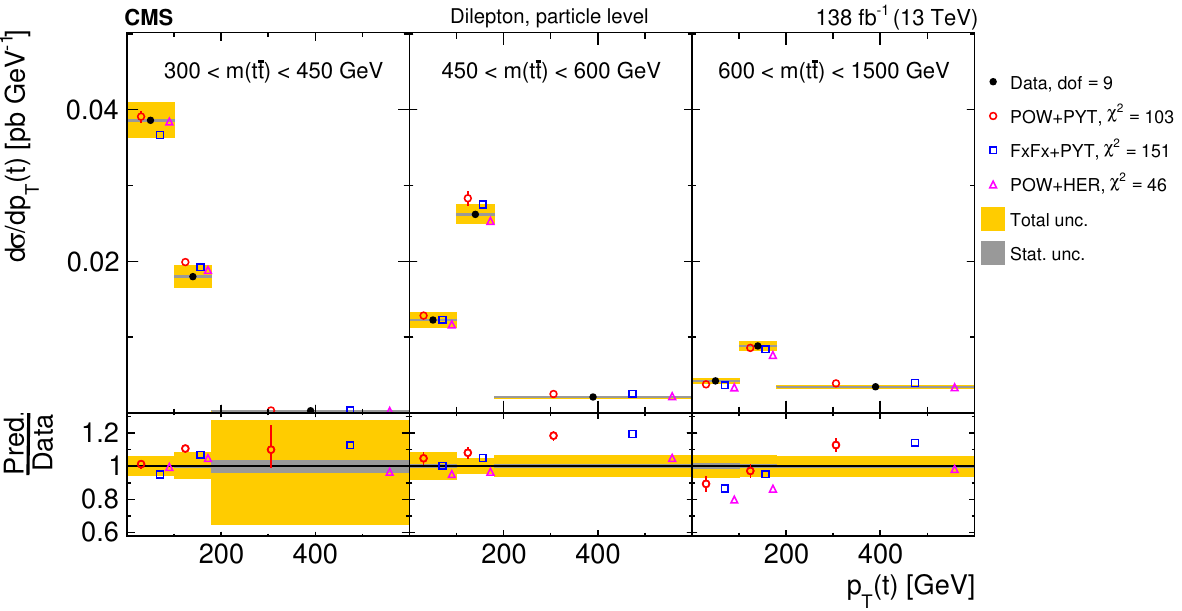}
\caption{Absolute \mttptt cross sections are shown for data (filled circles) and various MC predictions
(other points).
    Further details can be found in the caption of Fig.~\ref{fig:res_ytptt_abs}.}
    \label{fig:res_mttptt_abs}
\end{figure}

\begin{figure}
\centering
\includegraphics[width=0.99\textwidth]{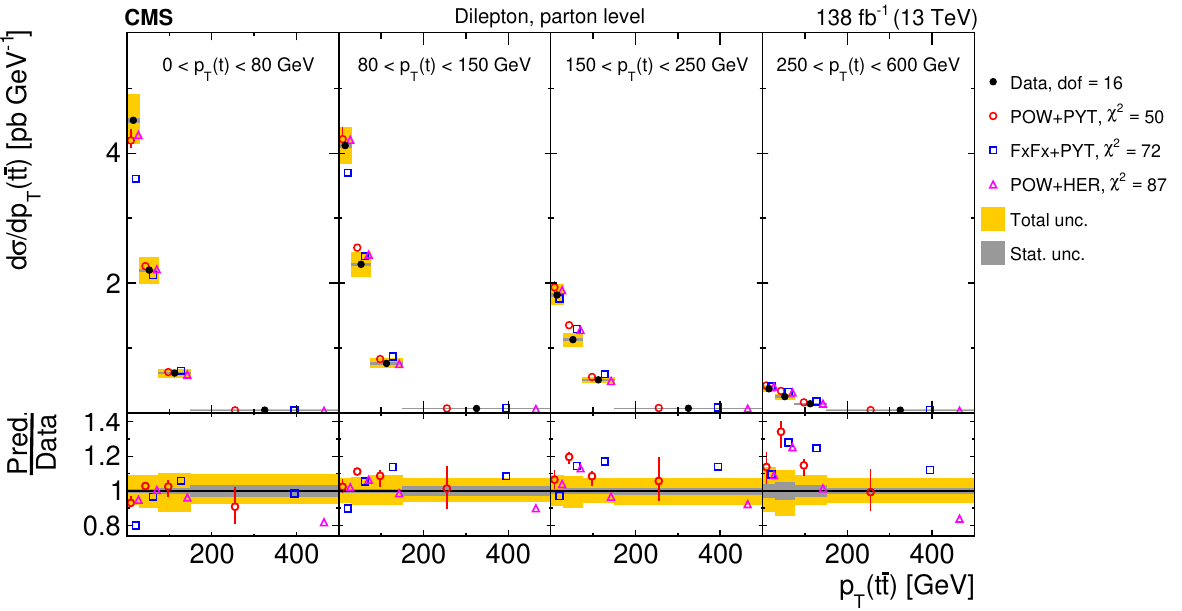}
\includegraphics[width=0.99\textwidth]{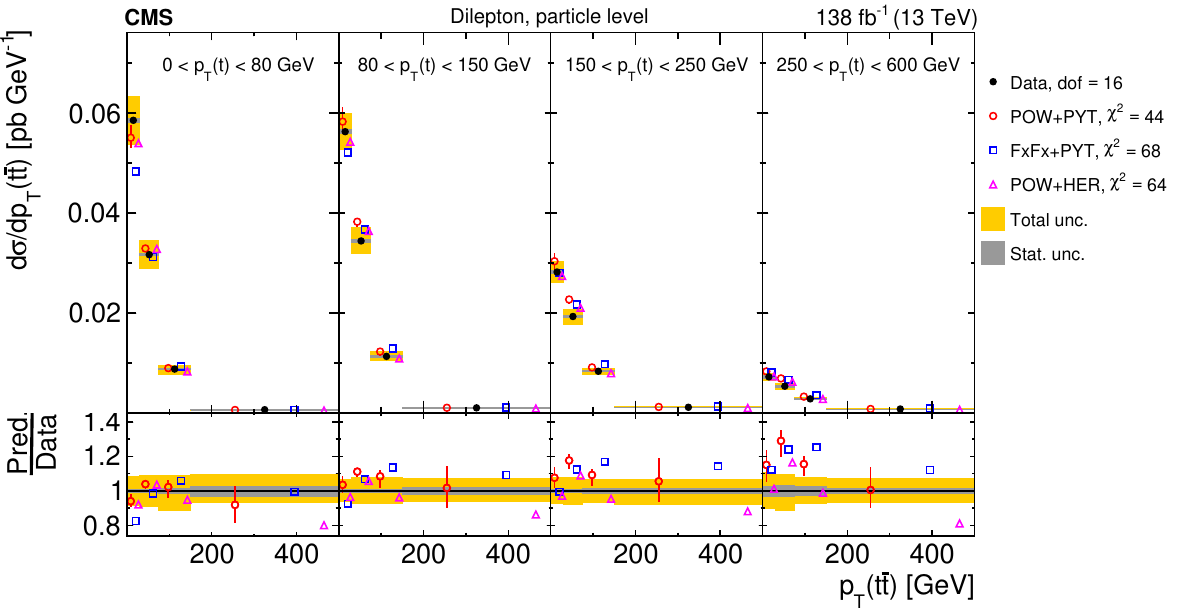}
\caption{Absolute \pttpttt cross sections are shown for data (filled circles) and various MC predictions
(other points).
     Further details can be found in the caption of Fig.~\ref{fig:res_ytptt_abs}.}
    \label{fig:res_pttpttt_abs}
\end{figure}

\begin{figure}
\centering
\includegraphics[width=0.99\textwidth]{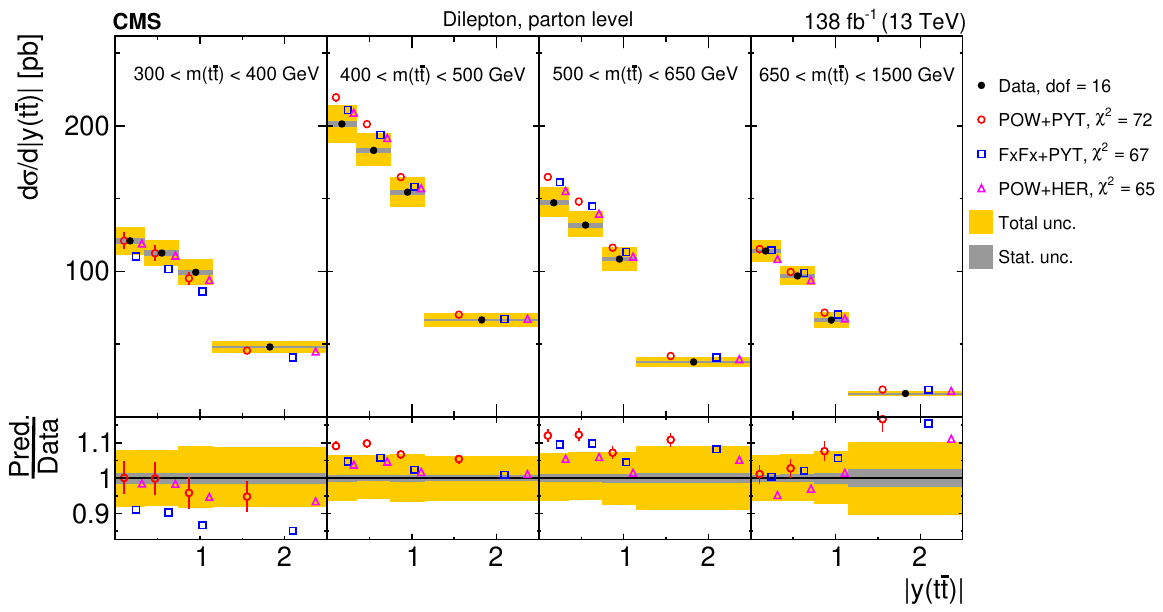}
\includegraphics[width=0.99\textwidth]{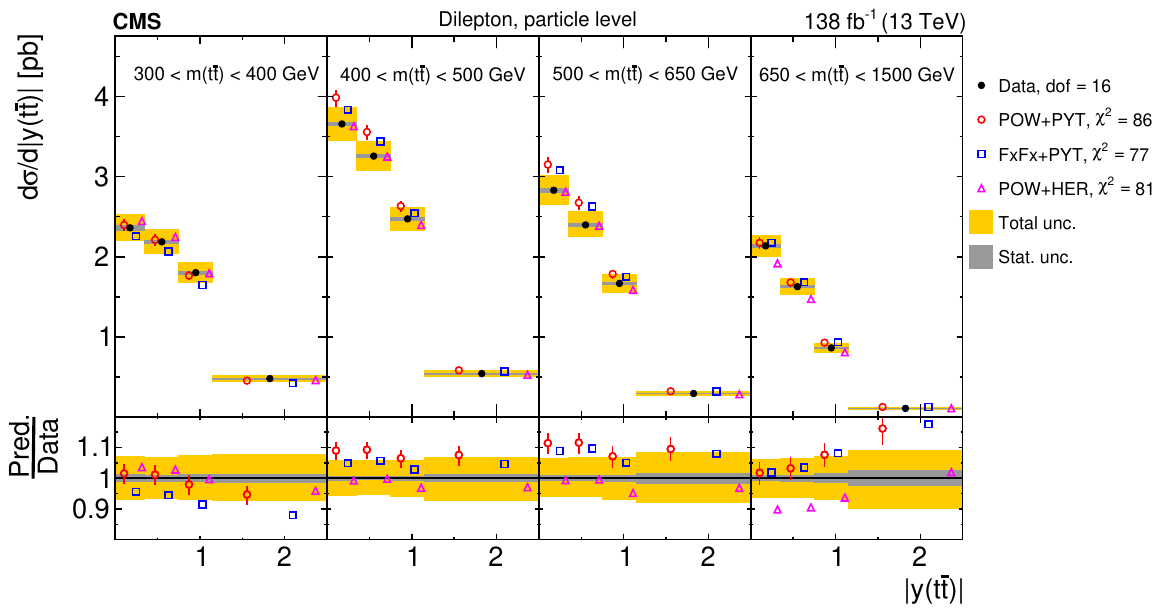}
\caption{Absolute \mttytt cross sections are shown for data (filled circles) and various MC predictions
(other points).
     Further details can be found in the caption of Fig.~\ref{fig:res_ytptt_abs}.}
    \label{fig:res_mttytt_abs}
\end{figure}

\begin{figure}
\centering
\includegraphics[width=0.99\textwidth]{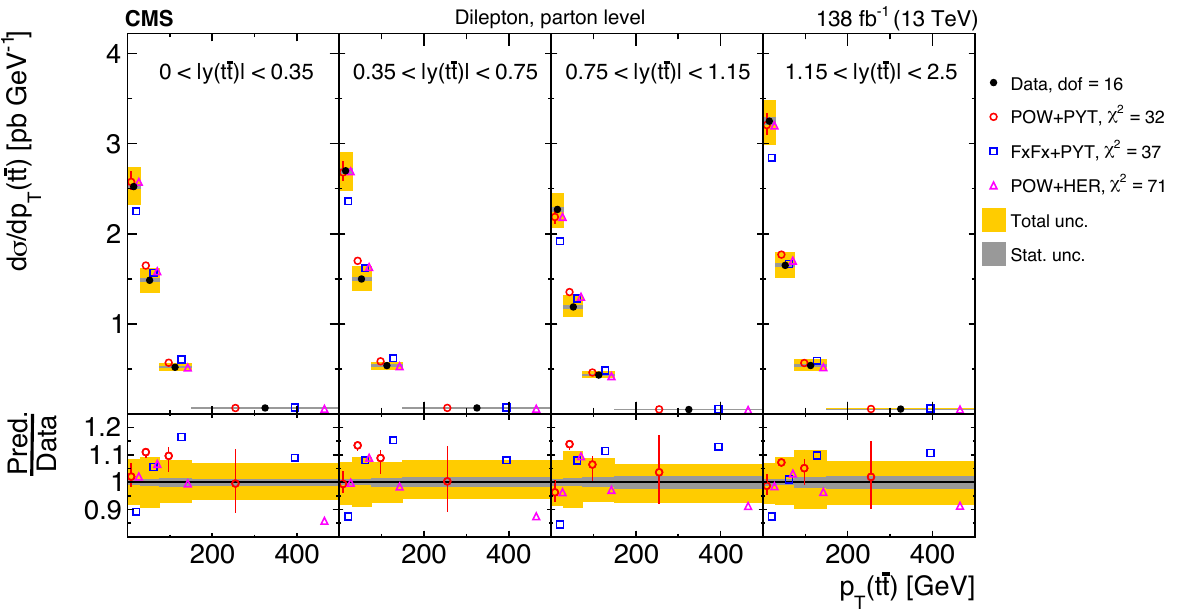}
\includegraphics[width=0.99\textwidth]{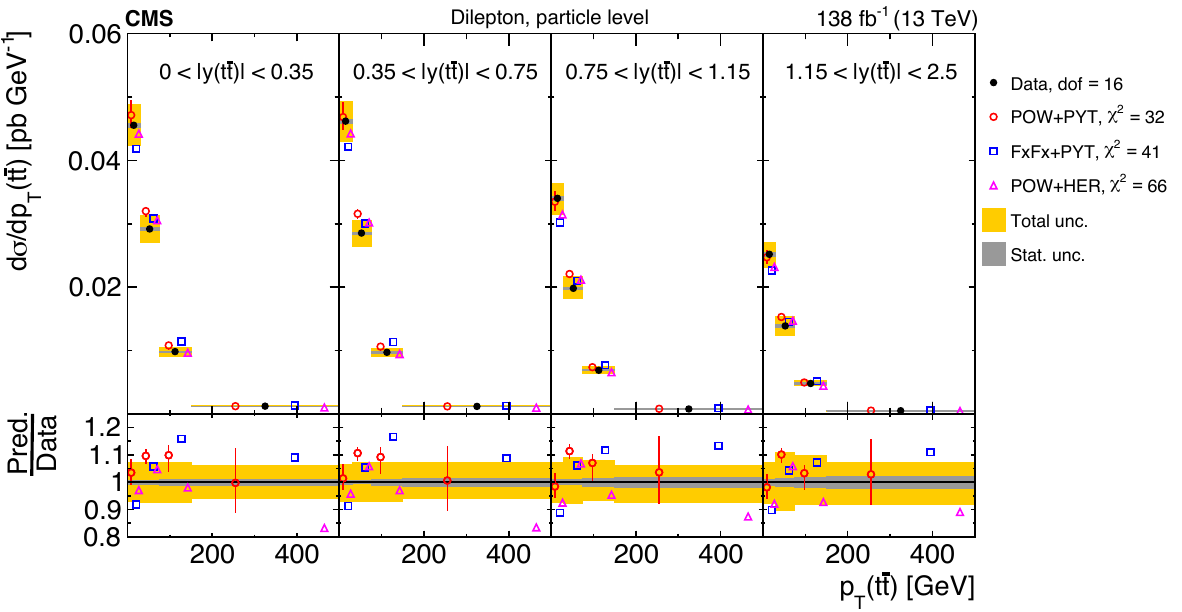}
\caption{Absolute \yttpttt cross sections are shown for data (filled circles) and various MC predictions
(other points).
     Further details can be found in the caption of Fig.~\ref{fig:res_ytptt_abs}.}
    \label{fig:res_yttpttt_abs}
\end{figure}

\begin{figure}
\centering
\includegraphics[width=0.99\textwidth]{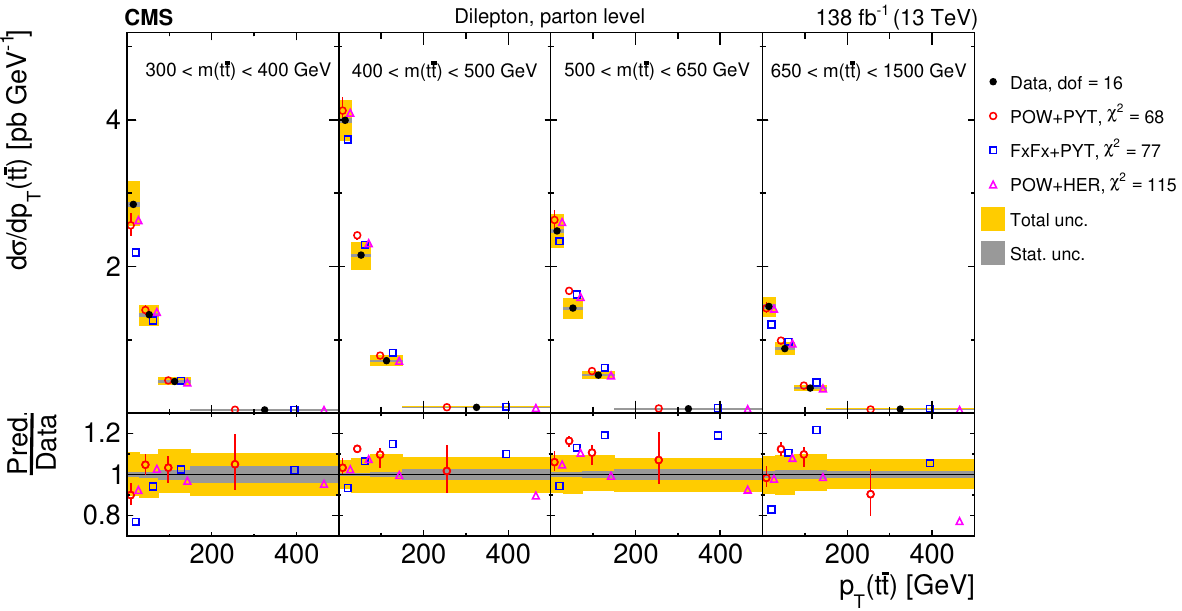}
\includegraphics[width=0.99\textwidth]{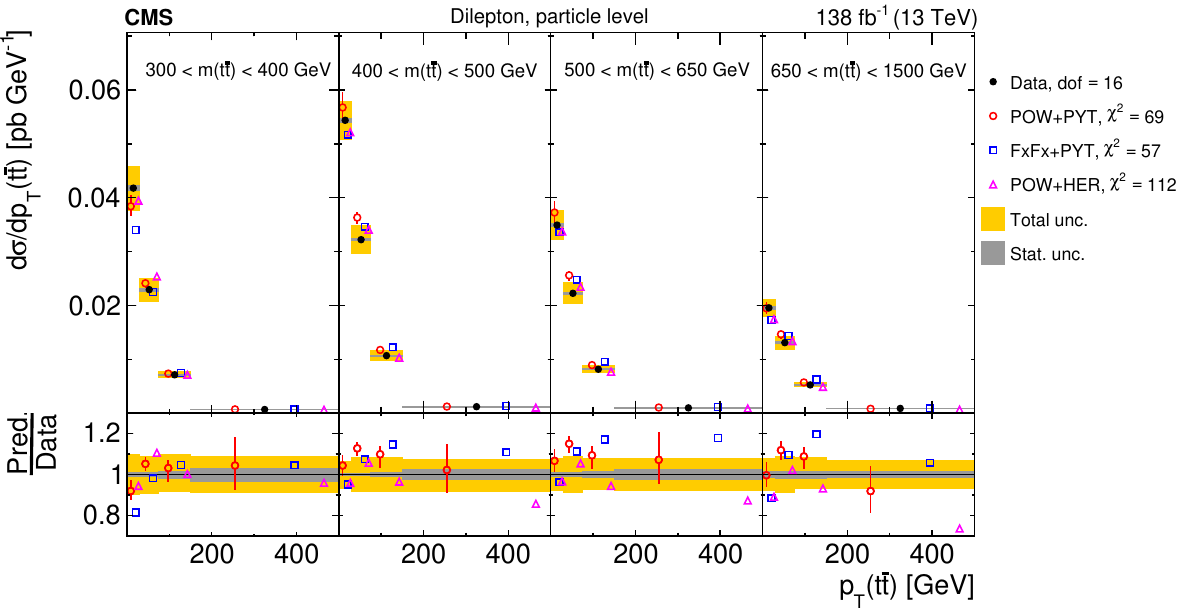}
\caption{Absolute \mttpttt cross sections are shown for data (filled circles) and various MC predictions
(other points).
     Further details can be found in the caption of Fig.~\ref{fig:res_ytptt_abs}.}
    \label{fig:res_mttpttt_abs}
\end{figure}

\begin{figure}
\centering
\includegraphics[width=0.99\textwidth]{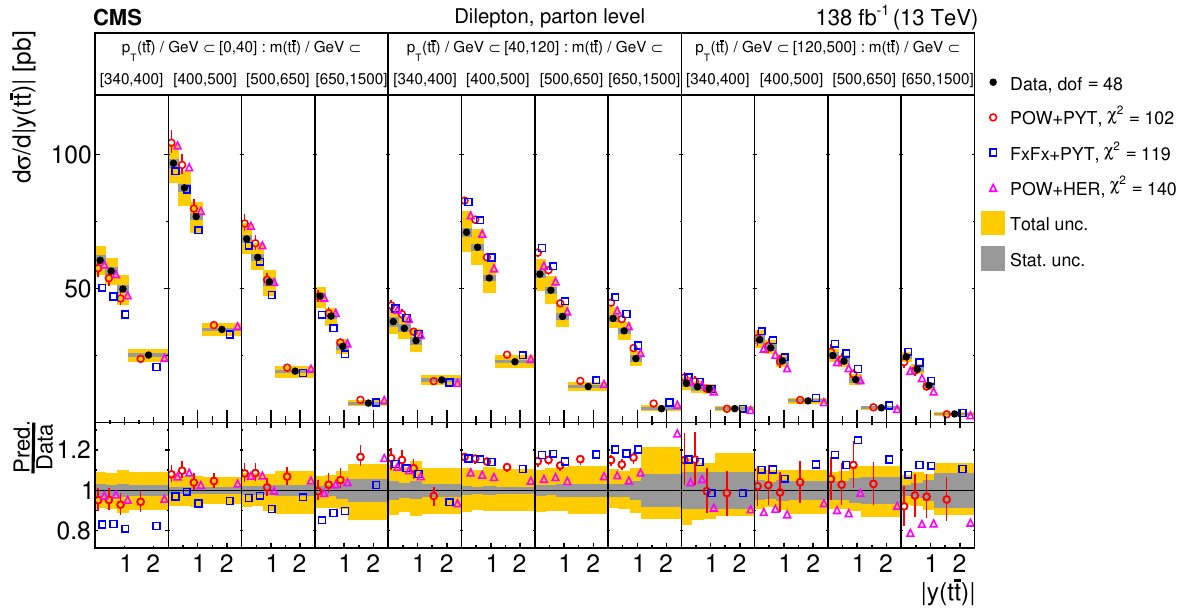}
\includegraphics[width=0.99\textwidth]{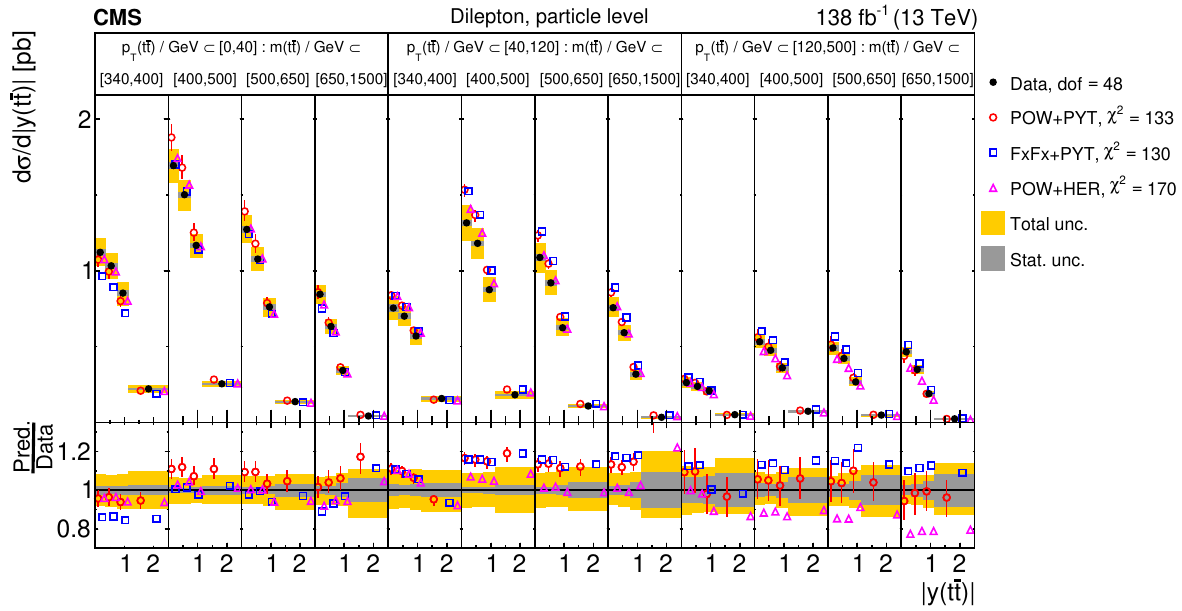}
\caption{Absolute \ptttmttytt cross sections are shown for data (filled circles) and various MC predictions
(other points).
     Further details can be found in the caption of Fig.~\ref{fig:res_ytptt_abs}.}
    \label{fig:res_ptttmttytt_abs}
\end{figure}

\begin{figure}
\centering
\includegraphics[width=0.99\textwidth]{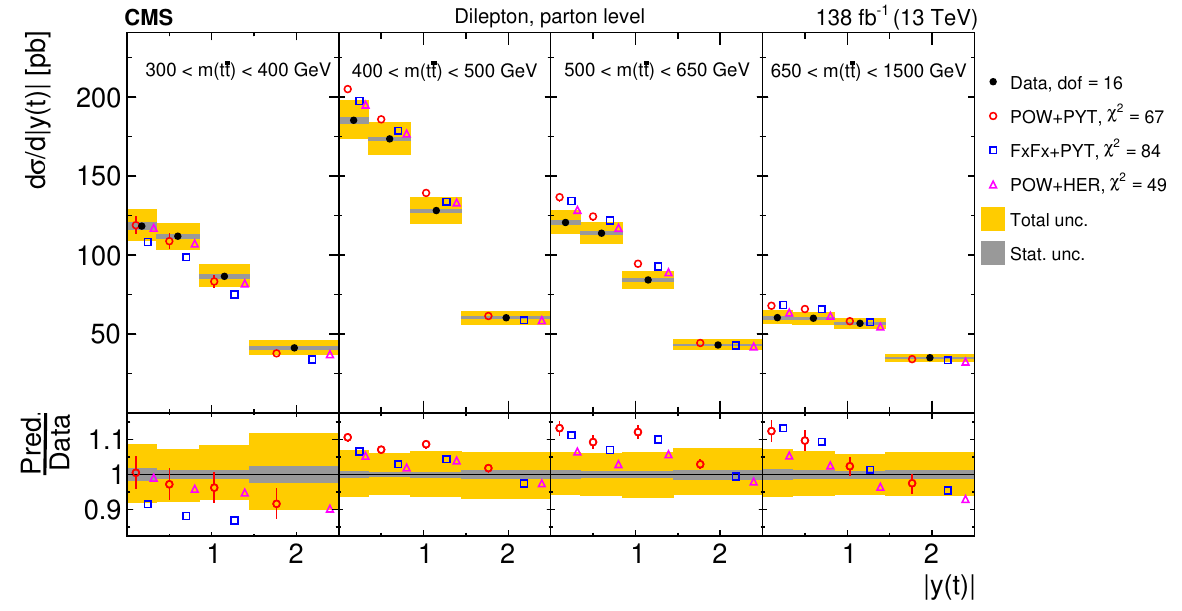}
\includegraphics[width=0.99\textwidth]{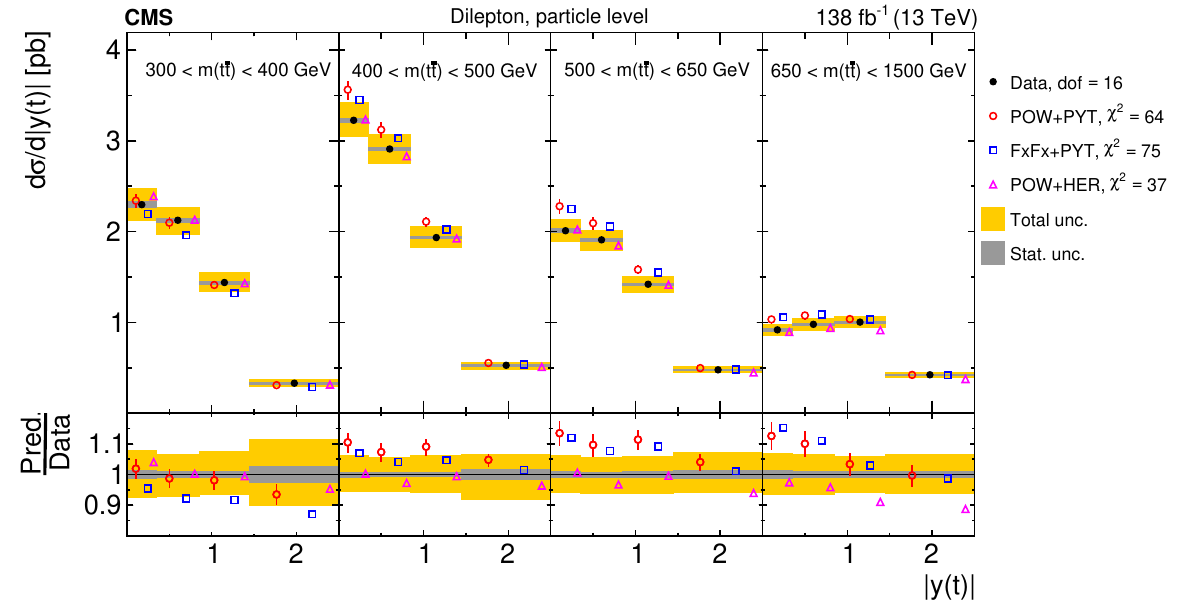}
\caption{Absolute \mttyt cross sections are shown for data (filled circles) and various MC predictions
(other points).
    Further details can be found in the caption of Fig.~\ref{fig:res_ytptt_abs}.}
   \label{fig:res_mttyt_abs}
\end{figure}

\begin{figure}
\centering
\includegraphics[width=0.99\textwidth]{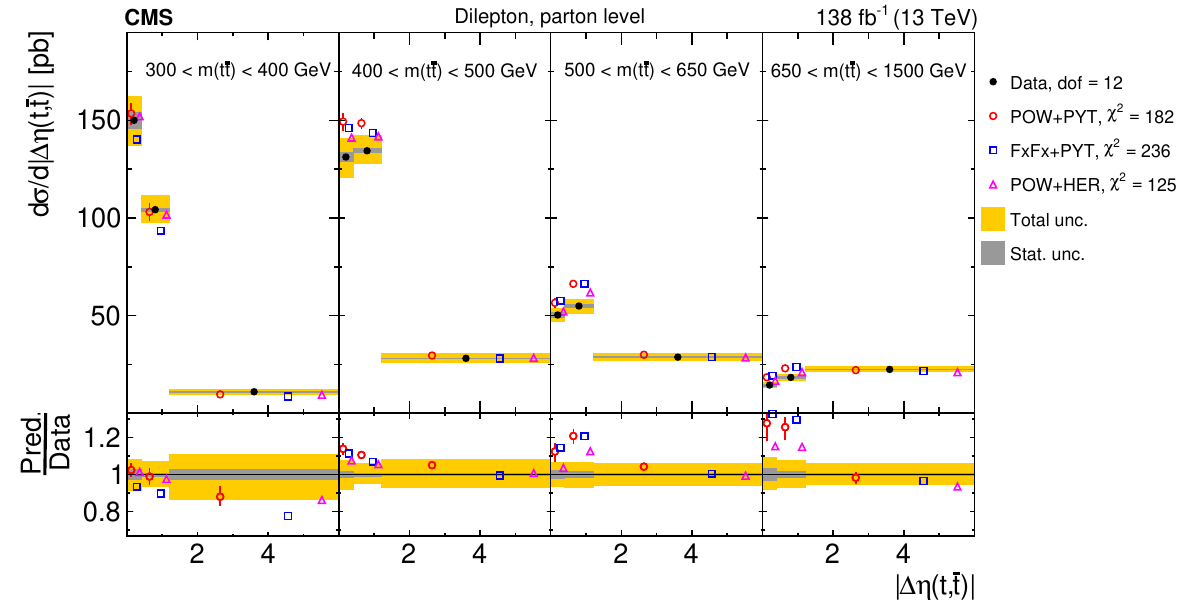}
\includegraphics[width=0.99\textwidth]{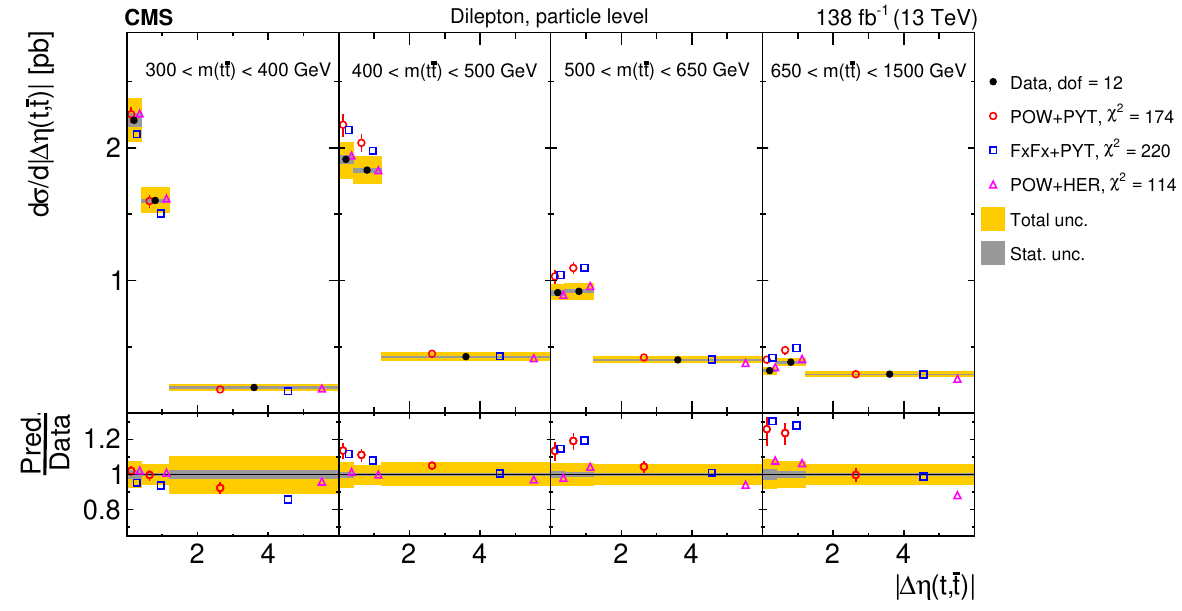}
\caption{Absolute \mttdetatt cross sections are shown for data (filled circles) and various MC predictions 
(other points).
    Further details can be found in the caption of Fig.~\ref{fig:res_ytptt_abs}.}
   \label{fig:res_mttdetatt_abs}
\end{figure}

\begin{figure}
\centering
\includegraphics[width=0.99\textwidth]{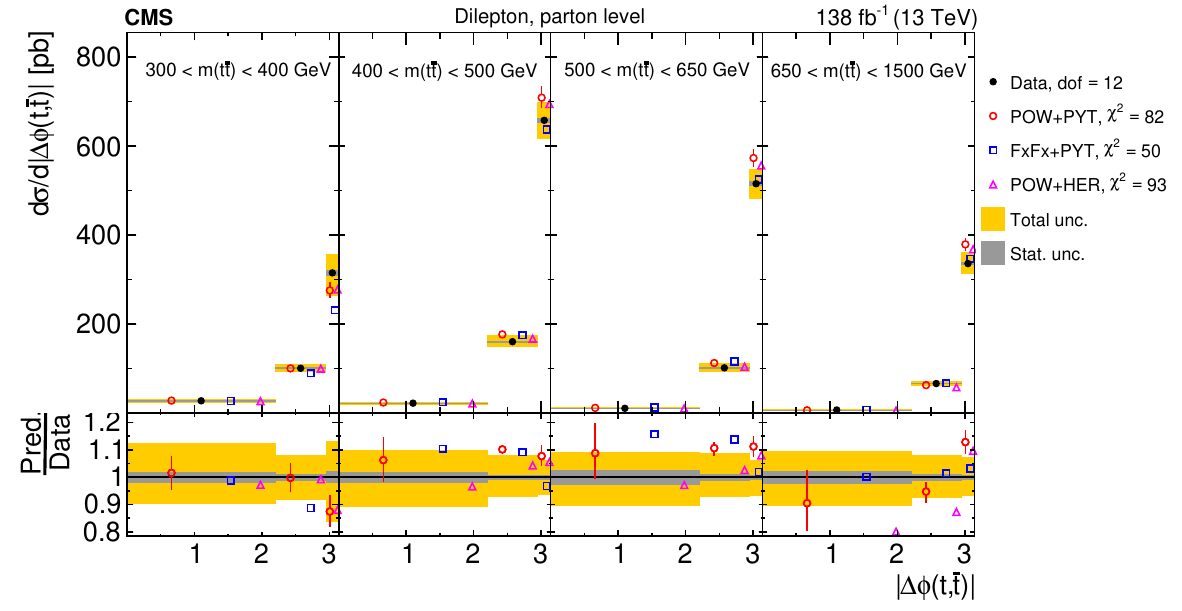}
\includegraphics[width=0.99\textwidth]{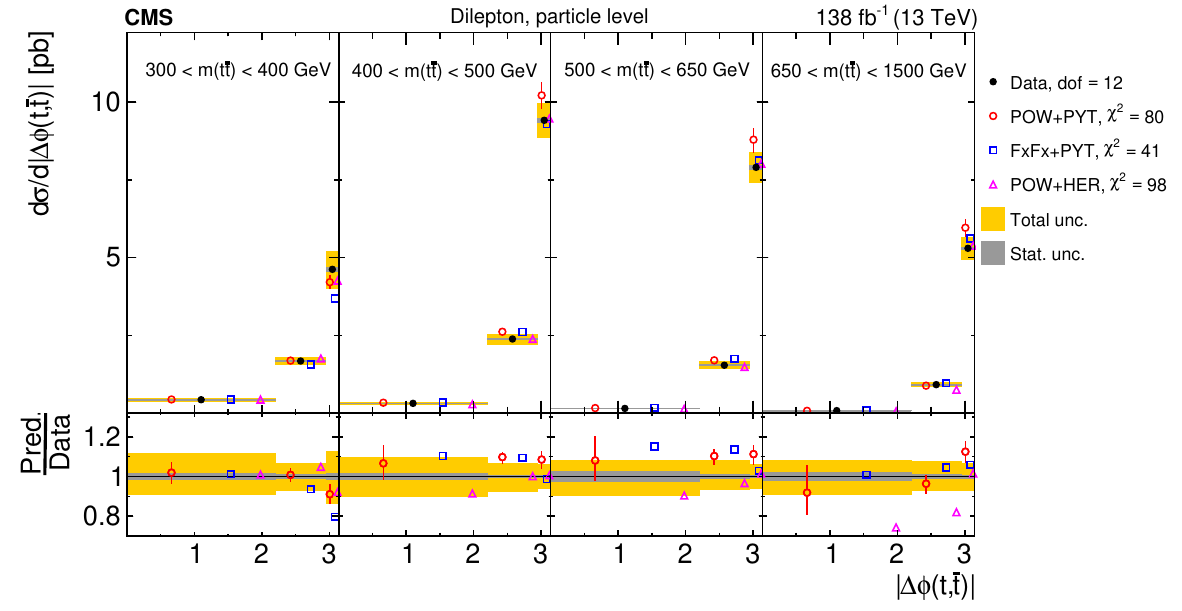}
\caption{Absolute \mttdphitt cross sections are shown for data (filled circles) and various MC predictions
(other points).
     Further details can be found in the caption of Fig.~\ref{fig:res_ytptt_abs}.}
    \label{fig:res_mttdphitt_abs}
\end{figure}

\begin{figure*}[!phtb]
\centering
\includegraphics[width=0.49\textwidth]{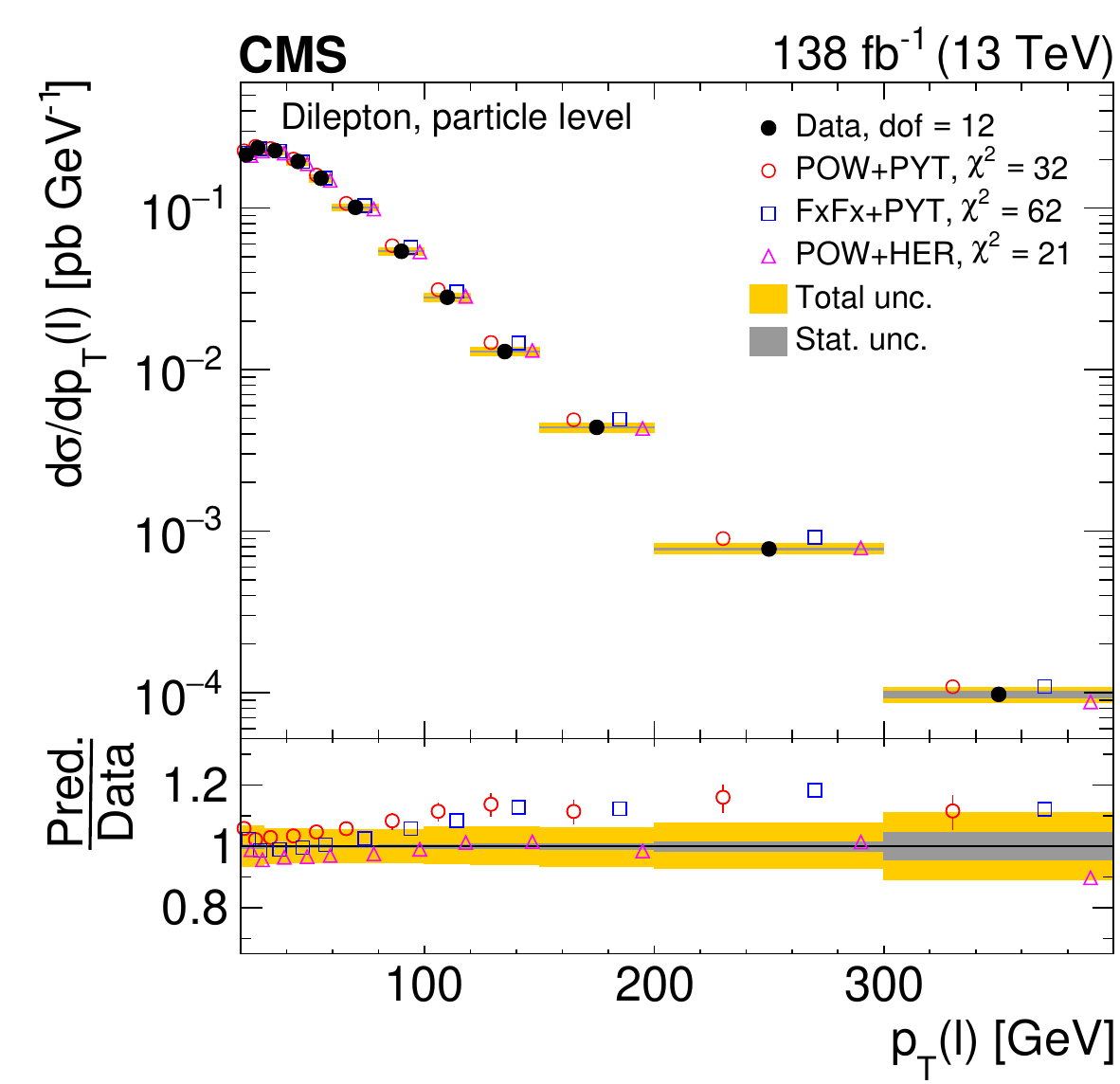}
\includegraphics[width=0.49\textwidth]{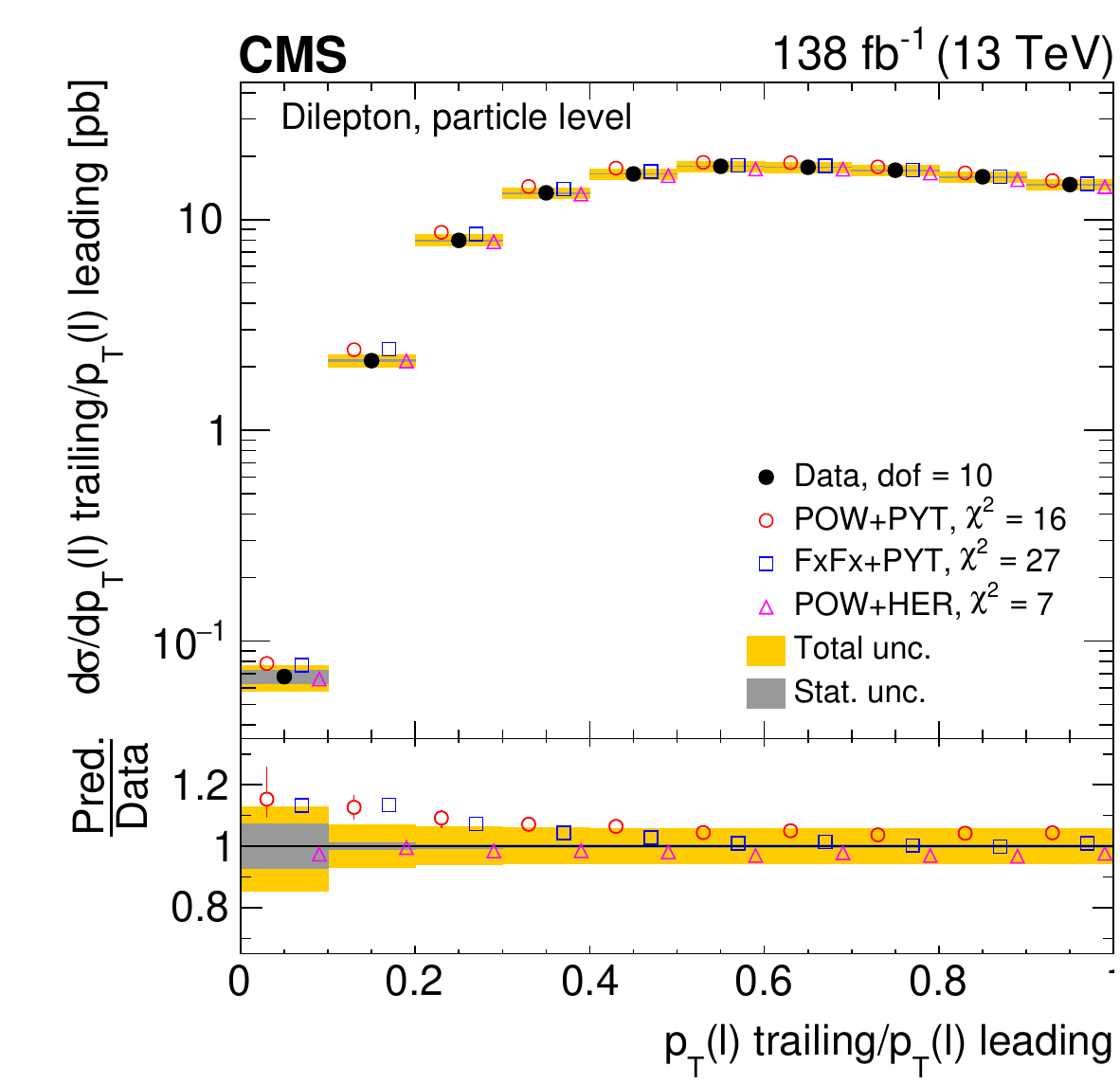}
\includegraphics[width=0.49\textwidth]{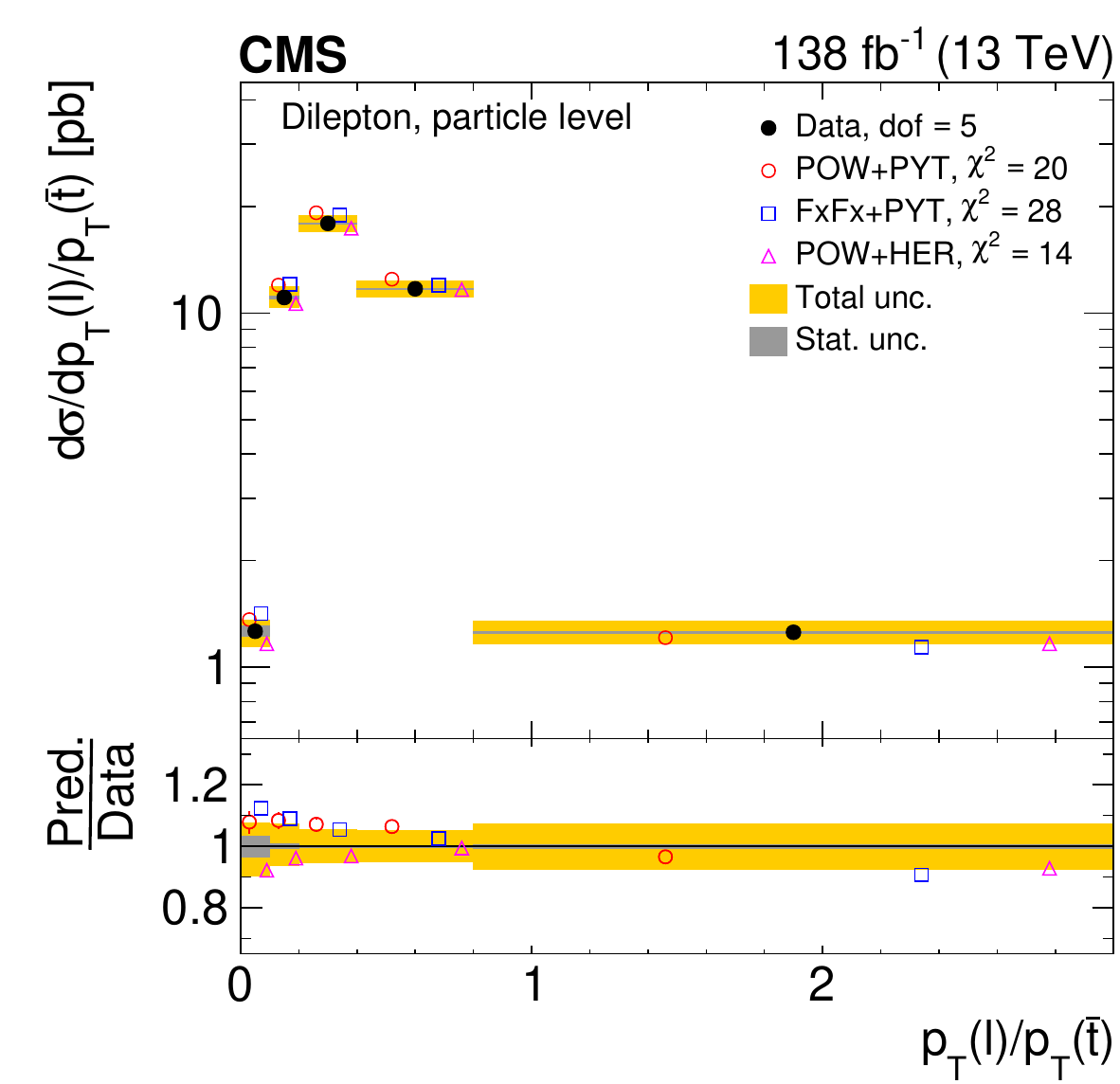}
\caption{Absolute differential \ttbar production cross sections as a function of \pt of the lepton (upper left),
of the ratio of the trailing and leading lepton \pt
(upper right), and of the ratio of lepton and top antiquark \pt (lower middle),
measured at the particle level in a fiducial phase space.
The data are shown as filled circles with grey and yellow bands indicating the statistical and total
uncertainties (statistical
and systematic uncertainties added in quadrature), respectively.
For each distribution, the number of degrees of freedom (dof) is also provided.
The cross sections are compared to various MC predictions (other points).
The estimated uncertainties in the \PowPyt (`POW-PYT') simulation are represented by vertical bars on
the corresponding points.
For each MC model, a value of \chisq is reported that takes into account the measurement uncertainties.
The lower panel in each plot shows the ratios of the predictions to the data.}
\label{fig:res_ptlep_abs}
\end{figure*}

\begin{figure*}[!phtb]
\centering
\includegraphics[width=0.49\textwidth]{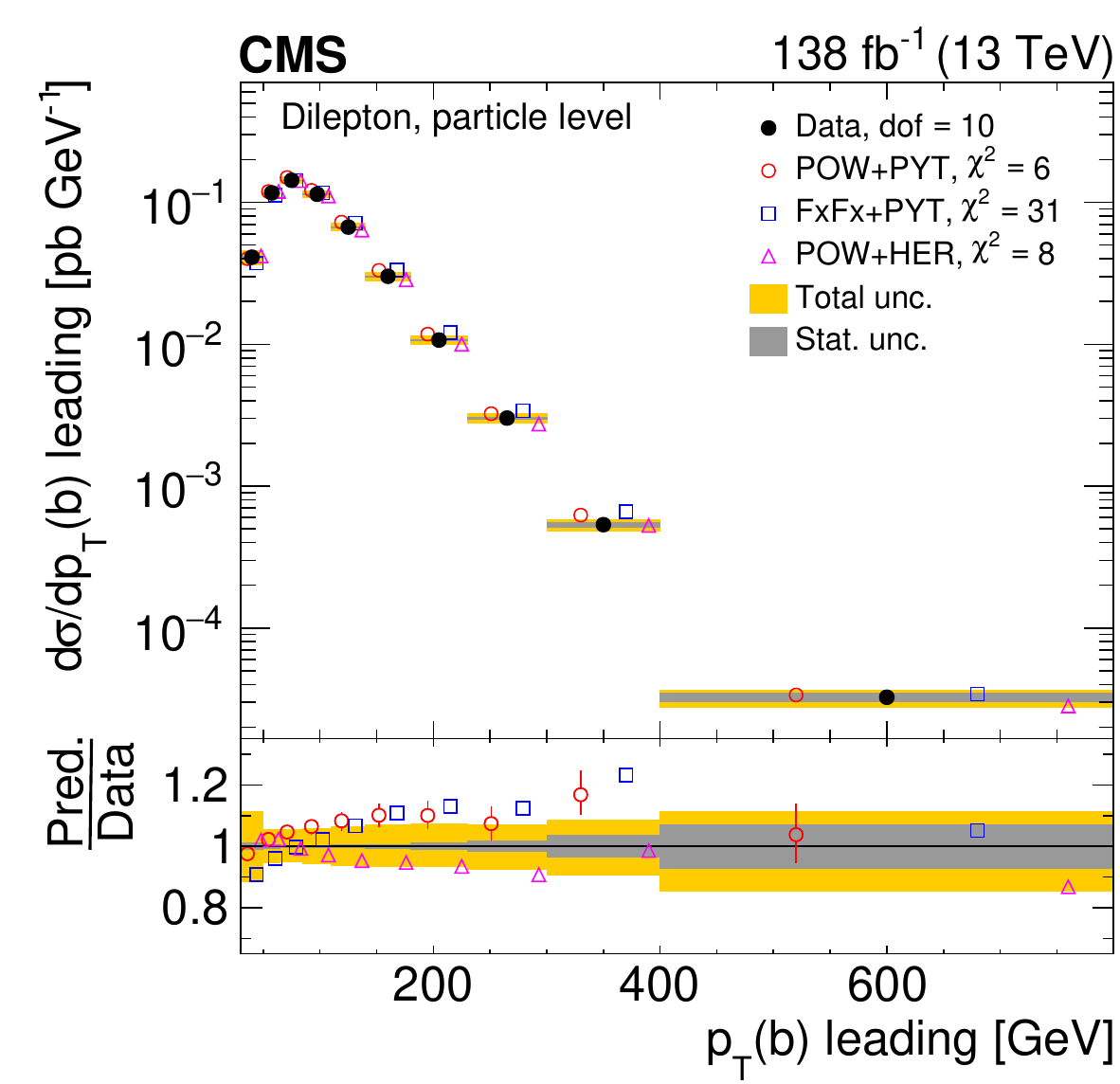}
\includegraphics[width=0.49\textwidth]{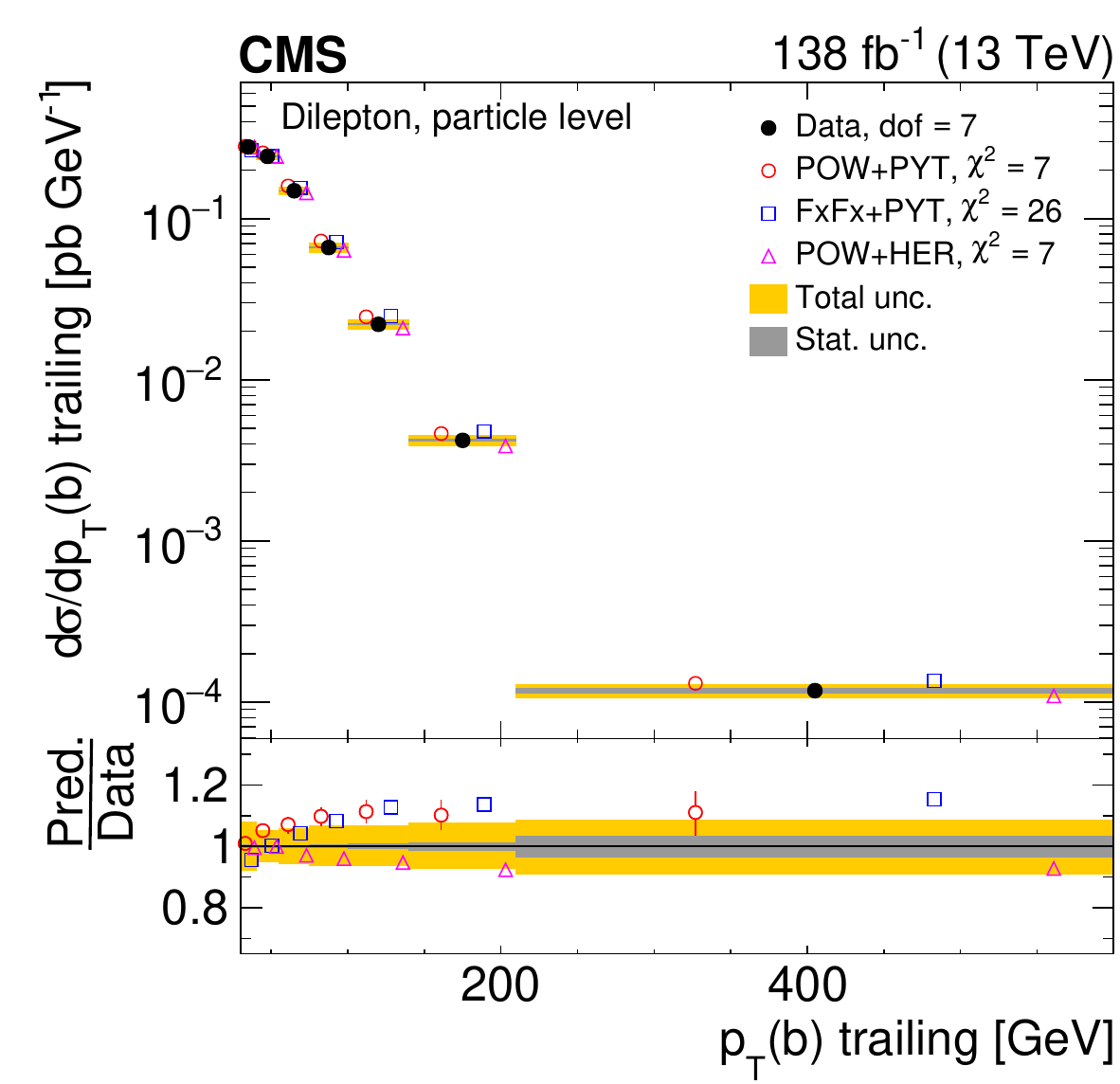}
\includegraphics[width=0.49\textwidth]{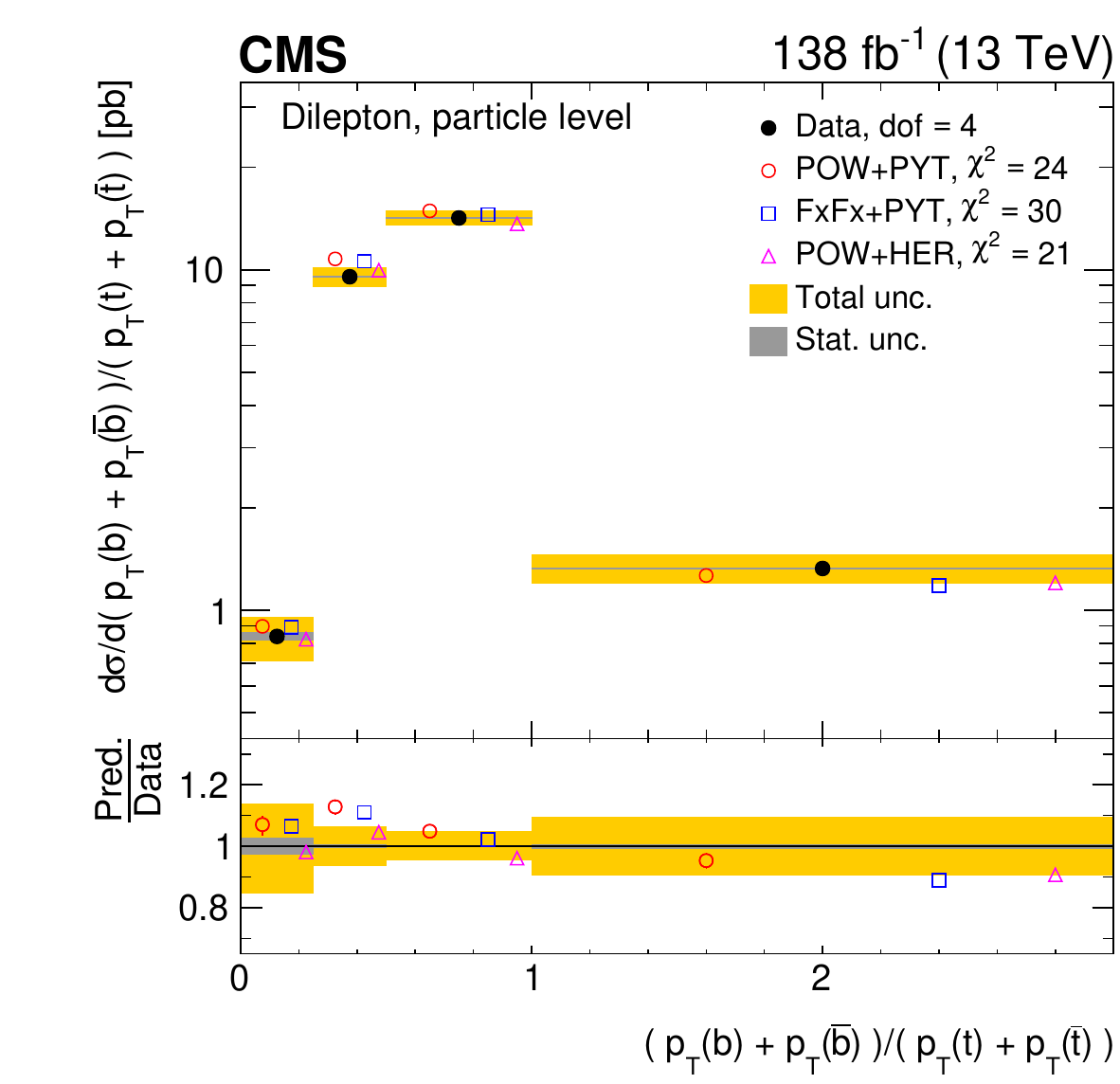}
\caption{Absolute differential \ttbar production cross sections as functions of the \pt of the
leading (upper left) and trailing (upper right) \PQb jet,
and \rptbsptts (lower).
Further details can be found in the caption of Fig.~\ref{fig:res_ptlep_abs}.}
\label{fig:res_ptb_abs}
\end{figure*}

\begin{figure*}[!phtb]
\centering
\includegraphics[width=0.49\textwidth]{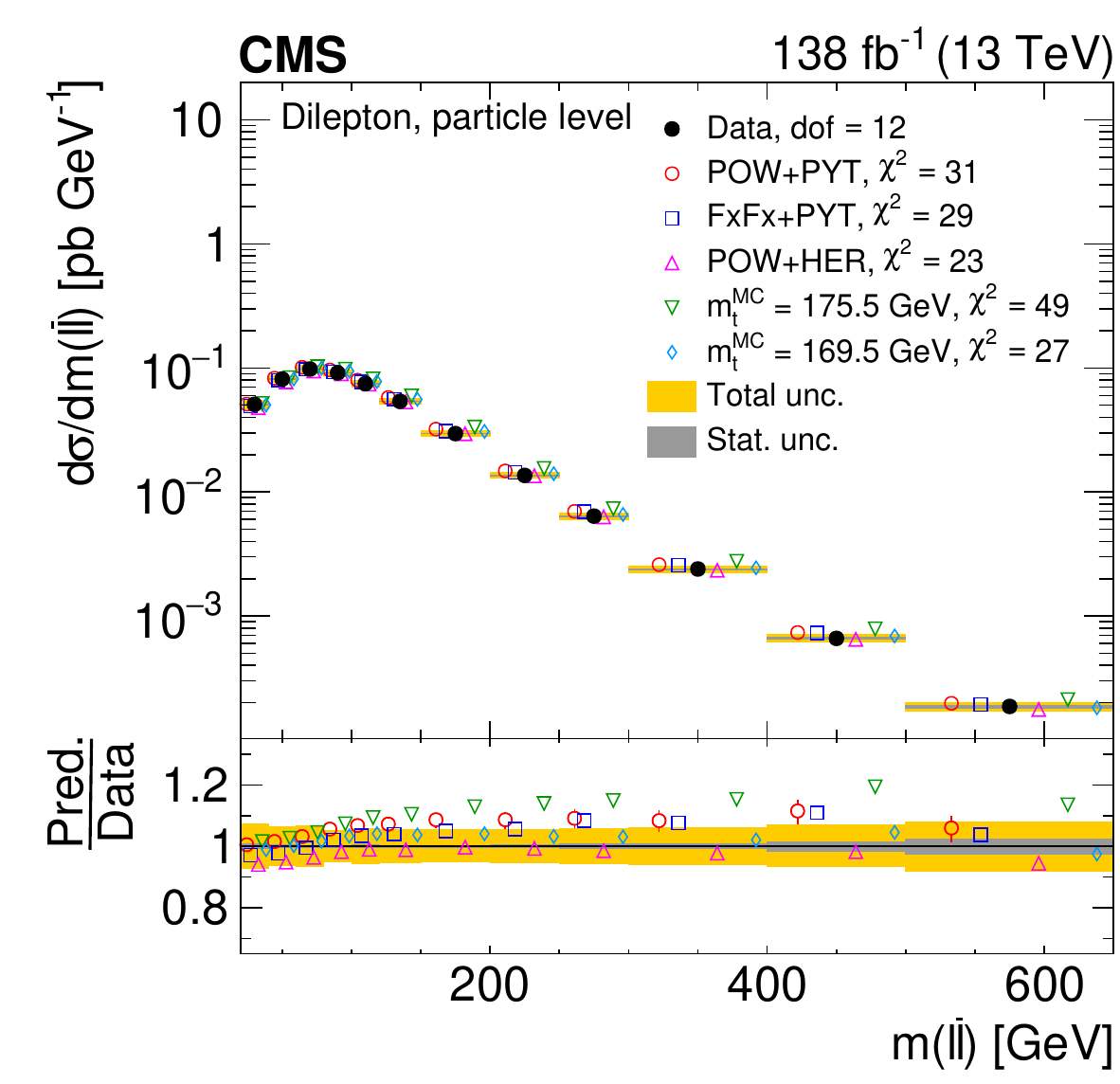}
\includegraphics[width=0.49\textwidth]{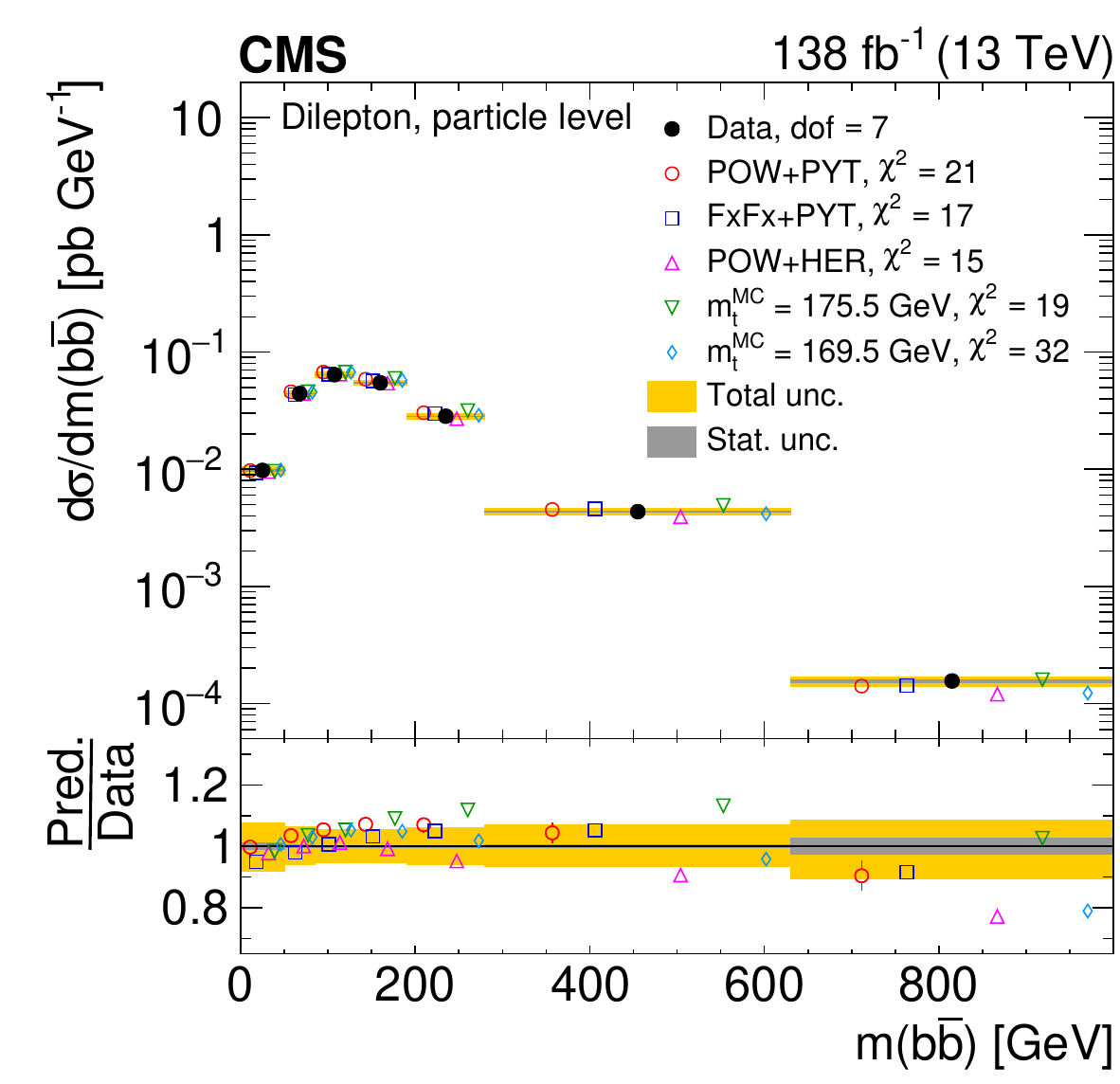}
\includegraphics[width=0.49\textwidth]{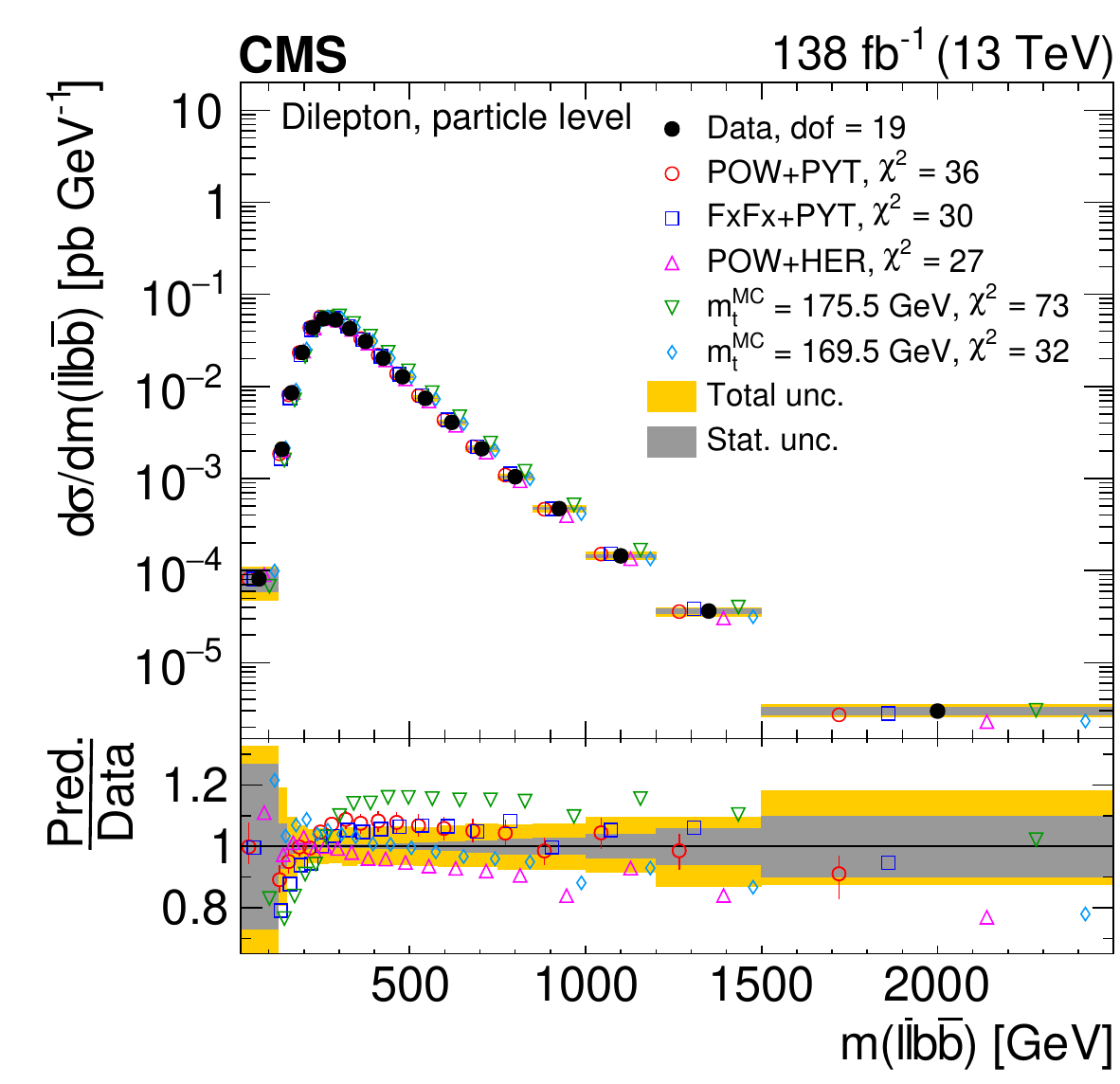}
\caption{Absolute differential \ttbar production cross sections as functions of \mll (upper left), \mbb
(upper right) and \mllbb (lower)
are shown for data (filled circles) and various MC predictions (other points).
Further details can be found in the caption of Fig.~\ref{fig:res_ptlep_abs}.}
\label{fig:res_mll_abs}
\end{figure*}

\begin{figure}
\centering
\includegraphics[width=0.49\textwidth]{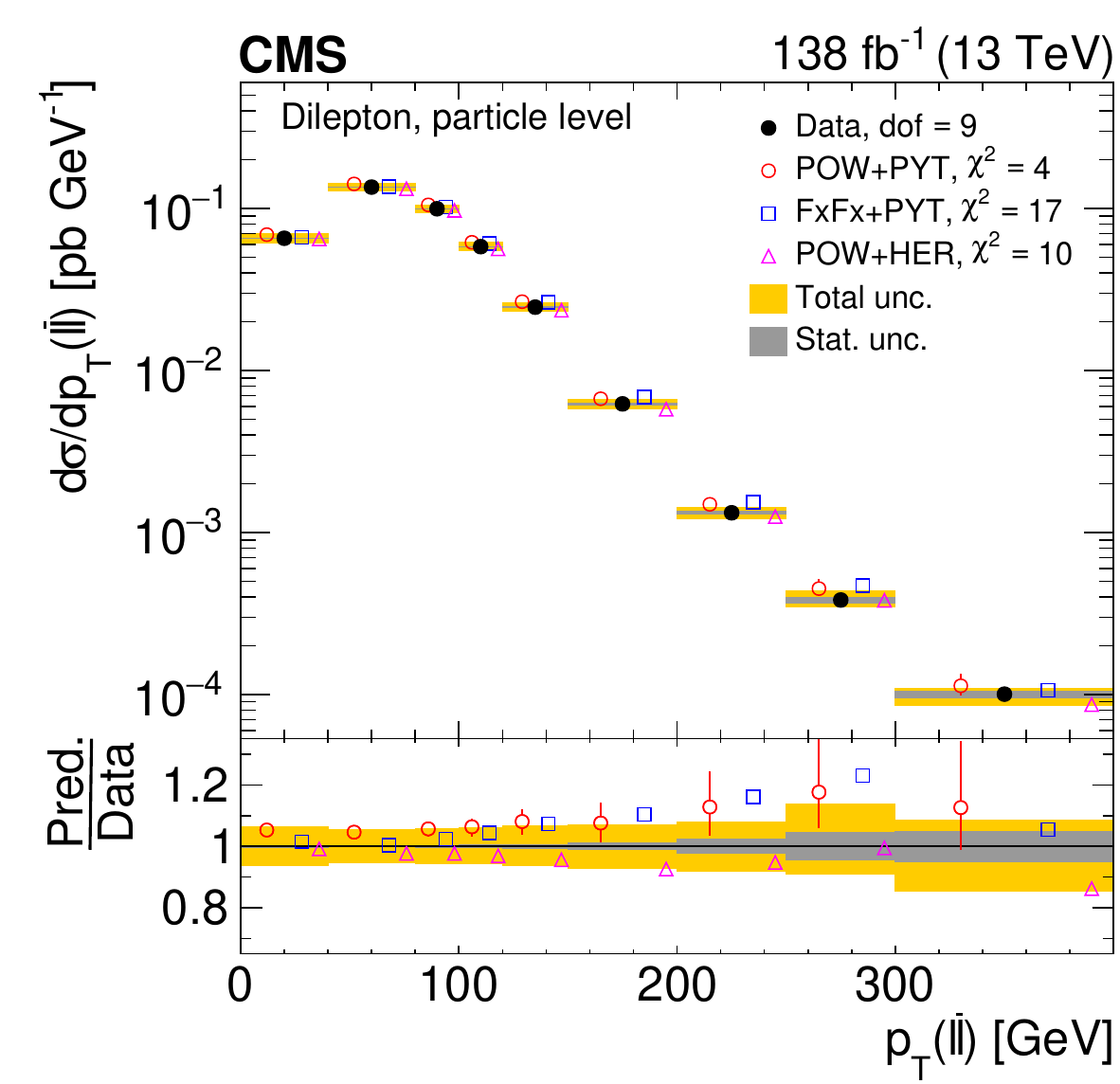}
\includegraphics[width=0.49\textwidth]{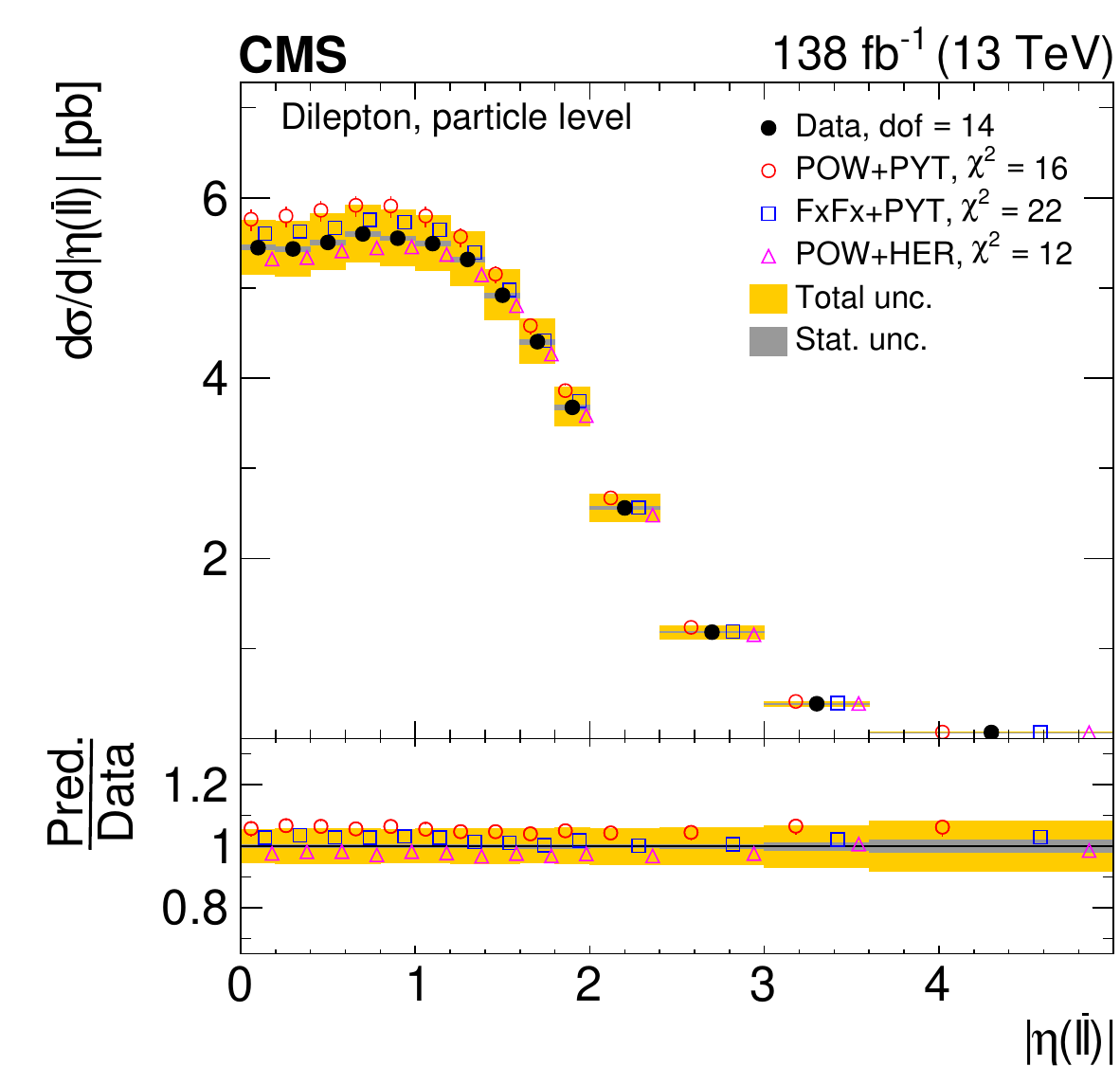}
\caption{Absolute differential \ttbar production cross sections as functions of \ptll (left) and \absetall (right)
are shown for data (filled circles) and various MC predictions (other points).
Further details can be found in the caption of Fig.~\ref{fig:res_ptlep_abs}.}
\label{fig:res_ptll_abs}
\end{figure}

\begin{figure}
\centering
\includegraphics[width=1.00\textwidth]{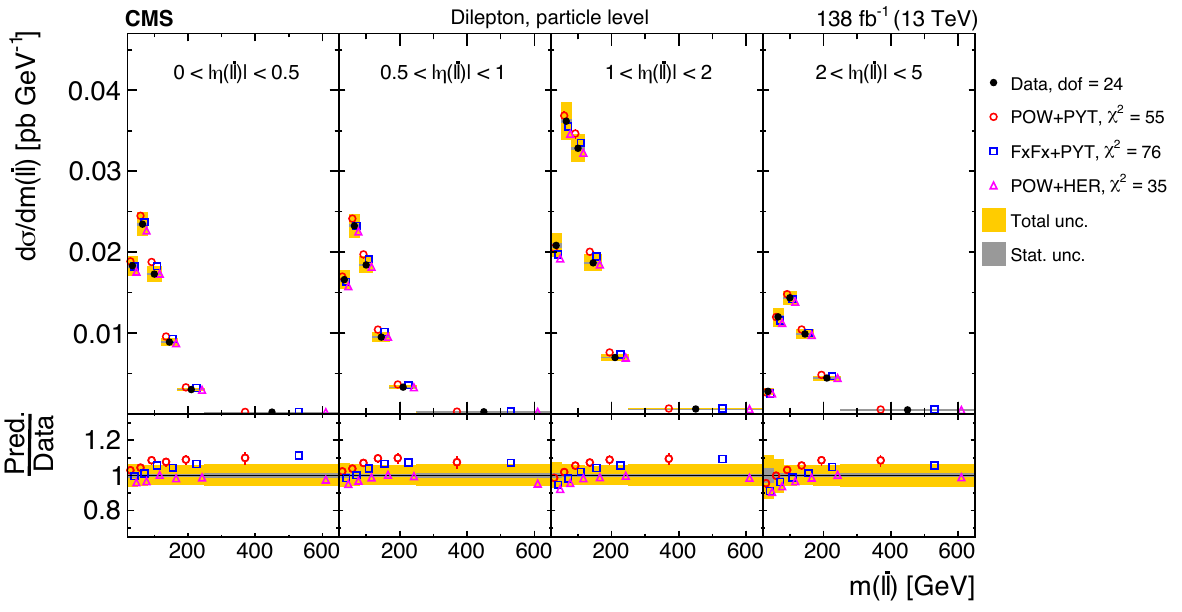}
\caption{Absolute \etallmll cross sections are shown for data (filled circles) and various MC predictions
(other points).
    Further details can be found in the caption of Fig.~\ref{fig:res_ptlep_abs}.}
    \label{fig:res-etallmll_abs}
\end{figure}

\begin{figure}
\centering
\includegraphics[width=1.00\textwidth]{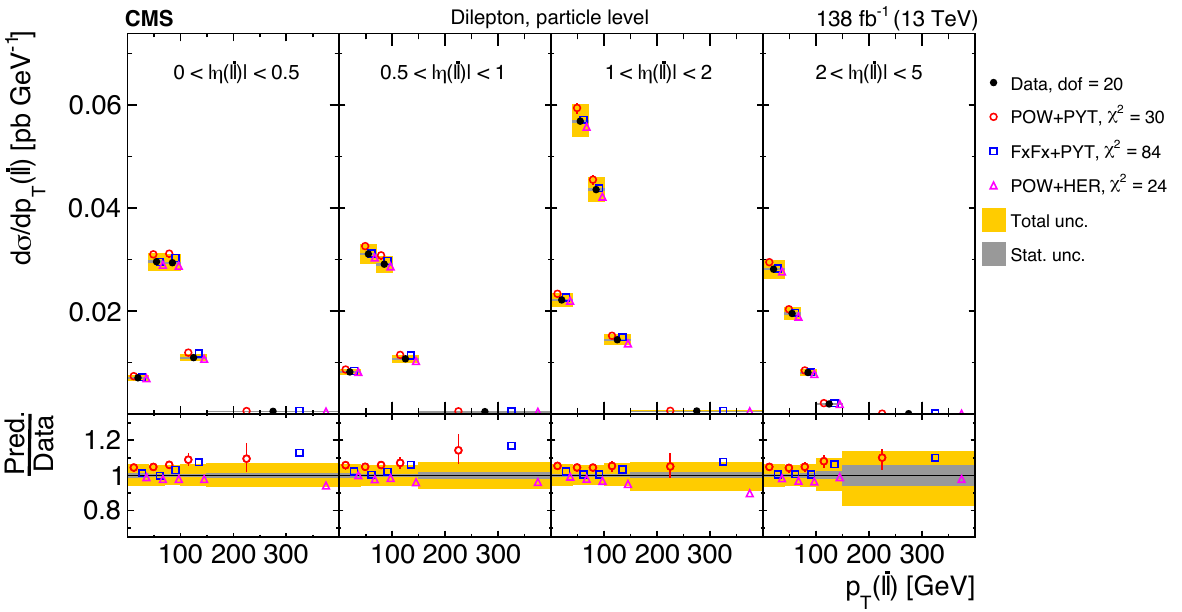}
    \caption{Absolute \etallptll cross sections are shown for data (filled circles) and various MC predictions
(other points).
    Further details can be found in the caption of Fig.~\ref{fig:res_ptlep_abs}.}
    \label{fig:res-etallptll_abs}
\end{figure}

\begin{figure}
\centering
\includegraphics[width=1.00\textwidth]{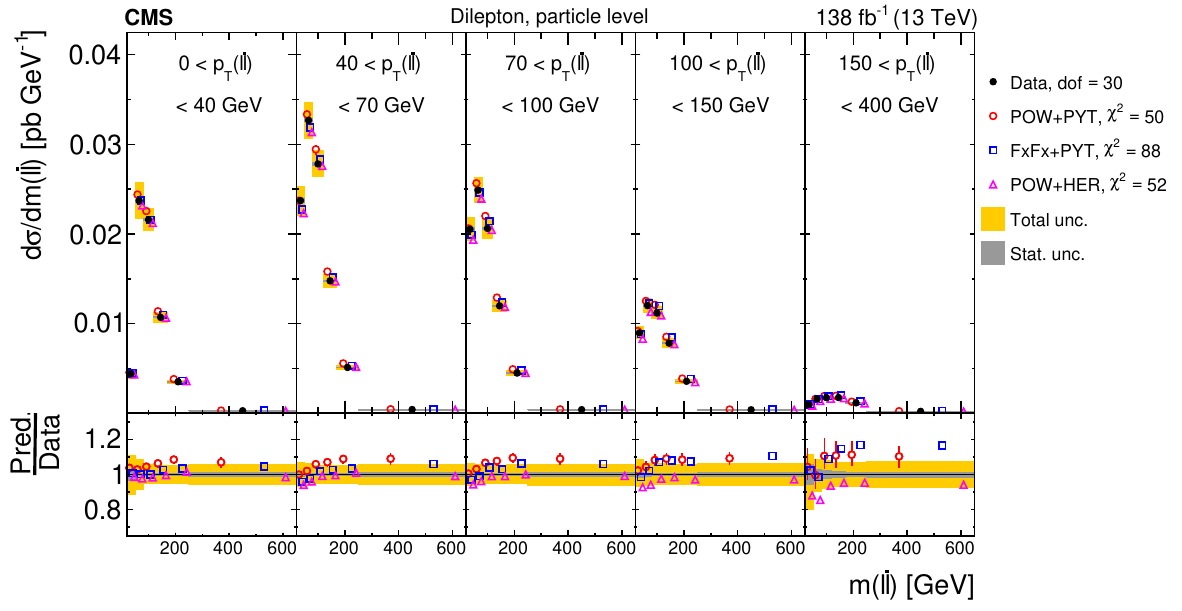}
\caption{Absolute \ptllmll cross sections are shown for data (filled circles) and various MC predictions
(other points).
    Further details can be found in the caption of Fig.~\ref{fig:res_ptlep_abs}.}
    \label{fig:res-ptllmll_abs}
\end{figure}

\begin{figure}
    \centering
\includegraphics[width=0.49\textwidth]{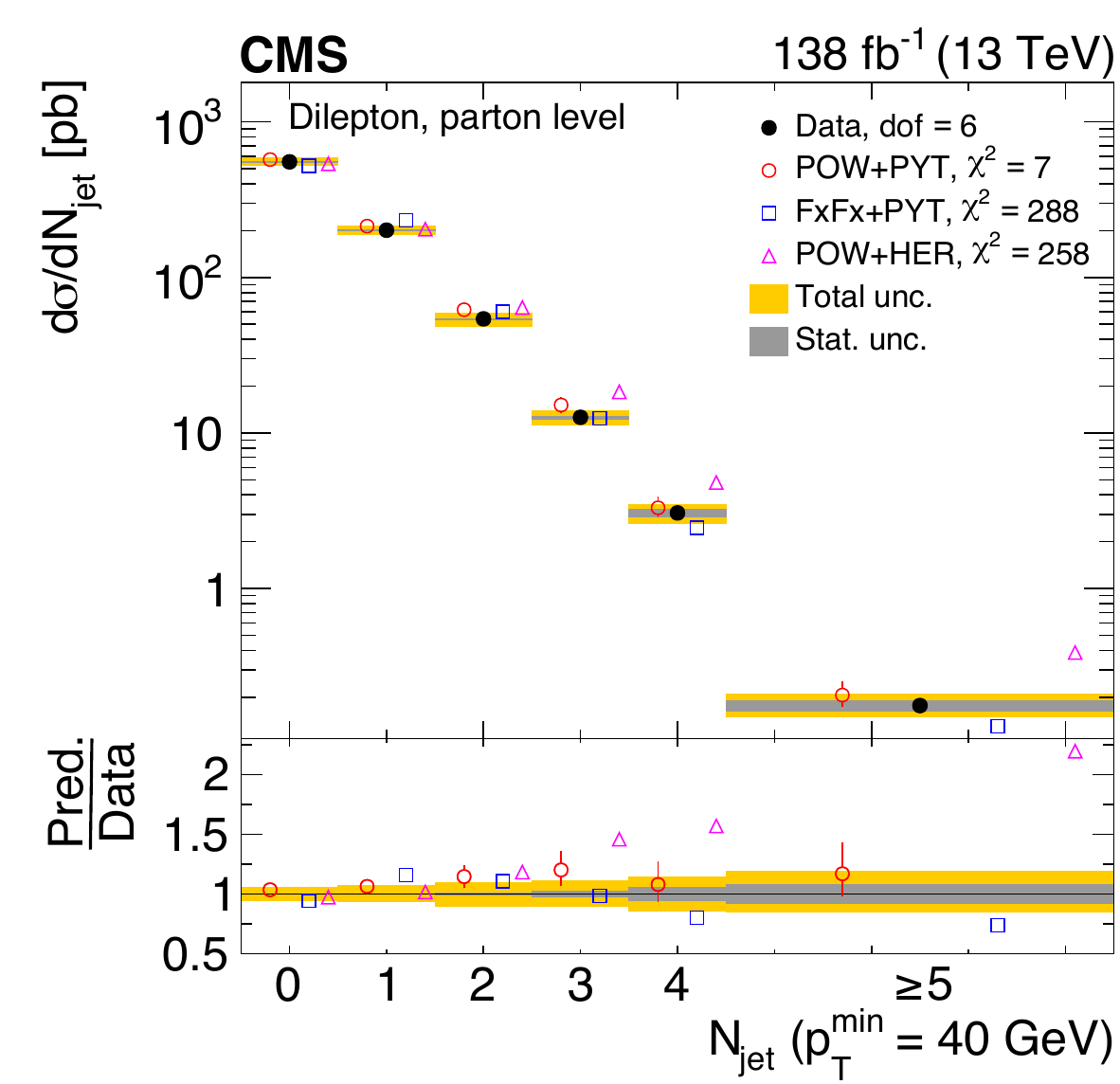}
\includegraphics[width=0.49\textwidth]{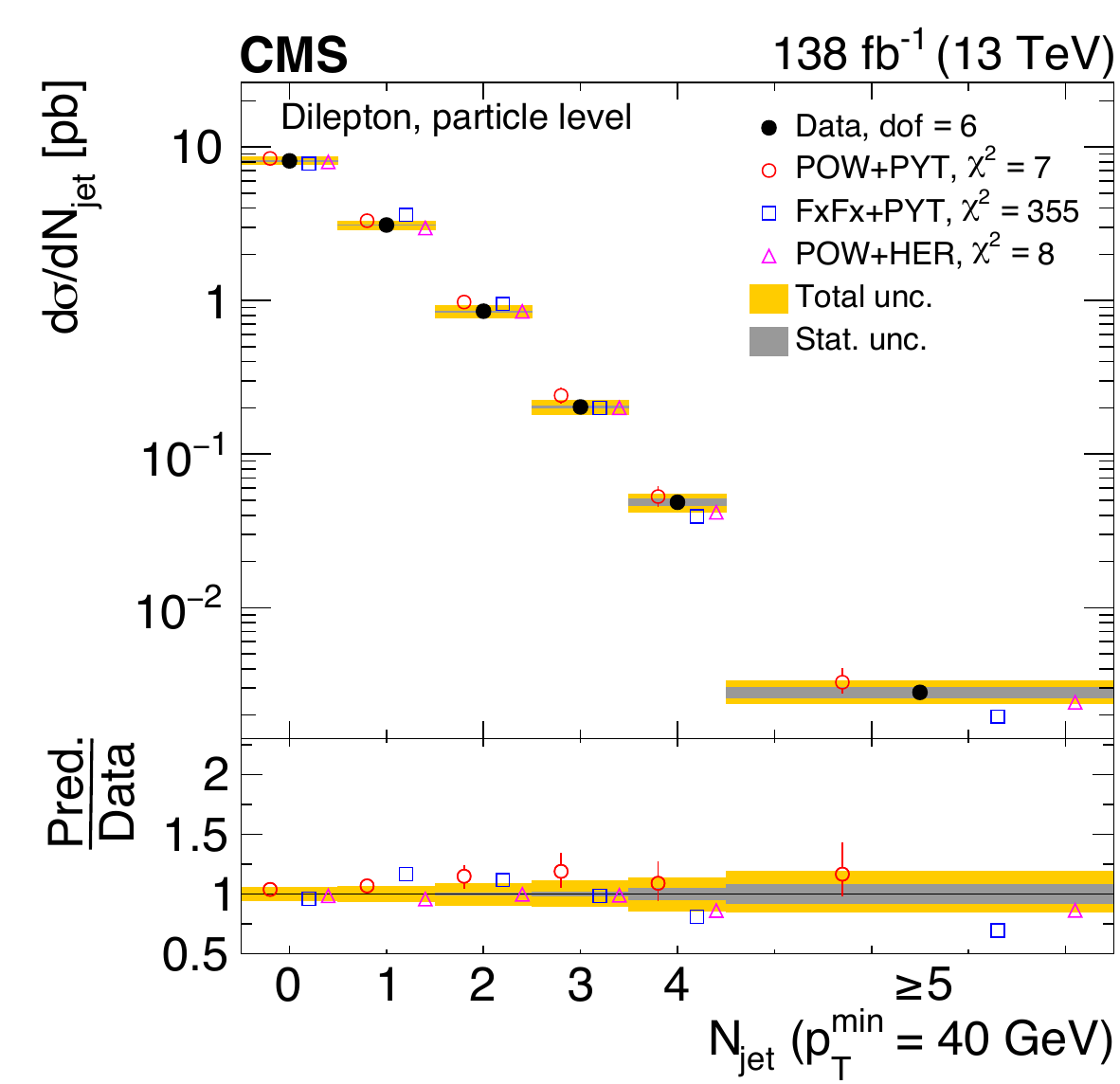}
\includegraphics[width=0.49\textwidth]{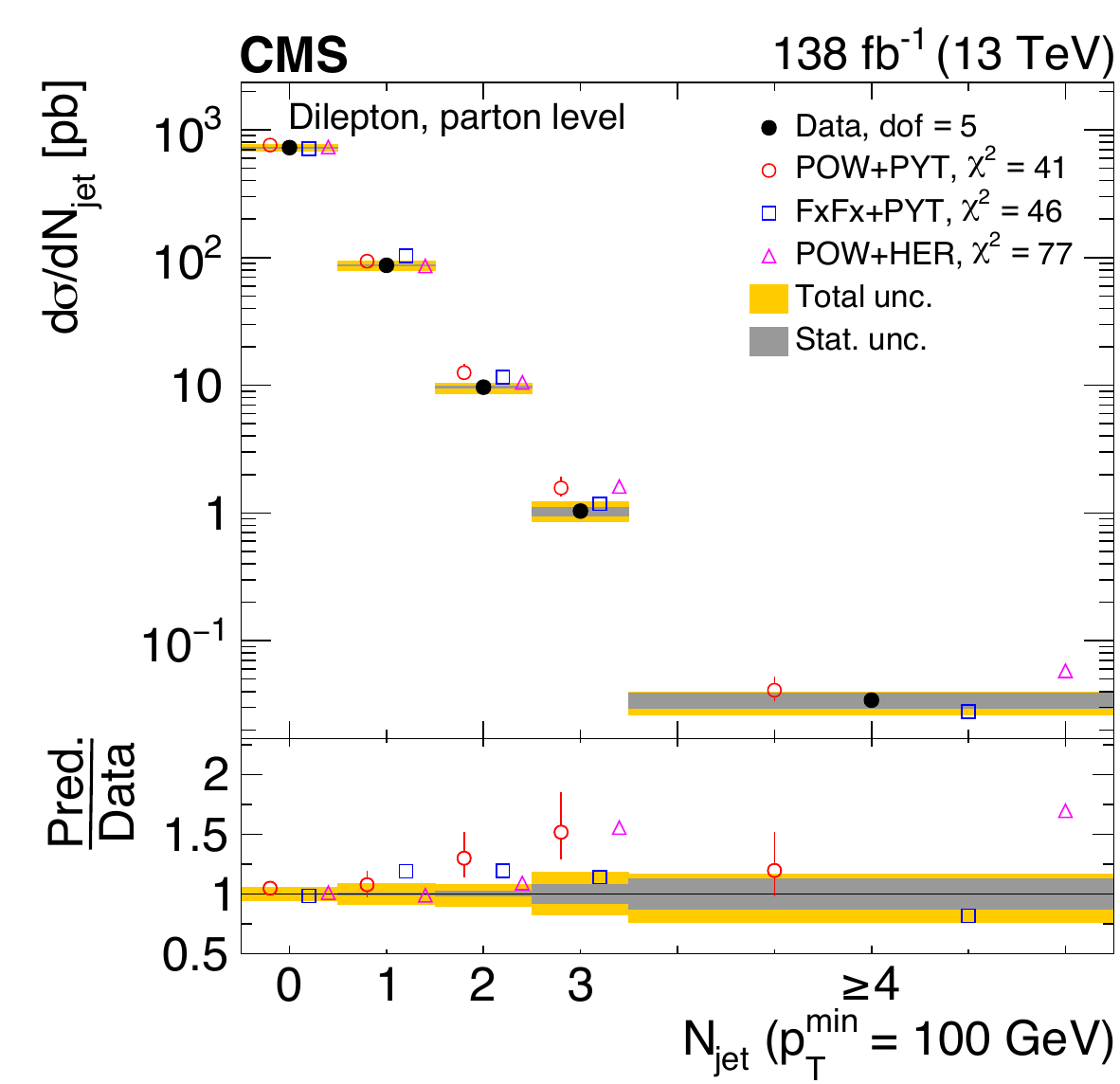}
\includegraphics[width=0.49\textwidth]{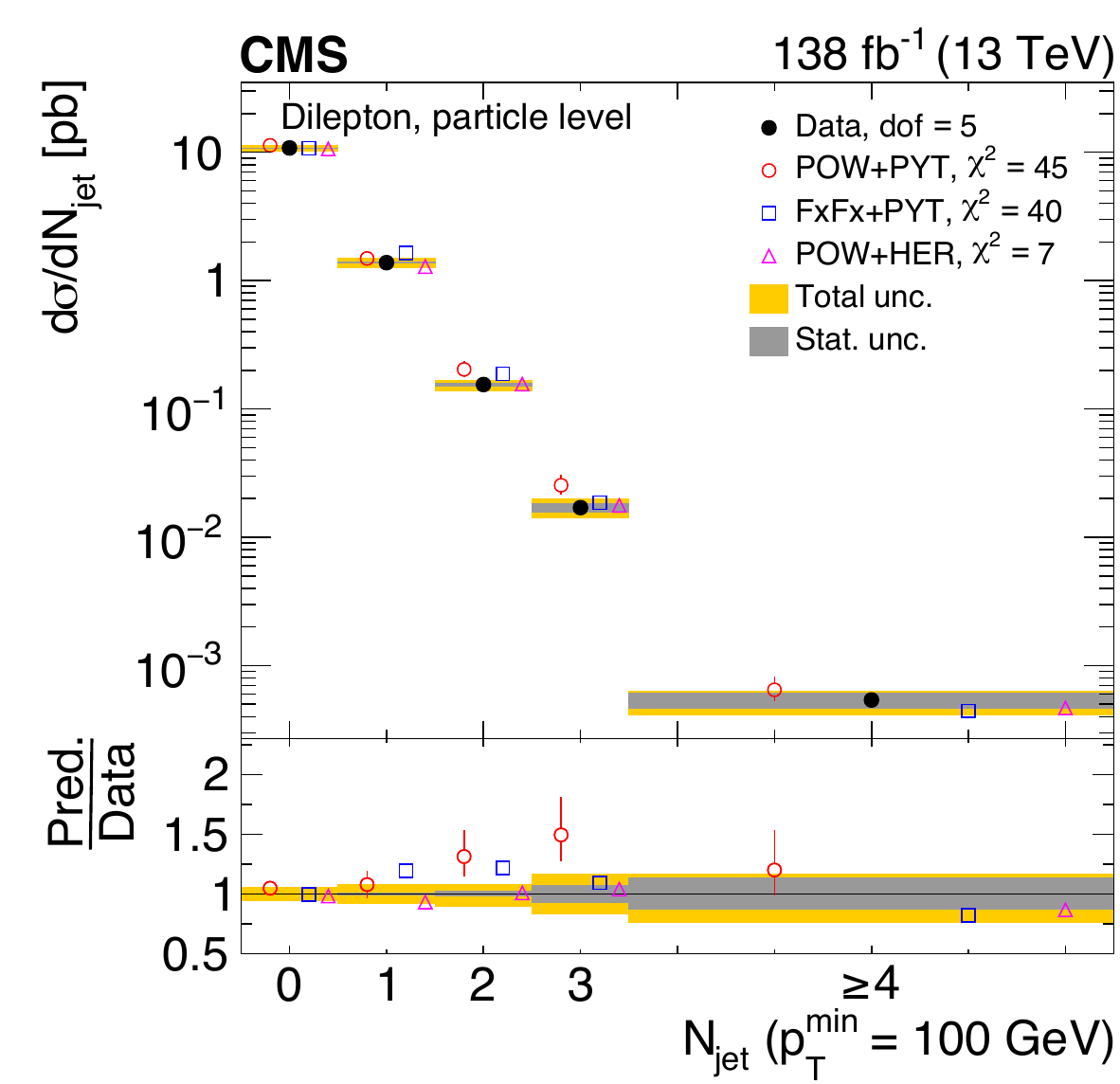}
\caption{
Absolute differential \ttbar production cross sections as a function of \nj, for a minimum jet \pt of 40\GeV
(upper) and 100\GeV (lower), measured at the parton level in the full phase space (left) and at the particle level in a
fiducial phase space (right).
The data are shown as filled circles with grey and yellow bands indicating the statistical and total
uncertainties (statistical and systematic uncertainties added in quadrature), respectively.
For each distribution, the number of degrees of freedom (dof) is also provided.
The cross sections are compared to various MC predictions (other points).
The estimated uncertainties in the \PowPyt (`POW-PYT') simulation are represented by vertical
bars on the corresponding points.
For each MC model, a value of \chisq is reported that takes into account the measurement uncertainties.
The lower panel in each plot shows the ratios of the predictions to the data.}
\label{fig:res_nj40_abs}
\end{figure}

\begin{figure}
\centering
\includegraphics[width=0.99\textwidth]{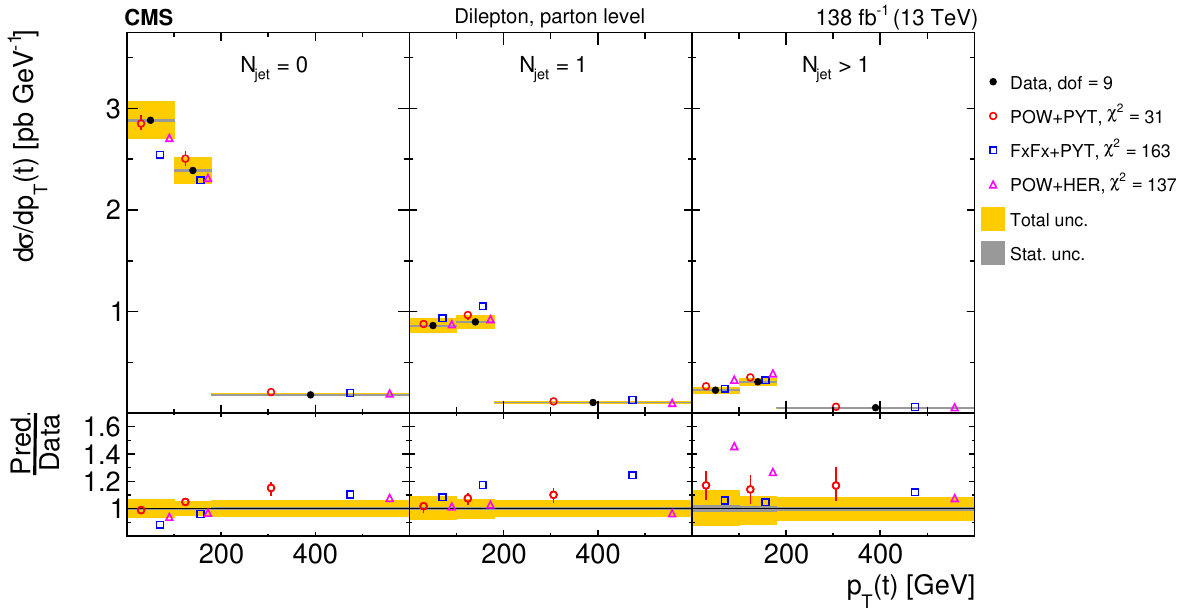}
\includegraphics[width=0.99\textwidth]{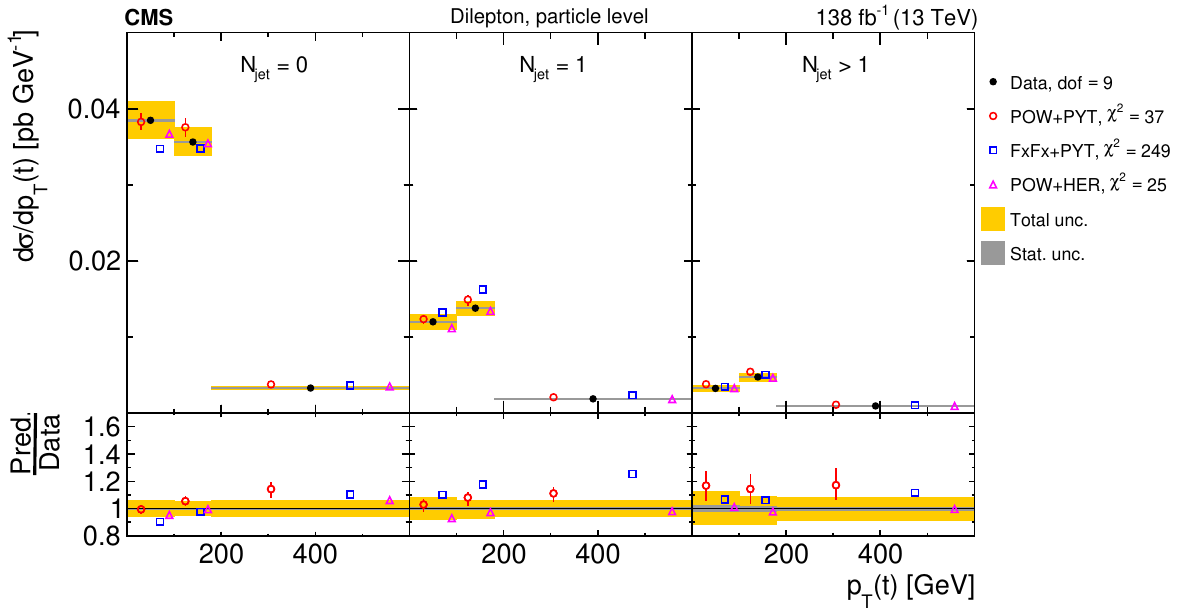}
\caption{Absolute \njptt cross sections measured at the parton level in the full phase space (upper) and
at the particle level in a fiducial phase space (lower). The data are shown as filled circles with grey and
yellow bands
indicating the statistical and total (sum in quadrature of statistical and systematic) uncertainties, respectively.
For each distribution, the number of degrees of freedom (dof) is also provided.
The cross sections are compared to various MC predictions (other points). The estimated uncertainties in the \PowPyt
(`POW-PYT') simulation
are represented by vertical bars on the corresponding points.
For each MC model, a value of \chisq is reported that takes into account the measurement uncertainties.
The lower panel in each plot shows the ratios of the predictions to the data.}
\label{fig:res_njptt_abs}
\end{figure}

\begin{figure}
\centering
\includegraphics[width=0.99\textwidth]{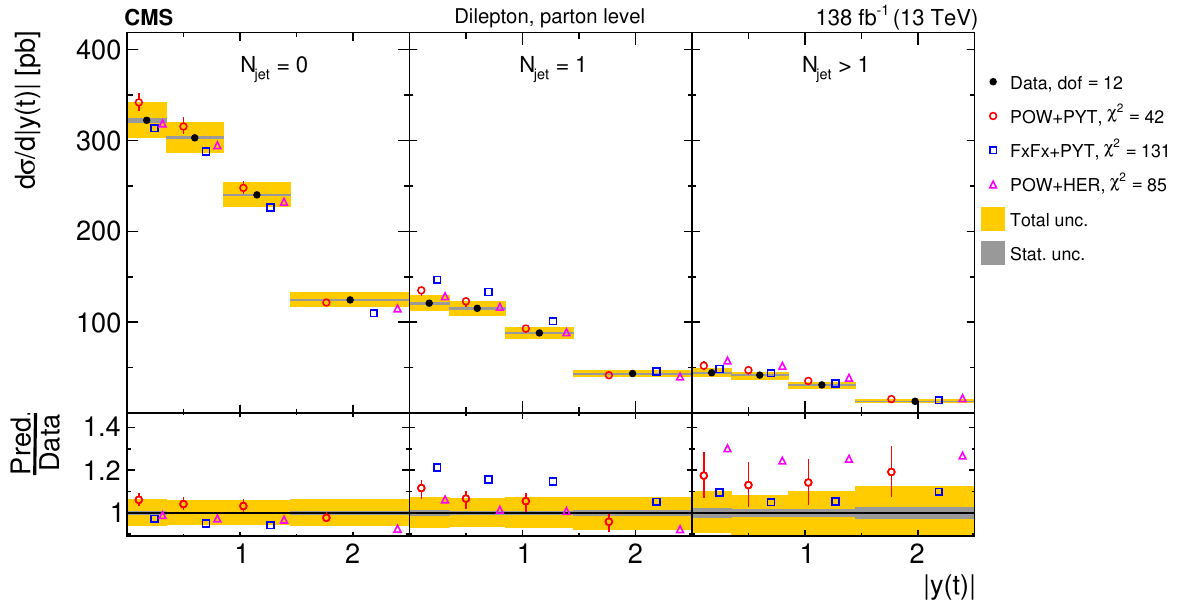}
\includegraphics[width=0.99\textwidth]{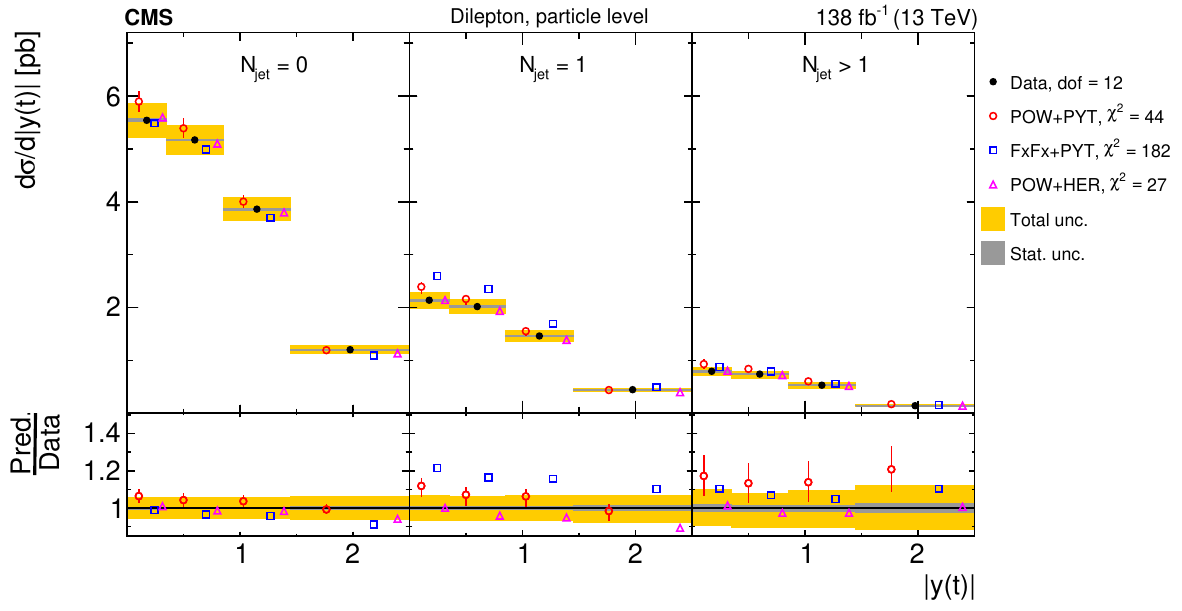}
\caption{Absolute \njyt cross sections are shown for data (filled circles) and various MC predictions
(other points).
Further details can be found in the caption of Fig.~\ref{fig:res_njptt_abs}.}
\label{fig:res_njyt_abs}
\end{figure}

\begin{figure}
\centering
\includegraphics[width=0.99\textwidth]{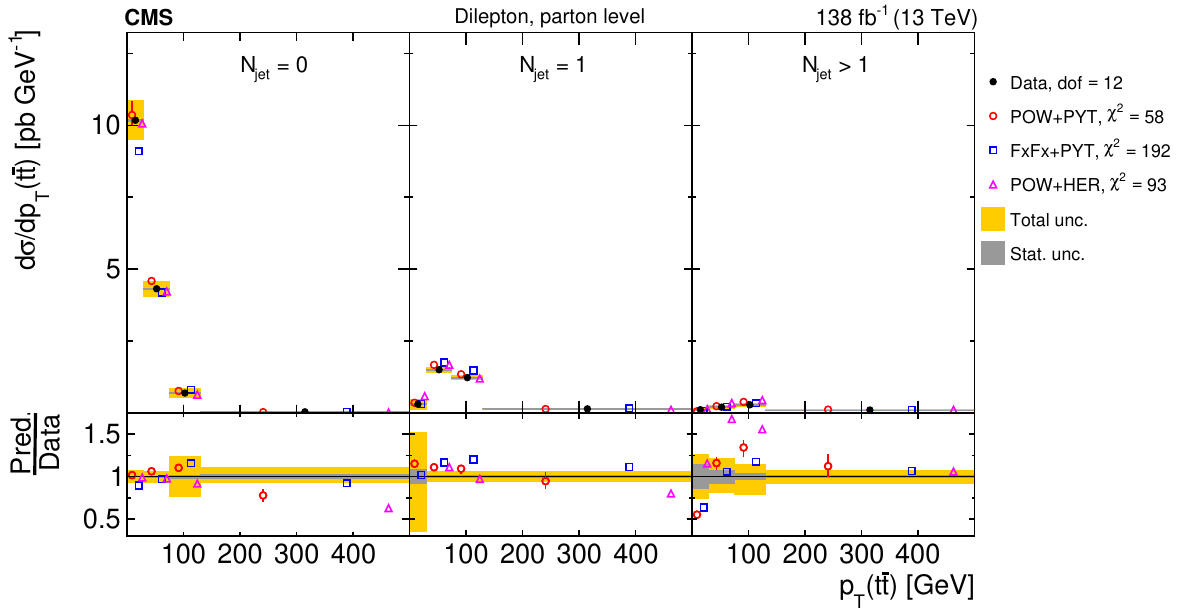}
\includegraphics[width=0.99\textwidth]{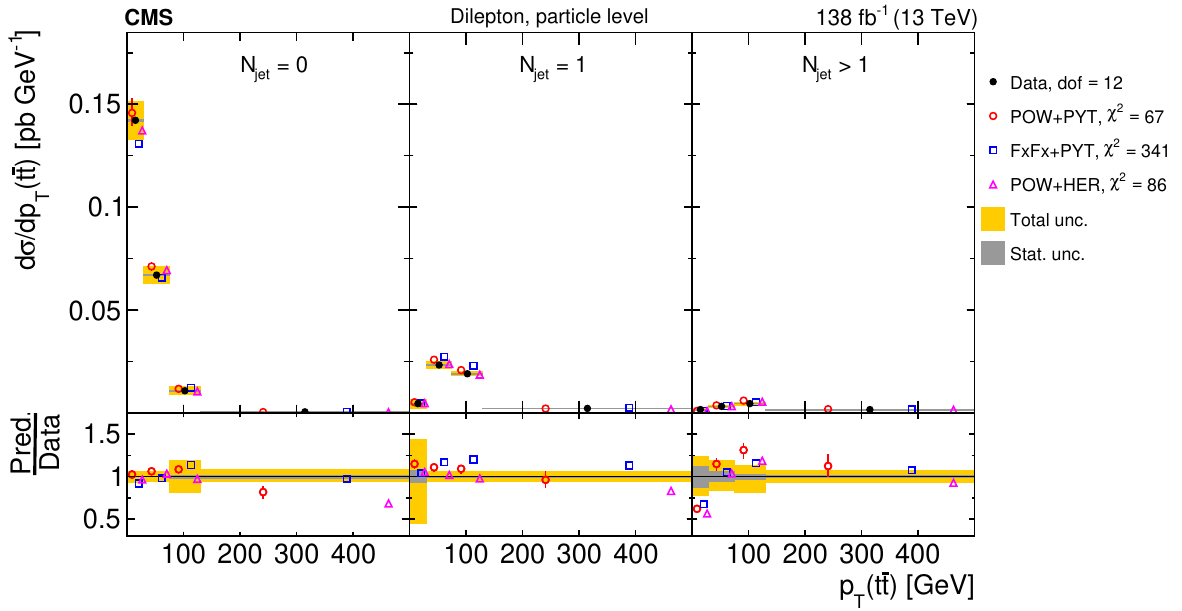}
\caption{Absolute $[\nj,\pttt]$
    cross sections are shown for data (filled circles) and various MC predictions (other points).
    Further details can be found in the caption of Fig.~\ref{fig:res_njptt_abs}.}
\label{fig:res_njpttt_abs}
\end{figure}

\begin{figure}
\centering
\includegraphics[width=0.99\textwidth]{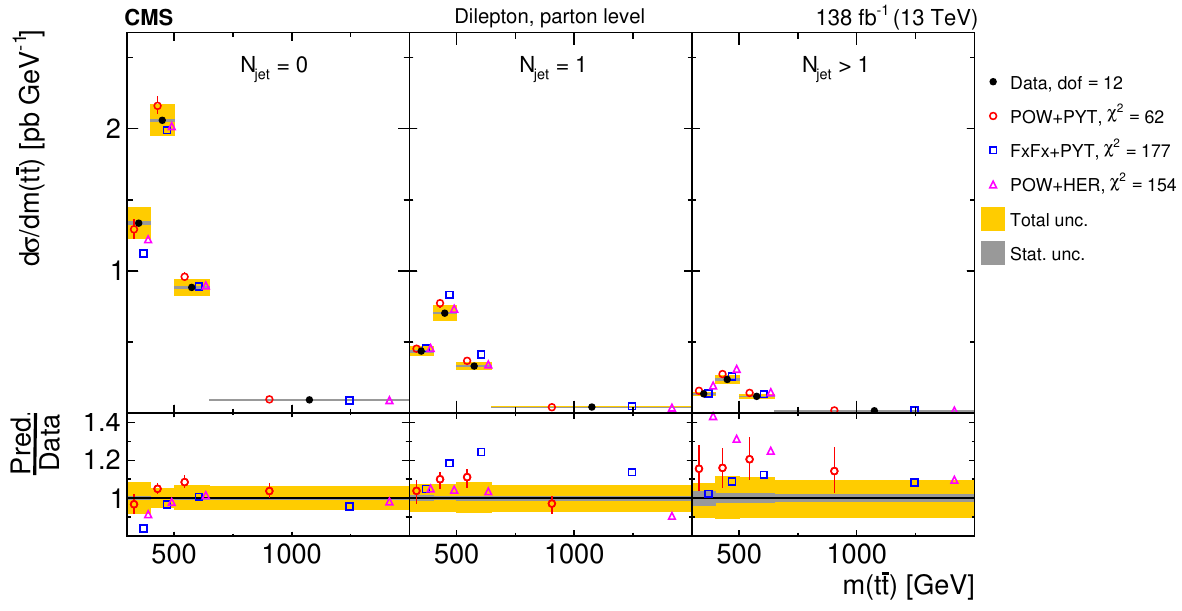}
\includegraphics[width=0.99\textwidth]{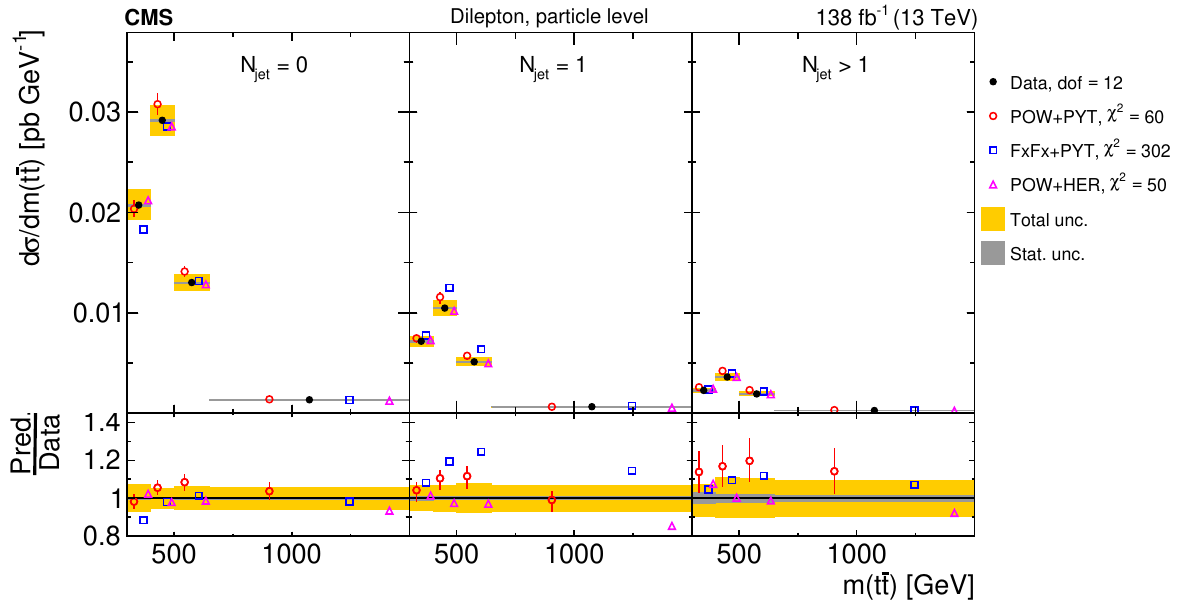}
\caption{Absolute \njmtt cross sections are shown for data (filled circles) and various MC predictions
(other points).
Further details can be found in the caption of Fig.~\ref{fig:res_njptt_abs}.}
\label{fig:res_njmtt_abs}
\end{figure}

\begin{figure}
\centering
\includegraphics[width=0.99\textwidth]{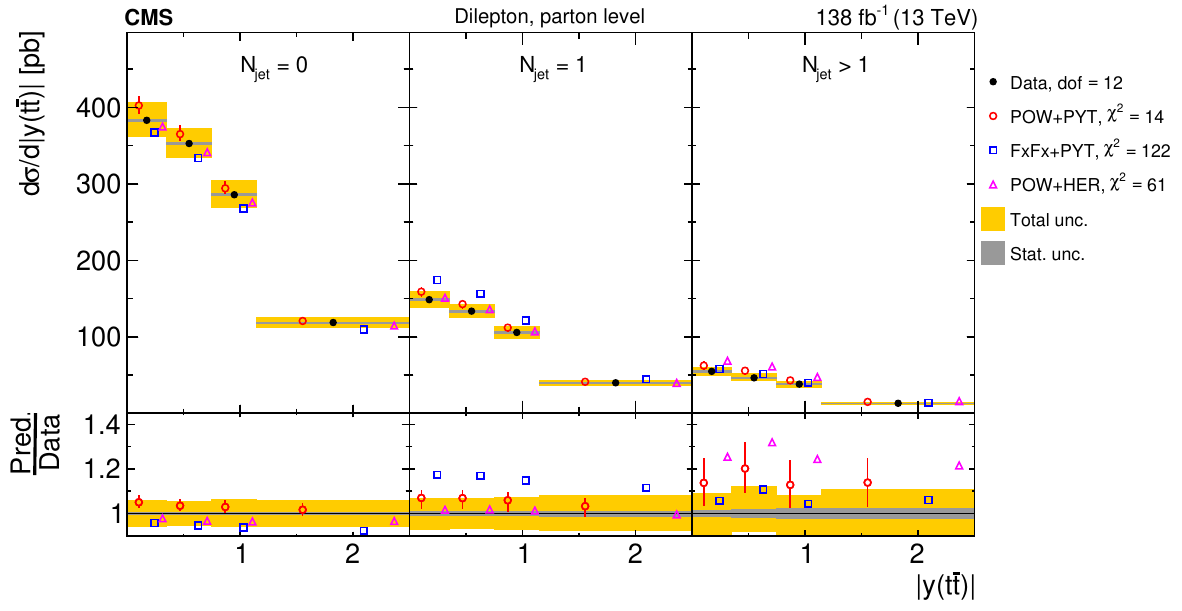}
\includegraphics[width=0.99\textwidth]{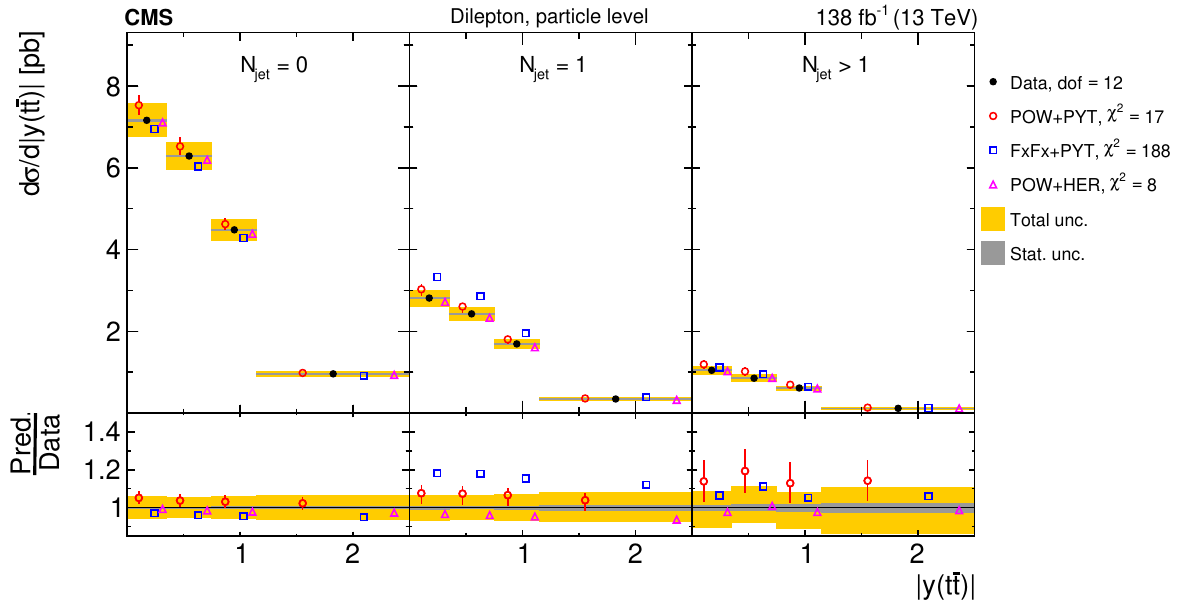}
\caption{Absolute \njytt cross sections are shown for data (filled circles) and various MC predictions
(other points).
Further details can be found in the caption of Fig.~\ref{fig:res_njptt_abs}.}
\label{fig:res_njytt_abs}
\end{figure}

\begin{figure}
\centering
\includegraphics[width=0.99\textwidth]{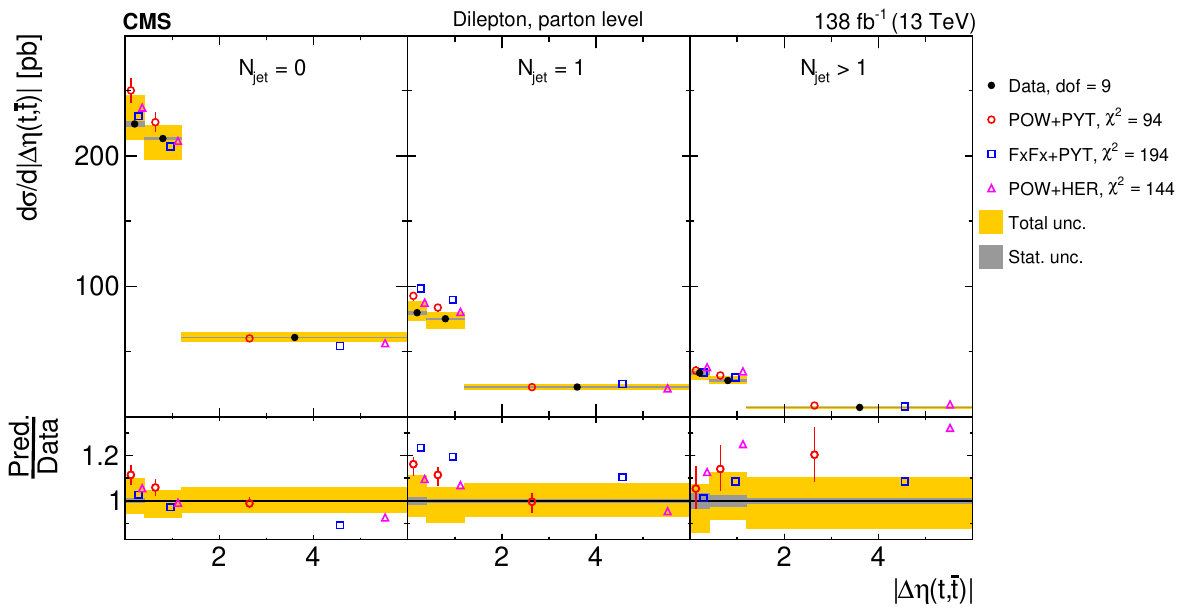}
\includegraphics[width=0.99\textwidth]{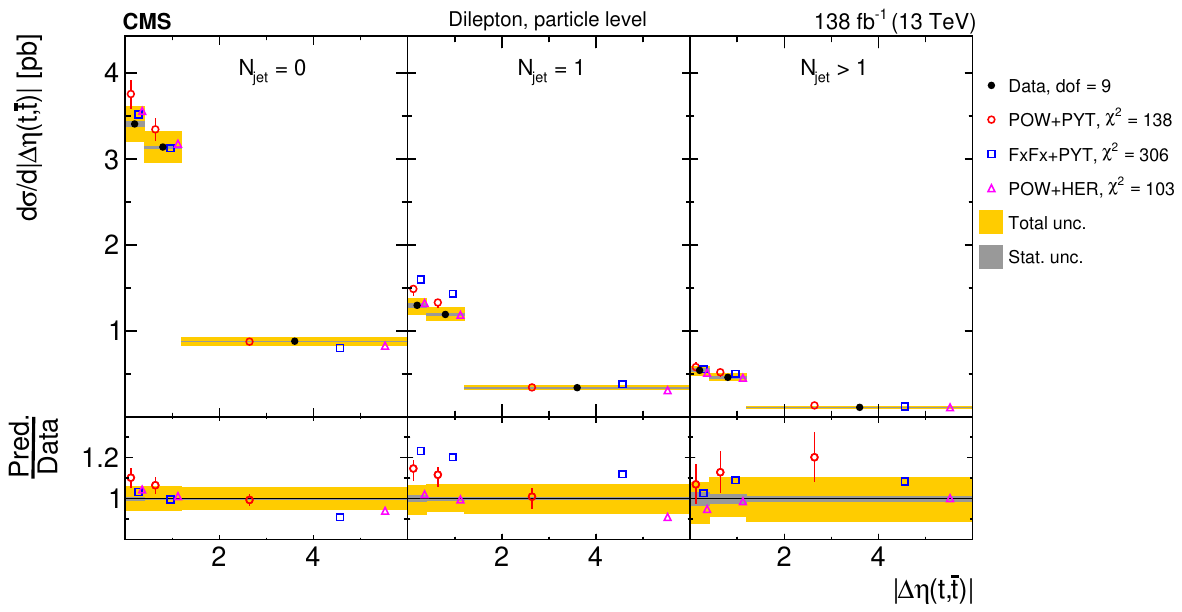}
\caption{Absolute \njdetatt cross sections are shown for data (filled circles) and various MC predictions 
(other points).
Further details can be found in the caption of Fig.~\ref{fig:res_njptt_abs}.}
\label{fig:res_njdeta_abs}
\end{figure}

\begin{figure}
\centering
\includegraphics[width=1.00\textwidth]{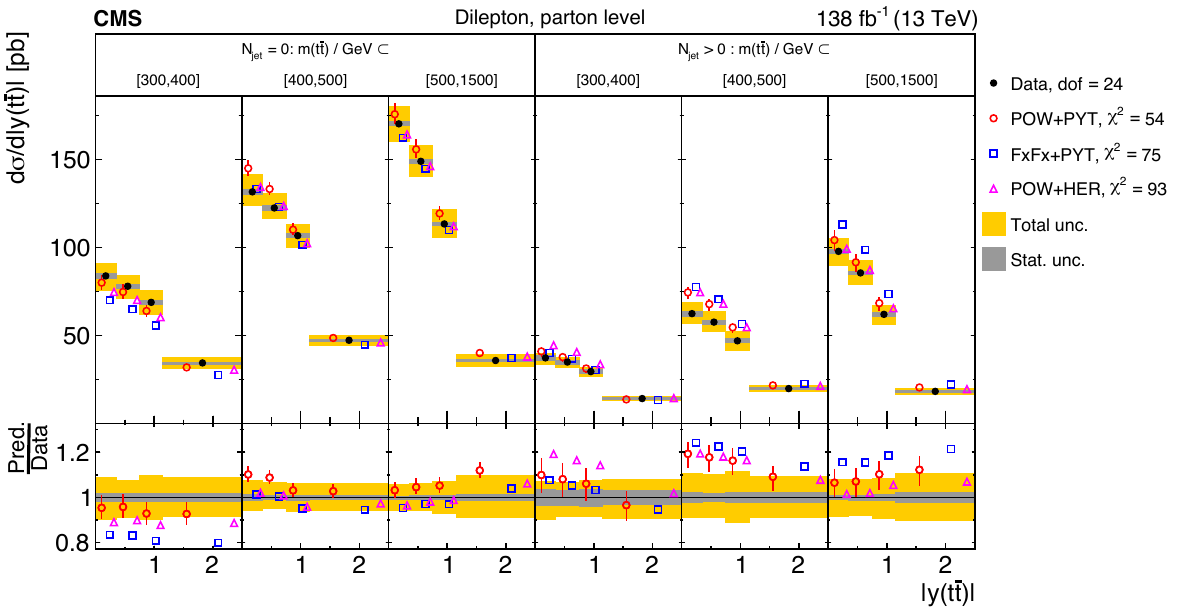}
\includegraphics[width=1.00\textwidth]{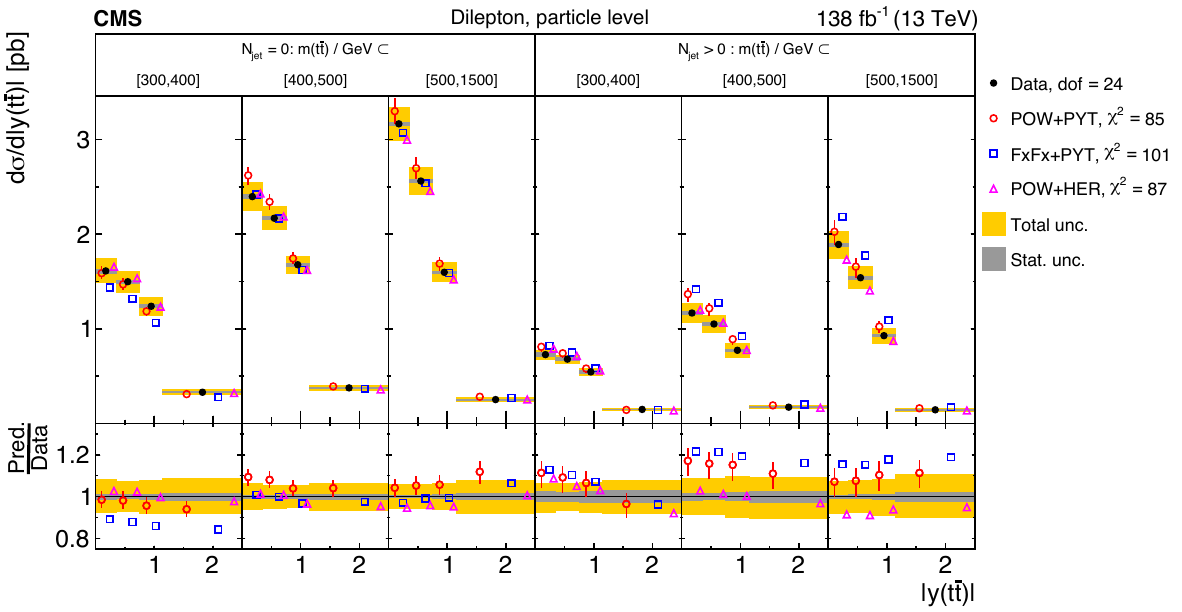}
\caption{Absolute \njmttytttwo cross sections are shown for data (filled circles) and various MC predictions
(other points).
Further details can be found in the caption of Fig.~\ref{fig:res_njptt_abs}.}
\label{fig:res_nj2mttytt_abs}
\end{figure}

\begin{figure}
\centering
\includegraphics[width=1.00\textwidth]{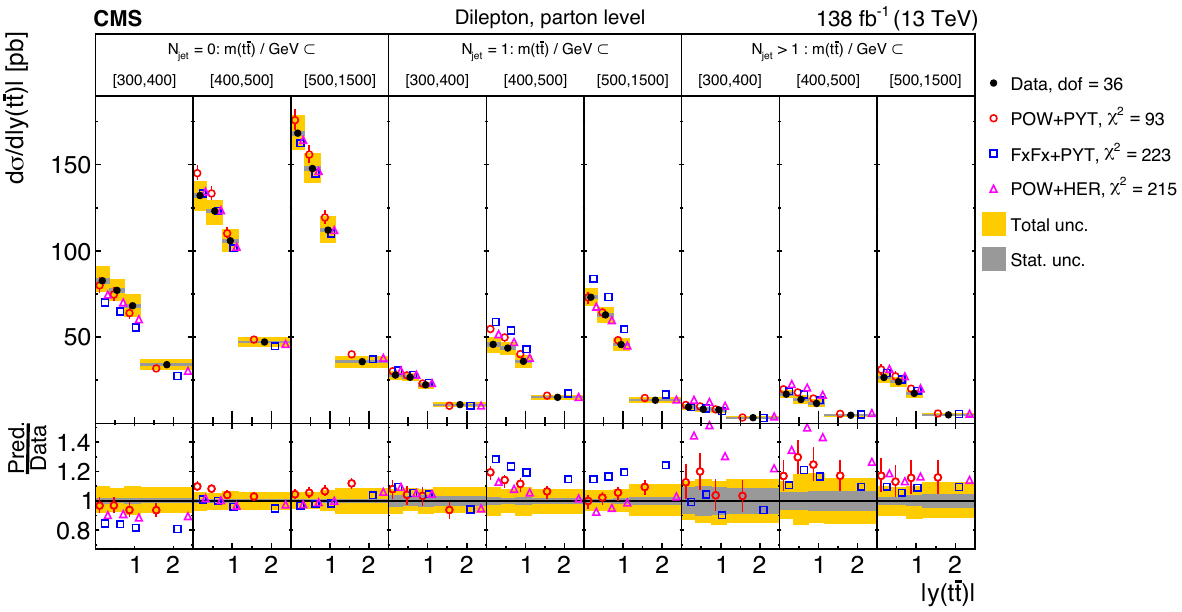}
\includegraphics[width=1.00\textwidth]{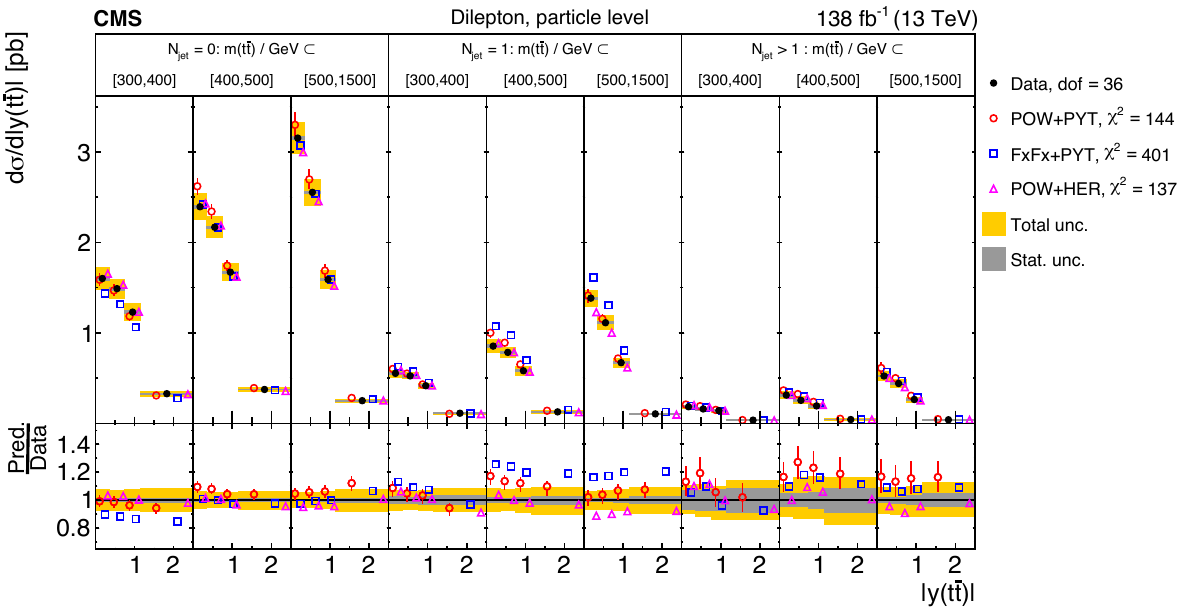}
\caption{Absolute \njmttyttthree cross sections are shown for data (filled circles) and various MC predictions
(other points).
Further details can be found in the caption of Fig.~\ref{fig:res_njptt_abs}.}
\label{fig:res_nj3mttytt_abs}
\end{figure}

\begin{figure}
\centering
\includegraphics[width=1.00\textwidth]{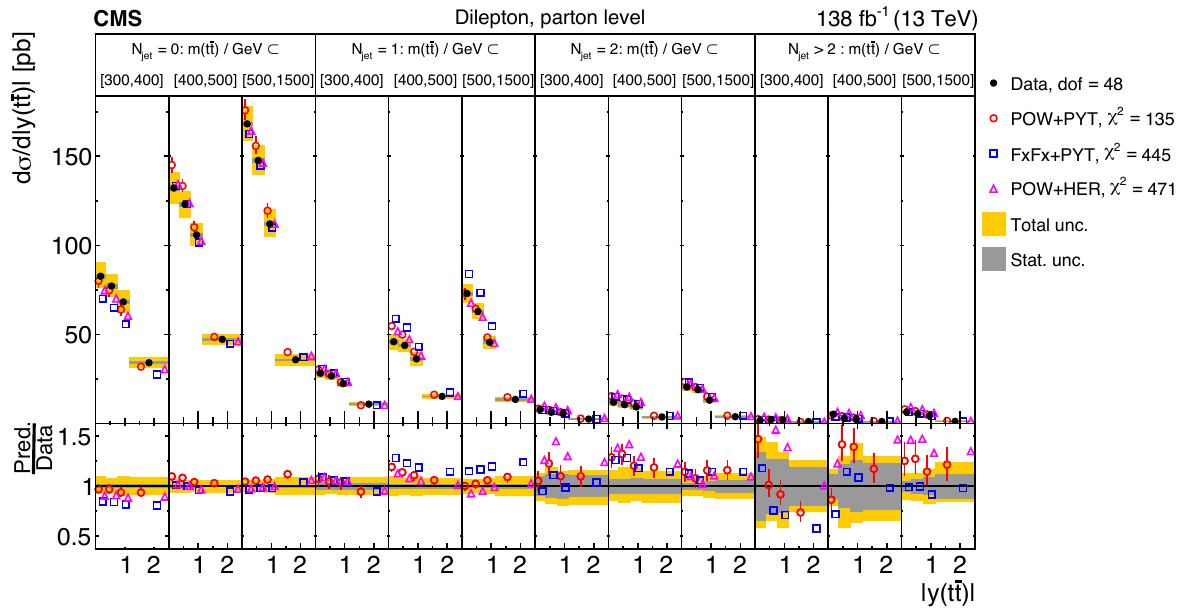}
\includegraphics[width=1.00\textwidth]{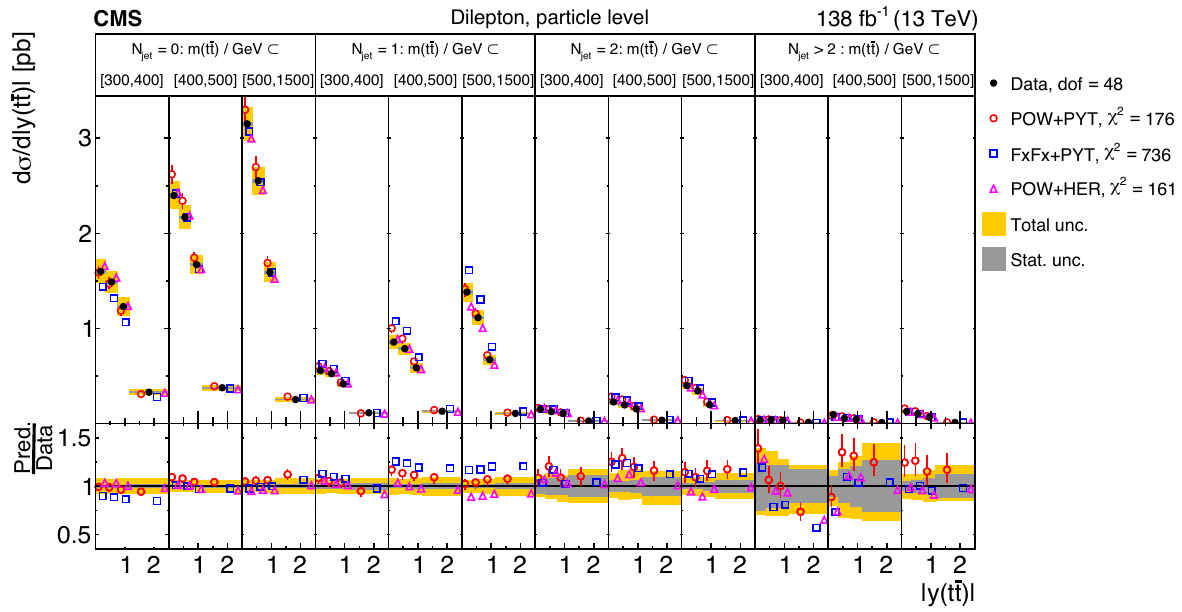}
\caption{Absolute \njmttyttfour cross sections are shown for data (filled circles) and various MC predictions
(other points).
Further details can be found in the caption of Fig.~\ref{fig:res_njptt_abs}.}
\label{fig:res_nj4mttytt_abs}
\end{figure}

The absolute differential cross sections comparing data to predictions based on MC
simulations are shown in Figs.~\ref{fig:res_ptt_abs}--\ref{fig:res_nj4mttytt_abs}, and the
corresponding \chisq values are presented in Tables~\ref{tab:chi2mc_1d_abs_parton}--\ref{tab:chi2mc_abs_particle_addjets}. The $p$-values of the \chisq tests are provided in Tables~\ref{tab:pvaluemc_1d_abs_parton}--\ref{tab:pvaluemc_abs_particle_addjets}.
The measurements are compared to the three MC simulations introduced in Section~\ref{sec:simulation}:
\PowPyt (\PowPytSh), \aMCPyt (\aMCPytSh), and 
\PowHer (\PowHerSh).
As detailed in Section~\ref{sec:simulation},
the \PowPyt simulation is normalized to the cross section calculated at NNLO+NNLL.
Comparing the absolute cross sections to the normalized ones presented in
Section~\ref{sec:res}, two effects are visible.
Firstly, the uncertainties in the measured absolute
cross sections are considerably larger than those of the corresponding normalized ones, which can be attributed to bin-to-bin correlated global normalization uncertainties.
Secondly, the absolute total cross section obtained by
integrating over the spectra of the three MC models is within 5\% of
that calculated from the data, indicating
a reasonable agreement.

\clearpage

\clearpage

\begin{table*}
\centering
 \topcaption{The \chisq values and \ndf of the measured absolute single-differential cross sections for \ttbar and top quark kinematic observables at the parton level are shown with respect to the predictions of various MC generators. 
The \chisq values are calculated taking only measurement uncertainties into account and excluding theory uncertainties.  
For \PowPytSh, the \chisq values including theory uncertainties are indicated with the brackets (w. unc.).}
 \label{tab:chi2mc_1d_abs_parton}
 \renewcommand{\arraystretch}{1.4}
 \centering
 \begin{tabular}{lccccc}
 \multirow{1}{*}{Cross section} & \hspace*{0.3 cm} \multirow{2}{*}{\ndf} \hspace*{0.3 cm} & \multicolumn{3}{c}{\chisq} \\
 \cline{3-5}
{variables} && \PowPytSh (w. unc.)  & \aMCPytSh  & \PowHerSh   \\
\hline
\ptt& 7 & 21 \: (13) & 43 & 5 \\
\ptat& 7 & 19 \: (12) & 43 & 6 \\
\yt& 10 & 28 \: (24) & 34 & 21 \\
\yat& 10 & 33 \: (28) & 40 & 25 \\
\pttt& 7 & 24 \: (8) & 39 & 35 \\
\ytt& 12 & 13 \: (9) & 19 & 8 \\
\mtt  & 7 & 6 \: (4) & 9 & 4 \\
\dphitt& 4 & 4 \: (2) & 7 & 11 \\
\dytt& 8 & 18 \: (10) & 19 & 14 \\
\rpttmtt& 5 & 39 \: (21) & 104 & 13 \\
\rptttmtt& 9 & 20 \: (7) & 32 & 39 \\
\logxone& 9 & 16 \: (12) & 18 & 12 \\
\logxtwo& 9 & 14 \: (9) & 24 & 7 \\
 \end{tabular} 
\end{table*}

\begin{table*}
\centering
 \topcaption{The \chisq values and \ndf of the measured absolute single-differential cross sections for \ttbar and top quark kinematic observables at the particle level are shown with respect to the predictions of various MC generators. 
The \chisq values are calculated taking only measurement uncertainties into account and excluding theory uncertainties.  
For \PowPytSh, the \chisq values including theory uncertainties are indicated with the brackets (w. unc.).}
 \label{tab:chi2mc_1d_abs_particle_ttbar}
 \renewcommand{\arraystretch}{1.4}
 \centering
 \begin{tabular}{lccccc}
 \multirow{1}{*}{Cross section} & \hspace*{0.3 cm} \multirow{2}{*}{\ndf} \hspace*{0.3 cm} & \multicolumn{3}{c}{\chisq} \\
 \cline{3-5}
{variables} && \PowPytSh (w. unc.)  & \aMCPytSh  & \PowHerSh   \\
\hline
\ptt& 7 & 22 \: (13) & 44 & 7 \\
\ptat& 7 & 20 \: (12) & 44 & 5 \\
\yt& 10 & 24 \: (18) & 32 & 19 \\
\yat& 10 & 28 \: (23) & 32 & 26 \\
\pttt& 7 & 23 \: (8) & 34 & 41 \\
\ytt& 12 & 13 \: (8) & 23 & 9 \\
\mtt  & 7 & 7 \: (4) & 5 & 7 \\
\dphitt& 4 & 4 \: (1) & 4 & 7 \\
\dytt& 8 & 17 \: (11) & 15 & 18 \\
\rpttmtt& 5 & 33 \: (23) & 71 & 14 \\
\rptttmtt& 9 & 21 \: (7) & 46 & 61 \\
\logxone& 9 & 16 \: (10) & 17 & 19 \\
\logxtwo& 9 & 12 \: (7) & 19 & 10 \\
 \end{tabular}
\end{table*}

\begin{table*}
 \centering
 \topcaption{The \chisq values and \ndf of the measured absolute multi-differential cross sections for \ttbar and top quark kinematic observables at the parton level are shown with respect to the predictions of various MC generators. 
The \chisq values are calculated taking only measurement uncertainties into account and excluding theory uncertainties.  
For \PowPytSh, the \chisq values including theory uncertainties are indicated with the brackets (w. unc.).}
 \label{tab:chi2mc_abs_parton_ttbar}
 \renewcommand{\arraystretch}{1.4}
 \centering
 \begin{tabular}{lccccc}
 \multirow{1}{*}{Cross section} & \hspace*{0.3 cm} \multirow{2}{*}{\ndf} \hspace*{0.3 cm} & \multicolumn{3}{c}{\chisq} \\
 \cline{3-5}
{variables} && \PowPytSh (w. unc.)  & \aMCPytSh  & \PowHerSh   \\
\hline
\ytptt& 16 & 48 \: (36) & 75 & 30 \\
\mttptt& 9 & 93 \: (36) & 156 & 42 \\
\pttpttt& 16 & 50 \: (25) & 72 & 87 \\
\mttytt  & 16 & 72 \: (46) & 67 & 65 \\
\yttpttt& 16 & 32 \: (17) & 37 & 71 \\
\mttpttt  & 16 & 68 \: (47) & 77 & 115 \\
\ptttmttytt  & 48 & 102 \: (71) & 119 & 140 \\
\mttyt& 16 & 67 \: (39) & 84 & 49 \\
\mttdetatt& 12 & 182 \: (34) & 236 & 125 \\
\mttdphitt& 12 & 82 \: (51) & 50 & 93 \\
 \end{tabular}
\end{table*}

\begin{table*}
 \centering
 \topcaption{The \chisq values and \ndf of the measured absolute multi-differential cross sections for \ttbar and top quark kinematic observables at the particle level are shown with respect to the predictions of various MC generators. 
The \chisq values are calculated taking only measurement uncertainties into account and excluding theory uncertainties.  
For \PowPytSh, the \chisq values including theory uncertainties are indicated with the brackets (w. unc.).}
 \label{tab:chi2mc_abs_particle_ttbar}
 \renewcommand{\arraystretch}{1.4}
 \centering
 \begin{tabular}{lccccc}
 \multirow{1}{*}{Cross section} & \hspace*{0.3 cm} \multirow{2}{*}{\ndf} \hspace*{0.3 cm} & \multicolumn{3}{c}{\chisq} \\
 \cline{3-5}
{variables} && \PowPytSh (w. unc.)  & \aMCPytSh  & \PowHerSh   \\
\hline
\ytptt& 16 & 44 \: (28) & 68 & 27 \\
\mttptt& 9 & 103 \: (37) & 151 & 46 \\
\pttpttt& 16 & 44 \: (21) & 68 & 64 \\
\mttytt  & 16 & 86 \: (41) & 77 & 81 \\
\yttpttt& 16 & 32 \: (19) & 41 & 66 \\
\mttpttt  & 16 & 69 \: (37) & 57 & 112 \\
\ptttmttytt  & 48 & 133 \: (69) & 130 & 170 \\
\mttyt& 16 & 64 \: (27) & 75 & 37 \\
\mttdetatt& 12 & 174 \: (32) & 220 & 114 \\
\mttdphitt& 12 & 80 \: (44) & 41 & 98 \\
 \end{tabular}
\end{table*}

\begin{table*}
\centering
 \topcaption{The \chisq values and \ndf of the measured absolute single-differential cross sections for lepton and \PQb-jet kinematic observables at the particle level are shown with respect to the predictions of various MC generators. 
The \chisq values are calculated taking only measurement uncertainties into account and excluding theory uncertainties.  
For \PowPytSh, the \chisq values including theory uncertainties are indicated with the brackets (w. unc.).}
 \label{tab:chi2mc_1d_abs_particle_lepb}
 \renewcommand{\arraystretch}{1.4}
 \centering
 \begin{tabular}{lccccc}
 \multirow{1}{*}{Cross section} & \hspace*{0.3 cm} \multirow{2}{*}{\ndf} \hspace*{0.3 cm} & \multicolumn{3}{c}{\chisq} \\
 \cline{3-5}
{variables} && \PowPytSh (w. unc.)  & \aMCPytSh  & \PowHerSh   \\
\hline
\ptlep& 12 & 32 \: (19) & 62 & 21 \\
\ptlep trailing/\ptlep leading& 10 & 16 \: (11) & 27 & 7 \\
\ptlep/\ptat& 5 & 20 \: (17) & 28 & 14 \\
\ptb leading& 10 & 6 \: (5) & 31 & 8 \\
\ptb trailing& 7 & 7 \: (5) & 26 & 7 \\
\rptbsptts& 4 & 24 \: (19) & 30 & 21 \\
\mll& 12 & 31 \: (25) & 29 & 23 \\
\mbb& 7 & 21 \: (16) & 17 & 15 \\
\mllbb& 19 & 36 \: (19) & 30 & 27 \\
\ptll& 9 & 4 \: (3) & 17 & 10 \\
\absetall& 14 & 16 \: (10) & 22 & 12 \\
\etallmll& 24 & 55 \: (29) & 76 & 35 \\
\etallptll& 20 & 30 \: (15) & 84 & 24 \\
\ptllmll& 30 & 50 \: (39) & 88 & 52 \\
 \end{tabular}
\end{table*}

\begin{table*}
 \centering
 \topcaption{The \chisq values and \ndf of the measured absolute differential cross sections as a function of the additional-jet multiplicity in the events, at the parton level of the top quark and antiquark, are shown with respect to the predictions of various MC generators. 
The \chisq values are calculated taking only measurement uncertainties into account and excluding theory uncertainties.  
For \PowPytSh, the \chisq values including theory uncertainties are indicated with the brackets (w. unc.).}
 \label{tab:chi2mc_abs_parton_addjets}
 \renewcommand{\arraystretch}{1.4}
 \centering
 \begin{tabular}{lccccc}
 \multirow{1}{*}{Cross section} & \hspace*{0.3 cm} \multirow{2}{*}{\ndf} \hspace*{0.3 cm} & \multicolumn{3}{c}{\chisq} \\
 \cline{3-5}
{variables} && \PowPytSh (w. unc.)  & \aMCPytSh  & \PowHerSh   \\
\hline
\njforty & 6 & 7 \: (5) & 288 & 258 \\
\njhundred & 5 & 41 \: (11) & 46 & 77 \\
\njptt& 9 & 31 \: (17) & 163 & 137 \\
\njyt& 12 & 42 \: (32) & 131 & 85 \\
\njpttt& 12 & 58 \: (43) & 192 & 93 \\
\njmtt  & 12 & 62 \: (48) & 177 & 154 \\
\njytt& 12 & 14 \: (7) & 122 & 61 \\
\njdetatt& 9 & 94 \: (40) & 194 & 144 \\
\njmttytttwo  & 24 & 54 \: (39) & 75 & 93 \\
\njmttyttthree  & 36 & 93 \: (63) & 223 & 215 \\
\njmttyttfour  & 48 & 135 \: (92) & 445 & 471 \\
 \end{tabular}
\end{table*}

\begin{table*}
 \centering
 \topcaption{The \chisq values and \ndf of the measured absolute differential cross sections as a function of the additional-jet multiplicity in the events, at the particle level of the top quark and antiquark, are shown with respect to the predictions of various MC generators. 
The \chisq values are calculated taking only measurement uncertainties into account and excluding theory uncertainties.  
For \PowPytSh, the \chisq values including theory uncertainties are indicated with the brackets (w. unc.).}
 \label{tab:chi2mc_abs_particle_addjets}
 \renewcommand{\arraystretch}{1.4}
 \centering
 \begin{tabular}{lccccc}
 \multirow{1}{*}{Cross section} & \hspace*{0.3 cm} \multirow{2}{*}{\ndf} \hspace*{0.3 cm} & \multicolumn{3}{c}{\chisq} \\
 \cline{3-5}
{variables} && \PowPytSh (w. unc.)  & \aMCPytSh  & \PowHerSh   \\
\hline
\njforty & 6 & 7 \: (4) & 355 & 8 \\
\njhundred & 5 & 45 \: (11) & 40 & 7 \\
\njptt& 9 & 37 \: (15) & 249 & 25 \\
\njyt& 12 & 44 \: (26) & 182 & 27 \\
\njpttt& 12 & 67 \: (41) & 341 & 86 \\
\njmtt  & 12 & 60 \: (40) & 302 & 50 \\
\njytt& 12 & 17 \: (6) & 188 & 8 \\
\njdetatt& 9 & 138 \: (43) & 306 & 103 \\
\njmttytttwo  & 24 & 85 \: (46) & 101 & 87 \\
\njmttyttthree  & 36 & 144 \: (71) & 401 & 137 \\
\njmttyttfour  & 48 & 176 \: (97) & 736 & 161 \\
 \end{tabular}

\end{table*}

\clearpage

\subsection{Comparisons to higher-order theoretical predictions}
\label{sec:res_th_abs}

\begin{figure*}[!phtb]
\centering
\includegraphics[width=0.49\textwidth]{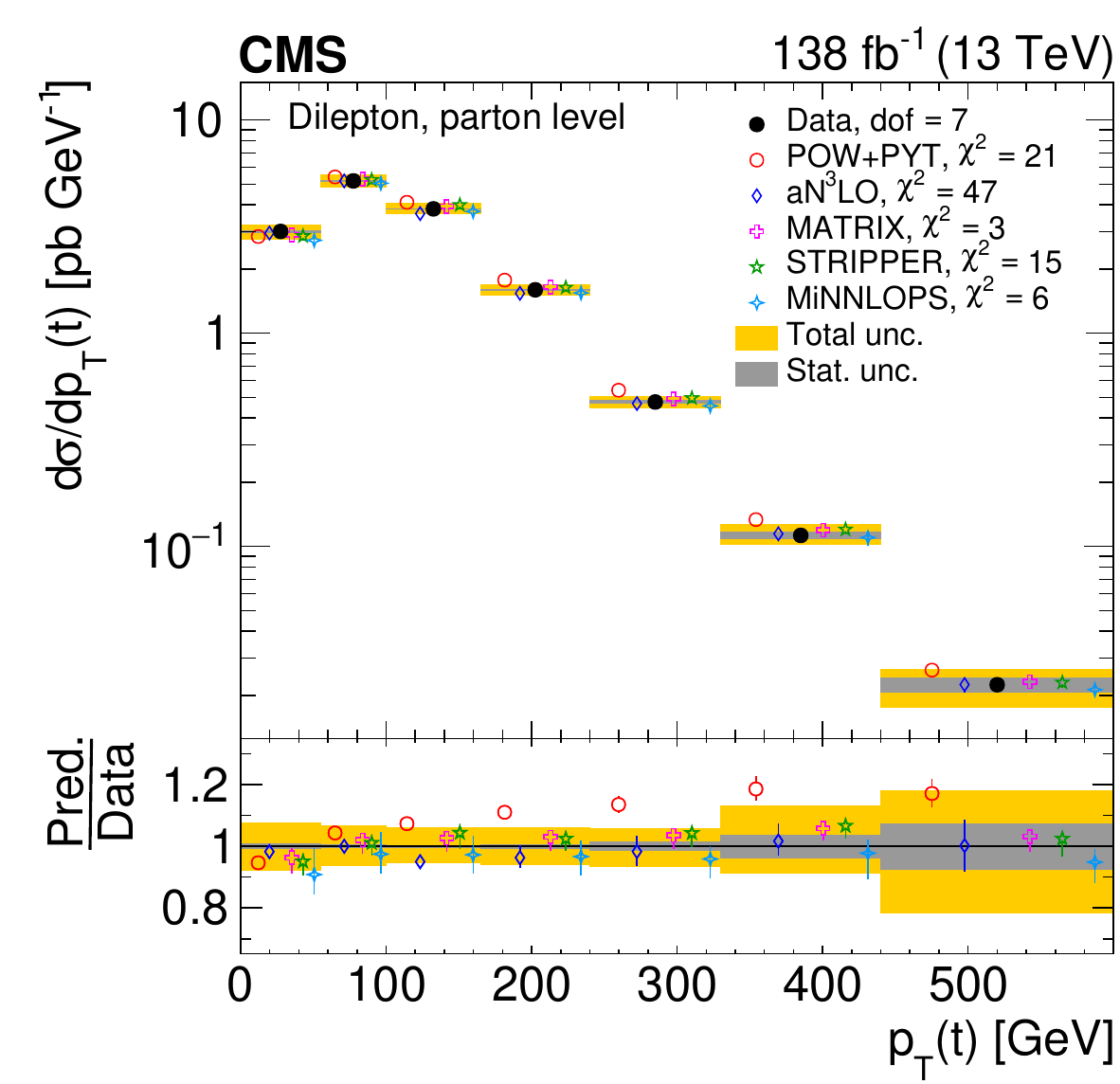}
\includegraphics[width=0.49\textwidth]{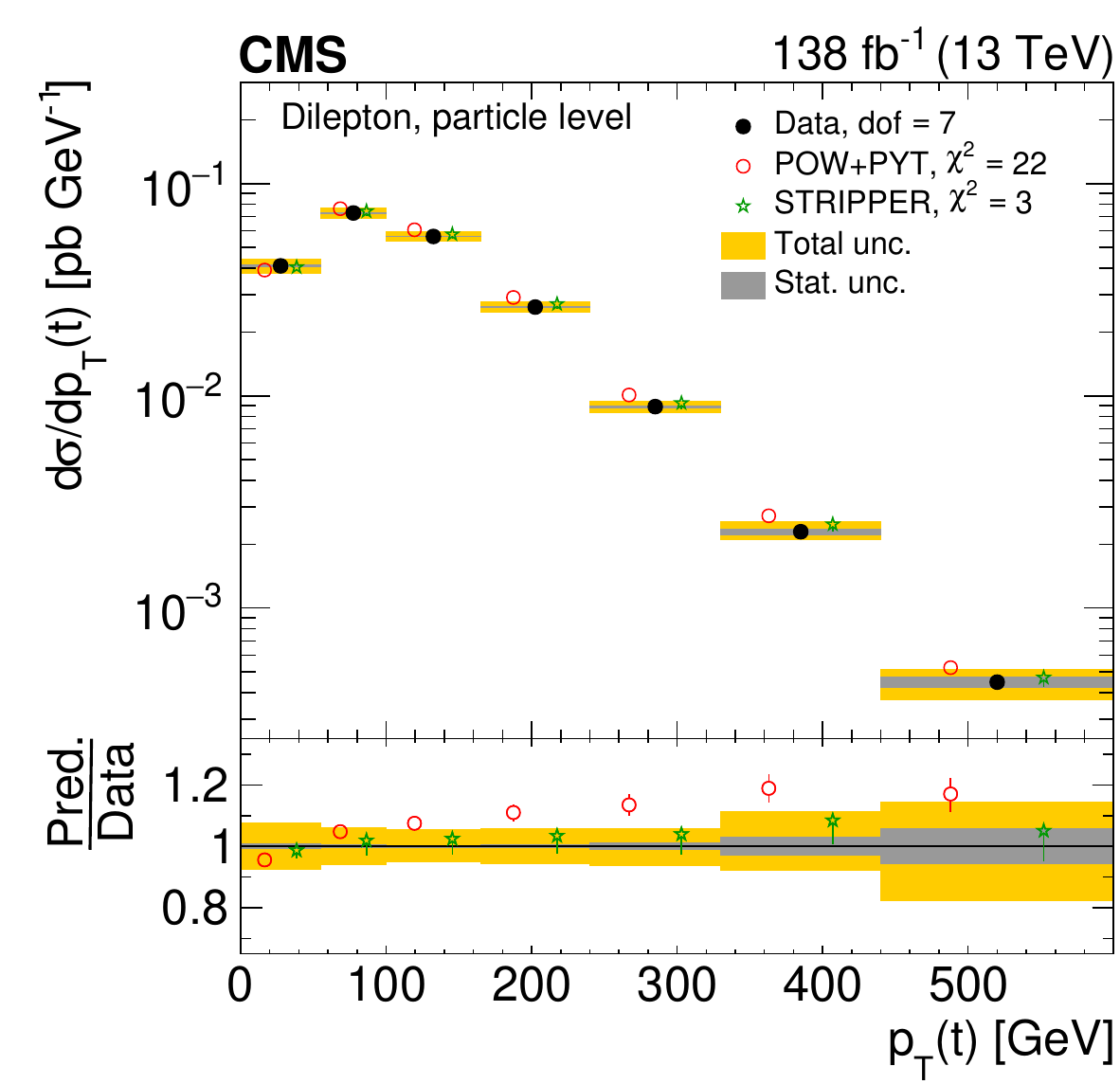}
\includegraphics[width=0.49\textwidth]{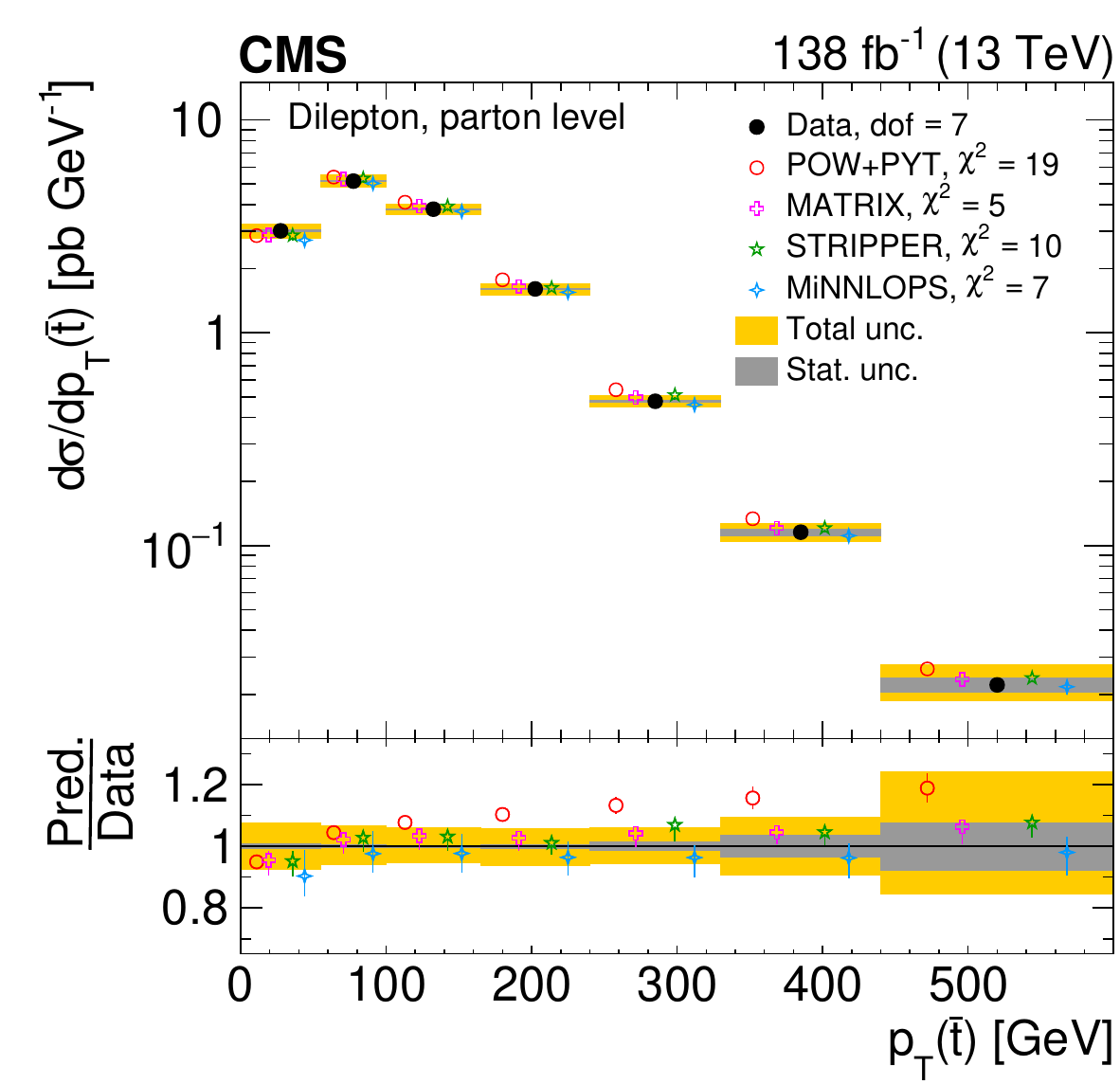}
\includegraphics[width=0.49\textwidth]{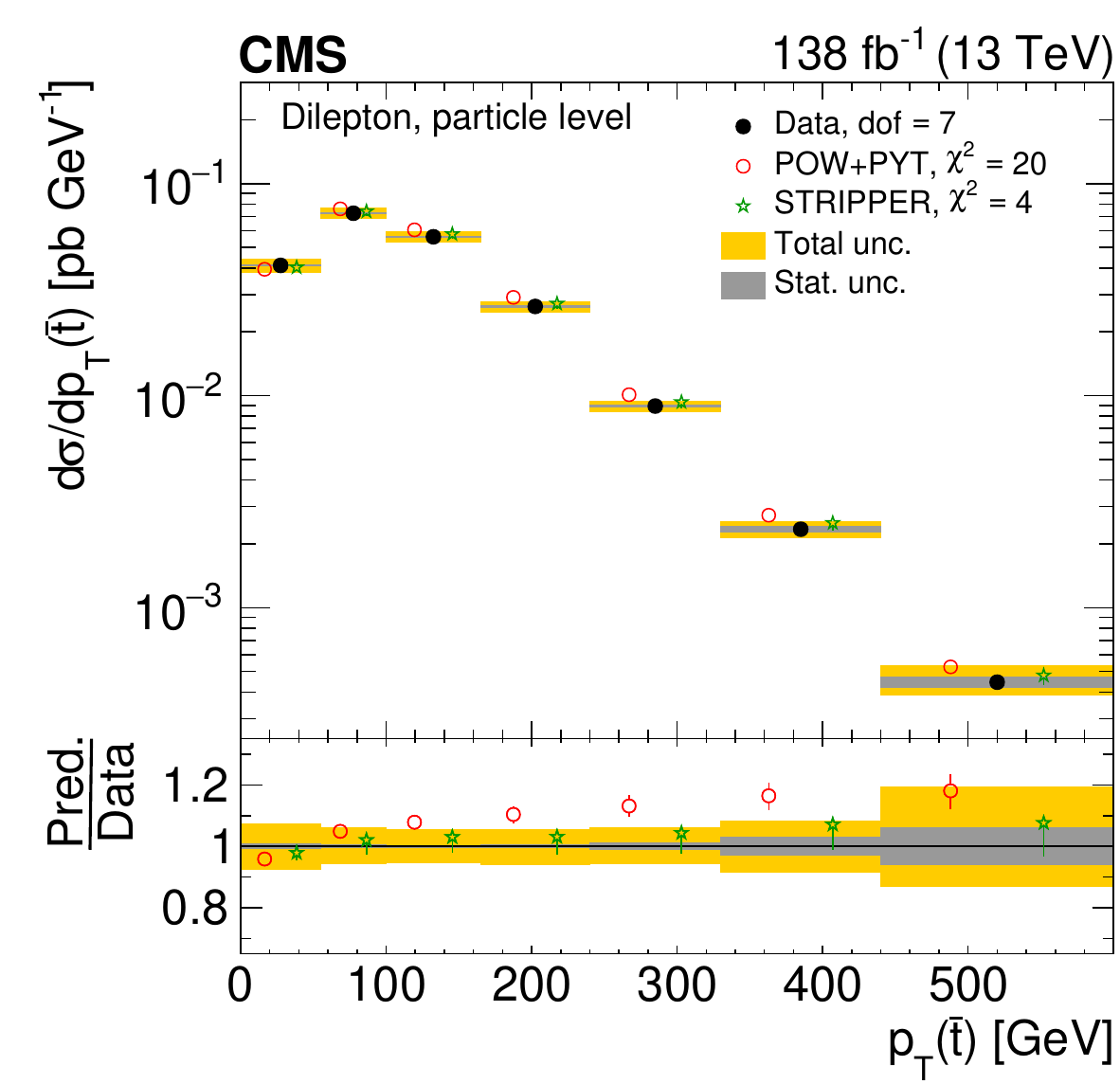}
\caption {Absolute differential \ttbar production cross sections as functions of \ptt (upper) and \ptat (lower), 
measured
at the parton level in the full phase space (left) and at the particle level in a fiducial phase space (right).
The data are shown as filled circles with grey and yellow bands indicating the statistical and total uncertainties
(statistical and systematic uncertainties added in quadrature), respectively.
For each distribution, the number of degrees of freedom (dof) is also provided.
The cross sections are compared to predictions from the \PowPyt (`POW-PYT', open circles) simulation and various theoretical
predictions with beyond-NLO precision (other points).
The estimated uncertainties in the predictions are represented by vertical bars on the corresponding points.
For each model, a value of \chisq is reported that takes into account the measurement uncertainties.
The lower panel in each plot shows the ratios of the predictions to the data.}
\label{fig:xsec-1d-theory-abs-ptt-ptat}
\end{figure*}

\begin{figure*}[!phtb]
\centering
\includegraphics[width=0.49\textwidth]{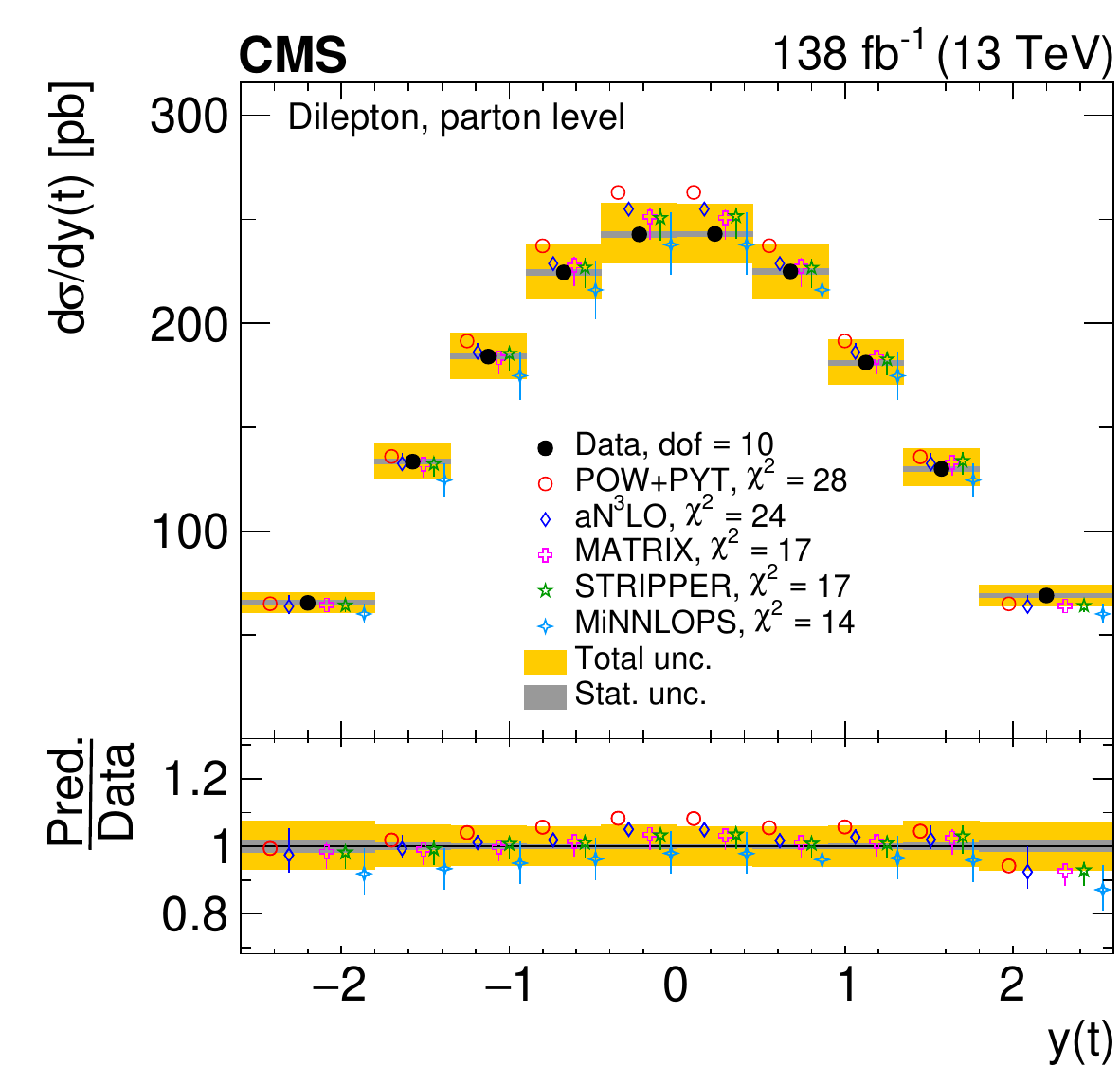}
\includegraphics[width=0.49\textwidth]{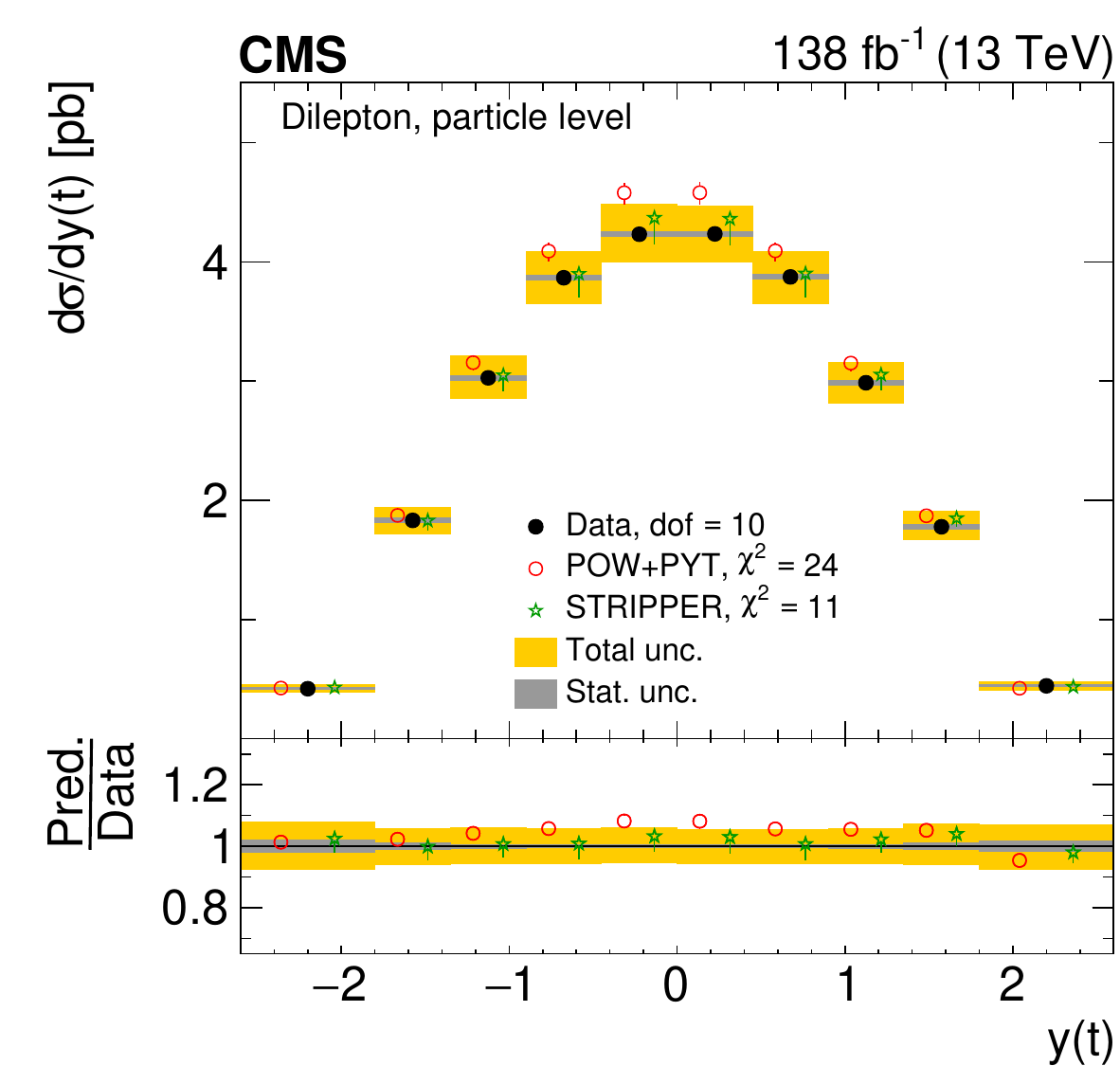}
\includegraphics[width=0.49\textwidth]{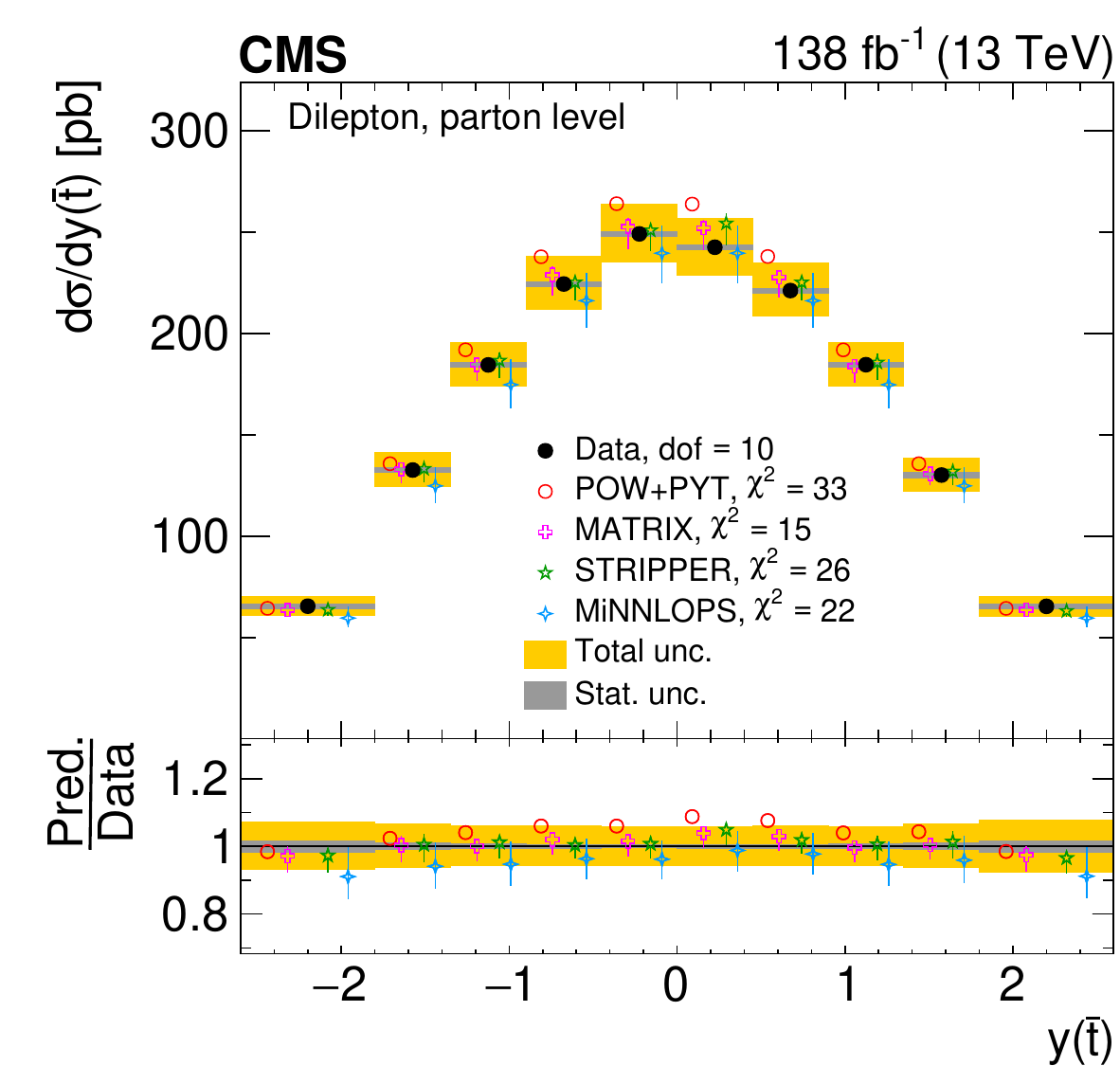}
\includegraphics[width=0.49\textwidth]{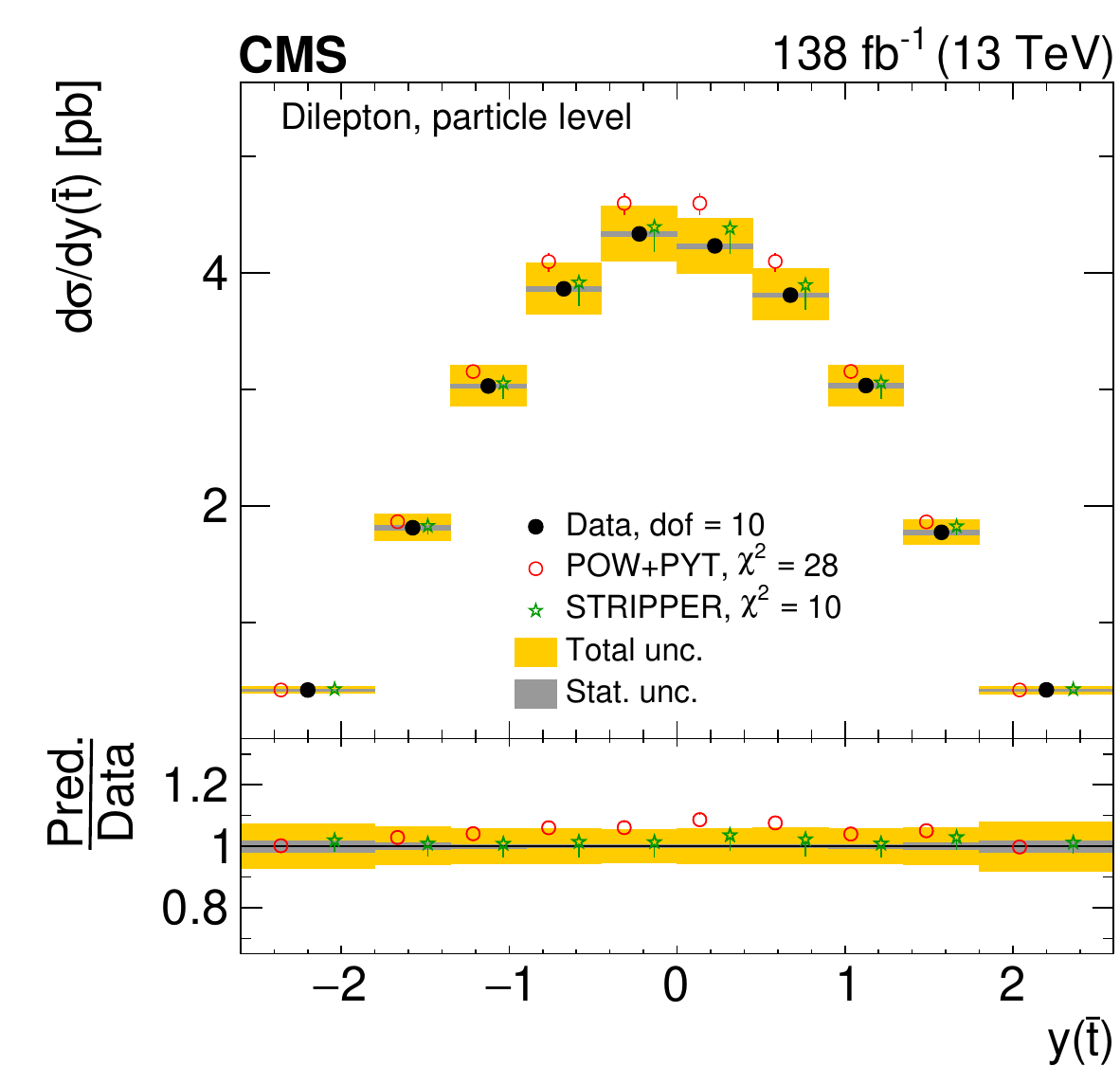}
\caption{Absolute differential \ttbar production cross sections as functions of \yt (upper) and \yat (lower)
are shown for data (filled circles), \PowPyt (`POW-PYT', open circles) simulation, and various theoretical predictions with
beyond-NLO precision (other points).
Further details can be found in the caption of Fig.~\ref{fig:xsec-1d-theory-abs-ptt-ptat}.}
\label{fig:xsec-1d-theory-abs-yt-yat}
\end{figure*}

\begin{figure*}[!phtb]
\centering
\includegraphics[width=0.49\textwidth]{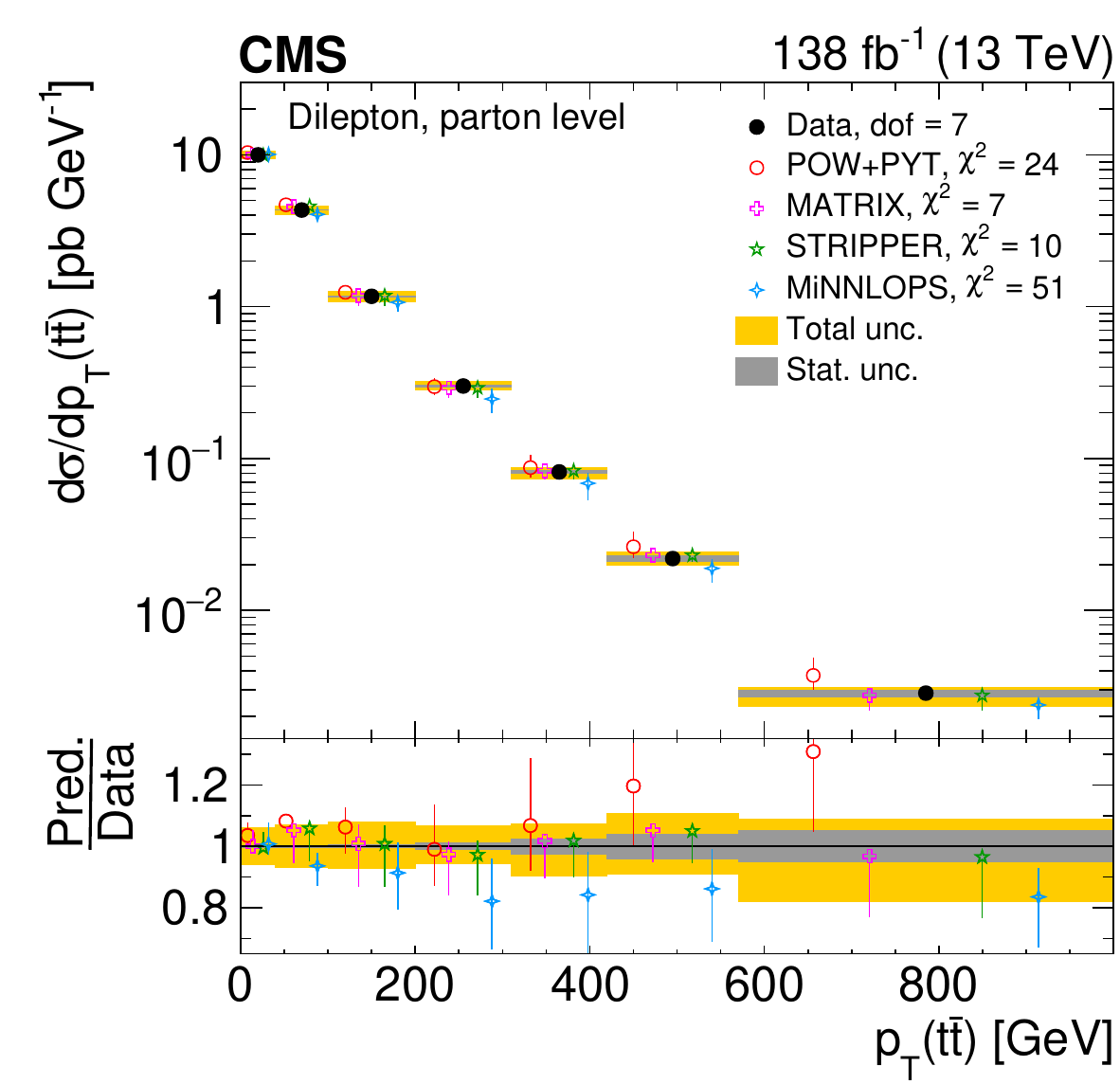}
\includegraphics[width=0.49\textwidth]{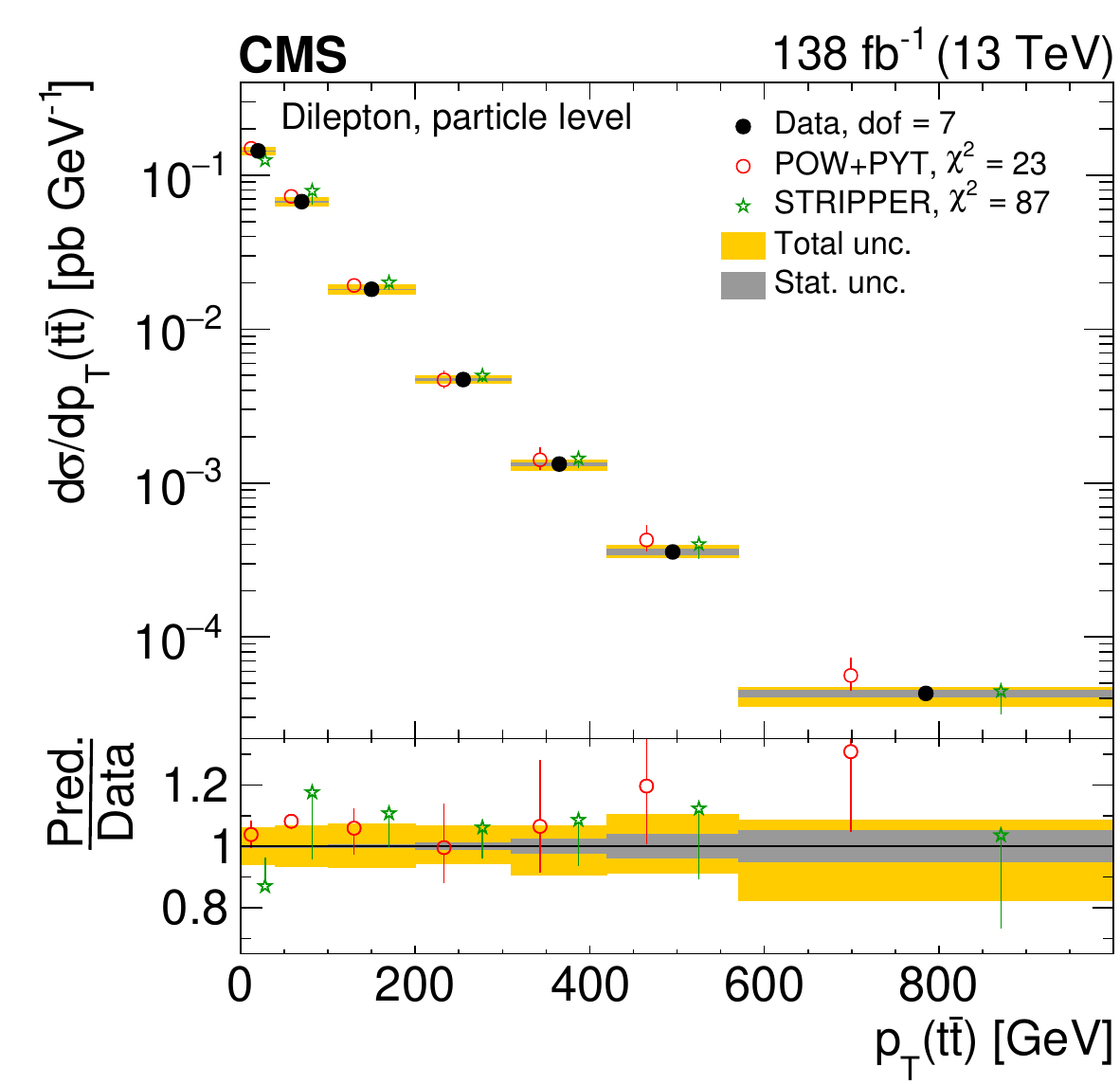}
\includegraphics[width=0.49\textwidth]{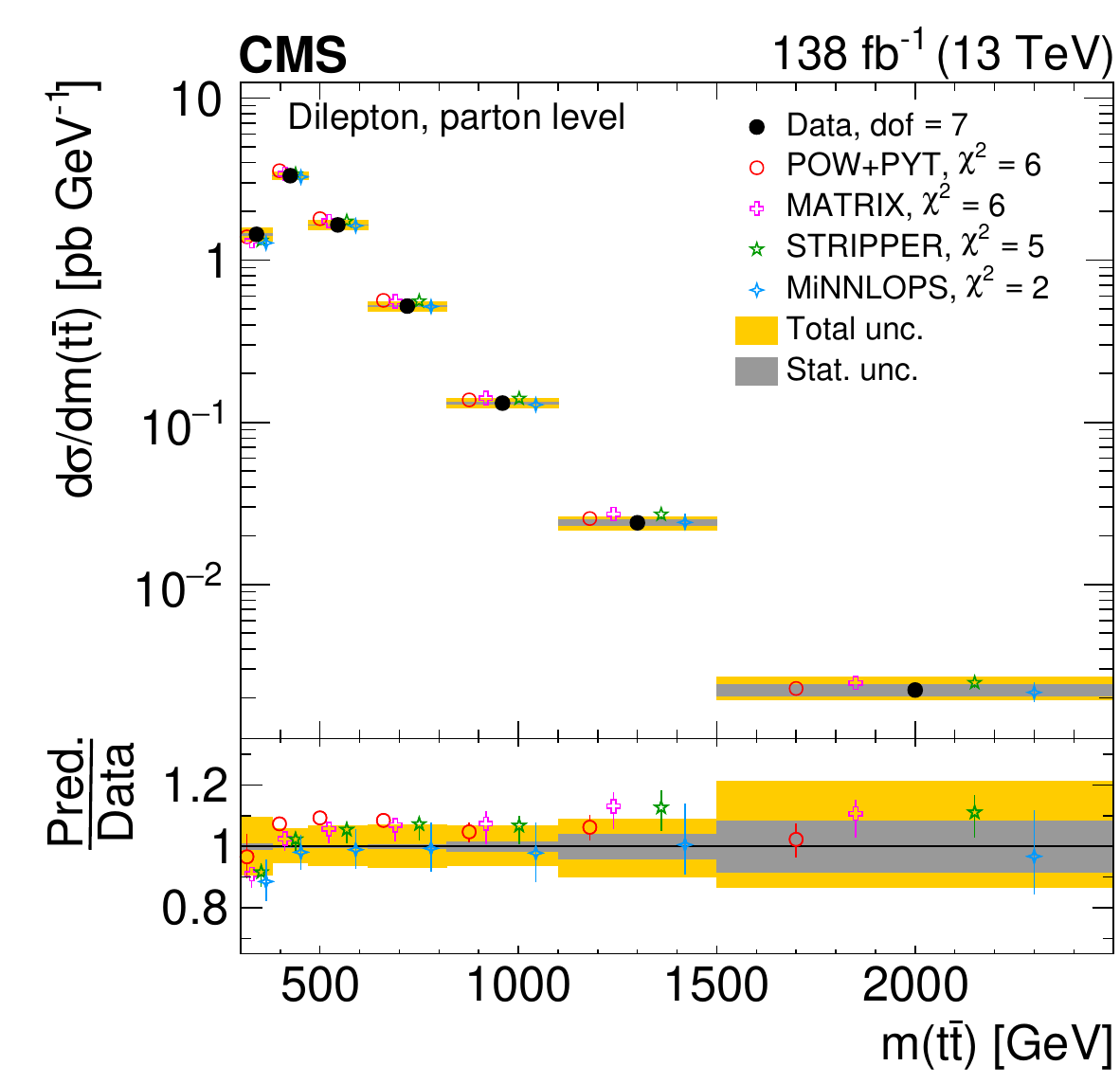}
\includegraphics[width=0.49\textwidth]{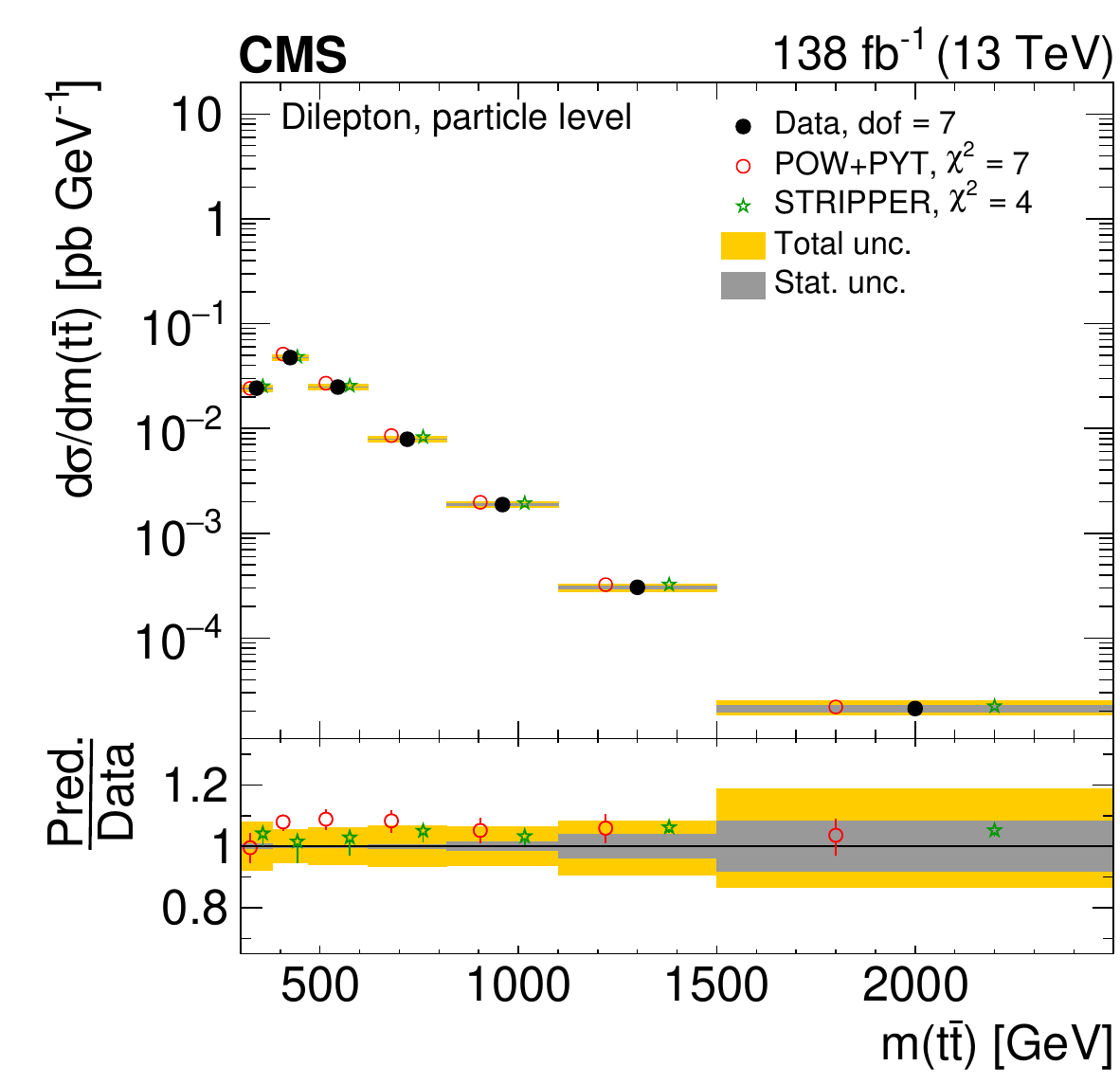}
\includegraphics[width=0.49\textwidth]{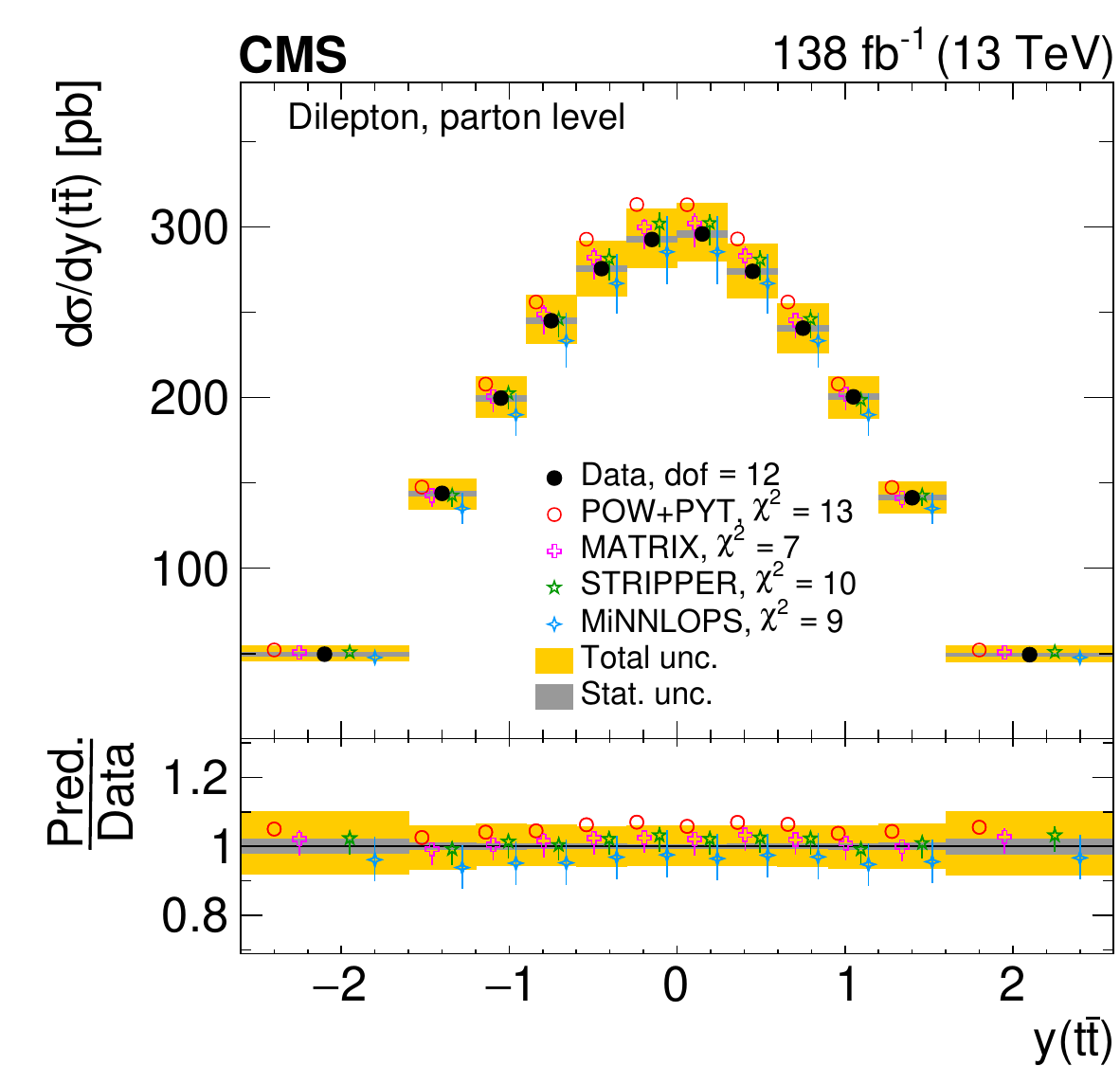}
\includegraphics[width=0.49\textwidth]{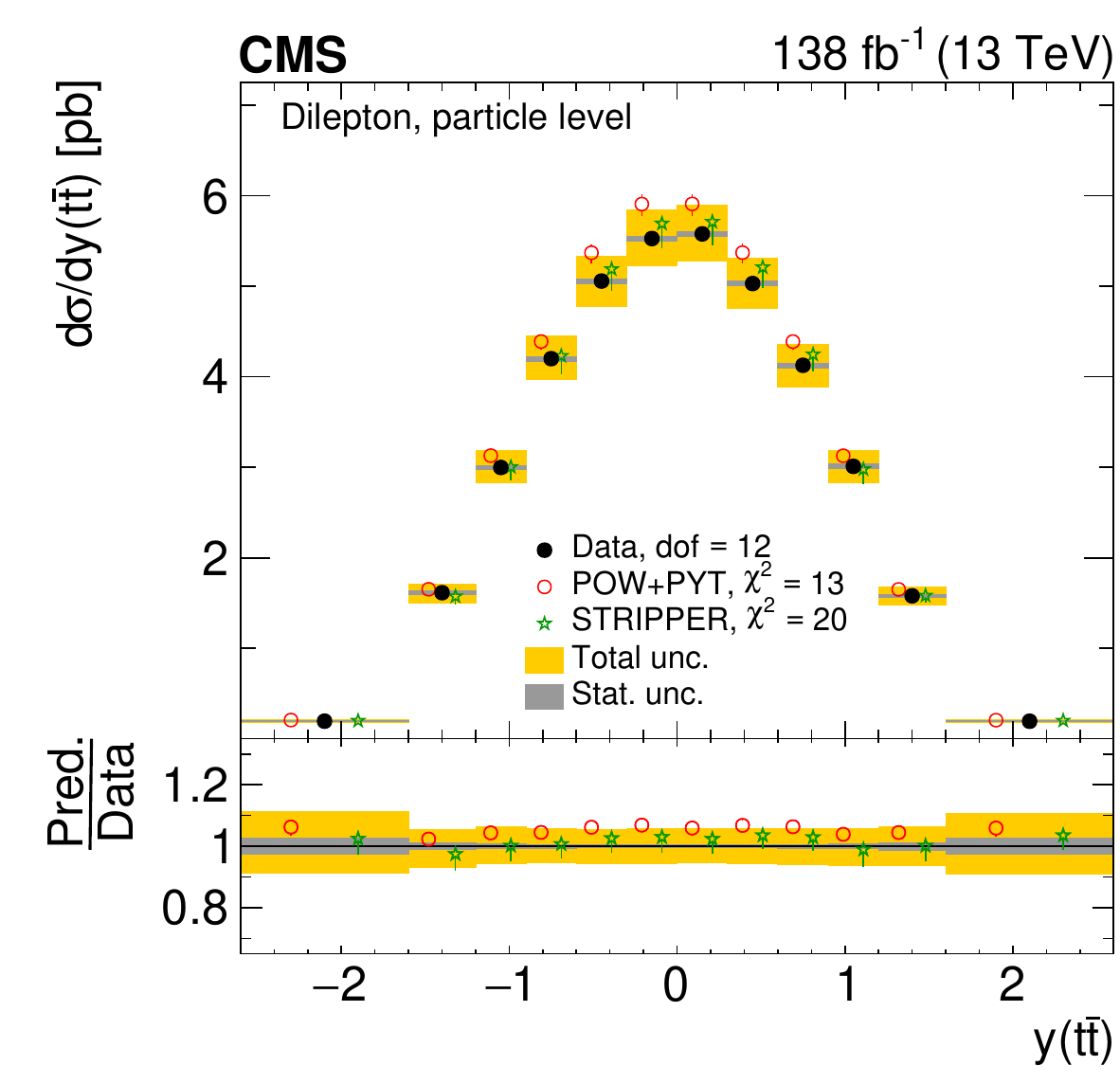}
\caption{Absolute differential \ttbar production cross sections as functions of \pttt (upper), \mtt (middle), and \ytt
(lower) are shown for data (filled circles), \PowPyt (`POW-PYT', open circles) simulation, and various theoretical
predictions with beyond-NLO precision (other points).
Further details can be found in the caption of Fig.~\ref{fig:xsec-1d-theory-abs-ptt-ptat}.}
\label{fig:xsec-1d-theory-abs-pttt-mtt-ytt}
\end{figure*}

\begin{figure*}[!phtb]
\centering
\includegraphics[width=0.49\textwidth]{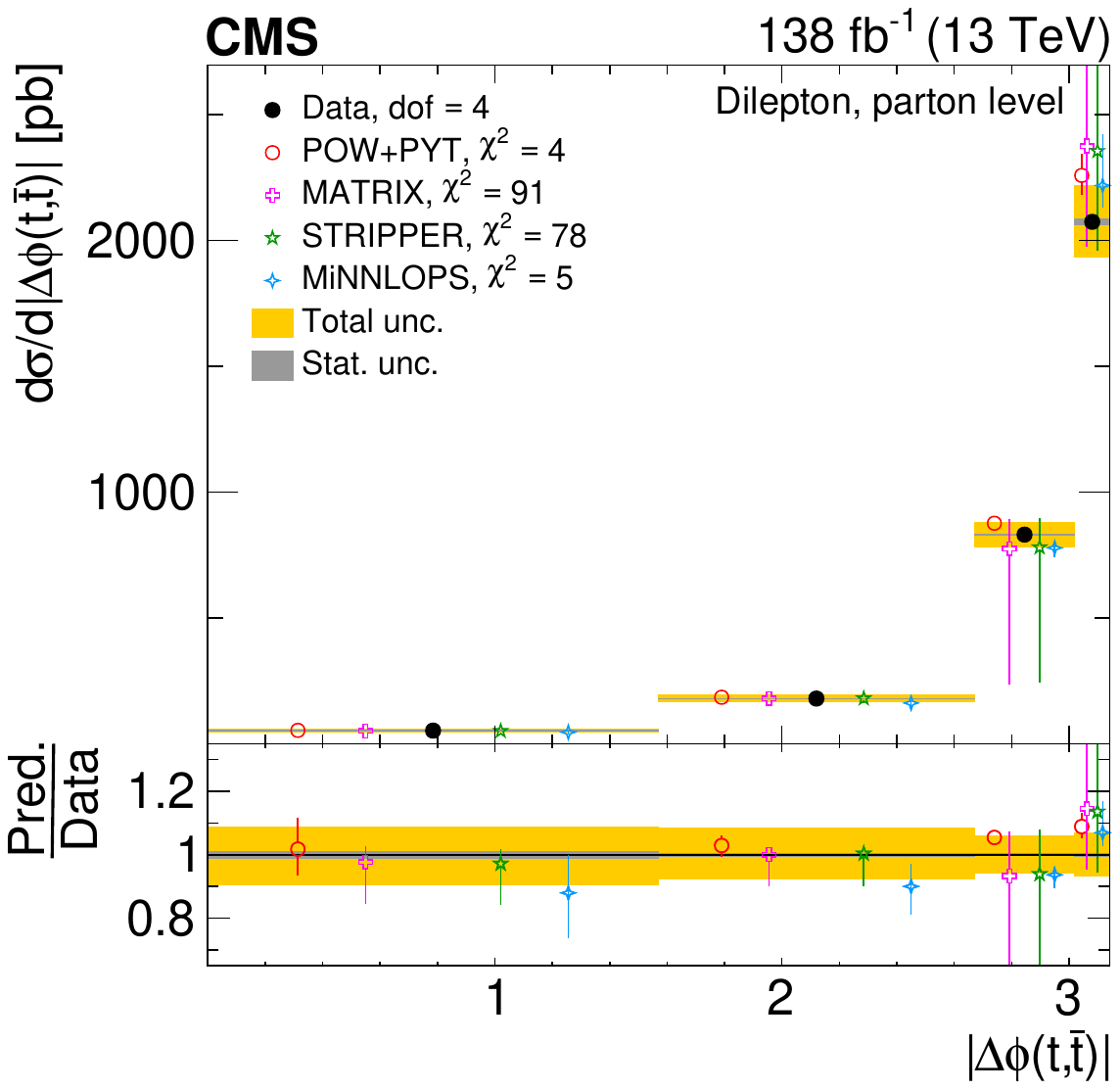}
\includegraphics[width=0.49\textwidth]{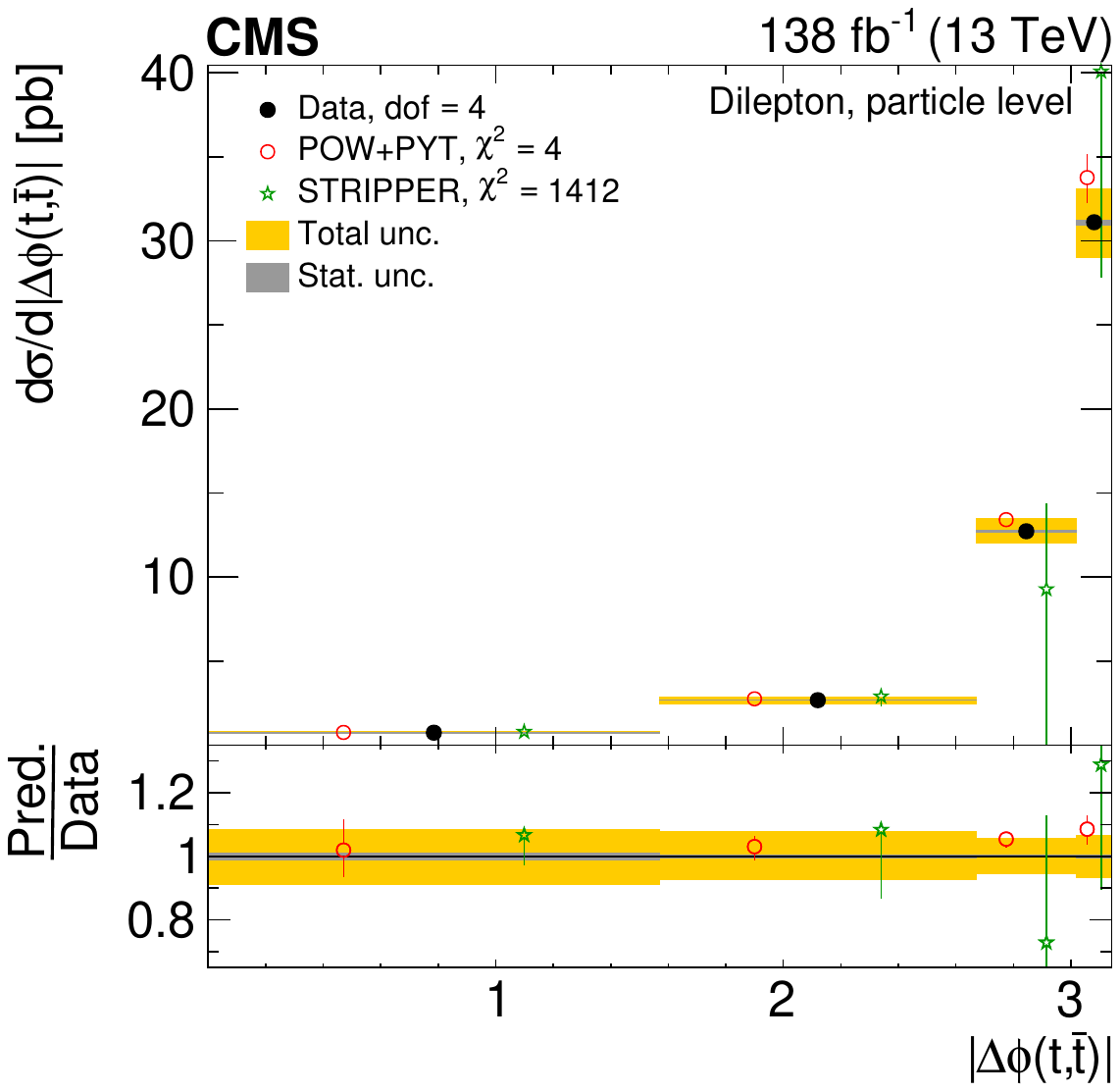}
\includegraphics[width=0.49\textwidth]{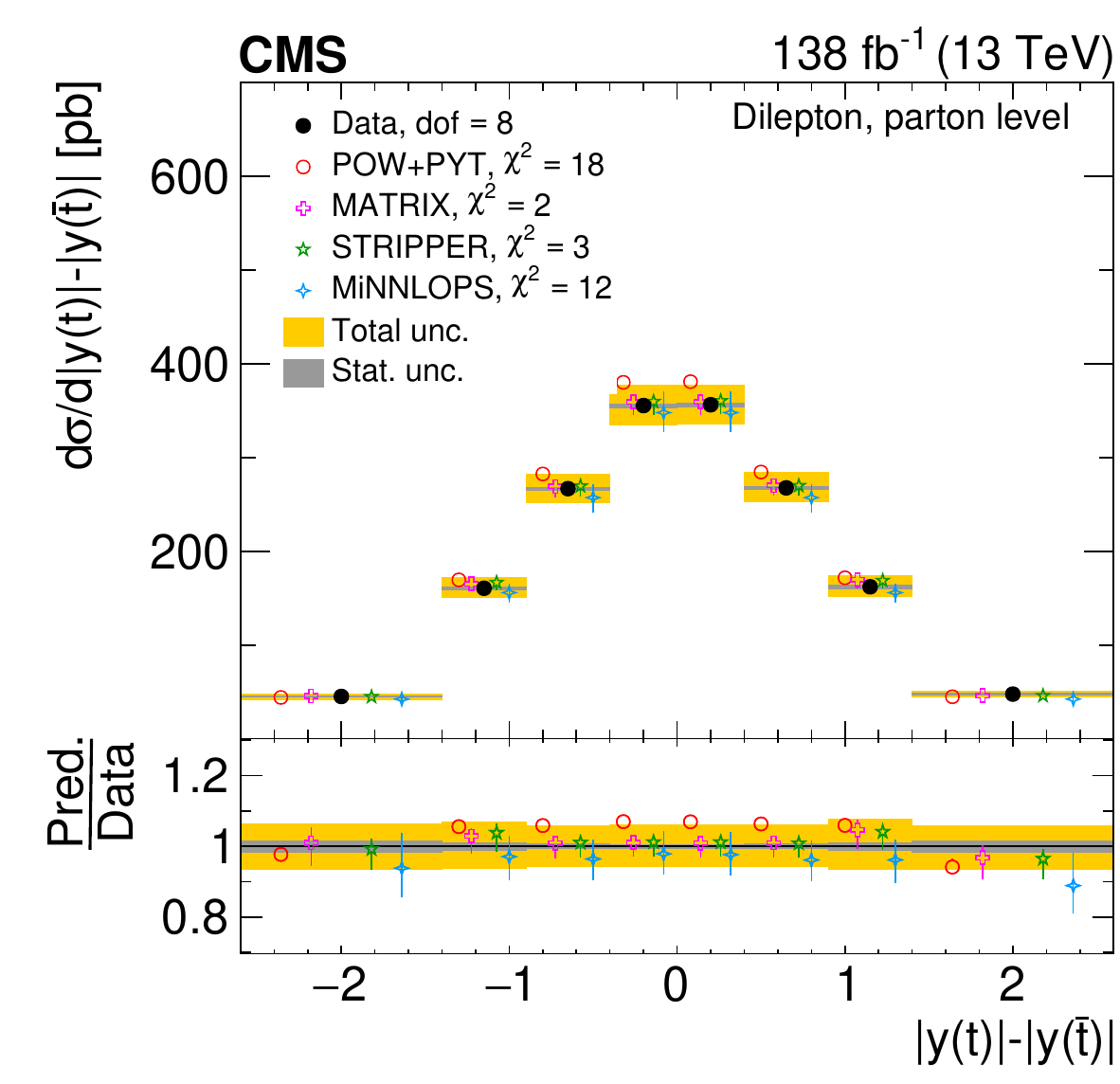}
\includegraphics[width=0.49\textwidth]{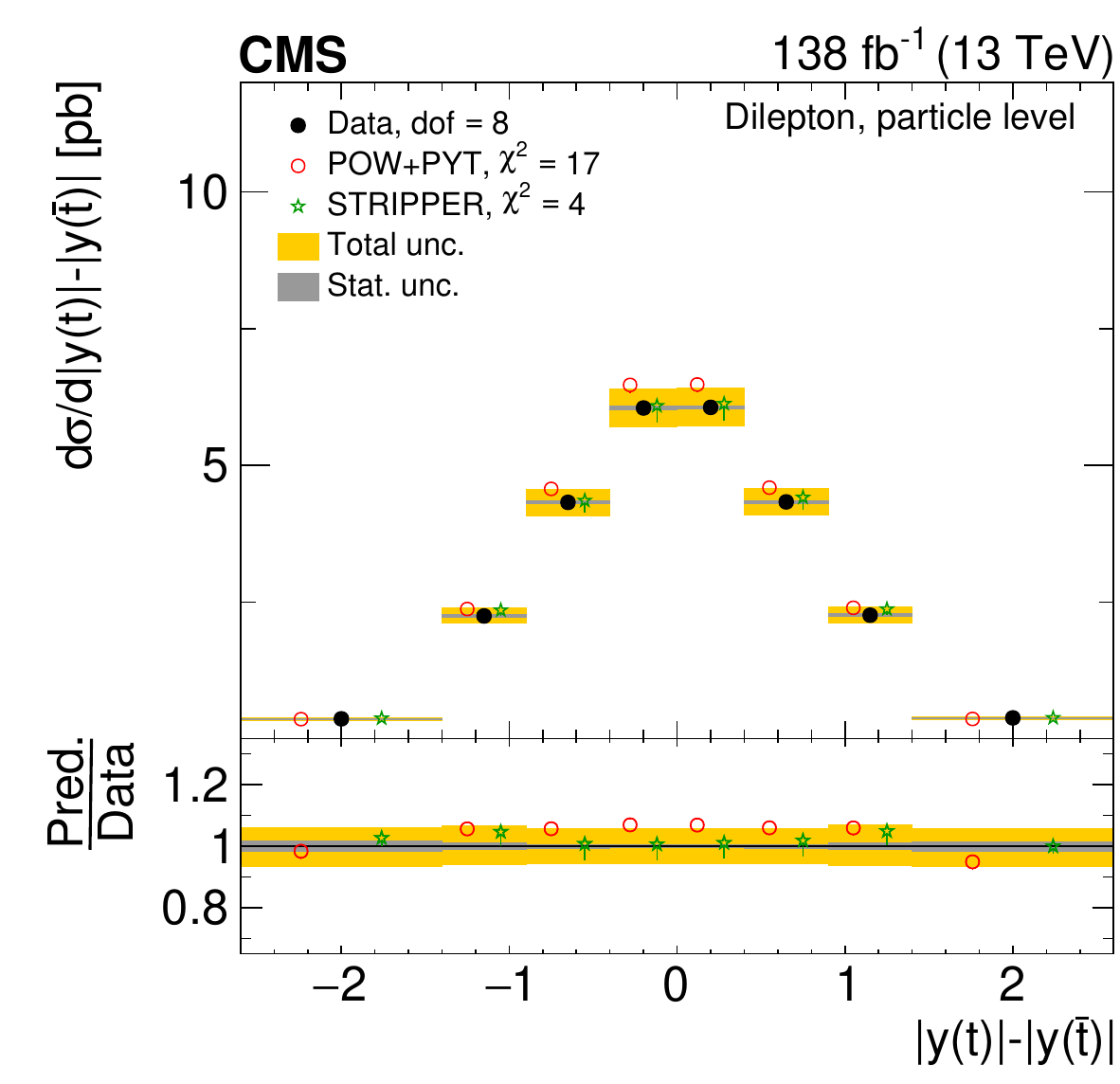}
\caption{Absolute differential \ttbar production cross sections as functions of \dphitt (upper) and \dytt (lower)
are shown for data (filled circles), \PowPyt (`POW-PYT', open circles) simulation, and various theoretical predictions with
beyond-NLO precision (other points).
Further details can be found in the caption of Fig.~\ref{fig:xsec-1d-theory-abs-ptt-ptat}.}
\label{fig:xsec-1d-theory-abs-dphitt-dytt}
\end{figure*}

\begin{figure*}[!phtb]
\centering
\includegraphics[width=0.49\textwidth]{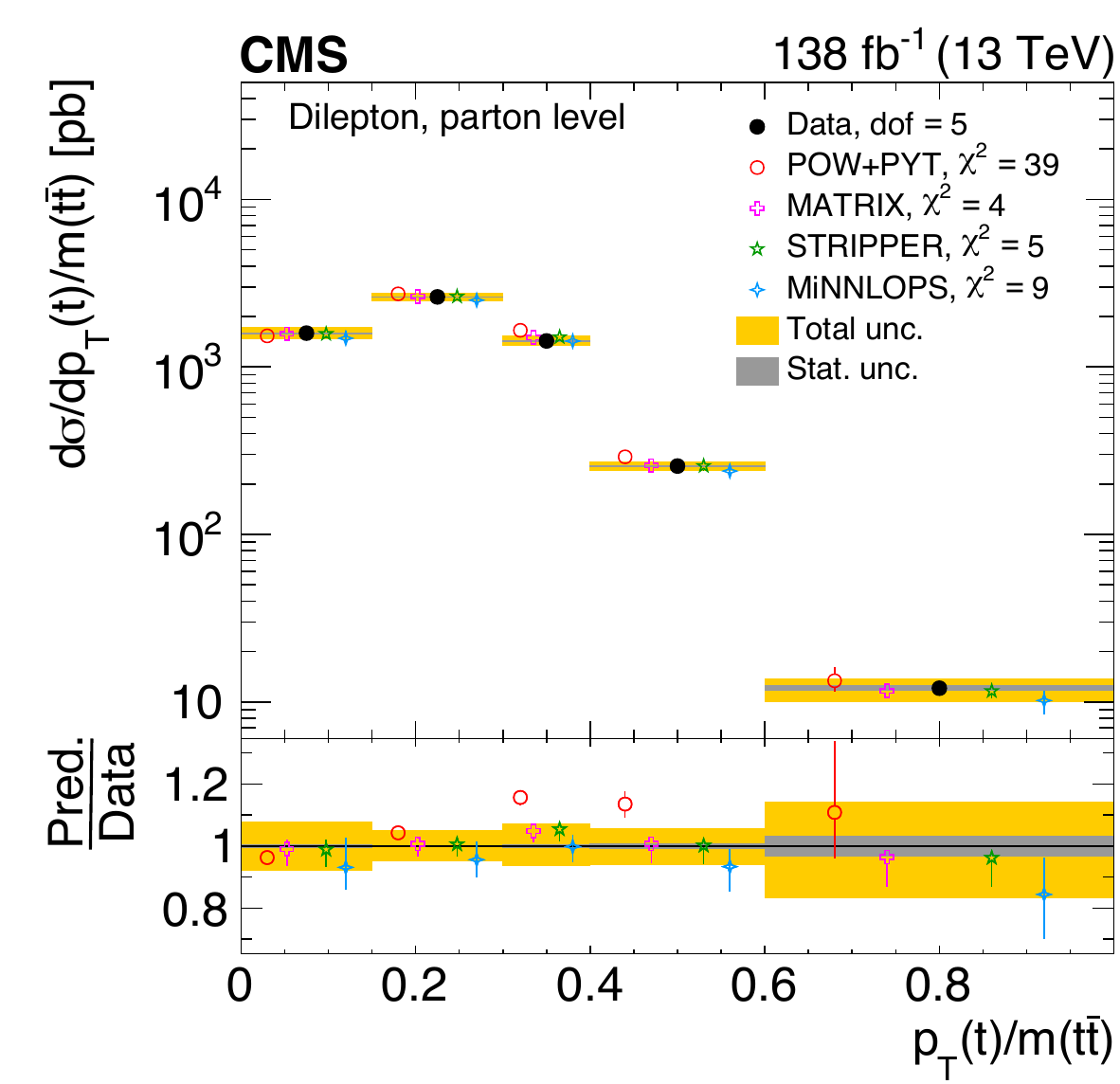}
\includegraphics[width=0.49\textwidth]{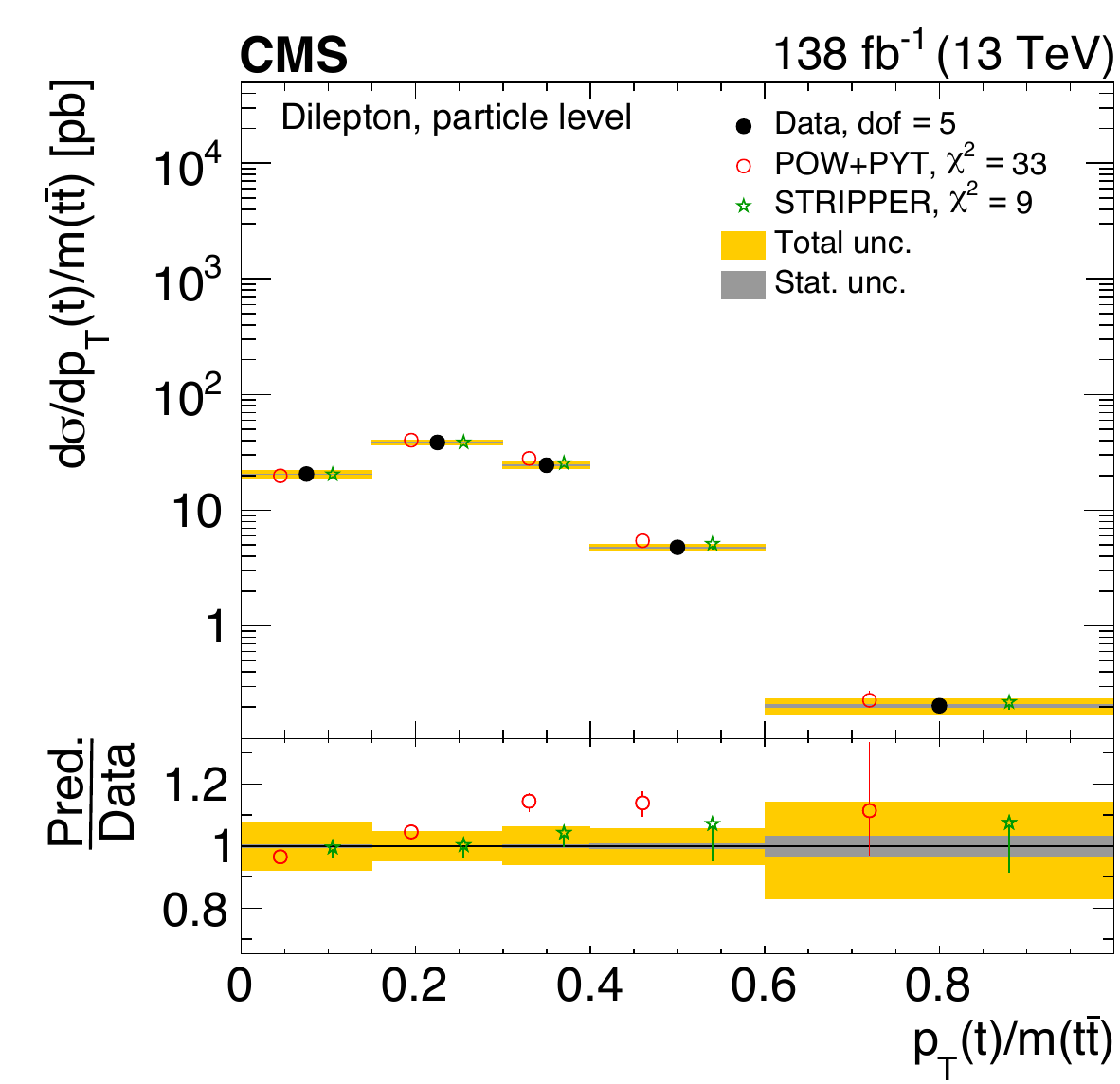}
\includegraphics[width=0.49\textwidth]{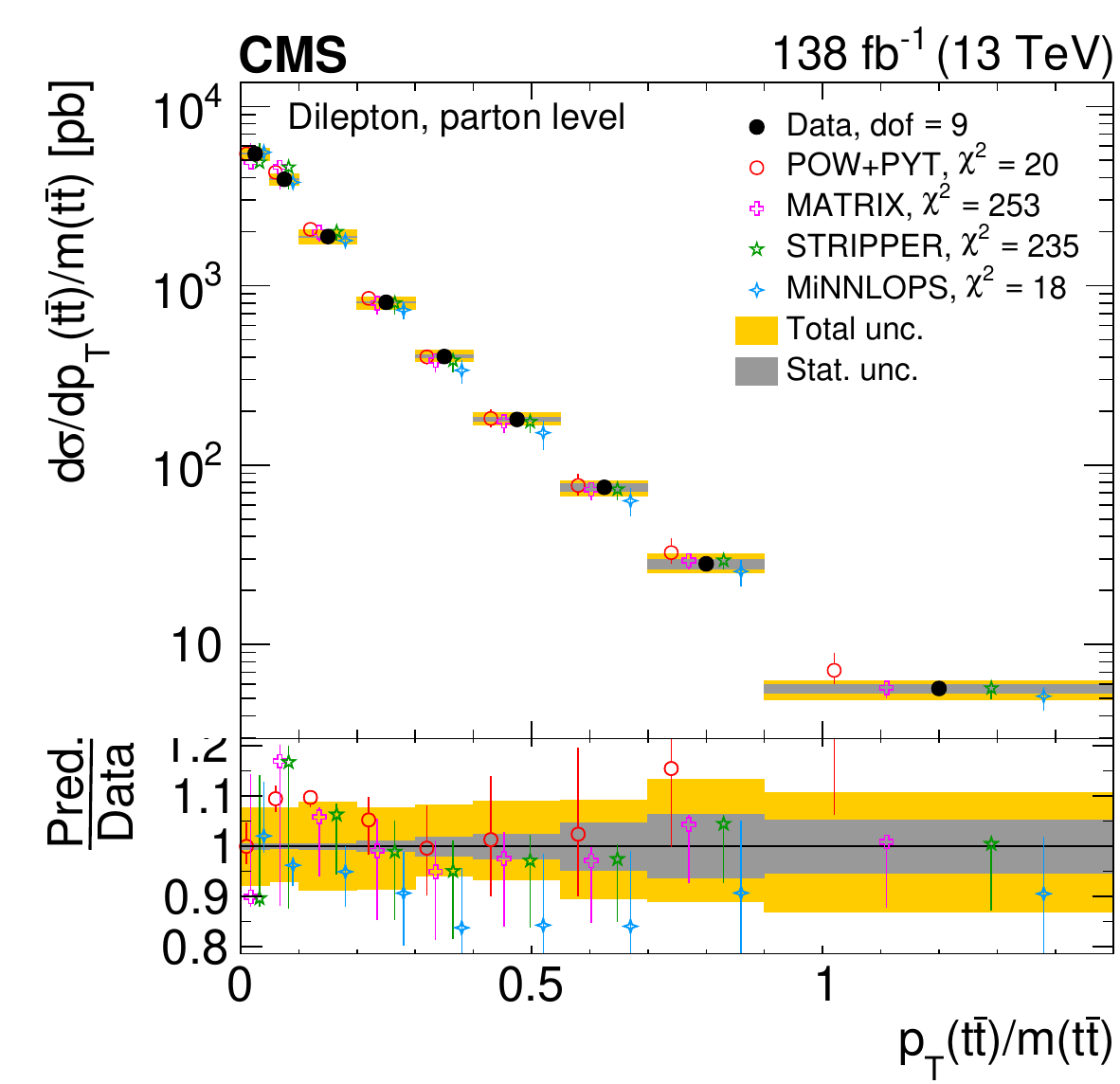}
\includegraphics[width=0.49\textwidth]{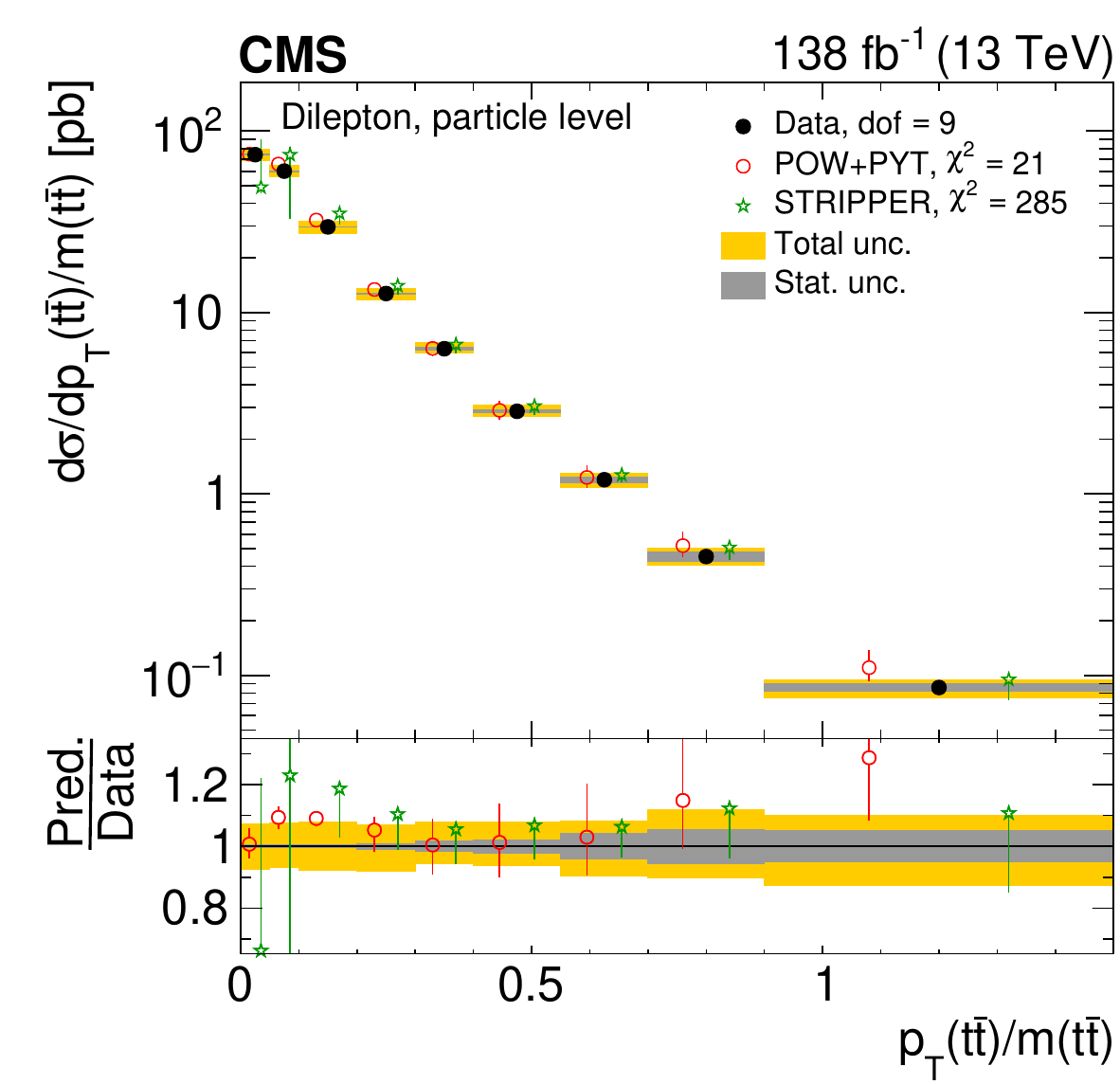}
\caption{Absolute differential \ttbar production cross sections as functions of \rpttmtt (upper) and \rptttmtt (lower)
are shown for data (filled circles), \PowPyt (`POW-PYT', open circles) simulation, and various theoretical predictions with
beyond-NLO precision (other points).
Further details can be found in the caption of Fig.~\ref{fig:xsec-1d-theory-abs-ptt-ptat}.}
\label{fig:xsec-1d-theory-abs-rpttmtt-rptttmtt}
\end{figure*}

\begin{figure*}[!phtb]
\centering
\includegraphics[width=0.49\textwidth]{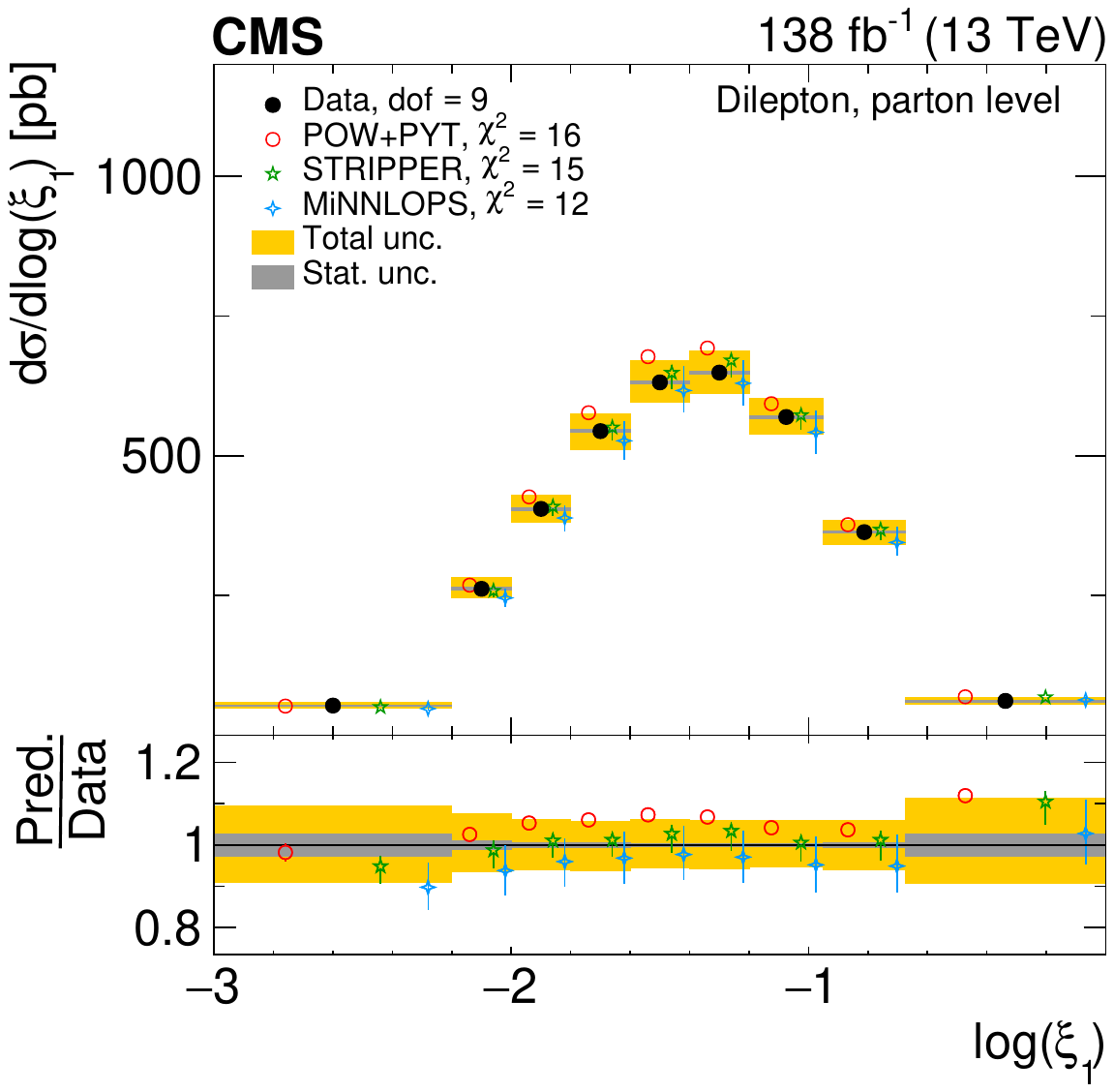}
\includegraphics[width=0.49\textwidth]{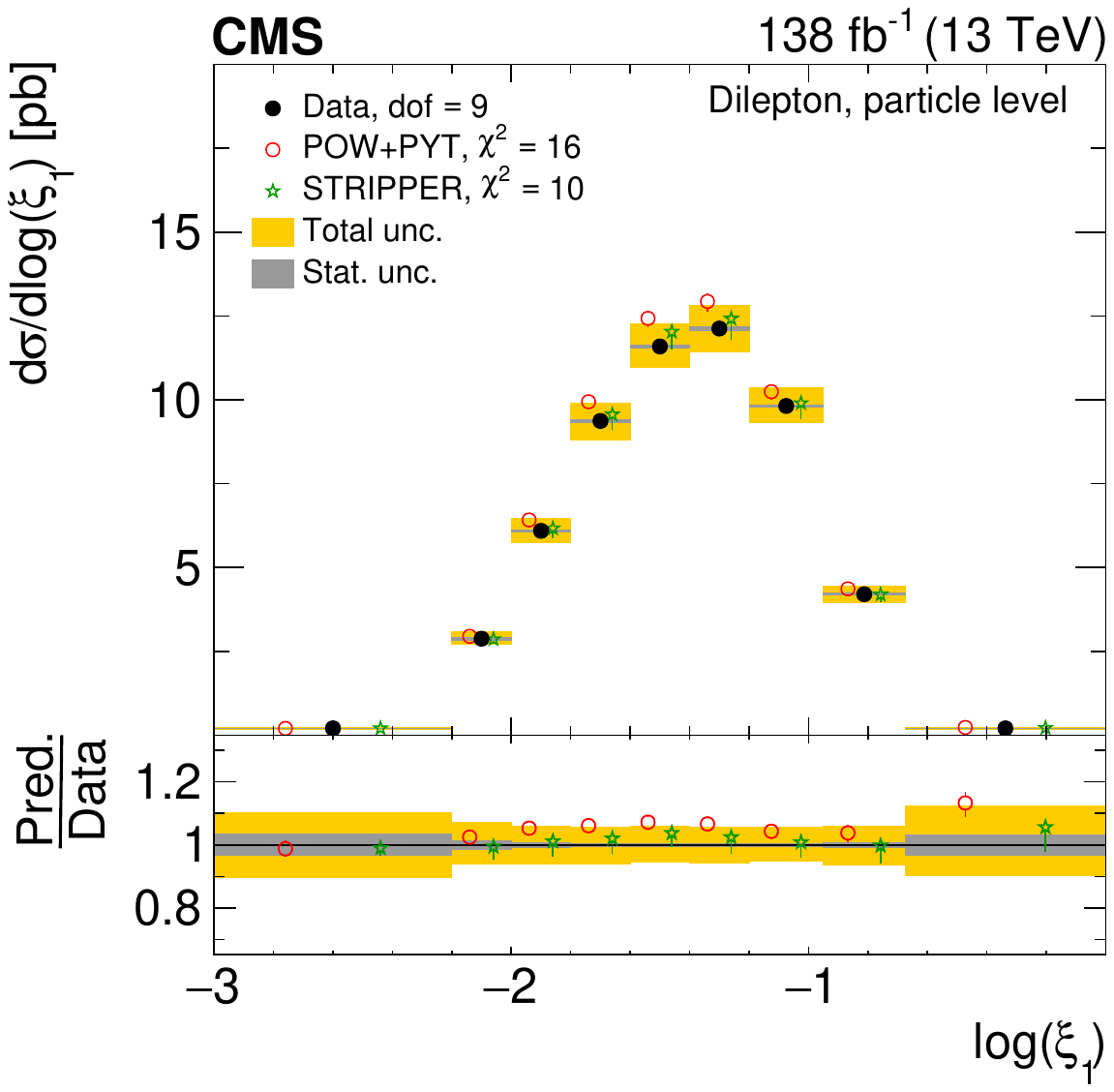}
\includegraphics[width=0.49\textwidth]{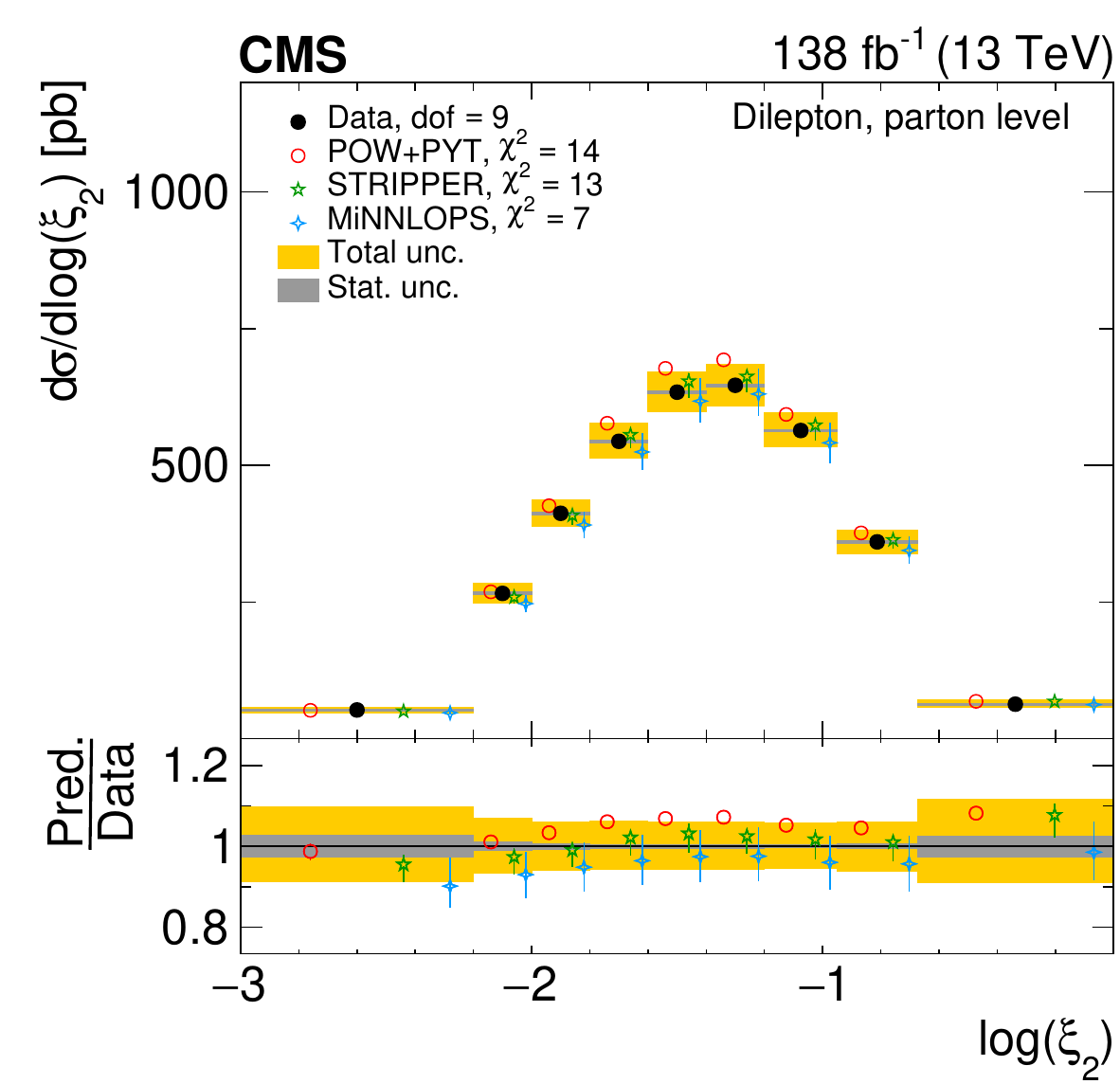}
\includegraphics[width=0.49\textwidth]{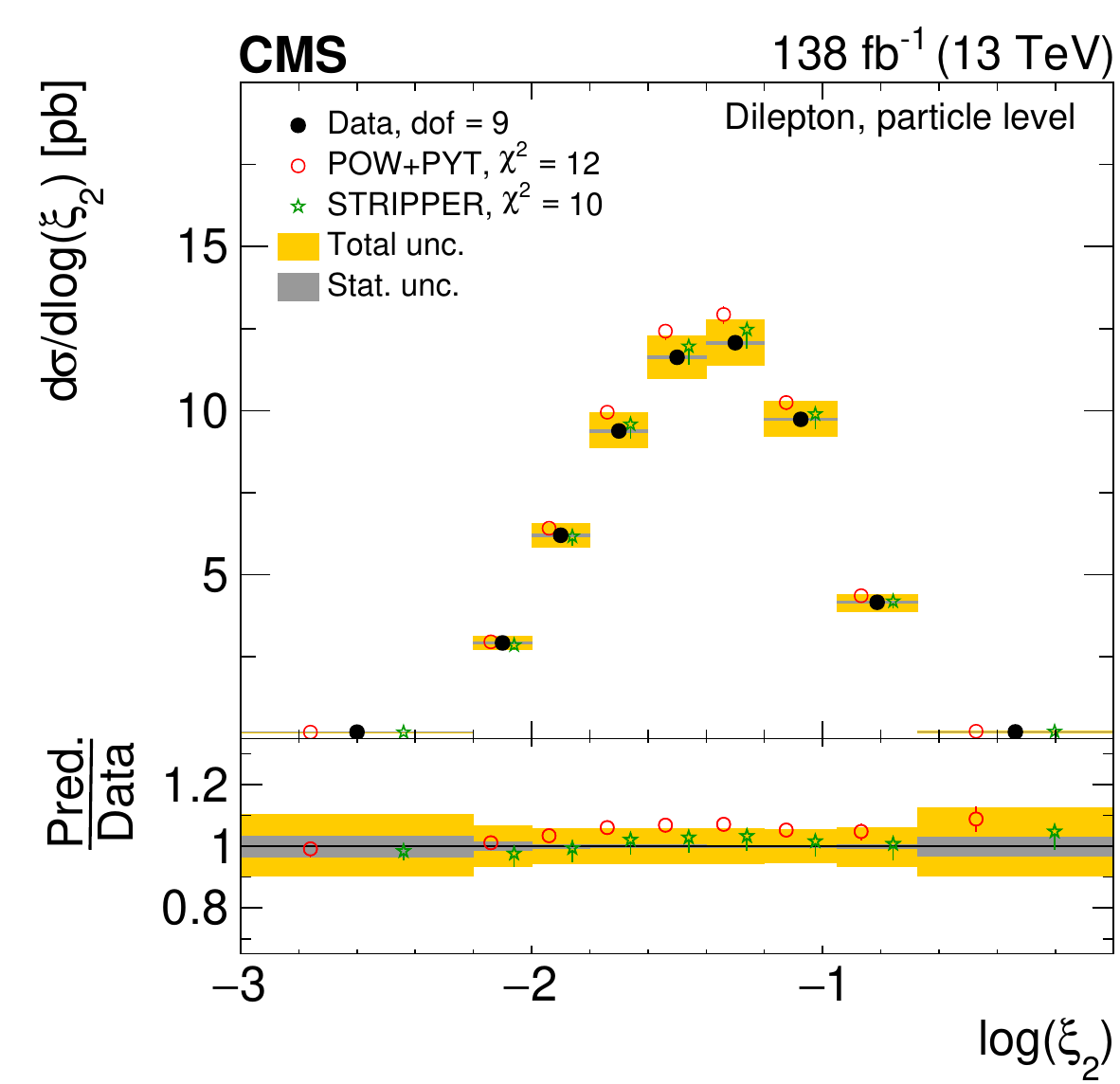}
\caption{Absolute differential \ttbar production cross sections as functions of \logxone (upper) and \logxtwo (lower)
are shown for data (filled circles), \PowPyt (`POW-PYT', open circles) simulation, and \StripperOnly
NNLO calculation (stars).
Further details can be found in the caption of Fig.~\ref{fig:xsec-1d-theory-abs-ptt-ptat}.}
\label{fig:xsec-1d-theory-abs-logxone-logxtwo}
\end{figure*}

\begin{figure}
\centering
\includegraphics[width=0.99\textwidth]{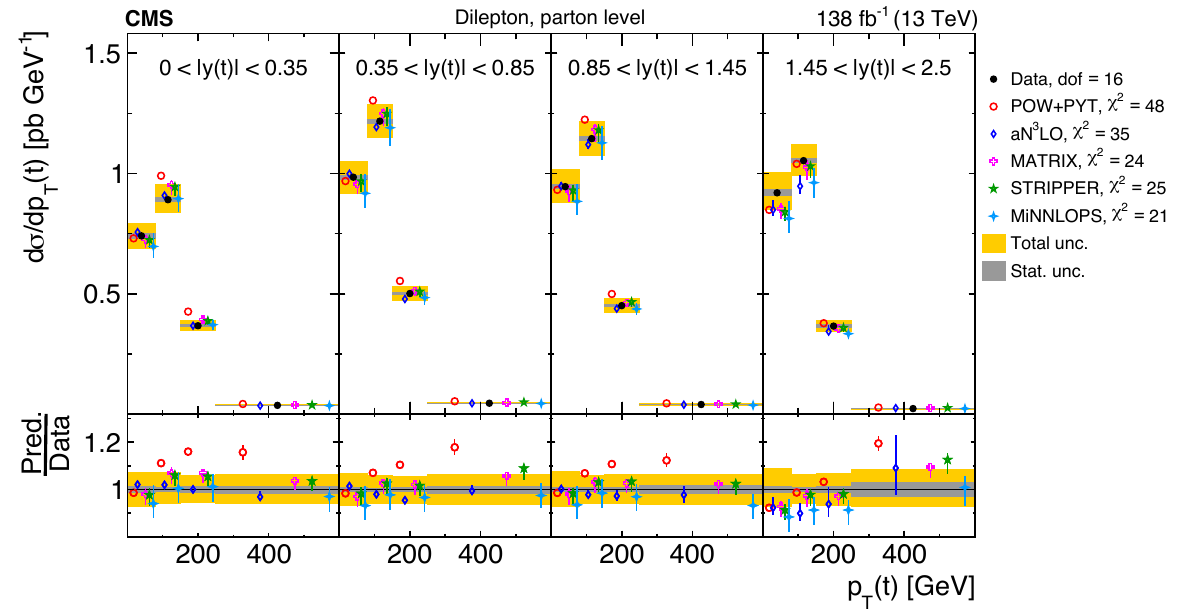}
\includegraphics[width=0.99\textwidth]{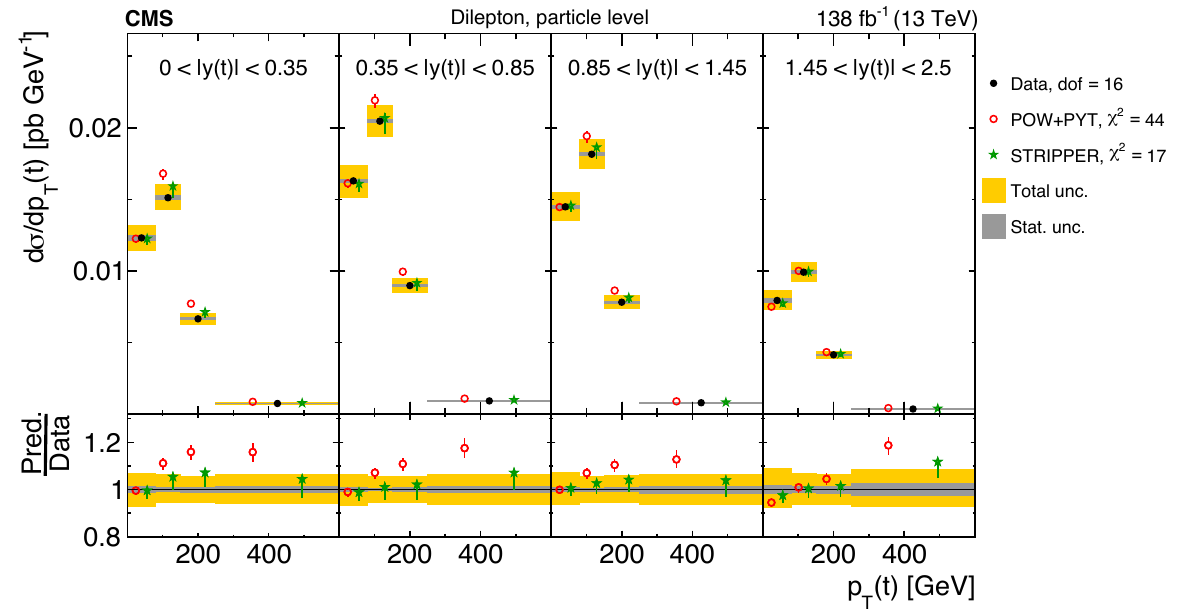}
\caption{Absolute \ytptt cross sections measured at the parton level in the full phase space (upper) and at the
particle level in a fiducial phase space (lower). The data are shown as filled circles with grey and yellow bands
indicating the statistical and total uncertainties (statistical and systematic uncertainties added in quadrature),
respectively.
For each distribution, the number of degrees of freedom (dof) is also provided.
The cross sections are compared to predictions from the \PowPyt (`POW-PYT', open circles) simulation and various theoretical
predictions with beyond-NLO precision (other points). The estimated uncertainties in the predictions are represented by
vertical bars on the corresponding points. For each model, a value of \chisq is reported that takes into account the
measurement uncertainties. The lower panel in each plot shows the ratios of the predictions to the data.}
\label{fig:xsec-md-theory-abs-ytptt}
\end{figure}

\begin{figure}
\centering
\includegraphics[width=0.99\textwidth]{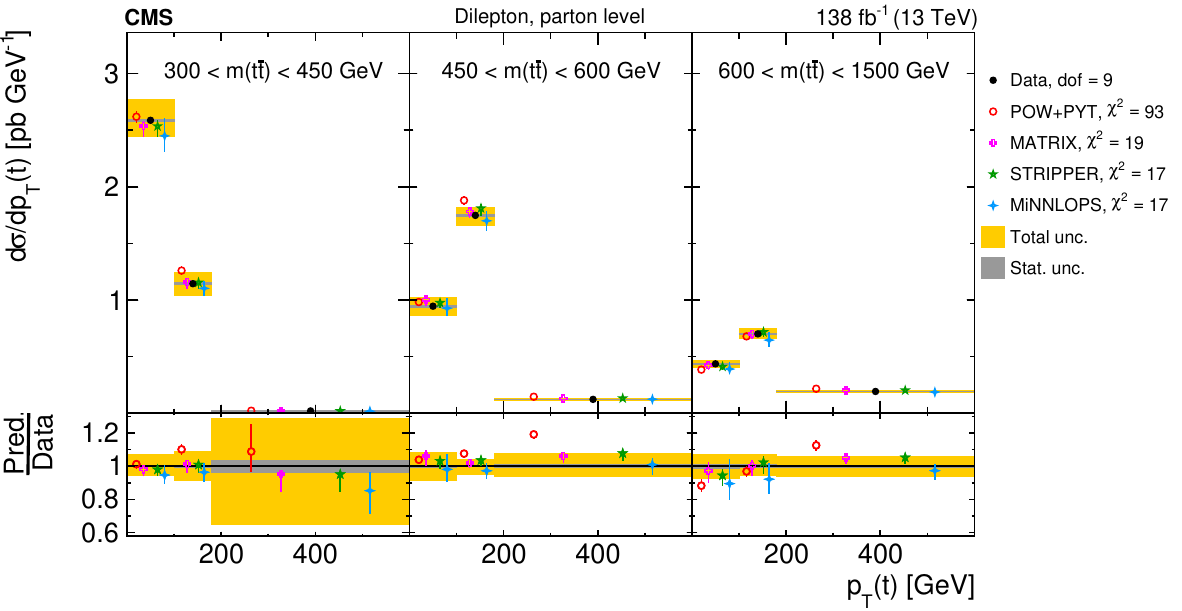}
\includegraphics[width=0.99\textwidth]{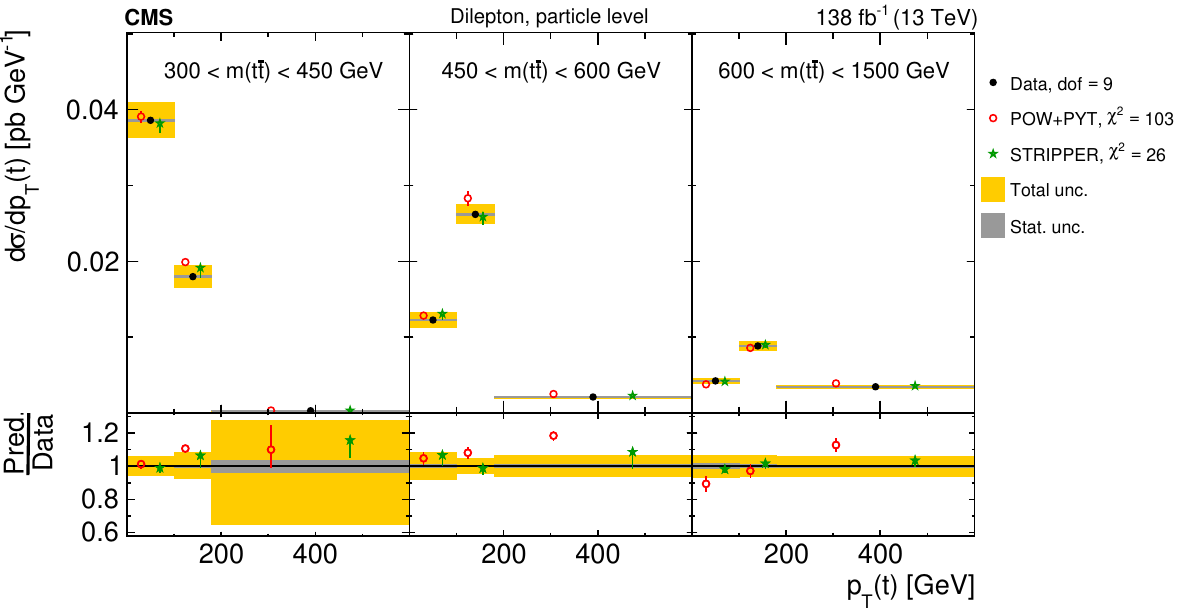}
\caption{Absolute \mttptt cross sections are shown for data (filled circles), \PowPyt (`POW-PYT', open circles)
simulation, and various theoretical predictions with beyond-NLO precision (other points).
    Further details can be found in the caption of Fig.~\ref{fig:xsec-md-theory-abs-ytptt}.}
\label{fig:xsec-md-theory-abs-mttptt}
\end{figure}

\begin{figure}
\centering
\includegraphics[width=0.99\textwidth]{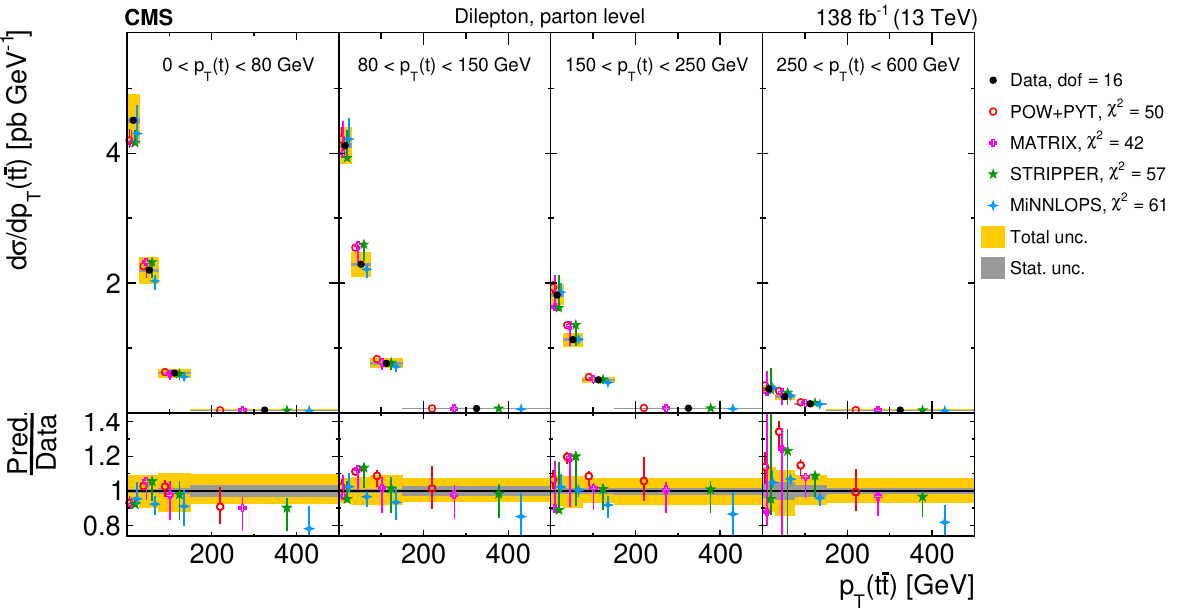}
\includegraphics[width=0.99\textwidth]{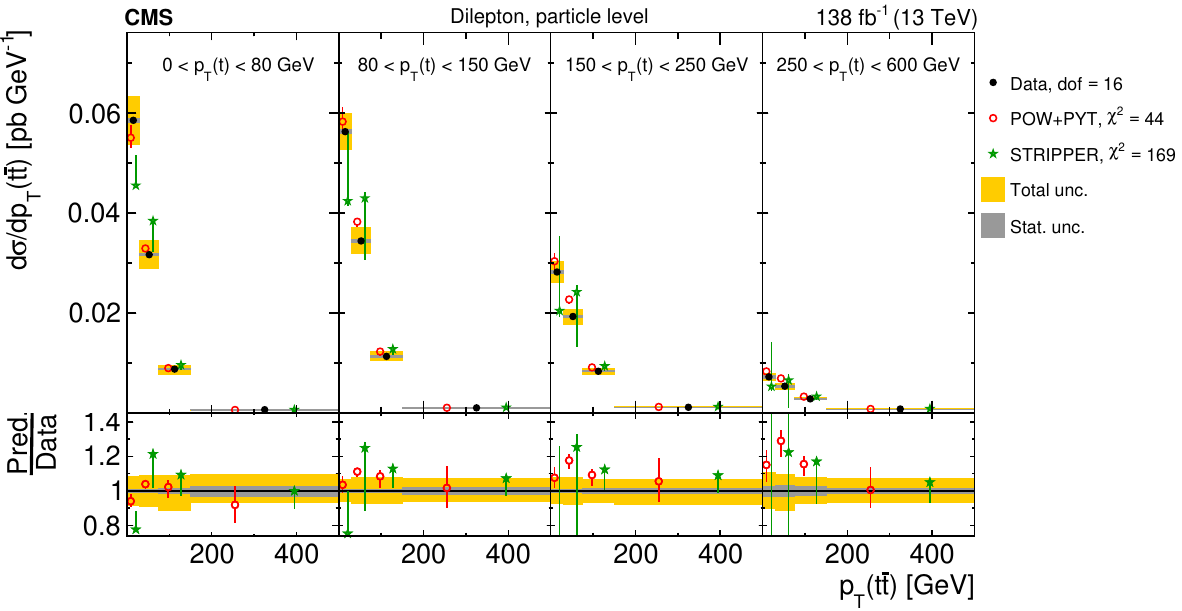}
\caption{Absolute \pttpttt cross sections are shown for data (filled circles), \PowPyt (`POW-PYT', open circles)
simulation, and various theoretical predictions with beyond-NLO precision (other points).
    Further details can be found in the caption of Fig.~\ref{fig:xsec-md-theory-abs-ytptt}.}
\label{fig:xsec-md-theory-abs-pttpttt}
\end{figure}

\begin{figure}
\centering
\includegraphics[width=0.99\textwidth]{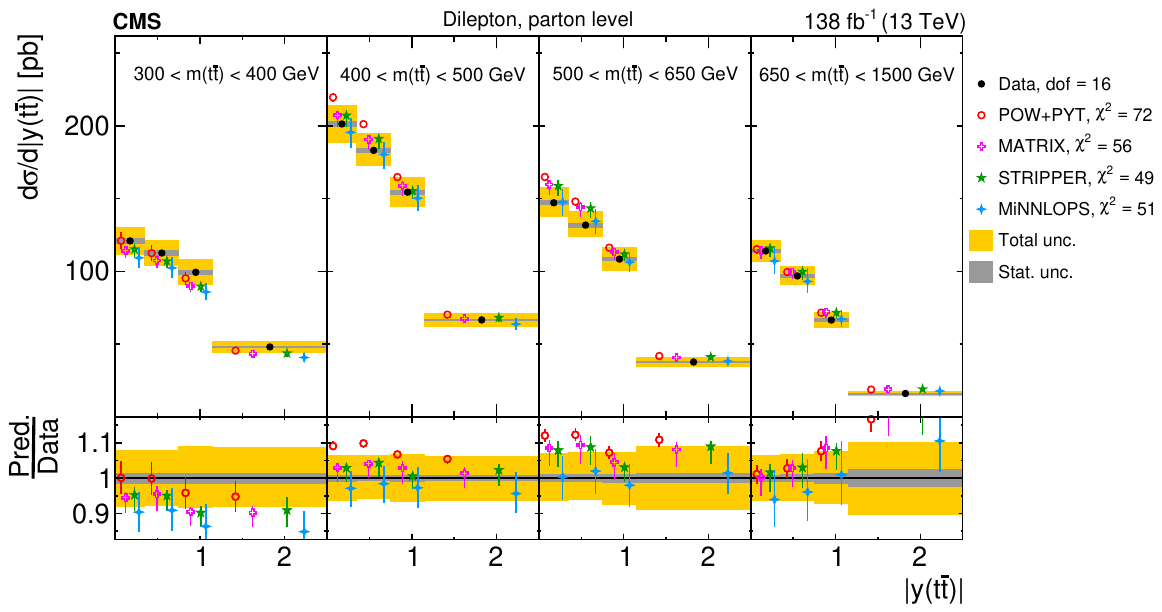}
\includegraphics[width=0.99\textwidth]{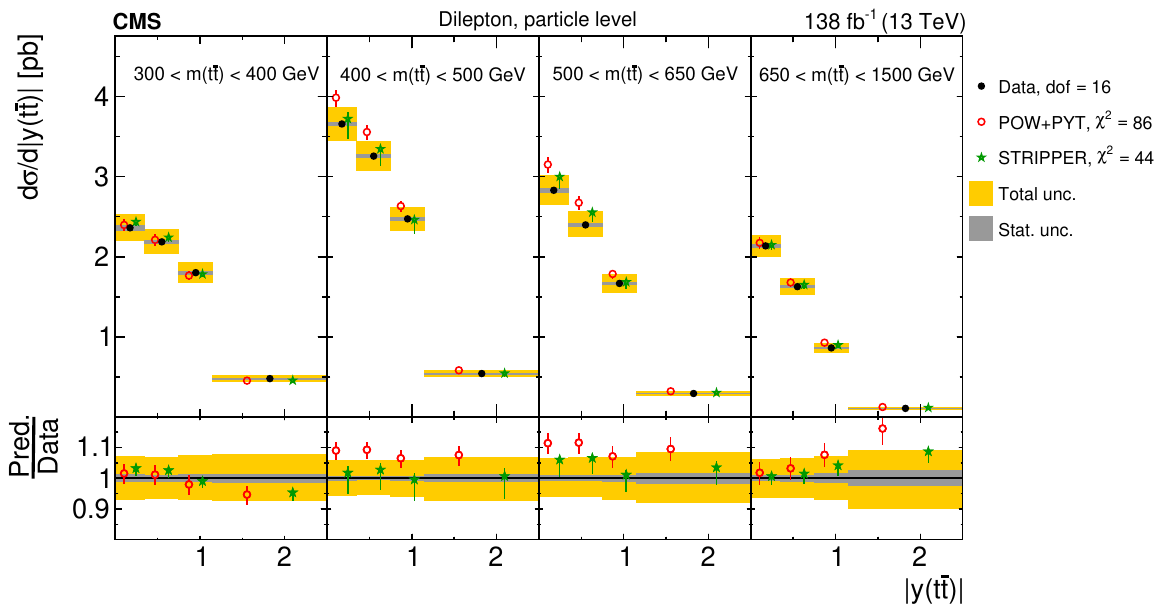}
\caption{Absolute \mttytt cross sections are shown for data (filled circles), \PowPyt (`POW-PYT', open circles)
simulation, and various theoretical predictions with beyond-NLO precision (other points).
    Further details can be found in the caption of Fig.~\ref{fig:xsec-md-theory-abs-ytptt}.}
\label{fig:xsec-md-theory-abs-mttytt}
\end{figure}

\begin{figure}
\centering
\includegraphics[width=0.99\textwidth]{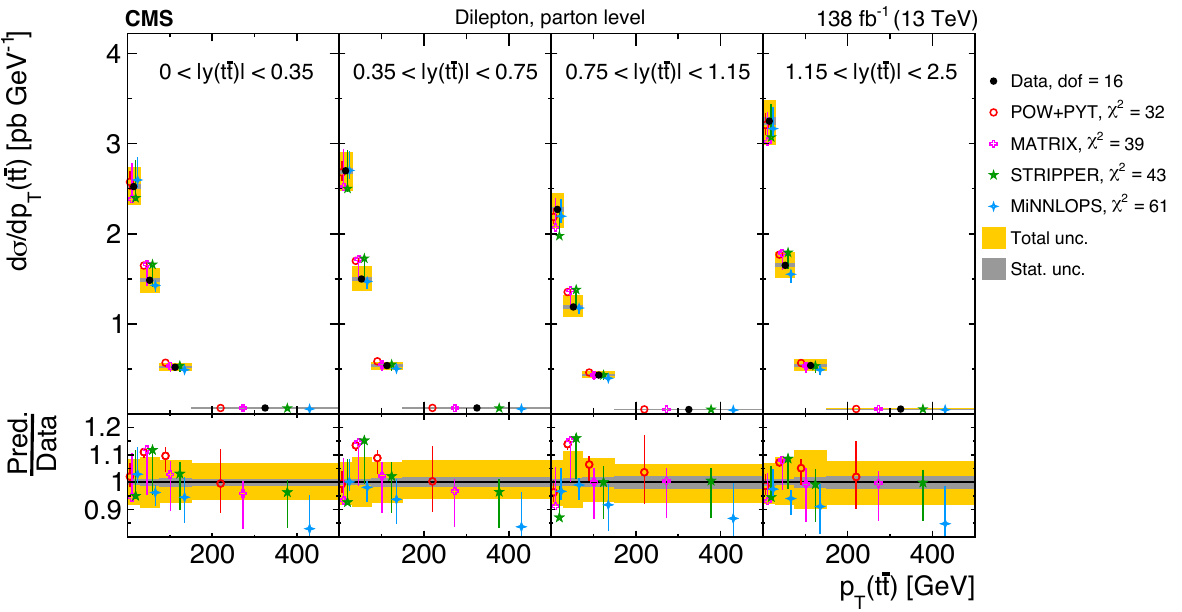}
\includegraphics[width=0.99\textwidth]{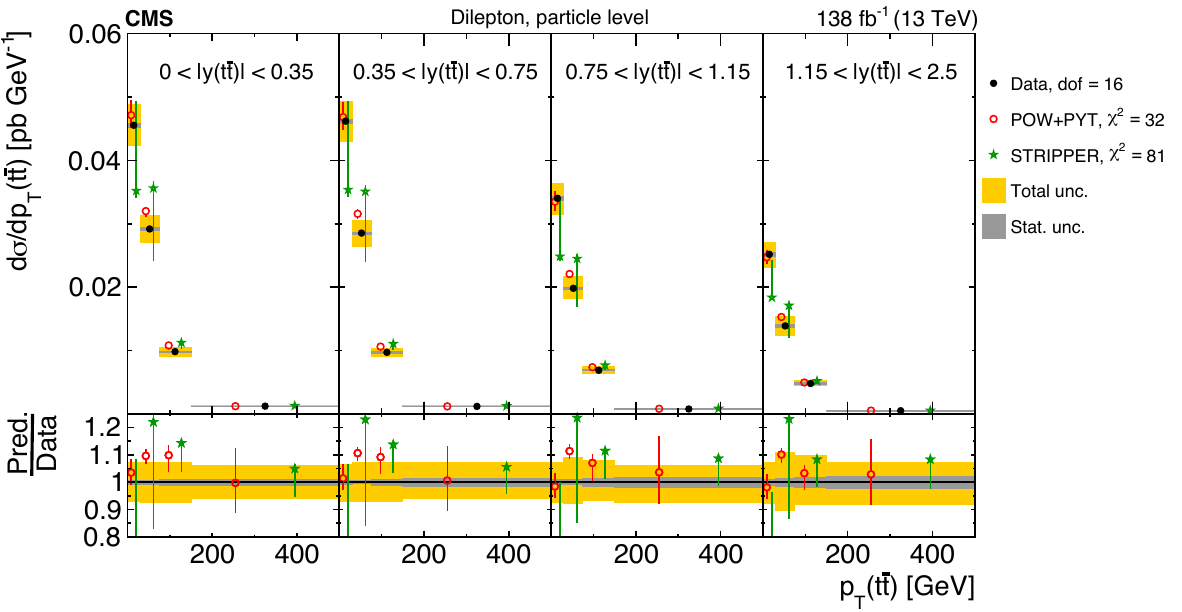}
\caption{Absolute \yttpttt cross sections are shown for data (filled circles), \PowPyt (`POW-PYT', open circles)
simulation, and various theoretical predictions with beyond-NLO precision (other points).
    Further details can be found in the caption of Fig.~\ref{fig:xsec-md-theory-abs-ytptt}.}
\label{fig:xsec-md-theory-abs-yttpttt}
\end{figure}

\begin{figure}
\centering
\includegraphics[width=0.99\textwidth]{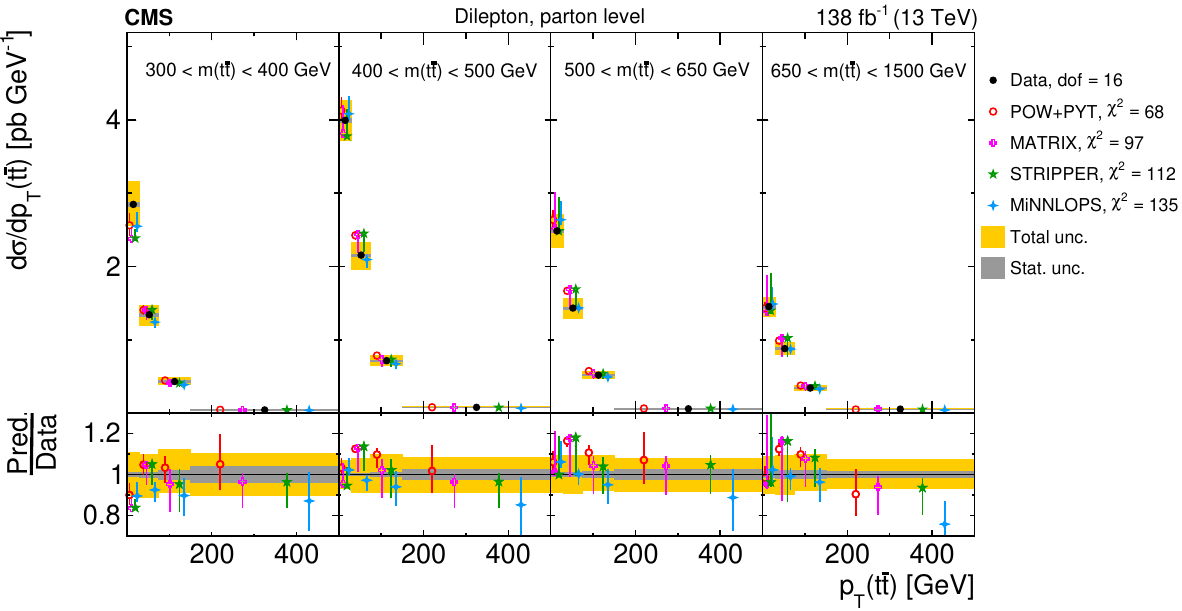}
\includegraphics[width=0.99\textwidth]{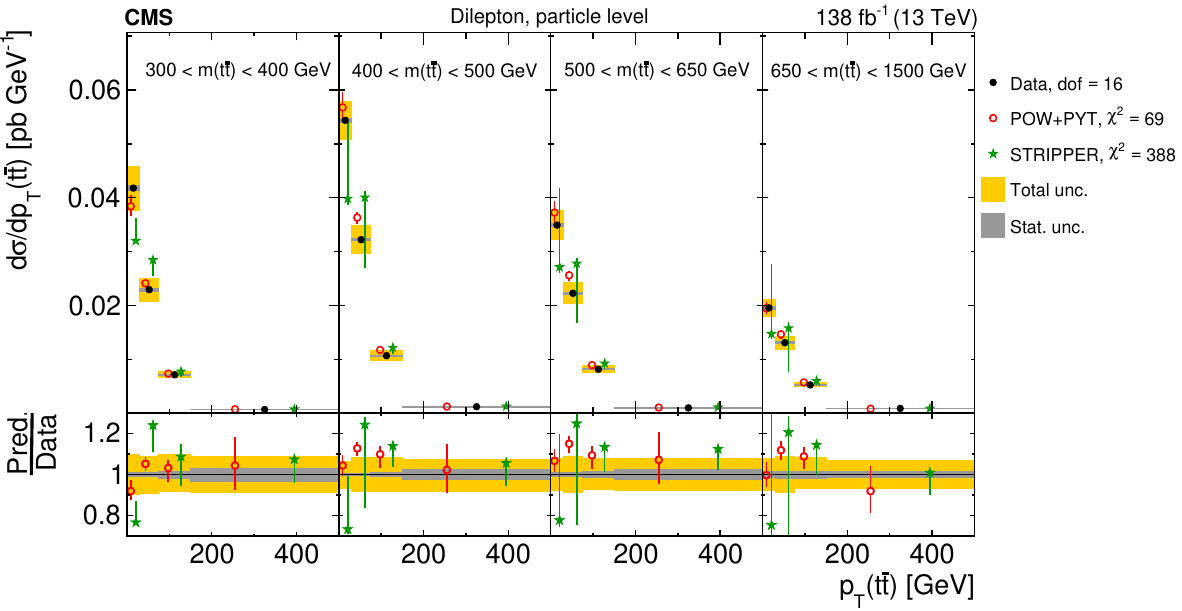}
\caption{Absolute \mttpttt cross sections are shown for data (filled circles), \PowPyt (`POW-PYT', open circles)
    simulation, and various theoretical predictions with beyond-NLO precision (other points).
    Further details can be found in the caption of Fig.~\ref{fig:xsec-md-theory-abs-ytptt}.}
\label{fig:xsec-md-theory-abs-mttpttt}
\end{figure}

\begin{figure}
\centering
\includegraphics[width=0.99\textwidth]{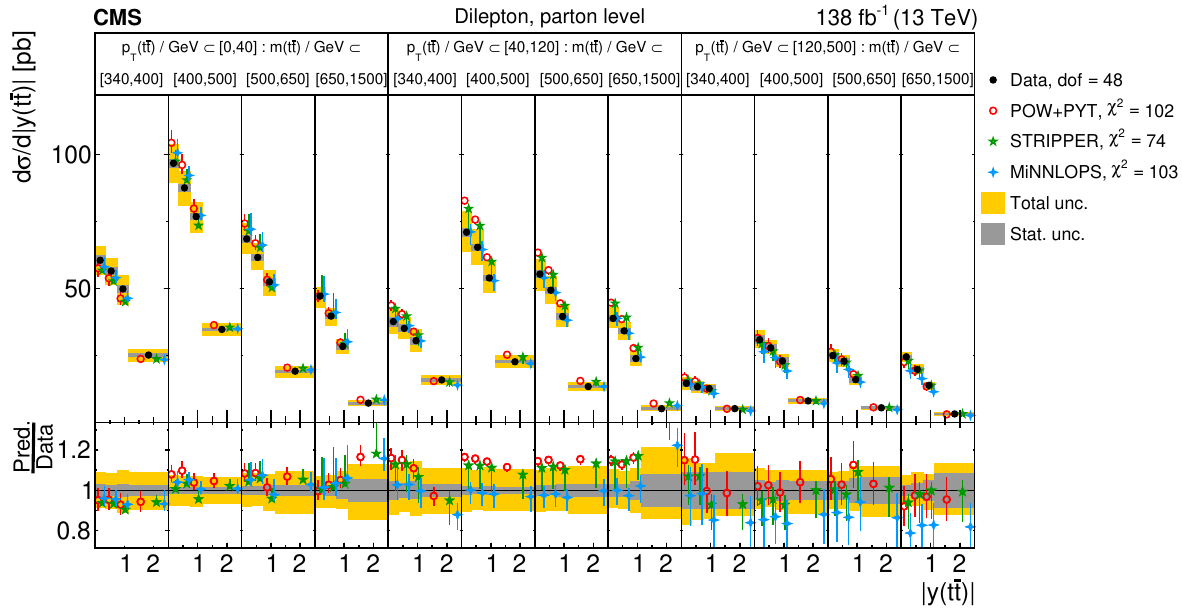}
\includegraphics[width=0.99\textwidth]{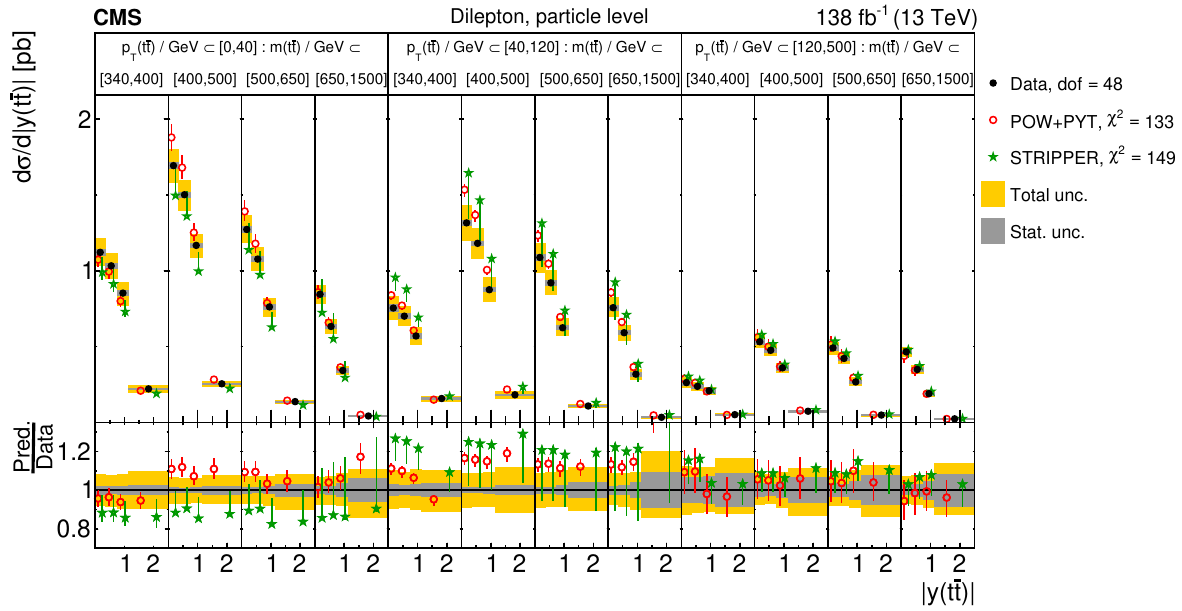}
\caption{Absolute \ptttmttytt cross sections are shown for data (filled circles),
    \PowPyt (`POW-PYT', open circles) simulation, and various theoretical predictions with beyond-NLO precision (other points).
    Further details can be found in the caption of Fig.~\ref{fig:xsec-md-theory-abs-ytptt}.}
\label{fig:xsec-md-theory-abs-ptttmttytt}
\end{figure}

\begin{figure}
\centering
\includegraphics[width=0.99\textwidth]{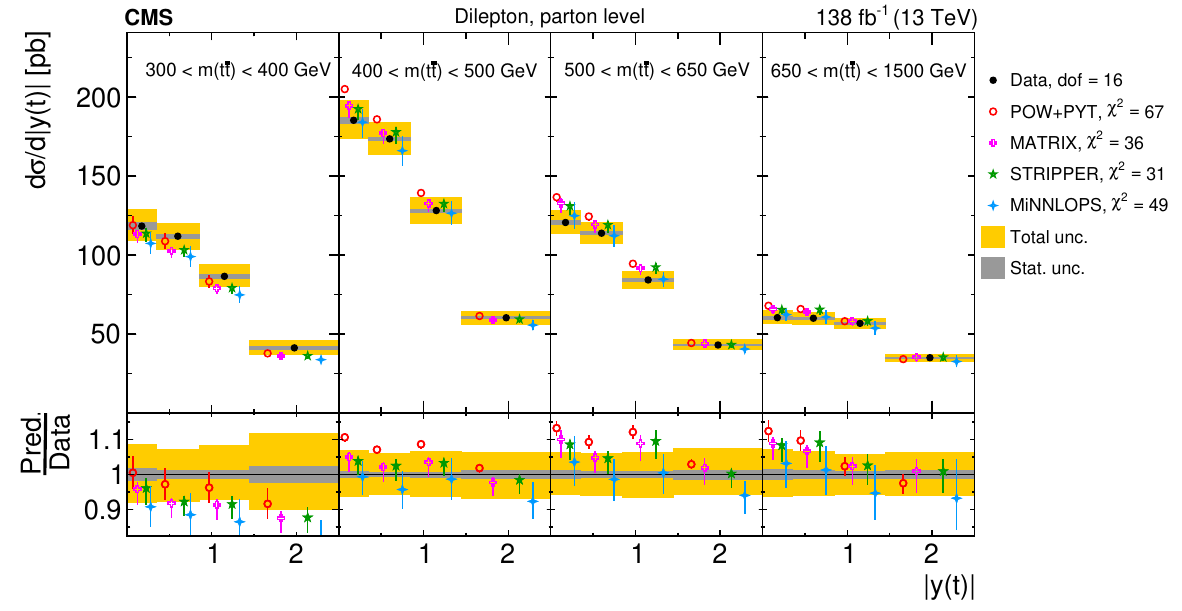}
\includegraphics[width=0.99\textwidth]{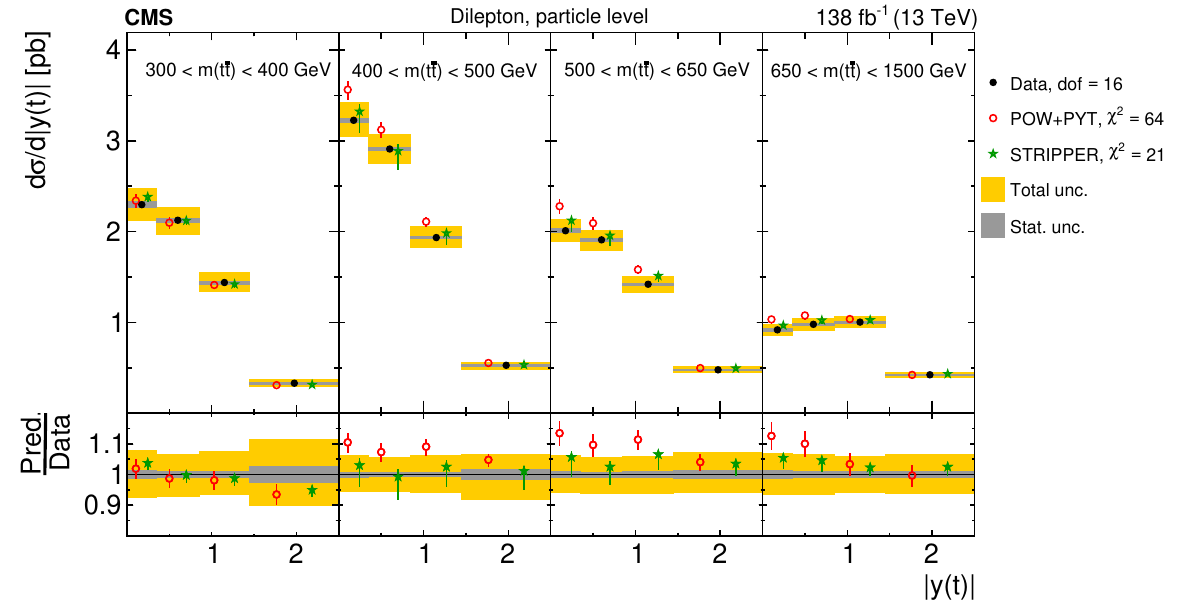}
\caption{Absolute \mttyt cross sections are shown for data (filled circles), \PowPyt (`POW-PYT', open circles)
simulation, and various theoretical predictions with beyond-NLO precision (other points).
    Further details can be found in the caption of Fig.~\ref{fig:xsec-md-theory-abs-ytptt}.}
\label{fig:xsec-md-theory-abs-mttyt}
\end{figure}

\begin{figure}
\centering
\includegraphics[width=0.99\textwidth]{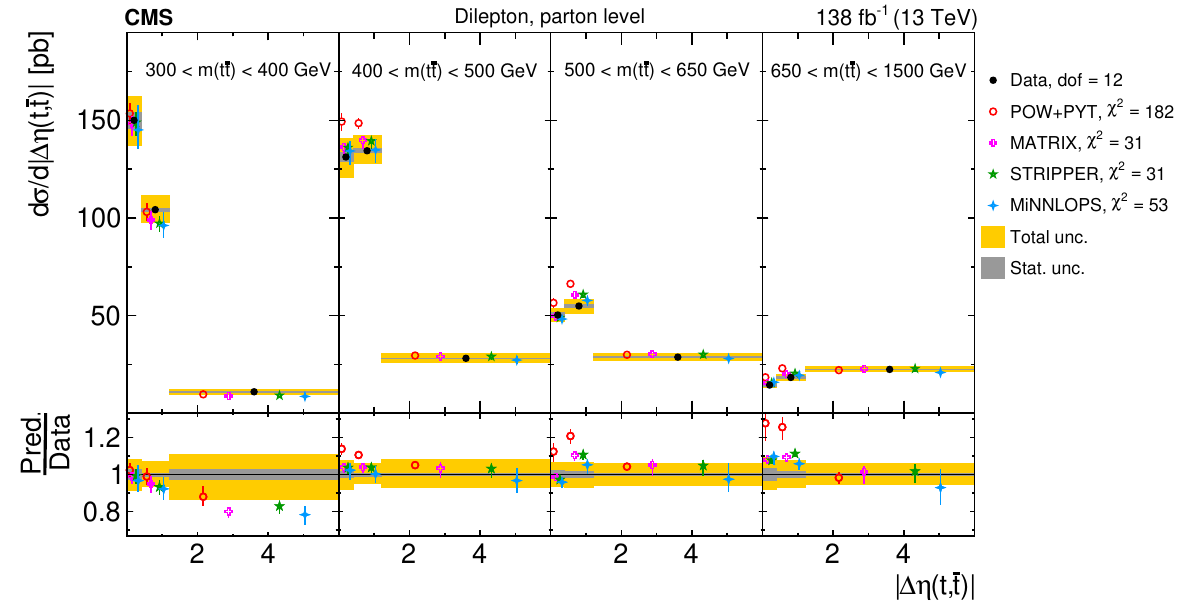}
\includegraphics[width=0.99\textwidth]{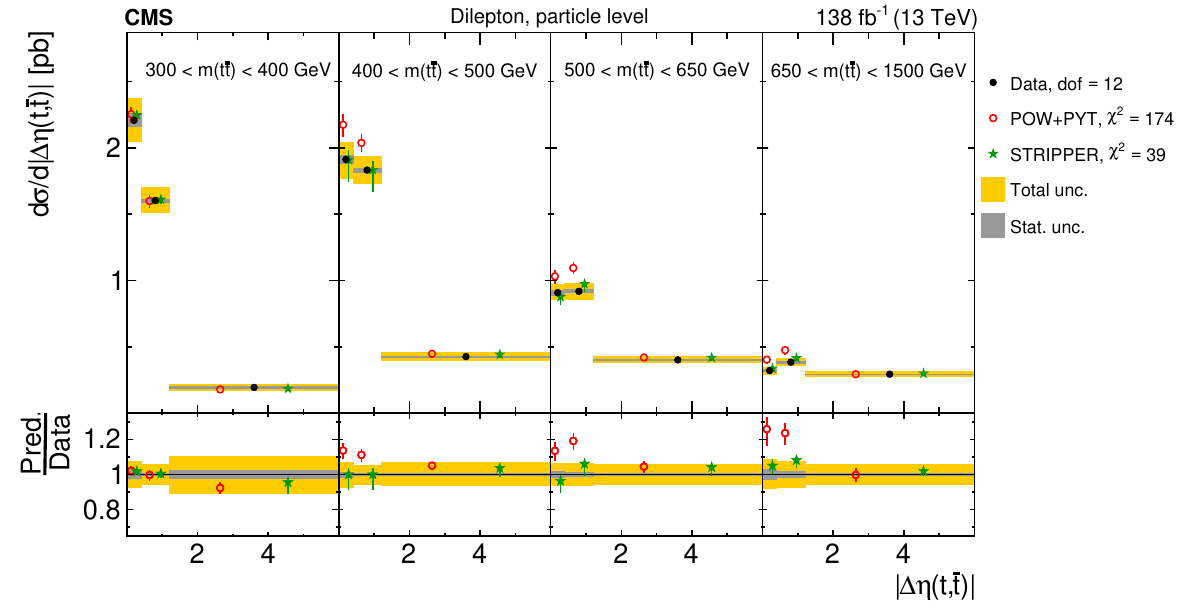}
\caption{Absolute \mttdetatt cross sections are shown for data (filled circles), \PowPyt (`POW-PYT', open circles)
simulation, and various theoretical predictions with beyond-NLO precision (other points).
    Further details can be found in the caption of Fig.~\ref{fig:xsec-md-theory-abs-ytptt}.}
\label{fig:xsec-md-theory-abs-mttdetatt}
\end{figure}

\begin{figure}
\centering
\includegraphics[width=0.99\textwidth]{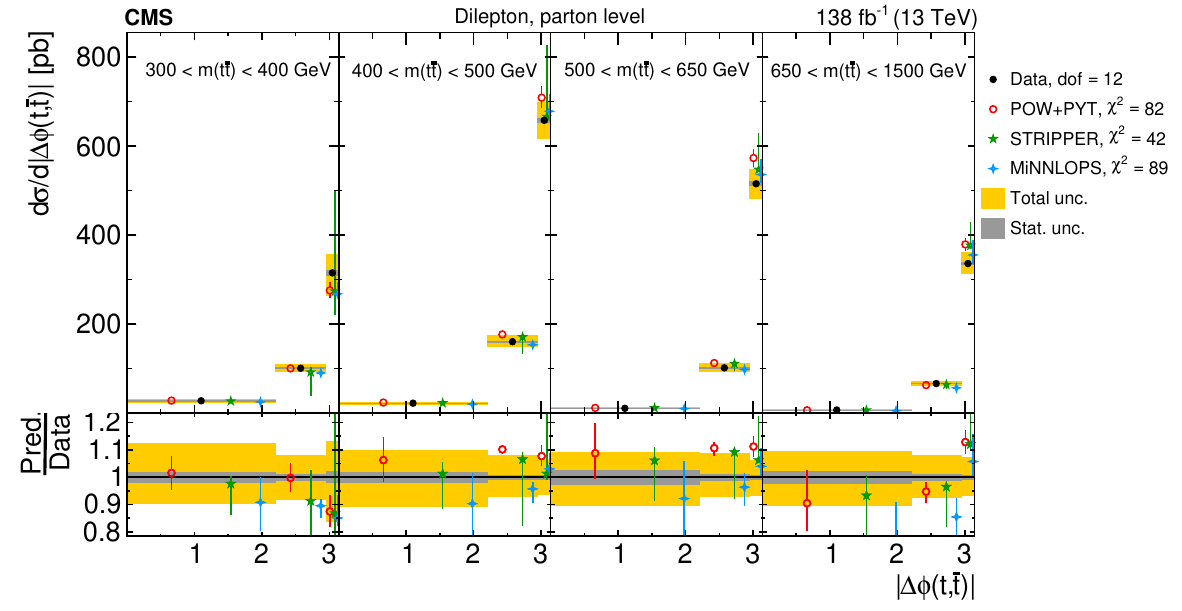}
\includegraphics[width=0.99\textwidth]{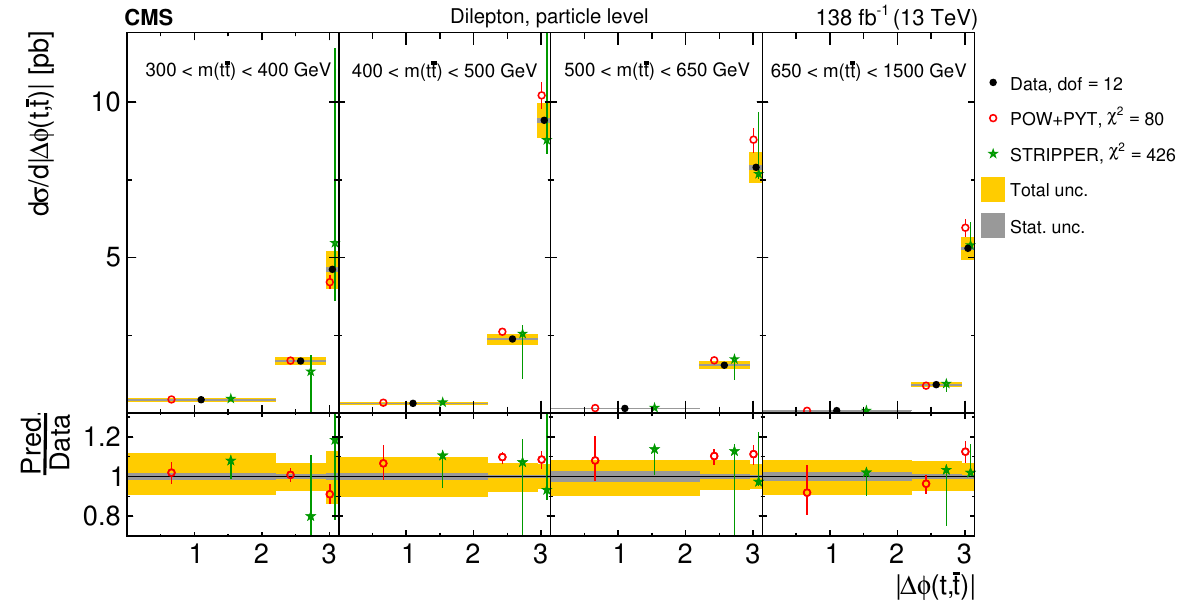}
\caption{Absolute \mttdphitt cross sections are shown for data (filled circles), \PowPyt (`POW-PYT', open circles)
simulation, and various theoretical predictions with beyond-NLO precision (other points).
    Further details can be found in the caption of Fig.~\ref{fig:xsec-md-theory-abs-ytptt}.}
\label{fig:xsec-md-theory-abs-mttdphitt}
\end{figure}

\begin{figure*}[!phtb]
\centering
\includegraphics[width=0.49\textwidth]{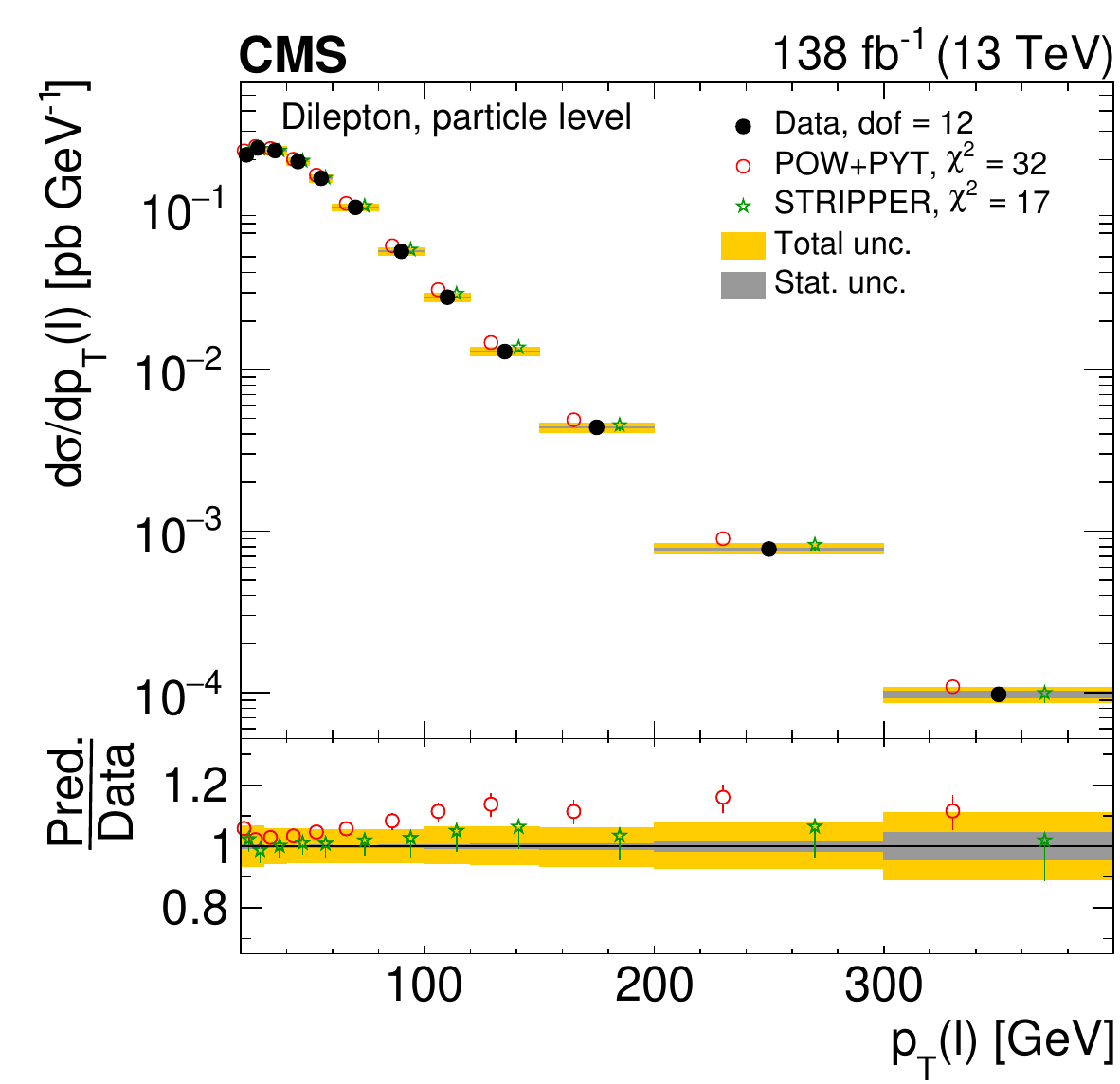}
\includegraphics[width=0.49\textwidth]{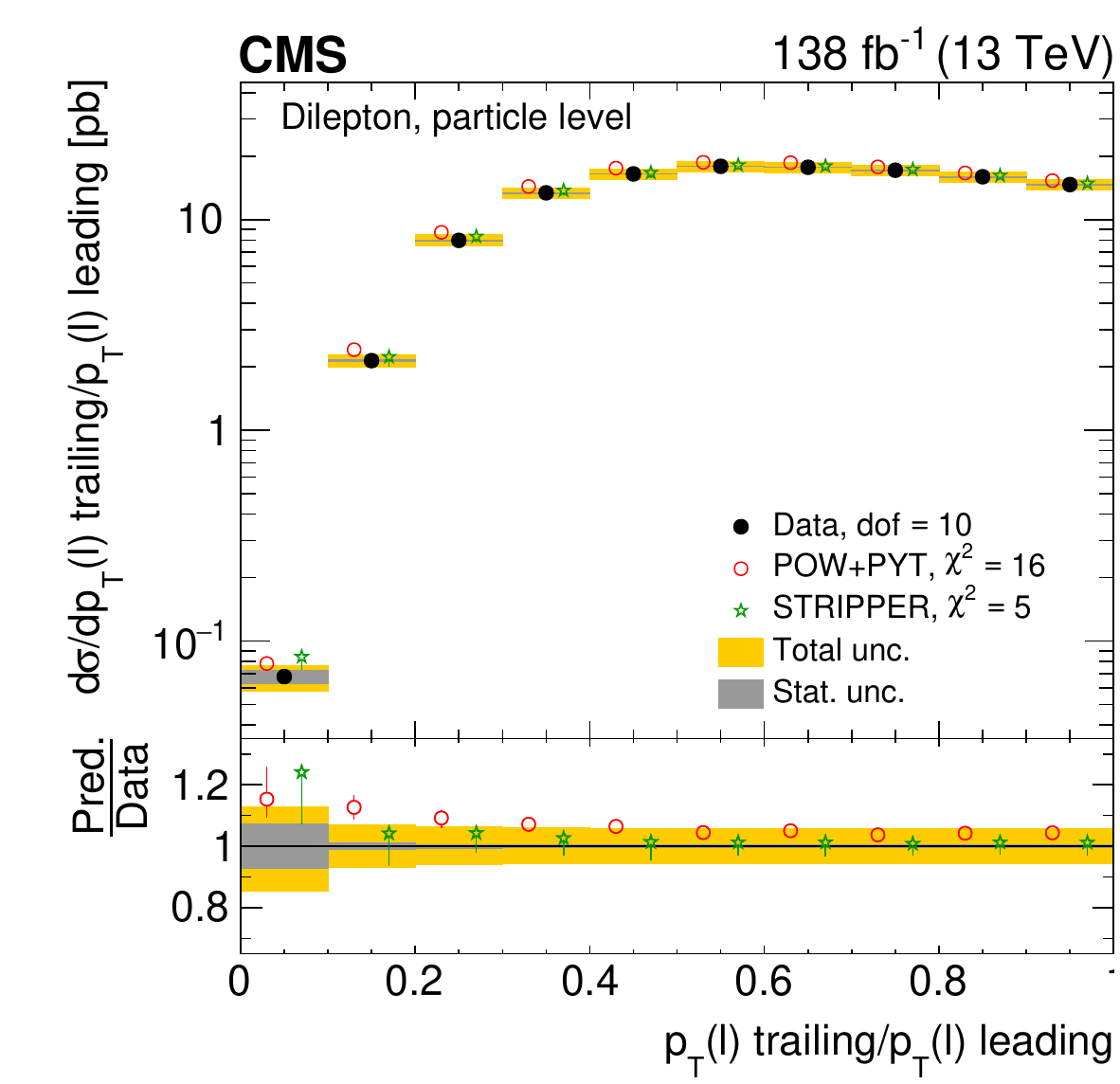}
\includegraphics[width=0.49\textwidth]{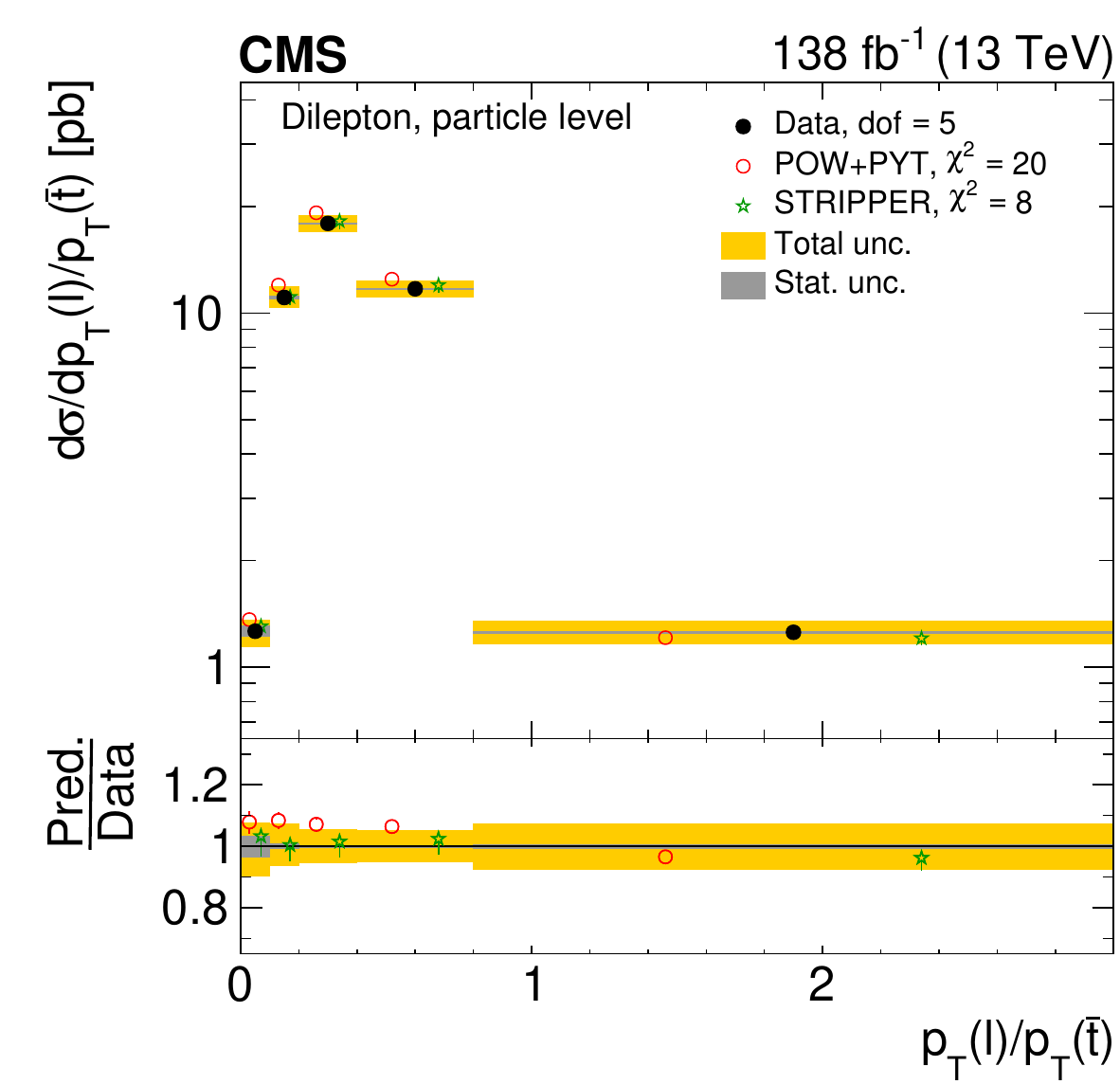}
\caption{Absolute differential \ttbar production cross sections as functions of \pt of the lepton (upper left), of the
ratio of the trailing and leading lepton \pt
(upper right), and of the ratio of lepton and top antiquark \pt (lower), measured at the particle level in a fiducial
phase space.
The data are shown as filled circles with grey and yellow bands indicating the statistical and total uncertainties
(statistical and systematic uncertainties added in quadrature), respectively.
For each distribution, the number of degrees of freedom (dof) is also provided.
The cross sections are compared to predictions from the \PowPyt (`POW-PYT', open circles) simulation and \StripperOnly NNLO
calculation (stars).
The estimated uncertainties in the predictions are represented by vertical bars on the corresponding points.
For each model, a value of \chisq is reported that takes into account the measurement uncertainties.
The lower panel in each plot shows the ratios of the predictions to the data.}
\label{fig:xsec-1d-theory-abs-ptlep-rptleptonic-rptlepptt}
\end{figure*}

\begin{figure*}[!phtb]
\centering
\includegraphics[width=0.49\textwidth]{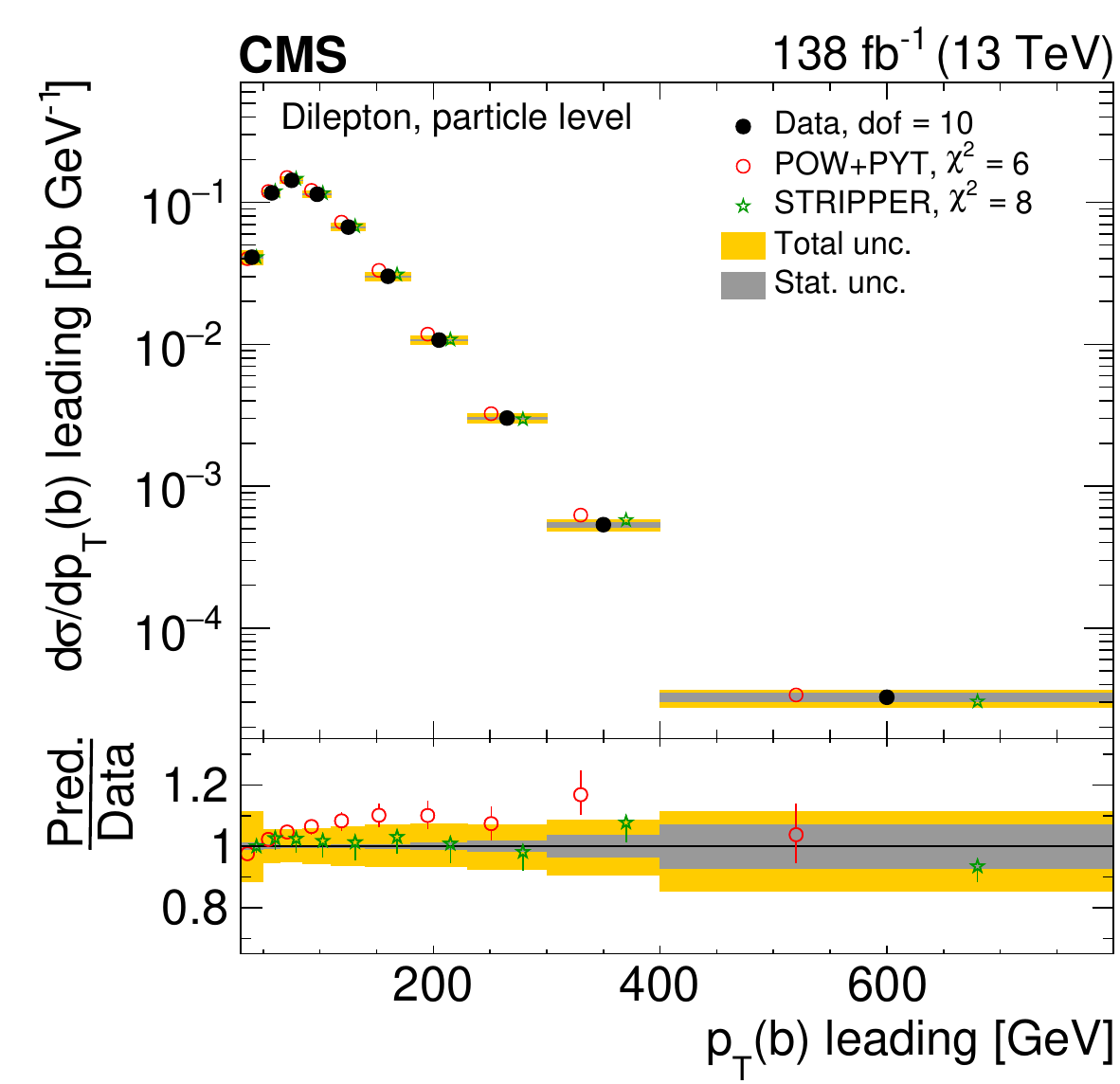}
\includegraphics[width=0.49\textwidth]{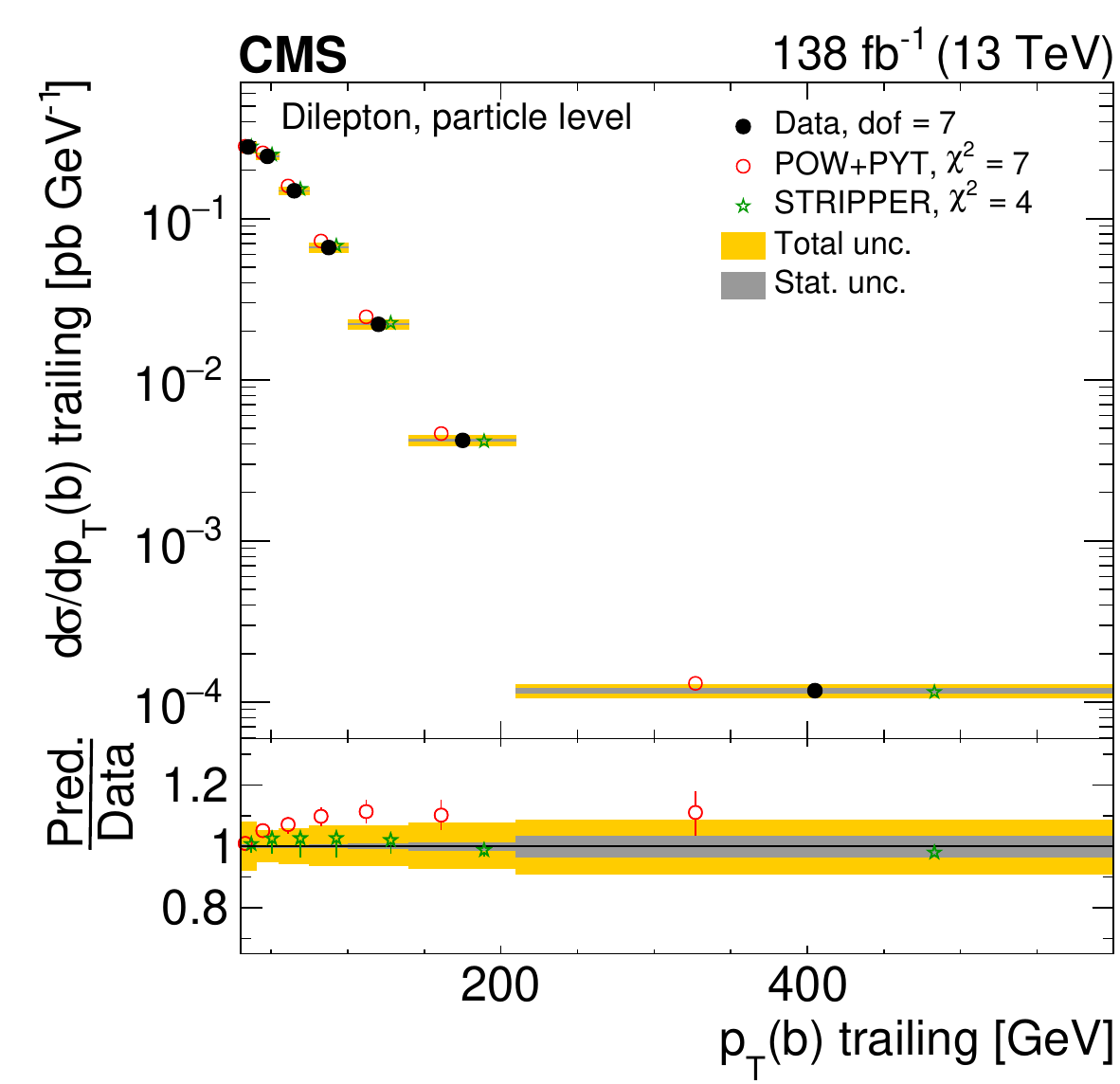}
\includegraphics[width=0.49\textwidth]{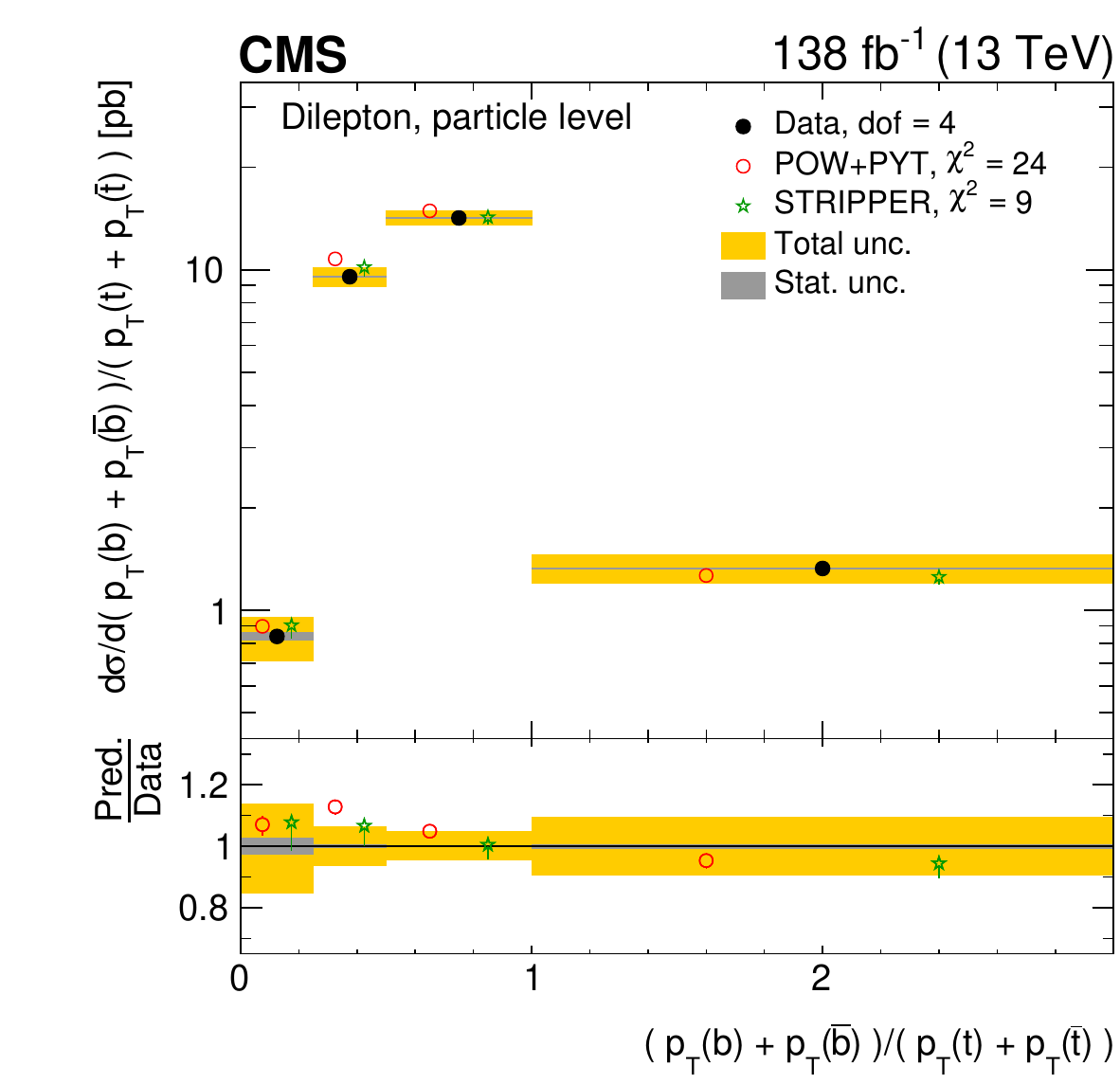}
\caption{Absolute differential \ttbar production cross sections as functions of the \pt of
the leading (upper left) and trailing (upper right) \PQb jet, and \rptbsptts (lower)
d-NLO precision (other points).
are shown for data (filled circles), \PowPyt (`POW-PYT', open circles) simulation, and \StripperOnly NNLO calculation (stars).
Further details can be found in the caption of Fig.~\ref{fig:xsec-1d-theory-abs-ptlep-rptleptonic-rptlepptt}.}
\label{fig:xsec-1d-theory-abs-ptb-ptnb-rptbsptts}
\end{figure*}

\begin{figure*}[!phtb]
\centering
\includegraphics[width=0.49\textwidth]{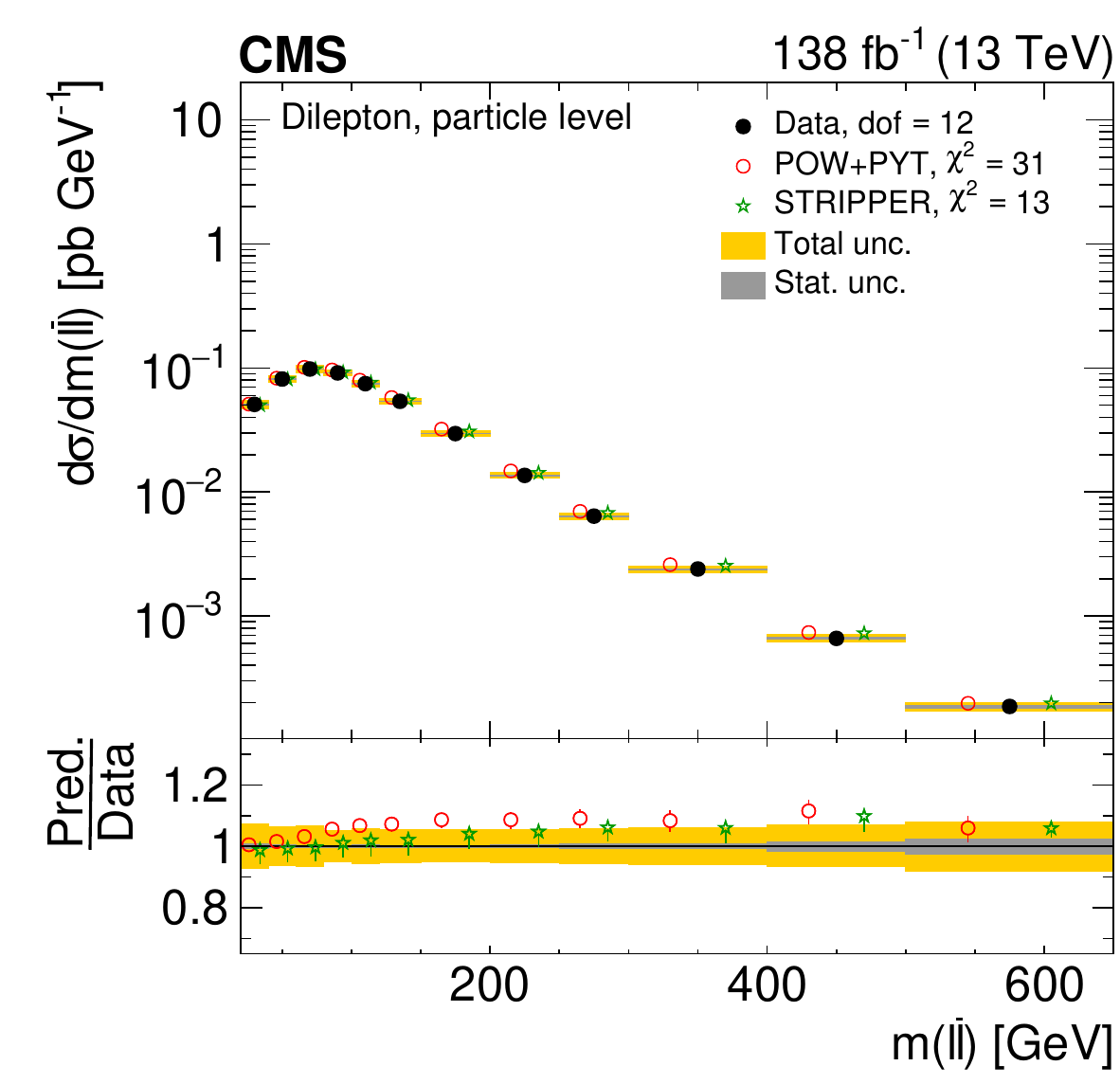}
\includegraphics[width=0.49\textwidth]{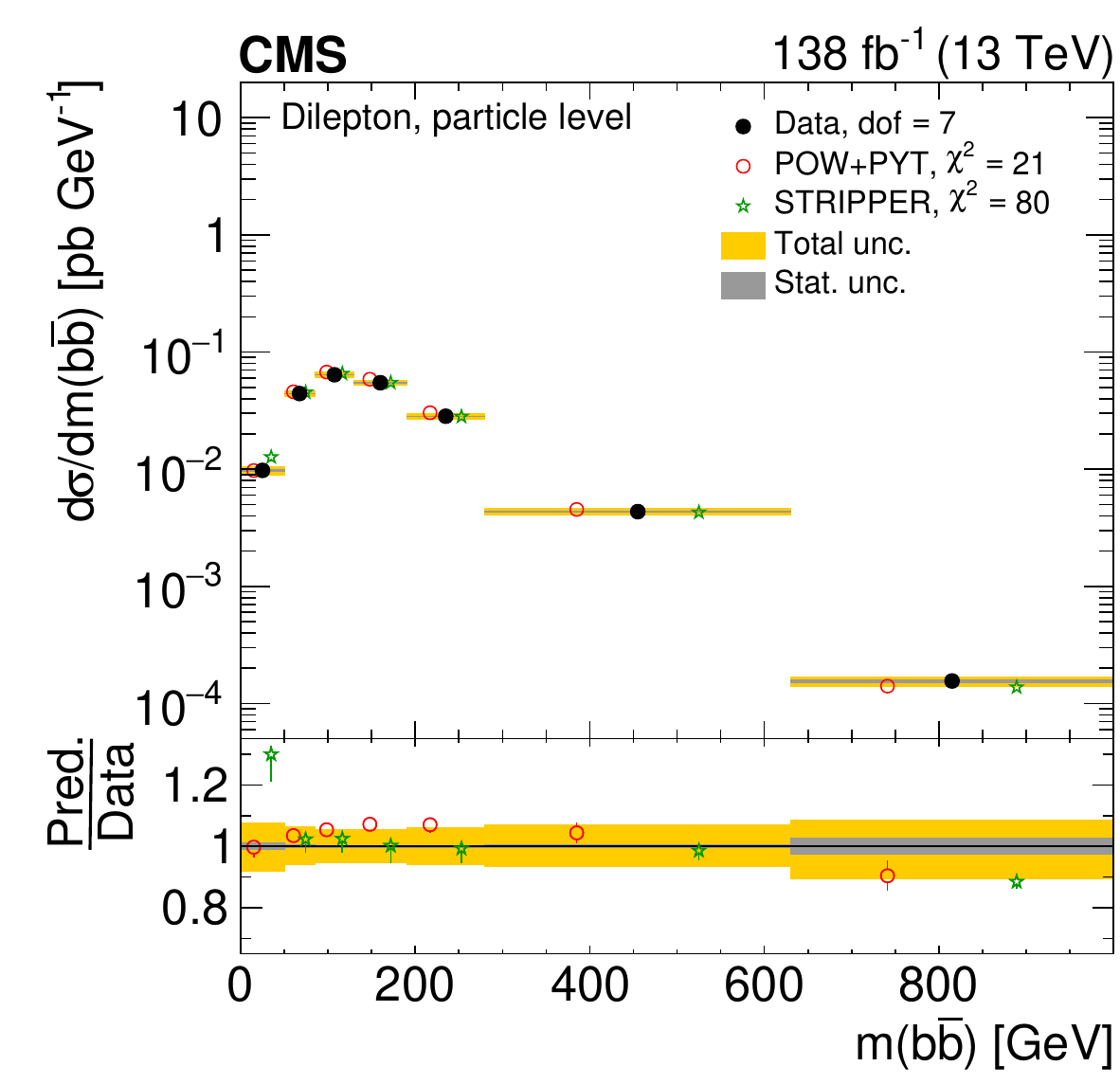}
\includegraphics[width=0.49\textwidth]{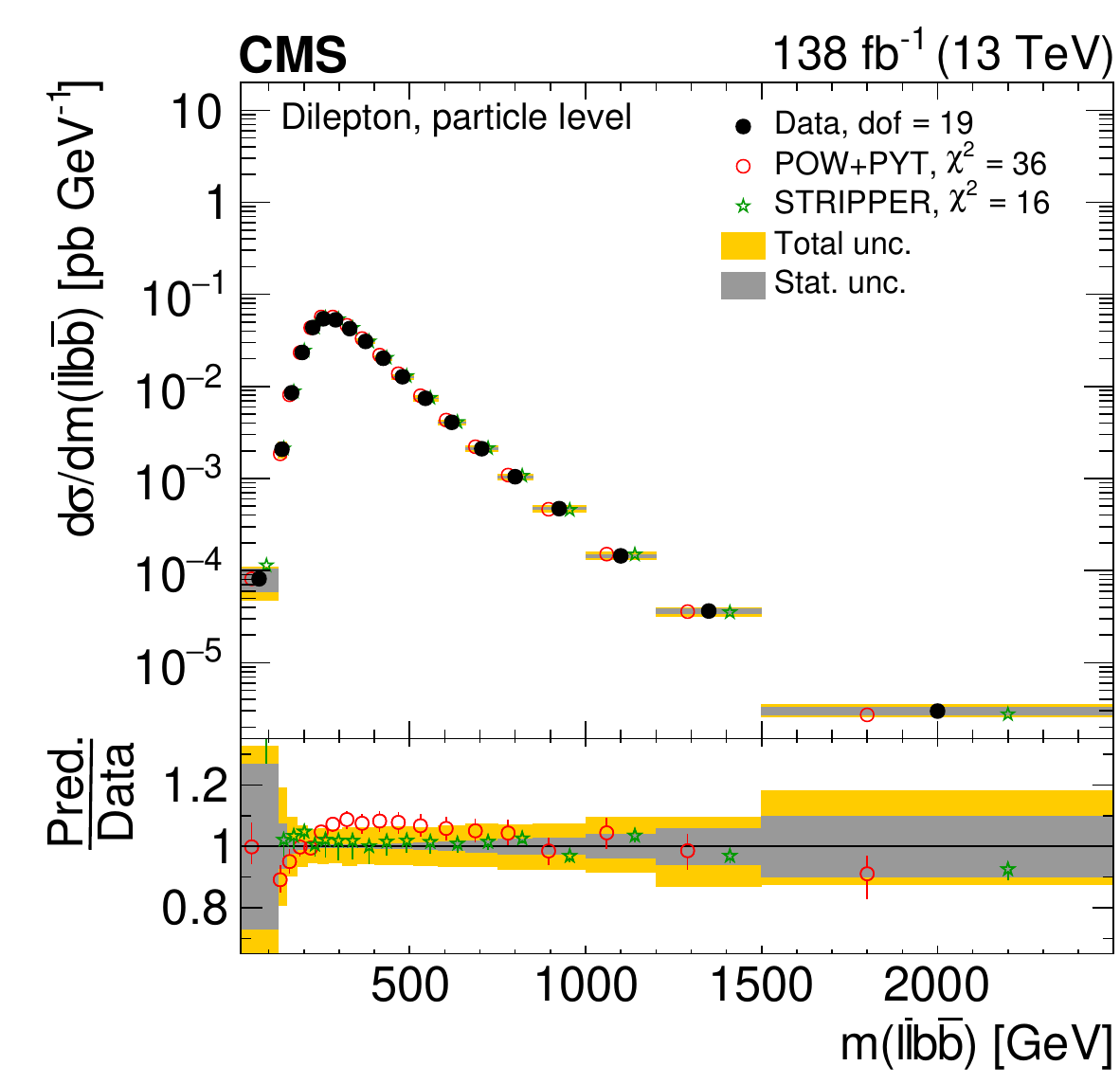}
\caption{Absolute differential \ttbar production cross sections as functions of \mll (upper left), \mbb (upper right),
and \mllbb (lower)
are shown for data (filled circles), \PowPyt (`POW-PYT', open circles) simulation, and \StripperOnly NNLO calculation (stars).
Further details can be found in the caption of Fig.~\ref{fig:xsec-1d-theory-abs-ptlep-rptleptonic-rptlepptt}.}
\label{fig:xsec-1d-theory-abs-mll-mbb-mllbb}
\end{figure*}

\begin{figure}
\centering
\includegraphics[width=0.49\textwidth]{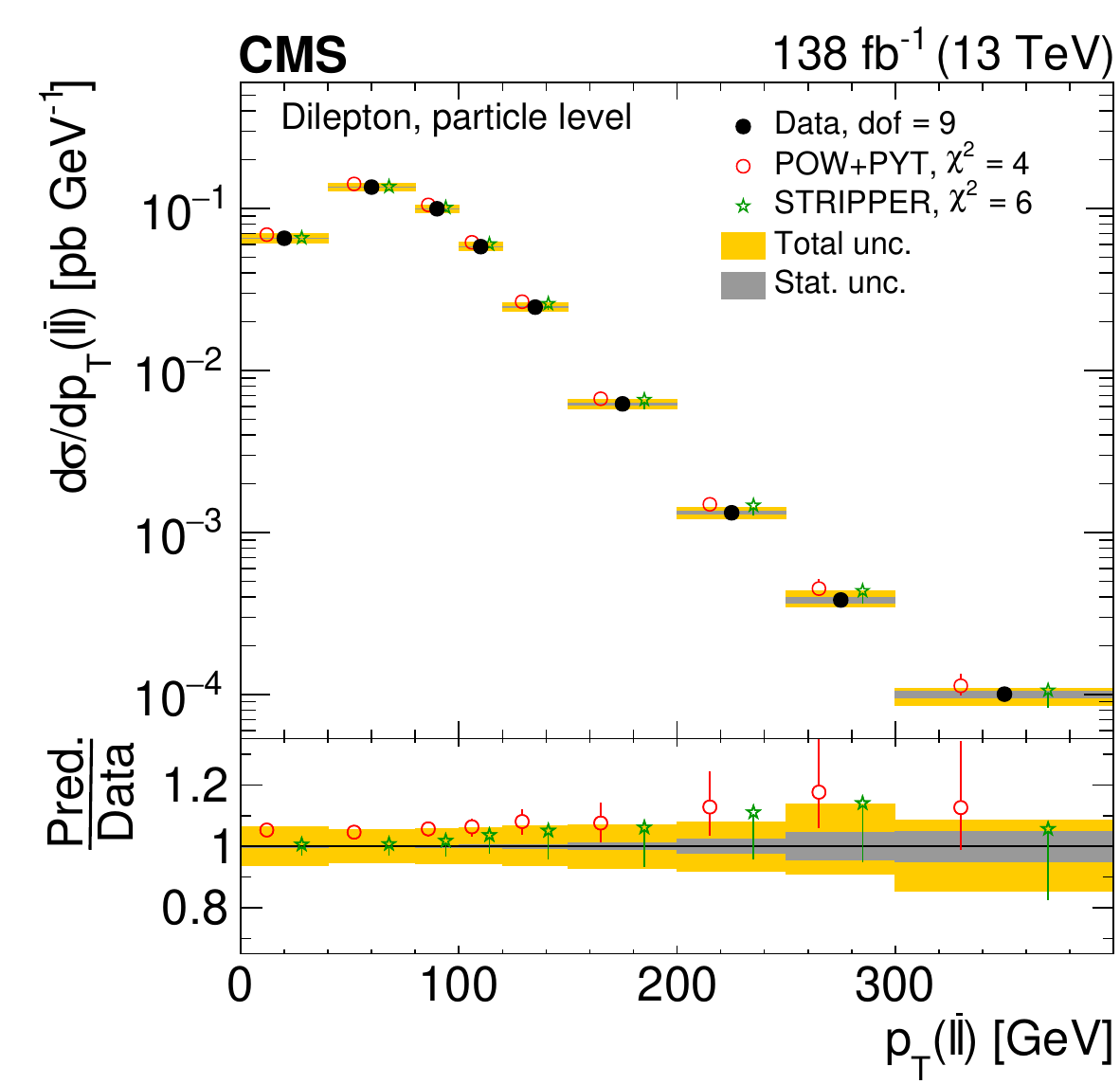}
\includegraphics[width=0.49\textwidth]{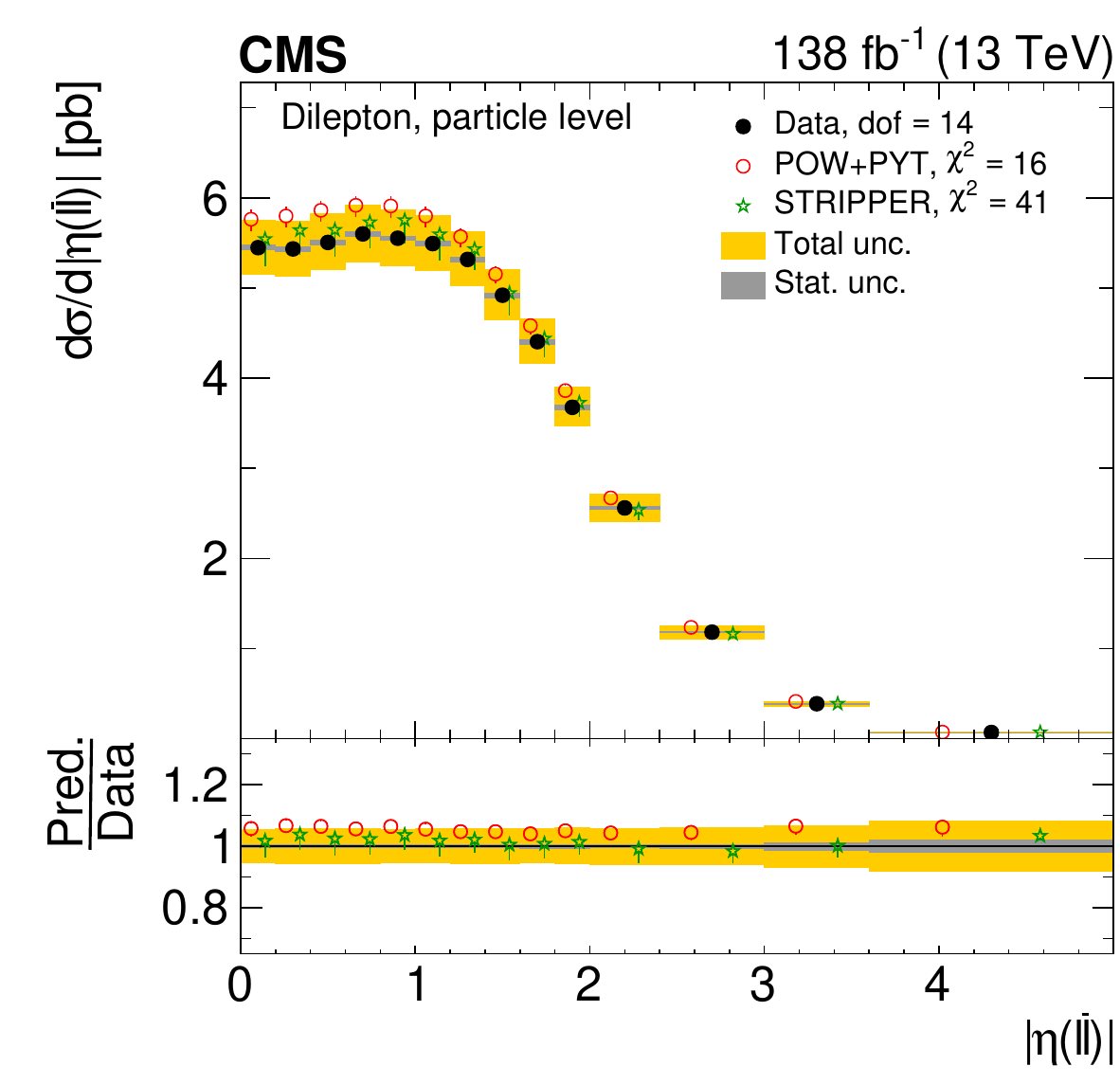}
\caption{Absolute differential \ttbar production cross sections as functions of \ptll (left) and \absetall (right)
are shown for data (filled circles), \PowPyt (`POW-PYT', open circles) simulation, and \StripperOnly NNLO calculation (stars).
Further details can be found in the caption of Fig.~\ref{fig:xsec-1d-theory-abs-ptlep-rptleptonic-rptlepptt}.}
\label{fig:xsec-1d-theory-abs-ptll-absetall}
\end{figure}

\begin{figure}
\centering
\includegraphics[width=1.00\textwidth]{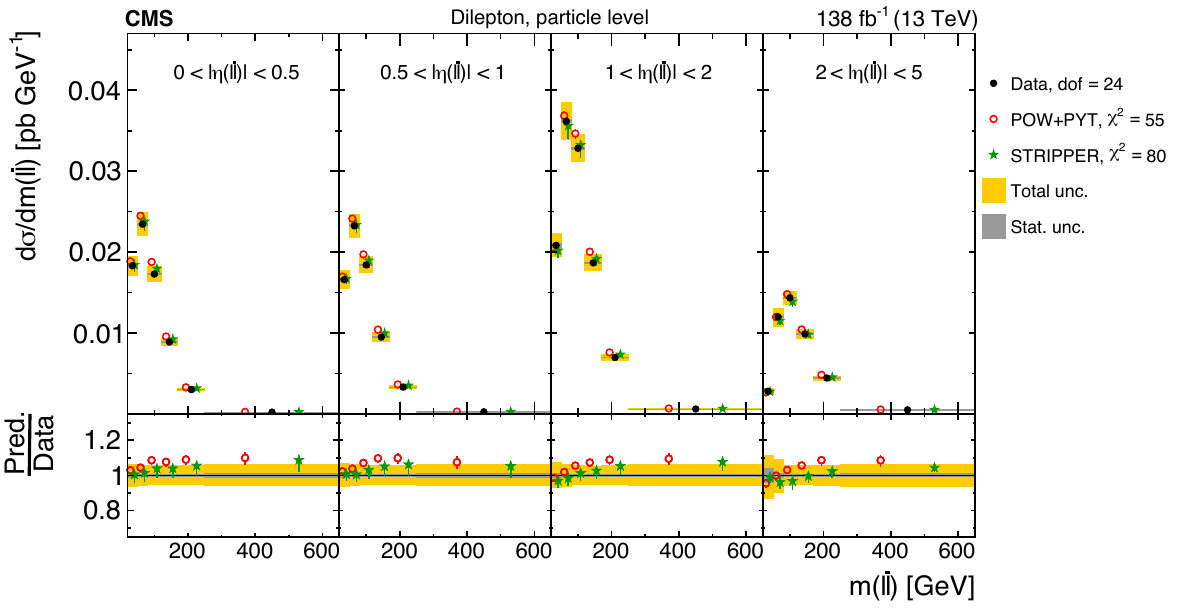}
\caption{Absolute \etallmll cross sections are shown for data (filled circles), \PowPyt (`POW-PYT', open circles)
simulation, and \StripperOnly NNLO calculation (stars).
    Further details can be found in the caption of Fig.~\ref{fig:xsec-1d-theory-abs-ptlep-rptleptonic-rptlepptt}.}
\label{fig:xsec-2d-theory-abs-etallmll}
\end{figure}

\begin{figure}
\centering
\includegraphics[width=1.00\textwidth]{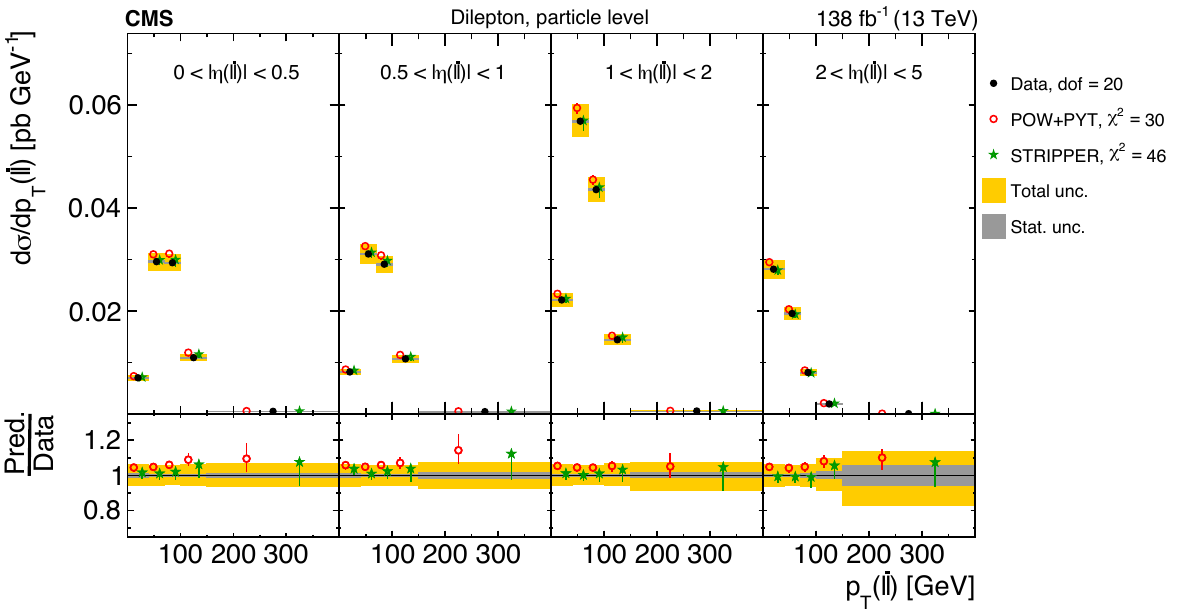}
\caption{Absolute \etallptll cross sections are shown for data (filled circles), \PowPyt (`POW-PYT', open circles)
simulation, and \StripperOnly NNLO calculation (stars).
    Further details can be found in the caption of Fig.~\ref{fig:xsec-1d-theory-abs-ptlep-rptleptonic-rptlepptt}.}
\label{fig:xsec-2d-theory-abs-etallptll}
\end{figure}

\begin{figure}
\centering
\includegraphics[width=1.00\textwidth]{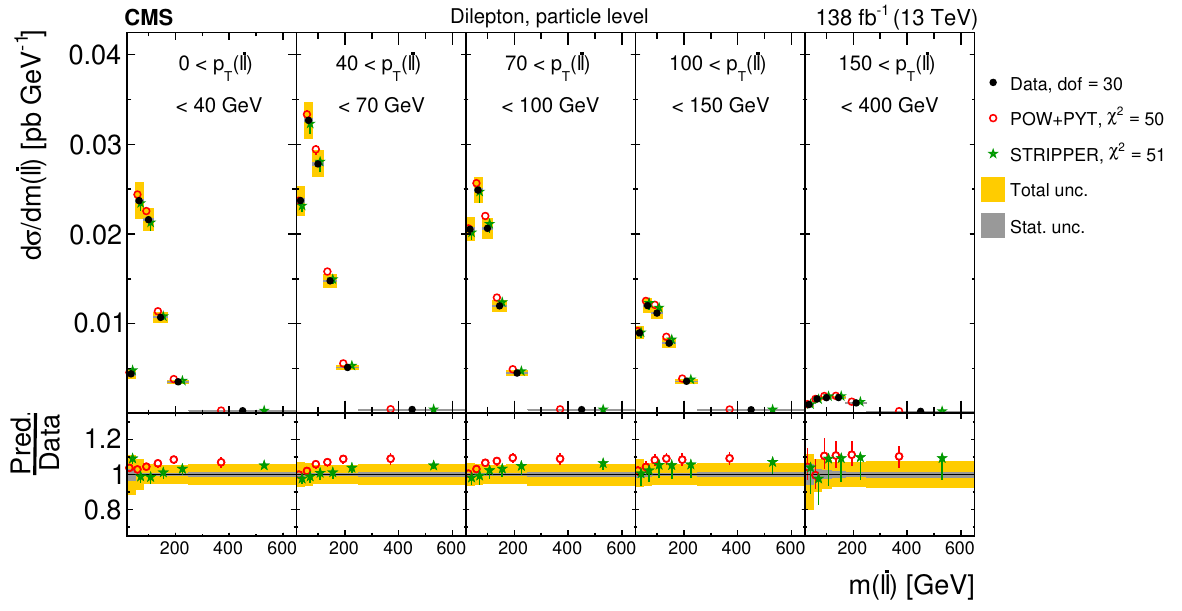}
\caption{Absolute \ptllmll cross sections are shown for data (filled circles), \PowPyt (`POW-PYT', open circles)
simulation, and \StripperOnly NNLO calculation (stars).
    Further details can be found in the caption of Fig.~\ref{fig:xsec-1d-theory-abs-ptlep-rptleptonic-rptlepptt}.}
\label{fig:xsec-2d-theory-abs-ptllmll}
\end{figure}

The absolute differential cross sections comparing data to theoretical predictions of beyond-NLO precision
are shown in Figs.~\ref{fig:xsec-1d-theory-abs-ptt-ptat}--\ref{fig:xsec-2d-theory-abs-ptllmll},
and the corresponding \chisq values are given in 
Tables~\ref{tab:chi2fixTheo_1d_abs_parton}--\ref{tab:chi2fixTheo_1d_abs_particle_lepb}. The $p$-values of the \chisq tests are presented in Tables~\ref{tab:pvaluefixTheo_1d_abs_parton}--\ref{tab:pvaluefixTheo_1d_abs_particle_lepb}.
The theoretical calculations are those discussed in Section~\ref{sec:res_theory_comp}:
\appNTLO, \Stripper, \Matrix, and \MiNNLOPS.
Comparing the absolute cross sections to the normalized ones in Section~\ref{sec:res_theory_comp},
one can see a similar level of agreement between data and predictions.
We conclude that the calculations provide overall reasonable predictions
of the total normalization of the data.

\clearpage

\clearpage

\clearpage

\clearpage

\begin{sidewaystable*}
\centering
 \topcaption{The \chisq values and \ndf of the measured absolute single-differential cross sections for \ttbar and top quark kinematic observables at the parton level are shown with respect to the \PowPyt (`POW-PYT') simulation and various theoretical predictions with beyond-NLO precision. 
The \chisq values are calculated taking only measurement uncertainties into account and excluding theory uncertainties.  
For \PowPytSh, the \chisq values including theory uncertainties are indicated with the brackets (w. unc.).}
 \label{tab:chi2fixTheo_1d_abs_parton}
 \renewcommand{\arraystretch}{1.4}
 \centering
 \begin{tabular}{lcccccc}
 \multirow{1}{*}{Cross section} & \hspace*{0.3 cm} \multirow{2}{*}{\ndf} \hspace*{0.3 cm} & \multicolumn{5}{c}{\chisq} \\
 \cline{3-7}
{variables} && \PowPytSh (w. unc.)  & \appNTLOOnly & \MatrixOnly  & \StripperOnly  & \MiNNLOPSOnly   \\
\hline
\ptt& 7 & 21 \: (13) & 47 & 3  & 15 & 6 \\
\ptat& 7 & 19 \: (12) & \NA & 5  & 10  & 7  \\
\yt& 10 & 28 \: (24) & 24  & 17  & 17 & 14  \\
\yat& 10 & 33 \: (28) & \NA & 15  & 26  & 22  \\
\pttt& 7 & 24 \: (8) & \NA & 7  & 10  & 51  \\
\ytt& 12 & 13 \: (9) & \NA & 7  & 10  & 9  \\
\mtt  & 7 & 6 \: (4) & \NA & 6  & 5  & 2  \\
\dphitt& 4 & 4 \: (2) & \NA & 91 & 78  & 5  \\
\dytt& 8 & 18 \: (10) & \NA & 2  & 3 & 12  \\
\rpttmtt& 5 & 39 \: (21) & \NA & 4  & 5  & 9  \\
\rptttmtt& 9 & 20 \: (7) & \NA & 253  & 235 & 18  \\
\logxone& 9 & 16 \: (12) & \NA & \NA & 15 & 12  \\
\logxtwo& 9 & 14 \: (9) & \NA & \NA & 13  & 7  \\
 \end{tabular}
\end{sidewaystable*}

\begin{table*}
\centering
 \topcaption{The \chisq values and \ndf of the measured absolute single-differential cross sections for \ttbar and top quark kinematic observables at the particle level are shown with respect to the \PowPyt (`POW-PYT') simulation and the \StripperOnly NNLO calculation. 
The \chisq values are calculated taking only measurement uncertainties into account and excluding theory uncertainties.  
For \PowPytSh, the \chisq values including theory uncertainties are indicated with the brackets (w. unc.).}
\label{tab:chi2fixTheo_1d_abs_particle_ttbar}
 \renewcommand{\arraystretch}{1.4}
 \centering
 \begin{tabular}{lccc}
 \multirow{1}{*}{Cross section} & \hspace*{0.3 cm} \multirow{2}{*}{\ndf} \hspace*{0.3 cm} & \multicolumn{2}{c}{\chisq} \\
 \cline{3-4}
{variables} && \PowPytSh (w. unc.)  & \StripperOnly  \\
\hline
\ptt& 7 & 22 \: (13) & 3  \\
\ptat& 7 & 20 \: (12) & 4  \\
\yt& 10 & 24 \: (18) & 11 \\
\yat& 10 & 28 \: (23) & 10  \\
\pttt& 7 & 23 \: (8) & 87 \\
\ytt& 12 & 13 \: (8) & 20 \\
\mtt  & 7 & 7 \: (4) & 4  \\
\dphitt& 4 & 4 \: (1) & 1412  \\
\dytt& 8 & 17 \: (11) & 4  \\
\rpttmtt& 5 & 33 \: (23) & 9  \\
\rptttmtt& 9 & 21 \: (7) & 285  \\
\logxone& 9 & 16 \: (10) & 10  \\
\logxtwo& 9 & 12 \: (7) & 10  \\
 \end{tabular}
\end{table*}

\begin{sidewaystable*}
 \centering
 \topcaption{The \chisq values and \ndf of the measured absolute multi-differential cross sections for \ttbar and top quark kinematic observables at the parton level are shown with respect to the \PowPyt (`POW-PYT') simulation and various theoretical predictions with beyond-NLO precision. 
The \chisq values are calculated taking only measurement uncertainties into account and excluding theory uncertainties.  
For \PowPytSh, the \chisq values including theory uncertainties are indicated with the brackets (w. unc.).}
 \label{tab:chi2fixTheo_abs_parton_ttbar}
 \renewcommand{\arraystretch}{1.4}
 \centering
 \begin{tabular}{lcccccc}
 \multirow{1}{*}{Cross section} & \hspace*{0.3 cm} \multirow{2}{*}{\ndf} \hspace*{0.3 cm} & \multicolumn{5}{c}{\chisq} \\
 \cline{3-7}
{variables} && \PowPytSh (w. unc.) & \appNTLOOnly  & \MatrixOnly & \StripperOnly & \MiNNLOPSOnly  \\
\hline
\ytptt& 16 & 48 \: (36) & 35 & 24  & 25  & 21  \\
\mttptt& 9 & 93 \: (36) & \NA & 19  & 17 & 17  \\
\pttpttt& 16 & 50 \: (25) & \NA & 42  & 57  & 61 \\
\mttytt  & 16 & 72 \: (46) & \NA & 56 & 49 & 51 \\
\yttpttt& 16 & 32 \: (17) & \NA & 39 & 43 & 61 \\
\mttpttt  & 16 & 68 \: (47) & \NA & 97  & 112 & 135 \\
\ptttmttytt  & 48 & 102 \: (71) & \NA & \NA & 74 & 103 \\
\mttyt& 16 & 67 \: (39) & \NA & 36 & 31 & 49  \\
\mttdetatt& 12 & 182 \: (34) & \NA & 31 & 31 & 53 \\
\mttdphitt& 12 & 82 \: (51) & \NA & \NA & 42  & 89 \\
 \end{tabular}
\end{sidewaystable*}

\begin{table*}
 \centering
 \topcaption{The \chisq values and \ndf of the measured absolute multi-differential cross sections for \ttbar and top quark kinematic observables at the particle level are shown with respect to the \PowPyt (`POW-PYT') simulation and the \StripperOnly NNLO calculation. 
The \chisq values are calculated taking only measurement uncertainties into account and excluding theory uncertainties.  
For \PowPytSh, the \chisq values including theory uncertainties are indicated with the brackets (w. unc.).}
 \label{tab:chi2fixTheo_abs_particle_ttbar}
 \renewcommand{\arraystretch}{1.4}
 \centering
 \begin{tabular}{lccc}
 \multirow{1}{*}{Cross section} & \hspace*{0.3 cm} \multirow{2}{*}{\ndf} \hspace*{0.3 cm} & \multicolumn{2}{c}{\chisq} \\
 \cline{3-4}
{variables} && \PowPytSh (w. unc.)  & \StripperOnly  \\
\hline
\ytptt& 16 & 44 \: (28) & 17  \\
\mttptt& 9 & 103 \: (37) & 26  \\
\pttpttt& 16 & 44 \: (21) & 169 \\
\mttytt  & 16 & 86 \: (41) & 44 \\
\yttpttt  & 16 & 32 \: (19) & 81 \\
\mttpttt& 16 & 69 \: (37) & 388 \\
\ptttmttytt  & 48 & 133 \: (69) & 149  \\
\mttyt& 16 & 64 \: (27) & 21 \\
\mttdetatt& 12 & 174 \: (32) & 39 \\
\mttdphitt& 12 & 80 \: (44) & 426 \\
 \end{tabular}
\end{table*}

\begin{table*}
\centering
 \topcaption{The \chisq values and \ndf of the measured absolute single-differential cross sections for lepton and \PQb-jet kinematic observables at the particle level are shown with respect to the \PowPyt (`POW-PYT') simulation and the \StripperOnly NNLO calculation. The \chisq values are calculated taking only measurement uncertainties into account and excluding theory uncertainties.  
For \PowPytSh, the \chisq values including theory uncertainties are indicated with the brackets (w. unc.).}
 \label{tab:chi2fixTheo_1d_abs_particle_lepb}
 \renewcommand{\arraystretch}{1.4}
 \centering
 \begin{tabular}{lccc}
 \multirow{1}{*}{Cross section} & \hspace*{0.3 cm} \multirow{2}{*}{\ndf} \hspace*{0.3 cm} & \multicolumn{2}{c}{\chisq} \\
 \cline{3-4}
{variables} && \PowPytSh (w. unc.)  & \StripperOnly   \\
\hline
\ptlep& 12 & 32 \: (19) & 17 \\
\ptlep trailing/\ptlep leading& 10 & 16 \: (11) & 5 \\
\ptlep/\ptat& 5 & 20 \: (17) & 8 \\
\ptb leading& 10 & 6 \: (5) & 8 \\
\ptb trailing& 7 & 7 \: (5) & 4 \\
\rptbsptts& 4 & 24 \: (19) & 9 \\
\mll& 12 & 31 \: (25) & 13 \\
\mbb& 7 & 21 \: (16) & 80 \\
\mllbb& 19 & 36 \: (19) & 16 \\
\ptll& 9 & 4 \: (3) & 6 \\
\absetall& 14 & 16 \: (10) & 41 \\
\etallmll& 24 & 55 \: (29) & 80 \\
\etallptll& 20 & 30 \: (15) & 46 \\
\ptllmll& 30 & 50 \: (39) & 51 \\
 \end{tabular}
\end{table*}

\clearpage

\section{Tables with \texorpdfstring{$p$-values}{p values}  of 
\texorpdfstring{\chisq tests}{chi2 tests}}
\label{sec:res_pval}

Tables~\ref{tab:pvaluemc_1d_nor_parton}--\ref{tab:pvaluefixTheo_1d_abs_particle_lepb} present the $p$-values of the performed \chisq tests between the measured
differential cross sections for \ttbar production and various predictions.
The corresponding \chisq values and number of degrees of freedom
can be found in the Tables~\ref{tab:chi2mc_1d_nor_parton}--\ref{tab:chi2fixTheo_1d_abs_particle_lepb}. 

\begin{table*}
\centering
\topcaption{The $p$-values are shown for the \chisq tests of the measured normalized single-differential cross sections for \ttbar and top quark kinematic observables at the parton level with respect to the predictions of various MC generators. The \chisq values are calculated taking only measurement uncertainties into account and excluding theory uncertainties. For \PowPytSh, the $p$-values of the \chisq tests including theory uncertainties are indicated with the brackets (w. unc.).}  
 \label{tab:pvaluemc_1d_nor_parton}
 \renewcommand{\arraystretch}{1.4}
 \centering
 \begin{tabular}{lccc}
 \multirow{1}{*}{Cross section} & \multicolumn{3}{c}{$p$-values of \chisq (in \%)} \\
 \cline{2-4}
{variables}& \PowPytSh (w. unc.)  & \aMCPytSh  & \PowHerSh   \\
\hline
\ptt& 2 \: (10) & $<$1 & 51 \\
\ptat& 5 \: (16) & $<$1 & 41 \\
\yt& $<$1 \: (2) & $<$1 & 1 \\
\yat& $<$1 \: ($<$1) & $<$1 & $<$1 \\
\pttt& $<$1 \: (32) & $<$1 & $<$1 \\
\ytt& 51 \: (74) & 7 & 73 \\
\mtt& 56 \: (77) & 30 & 70 \\
\dphitt& 82 \: (97) & 15 & 7 \\
\dytt& 3 \: (22) & $<$1 & 6 \\
\rpttmtt& $<$1 \: ($<$1) & $<$1 & 3 \\
\rptttmtt& 3 \: (62) & $<$1 & $<$1 \\
\logxone& 9 \: (26) & 3 & 15 \\
\logxtwo& 24 \: (54) & $<$1 & 51 \\
 \end{tabular}
\end{table*}

\begin{table*}
\centering
\topcaption{The $p$-values are shown for the \chisq tests of the measured normalized single-differential cross sections for \ttbar and top quark kinematic observables at the particle level with respect to the predictions of various MC generators. The \chisq values are calculated taking only measurement uncertainties into account and excluding theory uncertainties. For \PowPytSh, the $p$-values of the \chisq tests including theory uncertainties are indicated with the brackets (w. unc.).}  
 \label{tab:pvaluemc_1d_nor_particle_ttbar}
 \renewcommand{\arraystretch}{1.4}
 \centering
 \begin{tabular}{lccc}
 \multirow{1}{*}{Cross section} & \multicolumn{3}{c}{$p$-values of \chisq (in \%)} \\
 \cline{2-4}
{variables}& \PowPytSh (w. unc.)  & \aMCPytSh  & \PowHerSh   \\
\hline
\ptt& $<$1 \: (7) & $<$1 & 47 \\
\ptat& 3 \: (15) & $<$1 & 59 \\
\yt& 3 \: (7) & $<$1 & 2 \\
\yat& $<$1 \: (2) & $<$1 & $<$1 \\
\pttt& $<$1 \: (33) & $<$1 & $<$1 \\
\ytt& 51 \: (81) & 3 & 64 \\
\mtt& 59 \: (81) & 60 & 28 \\
\dphitt& 82 \: (97) & 36 & 10 \\
\dytt& 4 \: (20) & 4 & 1 \\
\rpttmtt& $<$1 \: ($<$1) & $<$1 & 2 \\
\rptttmtt& 2 \: (57) & $<$1 & $<$1 \\
\logxone& 8 \: (34) & 3 & 2 \\
\logxtwo& 32 \: (67) & 3 & 29 \\
 \end{tabular}
\end{table*}

\begin{table*}
 \centering
 \topcaption{The $p$-values are shown for the \chisq tests of the measured normalized multi-differential cross sections for \ttbar and top quark kinematic observables at the parton level with respect to the predictions of various MC generators.  The \chisq values are calculated taking only measurement uncertainties into account and excluding theory uncertainties. For \PowPytSh, the $p$-values of the \chisq tests including theory uncertainties are indicated with the brackets (w. unc.).}   
 \label{tab:pvaluemc_nor_parton_ttbar}
 \renewcommand{\arraystretch}{1.4}
 \centering
 \begin{tabular}{lccc}
 \multirow{1}{*}{Cross section} & \multicolumn{3}{c}{$p$-values of \chisq (in \%)} \\
 \cline{2-4}
{variables}& \PowPytSh (w. unc.)  & \aMCPytSh  & \PowHerSh   \\
\hline
\ytptt& $<$1 \: ($<$1) & $<$1 & 1 \\
\mttptt& $<$1 \: ($<$1) & $<$1 & $<$1 \\
\pttpttt& $<$1 \: (14) & $<$1 & $<$1 \\
\mttytt & $<$1 \: ($<$1) & $<$1 & $<$1 \\
\yttpttt& 2 \: (44) & $<$1 & $<$1 \\
\mttpttt & $<$1 \: ($<$1) & $<$1 & $<$1 \\
\ptttmttytt & $<$1 \: (5) & $<$1 & $<$1 \\
\mttyt& $<$1 \: ($<$1) & $<$1 & $<$1 \\
\mttdetatt& $<$1 \: ($<$1) & $<$1 & $<$1 \\
\mttdphitt& $<$1 \: ($<$1) & $<$1 & $<$1 \\
 \end{tabular}
\end{table*}

\begin{table*}
 \centering
 \topcaption{The $p$-values are shown for the \chisq tests of the measured normalized multi-differential cross sections for \ttbar and top quark kinematic observables at the particle level with respect to the predictions of various MC generators. The \chisq values are calculated taking only measurement uncertainties into account and excluding theory uncertainties. For \PowPytSh, the $p$-values of the \chisq tests including theory uncertainties are indicated with the brackets (w. unc.).}   
 \label{tab:pvaluemc_nor_particle_ttbar}
 \renewcommand{\arraystretch}{1.4}
 \centering
 \begin{tabular}{lccc}
 \multirow{1}{*}{Cross section} & \multicolumn{3}{c}{$p$-values of \chisq (in \%)} \\
 \cline{2-4}
{variables}& \PowPytSh (w. unc.)  & \aMCPytSh  & \PowHerSh   \\
\hline
\ytptt& $<$1 \: (5) & $<$1 & 2 \\
\mttptt& $<$1 \: ($<$1) & $<$1 & $<$1 \\
\pttpttt& $<$1 \: (21) & $<$1 & $<$1 \\
\mttytt & $<$1 \: ($<$1) & $<$1 & $<$1 \\
\yttpttt& 3 \: (28) & $<$1 & $<$1 \\
\mttpttt & $<$1 \: ($<$1) & $<$1 & $<$1 \\
\ptttmttytt & $<$1 \: (3) & $<$1 & $<$1 \\
\mttyt& $<$1 \: (4) & $<$1 & $<$1 \\
\mttdetatt& $<$1 \: ($<$1) & $<$1 & $<$1 \\
\mttdphitt& $<$1 \: ($<$1) & $<$1 & $<$1 \\
 \end{tabular}
\end{table*}

\begin{table*}
\centering
 \topcaption{The $p$-values are shown for the \chisq tests of the measured normalized single-differential cross sections for lepton and \PQb-jet kinematic observables at the particle level with respect to the predictions of various MC generators. The \chisq values are calculated taking only measurement uncertainties into account and excluding theory uncertainties. For \PowPytSh, the $p$-values of the \chisq tests including theory uncertainties are indicated with the brackets (w. unc.).}
 \label{tab:pvaluemc_1d_nor_particle_lepb}
 \renewcommand{\arraystretch}{1.4}
 \centering
 \begin{tabular}{lccc}
 \multirow{1}{*}{Cross section} & \multicolumn{3}{c}{$p$-values of \chisq (in \%)} \\
 \cline{2-4}
{variables}& \PowPytSh (w. unc.)  & \aMCPytSh  & \PowHerSh   \\
\hline
\ptlep& $<$1 \: (8) & $<$1 & 9 \\
\ptlep trailing/\ptlep leading& 10 \: (30) & $<$1 & 66 \\
\ptlep/\ptat& 4 \: (6) & $<$1 & 2 \\
\ptb leading& 86 \: (92) & $<$1 & 55 \\
\ptb trailing& 48 \: (70) & $<$1 & 35 \\
\rptbsptts& $<$1 \: ($<$1) & $<$1 & $<$1 \\
\mll& 2 \: (5) & $<$1 & 2 \\
\mbb& 2 \: (7) & 2 & 3 \\
\mllbb& 2 \: (45) & 6 & 6 \\
\ptll& 86 \: (96) & 9 & 31 \\
\absetall& 40 \: (77) & 7 & 58 \\
\etallmll& $<$1 \: (23) & $<$1 & 4 \\
\etallptll& 10 \: (78) & $<$1 & 19 \\
\ptllmll& 3 \: (15) & $<$1 & $<$1 \\
 \end{tabular}
\end{table*}

\begin{table*}
 \centering
 \topcaption{The $p$-values are shown for the \chisq tests of the measured normalized differential cross sections as a function of the additional-jet multiplicity in the events, at the parton level of the top quark and antiquark, with respect to the predictions of various MC generators. The \chisq values are calculated taking only measurement uncertainties into account and excluding theory uncertainties. For \PowPytSh, the $p$-values of the \chisq tests including theory uncertainties are indicated with the brackets (w. unc.).}
 \label{tab:pvaluemc_nor_parton_addjets}
 \renewcommand{\arraystretch}{1.4}
 \centering
 \begin{tabular}{lccc}
 \multirow{1}{*}{Cross section} & \multicolumn{3}{c}{$p$-values of \chisq (in \%)} \\
 \cline{2-4}
{variables}& \PowPytSh (w. unc.)  & \aMCPytSh  & \PowHerSh   \\
\hline
\njforty & 34 \: (64) & $<$1 & $<$1 \\
\njhundred & $<$1 \: (11) & $<$1 & $<$1 \\
\njptt& $<$1 \: (14) & $<$1 & $<$1 \\
\njyt& $<$1 \: ($<$1) & $<$1 & $<$1 \\
\njpttt& $<$1 \: ($<$1) & $<$1 & $<$1 \\
\njmtt & $<$1 \: ($<$1) & $<$1 & $<$1 \\
\njytt& 46 \: (94) & $<$1 & $<$1 \\
\njdetatt& $<$1 \: ($<$1) & $<$1 & $<$1 \\
\njmttytttwo & $<$1 \: (5) & $<$1 & $<$1 \\
\njmttyttthree & $<$1 \: (1) & $<$1 & $<$1 \\
\njmttyttfour & $<$1 \: ($<$1) & $<$1 & $<$1 \\
 \end{tabular}
\end{table*}

\begin{table*}
 \centering
 \topcaption{The $p$-values are shown for the \chisq tests of the measured normalized differential cross sections as a function of the additional-jet multiplicity in the events, at the particle level of the top quark and antiquark, with respect to the predictions of various MC generators.
The \chisq values are calculated taking only measurement uncertainties into account and excluding theory uncertainties. For \PowPytSh, the $p$-values of the \chisq tests including theory uncertainties are indicated with the brackets (w. unc.).}
 \label{tab:pvaluemc_nor_particle_addjets}
 \renewcommand{\arraystretch}{1.4}
 \centering
 \begin{tabular}{lccc}
 \multirow{1}{*}{Cross section} & \multicolumn{3}{c}{$p$-values of \chisq (in \%)} \\
 \cline{2-4}
{variables}& \PowPytSh (w. unc.)  & \aMCPytSh  & \PowHerSh   \\
\hline
\njforty & 30 \: (63) & $<$1 & 12 \\
\njhundred & $<$1 \: (8) & $<$1 & 19 \\
\njptt& $<$1 \: (12) & $<$1 & $<$1 \\
\njyt& $<$1 \: (1) & $<$1 & $<$1 \\
\njpttt& $<$1 \: ($<$1) & $<$1 & $<$1 \\
\njmtt & $<$1 \: ($<$1) & $<$1 & $<$1 \\
\njytt& 22 \: (94) & $<$1 & 65 \\
\njdetatt& $<$1 \: ($<$1) & $<$1 & $<$1 \\
\njmttytttwo & $<$1 \: ($<$1) & $<$1 & $<$1 \\
\njmttyttthree & $<$1 \: ($<$1) & $<$1 & $<$1 \\
\njmttyttfour & $<$1 \: ($<$1) & $<$1 & $<$1 \\
 \end{tabular}
\end{table*}

\clearpage

\begin{sidewaystable*}
\centering
\topcaption{The $p$-values are shown for the \chisq tests of the measured normalized single-differential cross sections for \ttbar and top quark kinematic observables at the parton level with respect to various fixed-order predictions. The \chisq values are calculated taking only measurement uncertainties into account and excluding theory uncertainties. For \PowPytSh, the $p$-values of the \chisq tests including theory uncertainties are indicated with the brackets (w. unc.).}
 \label{tab:pvaluefixTheo_1d_nor_parton}
 \renewcommand{\arraystretch}{1.4}
 \centering
 \begin{tabular}{lccccc}
 \multirow{1}{*}{Cross section} & \multicolumn{5}{c}{$p$-values of \chisq (in \%)} \\
 \cline{2-6}
{variables}& \PowPytSh (w. unc.)  & \appNTLOOnly & \MatrixOnly & \StripperOnly  & \MiNNLOPSOnly  \\
\hline
\ptt& 2 \: (10) & $<$1 & 78  & 22  & 62  \\
\ptat& 5 \: (16) & \NA & 69  & 14 & 49  \\
\yt& $<$1 \: (2) & $<$1  & 6 & 6 & 8  \\
\yat& $<$1 \: ($<$1) & \NA & 10 & $<$1 & $<$1  \\
\pttt& $<$1 \: (32) & \NA & 33 & 15  & $<$1  \\
\ytt& 51 \: (74) & \NA & 84 & 61 & 61  \\
\mtt& 56 \: (77) & \NA & 49 & 59  & 90 \\
\dphitt& 82 \: (97) & \NA & $<$1  & $<$1 & 14 \\
\dytt& 3 \: (22) & \NA & 94  & 86 & 10 \\
\rpttmtt& $<$1 \: ($<$1) & \NA & 58  & 41  & 28 \\
\rptttmtt& 3 \: (62) & \NA & $<$1 & $<$1  & 1 \\
\logxone& 9 \: (26) & \NA & \NA & 6 & 16  \\
\logxtwo& 24 \: (54) & \NA & \NA & 17 & 52 \\
 \end{tabular}
\end{sidewaystable*}

\begin{table*}
\centering
 \topcaption{The $p$-values are shown for the \chisq tests of the measured normalized single-differential cross sections for \ttbar and top quark kinematic observables at the particle level 
   with respect to the \PowPyt (`POW-PYT') simulation and the \StripperOnly NNLO calculation. The \chisq values are calculated taking only measurement uncertainties into account and excluding theory uncertainties. For \PowPytSh, the $p$-values of the \chisq tests including theory uncertainties are indicated with the brackets (w. unc.).}  
 \label{tab:pvaluefixTheo_1d_nor_particle_ttbar}
 \renewcommand{\arraystretch}{1.4}
 \centering
 \begin{tabular}{lcc}
 \multirow{1}{*}{Cross section} & \multicolumn{2}{c}{$p$-values of \chisq (in \%)} \\
 \cline{2-3}
{variables}& \PowPytSh (w. unc.)  & \StripperOnly  \\
\hline
\ptt& $<$1 \: (7) & 83 \\
\ptat& 3 \: (15) & 82 \\
\yt& 3 \: (7) & 29 \\
\yat& $<$1 \: (2) & 37 \\
\pttt& $<$1 \: (33) & $<$1  \\
\ytt& 51 \: (81) & 6 \\
\mtt& 59 \: (81) & 73 \\
\dphitt& 82 \: (97) & $<$1 \\
\dytt& 4 \: (20) & 81 \\
\rpttmtt& $<$1 \: ($<$1) & 9 \\
\rptttmtt& 2 \: (57) & $<$1 \\
\logxone& 8 \: (34) & 28 \\
\logxtwo& 32 \: (67) & 33 \\
 \end{tabular}
\end{table*}

\begin{sidewaystable*}
 \centering
 \topcaption{The $p$-values are shown for the \chisq tests of the measured normalized multi-differential cross sections for \ttbar and top quark kinematic observables at the parton level with respect to various fixed-order predictions. The \chisq values are calculated taking only measurement uncertainties into account and excluding theory uncertainties. For \PowPytSh, the $p$-values of the \chisq tests including theory uncertainties are indicated with the brackets (w. unc.).}
 \label{tab:pvaluefixTheo_nor_parton_ttbar}
 \renewcommand{\arraystretch}{1.4}
 \centering
 \begin{tabular}{lccccc}
 \multirow{1}{*}{Cross section} & \multicolumn{5}{c}{$p$-values of \chisq (in \%)} \\
 \cline{2-6}
{variables}& \PowPytSh (w. unc.)  & \appNTLOOnly  & \MatrixOnly   & \StripperOnly  & \MiNNLOPSOnly   \\
\hline
\ytptt& $<$1 \: ($<$1) & $<$1 & 7  & 6  & 9 \\
\mttptt& $<$1 \: ($<$1) & \NA & 7  & 4  & 6 \\
\pttpttt& $<$1 \: (14) & \NA & $<$1 & $<$1 & $<$1 \\
\mttytt& $<$1 \: ($<$1) & \NA &  & $<$1 & $<$1 \\
\yttpttt& 2 \: (44) & \NA & $<$1 & $<$1 & $<$1 \\
\mttpttt& $<$1 \: ($<$1) & \NA & $<$1 & $<$1 & $<$1 \\
\ptttmttytt& $<$1 \: (5) & \NA & \NA & 2 & $<$1 \\
\mttyt& $<$1 \: ($<$1) & \NA & $<$1  & 1 &   \\
\mttdetatt& $<$1 \: ($<$1) & \NA & $<$1 & $<$1  & $<$1 \\
\mttdphitt& $<$1 \: ($<$1) & \NA & \NA & $<$1 & $<$1 \\
 \end{tabular}
\end{sidewaystable*}

\begin{table*}
 \centering
 \topcaption{The $p$-values are shown for the \chisq tests of the measured normalized multi-differential cross sections for \ttbar and top quark kinematic observables at the particle level 
with respect to the \PowPyt (`POW-PYT') simulation and the \StripperOnly NNLO calculation. 
The \chisq values are calculated taking only measurement uncertainties into account and excluding theory uncertainties. For \PowPytSh, the $p$-values of the \chisq tests including theory uncertainties are indicated with the brackets (w. unc.).}  
 \label{tab:pvaluefixTheo_nor_particle_ttbar}
 \renewcommand{\arraystretch}{1.4}
 \centering
 \begin{tabular}{lcc}
 \multirow{1}{*}{Cross section} & \multicolumn{2}{c}{$p$-values of \chisq (in \%)} \\
 \cline{2-3}
{variables}& \PowPytSh (w. unc.)  & \StripperOnly \\
\hline
\ytptt& $<$1 \: (5) & 38 \\
\mttptt& $<$1 \: ($<$1) & 2 \\
\pttpttt& $<$1 \: (21) & $<$1 \\
\mttytt& $<$1 \: ($<$1) & $<$1 \\
\yttpttt& 3 \: (28) & $<$1 \\
\mttpttt& $<$1 \: ($<$1) & $<$1 \\
\ptttmttytt& $<$1 \: (3) & $<$1 \\
\mttyt& $<$1 \: (4) & 18 \\
\mttdetatt& $<$1 \: ($<$1) & $<$1 \\
\mttdphitt& $<$1 \: ($<$1) & $<$1 \\
 \end{tabular}
\end{table*}

\begin{table*}
\centering
 \topcaption{The $p$-values are shown for the \chisq tests of the measured normalized single-differential cross sections for lepton and \PQb-jet kinematic observables at the particle level 
with respect to the \PowPyt (`POW-PYT') simulation and the \StripperOnly NNLO calculation. 
The \chisq values are calculated taking only measurement uncertainties into account and excluding theory uncertainties. For \PowPytSh, the $p$-values of the \chisq tests including theory uncertainties are indicated with the brackets (w. unc.).}  
 \label{tab:pvaluefixTheo_1d_nor_particle_lepb}
 \renewcommand{\arraystretch}{1.4}
 \centering
 \begin{tabular}{lcc}
 \multirow{1}{*}{Cross section} & \multicolumn{2}{c}{$p$-values of \chisq (in \%)} \\
 \cline{2-3}
{variables}& \PowPytSh (w. unc.)  & \StripperOnly  \\
\hline
\ptlep& $<$1 \: (8) & 12  \\
\ptlep trailing/\ptlep leading& 10 \: (30) & 81 \\
\ptlep/\ptat& 4 \: (6) & 33 \\
\ptb leading& 86 \: (92) & 64 \\
\ptb trailing& 48 \: (70) & 81 \\
\rptbsptts& $<$1 \: ($<$1) & 4 \\
\mll& 2 \: (5) & 32 \\
\mbb& 2 \: (7) & $<$1 \\
\mllbb& 2 \: (45) & 68 \\
\ptll& 86 \: (96) & 73 \\
\absetall& 40 \: (77) & $<$1 \\
\etallmll& $<$1 \: (23) & $<$1 \\
\etallptll& 10 \: (78) & $<$1 \\
\ptllmll& 3 \: (15) & 2 \\
 \end{tabular}
\end{table*}

\begin{table*}
\centering
\topcaption{The $p$-values are shown for the \chisq tests of the measured absolute single-differential cross sections for \ttbar and top quark kinematic observables at the parton level with respect to the predictions of various MC generators. The \chisq values are calculated taking only measurement uncertainties into account and excluding theory uncertainties. For \PowPytSh, the $p$-values of the \chisq tests including theory uncertainties are indicated with the brackets (w. unc.).}  
 \label{tab:pvaluemc_1d_abs_parton}
 \renewcommand{\arraystretch}{1.4}
 \centering
 \begin{tabular}{lccc}
 \multirow{1}{*}{Cross section} & \multicolumn{3}{c}{$p$-values of \chisq (in \%)} \\
 \cline{2-4}
{variables}& \PowPytSh (w. unc.)  & \aMCPytSh  & \PowHerSh   \\
\hline
\ptt& $<$1 \: (7) & $<$1 & 61 \\
\ptat& $<$1 \: (9) & $<$1 & 50 \\
\yt& $<$1 \: ($<$1) & $<$1 & 2 \\
\yat& $<$1 \: ($<$1) & $<$1 & $<$1 \\
\pttt& $<$1 \: (35) & $<$1 & $<$1 \\
\ytt& 38 \: (67) & 9 & 78 \\
\mtt& 55 \: (78) & 28 & 76 \\
\dphitt& 40 \: (81) & 14 & 3 \\
\dytt& 2 \: (23) & 2 & 9 \\
\rpttmtt& $<$1 \: ($<$1) & $<$1 & 3 \\
\rptttmtt& 2 \: (66) & $<$1 & $<$1 \\
\logxone& 7 \: (23) & 4 & 20 \\
\logxtwo& 14 \: (47) & $<$1 & 60 \\
 \end{tabular}
\end{table*}

\begin{table*}
\centering
\topcaption{The $p$-values are shown for the \chisq tests of the measured absolute single-differential 
cross sections for \ttbar and top quark kinematic observables at the particle level with respect to the predictions of various MC generators. The \chisq values are calculated taking only measurement uncertainties into account and excluding theory uncertainties. For \PowPytSh, the $p$-values of the \chisq tests including theory uncertainties are indicated with the brackets (w. unc.).}
 \label{tab:pvaluemc_1d_abs_particle_ttbar}
 \renewcommand{\arraystretch}{1.4}
 \centering
 \begin{tabular}{lccc}
 \multirow{1}{*}{Cross section} & \multicolumn{3}{c}{$p$-values of \chisq (in \%)} \\
 \cline{2-4}
{variables}& \PowPytSh (w. unc.)  & \aMCPytSh  & \PowHerSh   \\
\hline
\ptt& $<$1 \: (7) & $<$1 & 46 \\
\ptat& $<$1 \: (11) & $<$1 & 68 \\
\yt& $<$1 \: (5) & $<$1 & 4 \\
\yat& $<$1 \: (1) & $<$1 & $<$1 \\
\pttt& $<$1 \: (37) & $<$1 & $<$1 \\
\ytt& 35 \: (78) & 3 & 74 \\
\mtt& 46 \: (80) & 67 & 39 \\
\dphitt& 40 \: (85) & 43 & 13 \\
\dytt& 3 \: (23) & 5 & 2 \\
\rpttmtt& $<$1 \: ($<$1) & $<$1 & 1 \\
\rptttmtt& 1 \: (59) & $<$1 & $<$1 \\
\logxone& 7 \: (35) & 4 & 3 \\
\logxtwo& 22 \: (67) & 3 & 36 \\
 \end{tabular}
\end{table*}

\begin{table*}
 \centering
 \topcaption{The $p$-values are shown for the \chisq tests of the measured absolute multi-differential cross sections for \ttbar and top quark kinematic observables at the parton level with respect to the predictions of various MC generators. The \chisq values are calculated taking only measurement uncertainties into account and excluding theory uncertainties. For \PowPytSh, the $p$-values of the \chisq tests including theory uncertainties are indicated with the brackets (w. unc.).}
 \label{tab:pvaluemc_abs_parton_ttbar}
 \renewcommand{\arraystretch}{1.4}
 \centering
 \begin{tabular}{lccc}
 \multirow{1}{*}{Cross section} & \multicolumn{3}{c}{$p$-values of \chisq (in \%)} \\
 \cline{2-4}
{variables}& \PowPytSh (w. unc.)  & \aMCPytSh  & \PowHerSh   \\
\hline
\ytptt& $<$1 \: ($<$1) & $<$1 & 2 \\
\mttptt& $<$1 \: ($<$1) & $<$1 & $<$1 \\
\pttpttt& $<$1 \: (6) & $<$1 & $<$1 \\
\mttytt & $<$1 \: ($<$1) & $<$1 & $<$1 \\
\yttpttt& $<$1 \: (37) & $<$1 & $<$1 \\
\mttpttt & $<$1 \: ($<$1) & $<$1 & $<$1 \\
\ptttmttytt & $<$1 \: (2) & $<$1 & $<$1 \\
\mttyt& $<$1 \: ($<$1) & $<$1 & $<$1 \\
\mttdetatt& $<$1 \: ($<$1) & $<$1 & $<$1 \\
\mttdphitt& $<$1 \: ($<$1) & $<$1 & $<$1 \\
 \end{tabular}
\end{table*}

\begin{table*}
 \centering
 \topcaption{The $p$-values are shown for the \chisq tests of the measured absolute multi-differential cross sections for \ttbar and top quark kinematic observables at the particle level with respect to the predictions of various MC generators. The \chisq values are calculated taking only measurement uncertainties into account and excluding theory uncertainties. For \PowPytSh, the $p$-values of the \chisq tests including theory uncertainties are indicated with the brackets (w. unc.).}
 \label{tab:pvaluemc_abs_particle_ttbar}
 \renewcommand{\arraystretch}{1.4}
 \centering
 \begin{tabular}{lccc}
 \multirow{1}{*}{Cross section} & \multicolumn{3}{c}{$p$-values of \chisq (in \%)} \\
 \cline{2-4}
{variables}& \PowPytSh (w. unc.)  & \aMCPytSh  & \PowHerSh   \\
\hline
\ytptt& $<$1 \: (3) & $<$1 & 4 \\
\mttptt& $<$1 \: ($<$1) & $<$1 & $<$1 \\
\pttpttt& $<$1 \: (17) & $<$1 & $<$1 \\
\mttytt & $<$1 \: ($<$1) & $<$1 & $<$1 \\
\yttpttt& 1 \: (27) & $<$1 & $<$1 \\
\mttpttt & $<$1 \: ($<$1) & $<$1 & $<$1 \\
\ptttmttytt & $<$1 \: (3) & $<$1 & $<$1 \\
\mttyt& $<$1 \: (5) & $<$1 & $<$1 \\
\mttdetatt& $<$1 \: ($<$1) & $<$1 & $<$1 \\
\mttdphitt& $<$1 \: ($<$1) & $<$1 & $<$1 \\
 \end{tabular}
\end{table*}

\begin{table*}
\centering
\topcaption{The $p$-values are shown for the \chisq tests of the measured absolute single-differential cross sections for lepton and \PQb-jet kinematic observables at the particle level with respect to the predictions of various MC generators. The \chisq values are calculated taking only measurement uncertainties into account and excluding theory uncertainties. For \PowPytSh, the $p$-values of the \chisq tests including theory uncertainties are indicated with the brackets (w. unc.).}
 \label{tab:pvaluemc_1d_abs_particle_lepb}
 \renewcommand{\arraystretch}{1.4}
 \centering
 \begin{tabular}{lccc}
 \multirow{1}{*}{Cross section} & \multicolumn{3}{c}{$p$-values of \chisq (in \%)} \\
 \cline{2-4}
{variables}& \PowPytSh (w. unc.)  & \aMCPytSh  & \PowHerSh   \\
\hline
\ptlep& $<$1 \: (9) & $<$1 & 6 \\
\ptlep trailing/\ptlep leading& 9 \: (34) & $<$1 & 70 \\
\ptlep/\ptat& $<$1 \: ($<$1) & $<$1 & 2 \\
\ptb leading& 85 \: (91) & $<$1 & 64 \\
\ptb trailing& 46 \: (64) & $<$1 & 41 \\
\rptbsptts& $<$1 \: ($<$1) & $<$1 & $<$1 \\
\mll& $<$1 \: (2) & $<$1 & 3 \\
\mbb& $<$1 \: (2) & 2 & 4 \\
\mllbb& $<$1 \: (48) & 6 & 10 \\
\ptll& 88 \: (97) & 6 & 34 \\
\absetall& 32 \: (77) & 7 & 63 \\
\etallmll& $<$1 \: (22) & $<$1 & 7 \\
\etallptll& 7 \: (80) & $<$1 & 24 \\
\ptllmll& 1 \: (12) & $<$1 & $<$1 \\
 \end{tabular}
\end{table*}

\begin{table*}
 \centering
 \topcaption{The $p$-values are shown for the \chisq tests of the measured absolute differential cross sections as a function of the additional-jet multiplicity in the events, at the parton level of the top quark and antiquark, with respect to the predictions of various MC generators. The \chisq values are calculated taking only measurement uncertainties into account and excluding theory uncertainties. For \PowPytSh, the $p$-values of the \chisq tests including theory uncertainties are indicated with the brackets (w. unc.).}  
 \label{tab:pvaluemc_abs_parton_addjets}
 \renewcommand{\arraystretch}{1.4}
 \centering
 \begin{tabular}{lccc}
 \multirow{1}{*}{Cross section} & \multicolumn{3}{c}{$p$-values of \chisq (in \$)} \\
 \cline{2-4}
{variables}& \PowPytSh (w. unc.)  & \aMCPytSh  & \PowHerSh   \\
\hline
\njforty & 30 \: (56) & $<$1 & $<$1 \\
\njhundred & $<$1 \: (5) & $<$1 & $<$1 \\
\njptt& $<$1 \: (5) & $<$1 & $<$1 \\
\njyt& $<$1 \: ($<$1) & $<$1 & $<$1 \\
\njpttt& $<$1 \: ($<$1) & $<$1 & $<$1 \\
\njmtt & $<$1 \: ($<$1) & $<$1 & $<$1 \\
\njytt& 32 \: (87) & $<$1 & $<$1 \\
\njdetatt& $<$1 \: ($<$1) & $<$1 & $<$1 \\
\njmttytttwo & $<$1 \: (3) & $<$1 & $<$1 \\
\njmttyttthree & $<$1 \: ($<$1) & $<$1 & $<$1 \\
\njmttyttfour & $<$1 \: ($<$1) & $<$1 & $<$1 \\
 \end{tabular}
\end{table*}

\begin{table*}
 \centering
 \topcaption{The $p$-values are shown for the \chisq tests of the measured absolute differential cross sections as a function of the additional-jet multiplicity in the events, at the particle level of the top quark and antiquark, with respect to the predictions of various MC generators. The \chisq values are calculated taking only measurement uncertainties into account and excluding theory uncertainties. For \PowPytSh, the $p$-values of the \chisq tests including theory uncertainties are indicated with the brackets (w. unc.).}
 \label{tab:pvaluemc_abs_particle_addjets}
 \renewcommand{\arraystretch}{1.4}
 \centering
 \begin{tabular}{lccc}
 \multirow{1}{*}{Cross section} & \multicolumn{3}{c}{$p$-values of \chisq (in \%)} \\
 \cline{2-4}
{variables}& \PowPytSh (w. unc.)  & \aMCPytSh  & \PowHerSh   \\
\hline
\njforty & 29 \: (66) & $<$1 & 22 \\
\njhundred & $<$1 \: (5) & $<$1 & 25 \\
\njptt& $<$1 \: (9) & $<$1 & $<$1 \\
\njyt& $<$1 \: (1) & $<$1 & $<$1 \\
\njpttt& $<$1 \: ($<$1) & $<$1 & $<$1 \\
\njmtt & $<$1 \: ($<$1) & $<$1 & $<$1 \\
\njytt& 13 \: (91) & $<$1 & 75 \\
\njdetatt& $<$1 \: ($<$1) & $<$1 & $<$1 \\
\njmttytttwo & $<$1 \: ($<$1) & $<$1 & $<$1 \\
\njmttyttthree & $<$1 \: ($<$1) & $<$1 & $<$1 \\
\njmttyttfour & $<$1 \: ($<$1) & $<$1 & $<$1 \\
 \end{tabular}
\end{table*}

\begin{sidewaystable*}
\centering
\topcaption{The $p$-values are shown for the \chisq tests of the measured absolute single-differential cross sections for \ttbar and top quark kinematic observables at the parton level with respect to various fixed-order predictions. The \chisq values are calculated taking only measurement uncertainties into account and excluding theory uncertainties. For \PowPytSh, the $p$-values of the \chisq tests including theory uncertainties are indicated with the brackets (w. unc.).}
 \label{tab:pvaluefixTheo_1d_abs_parton}
 \renewcommand{\arraystretch}{1.4}
 \centering
 \begin{tabular}{lccccc}
 \multirow{1}{*}{Cross section} & \multicolumn{5}{c}{$p$-values of \chisq (in \%)} \\
 \cline{2-6}
{variables}& \PowPytSh (w. unc.)  & \appNTLOOnly  & \MatrixOnly  & \StripperOnly  & \MiNNLOPSOnly \\
\hline
\ptt& $<$1 \: (7) & $<$1 & 84 & 4 & 55 \\
\ptat& $<$1 \: (9) & \NA & 71  & 17  & 47  \\
\yt& $<$1 \: ($<$1) & $<$1  & 8 & 7 & 17 \\
\yat& $<$1 \: ($<$1) & \NA & 13 & $<$1 & 2 \\
\pttt& $<$1 \: (35) & \NA & 42 & 21 & $<$1 \\
\ytt& 38 \: (67) & \NA & 86 & 65  & 72  \\
\mtt& 55 \: (78) & \NA & 51 & 61 & 94  \\
\dphitt& 40 \: (81) & \NA & $<$1 & $<$1 & 25  \\
\dytt& 2 \: (23) & \NA & 97 & 90  & 15  \\
\rpttmtt& $<$1 \: ($<$1) & \NA & 60  & 38 & 9  \\
\rptttmtt& 2 \: (66) & \NA & $<$1 & $<$1 & 3 \\
\logxone& 7 \: (23) & \NA & \NA & 8  & 23  \\
\logxtwo& 14 \: (47) & \NA & \NA & 18  & 66  \\
 \end{tabular}
\end{sidewaystable*}

\begin{table*}
\centering
 \topcaption{The $p$-values are shown for the \chisq tests of the measured absolute single-differential cross sections for \ttbar and top quark kinematic observables at the particle level 
with respect to the \PowPyt (`POW-PYT') simulation and the \StripperOnly NNLO calculation. 
The \chisq values are calculated taking only measurement uncertainties into account and excluding theory uncertainties.
For \PowPytSh, the $p$-values of the \chisq tests including theory uncertainties are indicated with the brackets (w. unc.).}
 \label{tab:pvaluefixTheo_1d_abs_particle_ttbar}
 \renewcommand{\arraystretch}{1.4}
 \centering
 \begin{tabular}{lcc}
 \multirow{1}{*}{Cross section} & \multicolumn{2}{c}{$p$-values of \chisq (in \%)} \\
 \cline{2-3}
{variables}& \PowPytSh (w. unc.)  & \StripperOnly  \\
\hline
\ptt& $<$1 \: (7) & 88 \\
\ptat& $<$1 \: (11) & 82 \\
\yt& $<$1 \: (5) & 33 \\
\yat& $<$1 \: (1) & 43 \\
\pttt& $<$1 \: (37) & $<$1 \\
\ytt& 35 \: (78) & 6  \\
\mtt& 46 \: (80) & 80 \\
\dphitt& 40 \: (85) & $<$1 \\
\dytt& 3 \: (23) & 87 \\
\rpttmtt& $<$1 \: ($<$1) & 11 \\
\rptttmtt& 1 \: (59) & $<$1 \\
\logxone& 7 \: (35) & 34 \\
\logxtwo& 22 \: (67) & 37 \\
 \end{tabular}
\end{table*}

\begin{sidewaystable*}
 \centering
 \topcaption{The $p$-values are shown for the \chisq tests of the measured absolute multi-differential cross sections for \ttbar and top quark kinematic observables at the parton level with respect to various fixed-order predictions. The \chisq values are calculated taking only measurement uncertainties into account and excluding theory uncertainties. For \PowPytSh, the $p$-values of the \chisq tests including theory uncertainties are indicated with the brackets (w. unc.).} 
 \label{tab:pvaluefixTheo_abs_parton_ttbar}
 \renewcommand{\arraystretch}{1.4}
 \centering
 \begin{tabular}{lccccc}
 \multirow{1}{*}{Cross section} & \multicolumn{5}{c}{$p$-values of \chisq (in \%)} \\
 \cline{2-6}
{variables}& \PowPytSh (w. unc.)  & \appNTLOOnly & \MatrixOnly & \StripperOnly & \MiNNLOPSOnly  \\
\hline
\ytptt& $<$1 \: ($<$1) & $<$1  & 8 & 6  & 19 \\
\mttptt& $<$1 \: ($<$1) & \NA & 2 & 5 & 4 \\
\pttpttt& $<$1 \: (6) & \NA & $<$1  & $<$1  & $<$1 \\
\mttytt& $<$1 \: ($<$1) & \NA & $<$1  & $<$1 & $<$1 \\
\yttpttt& $<$1 \: (37) & \NA & $<$1  & $<$1  & $<$1 \\
\mttpttt& $<$1 \: ($<$1) & \NA & $<$1  & $<$1  & $<$1 \ \\
\ptttmttytt& $<$1 \: (2) & \NA & \NA & $<$1  & $<$1 \\
\mttyt& $<$1 \: ($<$1) & \NA & $<$1  & 1  & $<$1  \\
\mttdetatt& $<$1 \: ($<$1) & \NA & $<$1 & $<$1 & $<$1 \\
\mttdphitt& $<$1 \: ($<$1) & \NA & \NA & $<$1 & $<$1  \\
 \end{tabular}
\end{sidewaystable*}

\begin{table*}
 \centering
 \topcaption{The $p$-values are shown for the \chisq tests of the measured absolute multi-differential cross sections for \ttbar and top quark kinematic observables at the particle level 
with respect to the \PowPyt (`POW-PYT') simulation and the \StripperOnly NNLO calculation. 
The \chisq values are calculated taking only measurement uncertainties into account and excluding theory uncertainties.
For \PowPytSh, the $p$-values of the \chisq tests including theory uncertainties are indicated with the brackets (w. unc.).}
 \label{tab:pvaluefixTheo_abs_particle_ttbar}
 \renewcommand{\arraystretch}{1.4}
 \centering
 \begin{tabular}{lcc}
 \multirow{1}{*}{Cross section} & \multicolumn{2}{c}{$p$-values of \chisq (in \%)} \\
 \cline{2-3}
{variables}& \PowPytSh (w. unc.)  & \StripperOnly  \\
\hline
\ytptt& $<$1 \: (3) & 39 \\
\mttptt& $<$1 \: ($<$1) & $<$1 \\
\pttpttt& $<$1 \: (17) & $<$1 \\
\mttytt& $<$1 \: ($<$1) & $<$1 \\
\yttpttt& 1 \: (27) & $<$1 \\
\mttpttt& $<$1 \: ($<$1) & $<$1 \\
\ptttmttytt& $<$1 \: (3) & $<$1 \\
\mttyt& $<$1 \: (5) & 20 \\
\mttdetatt& $<$1 \: ($<$1) & $<$1 \\
\mttdphitt& $<$1 \: ($<$1) & $<$1 \\
 \end{tabular}
\end{table*}

\begin{table*}
\centering
 \topcaption{The $p$-values are shown for the \chisq tests of the measured absolute single-differential
 cross sections for lepton and \PQb-jet kinematic observables at the particle level 
with respect to the \PowPyt (`POW-PYT') simulation and the \StripperOnly NNLO calculation. 
The \chisq values are calculated taking only measurement uncertainties into account and
excluding theory uncertainties. For \PowPytSh, the $p$-values of the \chisq tests including theory uncertainties are indicated with the brackets (w. unc.).}
 \label{tab:pvaluefixTheo_1d_abs_particle_lepb}
 \renewcommand{\arraystretch}{1.4}
 \centering
 \begin{tabular}{lcc}
 \multirow{1}{*}{Cross section} & \multicolumn{2}{c}{$p$-values of \chisq (in \%)} \\
 \cline{2-3}
{variables}& \PowPytSh (w. unc.)  & \StripperOnly  \\
\hline
\ptlep& $<$1 \: (9) & 14 \\
\ptlep trailing/\ptlep leading& 9 \: (34) & 86 \\
\ptlep/\ptat& $<$1 \: ($<$1) & 14 \\
\ptb leading& 85 \: (91) & 61 \\
\ptb trailing& 46 \: (64) & 78  \\
\rptbsptts& $<$1 \: ($<$1) & 7 \\
\mll& $<$1 \: (2) & 36 \\
\mbb& $<$1 \: (2) & $<$1 \\
\mllbb& $<$1 \: (48) & 66 \\
\ptll& 88 \: (97) & 78 \\
\absetall& 32 \: (77) & $<$1  \\
\etallmll& $<$1 \: (22) & $<$1 \\
\etallptll& 7 \: (80) & $<$1 \\
\ptllmll& 1 \: (12) & $<$1 \\
 \end{tabular}
\end{table*}

\end{appendix}
\cleardoublepage \section{The CMS Collaboration \label{app:collab}}\begin{sloppypar}\hyphenpenalty=5000\widowpenalty=500\clubpenalty=5000
\cmsinstitute{Yerevan Physics Institute, Yerevan, Armenia}
{\tolerance=6000
A.~Tumasyan\cmsorcid{0009-0000-0684-6742}
\par}
\cmsinstitute{Institut f\"{u}r Hochenergiephysik, Vienna, Austria}
{\tolerance=6000
W.~Adam\cmsorcid{0000-0001-9099-4341}, J.W.~Andrejkovic, T.~Bergauer\cmsorcid{0000-0002-5786-0293}, S.~Chatterjee\cmsorcid{0000-0003-2660-0349}, K.~Damanakis\cmsorcid{0000-0001-5389-2872}, M.~Dragicevic\cmsorcid{0000-0003-1967-6783}, A.~Escalante~Del~Valle\cmsorcid{0000-0002-9702-6359}, P.S.~Hussain\cmsorcid{0000-0002-4825-5278}, M.~Jeitler\cmsAuthorMark{1}\cmsorcid{0000-0002-5141-9560}, N.~Krammer\cmsorcid{0000-0002-0548-0985}, L.~Lechner\cmsorcid{0000-0002-3065-1141}, D.~Liko\cmsorcid{0000-0002-3380-473X}, I.~Mikulec\cmsorcid{0000-0003-0385-2746}, P.~Paulitsch, F.M.~Pitters, J.~Schieck\cmsAuthorMark{1}\cmsorcid{0000-0002-1058-8093}, R.~Sch\"{o}fbeck\cmsorcid{0000-0002-2332-8784}, D.~Schwarz\cmsorcid{0000-0002-3821-7331}, S.~Templ\cmsorcid{0000-0003-3137-5692}, W.~Waltenberger\cmsorcid{0000-0002-6215-7228}, C.-E.~Wulz\cmsAuthorMark{1}\cmsorcid{0000-0001-9226-5812}
% ...
\par}
\vskip\cmsinstskip
\dag:~Deceased\\
$^{1}$Also at TU Wien, Vienna, Austria\\
% ...
\end{sloppypar}
\end{document}